%% file: main.tex
\documentclass[a4paper]{book}
\pdfoutput=1 

\newenvironment{abstract}%
    {\cleardoublepage\thispagestyle{empty}\null\vfill\begin{center}%
    \bfseries\abstractname\end{center}}%
    {\vfill\null}
\usepackage{lipsum}
\usepackage{float}
\usepackage{feynmf}	
\usepackage{amssymb,graphicx, subcaption,wrapfig}
\usepackage{eucal}
\usepackage{listings}
\usepackage{xcolor}
\usepackage{multirow}
\usepackage{jheppub}  
\usepackage{dcolumn}
\usepackage{placeins}
\usepackage[english]{babel}
\usepackage[utf8x]{inputenc}
\usepackage[T1]{fontenc}
\usepackage{palatino}
\usepackage[switch*]{lineno}
\usepackage{tikz}

\usepackage{amsmath}
\usepackage{afterpage}
\usepackage[colorinlistoftodos]{todonotes}
\usepackage[colorlinks=true]{hyperref}
\usepackage{cleveref}
\usepackage{makeidx}
\usepackage{rotfloat}
\restylefloat{figure}
\def\ds{\displaystyle}
\def\bea{\begin{array}{c}}
\def\ea{\end{array}}
\def\be{\begin{equation}\bea\ds}
\def\eeq{\ea\end{equation}}
\def\bee{\begin{equation}\begin{array}{rcl}\ds}
\def\eee{\end{array}\end{equation}}
\newcommand{\bejamal}{\begin{equation}}  
\newcommand{\eejamal}{\end{equation}}  
\newcommand{\beajamal}{\begin{eqnarray}}  
\newcommand{\eeajamal}{\end{eqnarray}}  
\def\Rc{{\mathcal R}}
\def\tr{{\rm Tr}\,}
\def\nn{\nonumber}
\def\Lc{{\mathcal{L}}}
\def\Hc{{\mathcal{H}}}
\def\Rc{{\mathcal R}}
\def\p{\partial}
\def\tr{{\rm tr}\,}
\def\Tr{{\rm Tr}\,}
\newcommand{\nnjamal}{\nonumber\\}
\newcommand\Trule{\rule{0pt}{2.5ex}}
\newcommand\Brule{\rule[-1.ex]{0pt}{0pt}}

\newcommand{\dif}{\mathrm{d}}

\newcommand{\ord}[1]{\CMcal{O}\left(#1\right)}

\newcommand{\ADA}  {\rm{ADA}\xspace}
\newcommand{\ADC}  {\rm{ADC}\xspace}
\newcommand{\AD}   {\rm{AD}\xspace}

\newcommand{\drv}{{\rm d}}
\newcommand{\ee}           {\ensuremath{e^{+}e^{-}}} 
\newcommand{\eight}        {$\sqrt{s}~=~8$~Te\kern-.1emV\xspace}

\newcommand{\fd}   {f_{\rm D}}
\newcommand{\fI}   {f_{\rm I}}
\newcommand{\fiveExactly}  {$\sqrt{s}~=~5$~Te\kern-.1emV\xspace}
\newcommand{\fivenn}       {$\sqrt{s_{\mathrm{NN}}}~=~5.02$~Te\kern-.1emV\xspace}
\newcommand{\five}         {$\sqrt{s}~=~5.02$~Te\kern-.1emV\xspace}
\newcommand{\GeVmass}      {Ge\kern-.05emV/$c^2$\xspace}

\newcommand{\jets}{\textrm{jets}}
\newcommand{\jpsi} {\ensuremath{{\mathrm J}/\psi}\xspace}

\newcommand{\ktcolf}{{\boldsymbol k}}

\newcommand{\Lqcd}{\Lambda_{\mathrm QCD}}
\newcommand{\lt}{{\boldsymbol l}}

\newcommand{\MeVmass}      {Me\kern-.05emV/$c^2$\xspace}

\newcommand{\montecarlo}{Monte Carlo\xspace}

\newcommand{\nineH}        {$\sqrt{s}~=~0.9$~Te\kern-.1emV\xspace}

\newcommand{\plb}[3]{{\it Phys.~Lett.~}{\bf B #1} (#2) #3}
\newcommand{\psip} {\ensuremath{\psi'}\xspace}
\newcommand{\qt}{{\boldsymbol q}}
\newcommand{\rhozero} {\ensuremath{\rho^0}\xspace}

\newcommand{\seven}        {$\sqrt{s}~=~7$~Te\kern-.1emV\xspace}
\newcommand{\slcalP}{\raise.15ex\hbox{$/$}\kern-.63em\hbox{$\cal P$}}

\newcommand{\twoH}         {$\sqrt{s}~=~0.2$~Te\kern-.1emV\xspace}
\newcommand{\twohundrednn}       {$\sqrt{s_{\mathrm{NN}}}~=~200$~Ge\kern-.1emV\xspace}
\newcommand{\twosevensix}  {$\sqrt{s}~=~2.76$~Te\kern-.1emV\xspace}

\newcommand{\vm}[1]{\langle #1 \rangle}     

\newcommand{\vn}{\ensuremath{v_{\mathrm{n}}}\xspace}
\newcommand{\vtwo}{\ensuremath{v_{2}}\xspace}
\newcommand{\vthree}{\ensuremath{v_{3}}\xspace}
\newcommand{\vfour}{\ensuremath{v_{4}}\xspace}
\newcommand{\vtwotwo}{\ensuremath{\vtwo\{2\}}\xspace}
\newcommand{\vtwofour}{\ensuremath{\vtwo\{4\}}\xspace}
\newcommand{\vtwosix}{\ensuremath{\vtwo\{6\}}\xspace}
\newcommand{\vtwoeight}{\ensuremath{\vtwo\{8\}}\xspace}
\newcommand{\noff}  {\ensuremath{N_\text{trk}^\text{offline}}\xspace}
\newcommand{\rootsNN} {\ensuremath{\sqrt{\smash[b]{s_{_{\mathrm{NN}}}}}}\xspace}
\newcommand{\roots} {\ensuremath{\sqrt{\smash[b]{s}}}\xspace}
\newcommand{\PbPb}{\ensuremath{\mathrm{Pb}\mathrm{Pb}}\xspace}
\newcommand{\XeXe}{\ensuremath{\mathrm{Xe}\mathrm{Xe}}\xspace}
\newcommand{\pPb}{\ensuremath{\Pp\mathrm{Pb}}\xspace}
\newcommand{\pp}{\ensuremath{\Pp\Pp}\xspace}
\newcommand{\pn}{\ensuremath{\Pp\Pn}\xspace}
\newcommand{\ppbar}{\ensuremath{\Pp\PAp}\xspace}
\newcommand{\sigmaVis}{\ensuremath{\sigma_{\text{vis}}}\xspace}
\newcommand{\mub}{\ensuremath{\,\mu\text{b}}\xspace}
\newcommand{\xthree}{\PGccDoP{3872}}
\newcommand{\stt}{\ensuremath{\sigma_{\ttbar}}\xspace}
\newcommand{\twosevensixnn}{$\sqrt{s_{\mathrm{NN}}}~=~2.76$~Te\kern-.1emV\xspace}
\newcommand{\starlight}{STARlight\xspace}
\def\zt{z_t}
\def\xt{x_t}
\def\yt{y_t}
\def\ubar{\bar{u}}
\def\qbar{\bar{q}}
\newcommand{\slv}{\raise.15ex\hbox{$/$}\kern-.53em\hbox{$v$}}
\newcommand{\sln}{\raise.15ex\hbox{$/$}\kern-.53em\hbox{$n$}}
\newcommand{\slnbar}{\raise.15ex\hbox{$/$}\kern-.53em\hbox{$\bar{n}$}}
\newcommand{\slF}{\raise.15ex\hbox{$/$}\kern-.53em\hbox{$F$}}
\newcommand{\sll}{\raise.15ex\hbox{$/$}\kern-.40em\hbox{$l$}}
\newcommand{\sllbar}{\raise.15ex\hbox{$/$}\kern-.40em\hbox{$\bar{l}$}}
\newcommand{\slh}{\raise.15ex\hbox{$/$}\kern-.40em\hbox{$h$}}
\newcommand{\slP}{\raise.15ex\hbox{$/$}\kern-.53em\hbox{$P$}}
\newcommand{\slp}{\raise.15ex\hbox{$/$}\kern-.53em\hbox{$p$}}
\newcommand{\slq}{\raise.15ex\hbox{$/$}\kern-.53em\hbox{$q$}}
\newcommand{\slqbar}{\raise.15ex\hbox{$/$}\kern-.53em\hbox{$\bar{q}$}}
\newcommand{\slR}{\raise.15ex\hbox{$/$}\kern-.53em\hbox{$R$}}
\newcommand{\slz}{\raise.15ex\hbox{$/$}\kern-.53em\hbox{$Z$}}
\newcommand{\slzbar}{\raise.15ex\hbox{$/$}\kern-.53em\hbox{$\bar{Z}$}}
\newcommand{\slQ}{\raise.15ex\hbox{$/$}\kern-.53em\hbox{$Q$}}
\newcommand{\slK}{\raise.15ex\hbox{$/$}\kern-.53em\hbox{$K$}}
\newcommand{\slk}{\raise.15ex\hbox{$/$}\kern-.53em\hbox{$k$}}
\newcommand{\slkone}{\raise.15ex\hbox{$/$}\kern-.53em\hbox{$k_1$}}
\newcommand{\slktwo}{\raise.15ex\hbox{$/$}\kern-.53em\hbox{$k_2$}}
\newcommand{\slkthree}{\raise.15ex\hbox{$/$}\kern-.53em\hbox{$k_3$}}
\newcommand{\slkbar}{\raise.15ex\hbox{$/$}\kern-.53em\hbox{$\bar{k}$}}
\newcommand{\slkbarone}{\raise.15ex\hbox{$/$}\kern-.53em\hbox{$\bar{k}_1$}}
\newcommand{\slkbartwo}{\raise.15ex\hbox{$/$}\kern-.53em\hbox{$\bar{k}_2$}}
\newcommand{\slkbarthree}{\raise.15ex\hbox{$/$}\kern-.53em\hbox{$\bar{k}_3$}}
\newcommand{\slpone}{\raise.15ex\hbox{$/$}\kern-.53em\hbox{$p_1$}}
\newcommand{\slpbarone}{\raise.15ex\hbox{$/$}\kern-.53em\hbox{$\bar{p}_1$}}
\newcommand{\slptwo}{\raise.15ex\hbox{$/$}\kern-.53em\hbox{$p_2$}}
\newcommand{\slpbartwo}{\raise.15ex\hbox{$/$}\kern-.53em\hbox{$\bar{p}_2$}}
\newcommand{\slqone}{\raise.15ex\hbox{$/$}\kern-.53em\hbox{$q_1$}}
\newcommand{\slD}{\raise.15ex\hbox{$/$}\kern-.53em\hbox{$\!D$}}
\newcommand{\slC}{\raise.15ex\hbox{$/$}\kern-.53em\hbox{$C$}}
\newcommand{\slA}{\raise.15ex\hbox{$/$}\kern-.73em\hbox{$A$}}
\newcommand{\slSigma}{\raise.15ex\hbox{$/$}\kern-.53em\hbox{$\Sigma$}}
\newcommand{\slpartial}{\raise.15ex\hbox{$/$}\kern-.53em\hbox{$\partial$}}
\newcommand{\sleps}{\raise.15ex\hbox{$/$}\kern-.53em\hbox{$\epsilon$}}
\newcommand{\slepsbar}{\raise.15ex\hbox{$/$}\kern-.53em\hbox{$\overline{\epsilon}$}}
\newcommand{\slepsstar}{\raise.15ex\hbox{$/$}\kern-.53em\hbox{$\epsilon$}^\star}
\newcommand{\slS}{\raise.15ex\hbox{$/$}\kern-.73em\hbox{$S$}}

\usepackage{ptdr-definitions}
\usepackage{hepunits}
\usepackage{heppennames2}
\usepackage{xpatch}
\makeatletter
\xpatchcmd\@HepConStyle
 {\edef\@upcode{\updefault}}
 {\ifdefined\shapedefault\edef\@upcode{\shapedefault}\else\edef\@upcode{\updefault}\fi}
 {}{}
\makeatother

\makeindex
\makeatletter
\usepackage{titlesec}
\@addtoreset{section}{part}
\makeatother
\titleformat{\part}[display]
{\normalfont\LARGE\bfseries\centering}{}{0pt}{}
\usepackage{chngcntr}
\counterwithin{figure}{section}
\counterwithin{table}{section}

\begin{document}
\begin{titlepage}
	\centering
	\includegraphics[width=0.70\textwidth]{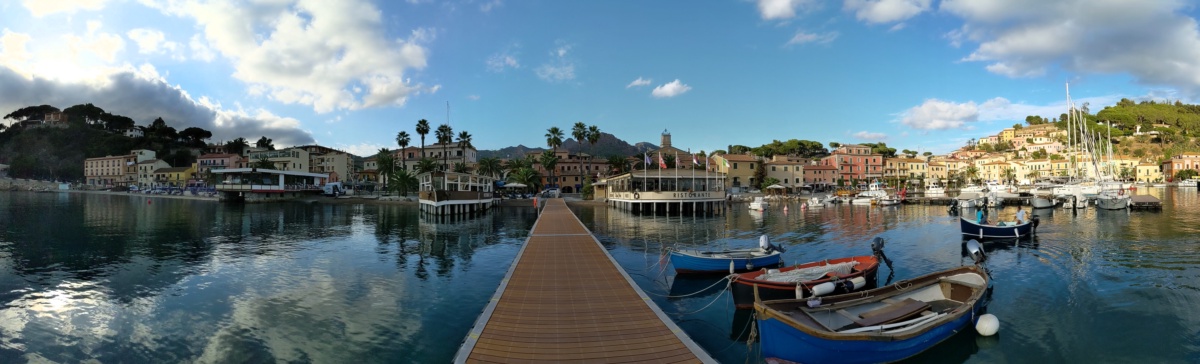}\par\vspace{1cm}
	{\scshape\Large  Elba Island, Italy \\ September 27--October 1 2021 \par}
	\vspace{1cm} 
	{\scshape\LARGE Proceedings of the \par}
	\vspace{1.5cm}
	{\huge\bfseries Low-$x$ 2021 International Workshop \par}
	\vspace{2cm}
	{\Large\itshape editors: L. Alcerro, G.~K. Krintiras, C. Royon \par}
	\vfill
	Organized and sponsored by the \par
	University of Kansas, INFN Pisa, and EMMI
	\vfill

	{\large \today\par}
\end{titlepage}

\newpage
\thispagestyle{empty}
\begin{center}

{\LARGE \bf Preface: Low-$x$ 2021 International Workshop}

\end{center}
\par\vspace*{7mm}\par

\bigskip

The 2021 edition of the  Low-$x$ International Workshop (\url{https://indico.cern.ch/event/1003281/}) took place from September 26 to October 1 in Elba, a Mediterranean island in Tuscany, Italy, 10 km from the coastal town of Piombino on the Italian mainland. This workshop has been the \textbf{XXVIII} edition in the series of the Workshop with previous editions held since 1993 at DESY, Saclay (May 1994), Cambridge (July 1995), Durham (June 1996), Madrid (June 1997), Berlin (June 1998), Tel Aviv (June 1999), Oxford (July 2000), Cracow (June 2001), Antwerpen (September 2002), Nafplio (June 2003), Prague (September 2004), Sinaia (June 2005), Lisbon (June 2006), Helsinki (September 2007), Crete (July 2008), Ischia (September 2009), Kavala (June 2010), Santiago de Compostela (June 2011), Cyprus (June 2012), Eilat (June 2013), Kyoto (June 2014), Sandomierz (September 2015), Gyongyos (June 2016), Bisceglie (June 2017), Reggio Calabria jointly with Diffraction 2018 (August 2018), Nicosia (August 2019). The purpose of the Low-$x$ Workshop series is to stimulate discussions between experimentalists and theorists in diffractive hadronic physics, QCD dynamics at low $x$, parton saturation, and exciting problems in QCD at HERA, Tevatron, LHC, RHIC and the future EIC.\\

The workshop, jointly organized by the University of Kansas and INFN Pisa and supported by EMMI, was attended by \textbf{88 participants} from \textbf{24 countries} both in person and remotely, including a large fraction of students and young researchers from universities and research institutes. During the entire workshop, more than \textbf{100 contributions} (\url{https://indico.cern.ch/event/1003281/contributions/contributions.pdf})  were made and lot of time was devoted to discussion sessions. The central topics of the workshop were 
\begin{itemize}
    \setlength\itemsep{-0.2em}
    \item Diffraction in $\Pe{}\Pp$ and $\Pe$-ion collisions (including EIC physics)
    \item Diffraction and photon-exchange in hadron-hadron, hadron-nucleus, and nucleus-nucleus collisions
    \item Spin Physics
    \item Low-$x$ PDFs, forward physics, and hadronic final states
\end{itemize}

The editors of these proceedings wish to thank all participants for
their highly valuable contributions. In particular we would like to
thank Michael Albrow, Valentina Avati, Irais Bautista Guzma, Dimitri Colferai, Mario Deile, Krzysztof Kutak, Cyrille Marquet, Leszek Motyka, and Kenneth Osterberg for their availability as discussion leaders as well as (following the order from the workshop page) Fernando Barreiro, Jochen Bartels, Andrzej Bialas, Irinel Caprini, Janusz Chwastowski, Jiri Chyla, Tamas Csorgo, Robin Devenish, Roberto Fiore, Konstantin Goulianos, Yoshitaka Hatta, Edmond Iancu, Valery Khoze, Michael Lublinsky, Cyrille Marquet, Alan Martin, Pierre Van Mechelen, Carlos Merino, Sasha Milov, Nicola Minafra, Al Mueller, Risto Orava, Alessandro Papa, Robi Peschanski, Michal Praszalowicz, Fotis Ptochos, Albert de Roeck, Angelo Scribano, Gregory Soyez, Eddi de Wolf for forming the
international advisory committee of the workshop.  We also want to
thank the University of Kansas library for their valuable support in publishing these proceedings. Last, but not least, we would like to thank Angelo Scribano from INFN Pisa for his indispensable help in the local organization of this workshop. \\
\\
L. Alcerro, G. K. Krintiras, and C. Royon. 
\clearpage
\setcounter{tocdepth}{0}
\tableofcontents
\clearpage
\begin{sidewaysfigure}
 \includegraphics[width=1.0\textwidth]{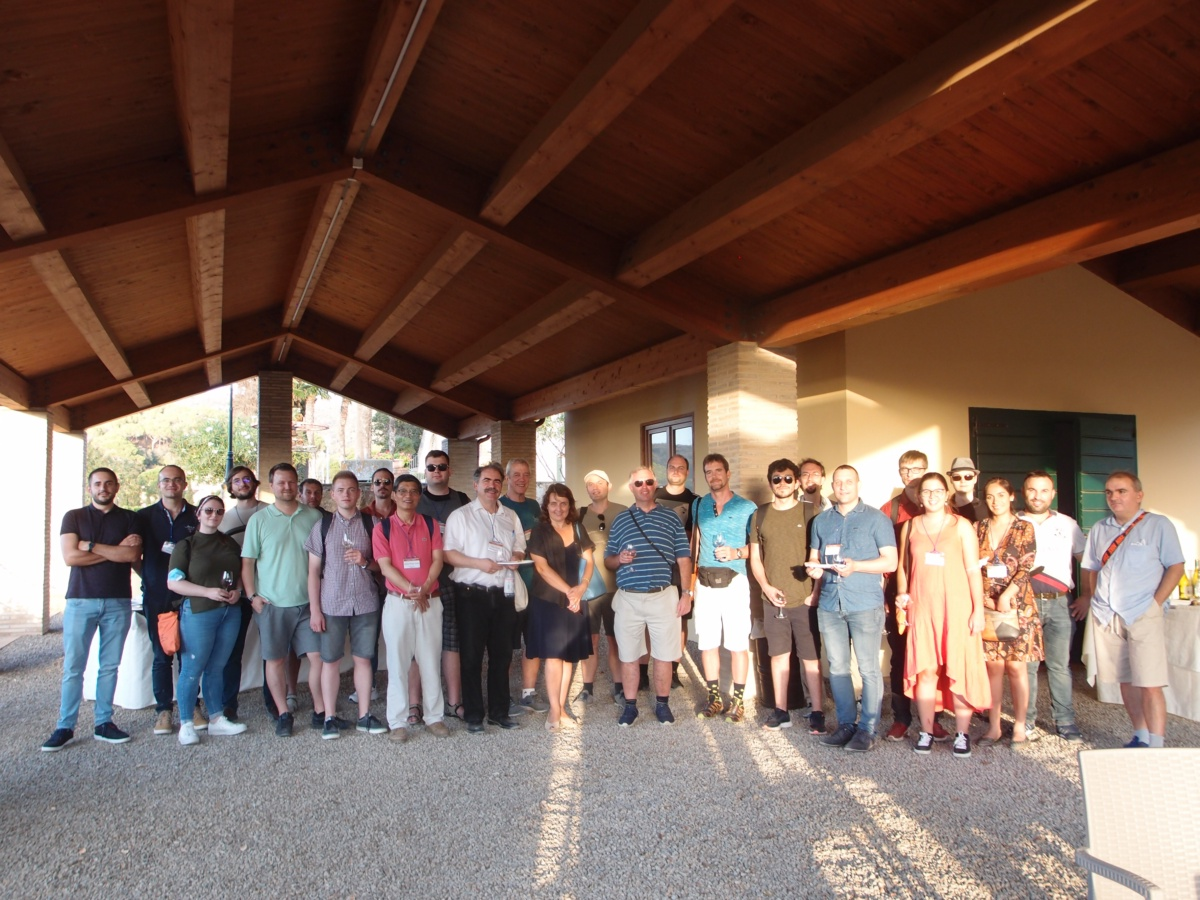} \\ 
\caption{Group photo of workshop participants during the excursion day around Elba Island.}
\clearpage
\end{sidewaysfigure}
\chapter{\LARGE{Diffraction and photon-exchange in hadron-hadron, hadron-nucleus, and nucleus-nucleus collisions}}
\graphicspath{ {albrowlowx/} }
\include{albrowlowx/albrow}
\graphicspath{ {Alcerro_Luis/} }
\include{Alcerro_Luis/Alcerro_Luis}
\graphicspath{ {Brodsky_proceedings_elba2021/Brodsky_proceedings_elba2021/Brodsky_proceedings_elba2021/} }
\include{Brodsky_proceedings_elba2021/Brodsky_proceedings_elba2021/Brodsky_proceedings_elba2021/sjb_Elba_contribution}

\graphicspath{ {low_x_2021} }
\include{low_x_2021/GKKrintiras}
\graphicspath{ {Maciej_Lewicki-Low-x_2021_proceedings/Low-x_2021_proceedings/} }
\include{Maciej_Lewicki-Low-x_2021_proceedings/Low-x_2021_proceedings/Lewicki}
\graphicspath{ {jamal/} }
\include{jamal/jamal}
\graphicspath{ {Low_X_proceedingsDMelnikov/} }
\include{Low_X_proceedingsDMelnikov/Low_X_proceedingsDMelnikov}
\graphicspath{ {Low_x_2021_F_Nemes/} }
\include{Low_x_2021_F_Nemes/Low_x_2021_F_Nemes}
\graphicspath{ {Odderon_lowx_Osterberg/} }
\include{Odderon_lowx_Osterberg/Osterberg}
\graphicspath{ {Petrov_article/} }
\include{Petrov_article/Petrov}
\graphicspath{ {proceedings_elba2021/proceedings_elba2021/} }
\include{proceedings_elba2021/proceedings_elba2021/Ribeiro}
\graphicspath{ {proceed_lowx2021/} }
\include{proceed_lowx2021/royon}

\chapter{\LARGE{Spin physics}}
\graphicspath{ {santimaria_proceedings_elba2021/} }
\include{santimaria_proceedings_elba2021/santimaria_proceedings_elba2021}

\chapter{\LARGE{QCD and saturation}}
\graphicspath{ {proceedings_elba2021_SalimCerci/proceedings_elba2021_SalimCerci/proceedings_elba2021/} }
\include{proceedings_elba2021_SalimCerci/proceedings_elba2021_SalimCerci/proceedings_elba2021/SalimCerci}
\graphicspath{ {chachamis_proceedings_elba2021/} }
\include{chachamis_proceedings_elba2021/chachamis_proceedings_elba2021}
\graphicspath{ {colferai/proceedings/} }
\include{colferai/proceedings/Colferai}

\chapter{\LARGE{Low-$x$ PDFs, forward physics, and hadronic final states}}
\graphicspath{ {Boettcher_Lowx2021_proceedings/proceedings/} }
\include{Boettcher_Lowx2021_proceedings/proceedings/Boettcher_Lowx2021_proceedings}
\graphicspath{ {proceeding_Lowx_2021_Celiberto/} }
\include{proceeding_Lowx_2021_Celiberto/proceeding_Lowx_2021_Celiberto}
\graphicspath{ {proceedings_elba2021_DenizSunarCerci/proceedings_elba2021_DenizSunarCerci/} }
\include{proceedings_elba2021_DenizSunarCerci/proceedings_elba2021_DenizSunarCerci/SunarCerci}
\graphicspath{ {draft_Weisong_proceedingslowx2021/} }
\include{draft_Weisong_proceedingslowx2021/Weisong}
\graphicspath{ {giugli_francesco/proceedings_elba2021/} }
\include{giugli_francesco/proceedings_elba2021/giugli_francesco}
\graphicspath{ {Lowxsubmission/Lowxsubmission/} }
\include{Lowxsubmission/Lowxsubmission/klein}
\graphicspath{ {LauraFabbri_Elba2021/LauraFabbri_Elba2021/} }
\include{LauraFabbri_Elba2021/LauraFabbri_Elba2021/LauraFabbri_Elba2021}
\graphicspath{ {Precision_QCD_measurements_from_CMS/Precision_QCD_measurements_from_CMS/} }
\include{Precision_QCD_measurements_from_CMS/Precision_QCD_measurements_from_CMS/Precision_QCD_measurements_from_CMS}
\graphicspath{ {proceedings-CSanchezGras/sanchez_cristina/} }
\include{proceedings-CSanchezGras/sanchez_cristina/Sanchez}
\graphicspath{ {proceedings_elba2021_ragoni/} }
\include{proceedings_elba2021_ragoni/proceedings_elba2021_ragoni}
\graphicspath{ {Zhang_ATLAS/} }
\include{Zhang_ATLAS/Zhang_ATLAS}
\end{document}

%% file: albrowlowx/albrow.tex
\vspace*{1.2cm}

\thispagestyle{empty}
\begin{center}
{\LARGE \bf{The FACET Project: Foward Aperture CMS ExTension\\
 to search for new Long-Lived Particles}
}
\par\vspace*{7mm}\par

{

\bigskip

\large \bf Michael G. Albrow}

\bigskip

{\large \bf  E-Mail: albrow@fnal.gov}

\bigskip

{Scientist Emeritus, Fermilab, USA}

\bigskip

{\it Presented at the Low-$x$ Workshop, Elba Island, Italy, September 27--October 1 2021}

\vspace*{15mm}

\end{center}
\vspace*{1mm}

\begin{abstract}
 FFACET is a proposed new subsystem for CMS to search for portals such as dark photons, dark higgs, heavy
neutral leptons and axion-like particles in the very forward direction at the High Luminosity LHC. 
Such particles can penetrate up to 50 m of iron and then decay inside a 14 m$^3$ vacuum pipe made by
enlarging an 18 m long section of the LHC pipe to a radius of 50 cm.
\end{abstract}

 \part[The FACET Project: Foward Aperture CMS ExTension to search for new Long-Lived Particles\\ \phantom{x}\hspace{4ex}\it{Michael G. Albrow}]{}

\section{Introduction}
FACET, short for \textbf{F}orward \textbf{A}perture \textbf{C}MS \textbf{E}x\textbf{T}ension,  
is a project under development to add a subsystem 
to CMS to search for beyond the standard model (BSM) long-lived particles (LLPs)
 in the high luminosity era
of the LHC, in Run 4 (2028) and beyond. 
The project was initiated with a two-day meeting in April 2020 \cite{april2020,albrowguan}, with one day discussing
a forward hadron spectrometer for strong interaction physics, and one day on searching for long-lived particles.
A description and more details of the physics potential are given in Ref. \cite{facetpaper}.

We can compare FACET to the pioneering FASER experiment \cite{faser} which is approved to search for LLPs in the very forward direction in Run 3,
and an upgrade FASER-2 \cite{faser2} which is being developed for Run 4. Major differences with FACET are (a) FASER-2 is
480 m from IR5 (with ATLAS) while FACET is 100 - 127 m from IR1 (with CMS). (b) FACET has 4$\times$ the solid
angle: 54.5 $\mu$sr cf. 13.6 $\mu$sr. (c) FASER-2 has a 5 m-long decay volume; FACET has 18 m which is evacuated to
eliminate background from particle interactions. (d) FASER-2 is centered at polar angle $\theta$ = 0$^\circ$ while
FACET covers 1 mrad $< \theta <$ 4 mrad. (e) FASER-2 is behind $\sim$ 100 m of rock absorber while FACET has $\sim$ 50 m of
iron. However FACET is located inside the main LHC tunnel where
radiation levels are much higher while FASER is located in a side tunnel. 

An important difference is that FASER is an independent experiment while FACET is not; it is proposed to be a new subsystem of CMS,
fully integrated and using the same advanced technology for its detectors. This has the added benefit of allowing
the study of correlations with the central event, and enables a standard model physics program especially in low pileup
$pp, pA$ and $AA$ collisions.

\begin{figure}[t]
  \vspace{-0.5 in}
 \begin{center}
\makebox[\textwidth][c]{\includegraphics[angle=270,origin=c,width=150mm]{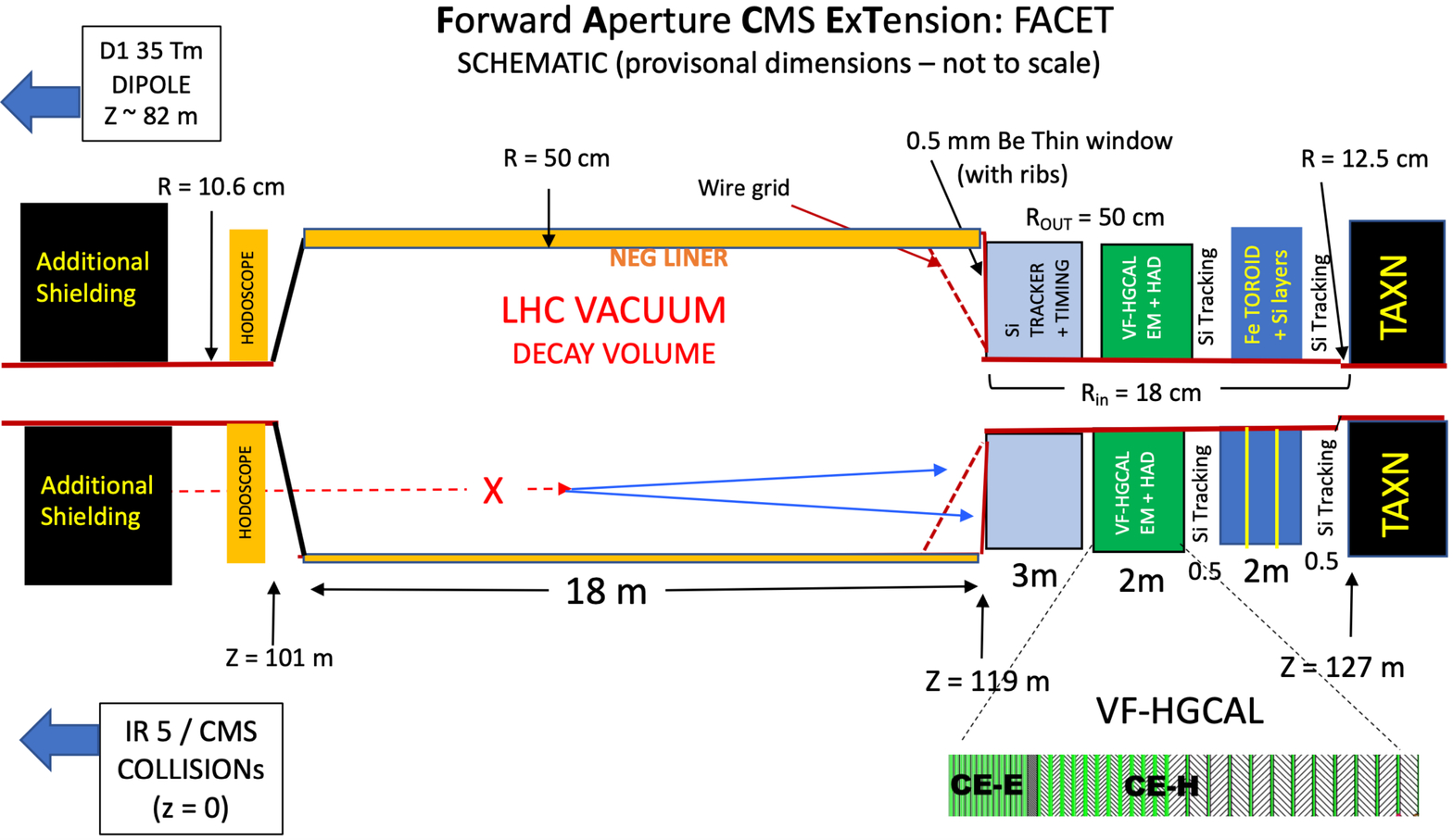}}
\end{center}
 \vspace{-1.5in}

\caption{Schematic layout of the proposed FACET spectrometer. The side view and top view are
the same since it is azimuthally symmetric. The IR5 collision region and the central CMS detectors
are 100 m to the left. An example of an LLP $X$ decaying inside the pipe is superimposed.}
\label{sketch}
 \end{figure}

 FACET will be located downstream of IR5 (at $z$ = 0) in an LHC straight section between the new (for Run 4) 
superconducting beam separation dipole D1 at $z$ = 80 m and the TAXN absorber at $z$ = 128 m. 
A schematic layout of the spectrometer is shown in Fig.~\ref{sketch}.
The beam pipe
between $z$= 101 m and 119 m will be enlarged to a radius of 50 cm. In front of the entrance window
will be a radiation-hard ``tagging'' hodoscope, with 2 - 3 planes of $\sim 1$ cm$^2$ quartz or radiation-hard scintillator
blocks. 
This must have very high efficiency with a precision time measurement for charged
particles entering the pipe. These are all background particles to be ignored in the subsequent analysis.
Excellent time resolution, $\sim$ 30 ps, together
with fast timing on the tracks from another plane between the tracker and the calorimeter 
will not only help the rejection of incoming background tracks but allows a study of their  momenta and composition.

Neutral LLPs produced with polar angle 
1 $< \theta <$ 4 mrad penetrate the iron of the LHC elements (quadrupoles Q1 - Q3  and dipole D1) and enter the big vacuum tank
where decays to SM particles can occur. The LHC-quality vacuum completely eliminates any background from interacting
particles inside a fiducial region starting behind the front window. The back window of the big pipe, where it transitions
from $R$ = 50 cm to $R$ = 18 cm, will be thin, e.g. 0.5 mm of Be with strengthening ribs,
 to minimise multiple scattering of the decay tracks\footnote{The front window may also need to be thin to minimize interactions
behind the tagging hodoscope; this is under study.} 
Behind that window, in air, the detector elements will be
3 m of silicon tracking (resolution $\sigma_x = \sigma_y \sim 30 \; \mu$m per plane) followed by a layer of
fast timing ($\sigma_t \sim$ 30 ps)\footnote{Since this Low-x Workshop 
we note that measuring the time-of-flight of these background tracks over the 22 m between the two 
hodoscopes, with a resolution $\delta \beta \lesssim 5 \times 10^{-4}$, together with the energy measured in the calorimeter,
will be very interesting. For example, consider particles with a delay relative to $\beta$ = 1 of 1 ns $\pm$ 50 ps  with a shower
of energy $E_{cal}$. These can be 0.63 GeV/c $\mu^\pm$, 0.83 GeV/c $\pi^\pm$, 3 GeV/c $K^\pm$, or 5.6 GeV/c $p$ or $\bar{p}$, easily
distinguished thereby measuring the identify and spectrum of these background tracks. That would be useful for testing and tuning
\textsc{fluka}, the LHC standard for machine protection etc. It also uniquely enables calibration of the HGCAL with hadrons of known
momenta up to tens of GeV with high  statistics even in short runs. The charge $Q$ is known from the Cherenkov light 
amplitude and track $dE/dx$
enabling measurements of light isotopes with lifetimes $\gtrsim 10^{-9}$ s in the showers, and to search for
objects such as strangelets (nuclei with extra strange quarks and therefore anomalous low charge:mass ratio).}.

The tracking and timing will be followed by a high granularity
electromagnetic and hadronic calorimeter, the HGCAL design. Muons that penetrate the HGCAL are detected in more
silicon tracking through an iron toroid. 

FACET is complementary to all other searches with unique 
access to regions of mass and coupling (or lifetime) for many \emph{portals}, hypothetical particles
that couple very weakly to both standard model particles (directly or through mixing) and to dark matter 
particles. Unlike most searches in the central detectors FACET  is sensitive to a wide variety of possible LLPs.
It has the potential to discover dark photons ($A'$), dark higgs ($h$ or $\phi$), heavy neutral 
leptons ($N_i$) and axion-like particles ($ALP$s or $a$) if they have large enough production cross section in the
very forward direction, small enough coupling to penetrate 300 $\lambda_{int}$ of iron, and lifetime
in the range $c\tau$ = 10 cm - 100 m before decaying to standard model charged particles and/or photons. 
A key feature is the high (LHC quality) vacuum tank for decays,
1 m diameter and 18 m long (14 m$^3$), made by enlarging a section of the LHC beam pipe. This allows some channels,
e.g. $X^0 \rightarrow$ multihadrons, $ \tau^+ \tau^-, c + \bar{c}$ and $b + \bar{b}$ to have zero background even in
3 ab$^{-1}$, while $e^+e^-$ and $\mu^+\mu^-$ decays may have very low backgrounds especially for
masses $\gtrsim$ 0.8 GeV. In 3 ab$^{-1}$ we expect to observe several thousand $K^0_L \rightarrow \mu^+ \mu^-$ and also
$K^0$ decays to 4 charged tracks, compromising the region around $M(X^0) =$ 0.5 GeV. 

Dark photons $A'$ are hypothetical neutral gauge bosons that do not have direct couplings with SM particles,
but they can interact indirectly by mixing with SM photons. If $M(A') < 1$ GeV their main production mechanism
is via the decays $\pi^0, \eta^0, \eta' \rightarrow \gamma \gamma$, the fluxes being highest at small polar angle $\theta$.
Fixed target experiments such as NA62 have higher luminosity, and the higher $\sqrt{s}$ of the LHC is not advantageous for
dark photons from these sources.
For  $M(A') > 1$ GeV the higher $\sqrt{s}$ of the LHC
is important, as additional sources such as Drell-Yan and quark- and proton-bremsstrahlung 
dominate. The LHC is essential if the source is a massive state such as a $Z'$ in the model of Ref.\cite{duaprime},
which would give FACET sensitivity up to $\sim$ 20 GeV.
The decay modes are the same as the final states in $e^+e^- \rightarrow \gamma^*$, with $\tau^+ \tau^-$,
$c + \bar{c}$ and multihadron decays being background-free above their thresholds. 
Measuring the relative rates of different channels could establish the identity of candidates as dark photons. 

Heavy neutral leptons $N_i$ (where $i$ represents flavor, perhaps with three different mass states to discover)
are present in many BSM theories; they may explain the light neutrino masses through the seesaw mechanism. Possible decay
modes are $N_{\mu} \rightarrow \mu^\pm W^{*\mp}$ with the virtual $W^*$ decaying to kinematically allowed leptonic
or hadronic channels, and the same modes but with $\mu^\pm$ replaced by $e^\pm, \tau^\pm$. If $N_i$ have masses in the few-GeV region
even a few good candidate events would be a discovery that would open a very rich new field of neutrino physics.
In the model of Ref.\cite{suchita} FACET has unique discovery reach up to $\sim$25 GeV.

\begin{figure}[t]
\vspace{-1.0in}
 \begin{center}
\makebox[\textwidth][c]{\includegraphics[angle=0,origin=c,width=120mm]{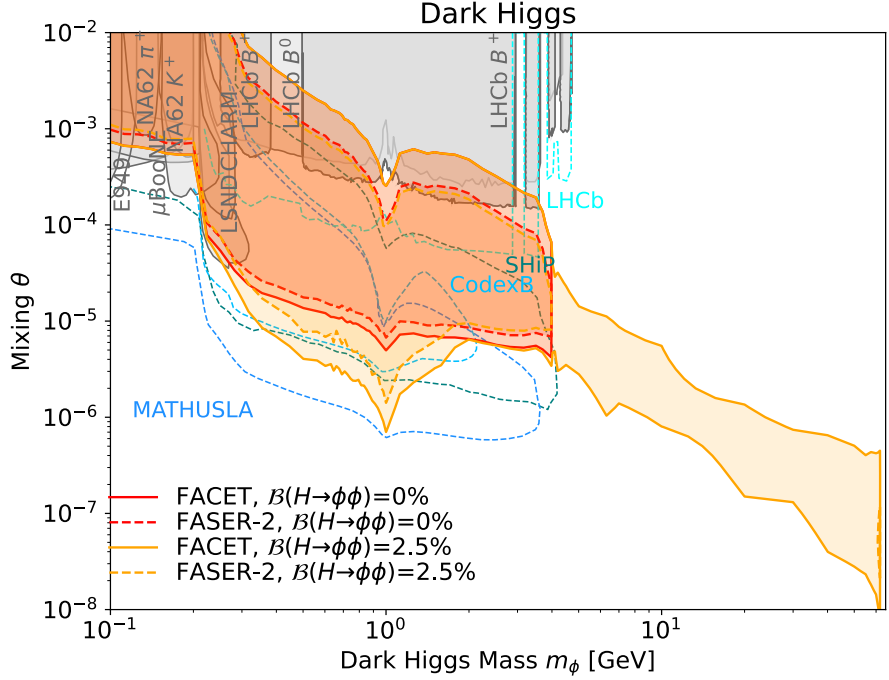}}
\end{center}
\vspace{-1.2in}
\caption{Reach of FACET and other existing and proposed experiments for a dark Higgs boson $\phi$
with the assumption of either 0\% (red lines) or 2.5\% (yellow lines) branching 
fraction for the  $H(125) \to \phi\phi$ decays. FACET offers a unique coverage all the way to half $M_H$ 
for a range of mixing angles. FACET and FASER-2 contours are calculated with \textsc{Foresee}~\protect\cite{foresee}.
Figure from Ref. \cite{facetpaper} which gives citations.}
\label{darkhiggs}
 \end{figure}

Also very exciting would be the discovery of another Higgs boson, a dark higgs, $h$ or in general a scalar $\phi$, 
having the same vacuum quantum numbers
as the H(125) but with mass possibly in the several GeV region. Present measurements of H(125) decays allow
an invisible decay fraction up to 5\%, which could be explained by an $h$ through mixing $H(125) \leftrightarrow h$ or 
decay $H(125) \rightarrow h + h$\footnote{If one $h$ is detected in FACET the other should be more central and give rise to
missing transverse energy $E_T$. Whether it is possible to detect this in a high pileup bunch crossing remains to be seen.}.
If $M(h) \lesssim$ 4.5 GeV rare $b$-decays are a potential source, with competition especially from B-factories and LHC-b. 
For 4.5 GeV $< M(h) <$ 60 GeV and a range of mixing angles FACET has unique coverage, as shown in Fig. \ref{darkhiggs}.
The most spectacular
decays are $h \rightarrow \tau^+ \tau^-, c \:+ \: \bar{c}$ and $b \: + \: \bar{b}$ if kinematically allowed, and with the heaviest
states favored; the scalar nature can be demonstrated by the relative decay fractions as well as the isotopic decay.
FACET has more sensitivity than FASER-2 due to its larger solid angle and longer decay volume, e.g.  
if there is no background and if 10 candidates were to be detected in FACET, FASER-2 would expect $<$ 1
\footnote{While the coverage of FACET is limited by LHC restrictions
in the horizontal direction, the solid angle could be increased nearly a factor $\times$2 in the vertical direction 
with a non-circular beam pipe.}.

Another possible portal is a heavy ALP, but the main decay mode to $\gamma + \gamma$ will have a high background
from random pairs of photons from $\pi^0$ and $\eta$ decay, etc. Even though the electromagnetic section of the HGCAL
measures the shower direction the vertex resolution is much worse than for charged tracks.

FACET will be live for every bunch crossing, with an expected pileup of $\sim$ 140 inelastic collisions, giving a total
integrated luminosity of $\sim$ 3 ab$^{-1}$. The \textsc{fluka} code, which is the LHC standard,  predicts about 25 charged particle tracks
with 18 cm $< R <$ 50 cm in each bunch crossing. Their origin (apart from any BSM signal!) is (a) from interactions
of beam halo and secondary particles with the beam pipe, collimators, magnets, etc. (b) from decays of neutral
hadrons, mainly $K^0_S, K^0_L$, and $\Lambda^0$. (c) Very small angle ($\theta \lesssim$ 1 mrad) charged particles that
pass through the D1 aperture, which deflects them to the left and right sides. The acceptance for the latter is 
limited to $\sim$ 2 TeV,
but they allow some standard model physics (e.g. measuring $\mu^+\mu^-$ pairs at Feynman-$x_F \sim$ 0.5).
 
In a fast Level-1 trigger the tracks will be projected upstream to the
2D  hodoscope in front of the front window. The main purpose of the hodoscope is to tag all entering charged
particles with very high efficiency (inefficiency $\lesssim 10^{-5}$) and ignore them; they are all background. 
Because of the high resolution of the tracker and because there is 
no significant magnetic field the uncertainty on the projected entrance point is $<$ 1 mm. 
Since in 3 ab$^{-1}$ there will be about $2 \times 10^{15}$ bunch crossings,
we still expect $\sim 10^5 - 10^6$ bunch crossings with two untagged tracks from different collisions entering the
decay volume, depending on the tagger inefficiency. However, the probability that these two background tracks intersect in space, i.e. have a distance of closest approach
$\lesssim 100 \; \mu$m inside the fiducial decay volume and matching in time effectively kills this pileup background.

Decays of $K^0_S, K^0_L$, and $\Lambda^0$ inside the pipe are a serious background for any LLPs with mass $M(X^0) \lesssim$
0.8 GeV decaying to hadrons. Their mass and momentum are reconstructed from the tracks and calorimeter energies (or muon momenta in the toroid),
and one can require pointing back to the IR, good timing and a flat distribution of decay distance (as it would be
for an LLP). However the background to a search for 2-body hadronic decays of an LLP is expected to be overwhelming
except for $M(X^0) \gtrsim$ 0.8 GeV. For higher masses 4-body decays become more probable, and a well-defined
vertex with $\geq$ 4 charged tracks should have zero background. The probability of two unrelated $K^0$ decays occurring
within the resolution in $x,y,z,t$ is very small but is being evaluated, as are all expected possible backgrounds. 

A Letter of Intent to CMS is being prepared to officially propose FACET as a new subsystem and initiate a technical design study.
The most critical item is the enlarged beam pipe, since that cannnot be installed in short technical 
stops, and the next planned long shutdown LS4 is in 2031. New sources of funding will be sought. 
The detectors required represent $\lesssim$ 5\% of the CMS forward upgrades, and could 
be installed (and upgraded if needed) in technical stops.

\section*{Acknowledgements}

The author thanks all the members of the FACET developing team, named as co-authors of Ref. \cite{facetpaper}. We acknowledge
valuable input from V. Baglin and P. Fessia (CERN) on the LHC pipe and infrastructure and V. Kashikhin (Fermilab) on the preliminary
toroid design. 

\nocite{*}
\bibliographystyle{auto_generated}
\bibliography{albrowlowx/albrow}

%% file: Alcerro_Luis/Alcerro_Luis.tex
\vspace*{1.2cm}

\thispagestyle{empty}
\begin{center}
{\LARGE \bf Results of top quark production in heavy ion collisions at $\sqrt{s_{NN}} = $ 5.02 TeV and 8.16 TeV with CMS}

\par\vspace*{7mm}\par

{

\bigskip

\large \bf Luis F. Alcerro} \\ On behalf of the CMS Collaboration

\bigskip

{\large \bf  E-Mail: l.alcerro@cern.ch}

\bigskip

{Department of Physics \& Astronomy, University of Kansas, United States}

\bigskip

{\it Presented at the Low-$x$ Workshop, Elba Island, Italy, September 27--October 1 2021}

\vspace*{15mm}

\end{center}
\vspace*{1mm}

\begin{abstract}

Droplets of a strongly interacting state of matter, known as Quark-Gluon Plasma (QGP) are constantly produced in high-energy collisions of heavy nuclei. Although various methods to study the properties of the QGP have been used, they all have in common the dependence on properties of the QGP integrated over its lifetime. For this reason, the time dependence of the QGP remains elusive so far. Unlike the probes used so far, the top quark has the particularity that it can decay even inside the QGP, then provide valuable information about the time evolution of the medium. In this proceeding we discuss the current status of top-antitop ($t\overline{t}$) quark pair production at the CMS experiment and the feasibility of using top quark to unveil the time evolution of the QGP.  
\end{abstract}
  \part[Results of top quark production in heavy ion collisions at $\sqrt{s_{NN}} = $ 5.02 TeV and 8.16 TeV with CMS\\ \phantom{x}\hspace{4ex}\it{Luis F. Alcerro on behalf of the CMS Collaboration}]{}
 \section{Introduction}
\begin{figure}
\begin{center}
\includegraphics[width=0.47\textwidth]{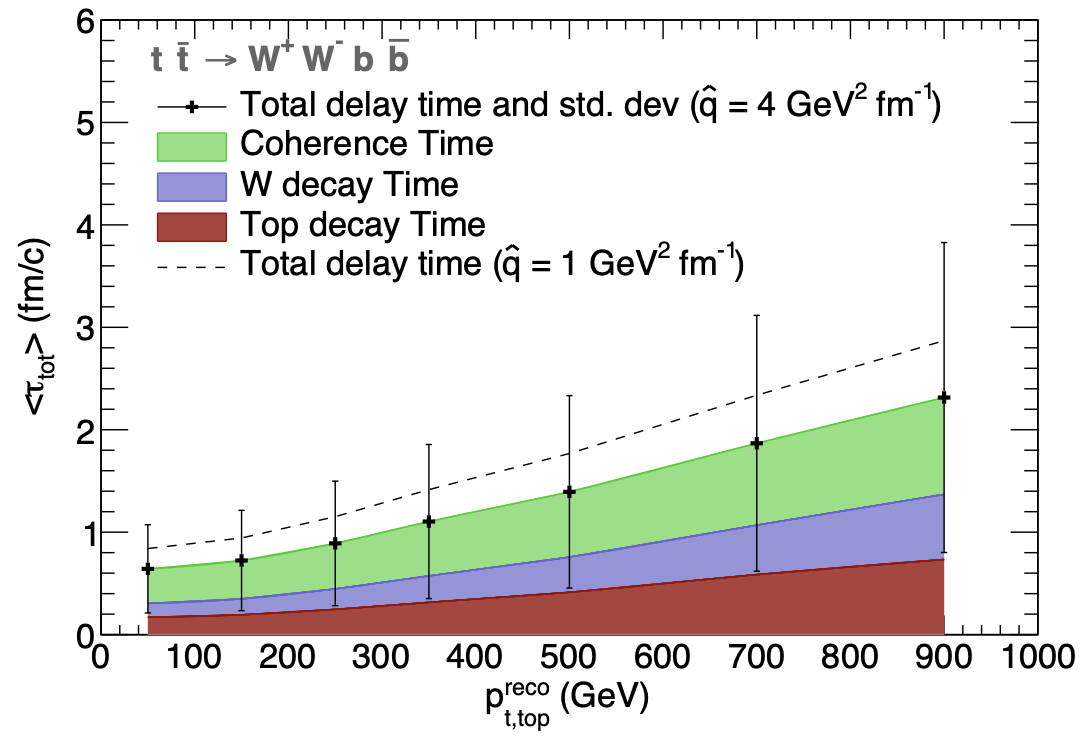}
\includegraphics[width=0.47\textwidth]{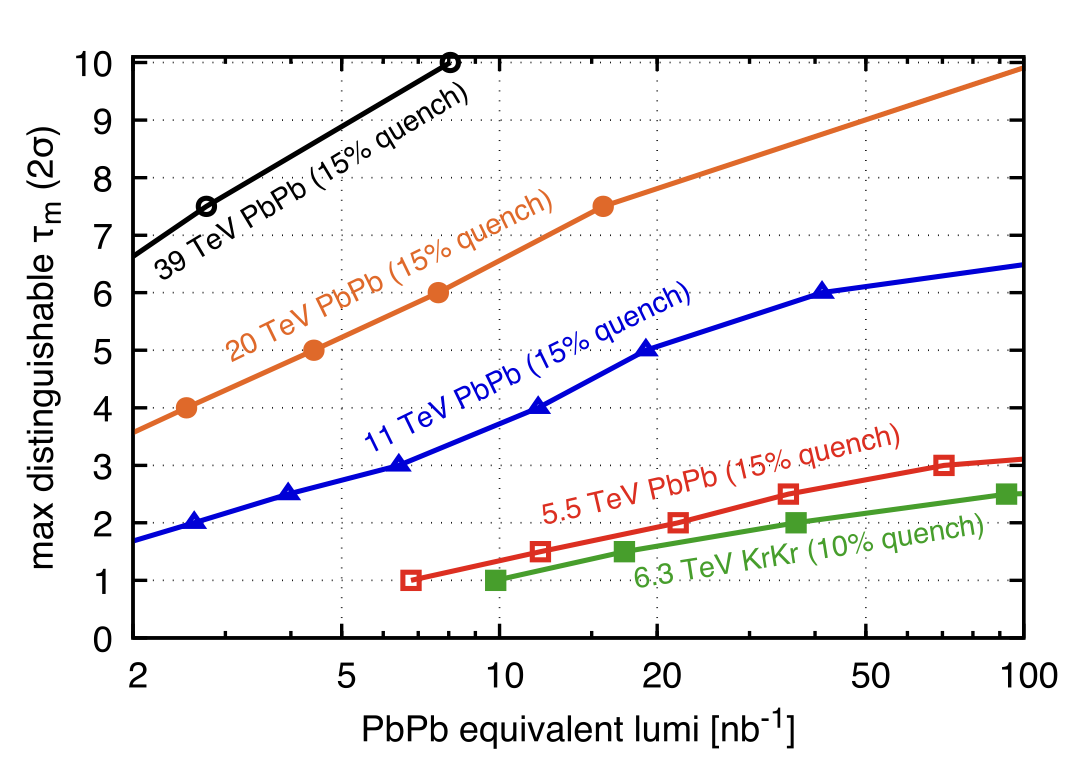}
\caption{Left: Total delay time $\tau_{tot}$ and the contributions from each component. Right: maximum medium quenching end-time, $\tau_m$, that
can be distinguished as a function of integrated luminosity. Plots taken from \cite{Apolinario:2017sob}. }
\label{pheno_1}
\end{center}
\end{figure}

The top quark was discovered in 1995 by the D0 \cite{D0:1995jca} and CDF \cite{CDF:1995wbb} Collaborations in the Tevatron experiment. With a mass of roughly $173$ GeV, it is mainly produced at LHC in $t\overline{t}$ pairs by gluon fusion and decays most of time in a b quark and a W boson, which eventually could decay either hadronically (leptonically) with a branching fraction of $\sim 66 \%$ ($\sim 33 \%$). Then we can classify the $t\overline{t}$ decay modes according to the decay products of the W bosons, namely:
\begin{itemize}
    \item $\ell +\text{jets}$ (semileptonic): \[t\overline{t}\rightarrow bb'W(\rightarrow \ell \nu)W'(\rightarrow q \overline{q}')  \]
    
    \item Dilepton (leptonic):
    \[t\overline{t} \rightarrow bb'W(\rightarrow \ell \nu)W'(\rightarrow \ell'\nu')\]
    
    \item All jets (hadronic):
    \[t\overline{t} \rightarrow bb'W(\rightarrow q\overline{q}')W'(\rightarrow q''\overline{q}''') \]
\end{itemize}
The semileptonic channel is characterized for its high branching ratio while the dilepton for its purity. The hadronic one represents the dirtiest and more challenging channel. \\
The study of top quark is key to understand various questions of QCD. In proton-proton collisions, top quark production is relevant to constrain proton PDF as well as to determine SM critical parameters, such as the $|V_{tb}|$ element of the CKM matrix. Moreover, heavy-ion collisions profits from proton-proton measurements at the same center-of-mass energies. In the case of proton-nucleus and nucleus-nucleus collisions, top quark production could serve as a probe to test nuclear PDFs as well as it paves the way for using the top quark as a probe to unveil the time structure of the QGP. \\
With a short lifetime with roughly $10^{-24}$ seconds, the top quark does not hadronize and decays before QCD mechanisms start acting. Unlike other jet quenching probes (eg. dijets, $Z/\gamma$ + jets), which are produced simultaneously with the collision, top quark can decay before of within the QGP, depending on its momentum. Taking "snapshots" at different times (or momentum) one could resolve the QGP time evolution. For this purpose, the semileptonic $t\overline{t}$ represents a golden channel due to its high branching fraction and signal/background discrimination. \\
A recent study shows the potential that hadronically decaying W bosons in top-antitop quark pair has to provide insights of the time structure of the QGP \cite{Apolinario:2017sob} . This is consequence of a "time delay" between the moment of the collision and that when the $q\overline{q}$ product of the W boson starts feeling the strong interaction of the QGP. This way, the $q\overline{q}$ pair propagates a in a certain decoherence time $\tau_d$ before start acting with the medium, so the $t \rightarrow b + W \rightarrow q \overline{q}$ decay does not see the full QGP, only the portion after \[\tau_{tot} = \gamma_{t, top} \tau_{top} + \gamma_{t,W} \tau_W + \tau_d ,\]
with $\gamma_{t,X} = (p_{t,X}^2/m_X^2 +1)^{1/2}$. In Fig. \ref{pheno_1} we can appreciate the significant contribution of the decoherence time to the total delay time. 
Using this approach, from Fig. \ref{pheno_1} we can have an idea of the current expectations of using the top quark as a probe to resolve the time dimension of the QGP in the context of future colliders and higher luminosities. We see that shorter QGP scenarios are potentially reachable during the High-Luminosity LHC era, while future colliders could resolve the full QGP evolution. \\
On the experimental side, the CMS Collaboration \cite{CMS:2008xjf} has observed top quark $t\overline{t}$ in proton-proton collisions at center-of-mass energies of 5 \cite{CMS:2017zpm}, 7, 8 \cite{CMS:2016yys,CMS:2016csa} and 13 TeV \cite{CMS:2017xrt, CMS:2016hbk}, testing different regions in Bjorken-x of the proton PDF. Top quark pair has been also measured at CMS in proton-lead (pPb) collisions at 8 TeV \cite{CMS:2017hnw} and there is evidence in lead-lead (PbPb) collisions at 5 TeV  \cite{CMS:2020aem}.
\section{$t\overline{t}$ in $pp$ at 5.02 TeV}
\begin{figure}
\begin{center}
\includegraphics[width=\textwidth]{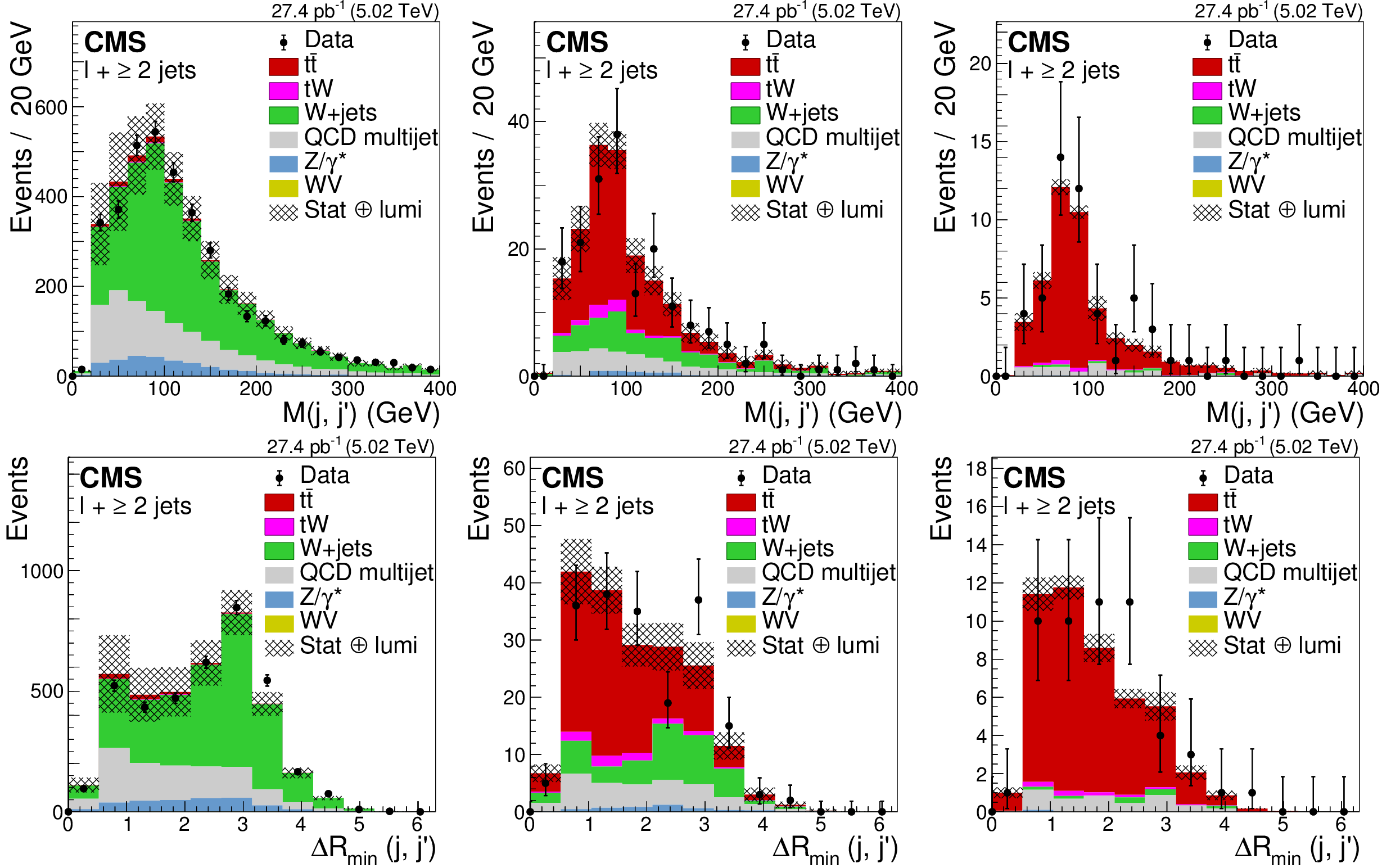}
\caption{Distributions of the invariant mass $M(j,j')$ and minimum angular distance $\Delta R_{min}(j,j')$ of the light jets $j,j'$. Events are classified into 0 b (left), 1 b (center) and $\geq 2$ b tagged jets categories. Plots taken from \cite{CMS:2017zpm}.}
\label{pp_1}
\end{center}
\end{figure}

\begin{figure}
\begin{center}
\includegraphics[width=0.47\textwidth]{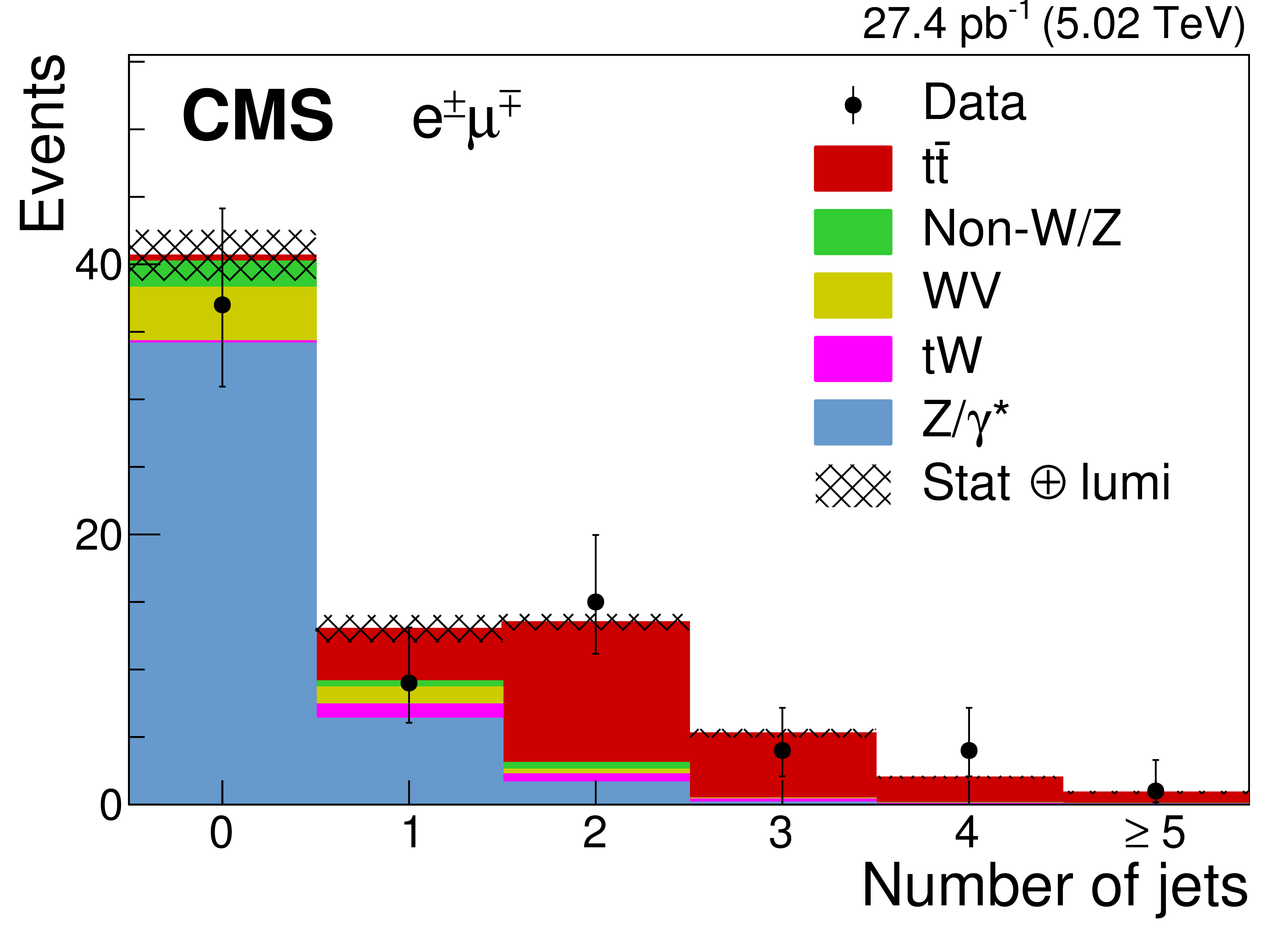}
\includegraphics[width=0.47\textwidth]{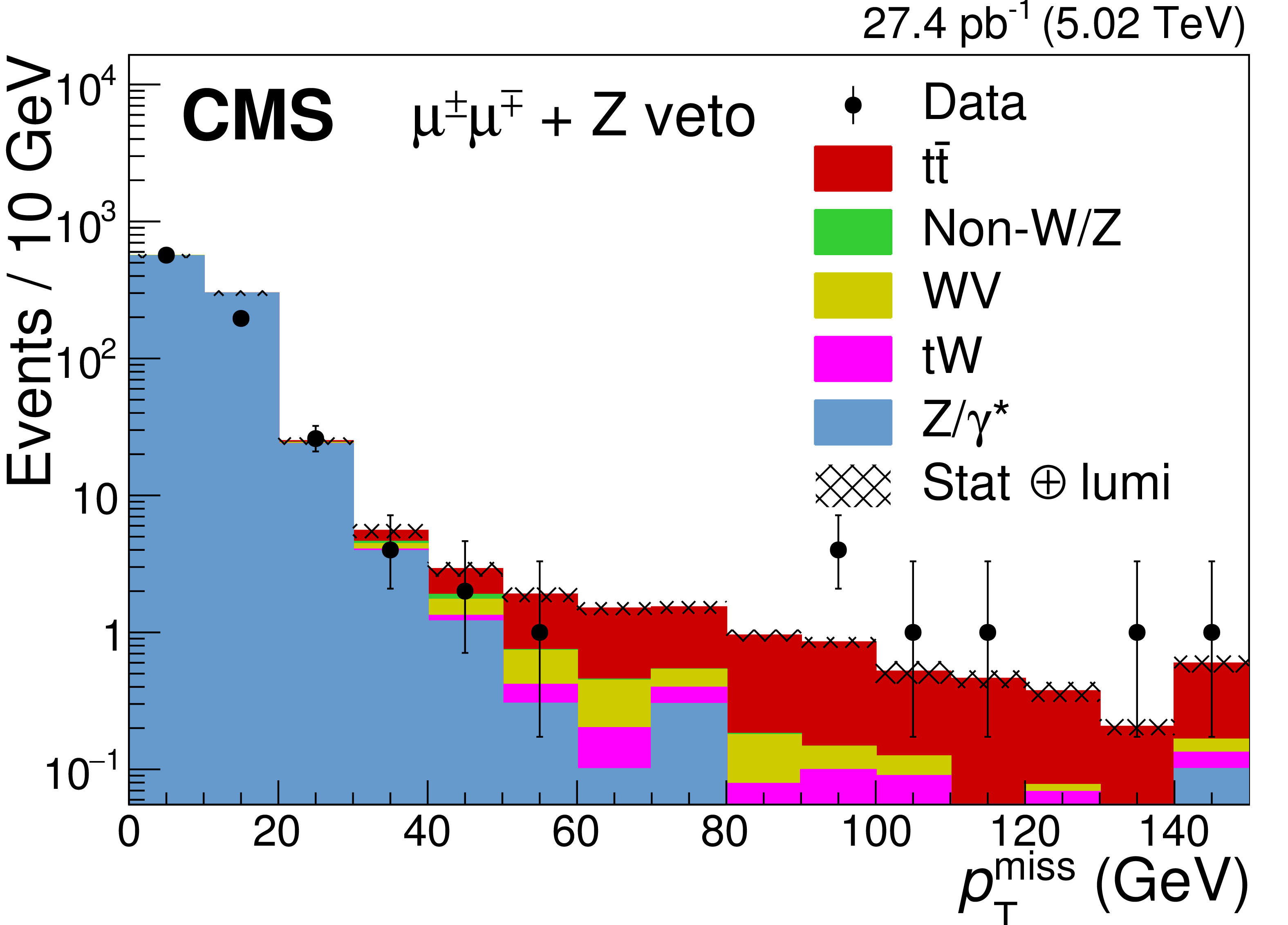}
\caption{Distributions of jet multiplicity (left) and missing transverse momentum (right) for events passing the dilepton criteria. Plots taken from \cite{CMS:2017zpm}.}
\label{pp_2}
\end{center}
\end{figure}

\begin{figure}
\begin{center}
\includegraphics[width=0.45\textwidth]{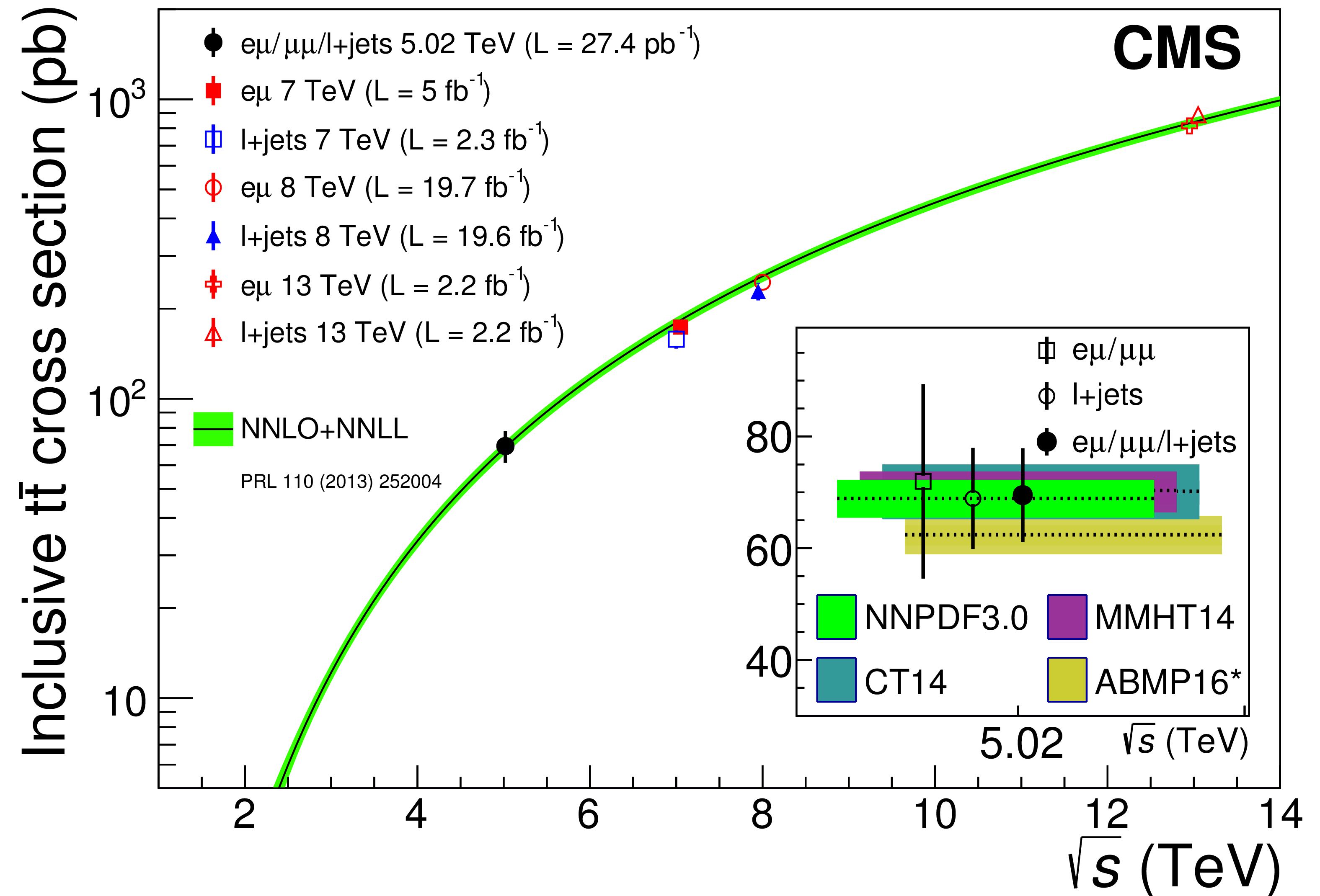}
\includegraphics[width=0.51\textwidth]{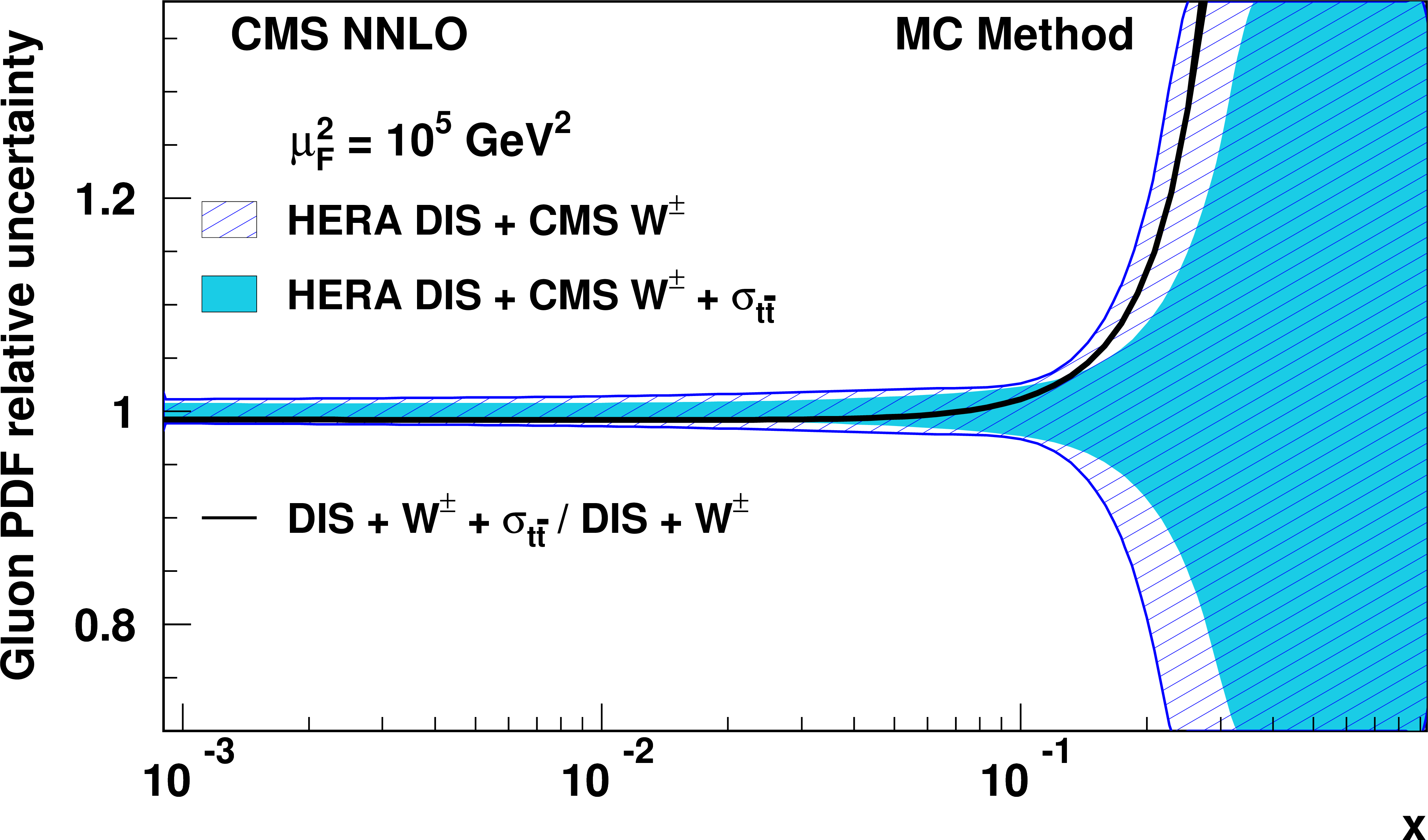}
\caption{Left: Inclusive $\sigma_{t\overline{t}}$ measurements at $\sqrt{s}=5,7,8,13$ TeV compared to theoretical predictions. Right: The relative uncertainties in the gluon distribution function of the proton as a function
of $x$ from a QCD analysis. Plots taken from \cite{CMS:2017zpm}.}
\label{pp_3}
\end{center}
\end{figure}

The first measurement of the total cross section of $t\overline{t}$, $\sigma_{t\overline{t}}$, at 5.02 TeV was performed by the CMS Collaboration using data recorded in November 2015 \cite{CMS:2017zpm} with a data sample that corresponds to an integrated luminosity of 27.4 pb$^{-1}$. For this measurement the $\ell$ + jets and dilepton channels were analyzed. The former is characterized by the presence in the final state of at least two b jets, two light jets, one lepton and momentum imbalance due two the undetected neutrino while the later involves two b jets, two light jets, two high energy leptons and momentum imbalance. \\
Control samples in data are used to estimate backgrounds coming from multijets (Drell-Yan, referred to as $Z/\gamma^*$) in $\ell$ + jets (dilepton) channels. All other contributions in both channels are estimated from simulations. \\
In the $\ell$ + jets channel, events are classified into 3 b jet multiplicity categories: 0b, 1b, $\geq 2$b (see Fig. \ref{pp_1}). The cross section is extracted by means of likelihood fits in the angular distance of the two light jets ($j,j'$) produced from the W boson hadronic decay, $\Delta R(j, j')$. In the dilepton analysis, the cross section is extracted with an event counting technique. Fig. \ref{pp_2} shows the jet multiplicity and missing transverse momentum distributions for events passing the dilepton criteria.
Cross sections measured in both channels are combined to determine an overall $t\overline{t}$ cross section:
\[\sigma_{t\overline{t}} = 69.5 \pm 6.1 \text{ }(stat)\pm 5.6 \text{ }(syst) \pm 1.6 \text{ }(lumi) \text{ pb.}\] 
This result is in agreement with theory and has been used in a QCD analysis showing a moderate reduction of the uncertainty in the gluon PDF of the proton, as shown in Fig. \ref{pp_3}.\\ It is worth to mention that this result has been recently updated with an increase in integrated luminosity of more
than an order of magnitude compared to the data set previously mentioned \cite{CMS:2021gwv}. This analysis takes into account the dilepton channel only, obtaining a cross section of $60.7\pm 5.0(stat) \pm 2.8(syst)\pm 1.1 (lumi)$ pb. In combination with the $\ell+$jets result from 2015 data \cite{CMS:2017zpm}, the updated cross section measurement is
\[\sigma_{t\overline{t}} = 63.0 \pm 4.1 \text{ }(stat)\pm 3.0 \text{ }(syst+lumi)  \text{ pb.}\] 

\section{$t\overline{t}$ in $pPb$ at 8.16 TeV}
\begin{figure}
\begin{center}
\includegraphics[width=\textwidth]{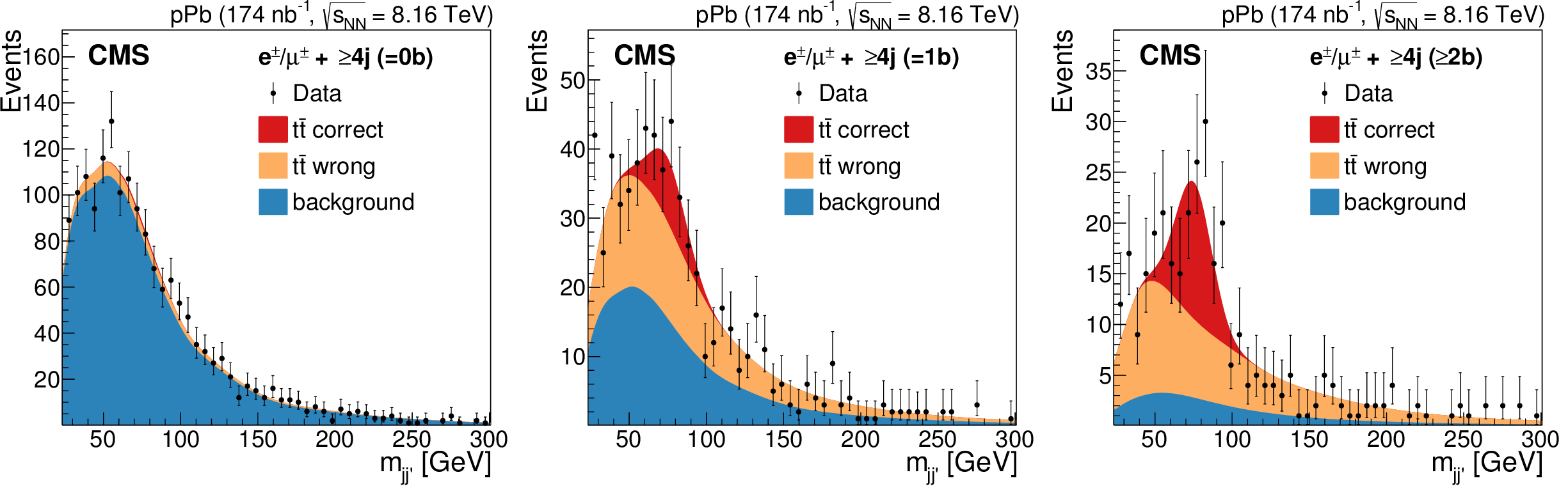}
\caption{Invariant mass distributions of the W candidate, $m_{jj^{′}}$, in the 0 (left), 1 (center), and 2 (right) b-tagged jet categories. Plots taken from \cite{CMS:2017hnw}.}
\label{ppb_1}
\end{center}
\end{figure}

\begin{figure}
\begin{center}
\includegraphics[width=\textwidth]{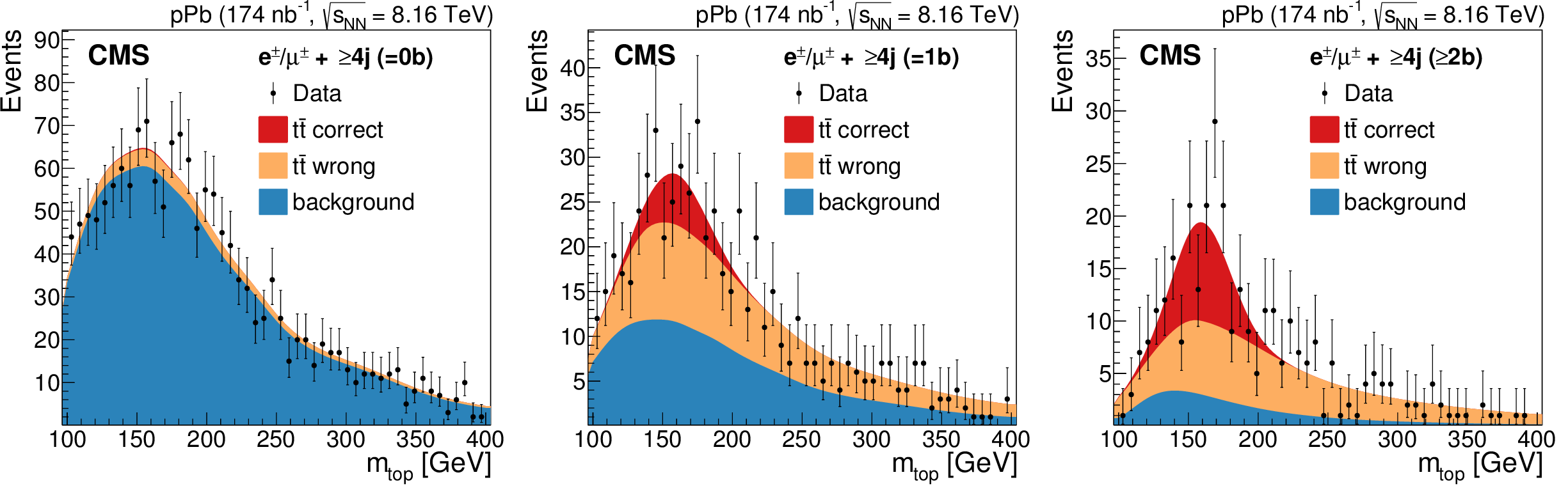}
\caption{Distributions of $m_{\text{top}}$, in the 0 (left), 1 (center), and 2 (right) b-tagged jet categories. Plots taken from \cite{CMS:2017hnw}. }
\label{ppb_2}
\end{center}
\end{figure}

\begin{figure}
\begin{center}
\includegraphics[width=0.5\textwidth]{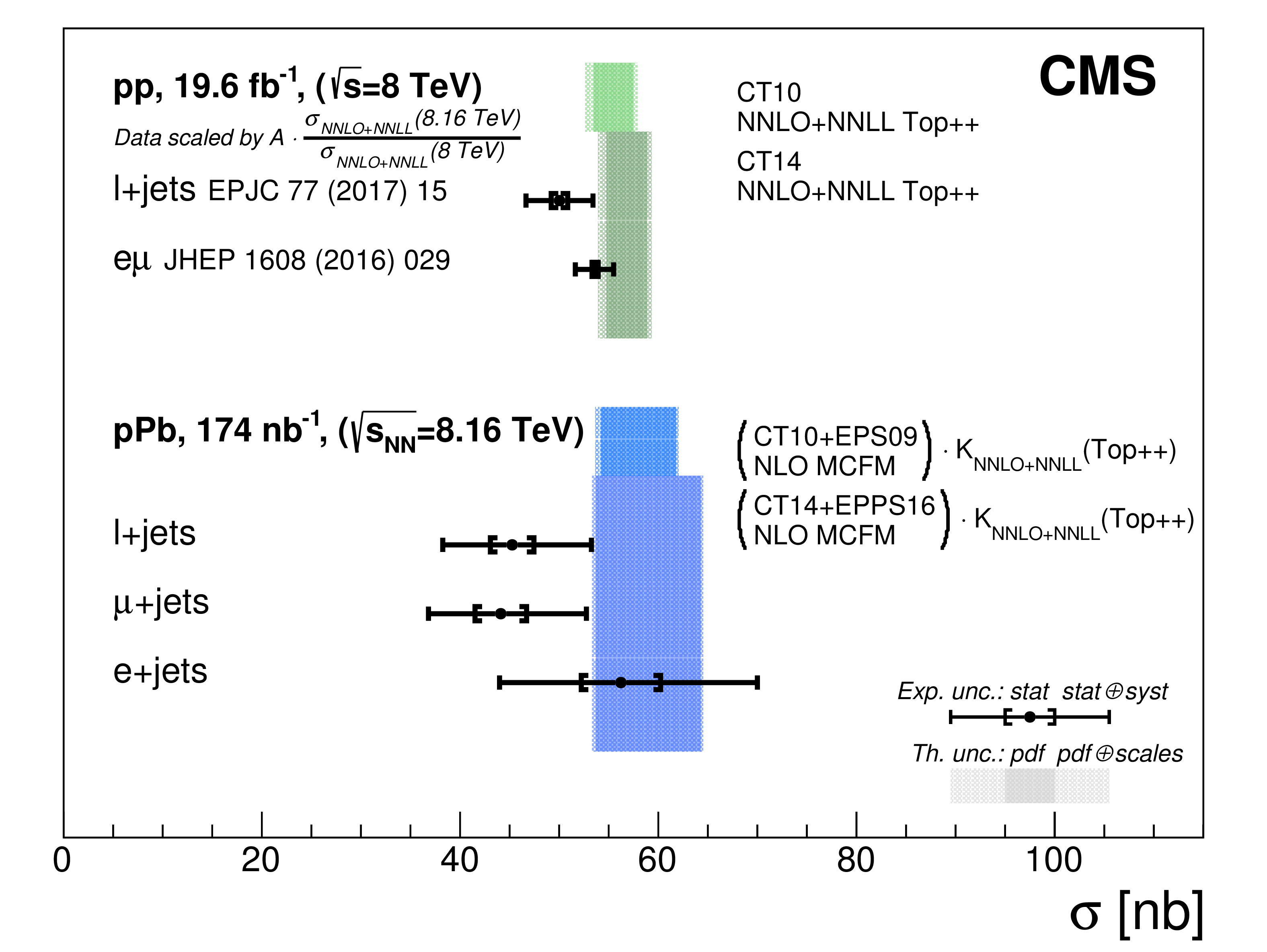}
\caption{Total $t\overline{t}$ cross sections in $e$+jets, $\mu$+jets and combined $\ell$+jets channels compared to theory and pp scaled data. Plots taken from \cite{CMS:2017hnw}.}
\label{ppb_3}
\end{center}
\end{figure}

The CMS Collaboration performed the first observation of top quark in proton-nucleus collisions using proton-lead at $\sqrt{s}=8.16$ TeV data taken in 2016 equivalent to integrated luminosity of 174 nb$^{-1}$ nb \cite{CMS:2017hnw}. For this analysis, due to its high branching fraction and moderate background contamination, only the $\ell$ + jets channel is considered.  Events are required to contain exactly one muon or
electron with $p_T>30$ GeV and $|\eta|<2.1$ (except in the electron case where there is the transition region $1.444<|\eta|<1.566$ between the barrel and endcap) and to be isolated from hadronic activity in both cases. Events are also required to have at least four anti-$k_T$ jets with cone size of 0.4, $p_T>25$ GeV and $|\eta|<2.5$. Jets coming from b quarks are identified based on the presence of a secondary vertex from B-hadron decays. \\
The main sources of background are QCD multijet and W + jets (collectively labeled as “non-top” background) which are taken from simulation. Since the presence of two b jets is very uncommon in non-top background, the number of jets passing a threshold of a b jet identification discriminant is used to categorize each event candidate into no (0b), exactly one (1b), or at least two (2b) tagged-jet categories. The invariant mass of the two light-flavor jets ($m_{jj^{'}}$) produced from the decay of the W boson is used as input for a likelihood fit for the cross section extraction (see Fig. \ref{ppb_1}). As a further examination of the hypothesis that the selected data are consistent with the production of top quarks, a proxy of the top quark mass, $m_{\text{top}}$, is constructed
as the invariant mass of candidates formed by pairing the W candidate with a b-tagged jet, $t \rightarrow b j j^{'}$. Figure \ref{ppb_2}  shows the distribution of $m_{\text{top}}$ reconstructed for events in the 0, 1, and 2 b-tagged jet categories.\\
The combined fit to both channels ($e/\mu$ + jets) and all three b-tagged jet categories yields the cross section:
\[\sigma_{t\overline{t}} = 45 \pm 8 \text{ nb,}\]
compatible with theory and proton-proton scaled data, as shown in Fig. \ref{ppb_3}.
\section{$t\overline{t}$ in PbPb at 5.02 TeV}
\begin{figure}
\begin{center}
\includegraphics[width=\textwidth]{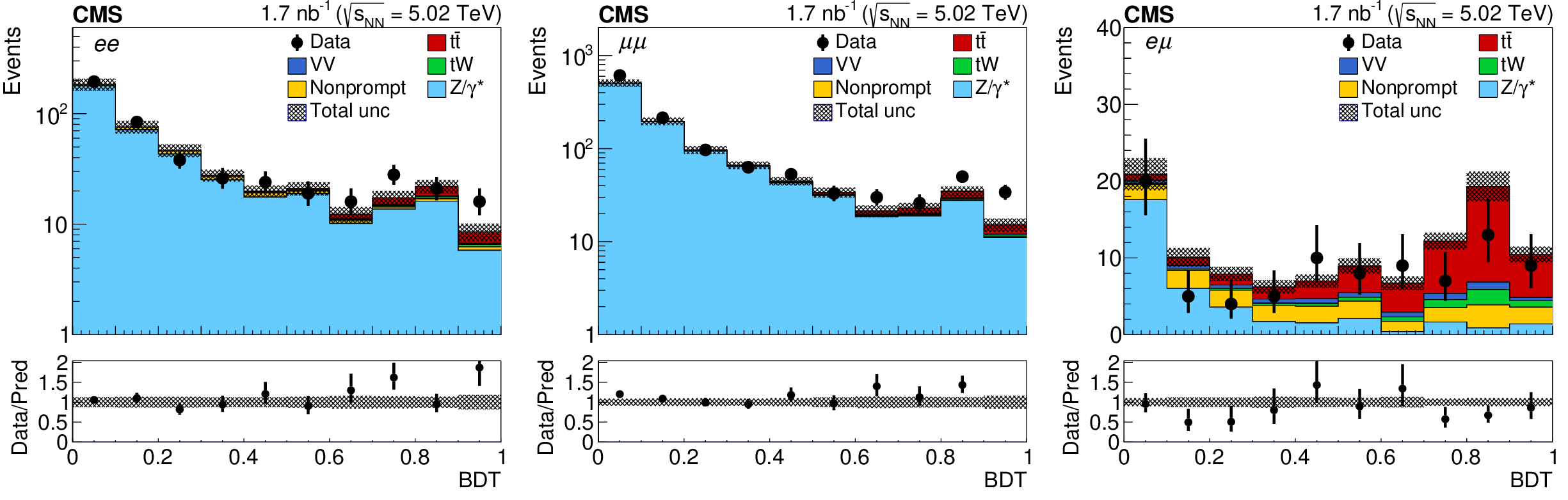}
\caption{BDT discriminator distributions in the $e^+e^−$ (left), $\mu^+ \mu^-$ (middle), and $e^\pm \mu^\mp$ (right) final states. Plots taken from \cite{CMS:2020aem}.}
\label{pbpb_1}
\end{center}
\end{figure}

\begin{figure}
\begin{center}
\includegraphics[width=\textwidth]{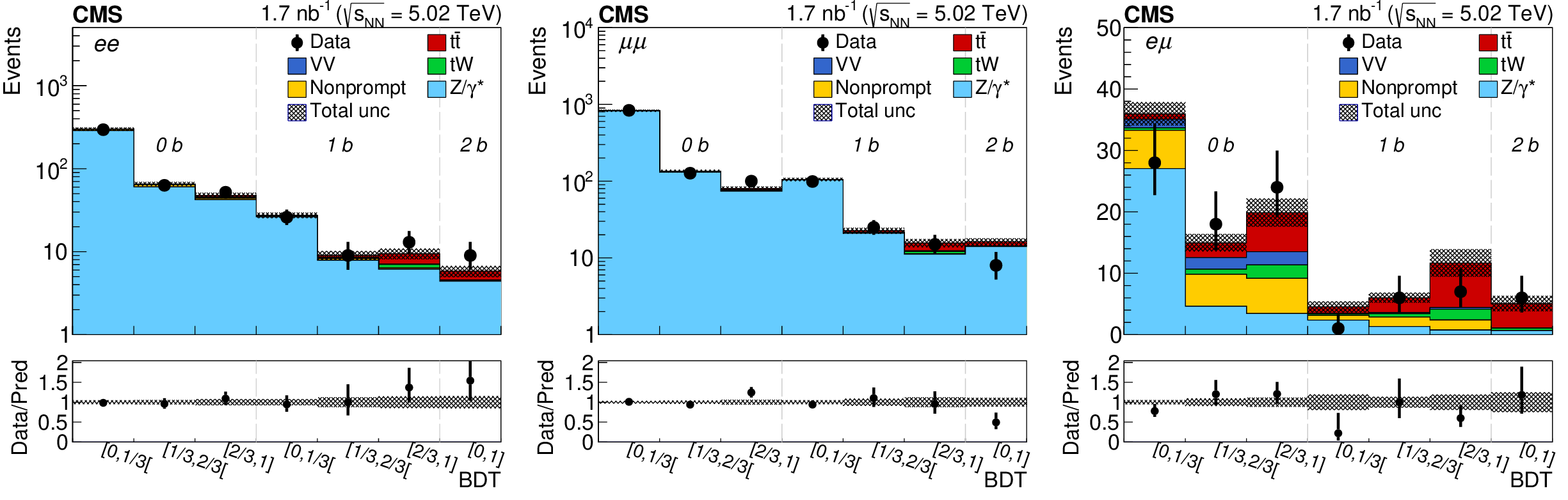}
\caption{ BDT discriminator distributions in the $e^+e^−$ (left), $\mu^+ \mu^-$ (middle), and $e^\pm \mu^\mp$ (right) final states separately for the 0b-, 1b-, and 2b-tagged jet multiplicity categories. Plots taken from \cite{CMS:2020aem}.}
\label{pbpb_2}
\end{center}
\end{figure}

\begin{figure}
\begin{center}
\includegraphics[width=0.5\textwidth]{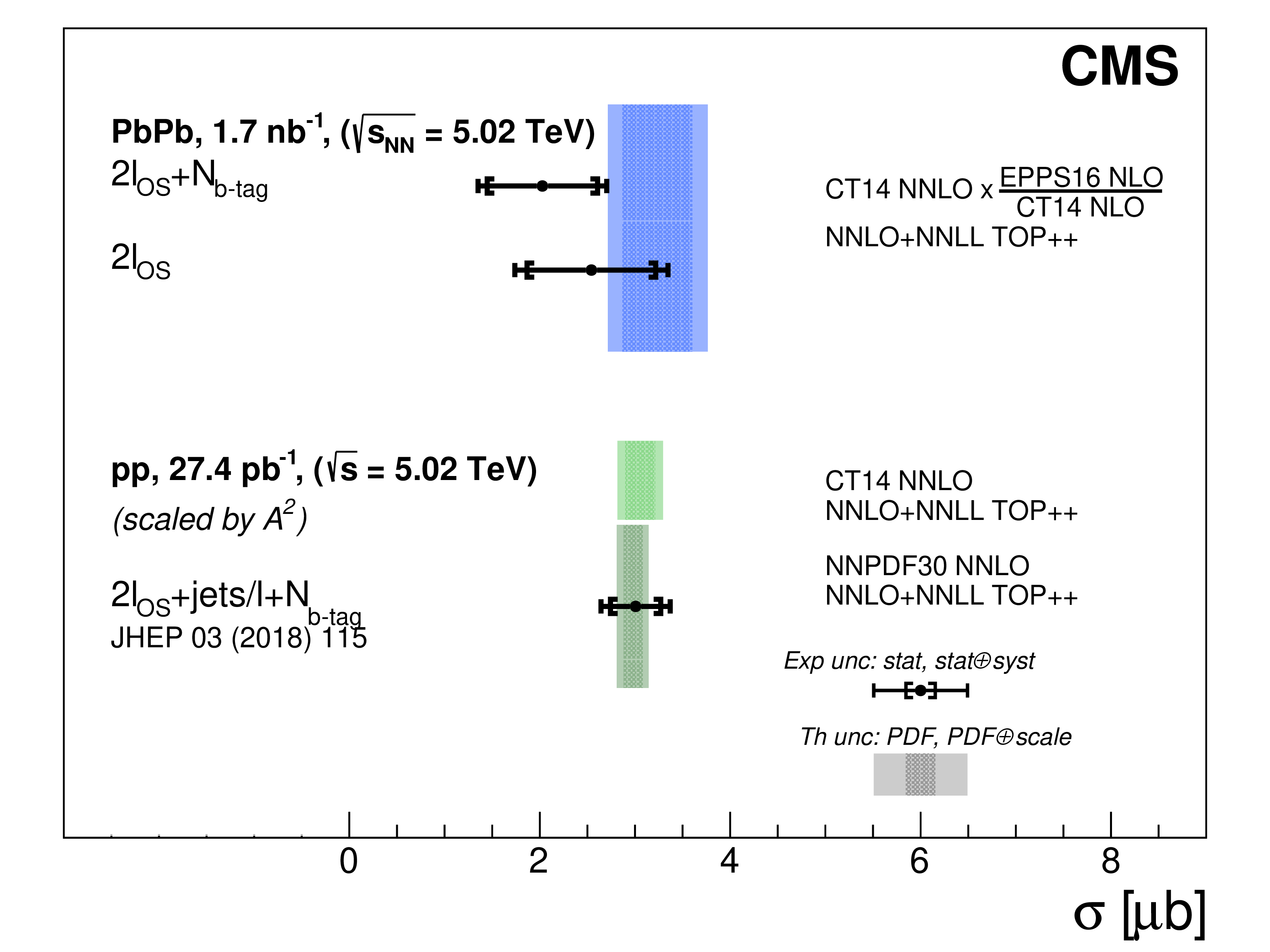}
\caption{Inclusive $t\overline{t}$ cross sections measured with two methods in the combined $e^+e^−$, $\mu^+ \mu^-$, and $e^\pm \mu^\mp$ final states in PbPb collisions at $\sqrt{s}$= 5.02 TeV, and pp results at the same energy. Plots taken from \cite{CMS:2020aem}.}
\label{pbpb_3}
\end{center}
\end{figure}

The very first evidence of top quark production in heavy nuclei was reported by the CMS Collaboration using data recorded in 2018, corresponding to an integrated luminosity of 1.7 nb$^{-1}$ of lead-lead collisions at center-of-mass energy of $\sqrt{s}=5.02$ TeV \cite{CMS:2020aem}. The purity of the dilepton channel is exploited in this analysis with and without inclusion of information coming from b tagged jets. \\
The data is filtered to contain two opposite sign (OS) leptons with $p_T>25$ (20) GeV and $|\eta|<2.1$ (2.4) for electrons (muons) with no nearby hadronic activity. The presence of b-tagged jets is further exploited in a second method to enhance the signal. Jets are tagged using information of secondary vertices. \\
The main sources of background are Drell-Yan (referred to as “$Z/\gamma∗$”) and W+jets and QCD multijets (referred as "nonprompt"). Drell-Yan processes are modeled from Monte Carlo with corrections obtained from data while nonprompt are directly derived from control regions in the data. \\
In both methods, a Boosted Decision Tree (BDT) is trained to discriminate genuine leptons with high $p_T$ between signal and background processes (see Figs. \ref{pbpb_1} and \ref{pbpb_2} ). In order to minimize effects of the imprecise knowledge of the jet properties in the heavy ion environment, BDTs use kinematic properties only. Likelihood fits to binned BDT distributions are performed separately for the two methods to extract the cross section, obtaining:
\[\sigma_{t\overline{t}} = 2.03^{+0.71}_{-0.64} \text{ }\mu \text{b}\]
with the $2\ell_{\text{OS}} +$ b-jets method and 
\[\sigma_{t\overline{t}} = 2.54^{+0.84}_{-0.74} \text{ }\mu \text{b}\]
with $2\ell_{\text{OS}}$. The results are compatible with theoretical calculations and pp at 5 TeV scaled by the number of binary nucleon-nucleon collisions in PbPb as well, as is shown in Fig. \ref{pbpb_3}.

\section{Summary}
The CMS experiment has shown the capability to perform top quark studies both in different systems and energies obtaining results in agreement with simulations. In particular, the evidence of top quark in nucleus-nucleus prepares the way to explore the time evolution of the QGP using top quark with LHC higher luminosities and future colliders.

\section*{Acknowledgements}

The work is supported in whole by the Nuclear Physics (NP) program of the U.S. Department of Energy (DOE) with number \href{https://pamspublic.science.energy.gov/WebPAMSExternal/Interface/Common/ViewPublicAbstract.aspx?rv=00d4fe0f-48a0-4d4a-baf1-c70867d9e499&rtc=24&PRoleId=10}{DE-FG02-96ER40981}.

\nocite{*}
\bibliographystyle{auto_generated.bst} 
\bibliography{Alcerro_Luis/Alcerro_Luis.bib}

%% file: Brodsky_proceedings_elba2021/Brodsky_proceedings_elba2021/Brodsky_proceedings_elba2021/sjb_Elba_contribution.tex
\vspace*{1.2cm}

\thispagestyle{empty}
\begin{center}
{\LARGE \bf  Novel Effects in QCD: Intrinsic Heavy Quarks, Color Transparency, and the Implications of Diffractive Contributions to Deep Inelastic Scattering for the Momentum Sum Rule} 
\par\vspace*{7mm}\par

{

\bigskip

\large \bf Stanley J. Brodsky}

\bigskip

{\large \bf  E-Mail: sjbth@slac.stanford.edu}

\bigskip

{SLAC National Accelerator Laboratory, Stanford, California}

\bigskip

{\it Presented at the Low-$x$ Workshop, Elba Island, Italy, September 27--October 1 2021}

\vspace*{15mm}

\end{center}
\vspace*{1mm}

\begin{abstract}

I discuss three novel features of QCD -- asymmetric intrinsic heavy-quark phenomena, color transparency, and the violation of the momentum sum rule for nuclear structure functions measured in deep inelastic lepton scattering.

\end{abstract}

  \part[Novel Effects in QCD: Intrinsic Heavy Quarks, Color Transparency, and the Implications of Diffractive Contributions to Deep Inelastic Scattering for the Momentum Sum Rules\\ \phantom{x}\hspace{4ex}\it{Stanley J. Brodsky}]{}

\section{Intrinsic Heavy Quarks}

Quantum Chromodynamics (QCD), the underlying theory of strong interactions, with quarks and gluons as the fundamental degrees of freedom, predicts that the heavy quarks in the nucleon-sea to have both perturbative ``extrinsic" and nonperturbative ``intrinsic" origins.  The extrinsic sea arises from gluon splitting which is triggered by a probe in the reaction. It can be calculated order-by-order in perturbation theory.  In contrast, the intrinsic sea is encoded in the nonperturbative wave functions of the nucleon eigenstate. 

The existence of nonperturbative intrinsic charm (IC) was originally proposed in the BHPS model~\cite{Brodsky:1980pb} and developed further in subsequent papers~\cite{Brodsky:1984nx,Harris:1995jx,Franz:2000ee}. The intrinsic contribution to the heavy quark  distributions of hadrons at high $x$ corresponds to Fock states such as  $|uud Q \bar Q>$ where the heavy quark 
pair is multiply connected to two or more valence quarks of the proton.   It is maximal at minimal off-shellness; i.e., when the constituents all have the same rapidity  $y_I$, and thus 
$x_i \propto \sqrt(m_i^2+ { \vec k_{\perp i}}^2 )$.  Here $x= {k^+\over P^+} = {k^0 + k^3\over P^0 + P^3}$ is the frame-independent light-front momentum fraction carried by the heavy quark in a hadron with momentum $P^\mu$. 
In the case of deep inelastic lepton-proton scattering, the LF momentum fraction variable $x$  in the proton structure functions can be identified with the Bjorken variable 
$x = {Q^2\over 2 p \cdot q}.$
These heavy quark contributions 
to the nucleon's PDF thus peak at large $x_{bj}$ and thus have important  implication for LHC and EIC collider phenomenology, including Higgs and heavy hadron production at high $x_F$~\cite{Royon:2015eya}.
It also opens up new opportunities to study heavy quark phenomena in fixed target experiments such as the proposed AFTER~\cite{Brodsky:2015fna} fixed target facility at CERN.  Other applications are presented in Refs.~\cite{Brodsky:2020zdq,Bednyakov:2017vck,Brodsky:2016fyh}.
The  existence of intrinsic heavy quarks also illuminates fundamental aspects of nonperturbative QCD.

In  light-front (LF) Hamiltonian theory, the intrinsic heavy quarks of the proton are associated with non-valence Fock states. 
such as $|uud Q \bar Q>$ in the hadronic eigenstate of the LF Hamiltonian; this implies that the heavy quarks are multi-connected to the valence quarks. The probability for the heavy-quark Fock states scales as $1/m^2_Q$ in non-Abelian QCD.  Since the LF wavefunction is maximal at minimum off-shell invariant mass; i.e., at equal rapidity, the intrinsic heavy quarks carry large momentum fraction $x_Q$.  A key characteristic is different momentum and spin distributions for the intrinsic $Q$ and $\bar Q$ in the nucleon; for example the charm-anticharm asymmetry, since the comoving quarks are sensitive to the global quantum numbers of the nucleon~\cite{Brodsky:2015fna}.  Furthermore, since all of the  intrinsic quarks in the $|u[ud] Q \bar Q>$  Fock state have similar rapidities as the valence quarks, they can re-interact, leading to significant $Q$ vs $\bar Q$ asymmetries.  The concept of intrinsic heavy quarks was also proposed in the context of  meson-baryon fluctuation models~\cite{Navarra:1995rq,Pumplin:2005yf} where intrinsic charm was identified with two-body state $\bar{D}^0(u\bar{c})\Lambda^+_c(udc)$ in the proton. This identification  predicts large asymmetries in the charm versus charm momentum and spin distributions,  Since these heavy quark distributions depend on the correlations determined by the valence quark distributions, they are referred to as {\it  intrinsic } contributions to the hadron's fundamental structure. A specific analysis of the intrinsic charm content of the deuteron is given in ref.~\cite{Brodsky:2018zdh}.
In contrast, the contribution to the heavy quark PDFS arising from gluon splitting are symmetric in $Q$ vs $\bar Q$. The contributions generated by DGLAP evolution at low $x$ can be considered as  {\it extrinsic} contributions since they only depend on the gluon distribution. The gluon splitting contribution to the heavy-quark degrees of freedom is  perturbatively calculable 
using  DGLAP
evolution.  To first approximation, the perturbative extrinsic heavy quark distribution falls as $(1-x)$ times the gluon distribution and is limited to low $x_{bj}.$
Thus, unlike the conventional $\log m^2_Q$ dependence of the low $x$  extrinsic gluon-splitting contributions, the probabilities for the intrinsic heavy quark Fock states at high $x$  scale as $1\over m_Q^2$  in non-Abelian QCD, and the relative probability of intrinsic bottom to charm is of order ${m^2_c\over m^2_b} \sim  {1\over 10}.$
In contrast, the probability for a higher Fock state containing heavy leptons in a QED atom  scales as $1\over m_\ell^4$, corresponding to the twist-8 Euler-Heisenberg light-by-light self-energy insertion.  Detailed derivations based on the OPE have been given in Ref. ~\cite{Brodsky:1984nx,Franz:2000ee}.

In an important recent development~\cite{Sufian:2020coz},  the difference of the charm and anticharm  quark distributions in the proton, $\Delta c(x) = c(x) -\bar c(x)$,  has been computed from first principles in QCD using lattice gauge theory.  A key  theoretical tool is the computation of the charm and anticharm quark contribution 
to the electromagnetic form factor of the proton which would vanish if $c(x) =\bar c(x).$    The exclusive-inclusive connection, together with the LFHQCD formalism, predicts the asymmetry of structure functions $c(x)- \bar c(x)$ which is also odd under charm-anticharm interchange.   
The predicted $c(x)- \bar c(x)$ distribution is large and nonzero at large at $x \sim 0.4$, consistent with the expectations of intrinsic charm. 

\begin{figure}[htp]
\begin{center}
\setlength\belowcaptionskip{-2pt}
\includegraphics[width=3.2in, height=2.3in]{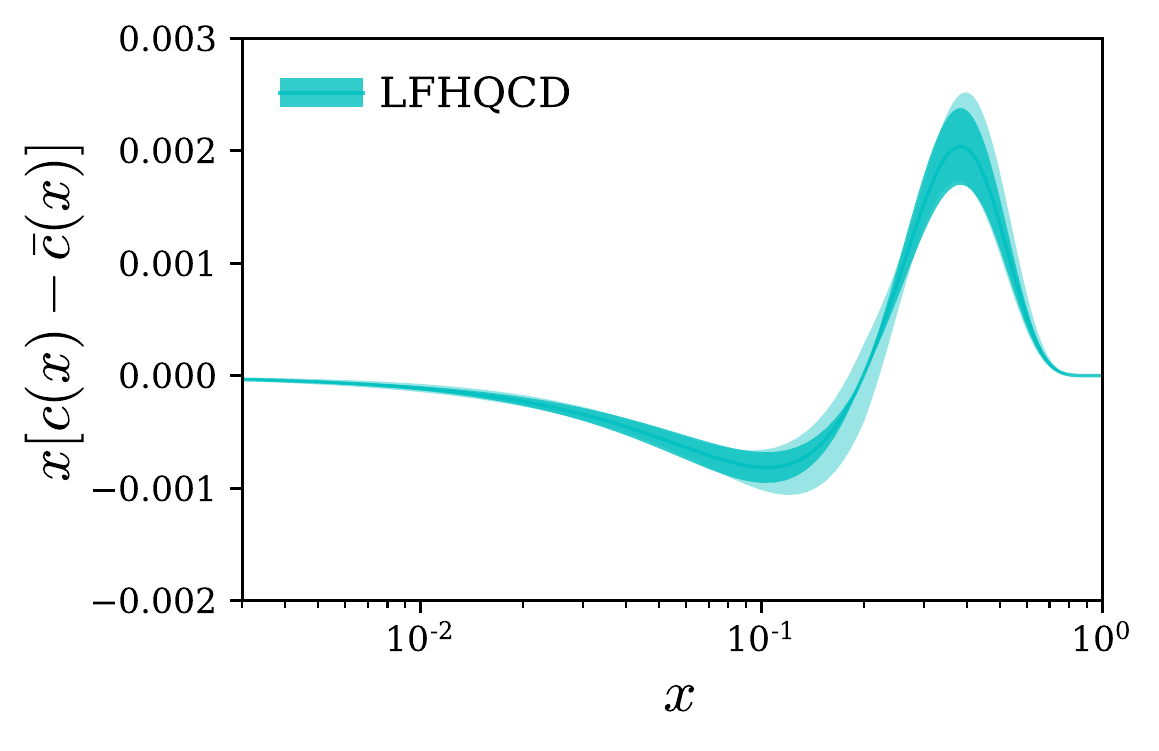}
\caption{The difference of charm and anticharm structure functions $x[c(x)-\bar{c}(x)]$ obtained from the LFHQCD formalism using the lattice QCD input of charm electromagnetic form factors $G^c_{E,M}(Q^2)$ \label{fig:ccbardis}. 
The outer cyan band indicates an estimate of systematic uncertainty in the $x[c(x)-\bar{c}(x)]$ distribution obtained from a variation of  the hadron scale $\kappa_c$ by 5\%.  From ref.~\cite{Sufian:2020coz}}.
\end{center}
\end{figure}

The $c(x)$ vs. $\bar c(x)$  asymmetry can also be understood physically by identifying the $ |uud c\bar c>$ Fock state with the $|\Lambda_{udc} D_{u\bar c}>$ off shell excitation of the proton.
See Fig.~\ref{fig:ccbardis}.   A related application of lattice gauge theory to the nonperturbative strange-quark sea from lattice QCD is given in ref.~\cite{Sufian:2018cpj}.

There have been many phenomenological calculations involving the existence of a non-zero IC  component which can explain anomalies in the experimental data and to predict  its novel signatures of IC in upcoming experiments~\cite{Brodsky:2015fna}.   A recent measurement by LHCb is shown in Fig. 10.
The observed spectrum exhibits a sizable enhancement at forward Z rapidities, consistent with the effect expected if the proton contains the $ |uud \bar c c>$ Fock state predicted by LFQCD.~\cite{LHCb:2021stx}

\begin{figure}
 \begin{center}
\includegraphics[height= 10cm,width=15cm]{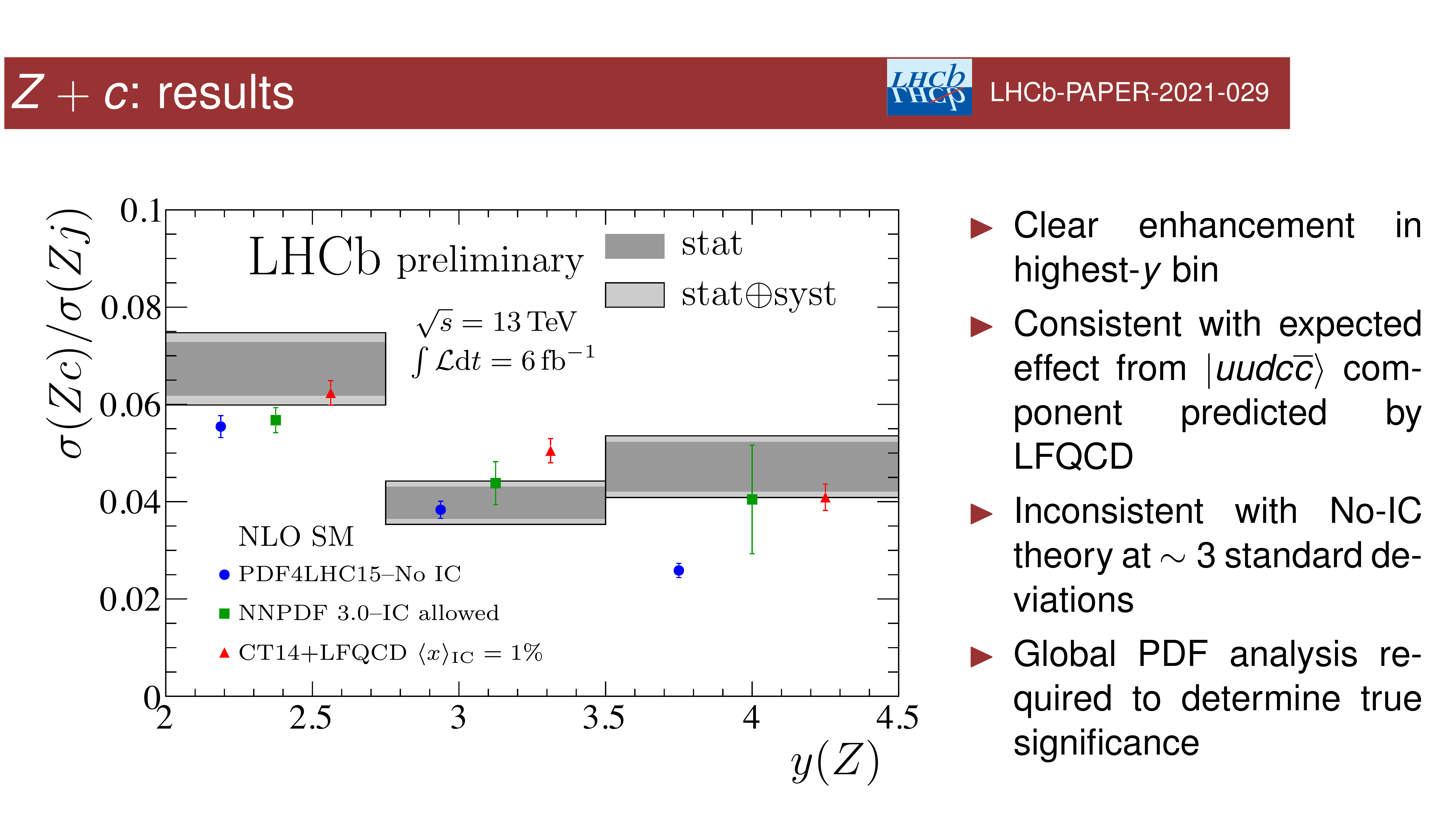}
\end{center}
\caption{The charm distribution in the proton determined from LHCb  measurements of 
$Z$ bosons produced in association with charm at forward rapidity~\cite{LHCb:2021stx}.}
\label{Bledslides1LHCb.pdf}
\end{figure}

Thus QCD predicts two separate and distinct contributions to the heavy quark distributions $q(x,Q^2)$ of  the nucleons at low and high $x$.
Here $x= {k^+\over P^+} = {k^0 + k^3\over P^0 + P^3}$ is the frame-independent light-front momentum fraction carried by the heavy quark in a hadron with momentum $P^\mu$. 
In the case of deep inelastic lepton-proton scattering, the LF momentum fraction variable $x$  in the proton structure functions can be identified with the Bjorken variable 
$x = {Q^2\over 2 p \cdot q}.$
At small $x$,  heavy-quark pairs are dominantly produced via the standard gluon-splitting subprocess $g \to Q \bar Q$.  

The presence of the heavy
quarks in nucleon  from this contribution is a result of the QCD DGLAP evolution of the light quark and gluon
PDFs.  Unlike the conventional $\log m^2_Q$ dependence of the low $x$  extrinsic gluon-splitting contributions, the probabilities for the intrinsic heavy quark Fock states at high $x$  scale as $1\over m_Q^2$  in non-Abelian QCD.   Thus the relative probability of intrinsic bottom to charm is of order ${m^2_c\over m^2_b} \sim  {1\over 10}.$
In contrast, the probability for a higher Fock state containing heavy leptons in a QED atom  scales as $1\over m_\ell^4$, corresponding to the twist-8 Euler-Heisenberg light-by-light self-energy insertion.  Detailed derivations based on the OPE have been given in Ref. ~\cite{Brodsky:1984nx,Franz:2000ee}.

\section{Color Transparency}

One of the most striking properties of QCD phenomenology is ``color transparency"~\cite{Brodsky:1988xz}, the reduced absorption of a hadron as it propagates through nuclear matter, if it is produced at high transverse momentum in a hard exclusive process, such as elastic lepton-proton scattering. The nuclear absorption reflects the size of the color dipole moment of the propagating hadron; i.e.,  the separation between its colored constituents.

The key quantity which measures the  transverse size of a scattered hadron in a given Fock state is~\cite{sjbGdT}
$$a_\perp = \sum_{i=1}^{n-1} x_i b_{\perp i}$$ 
The LF QCD formula for form factors can then be written compactly in impact space as 
$$F(Q^2) = \int^1_0 dx d^2 a_\perp  e^{i \vec q_\perp\cdot a_\perp} q(x, a_\perp)$$
and thus 
$$a^2_\perp(Q^2) =4{{d \over dQ^2} F(Q^2) \over F(Q^2)}$$
measures the slope of the hadron factor. 
We can use LF holography to predict for the valence Fock state
$$a^2_\perp = 4 {\tau-1\over Q^2}$$
which shows that, as expected, the hadronic size decreases with the momentum transfer $Q^2$, and that the size of the hadron increases with its twist $\tau$.

\begin{figure}
 \begin{center}
\includegraphics[height= 10cm,width=15cm]{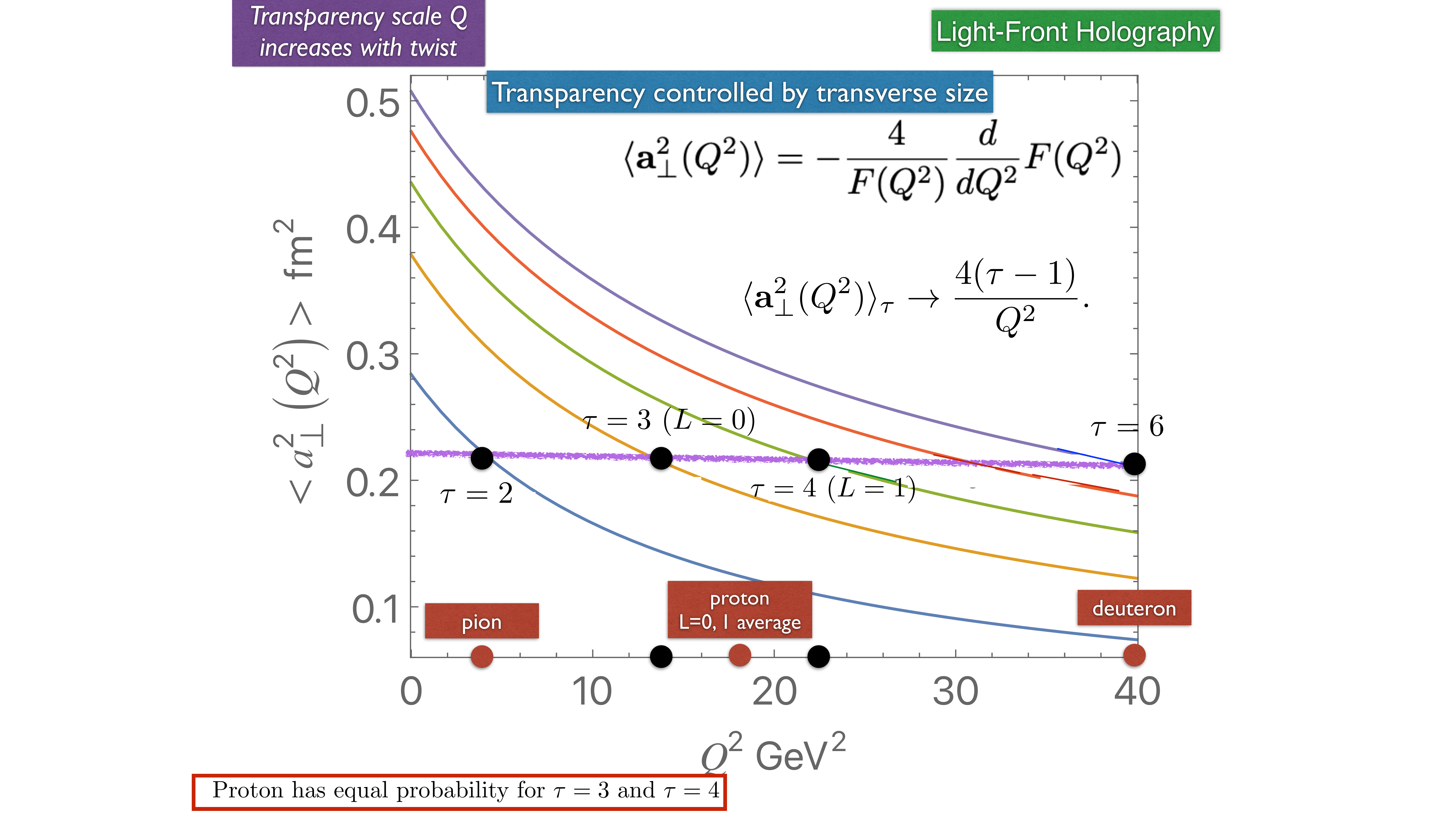}
\end{center}
\caption{Predictions from LF holography for the effective transverse size of hadrons}
\label{Bledslides10.pdf}
\end{figure}

A key prediction is that the size of $a_\perp$ is smaller for mesons $(\tau =2$  than for baryons with $\tau=3,4$, corresponding to the quark-diquark Fock states with $ L=0 $ and $ L=1 $ respectively.
In fact the proton is predicted to have ``two-stage" color transparency $Q^2>14~GeV^2$ for the $ |[ud] u>$  twist-3 Fock state with orbital angular momentum $ L=0$ 
 and $Q^2 > 16~GeV^2$ for the later onset of CT for its $L=1$ twist-4 component.  
 In fact, LF holography predicts equal  quark probability for the $L=0$  and $ L=1$ Fock states.
 
 \begin{figure}
 \begin{center}
\includegraphics[height= 10cm,width=15cm]{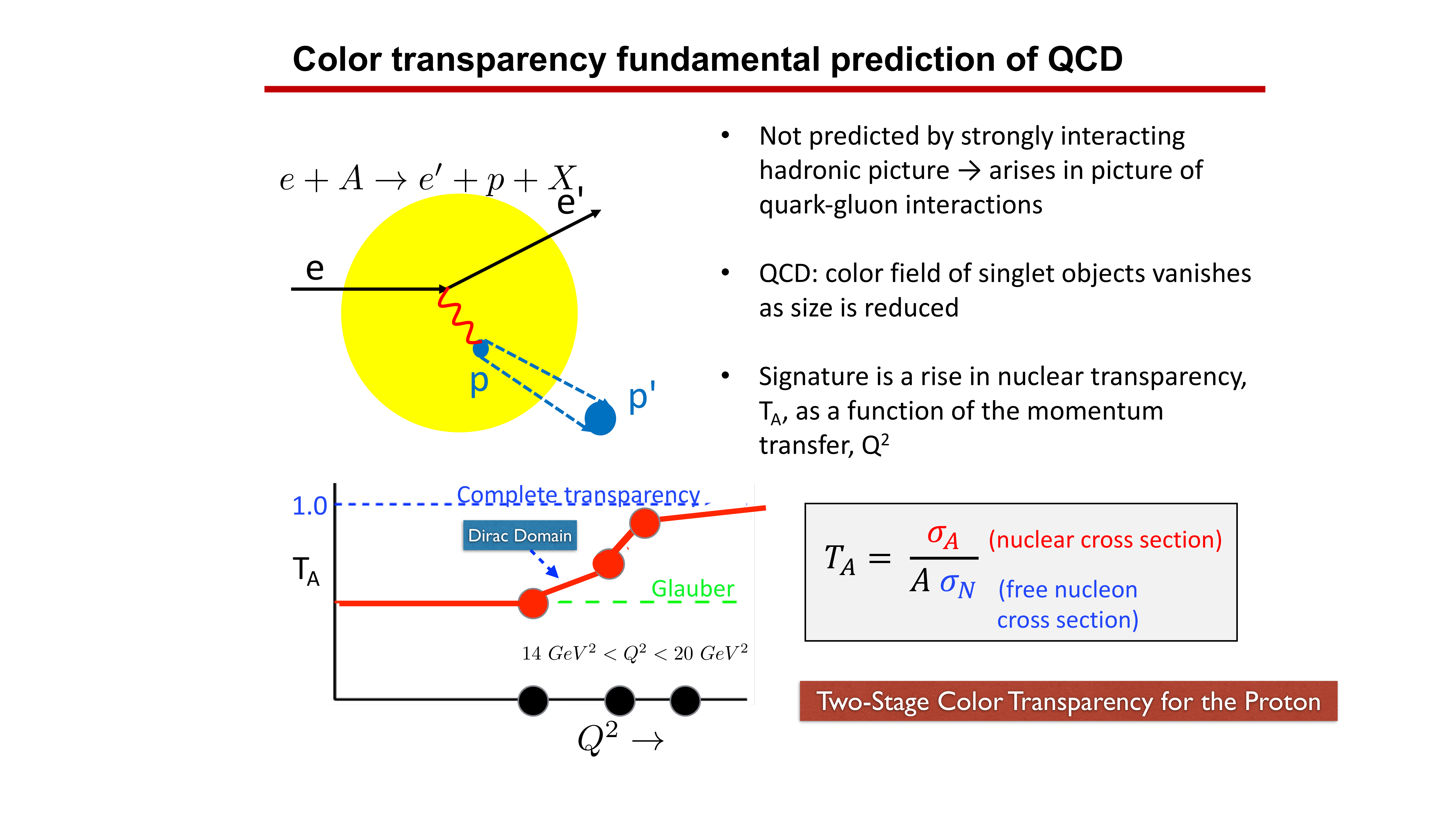}
\end{center}
\caption{Two-stage color transparency and transmission probability of the proton in a nuclear medium from LF Holography }
\label{Bledslides9.pdf}
\end{figure} 
 
 Color transparency is thus predicted to occur at a significantly  higher $Q^2$ for baryons  $(Q^2 > 14~GeV^2)$ than for mesons $(Q^2 > 4~GeV^2)$.
 This is consistent with a recent test of color transparency at JLab which has confirmed color transparency for the the $\pi$ and $\rho$ ~\cite{HallC:2020ijh}; however, the measurements
 in this experiment are limited to values below the range of $Q^2$ where proton color transparency is predicted to occur.

Remarkably, color transparency for the production of an intact deuteron nucleus in $e A \to d + X_{(A-2)}$  quasi-exclusive reactions should be observed at $Q^2 > 50~GeV^2$. This can be tested in $e d \to e d $ elastic scattering in a nuclear background.

It has been speculated~\cite{Caplow-Munro:2021xwi} that the ``Feynman mechanism", where the behavior of the struck quark at $x \sim 1$ in the proton LFWF plays a key role for hard exclusive hadronic processes
does not predict color transparency.  However, LF wavefunctions are functions of the invariant mass 
$\sum_i {\vec k^2_{\perp i }+ m^2_i \over x_i}$
so that their behavior at large $k_\perp$ and large $x$ are correlated.  Thus color transparency occurs for scattering amplitudes involving both the large transverse momentum and large $x$ domains. The three-dimensional LF spatial symmetry of LFWFs also leads to the 
exclusive-inclusive connection, relating the counting rules for the behavior of form factors at large $Q^2$ and structure functions at $x_{bj} \to 1$.

\section{Is the Momentum Sum Rule Valid for Nuclear Structure Functions? }

Sum rules for DIS  processes are analyzed using the operator product expansion of the forward virtual Compton amplitude, assuming it depends in the limit $Q^2 \to \infty$ on matrix elements of local operators such as the energy-momentum tensor.  The moments of the structure function and other distributions can then be evaluated as overlaps of the target hadron's light-front wavefunction,  as in the Drell-Yan-West formulae for hadronic form factors~\cite{Brodsky:1980zm,Liuti:2013cna,Mondal:2015uha,Lorce:2011dv}.
The real phase of the resulting DIS amplitude and its OPE matrix elements reflects the real phase of the stable target hadron's wavefunction.
The ``handbag" approximation to deeply virtual Compton scattering also defines the ``static"  contribution~\cite{Brodsky:2008xe,Brodsky:2009dv} to the measured parton distribution functions (PDF), transverse momentum distributions, etc.  The resulting momentum, spin and other sum rules reflect the properties of the hadron's light-front wavefunction.
However, final-state interactions which occur {\it after}  the lepton scatters on the quark, can give non-trivial contributions to deep inelastic scattering processes at leading twist and thus survive at high $Q^2$ and high $W^2 = (q+p)^2.$   For example, the pseudo-$T$-odd Sivers effect~\cite{Brodsky:2002cx} is directly sensitive to the rescattering of the struck quark. 
Similarly, diffractive deep inelastic scattering involves the exchange of a gluon after the quark has been struck by the lepton~\cite{Brodsky:2002ue}.  In each case the corresponding DVCS amplitude is not given by the handbag diagram since interactions between the two currents are essential.
These ``lensing" corrections survive when both $W^2$ and $Q^2$ are large since the vector gluon couplings grow with energy.  Part of the phase can be associated with a Wilson line as an augmented LFWF~\cite{Brodsky:2010vs} which do not affect the moments.  

The cross section for deep inelastic lepton-proton scattering (DIS) $\ell p \to \ell' p' X)$
includes a diffractive deep inelastic (DDIS) contribution 
in which the proton remains intact with a large longitudinal momentum fraction $x_F>0.9$
greater than 0.9 and small transverse momentum. The DDIS events, which can be identified with Pomeron exchange in the 
$t$-channel, account for approximately 10\%
of all of the DIS events. 
Diffractive DIS contributes at leading-twist (Bjorken scaling) and is the essential component of the two-step amplitude which causes shadowing and antishadowing of the nuclear PDF ~\cite{Brodsky:2021jmj}.  It is important to analyze whether the momentum and other sum rules derived from the OPE expansion in terms of local operators remain valid when these dynamical rescattering corrections to the nuclear PDF are included.   The OPE is derived assuming that the LF time separation between the virtual photons in the forward virtual Compton amplitude 
$\gamma^* A \to \gamma^* A$  scales as $1/Q^2$.
However, the propagation  of the vector system $V$ produced by the diffractive DIS interaction on the front face and its inelastic interaction with the nucleons in the nuclear interior $V + N_b \to X$ are characterized by a longer LF time  which scales as $ {1/W^2}$.  Thus the leading-twist multi-nucleon processes that produce shadowing and antishadowing in a nucleus are evidently not present in the $Q^2 \to \infty$ OPE analysis.

Thus, when one measures DIS, one automatically includes the leading-twist Bjorken-scaling DDIS events as a contribution to the DIS cross section, whether or not the final-state proton is explicitly detected.  In such events, the missing momentum fraction 
in the DDIS events could be misidentified with the light-front momentum fraction carried by sea quarks or gluons in the protons' Fock structure. The underlying QCD Pomeron-exchange amplitude which produces the DDIS events thus does not obey the operator product expansion nor satisfy momentum sum rules -- the quark and gluon distributions measured in DIS experiments will be misidentified, unless the measurements explicitly exclude the DDIS events~\cite{Brodsky:2019jla,Brodsky:2021bkt}

The Glauber propagation  of the vector system $V$ produced by the diffractive DIS interaction on the nuclear front face and its subsequent  inelastic interaction with the nucleons in the nuclear interior $V + N_b \to X$ occurs after the lepton interacts with the struck quark.  
Because of the rescattering dynamics, the DDIS amplitude acquires a complex phase from Pomeron and Regge exchange;  thus final-state  rescattering corrections lead to  nontrivial ``dynamical" contributions to the measured PDFs; i.e., they involve the physics aspects of the scattering process itself~\cite{Brodsky:2013oya}.  The $ I = 1$ Reggeon contribution to diffractive DIS on the front-face nucleon leads to flavor-dependent antishadowing~\cite{Brodsky:1989qz,Brodsky:2004qa}.  This could explain why the NuTeV charged current measurement $\mu A \to \nu X$ scattering does not appear to show antishadowing
 in contrast to deep inelastic electron nucleus scattering~\cite{Schienbein:2007fs}.
Again, the corresponding DVCS amplitude is not given by the handbag diagram since interactions between the two currents are essential to explain the physical phenomena.

It should be emphasized  that shadowing in deep inelastic lepton scattering on a nucleus  involves  nucleons at or near the front surface; i.e, the nucleons facing the incoming lepton beam. This  geometrical orientation is not built into the frame-independent nuclear LFWFs used to evaluate the matrix elements of local currents.  Thus the dynamical phenomena of leading-twist shadowing and antishadowing appear to invalidate the sum rules for nuclear PDFs.  The same complications occur in the leading-twist analysis of deeply virtual Compton scattering $\gamma^* A \to \gamma^* A$ on a nuclear target.

\section*{Acknowledgements}

Presented at the Low-$x$ Workshop, Elba Island, Italy, September 27--October 1, 2021.  I thank Christophe Royon for inviting me to make a contribution to this meeting.
I am also grateful to my collaborators on our recent work, especially Guy de T\'eramond, Alexandre Deur, Guenter Dosch, Sabbir Sufian, Simonetta Liuti, Ivan Schmidt, and Valery Lyubovitskij.  This work is supported by the Department of Energy, Contract DE--AC02--76SF00515. SLAC-PUB-17651

\nocite{*}
\bibliographystyle{auto_generated}
\bibliography{Brodsky_proceedings_elba2021/Brodsky_proceedings_elba2021/Brodsky_proceedings_elba2021/sjb_Elba_contribution}


%% file: low_x_2021/GKKrintiras.tex
\vspace*{1.2cm}

\thispagestyle{empty}
\begin{center}
{\LARGE \bf Collectivity in heavy ions at CMS}

\par\vspace*{7mm}\par

{

\bigskip

\large \bf \href{https://gkrintir.web.cern.ch}{Georgios Konstantinos Krintiras} \\ (on behalf of the CMS Collaboration)}

\bigskip

{\large \bf  E-Mail: gkrintir@cern.ch}

\bigskip

{The University of Kansas}

\bigskip

{\it Presented at the Low-$x$ Workshop, Elba Island, Italy, September 27--October 1 2021}

\vspace*{15mm}

\end{center}
\vspace*{1mm}

\begin{abstract}

Investigations of anisotropic flow can provide us with necessary information about the initial state and the evolution of the deconfined strongly interacting matter called quark-gluon plasma (QGP). CMS measurements of flow harmonics and their fluctuations and (de)correlations cover a wide range of transverse momenta, pseudorapidity and centrality or event activity in different collision systems. The flow harmonics are measured using two- and multiparticle correlations. Although most measurements focus on the flow in the transverse plane at $y = 0$, studying event-by-event fluctuations along the longitudinal direction is interesting as well since they put constraints on the complete three-dimensional modeling of the
initial state used in 3+1D models of the QGP evolution. Event-by-event correlations between harmonics of different orders, studied, \eg, with mixed-harmonic and symmetric cumulants, measure the nonlinear hydrodynamic response and probe the onset of long-range collective multiparticle correlations, respectively. Statistical properties, such as the non-Gaussianity of the flow fluctuations, can be extracted by careful inspection of the data showing that there are small differences
in the magnitudes of multiparticle
cumulants. Identified flow measurements (\eg, that of light and heavy flavor quarks, or \PZ bosons, etc) provide an important insight into the nature of collective phenomena in different collision systems.
\end{abstract}
 \part[Collectivity in heavy ions at CMS\\ \phantom{x}\hspace{4ex}\it{Georgios Konstantinos Krintiras on behalf of the CMS Collaboration}]{}
\section{Introduction}
\label{sec:intro_GKK}

One of the main purposes of the Large Hadron Collider (LHC) is to create a hot and dense, strongly interacting QCD medium, referred to as the quark-gluon plasma (QGP).
The study of ultrarelativistic heavy ion collisions~\cite{Krintiras:2020imy}, including the decomposition of azimuthal particle distributions into Fourier series, revealed QGP properties consistent with a collectively expanding (``flowing'') medium. The associated flow vectors $V_n \equiv{}$ \vn $e^{in\Psi_{n}}$, where \vn is the magnitude of the $n^{\text{th}}$-order Fourier harmonic and $\Psi_n$ its phase (also known as the $n^{\text{th}}$-or order ``symmetry plane angle''), reflect the hydrodynamic response of the medium to the transverse overlap region and its subnucleon fluctuations. Measurements of flow vectors, their event-by-event fluctuations and correlations between different orders of harmonics or symmetry planes provide input to QGP modeling, in particular, details about initial-state conditions and the dynamics of the subsequent deconfined phase (see Section~\ref{sec:AA}). 

Extensive measurements of light hadron azimuthal anisotropies are complemented by studies of heavy flavor (charm and bottom) quarks. In principle, their masses are much larger than the typical range of temperatures in the QGP, meaning thermal production of heavy quarks during the QGP phase is suppressed relative to that of light quarks. However, it is still postulated that charm quarks interact strongly enough to flow with the QGP. Experimental data at LHC (see Section~\ref{sec:AA}) so far reveal that charm quark hadrons have significant azimuthal anisotropies, suggesting that they participate in the overall collective flow of the medium. A nonthermalized probe is required to assess the interaction with the medium more thoroughly, with the heavier bottom quark being a natural candidate. Although a series of theoretical predictions for the azimuthal anisotropies of bottom quarks exist, only limited experimental data are currently available at LHC (see Section~\ref{sec:AA}). 

The observation of QGP-like phenomena in the small systems produced in proton-proton (\pp) and proton-lead (\pPb)
collisions indicates the possibility of final-state effects in such systems. In that sense, precision measurements of \PZ bosons in peripheral nucleus-nucleus collisions (see Section~\ref{sec:AA}) can provide an experimental reference for the expected yields of hard probes in the absence of final-state effects, which may lead to an improved understanding of the onset of collectivity in those ``small systems''. Given features similar to those observed in heavy ion collisions are revealed when the same observables are used in conjunction with high particle multiplicities in \pp and \pPb collisions (see Section~\ref{sec:SmallSystems}), it thus remains imperative that LHC experiments, like  CMS~\cite{Chatrchyan:2008aa}, pursue their quest for a medium of a similar origin as in measurements involving heavy ion collisions. It is not yet clear, however, to what extent the $V_n$ is driven by the initial spatial anisotropy and/or whether competing mechanisms, \eg, gluon field momentum correlations at the initial state, contribute to the seen final-state anisotropy. In all cases, utmost caution must be exercised when interpreting flow-related signatures, especially in small systems, given the potential contamination from nonflow effects. Similar to the heavy ion case, it is also of interest to measure heavy flavor anisotropies in such collision systems so that we obtain information about the interaction of heavy quarks with the medium in the smallest hadronic collision systems at LHC (see Section~\ref{sec:SmallSystems}).

\section{Flow related measurements in heavy ion collisions}
\label{sec:AA}

\begin{figure}[!htb]
\centering
\begin{minipage}{0.65\textwidth}
\centering
\includegraphics[width=0.99\textwidth]{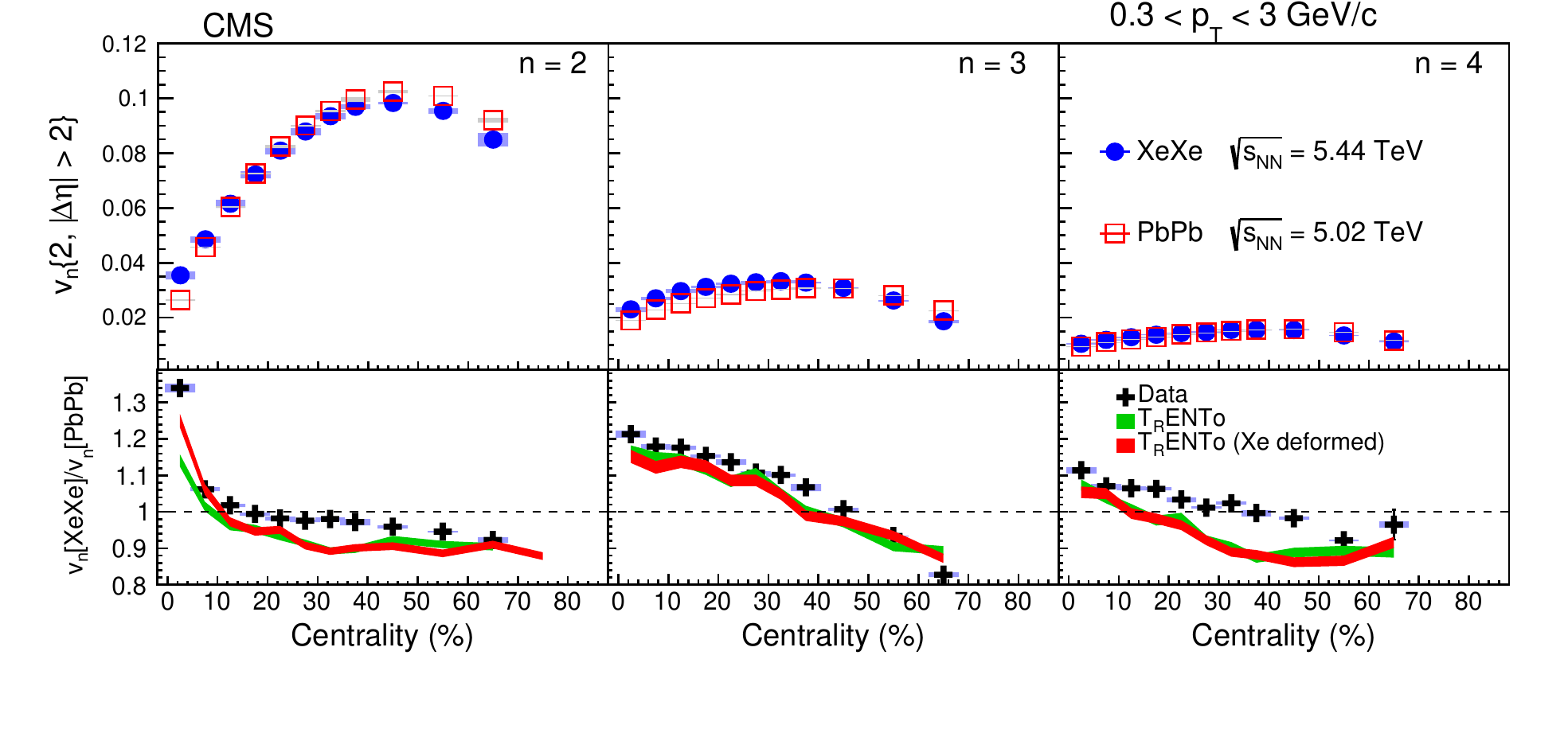}
\end{minipage}\hfill
\begin{minipage}{0.35\textwidth}
\centering
\includegraphics[width=0.99\textwidth]{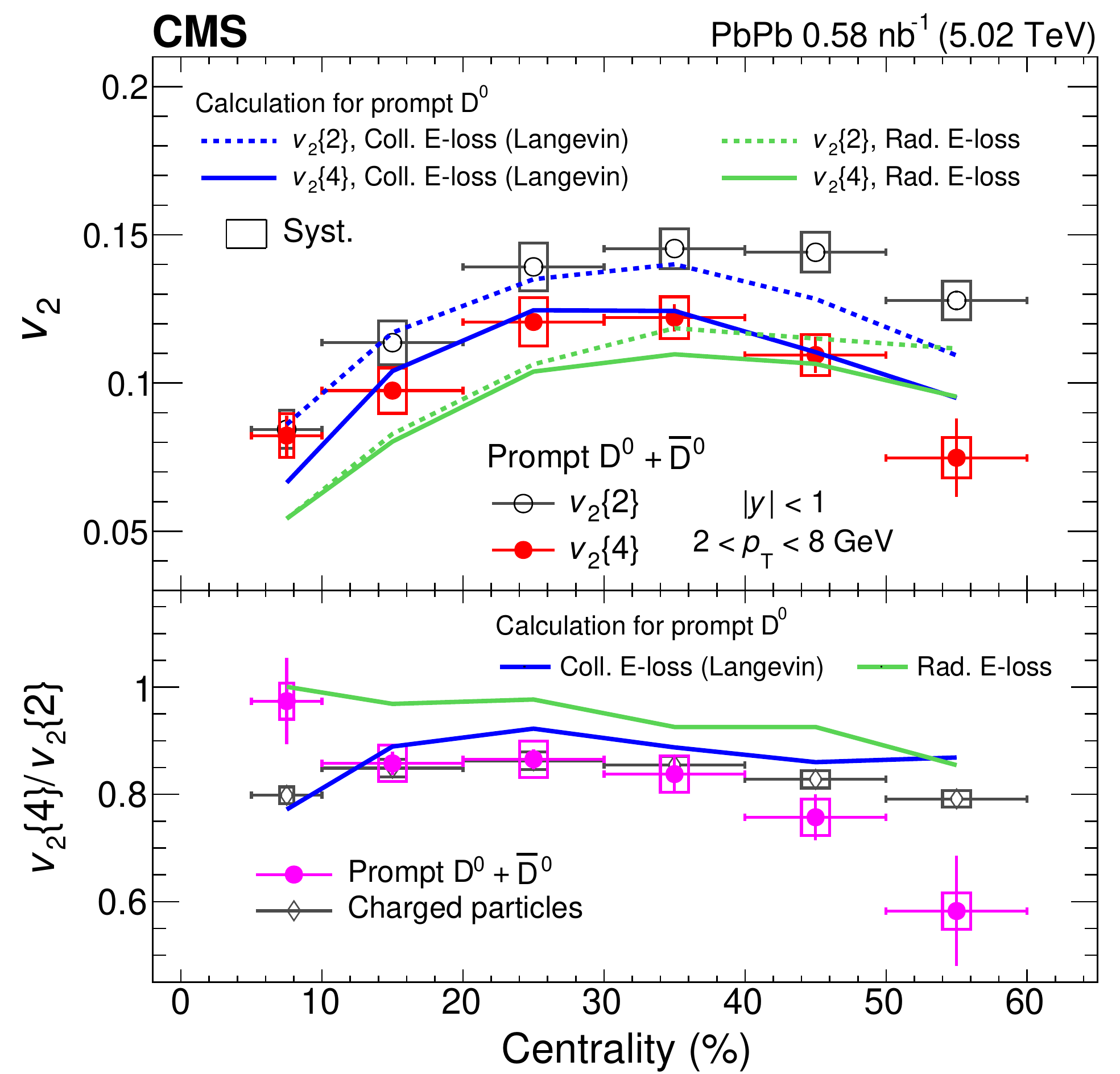}
\end{minipage}
\caption{Left: Centrality dependence of $\vtwo$, $\vthree$, and  $\vfour$  harmonic coefficients from two-particle correlations method for $0.3 < \pt < 3.0\,\GeVns$ for  \XeXe collisions at $5.44\,\TeVns$ and \PbPb collisions at $5.02\,\TeVns$~\cite{CMS:2019cyz}. The lower panels show the ratio of the results for the two systems. The bars and the shaded boxes represent statistical and systematic uncertainties, respectively. Theoretical predictions are compared to the data (shaded bands). Right: Prompt \PDz meson \vtwotwo~\cite{CMS:2020bnz} and \vtwofour~\cite{CMS:2021qqk}, and $\vtwofour/\vtwotwo$ compared to the same ratio for charged particles in the range $\abs{y}<1$ as a function of centrality for $2<\pt<8\,\GeVns$ in \PbPb collisions at $5.02\,\TeVns$. The vertical bars represent statistical uncertainties and open boxes denote the systematic uncertainties. The lines represent model calculations.}
\label{fig:fig1}
\end{figure}

The \vn measurements had led to ``the  discovery  of the  perfect  liquid''~\cite{Rafelski:2019twp} and contributed significantly to the understanding of initial-state effects and final-state evolution mechanisms of the QGP in various collisions systems~\cite{CMS:2019cyz}, as shown in Fig.~\ref{fig:fig1} (left), and extensive phase space regions in pseudorapidity~\cite{CMS:2017xnj} and transverse momentum~\cite{CMS:2017xgk}. Details of the initial-state conditions and the subsequent dynamics can be further probed by more complex observables: event-by-event flow fluctuations, decorrelations of flow vectors along the longitudinal direction, higher order $V_n$, in particular their linear and nonlinear components.

By unfolding statistical resolution effects from the event-by-event measured cumulant \vn distributions, higher order moments from the underlying probability distribution functions $p(\vn)$ can be determined. For example, non-Gaussian fluctuations in the initial-state energy density lead to differences in the higher order cumulant $\vtwo\{k\}$, and asymmetry about the mean of the $p(\vtwo)$. Parameters like the mean eccentricity in the reaction plane can be further extracted from elliptic power function fits to $p(\vtwo)$: the extracted eccentricities~\cite{CMS:2017glf} are actually found smaller than predictions based on models of initial-state conditions.

Flow vector decorrelations can be studied by forming a factorization ratio $r_n (\pt, \eta)$ such that when $\vn$ and/or $\Psi_n$ decorrelate the $r_n$ deviates from unity. These measurements can provide important constraints on the longitudinal structure, currently a challenge for three-dimensional hydrodynamic models. More specifically, CMS studies of $\eta$-dependent factorization breakdown~\cite{CMS:2015xmx} gave an indication of initial-state
fluctuations along the longitudinal direction, and subsequent studies at LHC comparing $r_n$ in \PbPb and \XeXe collisions~\cite{Aad:2020gfz} revealed that models tuned to describe the $\vn$ in both systems failed to reproduce the $r_n$. Higher order $V_n$ can be expressed in terms of linear and nonlinear modes, each being proportional to the same or lower order eccentricities and/or a combination of them. Constraints to hydrodynamic models can be imposed by measurements of the corresponding nonlinear response coefficients $\chi_{n(pk)}$, where $n$ represents the order of $V_n$ and $p$, $k$ with respect to lower-order symmetry plane angle or angles. Model calculations failed to describe the measured $\chi$~\cite{Sirunyan:2019izh}, in particular, $\chi_{7223}$.

At LHC, significant $\vtwo$ flow signal is observed for mesons containing a charm quark, \eg, prompt $\PJGy$~\cite{CMS:2016mah} and prompt $\PDz$~\cite{CMS:2017vhp}, while the first measurements with bottom quarks, \eg, nonprompt $\PJGy$~\cite{CMS:2016mah}, and $\PGU$(1S) and $\PGU$(2S)~\cite{CMS:2020efs}, are compatible with zero in the kinematic region studied so far. For the charm quark case, the prompt $\PDz$ meson $\vtwo$ has been so far measured
using two-particle correlation methods, while the first $\vtwofour$ measurement, \ie, using
multiparticle correlations, is presented most recently~\cite{CMS:2021qqk} (Fig.~\ref{fig:fig1}, right). The results can also discriminate between models of heavy quark energy loss and constrain heavy quark transport coefficients in the QGP, complementing measurements of the nuclear modification factor, \eg, Ref.~\cite{CMS:2018bwt}.

\begin{figure}[!htb]
\centering
\begin{minipage}{0.5\textwidth}
\centering
\includegraphics[width=0.85\textwidth]{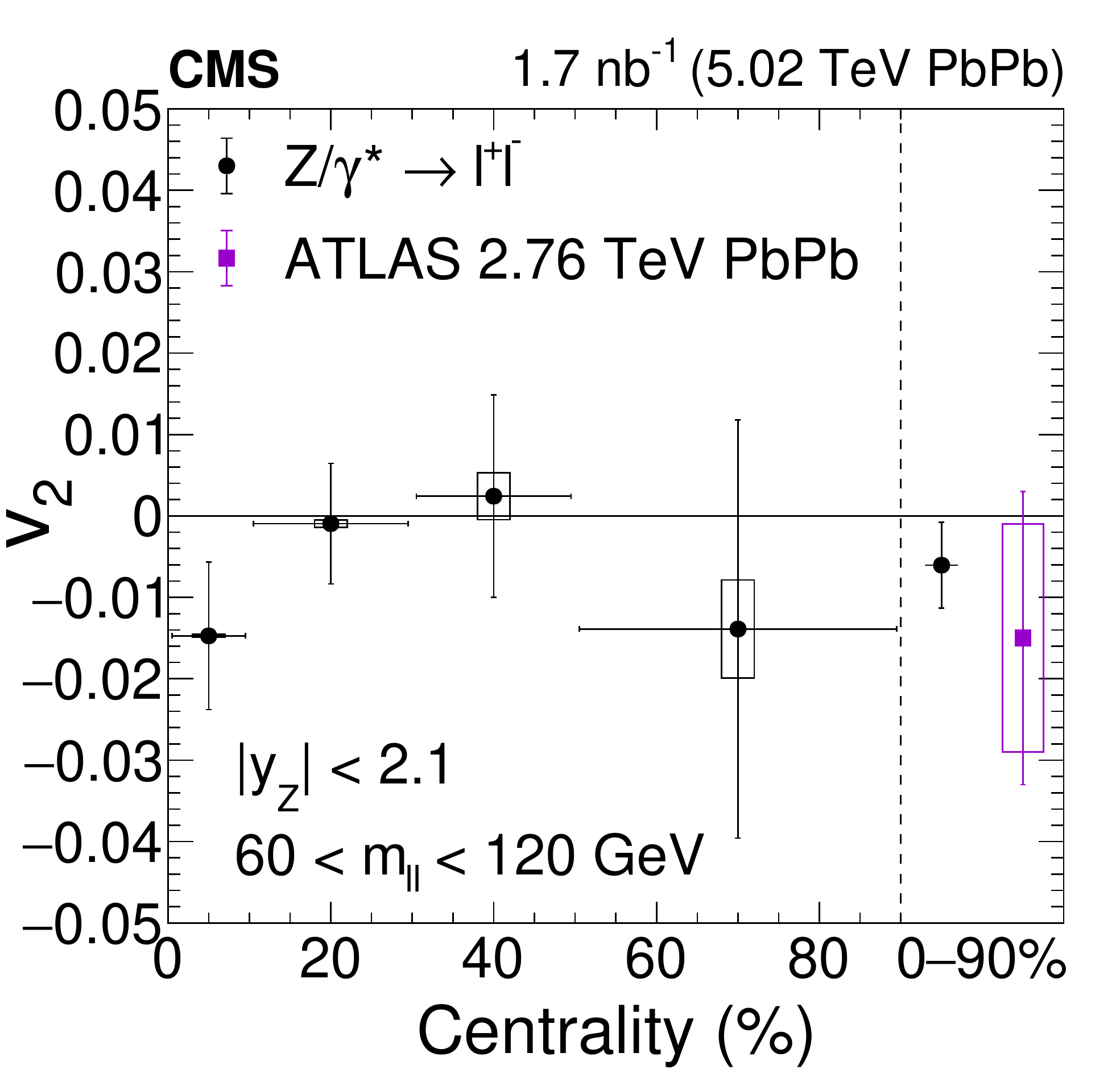}
\end{minipage}\hfill
\begin{minipage}{0.5\textwidth}
\centering
\includegraphics[width=0.85\textwidth]{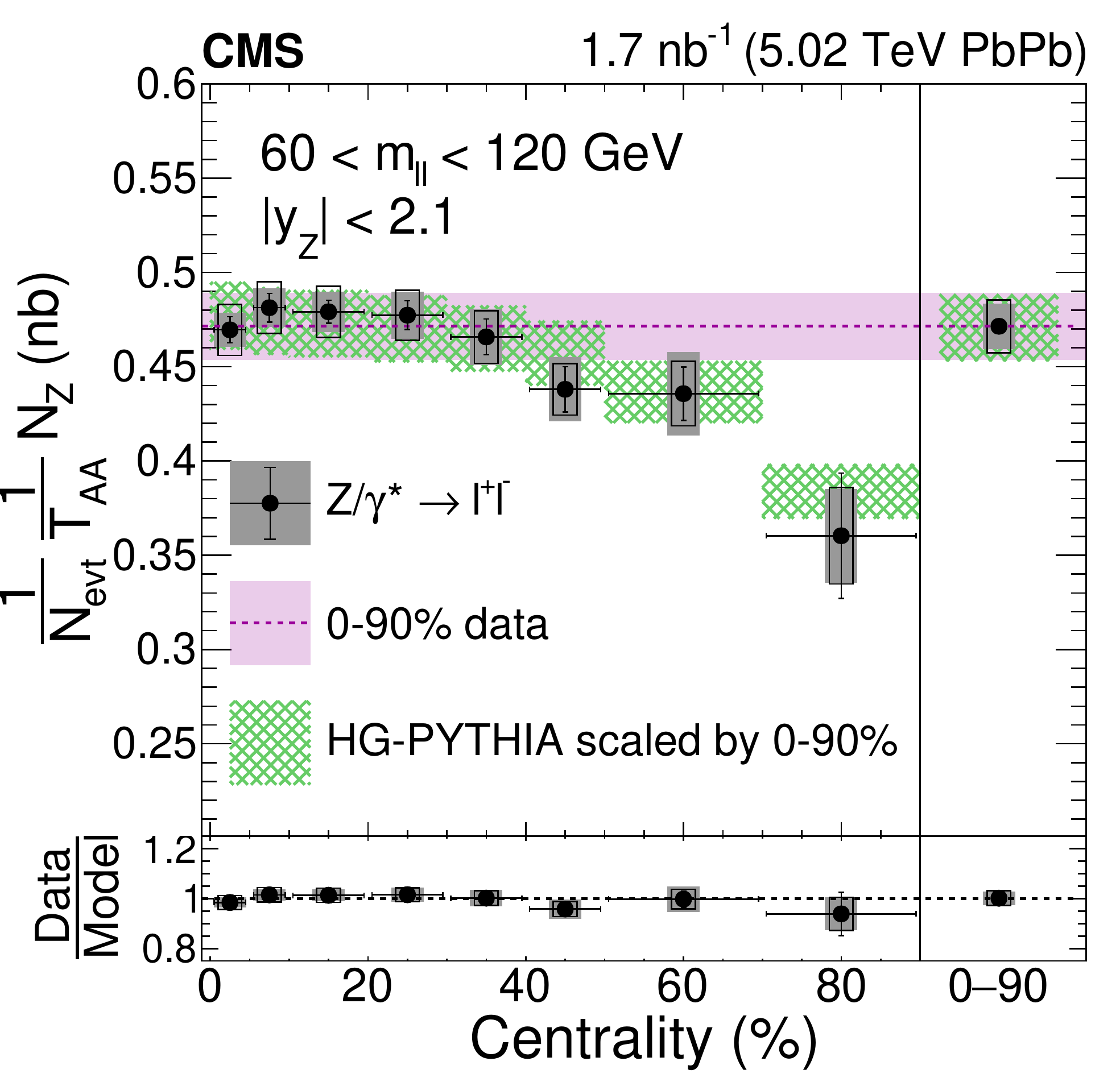}
\end{minipage}
\caption{Left: The $\vtwo$ of \PZ bosons in \PbPb collisions for various centrality bins.  The error bars represent statistical uncertainties. The boxes represent systematic uncertainties and may be smaller than the markers~\protect\cite{CMS:2021kvd}. A measurement from the ATLAS Collaboration at $2.76\,\TeVns$ is also shown~\protect\cite{ATLAS:2012qdj}. Right: Normalized yields of \PZ bosons as a function of centrality.  The error bars, hollow boxes, and solid gray boxes represent the statistical, systematic, and model uncertainties, respectively. The value of the 0--90\% data point, and the scaled  model prediction are shown for comparison, with the width of the bands representing the contribution from the total 0--90\% data point uncertainty.}
\label{fig:fig2}
\end{figure}

\PZ boson yields and the elliptic flow coefficient have been measured with high
precision as a function of centrality in lead-lead collisions~\cite{CMS:2021kvd}. The \PZ boson $\vtwo$ is compatible with zero (Fig.~\ref{fig:fig2}, left), consistent with the expectation of no significant final-state interactions in the QGP. Appropriately scaled \PZ boson yields ($R_{\mathrm{AA}}$) are found constant versus centrality, but a decreasing trend is seen for the first time for more peripheral events (Fig.~\ref{fig:fig2}, right). This is compatible with the \textsc{hg-pythia} model prediction that accounts for initial collision geometry and centrality selection effects~\cite{Loizides:2017sqq}, and in contrast to the findings in Ref.~\cite{ATLAS:2019maq} by $\approx{5}$\%. A slightly increasing $R_{\mathrm{AA}}$ with decreasing centrality was previously explained by arguing that the nucleon–nucleon
cross section may be shadowed in nucleus–nucleus collisions~\cite{Eskola:2020lee}, an interpretation with quite significant consequences for the understanding of heavy ion data, in particular in the context of the Glauber model. Instead, an alternative explanation of the data was recently provided~\cite{Jonas:2021xju} by assuming that there is a mild bias present in the centrality determination of
the measurement in Ref.~\cite{ATLAS:2019maq} about the size of the related systematic uncertainty. Overall, these results provide a new experimental proxy for estimating the average nucleon-nucleon integrated luminosity as a function of centrality in heavy ion collisions. The ratio of \PZ boson yields in \PbPb over \pp collisions can be used as an alternative to Glauber-model-based scaling for hard scattering processes, which also automatically accounts for potential effects related to event selection and centrality calibration. A centrality-determination independent measurement of the \PZ boson cross section is also possible in zero-bias \PbPb collisions but requires the larger data samples of Runs 3 and 4.

\section{Flow related measurements in proton-nucleus and proton-proton collisions}
\label{sec:SmallSystems}

A series of features in small systems are indicative of a collective behavior driven by the initial-state geometry and final-state effects: the near-side ridge in two-particle correlations, the \pt and event activity dependence of $\vn$, the $\vn$ dependence upon the hadron species and scaling with the number of valence quarks in the hadron, multiparticle cumulants and their ratios. In proton-proton collisions, it is even less clear what mechanism gives rise to the observed finite azimuthal anisotropy. Until now, hints towards a final-state description were given, \eg, the indication of a mass ordering with multiparticle angular correlations~\cite{Khachatryan:2016txc}.

\begin{figure}[!htb]
\centering
\begin{minipage}{0.3\textwidth}
\centering
\includegraphics[width=0.99\textwidth]{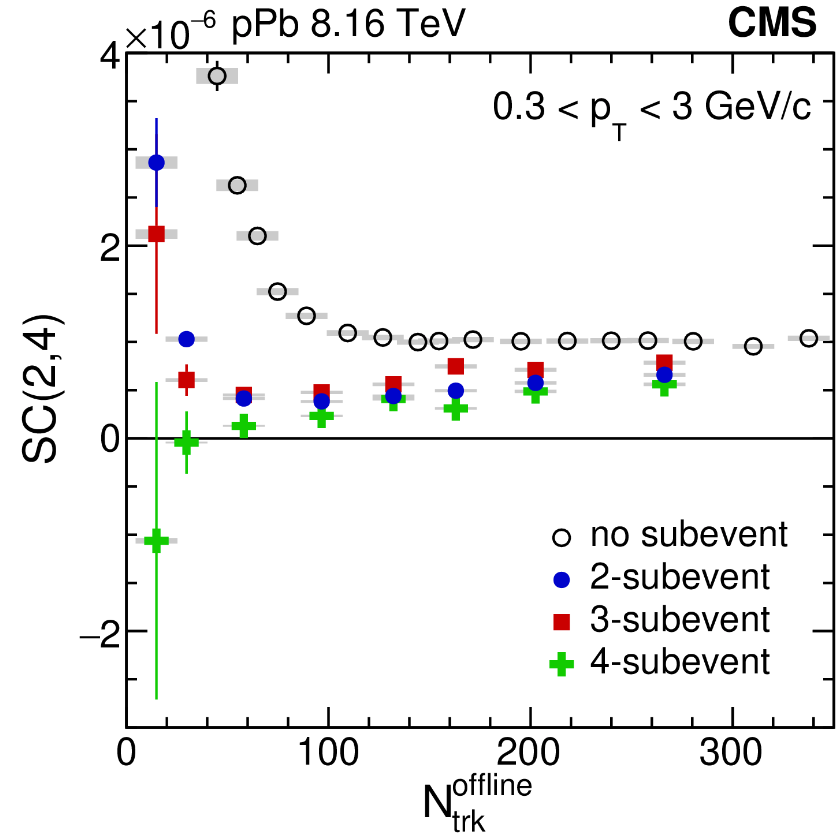}
\end{minipage}
\begin{minipage}{0.65\textwidth}
\centering
\includegraphics[width=0.99\textwidth]{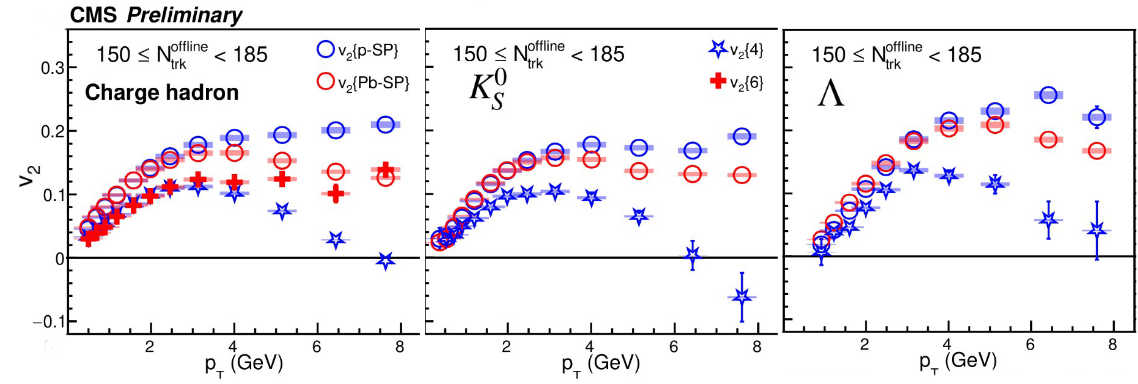}
\end{minipage}\hfill
\caption{Left: The $\text{SC}(2,4)$ distributions as a function of \noff from 2 subevents (full blue circles), 3 subevents (red squares), and 4 subevents (green crosses)~\cite{CMS:2019lin}.
For comparison, results from Ref.~\cite{CMS:2017kcs} with no subevents (open black circles), are also shown.
Bars represent statistical uncertainties while grey areas represent the systematic uncertainties. Right: Two- and multiparticle cumulant $\vtwo$ results of \pPb collisions at 8.16\,\TeVns for charged hadrons, \PKzS mesons, and \PgL\ baryons in different \noff ranges~\cite{CMS:2021fhf}. The two-particle results are based on event planes that are determined in either the proton-going (p-SP) or lead-going (Pb-SP) side of the forward hadron calorimeter. The shaded boxes show the total systematic uncertainties.}
\label{fig:fig3}
\end{figure}

Event-by-event correlations among the $\vtwo$, $\vthree$, and $\vfour$ Fourier harmonics have also been measured for small systems using the symmetric cumulant (SC) method~\cite{CMS:2019lin}. The correlation data reveal features similar to those observed in \PbPb collisions, \ie, a negative correlation is found between the $\vtwo$ and $\vthree$ harmonics (SC(2,3)), while the correlation is positive between the $\vtwo$ and $\vfour$ harmonics (SC(2,4)). First measurements of event-by-event SC(2, 3) and SC(2, 4) with two, three, or four subevents~\cite{CMS:2019lin} corroborate the findings using the SC technique without subevents. By significantly suppressing the nonflow
contribution, the four-subevent results for both SC(2, 3) and SC(2, 4) show a monotonically decreasing magnitude toward zero at an offline track multiplicity of $N_{\text{trk}}\approx{20}$, providing evidence for the onset of long-range collective particle correlations from high to low multiplicity events in \pPb collisions (Fig.~\ref{fig:fig3}, left). 

Using the multiparticle cumulant method, $\vtwofour$, $\vtwosix$, and $\vtwoeight$ values have been measured, for the first time, for \PKzS mesons and \PgL baryons in high-multiplicity \pPb collisions~\cite{CMS:2021fhf}. Nonflow effects are studied using jet veto and subevent methods.  A large difference between the $\vtwofour$ and $\vtwosix$ results (not present in \PbPb collisions) can be explained by jet-related correlations,
which are suppressed by rejecting events with at least one jet with transverse momentum $\pt > 20\,\GeVns$.
The subevent cumulant method is also performed to reduce short-range correlation effects, with the difference between the standard and the subevent cumulant methods attributed to the effect of event plane decorrelations. For both the jet suppression and subevent methods, the nonflow contribution is found to be particularly significant for at high track multiplicity.

\begin{figure}[!htb]
\centering
\begin{minipage}{0.33\textwidth}
\centering
\includegraphics[width=0.99\textwidth]{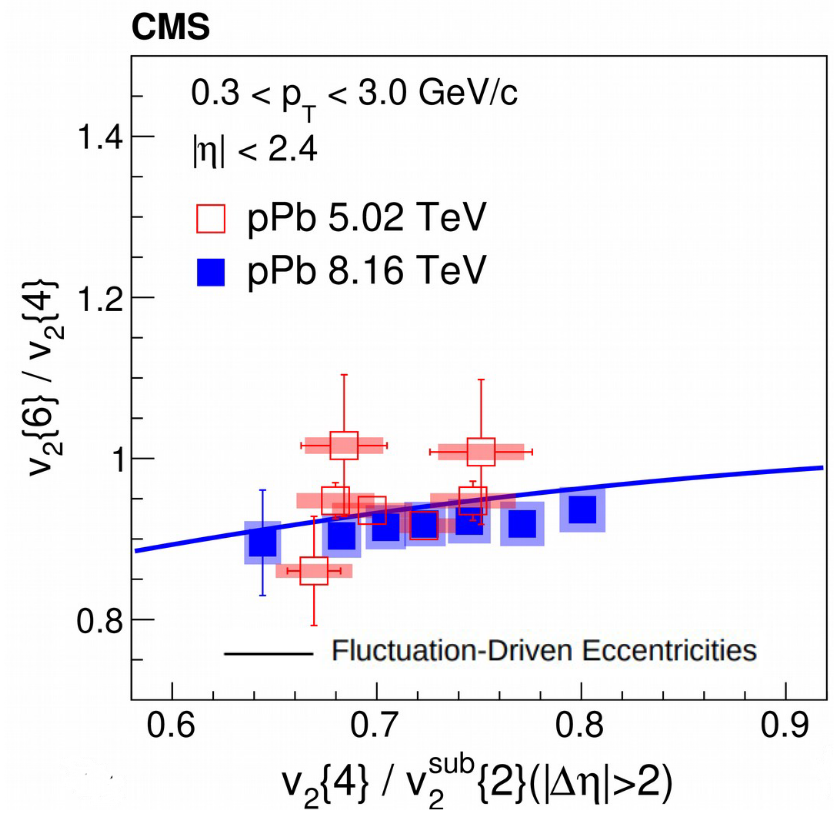}
\end{minipage}
\begin{minipage}{0.65\textwidth}
\centering
\includegraphics[width=0.99\textwidth]{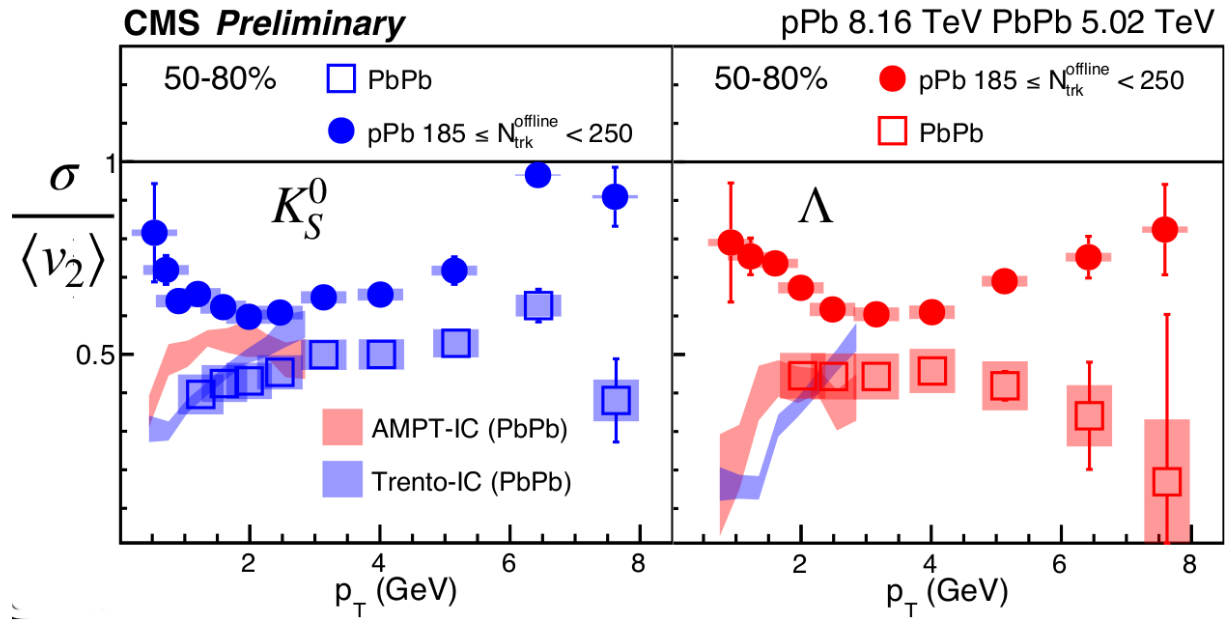}
\end{minipage}\hfill
\caption{Left:   The $\vtwo$ fluctuation results for \PbPb collisions at 5.02\,\TeVns in different centrality intervals and \pPb collisions at 8.16\,\TeVns with $185 \le \noff < 250$ for \PKzS mesons and \PgL baryons~\cite{CMS:2021fhf}. The shaded bands are hydrodynamic calculations of $\vtwo$ fluctuations in \PbPb collisions. The shaded boxes show the systematic uncertainties Right: 
Cumulant ratios $\vtwosix/\vtwofour$ as a function of $\vtwofour/v^{\text{sub}}_2\{2\}$ in \pPb collisions at $5.02$~\cite{CMS:2015yux} and 8.16\,\TeVns~\cite{CMS:2019wiy}.
Error bars and shaded areas denote statistical and systematic uncertainties, respectively. The solid curves show the expected behavior based on a hydrodynamics-motivated study of the role of initial-state fluctuations.}
\label{fig:fig4}
\end{figure}

The agreement of  calculations of purely fluctuation-driven origin with measurements of the ratios $\vtwosix/\vtwofour$ and $\vtwoeight/\vtwosix$~\cite{CMS:2015yux,CMS:2019wiy} shows that the differences found among the multiparticle cumulant results for the $\vtwo$ harmonic can be described by non-Gaussian initial-state fluctuations (Fig.~\ref{fig:fig4}, left). Similarly, the higher-order $\vthree\{4\}$ coefficient is reported for the first time for a small
system~\cite{CMS:2019wiy}. Both the \pPb and \PbPb systems have very similar $\vthree$ coefficients for the cumulant orders studied, indicating a similar, fluctuation-driven initial-state geometry. In addition, no obvious particle species dependence of the fluctuations in the $\vtwo$ values is observed (Fig.~\ref{fig:fig4}, right). for either the \PbPb or \pPb systems~\cite{CMS:2021fhf}. However, the flow fluctuations are observed to be larger in \pPb collisions compared to \PbPb collisions~\cite{CMS:2021fhf}, as expected if the overall collision geometry is a driving force. 

\begin{figure}[!htb]
\centering
\begin{minipage}{0.5\textwidth}
\centering
\includegraphics[width=0.85\textwidth]{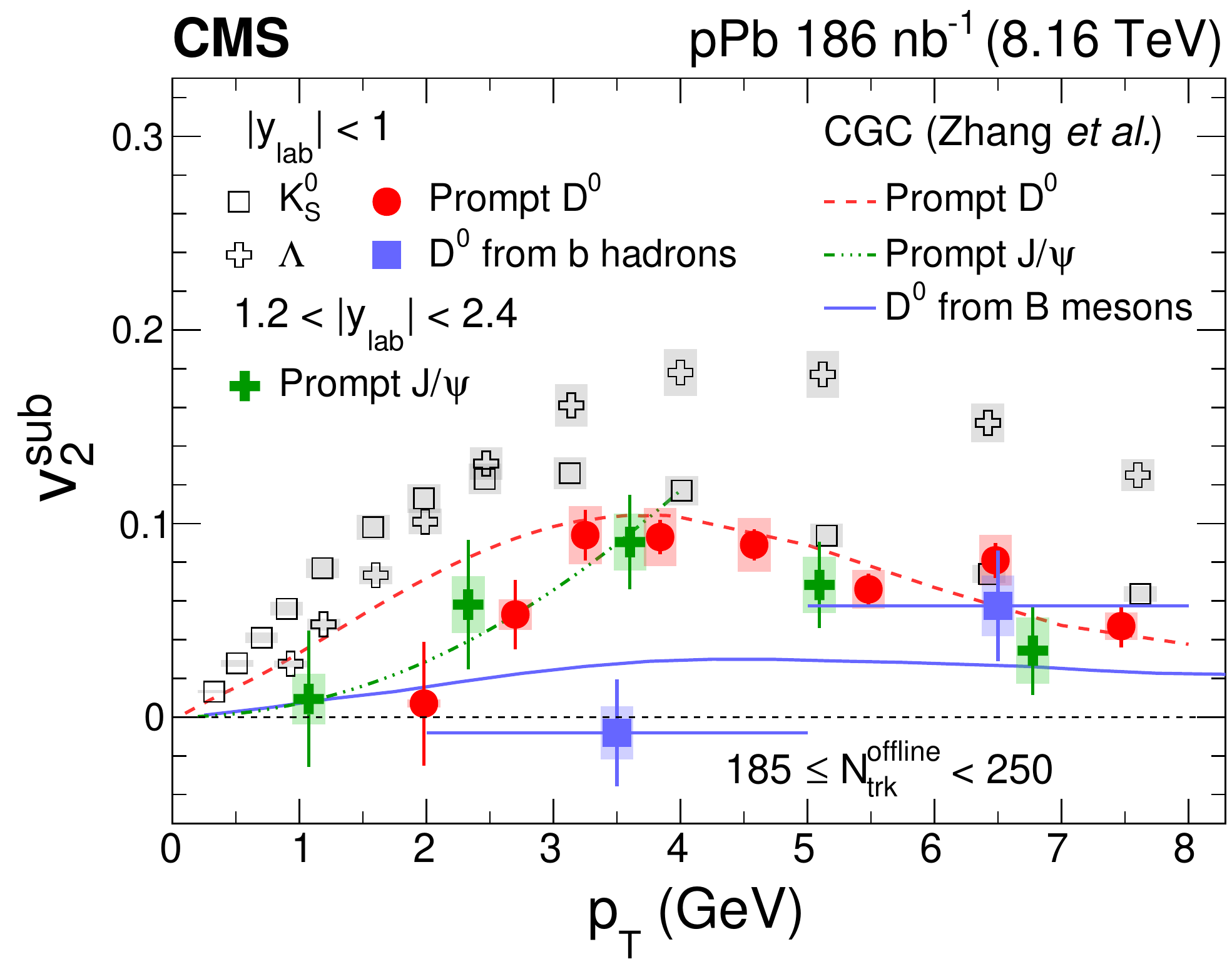}
\end{minipage}\hfill
\begin{minipage}{0.5\textwidth}
\centering
\includegraphics[width=0.99\textwidth]{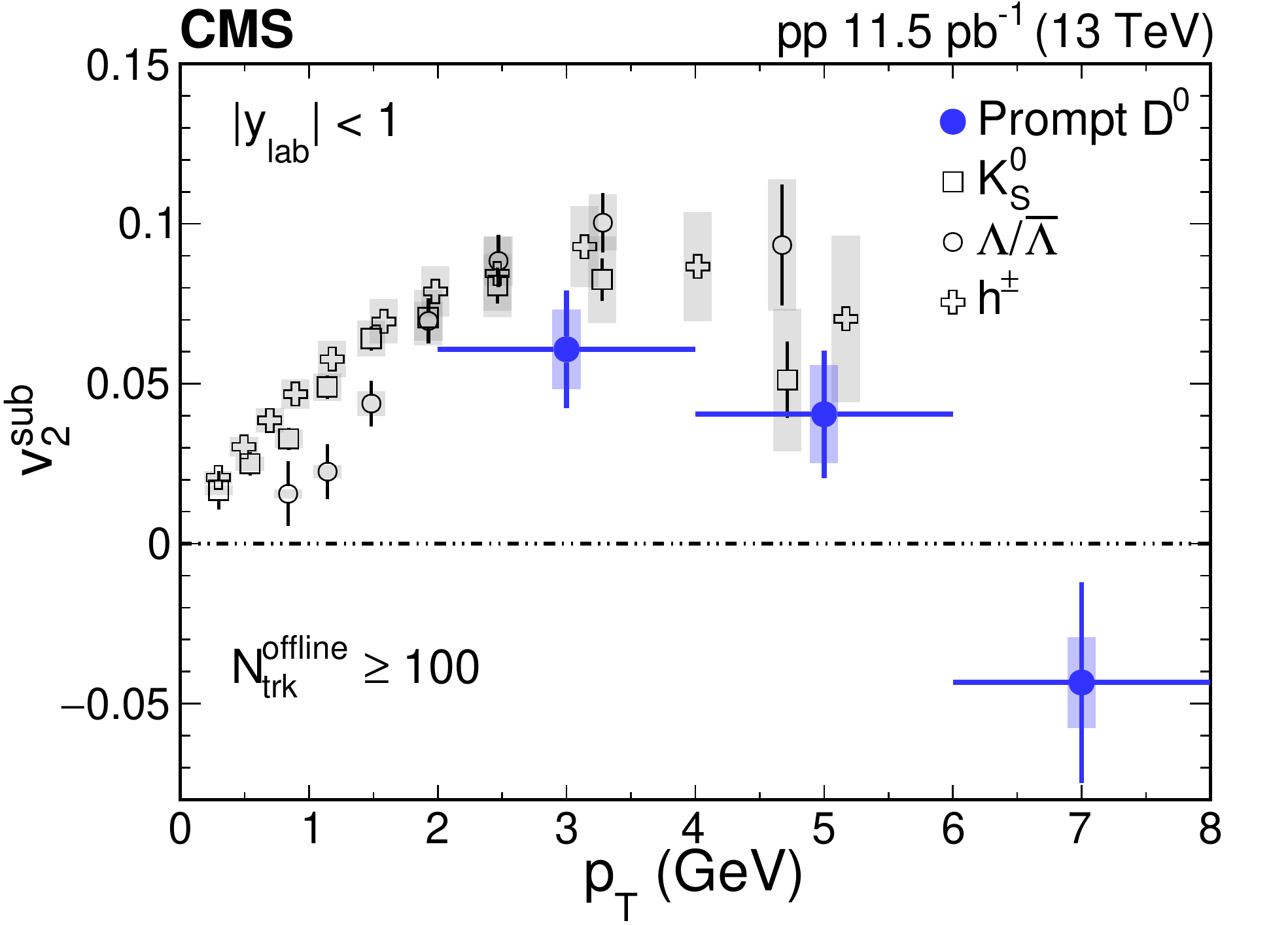}
\end{minipage}
\caption{Left: Results of $\vtwo$ for prompt~\protect\cite{CMS:2018loe} and nonprompt~\protect\cite{CMS:2020qul} \PDz mesons, as well as prompt \PJGy\ mesons~\cite{CMS:2018duw} and light flavor hadrons~\protect\cite{CMS:2018loe}, as a function of \pt in $\Pp{}\mathrm{Pb}$ collisions at $8.16$\,\TeVns. Lines show the theoretical calculations of prompt and nonprompt \PDz, and prompt \PJGy mesons, respectively. Right: Results of $\vtwo^{\text{sub}}$ for charged particles, \PKzS mesons, \PgL baryons, and prompt \PDz mesons as
a function of \pt for $\abs{y_{\text{lab}}}<1$, with $\noff \geq 100$ in  \pp collisions at $13\,\TeVns$~\protect\cite{CMS:2020qul}. The vertical bars correspond to the statistical uncertainties, while the shaded areas denote
the systematic uncertainties.}
\label{fig:fig5}
\end{figure}

In small colliding systems, the study of heavy flavor hadron collectivity has the potential to disentangle possible contributions from both initial- and final-state effects.  Recent observations of significant $\vtwo$ signal for prompt \PDz~\cite{CMS:2018loe}, and prompt \PJGy~\cite{CMS:2018duw} in $\Pp{}\mathrm{Pb}$ collisions (Fig.~\ref{fig:fig5}, left) provided the first evidence for charm quark collectivity in small systems. In spite of the mass differences, the observed $\vtwo$ signal for prompt \PJGy mesons is found to be comparable to that of prompt \PDz mesons and light-flavor hadrons at a given \pt range, possibly implying the existence of initial-state correlation effects. Further detailed investigations start addressing open questions for understanding the origin of heavy flavor quark collectivity in small
systems. These include the \pt and multiplicity dependence of charm quark collectivity in the $\Pp{}\Pp$ system (Fig.~\ref{fig:fig5}, right), and the details of collective behavior of beauty quarks in the $\Pp{}\mathrm{Pb}$ system~\cite{CMS:2020qul}.

Probing systems with even smaller interaction regions is important to understand the reach of a hydrodynamic description. The search for collectivity has been extended to electron-positron, electron-proton, and photon-nucleus interactions with none of these systems exhibiting evidence of the long-range correlations seen in hadronic collisions. High-energy \pPb ultraperipheral collisions, \ie, where the impact parameter is larger than the nucleus radius, provide a new system to extend the search of long-range correlations to photon-proton collisions. Two-particle $V_{n\Delta}$ Fourier coefficients and corresponding single-particle $\vn$ ($n=1\text{--}3$) azimuthal anisotropies are compared to models that do not incorporate collective effects: the data suggest the absence of collectivity in the \PGg{}\Pp system over the explored multiplicity range of up to $N_{\text{trk}}\approx{35}$~\cite{Aad:2021yhy,CMS:2020rfg}.

\section{Summary}
\label{sec:summary}

Although significant progress has been made on the level of precision achieved at large collision systems, the amount of data collected at LHC allow the measurement of more complex flow-related observables. More specifically, such measurements pose constraints on initial conditions, which in turn contribute to more precise modeling of the final-state dynamics. Previous  flow  measurements mostly focused on $V_n$ (overall  flow),  \ie, $\vn$ with respect to $\Psi_n$, but recently event-by-event flow fluctuations, longitudinal flow decorrelations, and measurements of nonlinear response coefficients demonstrate that model calculations cannot simultaneously describe all the aforementioned observables.

Similar observations further support the hydrodynamic origin of collective correlations in high-multiplicity events in small collision
systems. Multiparticle long-range correlations of inclusive charged and identified hadrons are observed down to the smallest collision systems, while paving the way for stringent tests of theory predictions on the grounds that heavy flavor particles are formed early and subsequently participate in the medium evolution. The origin of positive $\vn$ seen up to high \pt is however still not resolved. Whether this is a manifestation of a collective behavior of the system created in such collisions and/or a dilute system with parton scatterings requires experimental techniques for suppressing nonflow contamination. In parallel, precision measurements of \PZ bosons in peripheral nucleus-nucleus collisions can lead to an improved understanding of the onset of collectivity
in small systems, whereas finding ways to discriminate between the two different scenarios are developed too, \eg, new results provide information on the QCD dynamics of multiparticle production in high-energy photonuclear interactions.

\section*{Acknowledgements}

The work is supported in whole by the Nuclear Physics (NP) program of the U.S. Department of Energy (DOE) with number \href{https://pamspublic.science.energy.gov/WebPAMSExternal/Interface/Common/ViewPublicAbstract.aspx?rv=d1ddcae6-235b-4163-ae34-01fce58e5f90&rtc=24&PRoleId=10}{DOE DE-SC0019389} and \href{https://pamspublic.science.energy.gov/WebPAMSExternal/Interface/Common/ViewPublicAbstract.aspx?rv=00d4fe0f-48a0-4d4a-baf1-c70867d9e499&rtc=24&PRoleId=10}{DE-FG02-96ER40981}. 

\nocite{*}
\bibliographystyle{auto_generated.bst} 
\bibliography{low_x_2021/GKKrintiras.bib}


%% file: Maciej_Lewicki-Low-x_2021_proceedings/Low-x_2021_proceedings/Lewicki.tex
\vspace*{1.2cm}

\thispagestyle{empty}
\begin{center}
{\LARGE \bf Overview of ATLAS Forward Proton detectors for LHC Run 3 and plans for the HL-LHC}

\par\vspace*{7mm}\par

{

\bigskip

\large \bf Maciej P. Lewicki\footnote{on behalf of the ATLAS Collaboration}}

\bigskip

{\large \bf  E-Mail: maciej.lewicki@ifj.edu.pl}

\bigskip

{Institute of Nuclear Physics Polish Academy of Sciences, PL-31342 Krakow, Poland}

\bigskip

{\it Presented at the Low-$x$ Workshop, Elba Island, Italy, September 27--October 1 2021 \footnote{Copyright 2021 CERN for the benefit of the ATLAS Collaboration.
CC-BY-4.0 license.}}

\vspace*{15mm}

\end{center}
\vspace*{1mm}

\begin{abstract}

A key focus of the physics program at the LHC is the study of head-on proton-proton collisions. Among those, an important class of physics can be studied for cases where the protons narrowly miss one another and remain intact. In such cases, the electromagnetic fields surrounding the protons can interact producing high energy photon-photon collisions, for example. Alternatively, interactions mediated by the strong force can also result in intact forward scattered protons, providing probes of quantum chromodynamics. In order to aid identification and provide unique information about these rare interactions, instrumentation to detect and measure protons scattered through very small angles is installed in the beam-pipe far downstream of the interaction point. We describe the ATLAS Forward Proton `Roman Pot' Detectors, including their performance to date and expectations for the upcoming LHC Run 3, covering Tracking and Time-of-Flight Detectors as well as the associated electronics, trigger, readout, detector control and data quality monitoring. The physics interest, beam optics and detector options for extension of the programme into the High-Luminosity LHC era are also discussed. 
\end{abstract}
\part[Overview of ATLAS Forward Proton detectors for LHC Run 3 and plans for the HL-LHC\\ \phantom{x}\hspace{4ex}\it{Maciej P. Lewicki on behalf of the ATLAS Collaboration}]{}
 
\section{Introduction}

The predictions of forward proton scattering arise in a diverse range of physics, including the hard \cite{jl2,jl3} and nonperturbative QCD \cite{jl4}, interactions at electroweak scale \cite{jl7,jl8,jl9,jl10}, and searches for physics beyond Standard Model \cite{jl5,jl6,jl11,jl12,jl13,jl14}.
Such events, usually called diffractive, involve an exchange of a colourless object between interacting protons, one or both of which may remain intact.
Moreover, a \textit{rapidity gap} will be present -- an absence of particles produced into kinematic vicinity of the intact proton.
Historically, rapidity gap is a standard experimental signature of a diffractive event,  however, it is frequently outside the acceptance of detector, or is destroyed due to background, i.e. particles coming from \textit{pile-up} -- independent collisions happening in the same bunch crossing.
An alternative method of identifying diffractive events is a direct measurement (\textit{tagging}) of the scattered proton, which requires additional devices called \textit{forward detectors} far downstream from the interaction point. 


\begin{figure}[h]
\hspace*{-1.7cm}\includegraphics[width=1.1\textwidth]{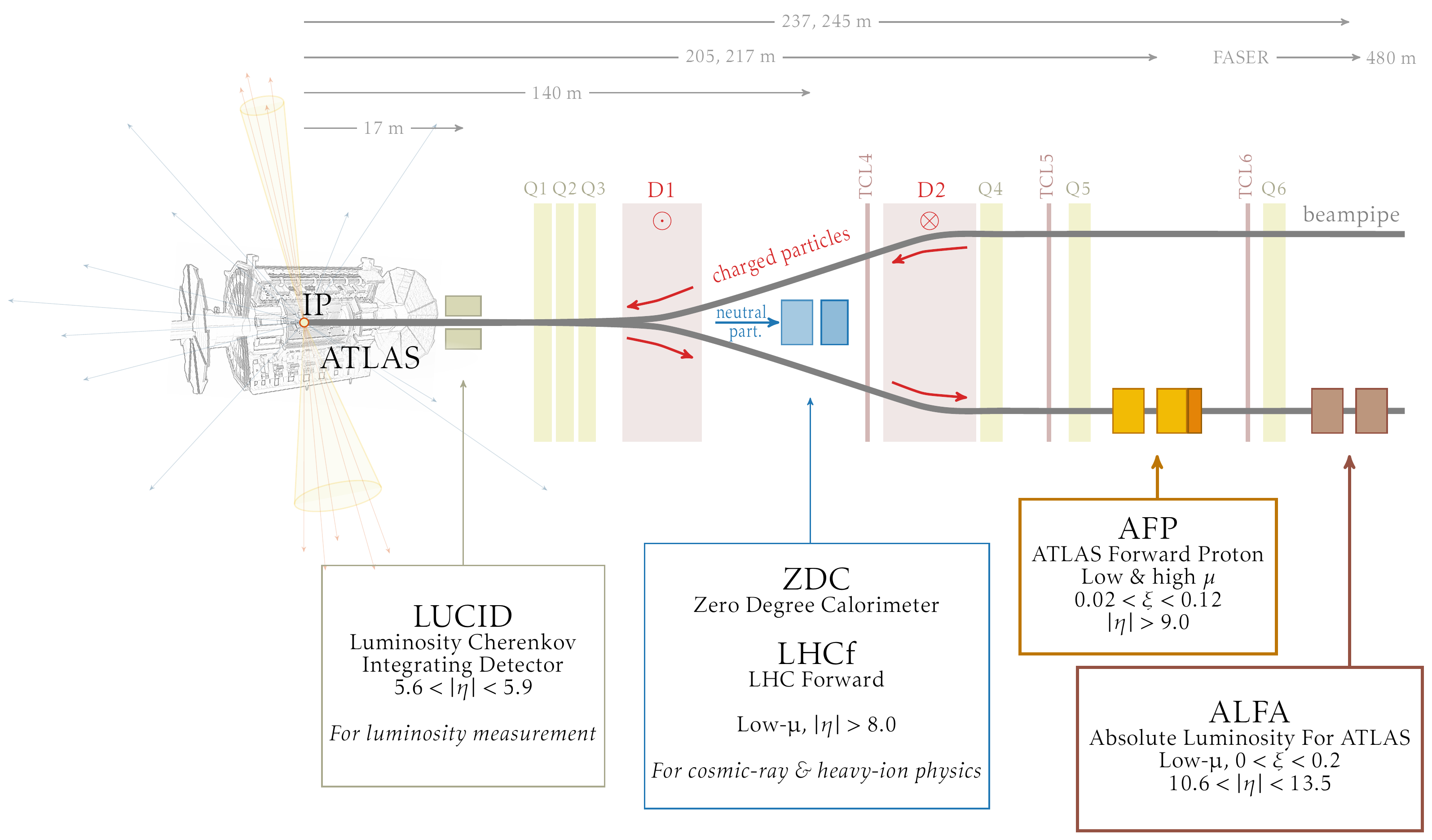}
\caption{A diagram of Forward Detectors in the ATLAS experiment showing their placement with respect to beam lines and optical instrumentation: dipoles (D1, D2), quadrupoles (Q1--6) and collimators (TCL4--6).}
\label{fig:fwd_det}
\end{figure}

\section{Forward Spectrometers in ATLAS}

ATLAS \cite{atlas_maciej} forward spectrometers are a set of instruments housed in Roman Pot devices registering the protons scattered at very small angles. A proton scattered at the interaction point (IP) is deflected outside the beam envelope by dipole and quadrupole magnets of the LHC~\cite{evans}. Its momentum can be determined by measuring points on its trajectory~\cite{tdr,optics}. The schematic layout of the forward spectrometers in ATLAS experiment with respect to the beam lines, optics instrumentation and other forward detectors is shown in Figure \ref{fig:fwd_det}.

\paragraph{Absolute Luminosity For ATLAS (ALFA)} performs measurements of soft diffraction and elastic scattering. It also provides an important input for Monte-Carlo generators, in particular, for modelling cosmic ray showers and simulation of the pile-up background.
The ALFA spectrometer system consists of four vacuum-sealed spectrometers housed in Roman Pots, which are inserted vertically (top and bottom) onto the beam line.
The NEAR and FAR stations are placed on each side of the ATLAS Interaction Point at 237 and 241/245\footnote{FAR station was initially installed at 241 m (Run 1) and then moved to 245 m (Run 2) to improve the reconstruction of proton kinematics} m respectively with the distance of the tracker's edge to the beam during normal operation at below 2 mm.
Each station houses a multi-layer scintillating fibre (SciFi) consisting of two main detectors (10 layers of 64 fibers each) used for tracking, and 4 outer layers for the purpose of precise alignment. Achieved tracking resolution is approximately $\sigma=30$ $\mu$m in both, vertical and horizontal direction.
The read-out is performed by the Multi-Anode-Photo-Multipliers and dedicated scintillators provide the triggering capability. ALFA detectors require special running conditions of low pile-up as well as high $\beta^*$ optics. 

\begin{figure}[h]
\centering
\includegraphics[width=1\linewidth]{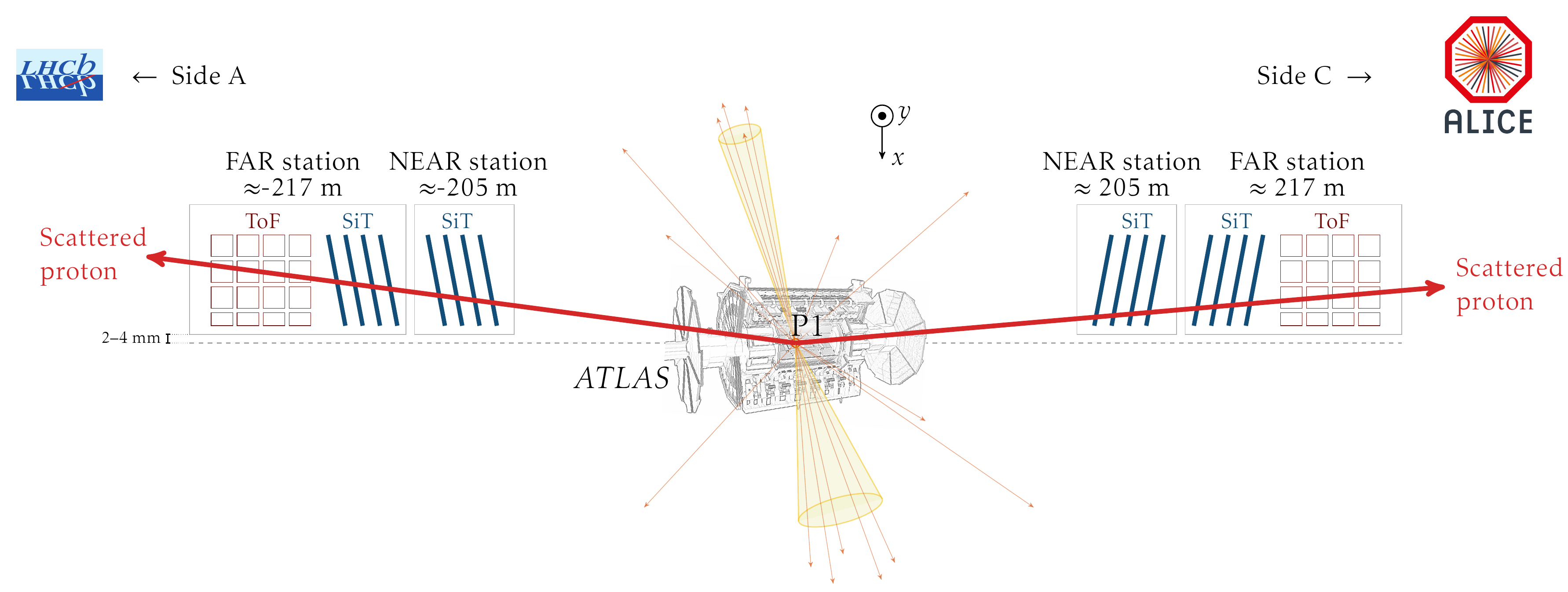}\\[-0.9cm]
\caption{A schematic diagram of the ATLAS Forward Proton detectors.}
\label{fig:sketch}
\end{figure}

\paragraph{ATLAS Forward Protons (AFP)}  spectrometer system consists of four Roman Pot stations. Their placement with respect to the beam lines is shown schematically in Figure~\ref{fig:sketch}. ``NEAR'' and ``FAR'' devices are placed at 205 m and 217 m on both sides of the IP and are inserted horizontally towards the beam.
Each station houses four planes of 3D silicon pixel sensors~\cite{sit1,sit2,sit3,sit4} forming the silicon tracker (SiT), which measures the trajectories of the scattered protons. The sensors have 336$\times$80 pixels with
50$\times$250 $\mu$m$^2$ area each, providing the combined spatial resolution of reconstructed proton tracks of 6 $\mu$m and 30 $\mu$m in $x$ and $y$ directions, respectively \cite{sit5}. Optimal resolution in $x$ coordinate is achieved with the sensors tilted by 14 degrees about the $x$-axis. 
\begin{figure}[h]
\centering
\hspace*{-0.7cm}
\includegraphics[height=0.3\linewidth]{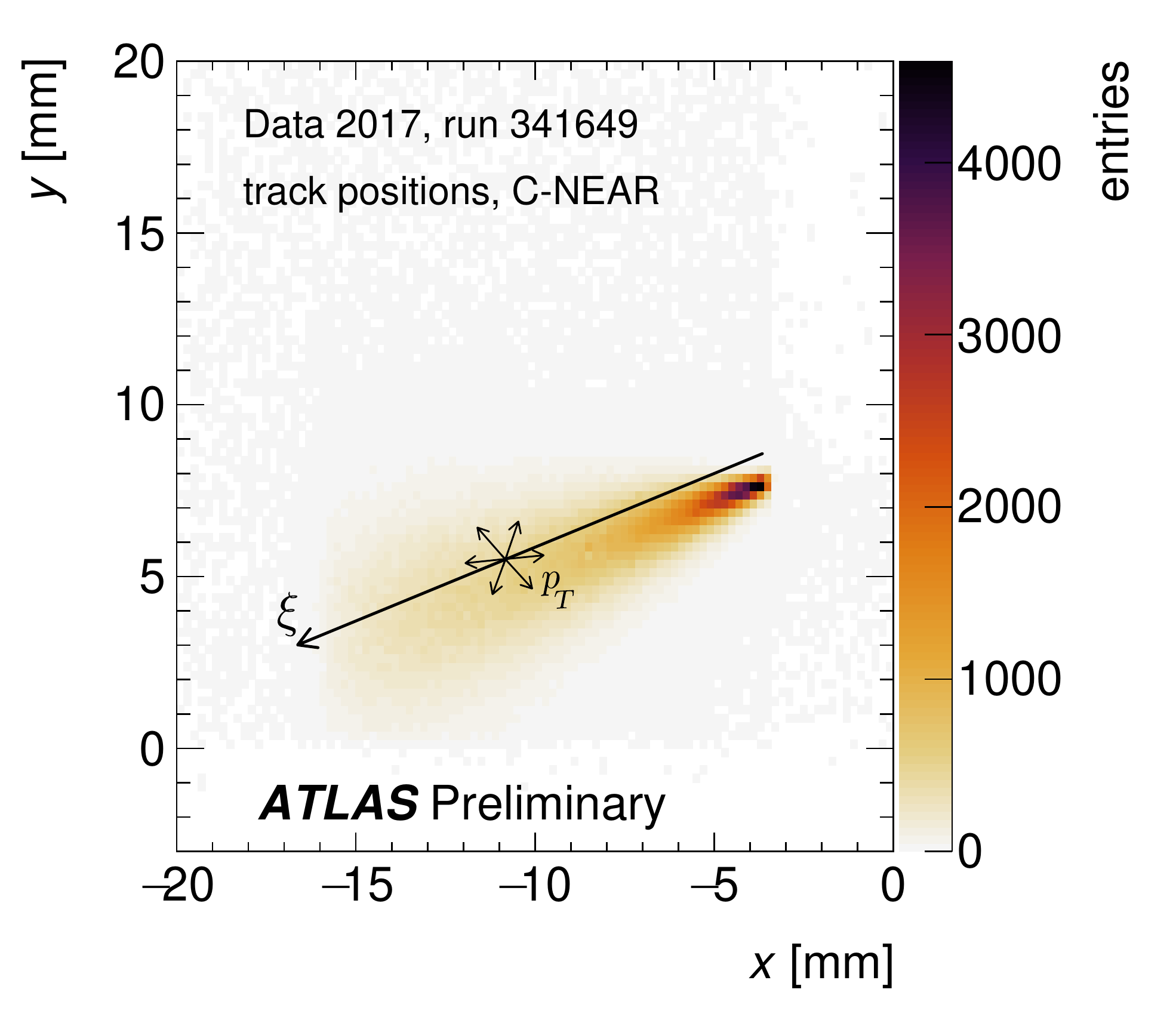}~~~~~~
\includegraphics[height=0.3\linewidth]{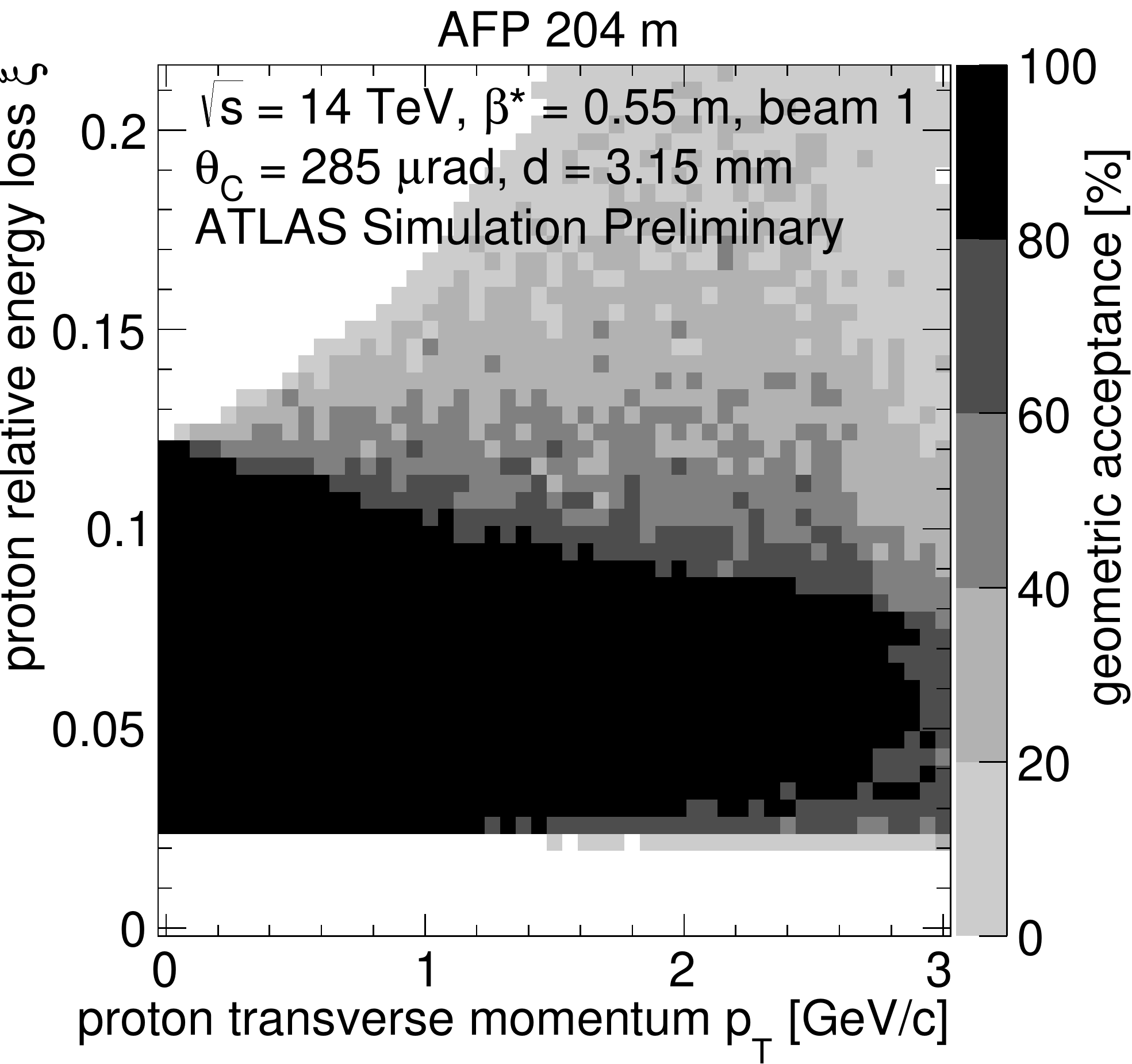}~~~~~~
\includegraphics[height=0.3\linewidth]{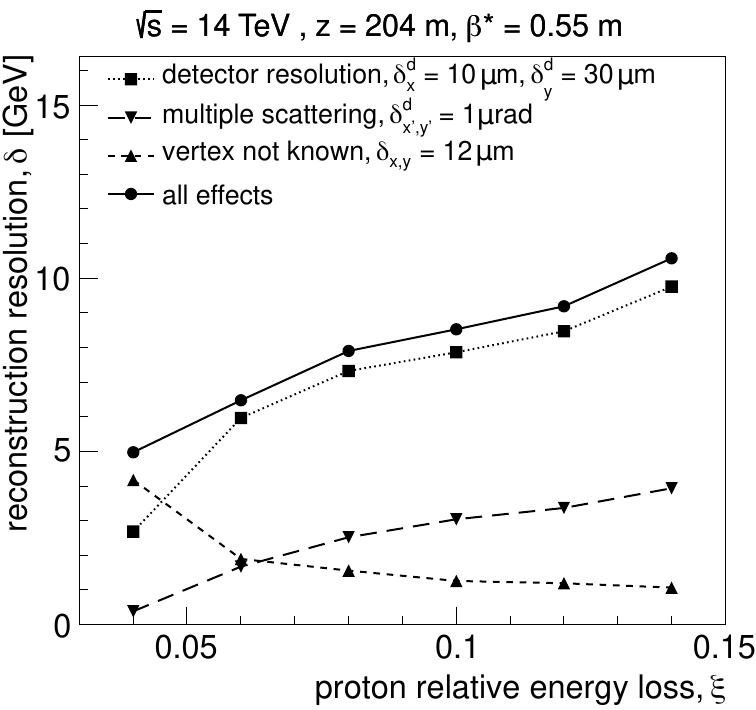}
\caption{Left: example distribution of reconstructed track positions ($x$ and $y$, transverse to the beam); the beam spot is approximately at (0,10) mm and deviations from this position are related to proton energy loss, as well as its transverse momentum (taken from \cite{proc_lhcp}).
Middle: Simulated geometric acceptance of the AFP detector as a function of the proton relative energy loss and its transverse momentum (from \cite{tdr}).
Right: AFP reconstruction resolution in dependence on proton relative energy loss $\xi$ calculated accounting also for multiple scattering and unknown position of the collision vertex \cite{tdr}. 
}
\label{fig:xy}
\end{figure}
The reconstruction of position of protons traversing the AFP detectors (see Figure \ref{fig:xy}, left panel), in a known magnetic field, allows the estimation of proton energy and transverse momentum \cite{unfold}. The main observable measured by the AFP is the proton fractional energy loss, defined as: $\xi=1-E_{\text{proton}}/E_{\text{beam}}$. The precision of unfolding proton kinematics based on the positions in AFP is directly affected by its spatial resolution. Figure \ref{fig:xy} (right panel) illustrates how the resolution changes with proton relative energy loss ($\xi$). Additional effects that affect the resolution of proton energy reconstruction include the unknown position of the primary vertex and multiple scattering.
The typical acceptance in $\xi$ and $p_\text{T}$ is illustrated in middle panel of Figure \ref{fig:xy}.

\begin{figure}[h]
\centering
	\includegraphics[height=0.36\linewidth]{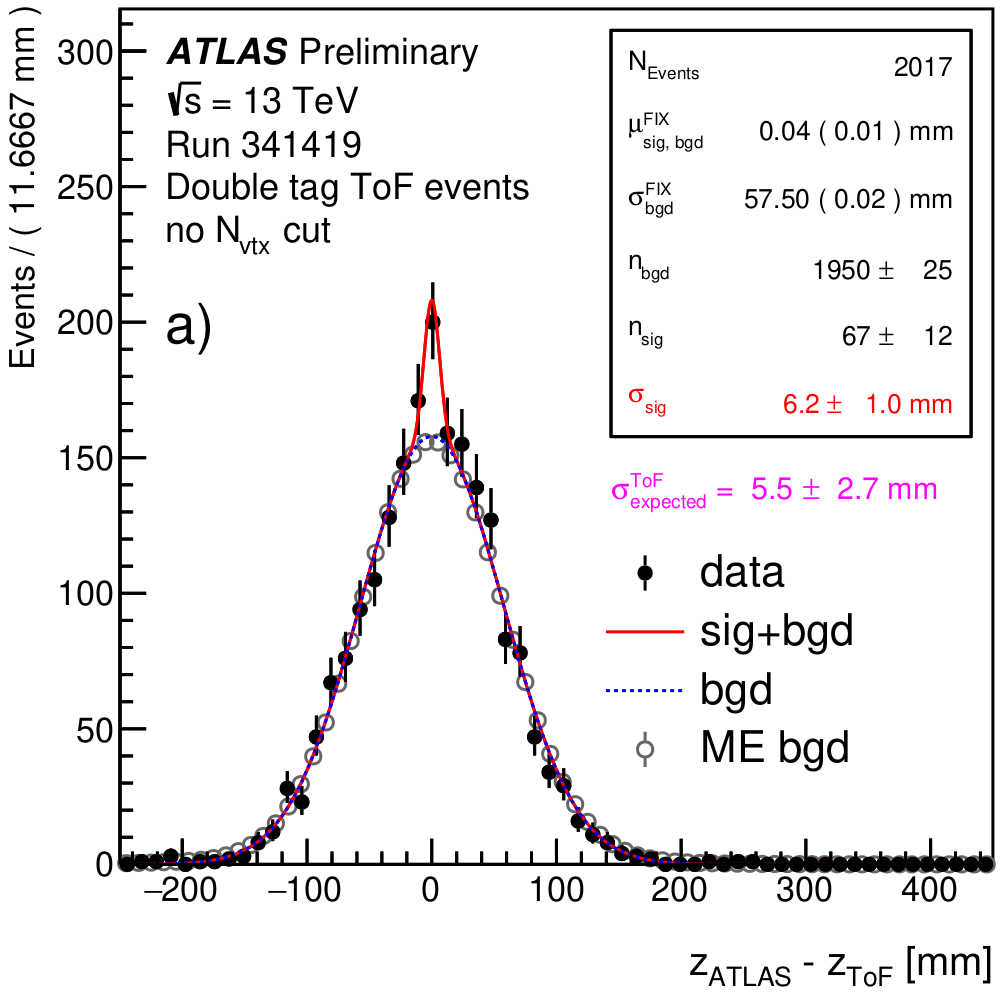}~~~~~~
	\includegraphics[height=0.36\linewidth]{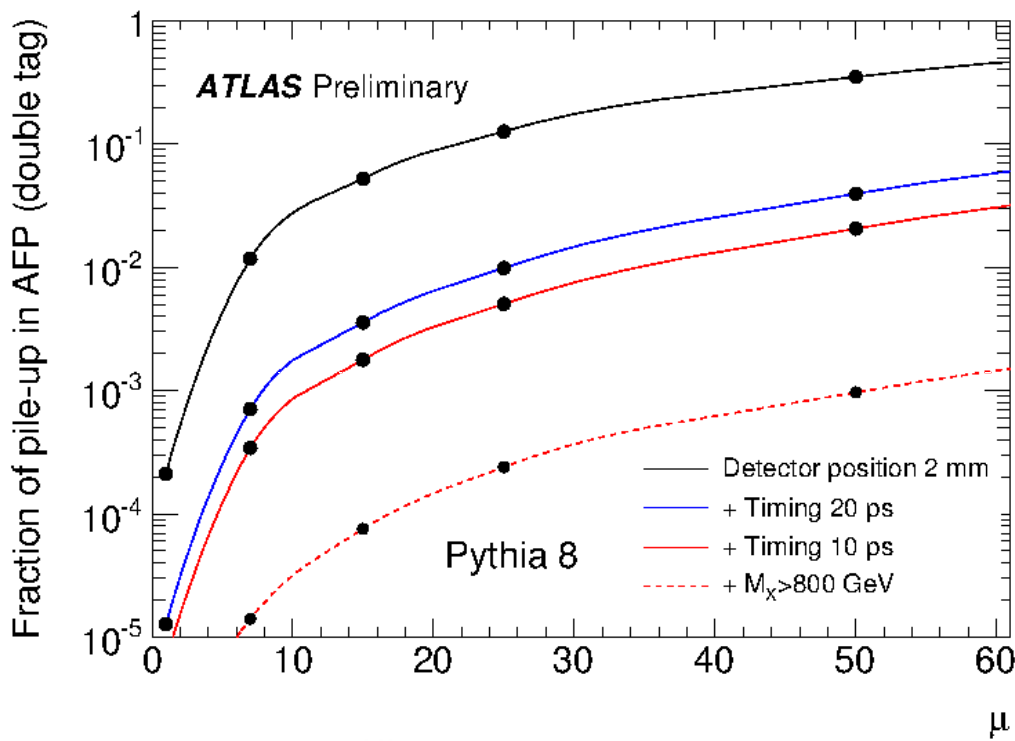}\\
	\caption{Left: Example distribution of $z_{\text{ATLAS}}-z_{\text{ToF}}$ measured in events with ToF signals on both sides of the interaction region, where $z_{\text{ATLAS}}$ stands for vertex $z$-positions reconstructed as primary ones by ATLAS (taken from \cite{tof0}).
	Right: A simulation of the fraction of pile-up events present in the data sample with a double AFP tag shown in dependence of the mean pile-up and for various timing resolutions (taken from Ref. \cite{tdr}).}
	\label{fig:tofres}
\end{figure}

\paragraph{Time-Of-Flight (ToF) detectors} are additional equipment present in Roman Pots at FAR stations. For processes in which protons are reconstructed on both sides of the IP, this allows rejection of background from pile-up by using the difference between the A and C-side ToF measurements to reconstruct the primary vertex position. 
The ToF detectors are based on Cherenkov radiation in quartz crystals, which leads to an excellent timing resolution. The performance of the ToF devices was measured for the data gathered in 2017  \cite{tof1,tof2} and obtained time resolution reaches values of 20 $\pm$ 4 ps and  26 $\pm$ 5 ps for sides A and C, respectively. Achieved time resolution translates to determination of primary vertex $z$-position with an accuracy of 5.5 $\pm$ 2.7 mm. Such level of precision allows for a substantial reduction of background in `double-tag' events, as shown in Figure \ref{fig:tofres} (right). However, the observed efficiency of ToF reconstruction was very low ($\approx$7\%) due to fast PMT degradation during the data taking. Recently, new PMTs were installed and preliminary tests show readiness for use in upcoming data-taking campaigns of Run 3.
\FloatBarrier

\begin{figure}[h]
\centering
\includegraphics[width=0.4\linewidth]{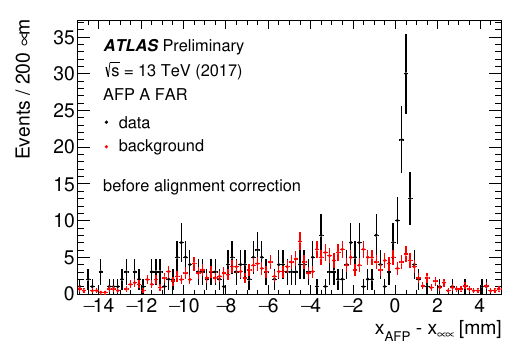}
\includegraphics[width=0.4\linewidth]{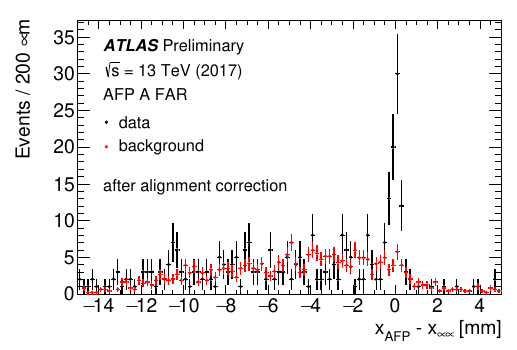}\\
\includegraphics[width=0.4\linewidth]{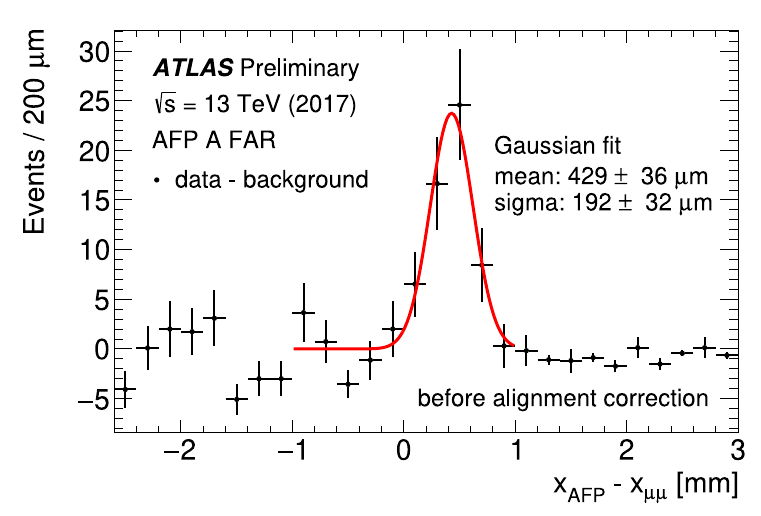}
\includegraphics[width=0.4\linewidth]{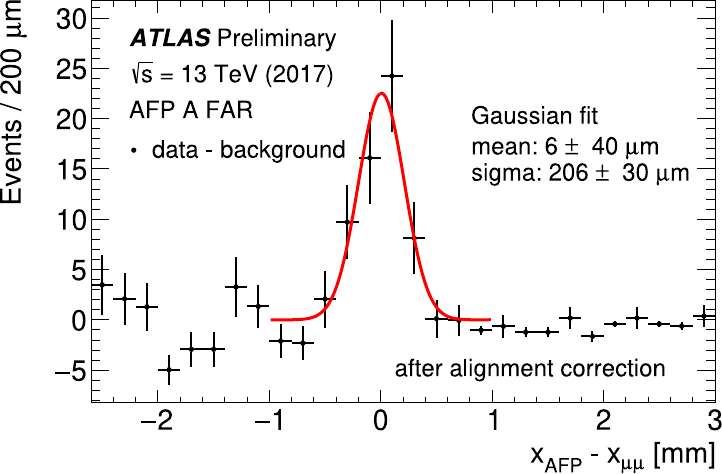}\\
\caption{The distribution of differences between measured track position $x_{\text{AFP}}$ and position expected based on dilepton system kinematics $x_{\mu\mu}$, compared with the background model based on event-mixing. The figures on the right show the same as those on the left after correcting the $x$ coordinates of all events by the alignment constant (taken from Ref. \cite{afp_align}).}
\label{fig:alignment}
\end{figure}

\paragraph{AFP global alignment} is performed by comparing the proton relative energy loss measured in the AFP $\xi_{\text{AFP}}$ with a corresponding value calculated based on the kinematics of produced lepton pair $\xi_{ll}$:
\begin{equation}
\xi_{\text{AFP}}=1-\frac{E_{\text{proton}}}{E_{\text{beam}}},~~~~~~~\xi_{ll}^{\text{A/C}}=\frac{m_{ll} e^{(+/-) y_{ll}}}{\sqrt{s}}
\label{eq:xill}
\end{equation}
The AFP alignment parameters are adjusted in such a way that the maximum of the distribution of $\xi_{\text{AFP}}-\xi_{\mu\mu}$ is at zero. Figure \ref{fig:alignment} illustrates the differences between projected track $x$ position based on lepton kinematics and the one measured by AFP, before and after alignment correction. A valuable advantage of such method is a low and well-modelled background, which allows achieving alignment precision currently quoted at 300 $\mu$m. Continued studies of data and simulation show promise that this value can be further improved.
	
\begin{figure}[h]
	\includegraphics[width=1\linewidth]{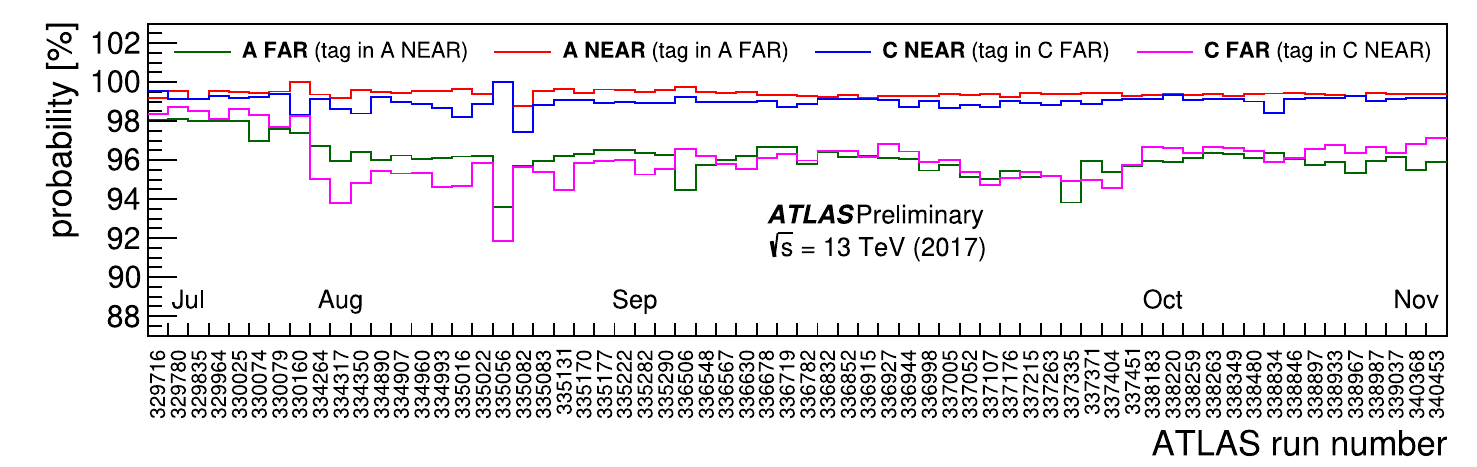}
	\caption{Track reconstruction efficiencies recorded in high-luminosity runs during 2017 data-taking (taken from Ref. \cite{afp_align}).}
	\label{fig:eff}
\end{figure}

\begin{figure}[h]
\centering
\includegraphics[height=0.35\linewidth]{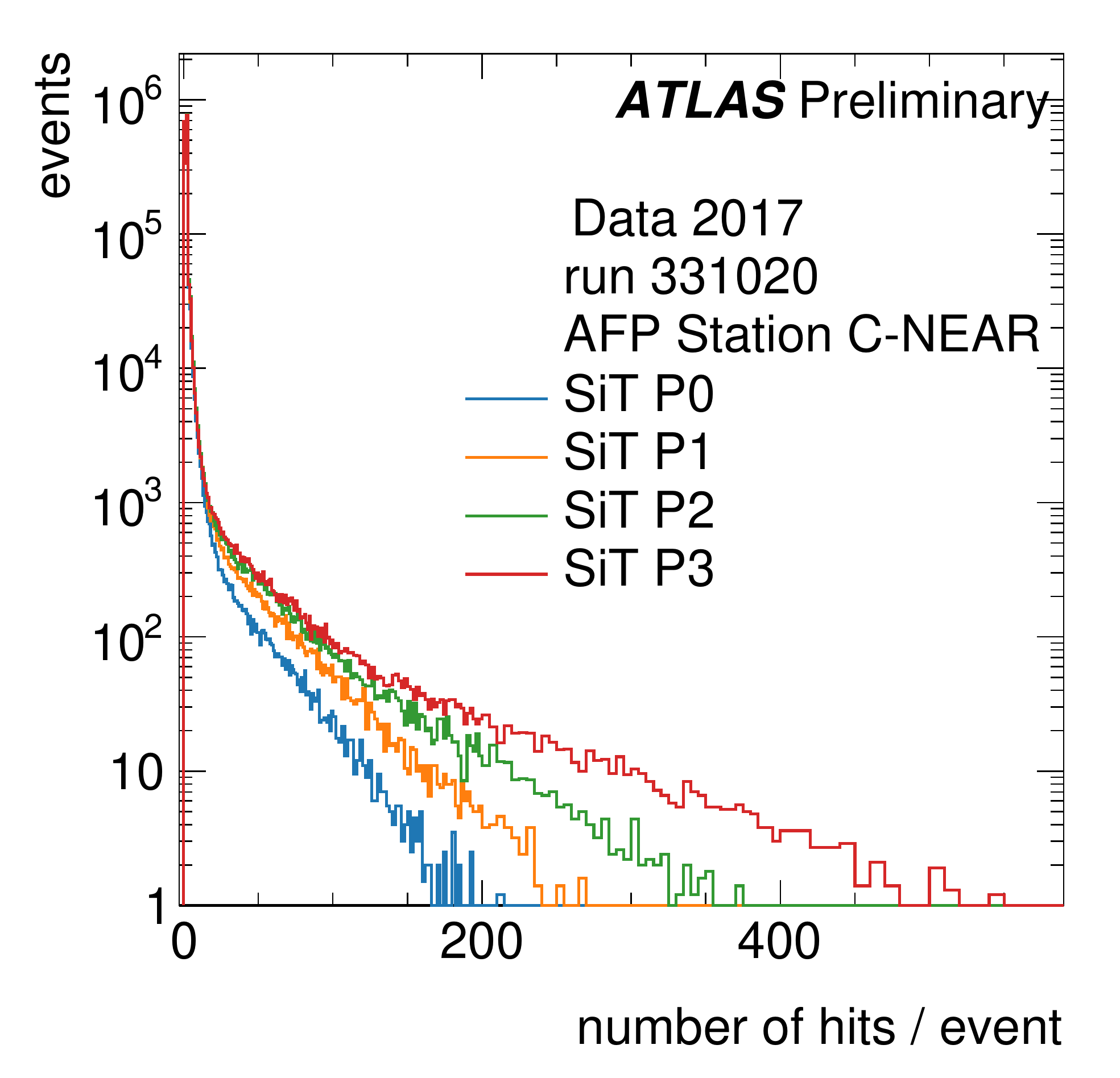}
\includegraphics[height=0.35\linewidth]{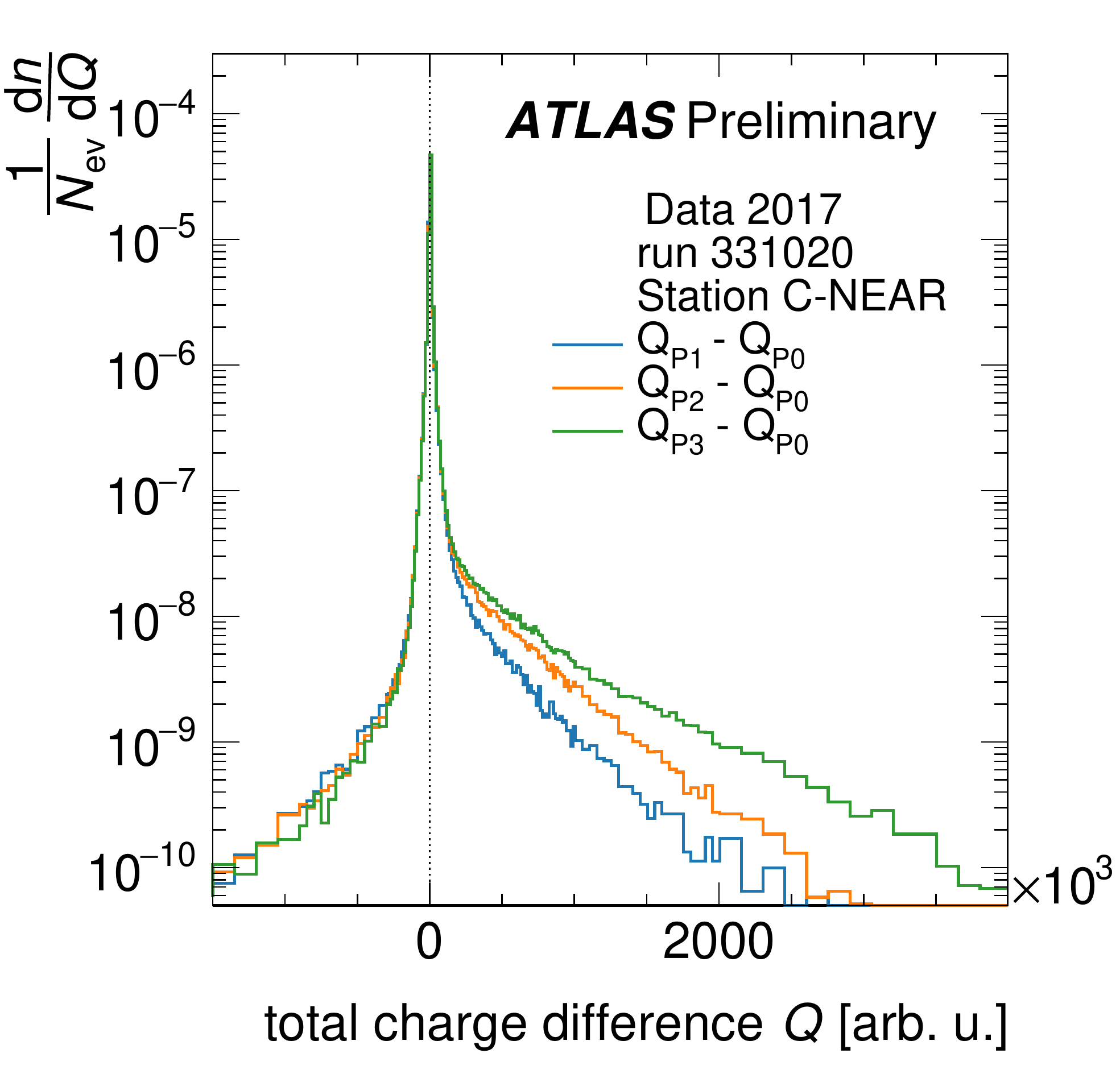}\\
\caption{Distribution of the number of hits per event (left) and the difference of recorded charge for each of silicon planes in C-NEAR station (right). Secondary interactions with detector material result in higher number of hits and higher recorded charge for consecutive SiT planes (taken from Ref. \cite{proc_lhcp}).}
\label{fig:charge}
\end{figure}

\paragraph{AFP track reconstruction efficiency} is calculated using a so called `tag-and-probe' method. The efficiency is defined as the ratio of events in which the track is recorded in one station (\textit{tag}) and not in the other (\textit{probe}) to the total number of events with tagged tracks. Measured track reconstruction efficiency during the 2017 data taking campaigns are shown in Figure \ref{fig:eff}.
NEAR stations record efficiency over 98\% for all studied datasets. A possible effect that might contribute to lower efficiencies of FAR stations (95\% -- 98\%) is the radiation degradation of the silicon tracker, as the FAR stations are inserted slightly closer (<1  mm) to the beam and are more exposed to the beam halo. However, lower efficiencies observed in FAR stations, are also a natural consequence of the `tag-and-probe' method used in this analysis, as the downstream stations are additionally affected by the showers created by interactions with detector material in the upstream station. The existence of showers is evident when examining the long non-Poisson tail in hit multiplicity per plane, which is higher for each consecutive pixel layer. Additionally, each consecutive plane registers on average higher number of hits (pixels that record a signal exceeding a threshold) and higher charge deposits, which is also expected under the presence of cascades created by interactions with SiT and Roman Pot floor. Both effects are illustrated in Figure \ref{fig:charge}.

\FloatBarrier

\section{Status and plans for Run 3}
	
	No major changes between Run 2 and Run 3 are planned in terms of detector technology, although numerous hardware and software updates were implemented.	
	The AFP tracker system was equipped with newly produced tracking modules and  new heat exchangers were installed to improve cooling capabilities.	
	Due to the repeated failures of MCP-PMTs in vacuum, the AFP ToF detector was redesigned with MCP-PMTs placed out of the detector secondary vacuum. The newly designed construction of the R2D2 based MCP-PMT back-end electronics was developed, successfully tested and used in the construction of the new ToF device. Additionally, a set of new, glue-less LQBars was installed, as well as picoTDC, single-channel pre-amplifiers, modified trigger and pulser modules.
	Updated systems were exposed to proton beams at DESY and CERN SPS and successfully underwent performance tests.
	
	The design of ALFA trackers remained unchanged as well and minor hardware updates include improvements to the cooling system and exchange of the readout electronics (due to radiation damage).
	All subsystems of both AFP and ALFA spectrometers are installed in LHC tunnels and are fully prepared for data-taking campaigns starting in 2022.
	
	Similarly, good progress is observed in development of software and simulation necessary for physics analyses of forward protons in ATLAS. Advances in studies of tracker performance and beam optics, as well as improvements in precision of detector alignment allow delivering a high-accuracy proton physics object to ATLAS. The properties of forward protons reconstructed with AFP and ALFA are used in several analyses across different working groups in ATLAS. A dedicated task force (Proton Combined Performance group) leads efforts to improve understanding of the proton object, including assessment of tracker efficiency or susceptibility to physical conditions, leading to a possible reduction of systematic uncertainties. Additionally, an ongoing work aiming at implementation of full \textsc{Geant4} simulation will allow to better understand possible effects related to detector geometry, alignment or interactions with detector material. Progresses in areas listed above contribute to advances with physics analysis, which are discussed in more detail in Section \ref{sec:analyses}.
	
	
	\begin{table}[h]
	{\renewcommand{\arraystretch}{1.4}
	\small
	\centering
	\begin{tabular}{m{4cm}ll}
 & \textbf{Run 2} & \textbf{Run 3 plans (requests)} \\
		beam and optics & $\sqrt{s}=13$ TeV, ~~$\beta^*=0.3$ m, 0.4 m & $\sqrt{s}=13$ TeV, ~~$0.2 < \beta^* < 1.1$ m \\
		AFP setup & one-arm (2016), two-arms (2017)& two-arms + TOF \\
		Standard runs & $\langle\mu\rangle$$\approx$35, int. lumi. 46.9 fb$^{-1}$ & $\langle\mu\rangle$$<$60, $\mathcal{O}(500~\text{fb}^{-1})$ \\
		Special runs at $\mu$$\approx$0 \newline (soft diffraction) & int. lumi.: $\approx$ 100 nb$^{-1}$ & $\mathcal{O}(100~\text{nb}^{-1})$ \\
		Special runs at 0.3$\lesssim$$\mu$$\lesssim$1 \newline (low $p_\text{T}$ jets) & int. lumi.: $\approx$ 1.15 pb$^{-1}$ & $\mathcal{O}(1~\text{pb}^{-1})$ \\
		Special runs at $\mu$$\approx$2 \newline (EW, hard diffr., SD $t\bar{t}$) & int. lumi.: $\approx$ 150 pb$^{-1}$ & $\mathcal{O}(100~\text{pb}^{-1})$ \\
	\end{tabular}\\
	}
	\caption{Comparison of most important properties of data taken by AFP in Run 2 and requests for Run 3 data taking.}
	\label{tab:data_afp}
	\end{table}
	
	\begin{table}[h]
	{\renewcommand{\arraystretch}{1.4}
	\small
	\centering
	\begin{tabular}{m{3cm}lm{5cm}}
& \textbf{Run 2} & \textbf{Run 3 plans (requests)} \\
		collision energy & $\sqrt{s}$=13 TeV & $\sqrt{s}\geq$ 13.5 TeV,  \\
		beam conditions & $\beta^*$=90 m, 2.5 km & $\beta^*$= 3, 6 km and/or \newline $\beta^*_x$= 3 km, $\beta^*_y$= 6 km \\
		& $\langle\mu\rangle$$\approx$35, $\langle\mu\rangle$$\approx$0 & only at $\langle\mu\rangle$$\approx$0 \\
	\end{tabular}\\
	}
	\caption{Comparison of most important properties of data taken by ALFA in Run 2 and possible plans for Run 3}
	\label{tab:data_alfa}
	\end{table}

\section{Recent ATLAS physics results with forward proton tag}
\label{sec:analyses}

	\subsection{First physics analysis with AFP proton tag}

\begin{figure}[h]
	{\centering
	\includegraphics[width=0.7\linewidth]{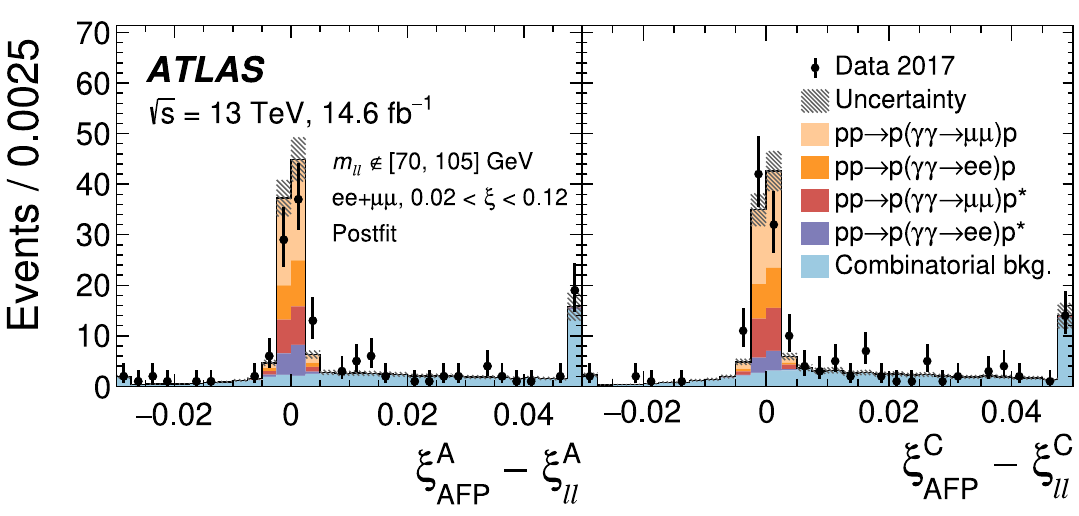}\\
	}
	\caption{Distributions of $\xi_{\text{AFP}}-\xi_{l\bar{l}}$ with $\xi_{l\bar{l}}$ and $\xi_{\text{AFP}}$ both in range [0.02, 0.12]. Black points with error bars illustrate data and its statistical uncertainty and coloured stacks represent model predictions for various processes contributing to the measured signal (p$^*$ denotes dissociated proton) with hatched area indicating their combined uncertainty (plots from Ref. \cite{dileptons}).}
	\label{fig:dilepton_xi}
	\end{figure}

	The analysis of the rich data collected by ATLAS Forward Protons detector is ongoing and recently the results on semi-exclusive dilepton production associated with forward proton scattering were published \cite{dileptons}, delivering the cross-sections measurements for $p \,p \rightarrow p\,(\gamma\gamma\rightarrow l\bar{l}) \,p \,p^*$ processes. Modelling of photon fusion in proton-proton interactions is poorly constrained, particularly at high $\gamma\gamma$ invariant masses. Direct proton measurement allows for a strong background suppression by means of kinematic matching of tagged forward protons and leptons measured in the central detector. The fractional proton energy loss measured in AFP, $\xi_{\text{AFP}}$, can be compared against the value of $\xi_{l\bar{l}}$ that can be derived from dilepton system kinematics as defined in Eq. (\ref{eq:xill}). With the criterium of $|\xi_{\text{AFP}}-\xi_{l\bar{l}}|<0.005$ a total of 57 and 123 candidates in the $ee+p$ and $\mu\mu+p$ final states were observed. With a background rejection on the level of $\approx$85\% and signal acceptance on the level of $\approx$95\% this corresponds to statistical significance of over 5$\sigma$ in each production channel, thus providing a direct evidence of forward proton scattering in association with electron and muon pairs produced via photon fusion. Table \ref{tab:dilepton} summarizes obtained cross sections for these processes in the detector fiducial region compared with the relevant theoretical predictions. Figure \ref{fig:dilepton_xi} shows a distribution of measured $\xi_{\text{AFP}}-\xi_{l\bar{l}}$ for both sides A and C together with predictions associated with various production channels.
	
	\begin{table}[h]
	\centering
	\begin{tabular}{lrr}
	$\sigma_{\text{\textsc{Herwig+Lpair}}} \times S_{\text{surv}}$ & $\sigma_{ee+p}^{\text{fid.}}$ (fb) & $\sigma_{\mu\mu +p}^{\text{fid.}}$ (fb) \\
	\hline
	$S_{\text{surv}}=1$ & 15.5 $\pm$ 1.2 & 13.5 $\pm$ 1.1 \\
	$S_{\text{surv}}$ using {\footnotesize$^{\text{EPJC 76 (2016) 9}}_{\text{PLB 741 (2015) 66}}$} & 10.9 $\pm$ 0.8 & 9.4 $\pm$ 0.7 \\
	\textsc{Superchic} & 12.2 $\pm$ 0.9 & 10.4 $\pm$ 0.7 \\
	\textbf{Measurement} & \textbf{11.0 $\pm$ 2.9} & \textbf{7.2 $\pm$ 1.8} \\
	\hline
	\end{tabular}\\
	\caption{Summary of model predictions on cross-sections for diffractive processes of di-lepton production compared with AFP measurements.}
	\label{tab:dilepton}
	\end{table}
	

	\subsection{Single diffraction results with ALFA proton tag}
	
\begin{figure}[h]
	{\centering
		\includegraphics[width=0.5\linewidth]{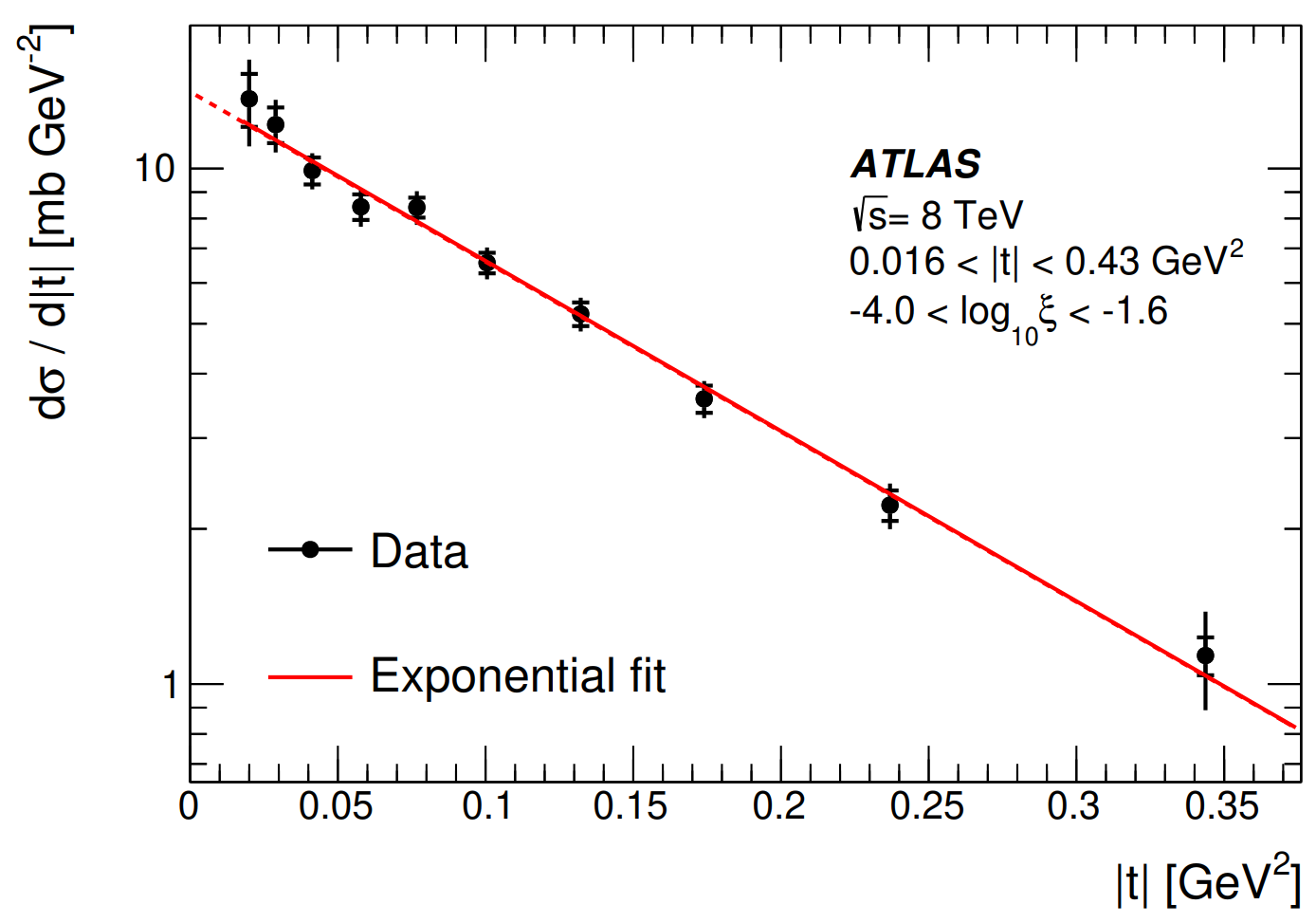}
	\\}
	\caption{The differential cross section as a function of $|t|$ with statistical and total uncertainties represented by inner and outer error bars, respectively. Red line shows fitted exponential function (plot taken from Ref. \cite{alfa_sd}).}
	\label{fig:alfa_sd}
	\end{figure}

	In a recent publication \cite{alfa_sd}, a dedicated sample of low-luminosity (mean pile-up $\langle\mu\rangle$$<$$0.08$) proton-proton collision data at $\sqrt{s}$ = 8 TeV is used to study of the dynamics of the inclusive single-diffractive dissociation process $p\, p \rightarrow X\,p$. Improving on previous related analyses, besides measurements of charged particles from the dissociated system $X$ performed by the central ATLAS detector components, the ALFA forward spectrometer provides reconstruction of the final-state intact protons. 
	The differential cross sections are measured as a function of the fractional proton
	energy loss ($-4.0$ $<$ $\log_{10}$$\xi$ $<$ $-1.6$), the squared four-momentum transfer ($0.016$ $<$ $|t|$ $<$ $0.43~\text{GeV}^2$), and the size of the rapidity gap $\Delta\eta$.
	The total cross section integrated across the fiducial range is shown in Table \ref{tab:xsec_alfa} with additional information on predictions of relevant theoretical models.	
	\begin{table}[h]
	\centering
		\begin{tabular}{lll}
		Distribution & $\sigma_{\text{SD}}^{\text{fiducial}(\xi,t)}$ [mb] &  $\sigma_{\text{SD}}^{t-\text{extrap}}$ [mb] \\ \hline
		\textsc{Pythia8} A2 (Schuler-Sj\"ostrand) & 3.69 & 4.35\\
		\textsc{Pythia8} A3 (Donnachie-Landshoff) & 2.52 & 2.98\\
		\textsc{Herwig7} & 4.96 & 6.11\\
		\textbf{Measurement} & \textbf{1.59 $\pm$ 0.13} & \textbf{1.88 $\pm$ 0.15} \\
		\hline
		\end{tabular}\\
		\caption{Summary of model predictions on cross-sections for single soft diffraction compared with the measurements by ALFA.}
		\label{tab:xsec_alfa}
	\end{table}
	As shown in Figure \ref{fig:alfa_sd} the data are consistent with an exponential $t$ dependence, $\text{d}\sigma/\text{d}t \propto e^{Bt}$ with slope parameter $B$ = 7.65 $\pm$ 0.34 GeV$^{-2}$.
	Interpreted in the framework of triple Regge phenomenology, the	$\xi$ dependence leads to a Pomeron intercept of $\alpha(0) = 1.07 \pm 0.09$.

\section{HL-LHC with AFP perspectives}


Predictions of forward proton scattering appear in a diverse range of topics, providing strong motivation for continuation of experimental efforts.
Diffractive processes were identified as a potential mechanism of top quark \cite{jl4}, as well as exclusive Higgs boson production \cite{jl5, jl6}.
Additionally, measurements of intact protons may provide a valuable input in the studies of two-photon processes in the context of SM electro-weak interactions \cite{jl9, jl10}, but also in high-mass sector beyond-SM \cite{jl11}. 
Extending the potential of new physics discoveries, a capability of direct proton tagging may be an important tool in the searches of sleptons and dark matter \cite{jl13,jl14}, as well as axion-like particles \cite{jl12}.
An important input to understanding the QCD sector of the SM and beyond may be also provided by novel studies of exclusive diffractive events \cite{jl2,jl3}.

While the rich physics opportunities open up with the capability of forward proton tagging, a number of challenges remain that keep the presence of AFP detectors in ATLAS operation after Long Shutdown 3 under discussion. The HL-LHC will feature a novel beamline that includes crab cavities, collimators and magnets at new positions and different settings. The beam instrumentation properties planned for Run 4 place Point 1 optics at a disadvantage comparing to conditions in Runs 2 and 3. The RP acceptance might be limited due to placement of beam optics devices (collimators, cavities), but more importantly, the beam crossing angle of $\phi=180^\circ$ renders the trajectories of diffractive protons closer to the beam, which in turn impairs the resolution of energy measurement.
The second major challenge is the rejection of background under the conditions of mean pile up at $\langle\mu\rangle$ $\approx$ 200. A distinction of individual vertices would only be reliable with a sub-10 ps precision of ToF detectors (Silicon/LGAD/Cherenkov technology) and additional timing devices in the central ATLAS detector. Similarly, novel solutions will be required in the area of data acquisition and analysis due to high event rates and consequently large data volumes. Parallel to technological challenges, an important practical issue is the acquisition and sustainability of manpower and resources for a time period of over 20 years.

\section{Summary}

The ATLAS Forward Proton spectrometers, ALFA and AFP, provide a capability of forward proton tagging and measuring its kinematics, thus delivering an important data in the studies of diffractive physics.
Both detector systems recorded rich datasets of standard and special, low-luminosity conditions during LHC Run 2.
The analyses of collected data are ongoing and active efforts are directed into improvement of data quality, including the accuracy of detector alignment or the estimation of trackers efficiencies.
First experimental results on dilepton production with AFP tag were recently published and many more measurements of diffractive and exclusive events from Run 2 data will come in a near future.
Similarly, first results on single diffraction with a tag in ALFA were published in 2020 and more analyses on diffractive and elastic processes are ongoing.

Both AFP and ALFA spectrometers underwent hardware improvements and, after successful tests with proton beams at DESY and SPS, were installed in the LHC tunnel and are ready for further tests preparing for Run 3 data-taking campaigns. The collection of physics data is expected to begin in Spring 2022 and in the course of next four years it is expected that the AFP will collect an order of magnitude more data for studies of diffractive physics.

The continuation of forward physics programmes in the HL-LHC era is currently under discussion within ATLAS. While a wide range of physics topics would benefit from forward proton tagging, a number of experimental challenges remain. The constraints on preferred detector localization and utilized technology are being discussed and corresponding feasibility studies are performed in parallel. Were the AFP to take data in Run 4, an optimization of beam optics must be considered in order to enhance the spectrometer acceptance.

\section*{Acknowledgements}

This work was partially supported by the Polish National Science Centre grant: 2019/34/E/ST2/00393.

\nocite{*}
\bibliographystyle{auto_generated}
\bibliography{Maciej_Lewicki-Low-x_2021_proceedings/Low-x_2021_proceedings/Lewicki}

%% file: jamal/jamal.tex
\vspace*{1.2cm}

\thispagestyle{empty}
\begin{center}
{\LARGE \bf From small to large $x$: toward a unified formalism for particle production in high energy collisions}

\par\vspace*{7mm}\par

{

\bigskip

\large \bf Jamal Jalilian-Marian}

\bigskip

{\large \bf  E-Mail: jamal.jalilian-marian@baruch.cuny.edu}

\bigskip

{Natural Sciences Department, Baruch College, the City University of New York 
\\
and 
\\
City University of New York Graduate Center, New York, NY}

\bigskip

{\it Presented online, Low-$x$ Workshop, Elba Island, Italy, September 27--October 1 2021}

\vspace*{15mm}

\end{center}
\vspace*{1mm}

\begin{abstract}
\noindent We propose and develop a new formalism which aims to unify the two main approaches to particle production in high energy QCD; that of collinear factorization applicable at high $Q^2$ with the Color Glass Condensate approach applicable at small $x$ (large rapidity). We use the new formalism to calculate double inclusive production of a photon and a quark (jet or hadron) in high energy proton-nucleus collisions. Latest progress in this direction is reported. 
\end{abstract}
  \part[From small to large $x$: toward a unified formalism for particle production in high energy collisionss\\ \phantom{x}\hspace{4ex}\it{Jamal Jalilian-Marian}]{}
\section{Introduction}
Collinear factorization formalism in perturbative QCD (pQCD) has been an extremely useful 
approach to particle production in high energy hadronic collisions. Very roughly and as applied to single inclusive hadron production it states that single inclusive hadron production in a proton-proton 
collision can be written as a convolution (in $x$) of three independent parts; parton distribution functions of the incoming hadrons, parton-parton scattering cross section, and parton-hadron fragmentation function. It guarantees a clean separation of short distance, perturbative physics 
from that of long-distance non-perturbative physics. The short-distance part, parton-parton scattering cross section is process dependent but can be calculated in perturbation theory, in principle, to
any order desired. The non-perturbative parts, parton distribution and fragmentation functions 
however are not amenable to weak-coupling methods but are universal, i.e. process independent, 
on which the predictive power of the approach depends. Despite its enormous success in predicting particle production yields in high energy collisions, collinear factorization formalism has severe 
limitations, mainly, it is applicable at asymptotically high  $Q^2 \rightarrow \infty$ limit 
dominated by leading twist operators. At any finite $Q^2$ there are corrections to this leading twist approximation which can be large, and worse, break collinear factorization. It is sobering to realize that high $Q^2$ processes occupy a very tiny corner of the QCD phase space and that particle production is dominated by low momentum processes due to the fast (power-like) fall off of the differential production cross section with transverse momentum $p_t$ of the produced particle.  Recalling that transverse momentum and rapidity of a produced particle (and the center of mass energy of the collision) determine the momentum fractions $x_1, x_2$ of the partons inside the projectile and target hadrons participating in a collision as in
\bejamal
x_{1,2} =  {p_t \over \sqrt{s}}\, e^{\pm y}
\eejamal
it becomes clear that processes with low momentum particles produced probe the small $x$ partons 
of the projectile and or target hadron/nucleus at high center of mass energy. Therefore as one increases the center of mass energy of a collison, for example proton-proton scattering at RHIC vs the LHC) at low to intermediate momenta one probes smaller and smaller $x$ region of a hadron/nucleus wave function. 
It is an experimental fact that parton (specially gluon) distribution functions grow very fast with decreasing $x$ due to the large radiation phase space ${d x \over x}$ becoming available for soft (carrying small longitudinal momentum fraction $x$) radiation. Therefore one expects that at very small values of $x$ a hadron or nucleus will be a state with a very large number of small $x$ gluons in it. Such a state is referred to as a Color Glass Condensate (CGC) due to the large gluon occupation number of the state and the fact that gluons are colored. Glass refers to the strikingly different time scales involved in dynamics of small vs large $x$ partons \cite{cgc-reviews,cgc-reviews2,cgc-reviews3,cgc-reviews4}. 

The large parton densities at small $x$ renders collinear factorization formalism useless as now one has many gluons that participate in a collision. In the language of operator product expansion this means that operators of all twist contribute comparably to the cross section and a twist expansion upon which the collinear factorization formalism is based is not valid. Therefore a new formalism that takes into account high gluon densities in high energy hadrons/nuclei is needed. Such a formalism was proposed by McLerran and Venugopalan who realized that high gluon densities lead to emergence a new semi-hard scale, called saturation scale $Q_s^2 \gg \Lambda_{QCD}$ which allows a weak-coupling yet non-perturbative approach to gluon saturation. Their formalism is a semi-classsical approach in which the high gluon occupancy state at small $x$ is described as a classical field radiated coherently by the large $x$ partons, collectively treated as sources of color charge $\rho$.  Distribution of the color charges at large $x$ is given by a weight functional and is non-perturbative and therefore modeled. However its dependence on (evolution with) $x$ can be calculated in perturbation theory. Due to the small $x$ kinematics the coupling of the color charges $\rho$ at large $x$ and the small $x$ gluons is taken to be eikonal which allows a re-summation of multiple scatterings, as required by the presence of large number of gluons in the projectile/target, into Wilson lines; path-ordered exponentials of gluon field along the direction of propagation. Quantum loop effects are included via a Wilsonian renrmalization group equation (RGE) that describes the dependence (evolution) of multi-point correlators of Wilson lines with $x$ (or equivalently, rapidity or energy). This Wilsonian RGE describing the evolution of the weight functional with $x$, or rapidity is the so-called JIMWLK evolution equation~\cite{jimwlk,jimwlk2,jimwlk3,jimwlk4,jimwlk5,jimwlk6,jimwlk7,jimwlk8,jimwlk9,jimwlk10}. It can be used to derive evolution equations for correlators of any number of Wilson lines which form the building blocks of the observables in this semi-classical approach. The JIMWLK equation is a functional renormalization group equation which reduces to a closed-form evolution equation for the two-point
function to the BK equation \cite{bk,bk2}.

While the semi-classical formalism of CGC is a well-motivated first principle approximation to QCD at small $x$ and has been very successful when applied to phenomenology it has some serious shortcomings. It is assumed that the gluon distribution function grows so fast that production cross sections are dominated by the lowest kinematic value of $x$ accessible called $x_{min}$. This ignores the contribution of all gluons with momentum fraction $x \ge x_{min}$~\cite{gsv}. Whereas this may not be important when one makes parametric estimates of physical observables, it is essential to go beyond this rather drastic approximation when precision calculations are required. 
As the dream of an Electron-Ion Collider comes closer to becoming reality it is important to develop a more general formalism that can encompass the the two main approaches to particle production in high energy collisions so that one can continuously map the QCD dynamics from small to large $x$ and from low to high transverse momenta.

\section{A new formalism}
In the Color Glass Condensate approach to particle production in the so called dilute-dense kinematics the first step is to consider scattering of a parton, a quark or gluon, from the classical color field of a target \cite{hybrid,hybrid2,hybrid3,hybrid4,hybrid5,hybrid6,hybrid7,hybrid8}. Due to the high gluon density of small $x$ gluons in the target one needs to resum multiple scatterings of the projectile from the target. This is possible only in the eikonal approximation, i.e. infinite energy limit \cite{eikonal,eikonal2}. In this approximation one ignores the deflection of the projectile parton so that it stays on a straight line. In momentum space this corresponds to the final state projectile having transverse momentum much less than its longitudinal momentum which is $O (\sqrt{s})$. In this limit the exchanged momenta are transverse only and no longitudinal momentum is exchanged. It is possible to include non-eikonal contributions as power suppressed corrections to the first order or two~\cite{aaa,aaa2,aaa3,aaa4,aaa5,kps} which extends the validity of the CGC approach to higher $p_t$ but not to the pQCD region. To go beyond the small $x$ approximation which necessarily limits the formalism to low $p_t$ one must therefore consider and allow exchange of large longitudinal momenta. This is indeed what happens in the collinear factorization approach where the large longitudinal momenta of the incoming partons are converted to transverse momenta of the outgoing partons. This however can not happen in the CGC approach since the small $x$ gluons do not carry large longitudinal momenta. 

To do this one must therefore include large $x$ gluons (and quarks in general) in the formalism. As a first step toward a particle production formalism valid at all $x$ and $p_t$ we consider generalizing the CGC approach valid at small $x$ to include the intermediate/large $x$ gluons of the target proton or nucleus from which a projectile parton can scatter \cite{jjm-largex,jjm-largex2}. We will therefore keep the kinematics of scattering from the large $x$ gluons of the target exact while treating scattering from the small $x$ gluons in the eikonal approximation. This allows us to resum multiple soft scatterings as in the CGC formalism while making a connection to pQCD and collinear factorization via inclusion of large $x$ gluons. Therefore in this new formalism, the first step is to consider multiple soft scatterings of a projectile from small $x$ gluons and one hard scattering from large $x$ gluons of the target. This is the analogue of a tree level pQCD and a classical CGC calculation. The next step then would be to do a one-loop correction to this leading order result. A one-loop calculation would then allow one to investigate the divergences that routinely appear in such calculations and to check whether they can be canceled or absorbed into physical quantities leading to their renormalization (evolution). 

\begin{figure}
\begin{center}
\epsfig{figure=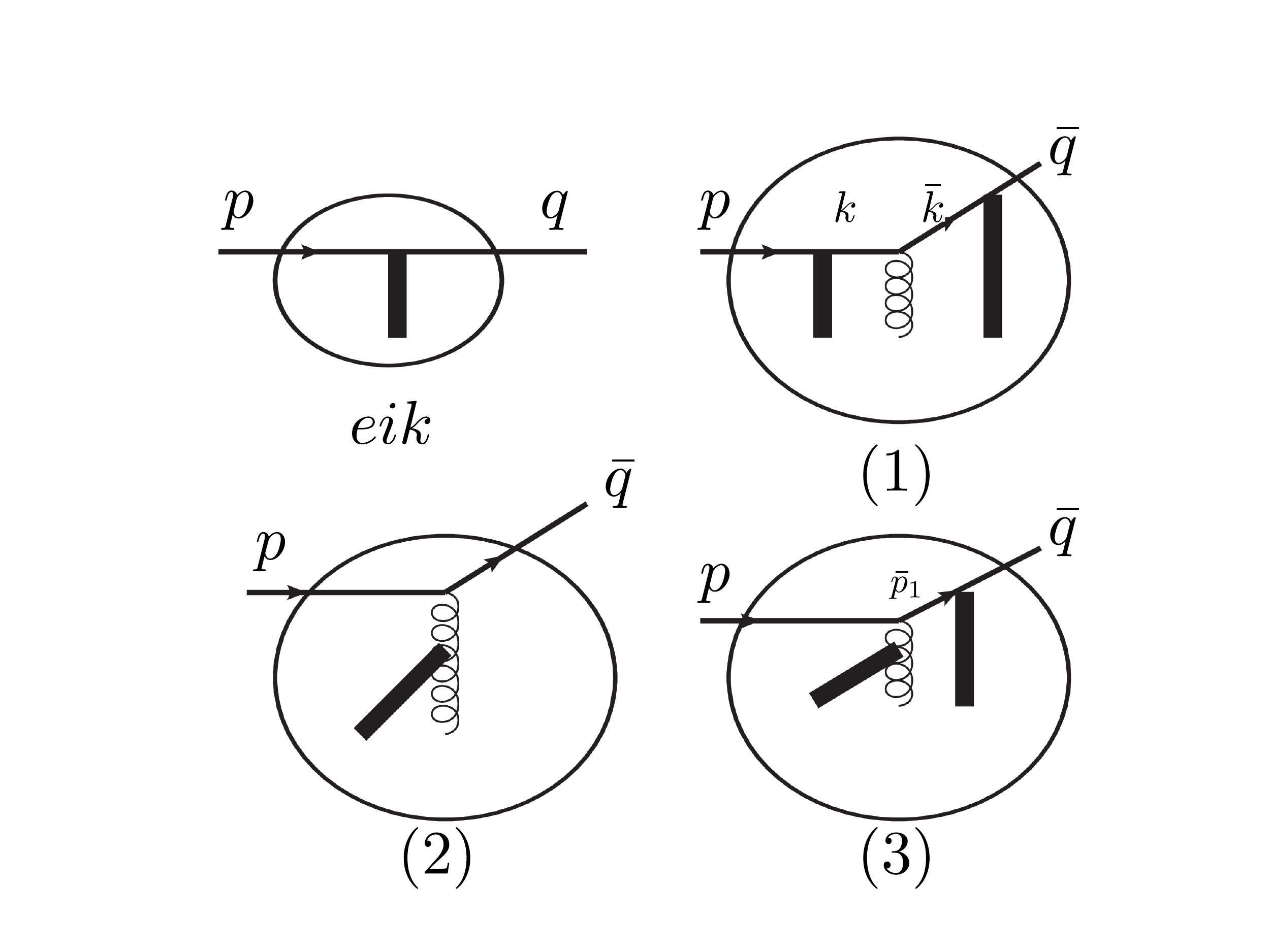,height=0.45\textwidth}
\caption{Scattering of a quark from the small and large $x$ gluons of a target.}
\label{fig:qA}
\end{center}
\end{figure}

We therefore start by considering scattering of a projectile quark from a target including both small $x$ and large $x$ gluons of the target. The amplitude for the scattering is shown in Fig. (\ref{fig:qA}) where the ellipse denotes the target and 
$p$ and $\bar{q}$ are the momenta of the incoming and outgoing projectile parton.  The thick solid line denotes Wilson lines which encode multiple soft scatterings of the projectile from the small $x$ gluons of the target. The wavy line represents single scattering from the large $x$ gluons of the target. The top left diagram represents the standard eikonal scattering while the top right diagram corresponds to having multiple soft scatterings from the target, then a single scattering from the large $x$ gluons, and then more soft scatterings from the small $x$ gluons of the target. The large $x$ gluons can themselves interact from the small $x$ gluons as in the bottom two diagrams.

The result for the full scattering amplitude at all $x$ (or any $p_t$) can thus be written 
as \cite{jjm-largex,jjm-largex2}
\bejamal 
 i \mathcal{M} = i \mathcal{M}_{eik} + i \mathcal{M}_1 + 
i \mathcal{M}_2 +  i \mathcal{M}_3
\label{eq:qA-lo}
\eejamal
where  $i \mathcal{M}_{eik}$ given by eq.~(\ref{eq:eik}) is the amplitude in the eikonal approximation used in CGC formalism while amplitudes $i \mathcal{M}_1$,  $i \mathcal{M}_2$ and  $i \mathcal{M}_3$ correspond to scattering from both small and large $x$ gluons of the target and
are given by eqs.~(\ref{eq:nsoft-hard-nsoft},\ref{eq:nsoft-on-hard},\ref{eq:nsoft-on-hard-nsoft-on-final}) respectively. 

\bejamal
i \mathcal{M}_{eik} (p,q) = 2 \pi \delta (p^+ - q^+)\,  
\ubar (q)\, \sln\, \int d^2 x_{t}\, e^{- i (q_t - p_t) \cdot x_{t}} \, 
\left[V (x_t) - 1\right]\, u(p)
\label{eq:eik}
\eejamal
where the infinite Wilson line $V (x_t)$ is defined as
\bejamal
V (x_t) \equiv \hat{P}\, 
\exp \left\{i g \int_{- \infty}^{+\infty} d x^+ \, S^-_a (x^+, x_t)\, t_a\right\}
\eejamal 

\beajamal
i \mathcal{M}_1 &=&  \int d^4 x\, d^2 z_t \, d^2 \bar{z}_t \, 
\int {d^2 k_t \over (2 \pi)^2} \, {d^2 \bar{k}_t \over (2 \pi)^2} \, 
e^{i (\bar{k} - k) x} \, 
e^{- i (\bar{q}_t - \bar{k}_t)\cdot \bar{z}_t}\, 
e^{- i (k_t - p_t)\cdot z_t}
 \nnjamal
&&
\ubar (\bar{q})\, \left[ 
\overline{V}_{AP} (x^+, \bar{z}_t) \, \slnbar \, {\slkbar \over 2 \bar{k}^+} \,
\left[ i g \, \slA (x)\right]\, 
{\slk \over 2 k^+} \, \sln \, V_{AP} (z_t, x^+)  
\right]\, u(p)
\label{eq:nsoft-hard-nsoft}
\eeajamal
with $k^+ = p^+, k^- = {k_t^2 \over 2 k^+}$, 
$\bar{k}^+ = \bar{q}^+, \bar{k}^- = {\bar{k}_t^2 \over 2 \bar{k}^+}$ 
and the semi-infinite, anti path-ordered Wilson lines in the fundamental representation are now defined 
as    
\bejamal
\overline{V}_{AP} (x^+, \bar{z}_t) \equiv \hat{P}\, 
\exp \left\{i g \int_{x^+}^{+\infty} d \bar{z}^+ \, \bar{S}^-_a 
(\bar{z}_t, \bar{z}^+)\, t_a\right\}
\label{eq:Wilsonbar-si-fundamental}
\eejamal
and
\bejamal
V_{AP} (z_t, x^+) \equiv \hat{P}\, 
\exp \left\{i g \int_{- \infty}^{x^+} d z^+ \, S^-_a (z_t, z^+)\, t_a\right\}.
\label{eq:Wilson-si-fundamental}
\eejamal 
where anti path-ordering (AP) in the amplitude means fields with the largest argument appear to the left.   

\beajamal
i \mathcal{M}_2 &=& {2\, i \over (p - \bar{q})^2} \, \int d^4 x\, 
e^{i (\bar{q} - p) x} \, 
\ubar (\bar{q})\, \bigg[ (i g \, t^a)\, 
\left[\partial_{x^+}\, U_{AP}^\dagger (x_t, x^+)\right]^{a b} \nonumber\\
&&
\left[n \cdot (p - \bar{q}) \, \slA_b (x) -  (p - \bar{q}) \cdot A_b (x) \, \sln\right]\, 
\bigg]\, u(p)
\label{eq:nsoft-on-hard}
\eeajamal
where the derivative acts on the $+$ coordinate of the adjoint Wilson line (anti path-ordered in the amplitude) 
and arises from the fact that
one can write the soft field at the last scattering point as a derivative on the Wilson line. 

\beajamal
i \mathcal{M}_3 &\!\!=\!\!& - 2\, i\, \int d^4 x\,  d^2 \bar{x}_t \, d \bar{x}^+\,  
 {d^2 \bar{p}_{1 t} \over (2 \pi)^2} \, 
e^{i (\bar{q}^+ - p^+) x^-} \, 
e^{- i (\bar{p}_{1 t} - p_t)\cdot x_t}\, 
e^{- i (\bar{q}_t - \bar{p}_{1 t})\cdot \bar{x}_t}
 \nnjamal
&&
\ubar (\bar{q})\, \bigg[
\left[\partial_{\bar{x}^+}\, \overline{V}_{AP} (\bar{x}^+, \bar{x}_t)\right]\, 
 \slnbar\, \slpbarone \,
(i g t^a)\, 
\left[\partial_{x^+}\, U^\dagger_{AP} (x_t, x^+)\right]^{a b}\nnjamal
&&
{
\left[ n \cdot (p - \bar{q}) \, \slA^b (x) -  (p - \bar{p}_1) \cdot A^b (x) \, \sln\right]
\over
\left[
2 n \cdot \bar{q} \, 2 n \cdot (p - \bar{q})\, p^- 
- 2 n \cdot (p - \bar{q})\, \bar{p}^2_{1 t} 
- 2 n \cdot \bar{q} \, (\bar{p}_{1 t} - p_t)^2
\right]
}
\bigg]\, u (p)
\label{eq:nsoft-on-hard-nsoft-on-final}
\eeajamal
The Leading Order quark-target scattering cross section can be obtained by squaring the 
amplitude in~(\ref{eq:qA-lo}) keeping in mind that there is no overlap between the eikonal 
and non-eikonal amplitudes as the non-eikonal amplitudes vanish in the eikonal limit by 
construction. 

In order to generalize the CGC formalism to large $x$ one must include the physics of
pQCD as contained in collinear factorization and its DGLAP evolution of parton distribution 
and fragmentation functions~\cite{dglap}. To do this one must do a one-loop correction 
to our one-loop
result which would involve radiation of gluons (and a quark anti-quark pair). One would 
then integrate out one of the final state partons which would lead to divergences. These 
divergences would then be either canceled by counter-terms or absorbed into running 
(evolution) of the ingredients of the cross section such as parton distribution and 
fragmentation functions as well as the $x$ evolution of the operators describing 
interactions with the target and result in a factorized cross section which would 
contain the JIMWLK evolution of the target at small $x$ and DGLAP evolution 
of the target at large $x$. As quarks and gluons interact strongly with the target 
this will be complicated. As a warm up we therefore consider radiation of a photon 
instead which is much simpler since photons do not interact strongly.
Furthermore, double inclusive production of a photon and quark (hadron or jet) is interesting 
on its own as it would allow investigation of two-particle correlations not only in the 
forward-forward rapidity region~\cite{photon,photon2,photon3,photon4} as is the case in the CGC formalism, but also in the 
forward-backward rapidity regions which is outside the domain of applicability of CGC. 

\subsection{Double inclusive quark-photon production}
\begin{figure}
\begin{center}
\epsfig{figure=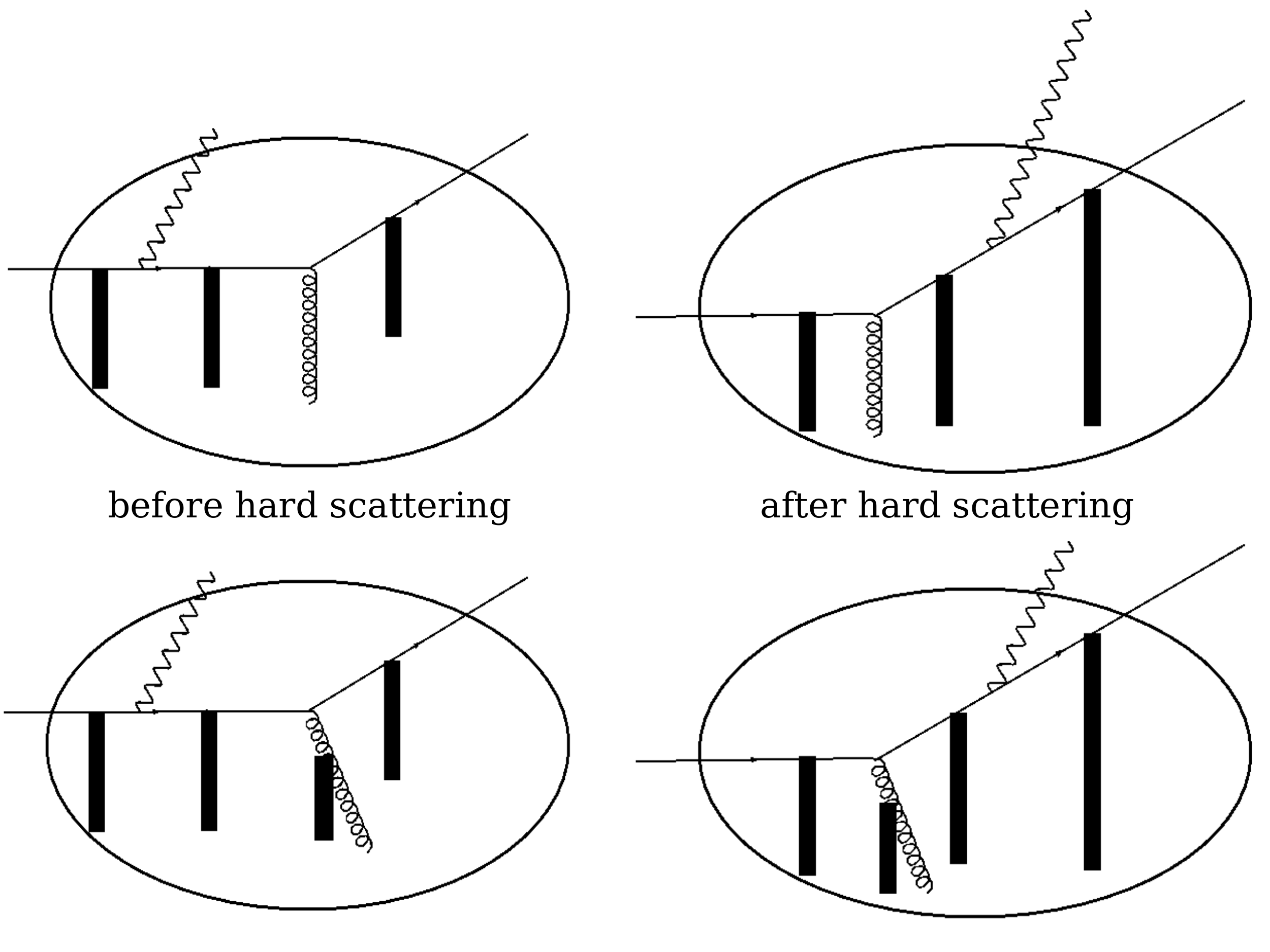,height=0.45\textwidth}
\caption{Radiation of a photon by a quark scattering from the small and large $x$ gluons of a target.}
\label{fig:qA-pho}
\end{center}
\end{figure}

We now consider scattering of a projectile quark from a target accompanied by radiation
of a photon by the quark. The diagrams are shown in Fig. (\ref{fig:qA-pho}) and correspond to
radiation of a photon by quark before, during and after multiple soft and single hard scatterings
of the quark from the target. The wavy line represents a photon while the springy line 
represents a large $x$ gluon (or a gluon exchange with full kinematics more accurately). 
As a reminder, in the eikonal limit there are only two diagrams corresponding to radiation 
of a photon either before or after multiple soft scatterings of the 
quark \cite{photon,photon2,photon3,photon4}. There is no radiation while the quark is traveling inside the target as the target is 
infinitely thin (shock wave) in the eikonal limit. The new diagrams correspond to radiation 
of a photon from a quark while the quark is traversing the target, as such photon radiation 
can happen anywhere. Furthermore, had we considered radiation of a gluon we would have 
needed to include radiation from the large $x$ gluons themselves and include multiple soft 
interactions of the radiated gluon with the target as well. 

Note that the thick solid lines denote Wilson lines which resum multiple soft scatterings of the quark from the target. As a Wilson line also include no-interactions (the $1$ term in the expansion of the exponential) it should be clear that the case of photon radiation before or after any interaction with the target is also included in these diagrams as can be verified by expanding the corresponding Wilson lines. The amplitudes can be written as 
\beajamal
i \mathcal{M}_1 (p,q,l) &=& e g\, \int {d^2 k_{2t} \over (2 \pi)^2} \, 
{d^2 k_{3t} \over (2 \pi)^2} \, 
{d^2 \bar{k}_{1 t} \over (2 \pi)^2} \, 
\int d^4 x\, d^2 y_{1t} \, d^2 y_{2 t}\, 
d^2 \bar{y}_{1 t} \, d z^+ \, \theta (x^+ - z^+) \, 
e^{i (l^+ + \bar{q}^+ - p^+) x^-} \nnjamal
&&
e^{- i (\bar{q}_t - \bar{k}_{1 t})\cdot \bar{y}_{1 t}}\, 
e^{- i (\bar{k}_{1 t} - k_{3t})\cdot x_t}\, 
e^{- i (k_{3 t} - k_{2 t})\cdot y_{2 t}}\,
e^{- i (l_t + k_{2t} - p_t)\cdot y_{1 t}}\,
\ubar (\bar{q})\, \overline{V} (\bar{y}_{1 t} ; x^+ , \infty) \,
{\slnbar \,\slkbarone \over 2 \bar{n}\cdot\bar{q}} \nnjamal
&&
\slA (x)\, 
\left[{\slkthree \over 2 n\cdot (p - l)} \, V (y_{2 t} ; z^+, x^+) \,
{\sln \, \slktwo \over 2 n\cdot (p - l)} \, + 
i \, {\delta (x^+ - z^+) \over 2 n\cdot (p -l)} \sln \right]\nnjamal
&&
\sleps (l) \, {\slkone \over 2 n \cdot p} \, 
V (y_{1 t}; - \infty , z^+)\, \sln \, u (p)
\eeajamal
corresponding to the first diagram at top left where photon radiation happens before hard scattering with large $x$ gluon non scattering. We also use the short-hand notation $k_1^+ = p^+ \, , \, k_{1t} = l_t + k_{2t} \, $,  $k_2^+ = k_3^+ = p^+ - l^+ $ and $\bar{k}_1^+ = \bar{q}^+$. The diagram at the top right corresponds to the case of photon radiation after hard scattering and non-scattering large $x$ gluon. The amplitude for this case is given by 

\beajamal
i \mathcal{M}_2 (p,q,l) &=& eg \int {d^2 k_{1 t} \over (2 \pi)^2} \, 
{d^2 \bar{k}_{1 t} \over (2 \pi)^2} \, 
{d^2 \bar{k}_{2 t} \over (2 \pi)^2} \, 
\int d^4 x\, d^2 y_{1 t} \, d^2 \bar{y}_{1 t}\, 
d^2 \bar{y}_{2 t} \, d \bar{z}^+ \, \theta (\bar{z} - x^+) \, 
e^{i (\bar{l}^+ + \bar{q}^+ - p^+) x^-} \nnjamal
&&
e^{- i (\bar{q}_t + \bar{l}_t - \bar{k}_{2 t})\cdot \bar{y}_{2 t}}\, 
e^{- i (\bar{k}_{2 t} - \bar{k}_{1 t})\cdot \bar{y}_{1 t} }\, 
e^{- i (k_{1 t} - p_{t})\cdot y_{1 t} }\,
e^{- i (\bar{k}_{1 t} - k_{1 t})\cdot x_t}\,
\ubar (\bar{q})\, \overline{V} (\bar{y}_{2 t} ; \bar{z}^+ , \infty) \nnjamal
&&
{\slnbar \, \slkbarthree \over 2 \bar{n} \cdot \bar{q}} \, 
\sleps (\bar{l})\,  
\left[{\slkbartwo \over 2 \bar{n} \cdot \bar{k}_2} \, 
\overline{V} (\bar{y}_{1 t} ; x^+, \bar{z}^+) \,
{\slnbar \, \slkbarone \over 2 \bar{n} \cdot \bar{k}_1} \, + 
i \, {\delta (x^+ - \bar{z}^+) \over 2 \bar{n} \cdot (\bar{q} + \bar{l}) }
\slnbar \right] 
\nnjamal
&&
\slA (x) \, 
{\slkone \over 2 n \cdot p} \, V ( y_{1 t} ; - \infty , x^+) \, 
\sln \, u (p) 
\eeajamal
with $k_1^+ = p^+ \, $, 
$\bar{k}_1^+ =  \bar{k}_2^+ = \bar{q}^+ + \bar{l}^+$, 
 $\bar{k}_3^+ = \bar{q}^+$ and 
$ \bar{k}_{3 t} = \bar{k}_{2 t} - \bar{l}_t$.

The bottom diagrams correspond to the case when the large $x$ gluon scatters from the small $x$ gluons of the target itself. Again there are two possibilities; the photon can be radiate before or after scattering from the large $x$ gluons. The amplitudes for these cases are given by
\beajamal
i \mathcal{M}_3 (p,q,l)\!\!\! &=&\!\!\!  e g  \, 
\int {d^2 k_{1 t} \over (2 \pi)^2} \, 
{d^2 k_{3 t} \over (2 \pi)^2} \, 
{d^2 \bar{k}_{1 t} \over (2 \pi)^2} \, 
\int \, d^4 x\, d^2 y_{1 t} \, d^2 y_{2 t}\, d^2 \bar{y}_{1 t}\, 
d z^+ \, d r^+ \, \theta (x^+ - r^+) \,\theta (r^+ - z^+) \nnjamal
\!\!\! &&\!\!\!
e^{i (\bar{q}^+  + l^+  - p^+) r^-} 
e^{- i (\bar{q}_t - \bar{k}_{1 t})\cdot \bar{y}_{1 t} } 
e^{- i (\bar{k}_{1 t} - k_{3 t})\cdot x_t}
e^{- i (k_{3 t} + l_t - k_{1 t} ) \cdot y_{2 t} }
e^{- i (k_{1 t} - p_t) \cdot y_{1 t}  }
\nnjamal
\!\!\! &&\!\!\!
\ubar (\bar{q})\, \overline{V} (\bar{y}_{1 t} ; r^+ , + \infty) \,
{\slnbar \,\slkbarone \over 2 \bar{n}\cdot\bar{k}_1 }
t^a \, 
\left[\partial_{x^+} \, U^\dagger_{AP} (x_t ; r^+ , x^+) \right]^{a b}\,
\left[\slA^b (x) -  
{(k_3 - \bar{k}_1) \cdot A^b (x) \over n \cdot (k_3 - \bar{k}_1)} \, 
\sln\right] \nnjamal
\!\!\! &&\!\!\!
\left[
{\slkthree \,\sln \, \slktwo \over 2 n\cdot k_3 \, 2 n\cdot k_2 } \, 
V (y_{2 t} ; z^+ , r^+) \, + 
i {\delta (z^+ - r^+)  \over 2 n\cdot k_2} \sln  \right]\,
{\sleps (l) \, \slkone \, \sln \over 2 n\cdot k_1 } \, V (y_{1 t} ; - \infty , z^+) \, u (p)
\eeajamal
with $n\cdot k_1 = n \cdot p$, 
$n \cdot k_2 = n \cdot k_3 = n\cdot (p - l)$, 
$\bar{n} \cdot \bar{k}_1 = \bar{n} \cdot \bar{q}$ and 
$k_{2 t} = k_{1 t} - l_t$
and
\beajamal
i \mathcal{M}_4 (p,q,l)\!\!\! &=&\!\!\!  e g  \, 
\int {d^2 k_{1t} \over (2 \pi)^2} \, 
{d^2 \bar{k}_{1t} \over (2 \pi)^2} \, 
{d^2 \bar{k}_{2t} \over (2 \pi)^2} \, 
\int \, d^4 x\, d^2 y_{1 t} \, d^2 \bar{y}_{1 t}\, d^2 \bar{y}_{2 t}\,  
d z^+ \, d r^+ \, 
\theta (\bar{z}^+ - r^+) \,\theta (x^+ - r^+) \nnjamal
\!\!\! &&\!\!\!
e^{i (\bar{q}^+  + \bar{l}^+  - p^+) r^-} 
e^{- i (\bar{q}_t + \bar{l}_t - \bar{k}_{2 t} ) \cdot \bar{y}_{2 t} }
e^{- i (\bar{k}_{2 t} - \bar{k}_{1 t})\cdot \bar{y}_{1 t} }
e^{- i (\bar{k}_{1 t} - k_{1 t})\cdot x_t}
e^{- i (k_{1 t} - p_t) \cdot y_{1 t}  }
\nnjamal
\!\!\! &&\!\!\!
\ubar (\bar{q})\, \overline{V} (\bar{y}_{2 t} ; \bar{z}^+ , + \infty) \,
{\slnbar \,\slkbarthree \, \sleps (\bar{l}) \over 2 \bar{n}\cdot\bar{k}_3 } 
\left[
{\slkbartwo \,\slnbar \, \slkbarone \over 2 \bar{n}\cdot \bar{k}_2 \, 2 \bar{n}\cdot \bar{k}_1 } \, 
\overline{V} (\bar{y}_{1 t} ; r^+ , \bar{z}^+) \, + 
i {\delta (\bar{z}^+ - r^+)  \over 2 \bar{n}\cdot \bar{k}_2} \slnbar  \right]
\nnjamal
\!\!\! &&\!\!\!
t^a  \left[\partial_{x^+} \, U^\dagger_{AP} (x_t ; r^+ , x^+) \right]^{a b}\,
\left[\slA^b (x) -  
{(k_1 - \bar{k}_1) \cdot A^b (x) \over n \cdot (k_1 - \bar{k}_1) } \, 
\sln\right] \nnjamal
\!\!\! &&\!\!\!
{\slkone \, \sln \over 2 n\cdot k_1 } \, V (y_{1 t} ; - \infty , r^+) \, u (p)
\eeajamal
where all $4$-momenta are on-shell with $n\cdot k_1 = n \cdot p$, 
$\bar{n} \cdot \bar{k}_1 = \bar{n} \cdot \bar{k}_2 = \bar{n} \cdot (\bar{q} + \bar{l})$, $\bar{n}\cdot\bar{k}_3 = \bar{n}\cdot\bar{q}$ and 
$\bar{k}_{3 t} = \bar{k}_{2 t} - \bar{l}_t$.

The next step in calculating the double inclusive production cross section would be to square this amplitude (sum of the $4$ amplitudes). Rather than doing this we use spinor helicity techniques~\cite{dixon} to
evaluate these amplitudes individually for a given helicity state. The advantage of this is that it allows us to consider spin asymmetries in the double photon-hadron production process which we intend to do in the future. For the details of the use of spinor helicity techniques in the CGC formalism we refer to~\cite{ahjt,ahjt2}.

Here we show our results for the helicity amplitudes for the first diagram. The other ones are similar and will be reported elsewhere. The Dirac numerators for the first diagram are defined as
\beajamal
\mathcal{N}_{1-1} &=&
\ubar (\bar{q})\, {\slnbar \,\slkbarone \over 2 \bar{n}\cdot\bar{q}} \, \,
\slA (x)\, 
{\slkthree \, \sln \, \slktwo \sleps (l) \, \slkone  \, \sln \over 
2 n \cdot p \, 2 n\cdot (p - l)\, 2 n\cdot (p - l)}  
\, u (p)\nnjamal
\mathcal{N}_{1-2} &=& 
\ubar (\bar{q})\, {\slnbar \,\slkbarone \over 2 \bar{n}\cdot\bar{q}} \, \, 
\slA (x)\, 
{ \sln \, \sleps (l) \, \slkone \, \sln \over 
2 n \cdot p \, 2 n\cdot (p -l)} \, u (p)
\eeajamal

\noindent and the corresponding helicity amplitudes are 
\beajamal
\mathcal{N}^{+ +}_{1-1} &=& \left(\mathcal{N}^{- -}_{1-1}\right)^\star = 
- \sqrt{{n\cdot p \over n\cdot (p - l)}} \, 
{\left[n\cdot l \, k_{2\perp}\cdot\epsilon_\perp^\star - 
n\cdot (p - l) \, l_\perp\cdot\epsilon_\perp^\star\right]
 \over n\cdot l \, n\cdot (p -l)}  \,
\langle \bar{k}_1^+ | \, \slA (x) | k_3^+ \rangle \nnjamal
\mathcal{N}^{+ +}_{1-2} &=& \left(\mathcal{N}^{- -}_{1-2}\right)^\star = 
- \sqrt{{n\cdot p \over n\cdot (p - l)}} \,
\langle \bar{k}_1^+ | \, \slA (x) | n^+ \rangle \nnjamal
\mathcal{N}^{+ -}_{1-1} &=& \left(\mathcal{N}^{- +}_{1-1}\right)^\star = 
- \sqrt{{n\cdot p \over n\cdot (p - l)}} \, 
{\left[n\cdot p \, l_\perp\cdot\epsilon_\perp - 
n\cdot l \, k_{1\perp}\cdot\epsilon_\perp\right]
 \over n\cdot p \, n\cdot l}  \,
\langle \bar{k}_1^+ | \, \slA (x) | k_3^+ \rangle \nnjamal
\mathcal{N}^{+ -}_{1-2} &=& \mathcal{N}^{- +}_{1-2} = 0
\eeajamal
where $\{\pm \pm\}$ refers to the helicity of the projectile quark and the real (transverse) photon.  
Using these helicity amplitudes it is clear that there will be a spin asymmetry in the cross section, the so called $A_{LL}$. 

There is much to gain from a formalism that can unify the pQCD and collinear factorization formalism with that valid at high $p_t$ (large $x$) with that of CGC applicable at low $p_t$ (small $x$). In addition to clarifying the kinematics where the CGC formalism is valid it will enable us to study spin asymmetries at intermediate-high $p_t$, forward-backward rapidity and azimuthal angular correlations and much more. It in principle can also be used to investigate jet energy loss from the earliest moments of a high energy heavy ion collision as well as the total neutrino-nucleon scattering cross sections~\cite{hj}.

Comments: Presented online at the Low-$x$ Workshop, Elba Island, Italy, September 27--October 1 2021.
 
\section*{Acknowledgements}

We acknowledge support by the DOE Office of Nuclear Physics
through Grant No.\ DE-SC0002307 and by PSC-CUNY through 
grant No. 63158-0051. 

\nocite{*}
\bibliographystyle{auto_generated}
\bibliography{jamal/jamal}


%% file: Low_X_proceedingsDMelnikov/Low_X_proceedingsDMelnikov.tex
\vspace*{1.2cm}

\thispagestyle{empty}
\begin{center}
{\LARGE \bf Glueballs. Updates from Lattice and Holography}

\par\vspace*{7mm}\par

{

\bigskip

\large \bf Dmitry Melnikov}

\bigskip

{\large \bf  E-Mail: dmitry.melnikov@ufrn.br}

\bigskip

{International Institute of Physics, Federal University of Rio Grande do Norte, \\ Campus Universit\'ario, Lagoa Nova, Natal-RN  59078-970, Brazil}

\bigskip

{Institute for Theoretical and Experimental Physics, \\ B.~Cheremushkinskaya 25, Moscow 117218, Russia}

\bigskip

{\it Presented at the Low-$x$ Workshop, Elba Island, Italy, September 27--October 1 2021}

\vspace*{15mm}

\end{center}
\vspace*{1mm}
 
\begin{abstract}

Recently there has been interesting progress in the study of glueball states in lattice gauge theories, in particular extrapolation of their spectrum to the limit of large number of colors. This progress coincided with the announcement of the discovery of the odderon in proton-proton and proton-anti-proton collisions, a colorless $C$-odd particle expected to be the $1^{--}$ glueball (more precisely, the corresponding Regge trajectory). In this note we review the holographic predictions for the spectrum of glueballs, focusing on the Klebanov-Strassler model. We demonstrate that thanks to the universal behavior of the glueball spectrum across a range of gauge theories, the holographic model can give reliable predictions for the structure of the spectrum in $SU(N)$ Yang-Mills theories with and without supersymmetry. This is especially important for the supersymmetric theories, for which no lattice predictions exist and the holographic model remains the most tractable approach. For the odderon, both lattice and holography give a consistent prediction of the relative position of the leading poles in the pomeron and odderon trajectories, that is $2^{++}$ and $1^{--}$ mass ratios.
\end{abstract}
 \part[Glueballs. Updates from Lattice and Holography\\ \phantom{x}\hspace{4ex}\it{Dmitry Melnikov}]{}
\vspace{1cm}

In~\cite{TOTEM:2020zzr} the D0 and TOTEM collaborations announced a $3.4\sigma$ divergence of the $pp$ and $p\bar p$ cross sections, compatible with a $t$-channel exchange of a colorless $C$-odd particle, an odderon~\cite{Lukaszuk:1973nt}. Further work is under way to improve the statistics and discover more properties of this particle.

The principle candidate for the odderon is a $C$-odd three-gluon bound state -- a glueball state $1^{--}$ in the $J^{PC}$ classification, and its Regge trajectory. Apart from this progress in hadron scattering the interest to purely gluonic states appeared recently in the context of models of physics beyond the Standard Model, e.g.~\cite{Kang:2008ea,Juknevich:2009ji,Juknevich:2009gg}. Despite the fact that the existence of such bound states was conjectured long ago, their experimental detection, let alone physical properties, turned out to be a very complicated task. The main problem is heavy mixing of such nonquarkyonic and apparently short-lived states with heavy and excited states of ordinary mesons. Most of the presently available information about glueballs comes from lattice simulations. 

Lattice studies focus on the lowest states, kinematically stable in the pure glue theory. The work of Morningstar and Peardon~\cite{Morningstar:1999rf} found 12 such lightest states with spins varying from $J=0$ to $J=3$ in the pure glue $SU(3)$ Yang-Mills theory.\footnote{See~\cite{Michael:1988jr,Michael:1989ry,Bali:1993fb} for yet earlier work on the lattice spectrum.} The ensuing lattice studies improved the original predictions of the glueball masses~\cite{Chen:2005mg,Meyer:2004gx,Athenodorou:2020ani}, observing additional (excited) states, studied the effects of introducing quarks~\cite{Bali:2000vr,Hart:2001fp,Hart:2006ps,Richards:2010ck,Gregory:2012hu,Sun:2017ipk,Brett:2019tzr}, investigated the dependence of the gauge group and its rank~\cite{Teper:1998kw,Lucini:2004my,Lucini:2010nv,Holligan:2019lma,Bennett:2020hqd,Bennett:2020qtj,Athenodorou:2021qvs} and attempted to estimate the decay constants of a few lightest states~\cite{Chen:2005mg}, cf.~\cite{Sexton:1994wg,Sexton:1996ed,Yamanaka:2019yek,Llanes-Estrada:2021evz}. 

One fact about glueballs, corroborated by the lattice analysis, is relatively light dependence of their masses of the number of colors,\footnote{In the $Sp(N_c)$ case the leading correction to the $N_c\to\infty $ value is $O(1/N_c)$.}
\be
\label{MassNc}
m \ \simeq \ m_\infty + \frac{c}{N_c^2}\,. 
\eeq
In table~\ref{tab:SUN} we demonstrate the $SU(N_c)$~\cite{Lucini:2004my} and $Sp(N_c)$~\cite{Bennett:2020qtj} lattice results for the mass of the lightest $0^{++}$ glueball and its ratio with the next in mass $2^{++}$ state. For $SU(N_c)$ the reported variation of the masses is within 14-16\%, while the variation of the ratios is even smaller, about 10\%. In the $Sp(N_c)$ case the variation of the ratio is even smaller, about 5\%.
\begin{table}[h]
    \centering
    \begin{tabular}{||c|cccccc||}
      \hline \Trule\Brule $G$  & $SU(2)$ & $SU(3)$ & $SU(4)$ & $SU(6)$ & $SU(8)$ & $SU(\infty)$ \\
        \hline\Trule\Brule $m_{0^{++}}$  & 3.78 & 3.55 & 3.36 & 3.25 & 3.55 & 3.31 \\ 
        \Trule\Brule$m_{2^{++}}/m_{0^{++}}$ & 1.44 & 1.35 & 1.45 & 1.46 & 1.32 & 1.46 \\
        \hline\hline \Trule\Brule $G$ & $Sp(1)$ & $Sp(2)$ & $Sp(3)$ & $Sp(4)$ & -- & $Sp(\infty)$ \\
        \hline \Trule\Brule $m_{2^{++}}/m_{0^{++}}$ & 1.41 & 1.41 & 1.48 & 1.41 & -- & 1.47 \\ \hline
    \end{tabular}
    \caption{Lattice masses (in appropriate units) of the $0^{++}$ and $2^{++}$ glueballs in the pure glue $SU(N_c)$~\cite{Lucini:2004my} and $Sp(N_c)$ theories~\cite{Bennett:2020qtj}.}
    \label{tab:SUN}
\end{table}

The second observation is a reasonably light effect of quark mixing in the unquenched version of the Yang-Mills theory. The phenomenological OZI rule~\cite{Okubo:1963fa,Zweig:1964jf,Iizuka:1966fk} does not favor quark mixing with purely gluonic states. Lattice simulations somewhat confirm this effect, as can be seen from the data in table~\ref{tab:unquenched}. The match is better for the lighter states (5\% for $0^{++}$) and the ratio $m_{2^{++}}/m_{0^{++}}\sim 1.46$ obtained for QCD with three flavors is compatible with the $SU(N_c)$ and $Sp(N_c)$ results~(universality of this ratio was acknowledged in~\cite{Bennett:2020hqd}). Mass ratios appears to be even more stable with respect to model variation, which might indicate that non only $m_\infty$, but also $c$ in equation~(\ref{MassNc}) is a universal quantity.

\begin{table}[h]
    \centering
    \begin{tabular}{||c|cccc||ccc||}
        \hline \Trule\Brule $J^{PC}$ & $0^{++}$ & $2^{++}$ & $1^{+-}$ & $0^{-+}$ & $m_{2^{++}}/m_{0^{++}}$ & $m_{1^{+-}}/m_{2^{++}}$ & $m_{0^{-+}}/m_{2^{++}}$ \\
        \hline \Trule\Brule  YM & 1710 & 2390 & 2980 & 3640 & 1.40 & 1.25 & 1.52 \\
        \Trule\Brule QCD$_3$ & 1795 & 2620 & 3270 & 4490 & 1.46 & 1.25 & 1.71 \\ \hline
    \end{tabular}
    \caption{Comparison of the lattice masses of glueballs (GeV) in the quenched Yang-Mills and QCD with three flavors~\cite{Gregory:2012hu}. The ratio of the masses of $m_{2^{++}}$ and $m_{0^{++}}$ for the two cases are $1.40$ and $1.46$ respectively.}
    \label{tab:unquenched}
\end{table}

The above features of the glueball spectrum make this data interesting from the point of view of the holographic models. Classical gravity limit of holography~\cite{Maldacena:1997re,Gubser:1998bc,Witten:1998qj} can only capture the limit of large number of colors $N_c\to\infty$ and large coupling constant $\lambda=g_{YM}^2N_c\to\infty$ of Yang-Mills theory. Without implementing quantum corrections to gravity one can only hope accessing some universal properties of hadrons. Lattice simulations indicate that the glueball spectrum is one such property.

Sparing most of the details, the holographic approach in the classical gravity limit consists of choosing a background (solution of an appropriate supergravity theory) that is expected to be equivalent to the gauge theory of interest and calculating the spectrum of linearized equations of motion over the background solution. If the background is truly equivalent to the sought gauge theory in the corresponding limit then the gravity spectrum is equivalent to spectrum of the gauge theory. Higher order correlation functions are also available through the calculations beyond linear order.

Early take on glueballs in holography~\cite{Csaki:1998qr,Brower:2000rp} invoked the so-called Witten's model~\cite{Witten:1998zw}. The background of this model is obtained from a new-horizon limit of the world-volume theory of D4 branes compactified on a circle. This background preserves no supersymmetry and gives a qualitative idea about the gauge theory spectrum, but also exhibits features incompatible with the Yang-Mills theory, such as mass degeneracy for different glueball families, which indicates to symmetries absent in the Yang-Mills. It also shows larger spacing of the mass eigenvalues, as demonstrated in table~\ref{tab:holomodels}. In Witten's model the ratio $m_{2^{++}}/m_{0^{++}}\sim 1.74$ and the discrepancy grows for heavier states.

\begin{table}[h]
    \centering
   \begin{tabular}{||c|ccccc||}
      \hline \Trule\Brule $J^{\,PC}$ & $SU(\infty)$ & $Sp(\infty)$ & BMT & BBC$_D$ & BBC$_N$ \\
      \hline
      \Trule\Brule ${0^{++}}$ & 1 & 1 & 1 & 1 & 1 \\
      \Trule\Brule ${2^{++}}$ & 1.49 & 1.47 & 1.74 & 1.48 & 1.56\\
      \Trule\Brule ${0^{-+}}$ & 1.53 & 1.54 & 2.09 & -- & --\\
      \Trule\Brule $1^{+-}$ & 1.88 & --  & 2.70 & -- & -- \\
      \Trule\Brule $1^{--}$ & 2.32 & --  & 3.37 & -- & -- \\
      \Trule\Brule $0^{+-}$ & 3.01 & --  & -- & -- & -- \\
      \hline\hline
      \Trule\Brule $0^{++\ast}$ & 1.89 & 1.94 & 2.53 & 1.63 & 1.83 \\
      \Trule\Brule $2^{++\ast}$ & 2.11 & -- & 2.76 & 2.15 & 2.49 \\
      \hline
    \end{tabular}	
    \caption{Spectra of the lightest glueballs in the Witten's (BMT~\cite{Brower:2000rp}) and hard wall models (BBC~\cite{Boschi-Filho:2005xct}) compared to the state of the art lattice extrapolations of $m_\infty$ for $SU(N)$~\cite{Athenodorou:2021qvs} and $Sp(N)$~\cite{Bennett:2020qtj}. The masses are normalized to the mass of the $0^{++}$ state. Models BBC$_N$ and BBC$_D$ used different boundary conditions (Dirichlet or Neumann) in the calculation of the spectrum.}
    \label{tab:holomodels}
\end{table}

A simpler approach towards the properties of the holographic hadrons can be taken via the class of bottom-up models. The so-called hard wall model~\cite{Erlich:2005qh,Polchinski:2001tt,Polchinski:2002jw} is its simplest representative. For the purpose of this note, Light-Front Holography models~\cite{Brodsky:2006uqa} can also be attributed to this class. In the hard wall model the background is five-dimensional anti de Sitter space, whose group of symmetries is precisely the conformal group in 3+1 dimensions, and a cutoff is introduced to break this symmetry explicitly. The dual gauge theory to such a background is expected to be approximately conformal (3+1)-dimensional gauge theory at high energies, with an IR scale defining the masses, as in Yang-Mills or QCD. One then considers wave equations of different spin matter in the $AdS_5$ background. This problem leads to the spectrum of the light states shown in table~\ref{tab:holomodels}, as calculated by~\cite{Boschi-Filho:2005xct}. 

Remarkably, the simplest hard wall model predicts the value $m_{2^{++}}/m_{0^{++}}\sim 1.48$, closest to the extrapolated ratio of $m_\infty$ on the lattice. This ``magic number" can be expressed as a ratio of the first non-trivial zeroes $x_{2,1}$ and $x_{4,1}$ of Bessel functions $J_2(x)$ and $J_4(x)$,
\be
\text{hard wall:}\quad \frac{m_{2^{++}}}{m_{0^{++}}} \ = \ \frac{x_{4,1}}{x_{2,1}} \ \simeq \ 1.47759\,.
\eeq 
At the same time, it is not possible to implement states with non-trivial parity and charge conjugation, since the simple model does not have a natural way to implement the corresponding symmetries. In contrast, top-down holographic constructions inherit these symmetries from string theory, as in the example of Witten's model. 

The Klebanov-Strassler model~\cite{Klebanov:2000hb} is another example of a top-down holographic model, which possesses rich physics. It is a model of D3 and D5 branes in type IIB string theory compactified on a six-dimensional cone (conifold). In the low-energy limit the compactification gives rise to a type IIB supergravity background on a space which is a warped product of $AdS_5$ and a pair of spheres $S^3\times S^2$~\cite{Klebanov:1998hh,Klebanov:1999rd,Klebanov:2000nc}. Smoothing out the singularity of the cone (deformation of the conifold) introduces a scale that breaks (3+1)-dimensional conformal invariance and provides an interesting example of a holographic model with a logarithmic running of the coupling constant. The background has $SU(2)\times SU(2)$ global symmetries, so its spectrum is organized in the irreducible representations of this group.

The Klebanov-Strassler background preserves ${\cal N}=1$ supersymmetry, so the dual gauge theory is a non-conformal ${\cal N}=1$ supersymmetric Yang-Mills theory with additional matter fields transforming under the global symmetry group, subject to a unique superpotential. In the IR the theory flows to a strongly coupled fixed point (it shows an interesting RG flow known as ``cascade'' of Seiberg dualities~\cite{Klebanov:2000hb,Strassler:2005qs}) which was initially expected to be in the universality class of the pure supersymmetric Yang-Mills theory. It was then understood that the IR Klebanov-Strassler theory has light operators and massless states due to spontaneous breaking of the baryon $U(1)$ symmetry by the baryonic operators~\cite{Gubser:2004qj,Gubser:2004tf,Benna:2006ib}. 

Despite being a modification of the pure ${\cal N}=1$ Yang-Mills, the Klebanov-Strassler gauge theory inherits many properties of the parent. Apart from the mentioned logarithmic running of the coupling, the holographic theory exhibits the same mechanism of breaking of the $U(1)$ R-symmetry~\cite{Klebanov:2002gr}. The singlet sector mostly contains the super Yang-Mills operators~\cite{Apreda:2003gc,Cassani:2010na}. In view of the universality of the spectrum, one can hope that $SU(2)\times SU(2)$-singlet states can reproduce the appropriate limit of the low-energy spectrum of the supersymmetric, or even pure bosonic Yang-Mills, since little is known about glueballs of the supersymmetric theory. Let us summarize the results of the study of the spectrum of the singlet states in the Klebanov-Strassler theory.

\begin{figure}
    \centering
    \includegraphics[width=0.45\linewidth]{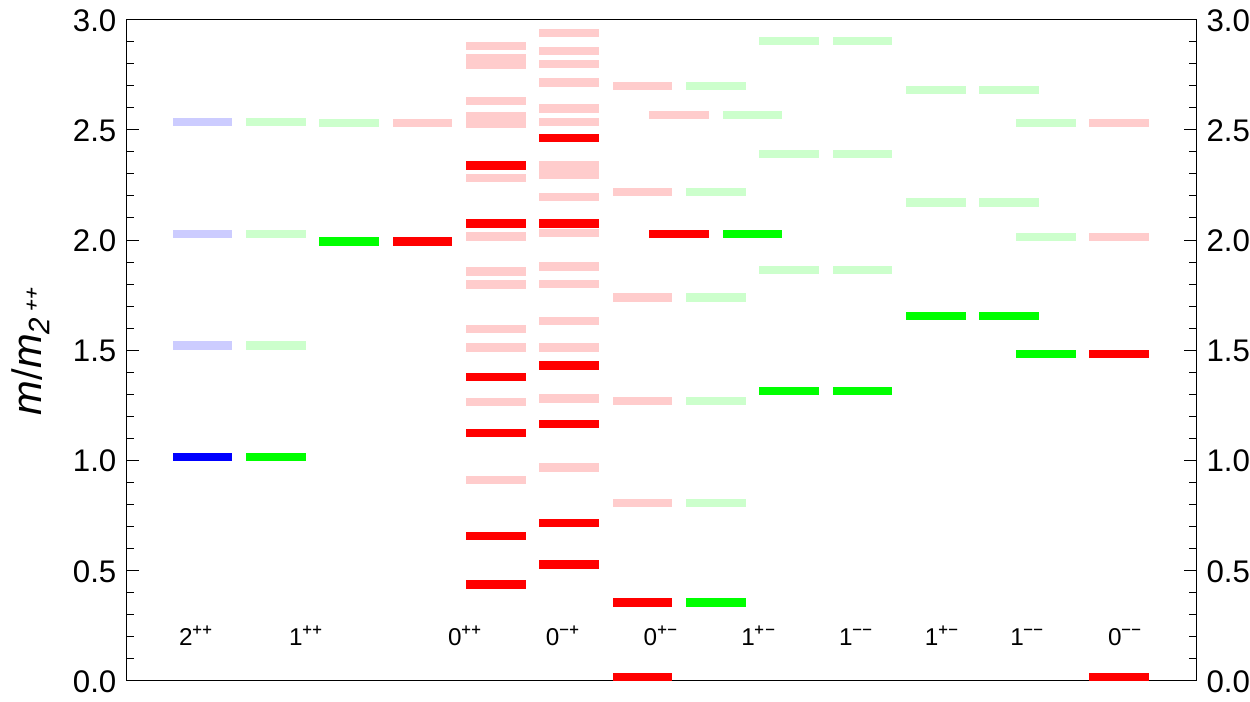} \quad
    \includegraphics[width=0.45\linewidth]{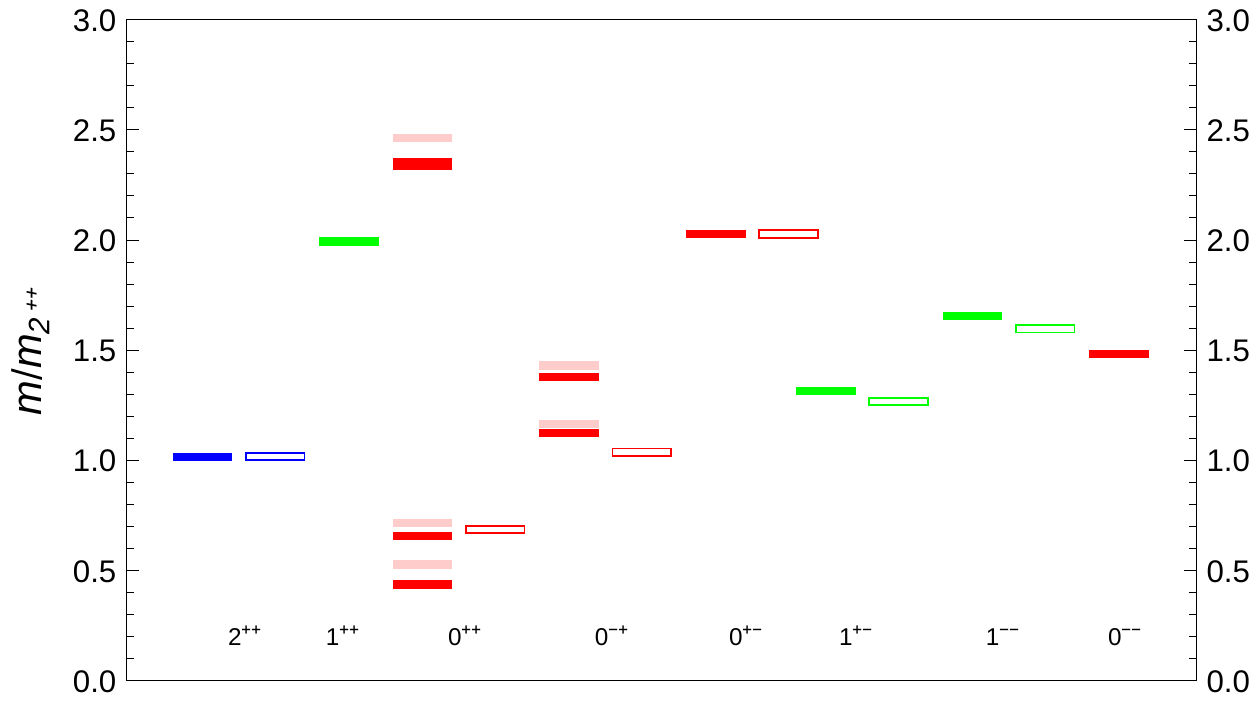}
    \caption{Full light $SU(2)\times SU(2)$-singlet spectrum of the bosonic states in the Klebanov-Strassler theory (left). The ground states of each supermultiplet are shown in full color, while the excited states are faded. All the mass values are computed in the units of the ground state of $2^{++}$. The discrepancy of the $0^{++}$~\cite{Berg:2006xy} and $0^{-+}$~\cite{Melnikov:2020cxe} sectors is visible. The right plot compares the masses of the ground states in the supersymmetric Yang-Mills sector (fill rectangles) and the ground states in $SU(N)$ bosonic Yang-Mills, extrapolated to $N=\infty$ (empty rectangles). In the $0^{++}/0^{-+}$ sector the results of~\cite{Berg:2006xy} are shown in full color, while that of~\cite{Melnikov:2020cxe} are faded. States that do not have pairs correspond to hybrid glueballs with no analog in the bosonic theory, except for the heaviest $0^{++}$ that has not been approached on the lattice.}
    \label{fig:fullspec}
\end{figure}

The spectrum of singlet fluctuations of the Klebanov-Strassler background was studied in a series of papers including~\cite{Amador:2004pz,Berg:2005pd,Dymarsky:2006hn,Berg:2006xy,Dymarsky:2007zs,Benna:2007mb,Dymarsky:2008wd,Gordeli:2009nw,Gordeli:2013jea,Melnikov:2020cxe}. (See also~\cite{Caceres:2005yx,Elander:2009bm,Bianchi:2010cy,Elander:2014ola,Elander:2017cle,Elander:2017hyr} for the studies of non-singlet states or deformations of the Klebanov-Strassler theory.) The structure of the spectrum in the supergravity limit was explained in~\cite{Gordeli:2009nw, Gordeli:2013jea,Melnikov:2020cxe}: it contains two massless scalar supermultiplets and thirteen massive supermultiplets classified as one graviton, two gravitino, four vector and six scalar supermultiplets, according to the highest spin of the contained fields (with the exception of the scalar multiplets that contain scalars and spin half fermions). The massless states are not part of the Yang-Mills spectrum, as well as the associated (light) massive vector supermultiplet, containing a $0^{+-}$ scalar~\cite{Benna:2007mb} and a $1^{+-}$ vector~\cite{Dymarsky:2008wd}. In the scalar sector there is one multiplet containing non-Yang-Mills degrees of freedom. The remaining eleven massive multiplets can be compared with the spectrum of the Yang-Mills. This is done in table~\ref{tab:KSsinglets} and illustrated in figure~\ref{fig:fullspec}.

\begin{table}[h]
    \centering
    \begin{tabular}{||c|c|c|cc|c||}
     \hline \Trule\Brule & $J^{\,PC}$ & Multiplet & $m^{\rm GS}$ & $m^{\ast}$  & Ref.\\
      \hline
      \Trule\Brule 1 & $1^{++},2^{++}$ & graviton & 1 & 1.51  & \cite{Dymarsky:2006hn}\\ \hline
      \Trule\Brule 2 & $1^{+-},1^{--}$ & gravitino & 1.30 & 1.85  & \cite{Dymarsky:2008wd} \\ 
      \Trule\Brule 3 & $1^{+-},1^{--}$ & gravitino & 1.64 & 2.15  & \cite{Dymarsky:2008wd} \\ \hline
      \Trule\Brule 4 & $0^{--},1^{--}$ & vector & 1.47 & 2.00  & \cite{Benna:2007mb} \\ 
       \Trule\Brule 5 & $0^{+-},1^{+-}$ & vector & 2.01 & 2.55  & \cite{Benna:2007mb} \\ \hline
       \Trule\Brule 6 & $0^{++},1^{++}$ & vector & 1.99 & 2.53  & \cite{Gordeli:2009nw} \\ \hline\hline
       \Trule\Brule 7 & $0^{++},0^{-+}$ & scalar & 0.421 & 0.894  & \cite{Berg:2006xy} \\ \hline
       \Trule\Brule 8 & $0^{++},0^{-+}$ & scalar & 0.640 & 1.25  & \cite{Berg:2006xy} \\ 
       \Trule\Brule 9 & $0^{++},0^{-+}$ & scalar & 1.11 & 1.58  & \cite{Berg:2006xy} \\ \hline\hline
        \Trule\Brule 10 & $0^{++},0^{-+}$ & scalar & 1.36 & 1.84  & \cite{Berg:2006xy} \\ 
         \Trule\Brule 11 & $0^{++},0^{-+}$ & scalar & 2.32 & 2.87  & \cite{Berg:2006xy} \\ \hline
    \end{tabular} \quad
    \begin{tabular}{||c|cc||}
     \hline \Trule\Brule $J^{\,PC}$  & $m^{\rm GS}$ & $m^{\ast}$  \\
      \hline
      \Trule\Brule $2^{++}$  & 1 & 1.43  \\ \hline
      \Trule\Brule $1^{+-}$  & 1.25 & 1.62   \\ 
      \Trule\Brule $1^{--}$  & 1.58 & $\geq 1.88$  \\ \hline
      \Trule\Brule    &  &       \\ 
      \Trule\Brule $0^{+-}$  & $\geq2.01$ & --  \\ \hline
      \Trule\Brule    &  &       \\  \hline\hline
      \Trule\Brule   &  &    \\  \hline
       \Trule\Brule $0^{++}$  & 0.668 & 1.27  \\ 
       \Trule\Brule $0^{-+}$  & 1.02 & 1.53  \\ \hline\hline
       \Trule\Brule   &  &    \\  
       \Trule\Brule  $0^{++}$ & -- & --   \\  \hline
    \end{tabular}
    \caption{Spectrum of the lightest supermultiplets of the supersymmetric Yang-Mills sector of the Klebanov-Strassler theory (left table). The bosonic members of the multiplets are indicated. The masses of the ground and first excited states, in units of the mass of the ground state of $2^{++}$, are shown, as well the references to the works, from which the spectra can be extracted. For the scalar sector, the assignment was made based on the analysis of~\cite{Melnikov:2020cxe,RodriguesFilho:2020rae}. The masses are compared with the masses recently obtained for the $SU(\infty)$ bosonic Yang-Mills on the lattice~\cite{Athenodorou:2021qvs} (right table). Lower bounds indicate states with known masses, but with a problem of confirming the $J^{PC}$ numbers in the continuum limit.}
    \label{tab:KSsinglets}
\end{table}

In table~\ref{tab:KSsinglets} we normalize all the masses to the mass of the ground state of the graviton multiplet, that is $2^{++}$ glueball. This state is particularly easy in the supergravity analysis, so it serves a natural reference for the remaining states. The $C$-odd sector is also relatively simple, although it involves some mixing between states with similar quantum numbers. The masses of the $1^{+-}$, $1^{--}$ and $0^{+-}$ glueballs can be compared with the $SU(N)$ bosonic Yang-Mills extrapolated to $N=\infty$~\cite{Athenodorou:2021qvs}. One can observe a very reasonable match of the masses of the ground states. For the excited states the correspondence is not so good: the holographic supersymmetric masses of excited states tend to grow more rapidly. Note that masses of $1^{--\ast}$ and $0^{+-}$ states appear with lower bounds in the $SU(\infty)$ part of the table. The continuum limit of the lattice result for this state does not allow to determine the $J^{PC}$ numbers unambiguously for the states with the indicated mass, although different arguments support the above identification~\cite{Athenodorou:2021qvs}. In case of $0^{+-}$ we can view the match with the holographic results as an additional argument in favor of the identification.

One vector multiplet in the $C$-odd sector contains no bosonic states, so there are no lattice results to compare. The same is true for the only vector multiplet in the $C$-even sector. The latter mode is also relatively simple.

The most complex is the scalar sector, originally studied in~\cite{Berg:2005pd,Berg:2006xy}. There are six scalars-pseudoscalar pairs with similar quantum numbers that mix together. This mix of scalars contains the bosonic $0^{++}$ and $0^{-+}$ modes and scalars dual to the gluino bilinear. The latter fermionic state is expected to be the lightest in the spectrum of the supersymmetric Yang-Mills. Higher gluon operators are also predicted by holography.

\begin{figure}
    \centering
    \includegraphics[width=0.45\linewidth]{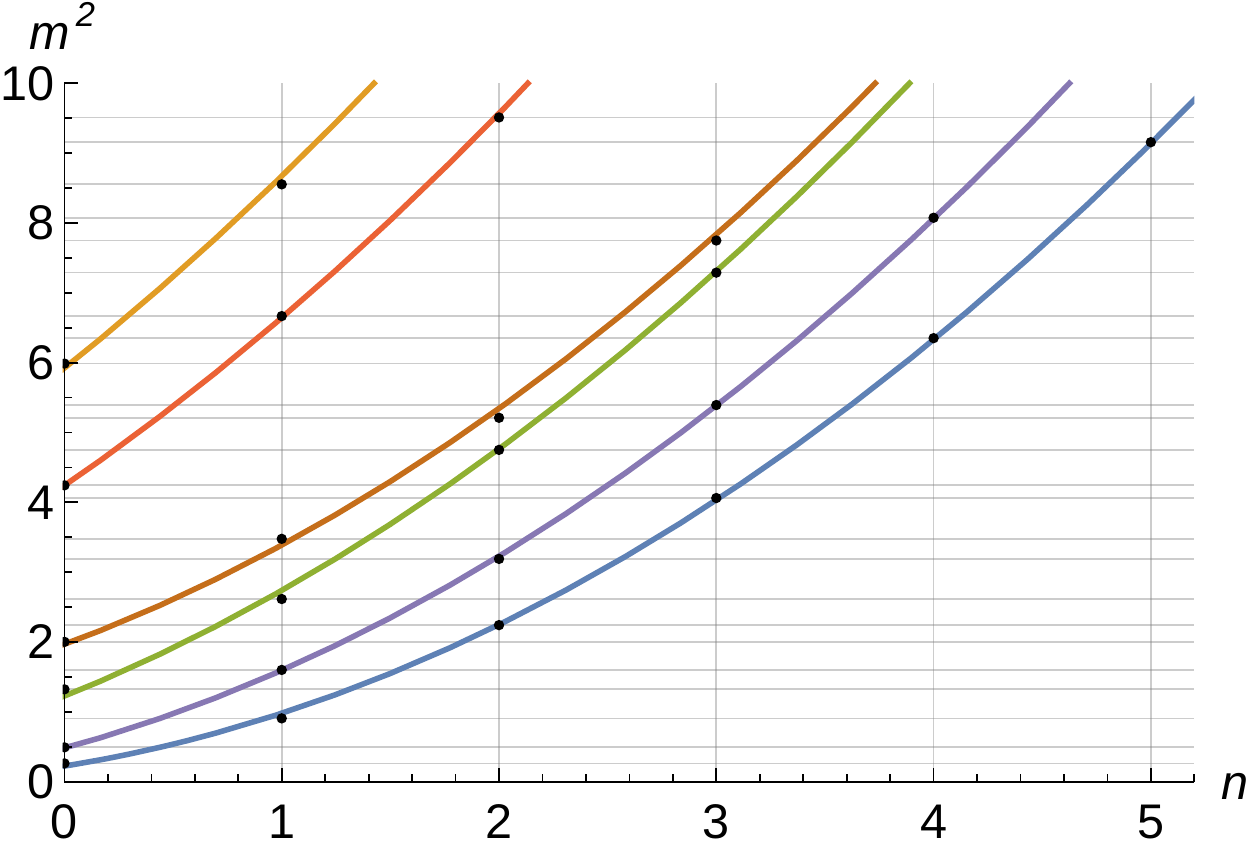}
    \quad
    \includegraphics[width=0.45\linewidth]{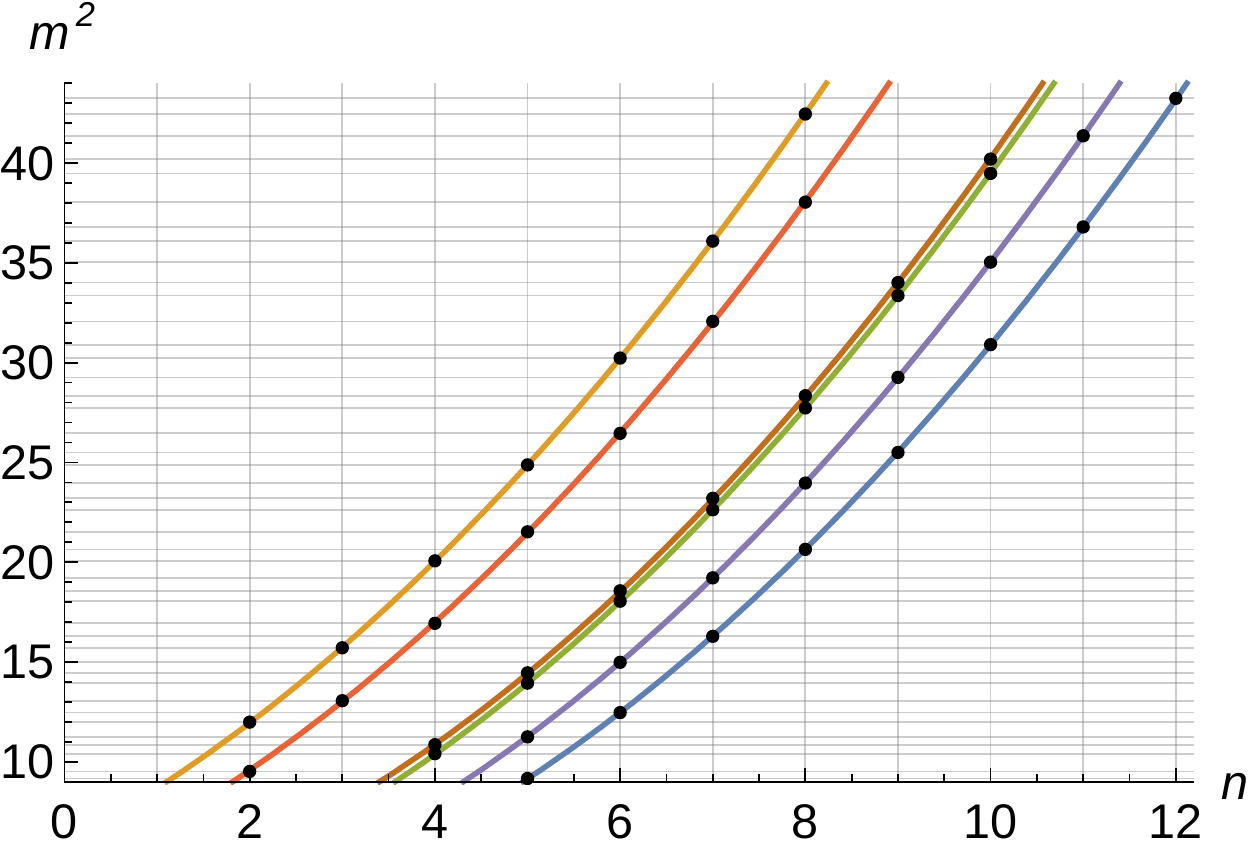}
    \caption{Quadratic fits of the $0^{-+}$ spectrum of $m^2$ found in~\cite{Melnikov:2020cxe}. The black dots (as well as the grids) indicate the mass eigenvalues and the curves are fits~(\ref{mll})-(\ref{m11}). The left plot shows the behavior of the fits for $m^2\leq 10$. The intercepts correspond to the ground states. The right plots illustrates the quality of the fits for $m^2\geq 10$. All mass values are shown in the units of the ground state mass of $2^{++}$.}
    \label{fig:fits}
\end{figure}

The spectrum of $0^{++}$, found in~\cite{Berg:2006xy} does not identify the operator origin of the mass eigenvalues. Identifying the six ground states and the corresponding excited states is a complicated task. In~\cite{Melnikov:2020cxe} the spectrum of $0^{-+}$ was calculated as a consistency check, since the two spectra are expected to be related by supersymmetry, yet the separation of the full scalar spectrum into six families was not completed. It was argued that the following ordering of states is consistent with the two approaches and with the lattice results~\cite{RodriguesFilho:2020rae},
\be
m_{\lambda\lambda} < m_{0^{++}} < m_{\lambda\lambda}^* < m_{0^{-+}} < m_{0^{++}}^* < \cdots\,,
\eeq
where $m_{\lambda\lambda}$ is the mass of the ground state of the gluino bilinear, $m_{\lambda\lambda}^\ast$ is its first excited state, $m_{0^{++}}$ and $m_{0^{-+}}$ are masses of the states in the spectrum of the two gluon operators $\tr F_{\mu\nu}F^{\mu\nu}$ and $\tr F_{\mu\nu}\tilde{F}^{\mu\nu}$, studied on the lattice. Here we also conjectured the masses of the ground states of the remaining two multiplets, including that of the heaviest four-gluon operator $\tr (F_{\mu\nu}F^{\mu\nu})^2$ to be seen on the lattice. This is done by applying quadratic fits to the values of mass squared. For the multiplets other than scalar the quadratic fits work quite well. The results of the fitting of the pseudoscalar sector~\cite{Melnikov:2020cxe} are shown in figure~\ref{fig:fits}. The following fits work very well for the heavy part of the spectrum, while for the light states we end up with a few noticeable deviations.
\begin{eqnarray}
m_{\lambda\lambda}^2 \ = \ 0.223 +0.499n + 0.257 n^2 \,, &&  m^{\rm GS} \ = \ 0.472\,, \label{mll}\\
m_{0^{++}}^2 \ = \ 0.480 + 0.854 n + 0.260 n^2\,, & &  m^{\rm GS} \ = \ 0.693\,,\\
m_{0{-+}}^2 \ = \ 1.22 + 1.26 n + 0.257 n^2\,, & & m^{\rm GS} \ = \ 1.11 \,, \\
m_{10}^2 \ = \ 1.97 + 1.15 n + 0.267 n^2\,, & & m^{\rm GS} \ = \ 1.40 \,, \\
m_{AB}^2 \ = \ 4.24 + 2.14 n + 0.261 n^2\,, & & m^{\rm GS} \ = \ 2.06 \,, \\
m_{11}^2 \ = \ 5.92 + 2.49 n + 0.260 n^2\,, & & m^{\rm GS} \ = \ 2.43 \,, \label{m11}
\end{eqnarray}
where all the values are given in units of the mass of the ground state of $2^{++}$. Here $m_{10}$ and $m_{11}$ are pseudoscalar predictions for the masses of the ground states in the entries 10 and 11 of table~\ref{tab:KSsinglets}. Eigenvalue $m_{AB}$ is expected to be absent from the spectrum of the pure supersymmetric Yang-Mills. We note that the fits slightly improve the convergence to the $SU(\infty)$ lattice values.

It is necessary to mention that the spectrum of pseudoscalars computed in~\cite{Melnikov:2020cxe}, although found to be consistent with the spectrum of $0^{++}$ in~\cite{Berg:2006xy}, contains a subset of eigenvalues exhibiting divergence, particularly significant at low mass. For example, the predictions for the masses of the three lightest scalar ground states were found to be $m_{\lambda\lambda}\simeq 0.511\,m_{2^{++}}$, $m_{0^{++}}\simeq 0.701\,m_{2^{++}}$ and $m_{0^{-+}}\simeq 1.15\,m_{2^{++}}$, cf. (\ref{mll})-(\ref{m11}), to be compared with the entries 7, 8 and 9 in table~\ref{tab:KSsinglets}. See also figure~\ref{fig:fullspec}. The spectrum of~\cite{Berg:2006xy} was independently confirmed in~\cite{Elander:2014ola}, so the status of the pseudoscalar calculation remains unclear. 

As a final comment we remind that no states of spin $J\geq 2$ are accessible to the holographic analysis in the supergravity limit, except for the $2^{++}$.

\bigskip

In conclusion we would like to discuss the relevance of the results presented in this note. Recent progress in lattice calculations largely confirmed the expectations of the universality of the glueball masses and allowed to extrapolate the results to the limit of large number of colors. The comparison of the lattice results for $SU(\infty)$ with the Yang-Mills subsector of the Klebanov-Strassler theory demonstrated consistency at the level expected from a holographic model. For the five lightest states with $J<2$ of the bosonic Yang-Mills theory computed on the lattice we observed the same hierarchy of the spectrum in the Klebanov-Strassler theory and a sub-ten-percent match for the masses, as summarized by table~\ref{tab:KSratios} for either $SU(\infty)$ and $SU(3)$ theories. With the support of the universality of the spectrum, these results give an independent check of the consistency of the lattice approach.

\begin{table}[h]
    \centering
    \begin{tabular}{||c|c|ccccc||}
\hline \Trule\Brule $J^{\,PC}$ & $2^{++}$ & $0^{++}$ & $0^{-+}$ &  $1^{+-}$ &  $1^{--}$ &  $0^{+-}$\\
      \hline
      \Trule\Brule Holography/$SU(\infty)$ & 1 & 0.959 & 1.081 & 1.037 & 1.038 & 0.999 \\ 
      \Trule\Brule Holography/$SU(3)$ & 1 & 0.920 & 1.027 & 1.048 & 0.967 & 1.057 \\ \hline
    \end{tabular}
    \caption{Comparison of the holographic (Klebanov-Strassler) and lattice predictions for the masses of the lightest glueball ground states in the $SU(\infty)$ and $SU(3)$ Yang-Mills theories~\cite{Athenodorou:2021qvs}.}
    \label{tab:KSratios}
\end{table}

More importantly, glueball spectrum turns out to be a rare guide to independent testing of the holographic correspondence. It shows that despite the unnaturalness of the classical holographic limit, first principle lattice calculations can be used to access physical observables in such a regime.

Despite a very reasonable match in table~\ref{tab:KSratios} and a possible room for improvement of convergence, the Klebanov-Strassler theory is not expected to give precisely the same results as the $SU(\infty)$ Yang-Mills. The results summarized in figure~\ref{fig:fullspec} should rather serve an approximation to the spectrum of the supersymmetric $SU(\infty)$ Yang-Mills, deformed by the presence of some matter. The consistency with the bosonic theory tells that the results presented here can give a similar, or even better match with the spectrum of the supersymmetric theory. The supersymmetric case gives a larger base for comparison with a multitude of additional states, but it unfortunately makes it a very complicated story for the lattice. We hope that such a comparison will be possible in the future.

We close the discussion with a prediction for the odderon, from the lattice and from holography. One can predict the value of the odderon to pomeron mass ratio to be close to
\be
\frac{m_{1^{--}}}{m_{2^{++}}} \ \simeq \  1.58 ~\text{(lattice)} \qquad \text{or} \qquad \frac{m_{1^{--}}}{m_{2^{++}}} \ \simeq \  1.64 ~\text{(holography)}\,.
\eeq

\paragraph{Acknowledgments.} These notes include the results of the collaboration with A.~Dymarsky and C.~Rodrigues Filho, whom the author would like to thank for discussions. This work was done with support of the Simons Foundation, award \#884966, via the Association International Institute of Physics (AIIF) and of the grant of the agency CNPq of the Brazilian Ministry of Science, Technology and Innovation \#433935/2018-9.

The material for these notes was originally prepared for the talk given at the Low X 2021 conference at Elba. While the talk was being prepared, a very timely paper~\cite{Athenodorou:2021qvs} appeared with new results on the $SU(\infty)$ spectrum on the lattice, which allowed to make a more detailed comparison of the spectrum in these notes, beyond $SU(3)$ and recent results on $Sp(\infty)$~\cite{Bennett:2020qtj} discussed in the talk.

\nocite{*}
\bibliographystyle{auto_generated}
\bibliography{Low_X_proceedingsDMelnikov/Low_X_proceedingsDMelnikov}

%% file: Low_x_2021_F_Nemes/Low_x_2021_F_Nemes.tex
\vspace*{1.2cm}

\thispagestyle{empty}
\begin{center}
{\LARGE \bf TOTEM results}

\par\vspace*{7mm}\par

{

\bigskip

\large \bf Frigyes Nemes}

\bigskip

{\large \bf  E-Mail: fnemes@cern.ch}

\bigskip

{CERN (also at Wigner RCP and MATE, Hungary)}

\bigskip

{\it Presented at the Low-$x$ Workshop, Elba Island, Italy, September 27--October 1 2021}

\vspace*{15mm}

\end{center}
\vspace*{1mm}

\begin{abstract}
TThe TOTEM experiment at the LHC has measured proton-proton (${\rm pp}$) elastic scattering in dedicated runs at $\sqrt{s}=2.76,~7,~8$ and 13~TeV centre-of-mass energies.
The total, elastic and inelastic ${\rm pp}$ cross-sections have been derived for each energy using the optical theorem and the luminosity independent method. At 13 TeV, the total ${\rm pp}$ cross-section has also been derived for the first time at LHC using the Coulomb amplitude for
the normalization of the elastic ${\rm d}\sigma/{\rm d}t$.
TOTEM has excluded using $\sqrt{s}=8$~TeV data a purely exponential nuclear differential cross-section ${\rm d}\sigma/{\rm d}t$ at low $|t|$ for elastic ${\rm pp}$ scattering. The effect has been confirmed at $\sqrt{s} = 13$~TeV.
The $\rho$ parameter, the real to imaginary ratio of the nuclear elastic scattering amplitude at $t = 0$, has been measured precisely at $\sqrt{s}=13$~TeV using the Coulomb-nuclear interference region.
In order to properly describe the measured $\rho$ value and all the TOTEM $\sigma_{\rm tot}$ measurements, in addition to the exchange of photons and colourless two-gluon compound, the so-called Pomeron, the exchange of a colourless C-odd three-gluon compound, also known as the Odderon, in the $t$-channel should be added for elastic ${\rm pp}$ scattering.
At all LHC energies, TOTEM has observed a diffractive minimum (``dip'') followed by a secondary maximum (``bump'') in the elastic ${\rm pp}$ ${\rm d}\sigma/{\rm d}t$. In the measurement of the D0 experiment at $\sqrt{s} = 1.96$~TeV, no dip or bump can be observed in the elastic proton-antiproton ${\rm d}\sigma/{\rm d}t$.
Under the assumption that possible effects due to the energy difference between the 2.76~TeV TOTEM and the 1.96~TeV D0 measurements can be neglected, the results provide evidence for the exchange
of a colourless C-odd 3-gluon compound in the $t$-channel of proton-(anti)proton elastic scattering. 
\end{abstract}
 \part[TOTEM results\\ \phantom{x}\hspace{4ex}\it{Frigyes Nemes}]{}
\section{INTRODUCTION}

\begin{figure}[H]
		\centering
    		\includegraphics[width=0.9\textwidth]{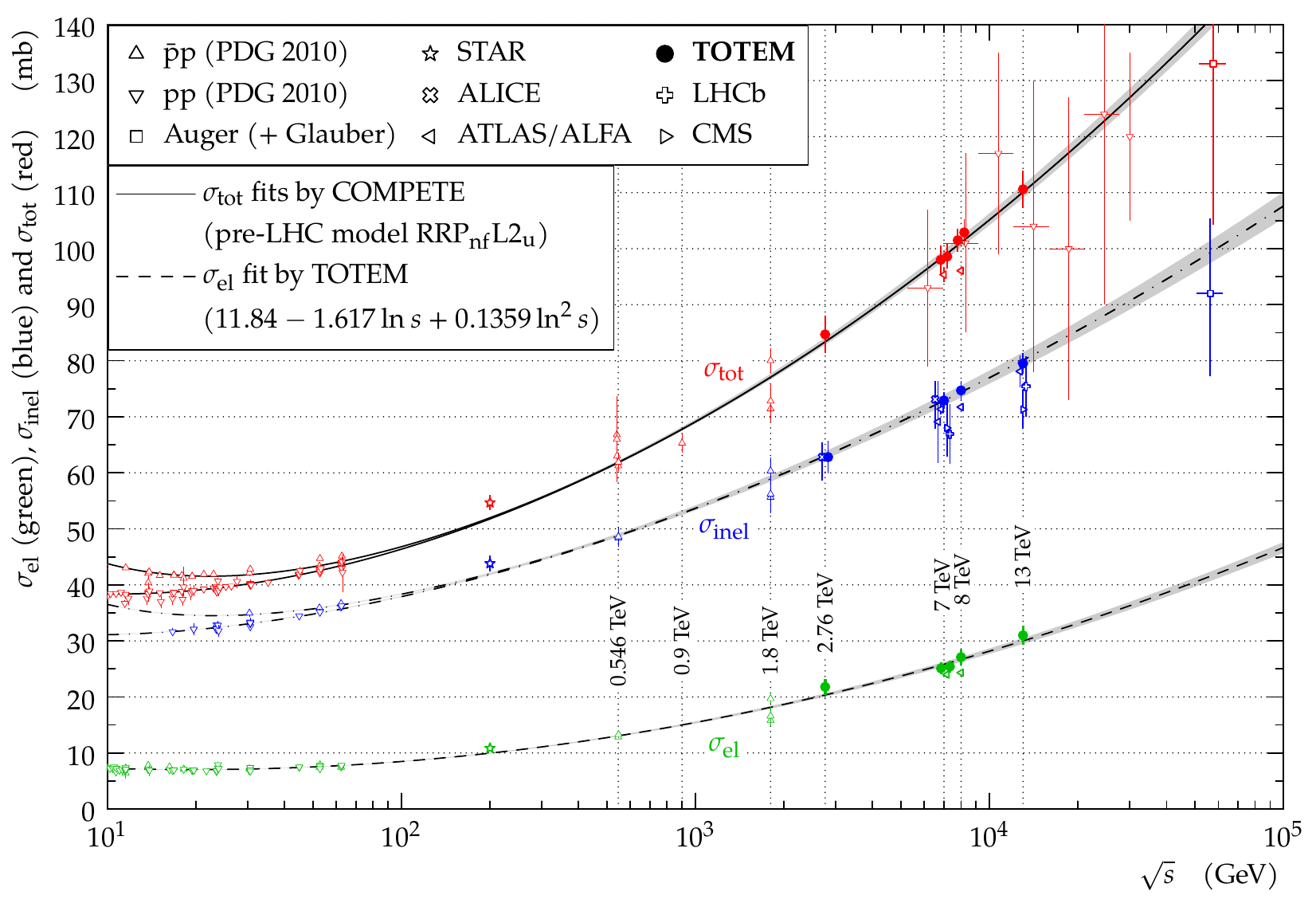}\vspace{-3mm}
		\caption{A compilation of total, inelastic and elastic pp cross-section measurements, see Ref.~\cite{Tanabashi2018pdg, Antchev:2017yns,STAR:2020phn} and references therein. The red points indicate the TOTEM total cross-section measurements at $\sqrt{s}=2.76, 7, 8$ and 13~TeV using the luminosity independent
		method.}
		\label{elasticscatteringsummary}
	\end{figure}

	The TOTEM (TOTal cross section, Elastic scattering and diffraction dissociation Measurement at the LHC) experiment has been designed to measure the total proton-proton (${\rm pp}$) cross-section, elastic scattering and
	diffractive processes at the LHC~\cite{Anelli:2008zza}, see e.g. Fig.~\ref{elasticscatteringsummary}.

	The experimental apparatus of TOTEM is composed of three subdetectors: the Roman Pots (RP), the T1 and the T2 inelastic forward telescopes. The detectors are placed symmetrically on both sides of the Interaction
	Point 5 (IP5), which is shared with the CMS experiment.

	The RPs are moveable beam-pipe insertions, hosting edgeless silicon detectors to detect leading protons scattered at very small angles.
 	In order to maximize the acceptance of the experiment for elastically scattered protons, the RPs are able to approach the beam center to a transverse
	distance as small as 1 mm. The alignment of the RPs is optimized by reconstructing common tracks going through the overlap between the vertical and
	horizontal RPs as well as by studying elastic events~\cite{TOTEM:2013vay}.
	
	The T1 telescope is based on cathode strip chambers placed at $\pm$9~m and covers the pseudorapidity range $3.1 \le |\eta| \le 4.7$; the T2
        telescope is based on gas electron multiplier chambers placed at $\pm$13.5~m and covers the pseudorapidity range $5.3 \le |\eta| \le 6.5$.
        The pseudorapidity coverage of the two telescopes at $\sqrt{s} = 2.76,~7,~8$ and 13 TeV allows the detection of about 96~\%, 95~\%, 94~\% and 92~\%, respectively, of the inelastic ${\rm pp}$ collisions, including collisions producing diffractive mass above about 2.1 GeV, 3.4 GeV, 3.6 GeV and 4.6 GeV, respectively~\cite{Antchev:2013haa, Antchev:2013paa, Antchev:2017dia}.

	Before the LHC long shutdown one (LS1) the RPs, used for the measurements, were located at distances of 215--220 m from IP5~\cite{Anelli:2008zza}. The actual layout, i.e., after the LHC LS1, is different in RP
	location and quantity. The RP stations previously installed at $\pm$147~m, from IP5, have been relocated to $\pm$210~m. Moreover, newly designed horizontal RPs have been installed
	between the two units of the $\pm220$~m station~\cite{TOTEM:2013iga,Albrow:2014lrm}.

\section{ELASTIC SCATTERING AND TOTAL CROSS-SECTION $\sigma_{\rm tot}$
          MEASUREMENTS}

        For each tagged elastic event the four-momentum transfer squared~$t$ is reconstructed using the LHC optical functions, characterized with the so-called
	betatron amplitude at IP5~$\beta^{*}$~\cite{Anelli:2008zza}. The TOTEM experiment developed a novel experimental method to estimate the optical functions at the RP locations,
	using the measured elastically scattered protons~\cite{Antchev:2014voa,Burkhardt:2012zza}.

	The total inelastic rate $N_{\rm inel}$, measured by the T1 and T2 telescopes, and the total nuclear elastic rate $N_{\rm el}$
	with its extrapolation to zero four-momentum transfer squared~$t=0$ are combined with the optical theorem to obtain the total cross-section
	in a luminosity, $\mathcal{L}$, independent way
	\begin{equation}
		\sigma_{\rm tot}=\frac{16\pi}{1+\rho^{2}}\cdot\left.\frac{{\rm d}N_{\rm el}}{{\rm d}t}\right|_{t=0}\cdot(N_{\rm el}+N_{\rm inel})^{-1}.
	\end{equation}
	The measured elastic $N_{\rm el}$ and inelastic rates $N_{\rm inel}$ allow for the determination of the elastic and inelastic cross-sections as well.

	The TOTEM experiment determined the total pp cross-section at $\sqrt{s}=7$~TeV using the luminosity independent method~\cite{Antchev:2013iaa}, which was shown to be consistent with the total cross-sections
	measured in independent ways, see Table~\ref{totalcrosssections}. The elastic and inelastic cross-sections were found to be $\sigma_{\rm el}=(25.1 \pm 1.1)$~mb and $\sigma_{\rm inel}=(72.9 \pm 1.5)$~mb, respectively.
	
\begin{table}[H]
		\centering
		\begin{tabular}{|c|c|c|c|c|} \hline
			Method				& $\mathcal{L}$ independent~\cite{Antchev:2013iaa}	& $N_{\rm inel}$ rate-indep.~\cite{Antchev:2011vs}		&$N_{\rm inel}$ rate-indep.~\cite{Antchev:2013gaa}	&$\rho$ indep.~\cite{Antchev:2013iaa}\\ \hline
			$\sigma_{\rm tot}$ [mb] 	& 98.0 $\pm$ 2.5					& 98.3 $\pm$ 2.8		& 98.6 $\pm$ 2.2		& 99.1 $\pm$ 4.3	\\ \hline
		\end{tabular}
		\caption{The total cross-section $\sigma_{\rm tot}$ results measured by the TOTEM experiment at $\sqrt{s}=7$~TeV with three different methods and two different data sets.}
		\label{totalcrosssections}
\end{table}

	The luminosity-independent measurements were repeated at $\sqrt{s}=2.76$, $8$ and $13$~TeV. At $\sqrt{s}=2.76$~TeV, the total cross-section was found to be $\sigma_{\rm tot}=(84.7 \pm 3.3)$~mb, while the elastic and inelastic cross-section were $\sigma_{\rm el}=(21.8 \pm 1.4)$~mb and $\sigma_{\rm inel}=(62.8 \pm 2.9)$~mb, respectively~\cite{Antchev:2017dia}. At $\sqrt{s}=8$~TeV, the total, elastic and inelastic cross-sections of $\sigma_{\rm tot}=(101.7\pm2.9)$~mb, $\sigma_{\rm el}=(27.1\pm1.4)$~mb and $\sigma_{\rm inel}=(74.7\pm1.7)$~mb, respectively, were obtained~~\cite{Antchev:2013paa}. Finally at $\sqrt{s}=13$~TeV, the total, elastic and inelastic cross-sections were found to be $\sigma_{\rm tot}=(110.6~\pm~3.4$)~mb, $\sigma_{\rm el}=(31.0~\pm~1.7)$~mb and $\sigma_{\rm inel}=(79.5~\pm~1.8)$~mb, respectively~~\cite{Antchev:2017dia}. 

In 2016 TOTEM took data during a special run with $\beta^{*}= 2500$~m optics at 13~TeV collision energy, which allowed to probe sufficiently low $|t|$-values to be sensitive to the Coulomb amplitude allowing a first total ${\rm pp}$ cross section measurement at LHC with Coulomb normalization $\sigma_{\rm tot}=(110.3\pm3.5$)~mb~\cite{Antchev:2017yns}. Combining the two uncorrelated TOTEM measurement at 13 TeV, luminosity independent and Coulomb normalization, yields $\sigma_{\rm tot}=(110.5 \pm 2.4)$~mb. Fig.~\ref{elasticscatteringsummary} shows a compilation of all the results together with other LHC measurements. The observed cross-sections are in agreement with the extrapolation of low-energy data to LHC and cosmic ray results as well.

Thanks to a high statistics $\beta^{*}=90$~m data set at $\sqrt{s}=8$~TeV energy, the TOTEM experiment excluded a purely exponential elastic pp differential cross-section~\cite{Antchev:2015zza}. The
significance of the exclusion is greater than 7$\sigma$ in the $|t|$ range from 0.027 to 0.2 GeV$^{2}$. Using refined parametrizations for the extrapolation
to the optical point, $t=0$, yields total cross-section values $\sigma_{\rm tot}=(101.5\pm2.1)$~mb and $\sigma_{\rm tot}=(101.9\pm2.1)$~mb, compatible with the previous measurements. The deviation from the purely exponential elastic pp differential cross-section has been confirmed 
at 13 TeV~\cite{Antchev:2018edk}.

	The TOTEM experiment performed its first measurement of elastic scattering in the Coulomb-nuclear interference (CNI) region~\cite{Antchev:2016vpy}.
	The data have been collected at $\sqrt{s}=8$~TeV with a special beam optics of $\beta^{*}=1000$~m in 2012.
	The $\rho$ parameter was for the first time at LHC extracted via the Coulomb-nuclear interference, and
	was found to be~$\rho = 0.12\pm0.03$. Taking the Coulomb-nuclear interference into account in the extrapolation
to the optical point, $t=0$, yields total cross-section values of  $\sigma_{\rm tot}=(102.9\pm2.3)$~mb and $\sigma_{\rm tot}=(103.0\pm2.3)$~mb for central and peripheral phase descriptions, respectively, compatible with the previous measurements. 

\begin{figure}[H]
    \centering
        \includegraphics[width=0.8\textwidth]{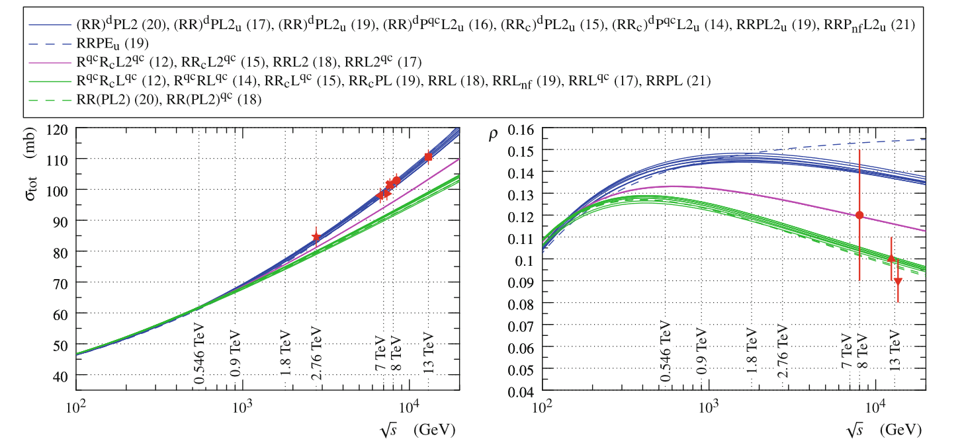}
    \caption{Predictions  of  COMPETE  models  for  pp interactions. Each model is represented by one line (see legend). The red points represent the reference TOTEM measurements.}
    \label{fig_4}
\end{figure}

The special run with $\beta^{*}= 2500$~m optics at 13~TeV collision energy with higher statistics allowed also for a precise measurement of the $\rho$ yielding $\rho = 0.09 \pm 0.01$ and $\rho = 0.10 \pm 0.01$, depending on different physics assumptions and mathematical modelling.
	This $\rho$ result combined with all the TOTEM $\sigma_{\rm tot}$ measurements indicate that it is not sufficient to include only photon and colourless C-even 2-gluon compound exchange, the so-called Pomeron, in the $t$-channel to properly describe elastic ${\rm pp}$ scattering. 
	A significantly better description is obtained both in the Regge-like frameworks and QCD by adding colourless C-odd 3-gluon compound exchange in the $t$-channel~\cite{Antchev:2017yns}, the so-called Odderon. On the contrary, if shown that the C-odd 3-gluon compound $t$-channel exchange is not of importance for the description of elastic ${\rm pp}$ scattering at low $|t|$, the $\rho$ value determined by TOTEM would represent a first evidence of a slowing down of the total cross-section growth at higher energies.

The $\rho$ and $\sigma_{\rm tot}$ results are incompatible with models with Pomeron exchange only and provide evidence of odderon exchange effects with significance between 3.4$\sigma$ and 4.6$\sigma$, see Fig.~\ref{fig_4}~\cite{Antchev:2017yns} and Ref.~\cite{TOTEM:2020zzr_nemes}.

\begin{figure}[H]
\centering
\includegraphics[width=0.7\textwidth]{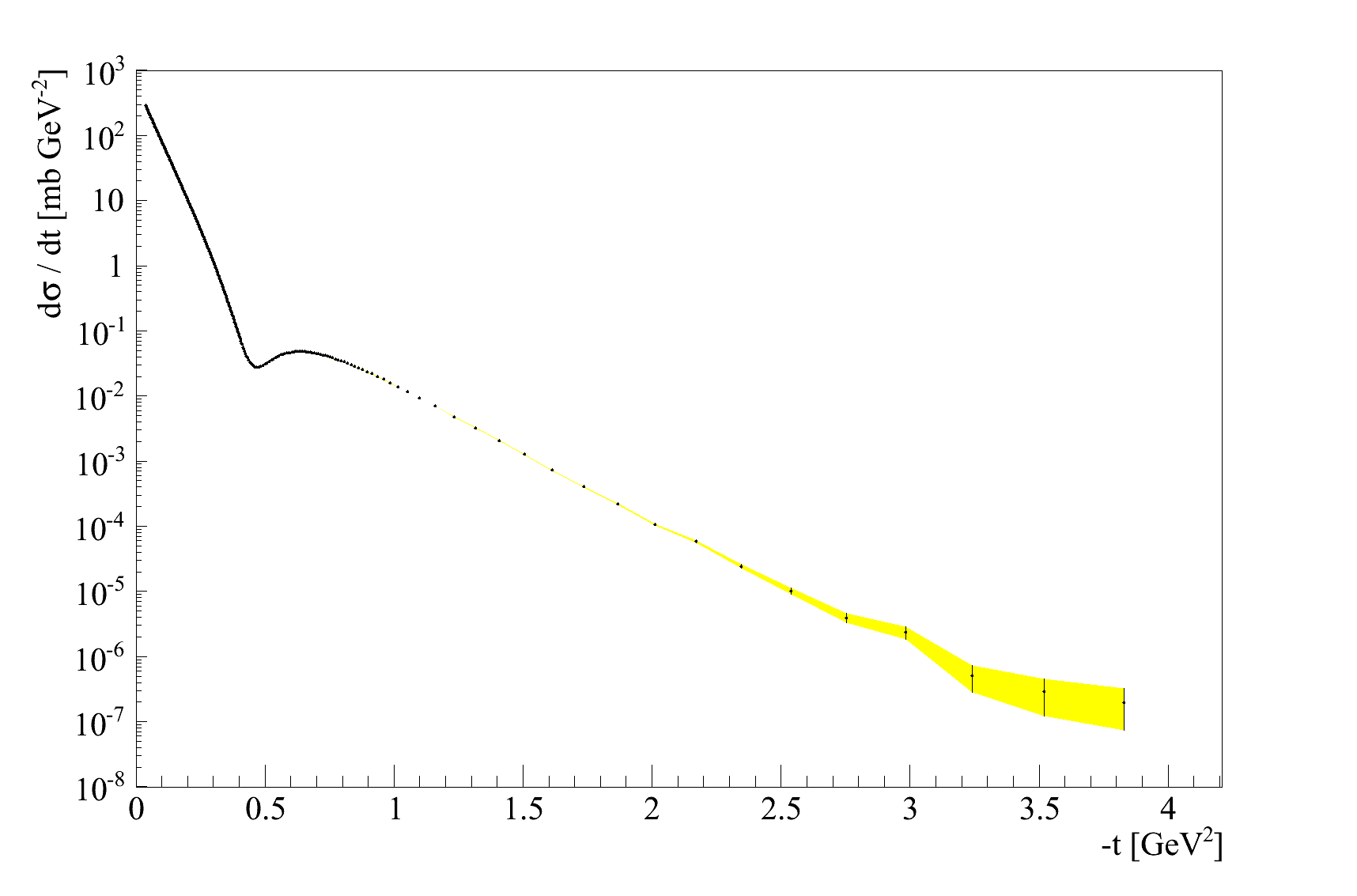}
\caption{(color) Differential elastic cross-section ${\rm d}\sigma/{\rm d}t$ at $\sqrt{s} = 13$~TeV. The statistical and $|t|$-dependent correlated systematic uncertainty envelope is shown
as a yellow band.}
\label{finalresult90m}
\end{figure}

\begin{figure}[H]
		\centering
    		\includegraphics[width=0.8\textwidth]{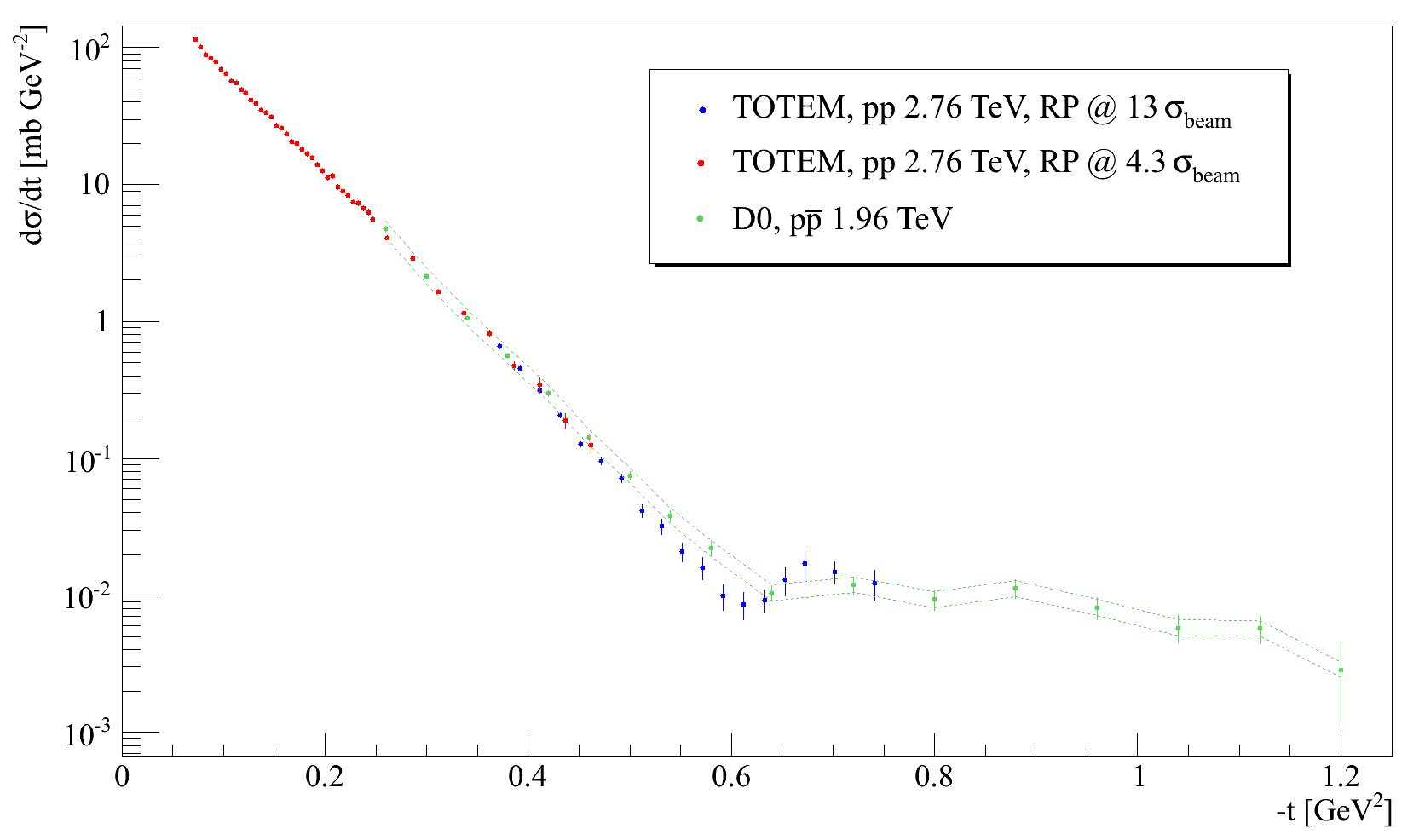}
		\caption{(color) The differential cross sections ${\rm d}\sigma/{\rm d}t$ at $\sqrt{s}=2.76$~TeV measured by the TOTEM experiment and the elastic $\rm p\bar{p}$ measurement of the D0 experiment at 1.96 TeV~\cite{Abazov:2012qb}. The
		green dashed line indicates the normalization uncertainty of the D0 measurement.}
		\label{totemd0}
	\end{figure}

\begin{figure}[H]
\begin{center}
\includegraphics[width=0.8\textwidth]{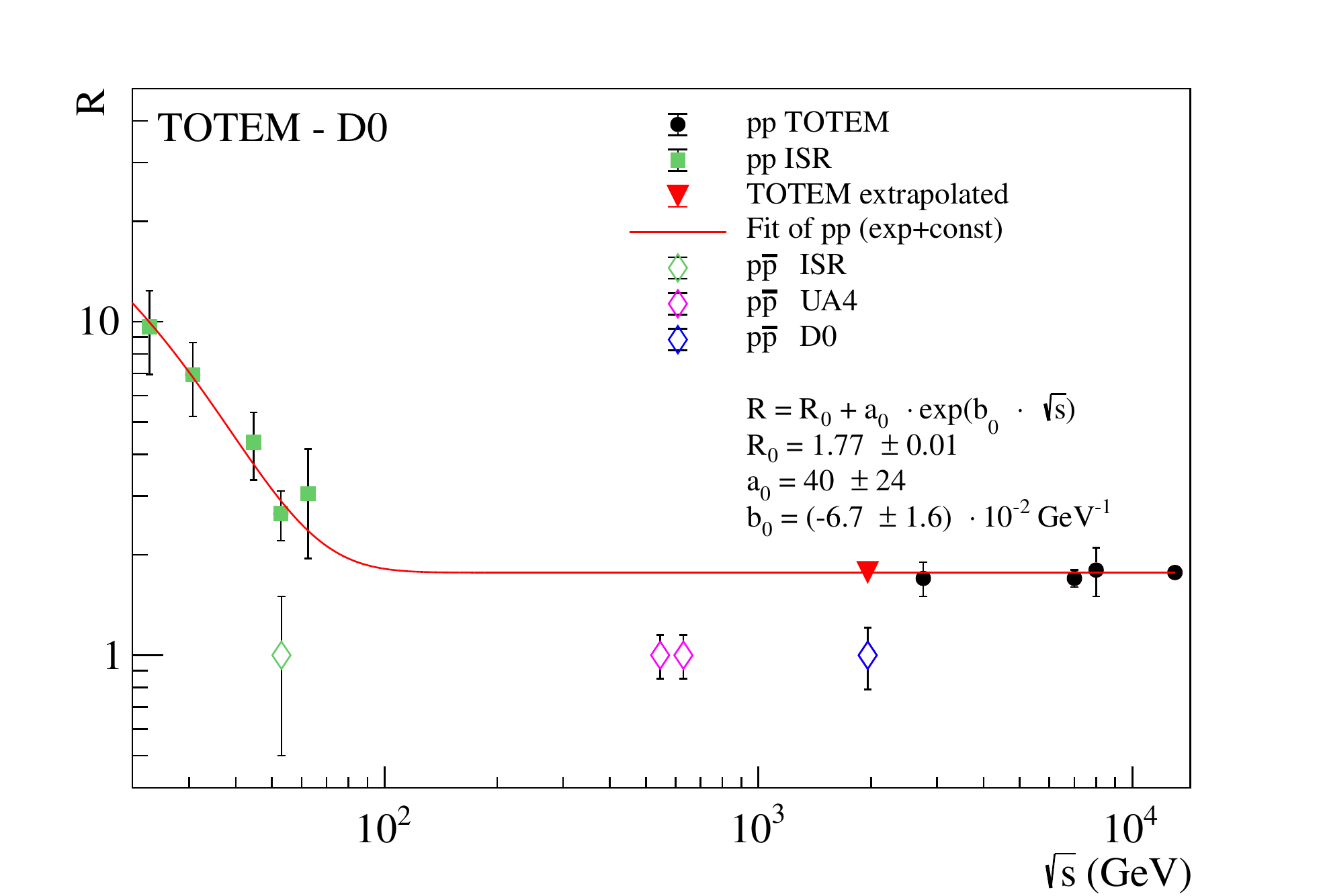}
\caption{The ratio $R$ of the cross sections at the bump and dip as a function of $\sqrt{s}$ for ${\rm pp}$ and ${\rm p\bar{p}}$. The pp data are fitted to the function noted in the legend.}
\label{elcross}
\end{center}
\end{figure}

At 13 TeV the differential cross-section has been measured in the [0.04; 4]~GeV$^{2}$ range of $|t|$ using a very-high statistics sample (more than 10$^9$) of elastic events taken in 2015 using a dedicated data acquisition system allowing an increased data taking rate by an order of magnitude. This sample allowed for a precise measurement of the non-exponential part that contains a diffractive minimum ``dip'' and a secondary maximum ``bump'', see Fig.~\ref{finalresult90m}~~\cite{Antchev:2018edk}. The dip position at 13 TeV was found to be $|t_{\rm dip}|=(0.47\pm0.004^{\rm stat}\pm0.01^{\rm syst})$~GeV$^{2}$ and the ratio of the ${\rm d}\sigma/{\rm d}t$ at the bump
	and at the dip $1.77\pm0.01^{\rm stat}$. Using $\beta^*$ = 11 m optics data taken in 2013, also the dip and bump at $\sqrt{s}=~$2.76~TeV could be observed; the position of the dip is $|t_{\rm dip}|=(0.61\pm0.03)$~GeV$^{2}$ and the bump-dip cross-section ratio $1.7\pm0.2$, as shown in Fig.~\ref{totemd0}~\cite{TOTEM:2018psk}. These new results confirm the ${\rm d}\sigma/{\rm d}t$ feature of dip and bump at TeV scale already observed at 7 TeV with a dip position of $|t_{\rm dip}|=(0.53\pm0.01^{\rm stat}\pm0.01^{\rm syst})$~GeV$^{2}$
	and a bump-dip cross-section ratio of $1.7\pm0.1$~\cite{Antchev:2011zz, Antchev:2017yns}. The result is confirmed at 8~TeV as well~\cite{TOTEM:2021imi}. The series of TOTEM elastic ${\rm pp}$ measurements show that the dip is a permanent feature of the ${\rm pp}$ differential cross-section at TeV scale. 
	This is expressed by a bump-to-dip cross section ratio R significantly larger than 1, see Fig.~\ref{elcross}~\cite{TOTEM:2020zzr_nemes}. However, for $\rm p\bar{p}$ at TeV scale, this R-value is close to 1, i.e. there is no dip and no bump in the differential cross section.

When the 2.76 TeV ${\rm d}\sigma/{\rm d}t$ measurement of TOTEM is compared directly to the proton-antiproton (${\rm p\bar{p}}$) measurement of the D0 experiment at $\sqrt{s} = 1.96$~TeV, a significant difference can be observed, see Fig.~\ref{totemd0}. Under the assumption that possible effects due to the energy difference between TOTEM and D0 can be neglected, the result provides evidence for a colourless C-odd 3-gluon compound exchange in the $t$-channel of
${\rm pp}$ and ${\rm p\bar{p}}$ elastic scattering, cf. also~\cite{TOTEM:2020zzr_nemes}. This conclusion has also been acknowledged by Ref.~\cite{Leader:2021zkf}.

	\section{CONCLUSIONS}
	The TOTEM experiment has measured elastic ${\rm pp}$ scattering at $\sqrt{s}=2.76, 7, 8$ and 13~TeV. The 
	total, elastic and inelastic ${\rm pp}$ cross-sections have been derived for all energies using
	the luminosity independent method and the optical theorem. At  $\sqrt{s}=8$~TeV, TOTEM has also excluded a purely
	exponential nuclear ${\rm pp}$ differential cross-section at low $|t|$. This deviation has been confirmed at 13 TeV. At 13 TeV, the $\rho$ parameter has been precisely measured and the total ${\rm pp}$ cross-section using the Coulomb amplitude has been derived for the first time at the LHC.
	The $\rho$ measurement combined with all the TOTEM $\sigma_{\rm tot}$ measurements indicate the necessity to add the exchange of a colourless C-odd 3-gluon compound in the $t$-channel of elastic ${\rm pp}$ scattering.
	
	At $\sqrt{s} = 2.76$~TeV, a diffractive minimum ``dip'' and a secondary maximum ``bump'' has been observed; when compared to the ${\rm p\bar{p}}$ measurement of the D0 experiment at
	$\sqrt{s} = 1.96$~TeV, a significant difference can be observed.  Under the assumption that possible effects due to the energy difference between TOTEM and D0 can be neglected, the result provides evidence for the exchange of a colourless C-odd 3-gluon compound in the $t$-channel of ${\rm pp}$ and ${\rm p\bar{p}}$ elastic scattering.
	At 13 TeV, the differential cross-section has been measured in the [0.04 GeV$^{2}$; 4 GeV$^{2}$] $|t|$-range allowing for the precise measurement of the dip. The series of TOTEM elastic ${\rm pp}$ measurements show that the dip is a permanent feature of the ${\rm pp}$
	differential cross-section at the TeV scale.\newline\vspace{2mm}

Comments: Presented at the Low-$x$ Workshop, Elba Island, Italy, September 27--October 1 2021.

\nocite{*}
\bibliographystyle{auto_generated}
\bibliography{Low_x_2021_F_Nemes/Low_x_2021_F_Nemes}

%% file: Odderon_lowx_Osterberg/Osterberg.tex
\vspace*{1.2cm}

\thispagestyle{empty}
\begin{center}
{\LARGE \bf Odderon observation: explanations and answers to questions/objections regarding the PRL publication}

\par\vspace*{7mm}\par

{

\bigskip

\large \bf Kenneth \"Osterberg on behalf of the D0 and TOTEM collaborations}

\bigskip

{\large \bf  E-Mail: kenneth.osterberg@helsinki.fi}

\bigskip

{Department of Physics and Helsinki Institute of Physics, University of Helsinki, Helsinki, Finland}

\bigskip

{\it Presented at the Low-$x$ Workshop, Elba Island, Italy, September 27--October 1 2021}

\vspace*{15mm}

\end{center}
\vspace*{1mm}

\begin{abstract}

The odderon observation recently published by the D0 and TOTEM collaborations has been widely accepted by a majority of the particle physics community and its importance recognized through dedicated physics seminars in the world major labs and physics institute. Naturally also some questions and objections have been raised, either privately or publicly, in discussion sessions and articles. In this proceedings article, a comprehensive list of these questions and objections are answered and supplementary material is provided. The methods and assumptions used in the extrapolation of the $pp$ elastic differential cross section to $\sqrt{s}$ = 1.96 TeV and its comparison to the D0 measurement in $p\bar{p}$ are shown to be valid and reasonable. Likewise, the methods and choices used for the $\rho$ measurement at LHC. Furthermore, objections against the odderon interpretation are demonstrated not to be valid. Finally, the combination of the different odderon significances, leading to the first experimental observation of odderon exchange, is shown to be well founded. 
\end{abstract}
  \part[Odderon observation: explanations and answers to questions/objections regarding the PRL publication\\ \phantom{x}\hspace{4ex}\it{Kenneth \"Osterberg on behalf of the D0 and TOTEM collaborations}]{}

 \section{Introduction}
 The D0 and TOTEM collaborations have recently published the observation of the odderon~\cite{Odderon-discovery}. The observation is based on combining two evidences for the odderon in complementary $|t|$-ranges using complementary TOTEM data sets: (1) a comparison of the proton-proton ($pp$) and proton-antiproton ($p\bar{p}$) elastic differential  cross sections ($d\sigma_{el}/dt$) in the $|t|$-range of the diffractive minumum ("dip") and the secondary maximum ("bump") of the $pp$ $d\sigma_{el}/dt$ at $\sqrt{s}$ = 1.96 TeV~\cite{Odderon-discovery} and (2) the total cross section ($\sigma_{tot}$) and $\rho$ measurements at \mbox{very low $|t|$ in $pp$ collisions at the LHC~\cite{TOTEM-rho-13TeV}.} The methods, assumptions and choices used in the analyses have raised questions that are answered in detail and supplementary material is provided in this proceedings contribution. Furthermore, the objections raised to the odderon interpretation of the evidences are shown not to be valid.

The explanations and answers are organized as follows. First the comparison of the $pp$ and $p\bar{p}$ $d\sigma_{el}/dt$ is briefly presented, then explanations regarding the $pp$ and $p\bar{p}$ comparison are provided and questions and objections raised are answered. Next the odderon evidence from the TOTEM $\rho$ and $\sigma_{tot}$ measurements is introduced and afterwards replies to the questions and objections raised regarding the analysis and interpretation are given. Finally the combination of the odderon signatures is discussed and responses to issues raised concerning the combination are provided.

\section{The comparison of elastic $pp$ and $p\bar{p}$ cross sections}
\begin{figure}
\begin{minipage}{0.55\linewidth}
\centerline{\includegraphics[width=0.925\linewidth]{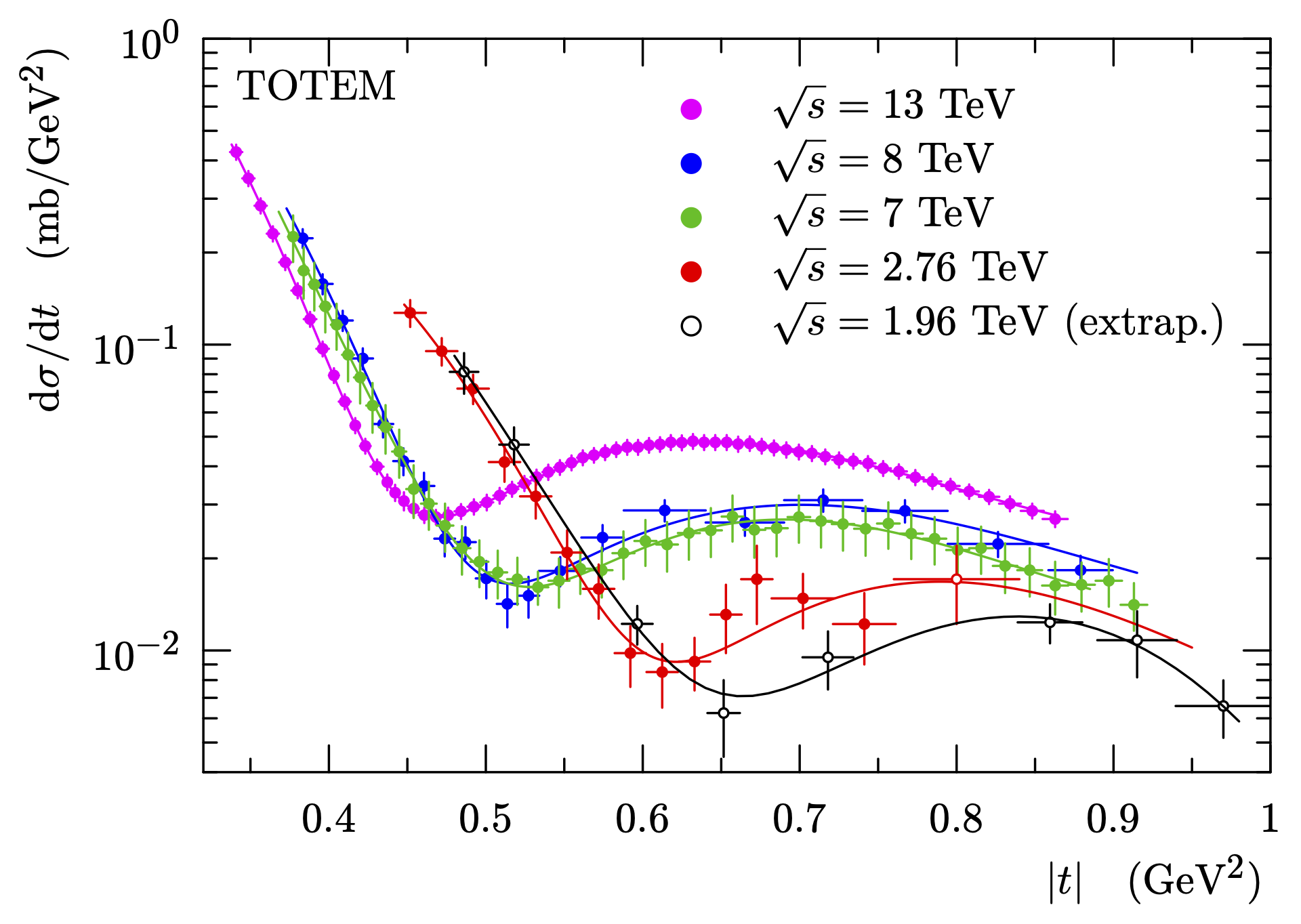}}
\end{minipage}
\hfill
\begin{minipage}{0.45\linewidth}
\centerline{\includegraphics[width=0.925\linewidth]{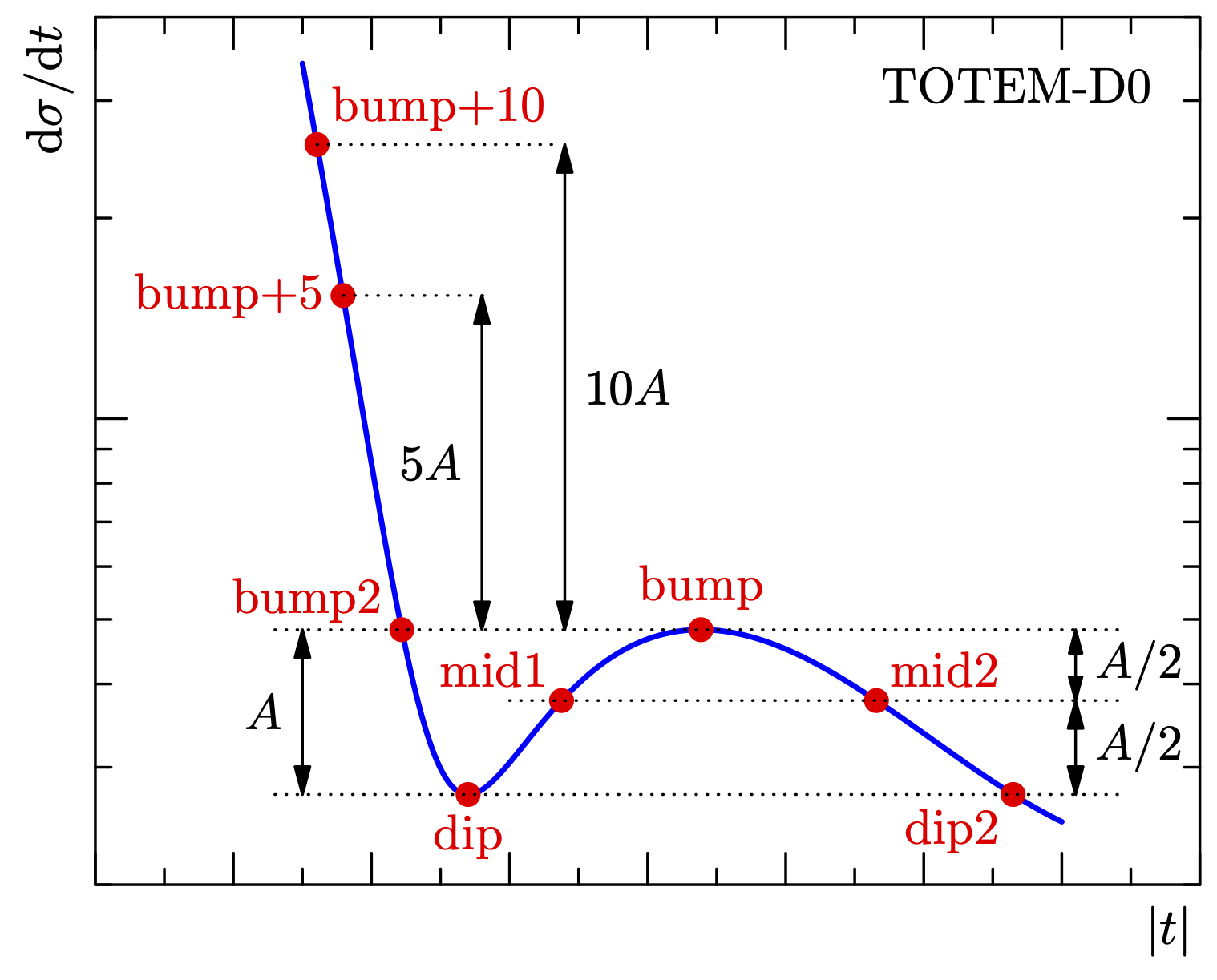}}
\end{minipage}
\hfill
\caption[]{Left: The TOTEM $pp$ elastic cross sections at 2.76, 7, 8, and 13 TeV (full circles), and
the extrapolation to 1.96 TeV (empty circles). Right: Schematic definition of the characteristic points in the TOTEM differential cross section data. $A$ represents the vertical bump to dip distance.}
\label{fig:dsigmadt}
\end{figure}

Each $pp$ $d\sigma_{el}/dt$ measurement at TeV energy scale shows a characteristic dip, followed by a bump, as illustrated by Fig.~\ref{fig:dsigmadt} (left), whereas the $p\bar{p}$ $d\sigma_{el}/dt$ at TeV energy scale only exhibits a flat behaviour in the region of the expected positions of the dip and bump. This difference in the $pp$ and $p\bar{p}$ $d\sigma_{el}/dt$ would naturally occur for $t$-channel odderon exchange, since at the dip the dominant pomeron exchange is largely suppressed, and the odderon amplitude can play a significant role. Contrary to the pomeron amplitude, the odderon amplitude has a different sign for $pp$ and $p\bar{p}$. 

\begin{figure}
\begin{center}
\centerline{\includegraphics[width=0.60\linewidth]{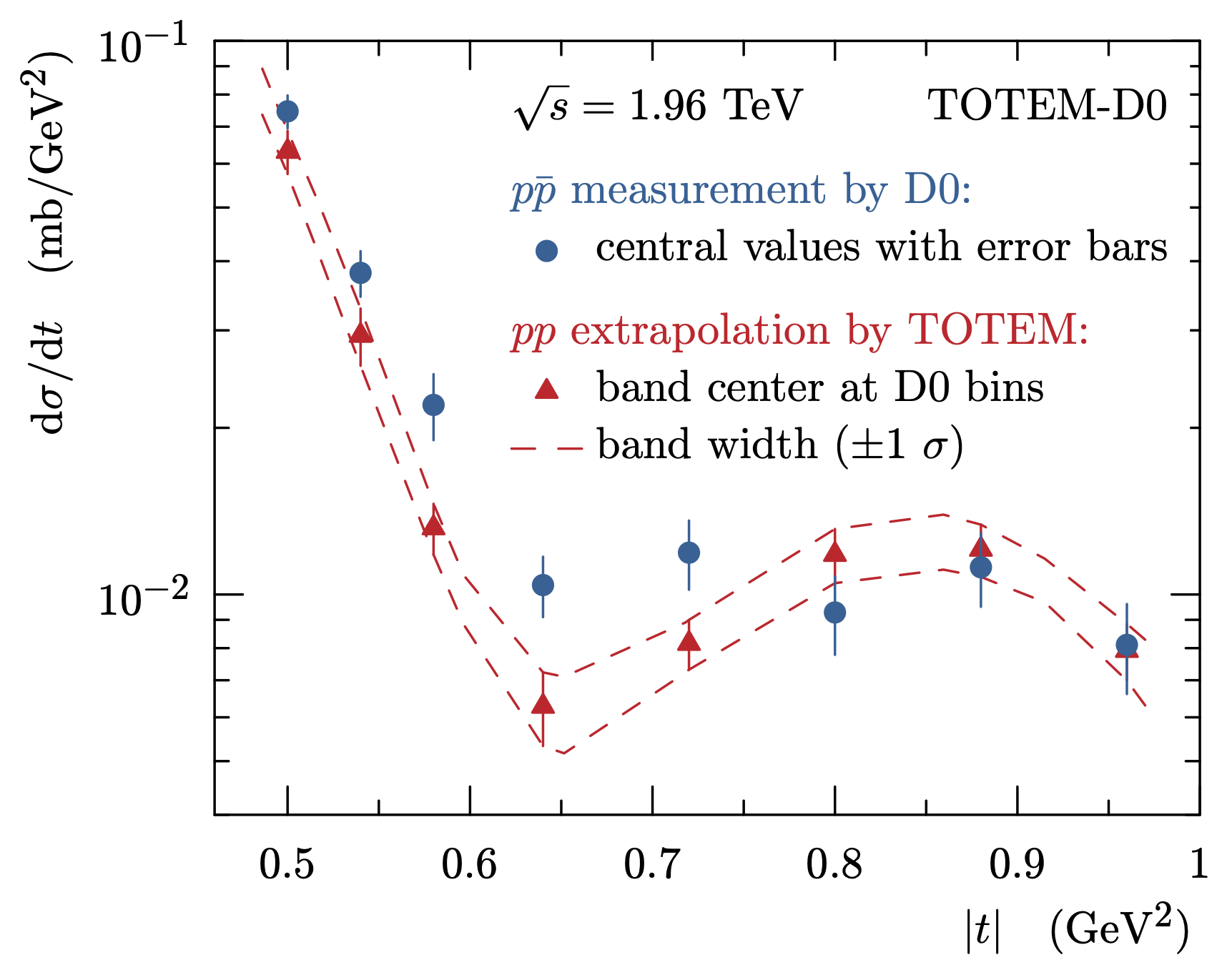}}
\caption{Comparison between the D0 $p\bar{p}$ measurement at 1.96 TeV and the extrapolated TOTEM $pp$ cross section (with its 1$\sigma$ uncertainty band), rescaled to match the D0 optical point. Note that the uncertainties at different $|t|$ values in the 1$\sigma$ uncertainty band are strongly correlated.}
\label{fig:dsigmadt_comp}
\end{center}
\end{figure}

To quantify the difference, eight characteristic points in the region of the dip and the bump, shown in Fig.~\ref{fig:dsigmadt} (right), of the TOTEM 2.76, 7, 8, and 13 TeV $pp$  $d\sigma_{el}/dt$ are extrapolated using a data-driven approach to obtain the 1.96 TeV $pp$ $d\sigma_{el}/dt$. The observed difference of 3.4$\sigma$ significance between the extrapolated $pp$ and the D0 $p\bar{p}$ $d\sigma_{el}/dt$ at 1.96 TeV in the region of the dip and the bump of the $pp$  $d\sigma_{el}/dt$, as shown by Fig.~\ref{fig:dsigmadt_comp}, is interpreted as evidence for odderon exchange. Note that the comparison is made in a common $t$-range (0.50 $\leq |t| \leq$ 0.96 GeV$^2$) of the $pp$ and $p\bar{p}$ $d\sigma_{el}/dt$.

\subsection{Questions and objections raised regarding the analysis and the interpretation}

\begin{figure}
\begin{center}
\centerline{\includegraphics[width=0.85\linewidth]{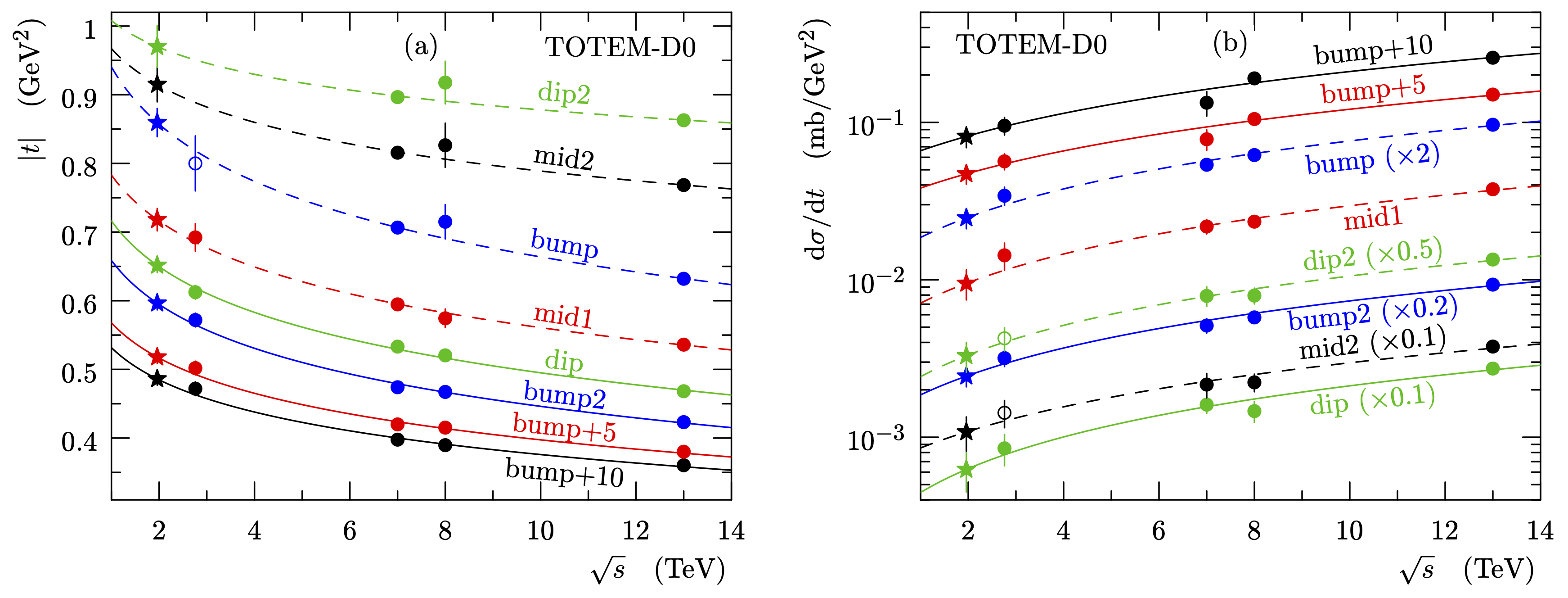}}
\caption{Characteristic points in (a) |t| and (b) $d\sigma / dt$ from TOTEM measurements at 2.76, 7, 8 and 13 TeV (circles) as functions of $\sqrt{s}$ extrapolated to 1.96 TeV (stars). Filled symbols are from measured pints; open symbols are from extrapolations or definitions of the characteristic points.}
\label{fig:dsigmadt_extra}
\end{center}
\end{figure}

A first objection that has been raised is a possible model dependence introduced by the formulas, $|t| = a \log (\sqrt{s} {\rm [TeV]}) + b$ and $d\sigma /dt = c \sqrt{s} {\rm [TeV]} + d$, used to extrapolate the TOTEM measured $|t|$  and differential cross section ($d\sigma /dt$)  values at 2.76, 7, 8 and 13 TeV to 1.96 TeV to obtain the characteristic points of the $pp$ $d\sigma /dt$ at 1.96 TeV, see Fig.~\ref{fig:dsigmadt_extra}. Firstly, it should be noted that the $\sqrt{s}$ range of the extrapolation from 2.76 TeV is small, only about 8 \%, compared to the $\sqrt{s}$ range that the validity of formulas are tested with the fits. Secondly, for each characteristic point, the closest measured point to the characteristic point in terms of $d\sigma /dt$ is used as measured and if two adjacent points have about equal $d\sigma /dt$, the two bins are merged avoiding any model-dependent extrapolation between bins. Thirdly, having 3-4 data points limits the extrapolation formulas to ones with maximally two parameters. Alternative functional forms with other $\log \sqrt{s}$ or $\sqrt{s}$ powers yield extrapolated values at 1.96 TeV well within the uncertainties of the extrapolated values given by the fits using the above $\sqrt{s}$ dependence for $|t|$ and $d\sigma /dt$. Fourthly, it is not obvious that the same functional form would give good fits for all characteristics points both in $|t|$ and $d\sigma /dt$ (majority of $\chi^2$ values $\sim$1 per degree of freedom (d.o.f.)) that probably is related to some general energy independent properties of elastic scattering, see e.g. Refs.~\cite{Martynov_Nicolescu, Durham}. So if there is any model dependence at all, it is largely contained in the quoted uncertainties, in particular due to short extrapolation range and the generality of the functional form used for extrapolating the characteristic points. Note also that the shape and hierarchy of the extrapolated $pp$ $d\sigma /dt$ w.r.t. the measured $pp$ $d\sigma /dt$ is preserved as shown by Fig.~\ref{fig:dsigmadt} (left), i.e. a constant bump-to-dip $d\sigma /dt$ ratio with energy, a descreasing $|t|$ of the diffractive cone, dip and bump position with energy and decreasing values of the $d\sigma /dt$'s in the dip-bump region with energy. Extrapolating the measured cross sections is more robust that fitting the $pp$ $d\sigma /dt$ at each energy and extrapolating the fit parameters, which tend to compensate each other and whose correlations might be different at different energies. 

\begin{figure}
\begin{center}
\centerline{\includegraphics[width=0.55\linewidth]{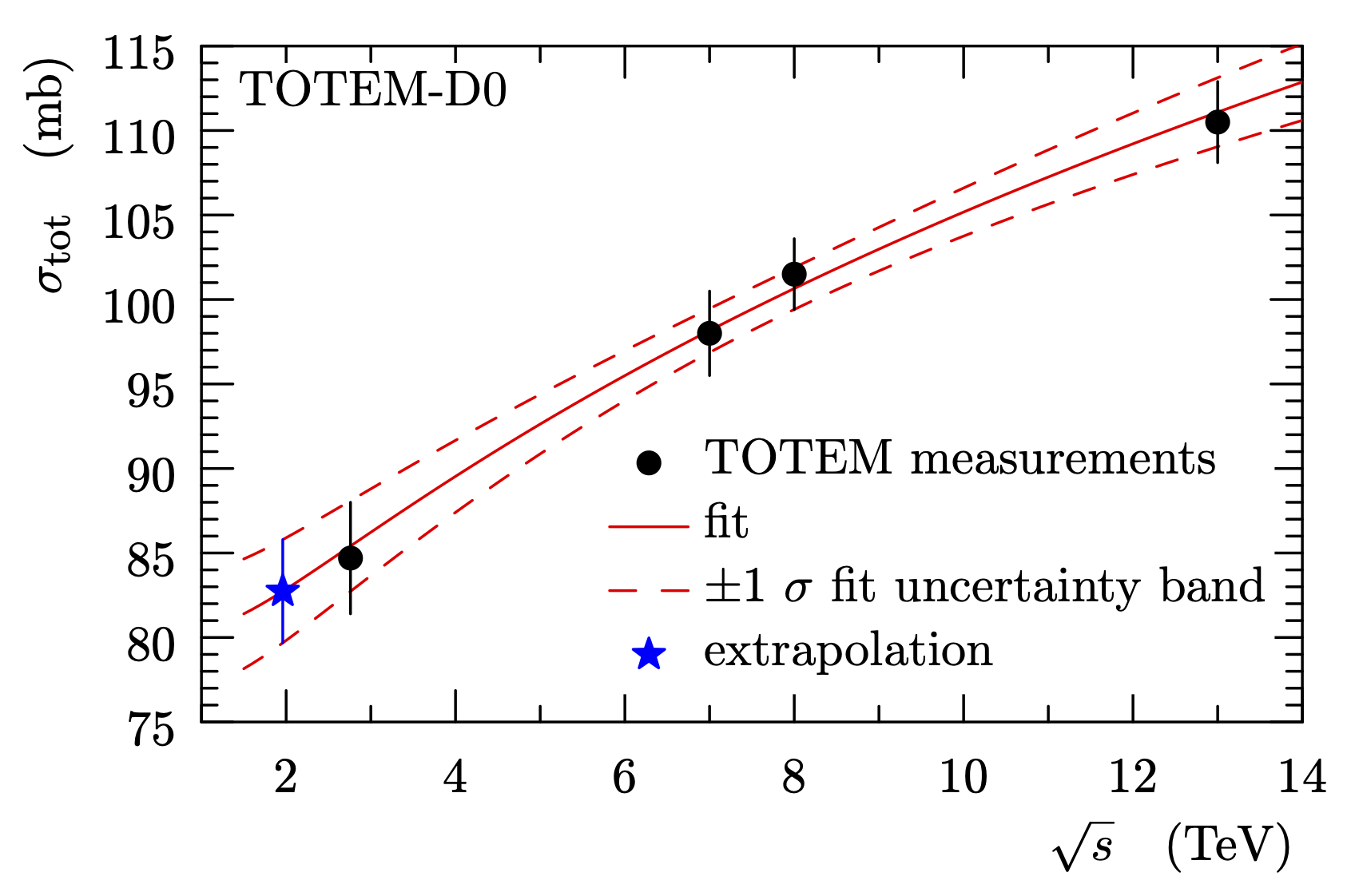}}
\caption{The $\sigma_{tot}$ from TOTEM measurements at 2.76, 7, 8 and 13 TeV (circles) as a functions of $\sqrt{s}$ extrapolated to the center-of-mass energy of the D0 measurement (star).}
\label{fig:sigmatot_extra}
\end{center}
\end{figure}

A similar objection has been raised concerning a possible model dependence introduced by the formula, $\sigma_{tot} = b_1 \log^2 (\sqrt{s} {\rm [TeV]}) + b_2$, used to extrapolate the TOTEM measured total cross section ($\sigma_{tot}$)  values at 2.76, 7, 8 and 13 TeV to 1.96 TeV as shown Fig.~\ref{fig:sigmatot_extra}, obtaining $\sigma_{tot} (pp)$ = 82.7 $\pm$ 3.7 mb at 1.96 TeV. Here the argumentation is similar to the one for the $|t|$ and $d\sigma/dt$ values. Firstly, the $\sqrt{s}$ range of the extrapolation is small, only about 8 \%, compared to the $\sqrt{s}$ range that the validity of formula is tested with the fit. Secondly, having four data points limits the extrapolation formulas to ones with maximally three parameters. Alternative functional forms such as $\log^2\sqrt{s} + \log\sqrt{s} + C$, $s + \sqrt{s} + C$ or  $s^{1/4} + C$ gave extrapolated values at 1.96 TeV well within the quoted $\sigma_{tot} (pp)$ uncertainty. Thirdly, the fit to the TOTEM $\sigma_{tot}$ measurements gives a $\chi^2$ per d.o.f. smaller than 1. So in conclusion, if there is any model dependence, it is well within the quoted uncertainty. Note that 1.96 TeV is in a boundary region for $\sigma_{tot}$, dominated by a $\log \sqrt{s}$ dependence for lower energies and a $\log^2 \sqrt{s}$ dependence for higher energies. Therefore the extrapolation of the TOTEM $\sigma_{tot}$ measurements is only valid for $\sqrt{s} \ge $ 1 TeV, which is sufficient for the purpose above. 

Also a somewhat similar objection has been raised concerning the interpolation of the characteristic points of the $pp$ $d\sigma_{el} /dt$ at 1.96 TeV to the $|t|$ values of the measured D0 $p\bar{p}$ $d\sigma_{el} /dt$ in the range 0.50 $\le |t| \le$ 0.96 GeV$^2$ using the double exponential:
\begin{eqnarray}
h (t) & = &  a_1 e^{-a_2 |t|^2 - a_3 |t|} +  a_4 e^{-a_5 |t|^3 -a_6 |t|^2 - a_7 |t|} \, \,  , 
\label{eq:double_exp}
\end{eqnarray}
where the first exponential describes the diffractive cone (with a steeper slope towards the dip) and the second exponential the asymmetric bump structure and subsequent falloff. The fit to the characteristic points of the $pp$ $d\sigma_{el} /dt$ at 1.96 TeV using Eq.~\ref{eq:double_exp},  gives a $\chi^2$ per d.o.f. smaller than 1. The same functional form describes well the measured $pp$ $d\sigma_{el} /dt$ in the dip and bump region for at 2.76, 7, 8 and 13 TeV, as illustrated in Fig.~\ref{fig:dsigmadt} (left), with a $a_4$ term i.e. bump term significantly different from zero. This reassures that Eq.~\ref{eq:double_exp} can be safely used for the interpolation given that the functional form corresponds to a very distinct shape of the $d\sigma_{el} /dt$. The interpolation uncertainty is evaluated using a MC simulation where the cross section values of the eight uncorrelated characteristic points at 1.96 TeV are varied within their Gaussian uncertainties and new fits given by Eq.~\ref{eq:double_exp} are performed. This provides a $pp$ cross section value at each $|t|$ value that was checked to correspond to a Gaussian distribution with the quoted uncertainty. All of this suggests that the model dependence due to the interpolation must be well within the quoted uncertainty.

Another objection is the assumption that the optical points (OP) (${d\sigma_{el}/dt}\left \vert_{t = 0} \right.$) of $pp$ and $p\bar{p}$ are equal. The basis is the Pomeranchuk theorem~\cite{Pomeranchuk} stating that the ratio of the $pp$ and $p\bar{p}$ $\sigma_{tot}$ is 1, when $\sqrt{s}$ approaches infinity. Using the optical theorem, this leads to the ratio of the OPs of  $pp$ and $p\bar{p}$ to be 1, when $\sqrt{s}$ approaches infinity. This doesn't imply that they are necessarily equal, however any possible difference between them must be due to the C-odd amplitude, which in the TeV-range is due to the odderon, since secondary reggeons can safely be ignored due to the decrease of their amplitude with $\sqrt{s}$, whereas the odderon amplitude is expected to increase with $\sqrt{s}$~\cite{Reggeons}. Therefore the assumption of equal $pp$ and $p\bar{p}$ OP is valid as long as the maximal possible odderon effect on the $\sigma_{tot}$ and hence on the OP is included as a systematic uncertainty for the OP. 

The assumption of equal $pp$ and $p\bar{p}$ OP can be tested comparing the extrapolated ${d\sigma_{el}/dt}\left \vert_{t = 0} \right.$ = 357 $\pm$ 26 mb/GeV$^2$ at 1.96 TeV with the extrapolation of the D0 $d\sigma^{p\bar{p}}_{el}/dt$ measurement to $|t| = 0$ obtaining  ${d\sigma_{el}/dt}\left \vert_{t = 0} \right.$ = 341 $\pm$ 49 mb/GeV$^2$. As can be noted they numerically agree well within the uncertainties, in fact the $p\bar{p}$ OP and its uncertainty encompasses the $pp$ OP and its uncertainty. 

Since the $pp$ and $p\bar{p}$ OP measurements measure the same physics quantity in the assumption of equal $pp$ and $p\bar{p}$ OP, one can estimate a weighted average from them and conclude that the precision on the common OP is determined by the measurement with the better precision, i.e. the $pp$ OP. Therefore the uncertainty on the $p\bar{p}$ OP can be ignored, since the uncertainty of two independent measurement of the same quantity never can be larger than the smaller of the two uncertainties. This procedure is still valid even if the $pp$ and $p\bar{p}$ OP would correspond to two different physics quantities with a known difference as long as the difference is included in the overall uncertainty. The maximal possible difference due to odderon exchange on the OP is estimated from the maximal odderon model to be 2.9 \% at 1.96 TeV that is added in quadrature to the uncertainty of the experimental $pp$ OP to give an overall 7.4 \% relative uncertainty on the common OP. Effects on the OP from secondary reggeons and from differences between the $pp$ and $p\bar{p}$ $\rho$ values are negligible.

Also the ability to extrapolate the D0 $d\sigma^{p\bar{p}}_{el}/dt$ to the OP has been questioned, since the measurement only covers $|t|$-values down to 0.26 GeV$^2$. In particular, since the $B$-slope measurements in $p\bar{p}$ at 0.546 TeV seems to indicate that the $B$-slope is 10-15 \% steeper for low $|t|$-values ($\lesssim$ 0.15 GeV$^2$) than higher $|t|$-values~\cite{UA4_slope}. However neither CDF~\cite{CDF_slope} nor E710~\cite{E710_slope1} observe any indication of a change of $B$-slope of that size below $|t|$ = 0.25 GeV$^2$ at 1.8 TeV.  
Even if the difference between the central values of the two E710 $B$-slope measurements~\cite{E710_slope1, E710_slope2} would be interpreted as an actual $B$-slope difference as a function of $|t|$, the change on the OP would be much smaller ($\sim$ 4 \%) than the luminosity uncertainty of 14.4 \% that dominates the D0 $p\bar{p}$ OP. Comparing TOTEM $\sigma_{tot}$ measurements at $\sqrt{s}$ = 8 and 13 TeV in $pp$ based on $B$-slopes extracted from data with and without acceptance in the Coulomb Nuclear Interference (CNI)-region, the ones with CNI-region data give about 1 \% higher $\sigma_{tot}$ thus about 2 \% higher OP (and steeper $B$-slope). So there is no indication that the D0 $p\bar{p}$ OP cannot be trusted. Note that if the pattern from 1.8 TeV $p\bar{p}$ and 8 and 13 TeV $pp$ measurements at low $|t|$ would be used to correct the D0 $p\bar{p}$ OP at $\sqrt{s}$ = 1.96 TeV for possible bias due to lack of such data, the central values of the D0 $p\bar{p}$ and TOTEM $pp$ OP's would be even closer.   

As a result of the interpolation from the characteristic points of the extrapolated $pp$ $d\sigma /dt$ to the $|t|$ values of the D0 $p\bar{p}$ $d\sigma /dt$, the $pp$ $d\sigma /dt$ at the $|t|$ values of the $p\bar{p}$ $d\sigma /dt$ are strongly correlated implying that the full covariance matrix of the $pp$ data points must be included in the $\chi^2$ for the comparison of the $pp$ and $p\bar{p}$ $d\sigma /dt$. The $\chi^2$-formula used:

\begin{eqnarray}
\chi^2 & = &  \sum_{i,j=1}^8 \left\{ \left( \frac{d\sigma^{pp, norm}_{el,i}}{dt} - \frac{d\sigma^{p\bar{p}}_{el,i}}{dt} \right) C^{-1}_{i,j} \left( \frac{d\sigma^{pp, norm}_{el,j}}{dt} - \frac{d\sigma^{p\bar{p}}_{el,j}}{dt} \right) \right\} + \frac{(A-A_0)^2}{\sigma_A^2} + \frac{(B-B_0)^2}{\sigma_B^2} \, \,  , 
\label{eq:chi2}
\end{eqnarray}

where $C_{i,j}$ is the covariance matrix, $A$ and $B$ are the two constraints and $d\sigma^{pp, norm}_{el,i} /dt$ is the $pp$ $d\sigma_{el}/dt$ normalized to the $p\bar{p}$ integral elastic cross section ($\sigma_{el}$) in the $|t|$ range of the comparison. The first constraint ($A$) is the normalization due to the matching of the $pp$ and $p\bar{p}$ OPs. The second constraint ($B$) is the matching of the $pp$ and $p\bar{p}$ $B$-slopes in the diffractive cone. The Pomeranchuk and the optical theorem infer that the ratio of the $pp$ and $p\bar{p}$ total $\sigma_{el}$ should be 1, when $\sqrt{s}$ goes to infinity. From this, one can deduce that the ratio of the $pp$ and $p\bar{p}$ elastic $B$-slopes should be 1, when $\sqrt{s}$ approaches infinity, since the $\sigma_{el}$ in the Coulomb region and in the region beyond the dip is negligible compared the one in the diffractive cone and the $d\sigma_{el}/dt$ in the diffractive cone is described by $e^{-B|t|}$~\cite{Cornille_Martin}. This doesn't imply that they are exactly equal but any difference between the $pp$ and $p\bar{p}$ elastic $B$-slopes at the TeV-scale is due to the odderon. Since the pomeron dominates in the diffractive cone region at 1.96 TeV, the $B$-slopes of $pp$ and $p\bar{p}$ are expected to be equal. This is verified to be true within the experimental uncertainties for the D0 $p\bar{p}$ and the TOTEM $pp$ $B$-slopes.

Therefore Eq.~\ref{eq:chi2} expresses the complete $\chi^2$, including the covariance matrix and the terms for the fully correlated uncertainties, thus also expressing the exact number of d.o.f. Eq.~\ref{eq:chi2} gives for six d.o.f. a significance of 3.4$\sigma$ for the difference between the TOTEM $pp$ and the D0 $p\bar{p}$ $d\sigma_{el}/dt$ at 1.96 TeV using the eight points in the region of the dip and the bump. The $\chi^2$ and therefore the significance is largely dominated by the first term of Eq.~\ref{eq:chi2} related to the shape of the $d\sigma_{el}/dt$. The obtained significance is confirmed by a Kolmogorov-Smirnov test of the difference between the $pp$ and $p\bar{p}$ $d\sigma_{el}/dt$ in the same $|t|$ range, where the correlations of the data points are included using Cholesky decomposition~\cite{Cholesky} and the normalisation difference via Stouffer's method~\cite{Stouffer}. 

\section{The TOTEM $\rho$ and $\sigma_{tot}$ measurements}

The second evidence of odderon exchange in elastic scattering is from the measurements of $\rho$, the ratio of the real and imaginary part of the elastic hadronic amplitude at $t$ = 0, and $\sigma_{tot}$ in $pp$ collisions at the LHC~\cite{TOTEM-rho-13TeV}. Models~\cite{Compete, Durham, Block_Halzen} are unable describe both the TOTEM $\sigma_{tot}$ and $\rho$ measurements without including odderon exchange. The disagreement between the measurements and the models is between 3.4$\sigma$ and 4.6$\sigma$ depending on the model. Comparison between the predictions of the COMPETE models~\cite{Compete} and the TOTEM $\sigma_{tot}$ and $\rho$ measurements is shown in Fig.~\ref{fig:compete_sigmatot_rho}. Note that the COMPETE~\cite{Compete} and Block-Halzen~\cite{Block_Halzen} models include secondary Reggeon-like C-odd terms proportional to $\sim 1/\sqrt{s}$ to describe the difference of $pp$ and $p\bar{p}$ scattering below 0.1 TeV that should not be confused with odderon-like C-odd terms that are expected to increase with $\sqrt{s}$.

When comparing different $\rho$ measurements, it is important to make sure that the prescriptions (functional form for the hadronic amplitude and the phase, CNI formula and $|t|$-range) used in the extraction are as similar as possible, otherwise it doesn't necessarily lead to the same physics quantity. This is especially true in the comparison with previous $\rho$ measurements. The $\sqrt{s}$ trend in the TeV range predicted by odderon exchange~\cite{Martynov_Nicolescu, Durham13TeV} is observed for the most precise $\rho$ measurements for $pp$ and $p\bar{p}$ in the TeV range, when extracted using the same prescription:  $\rho$ = 0.135 $\pm$ 0.015 at 0.546 TeV in $p\bar{p}$~\cite{UA4_rho} and $\rho$ = 0.09 $\pm$ 0.01 at 13 TeV in $pp$~\cite{TOTEM-rho-13TeV}. Note also that several groups, including A. Donnachie and P.V. Landshoff~\cite{Donnachie_Landshoff} and J.R. Cudell and O.V. Selyugin~\cite{Cudell_Selyugin}, have obtained compatible $\rho$ values (in the range 0.08-0.10), when taking the TOTEM 13 TeV CNI data as given and using a similar prescription as TOTEM~\cite{TOTEM-rho-13TeV}, contrary to the results they quote when they misinterpret or allow themselves the freedom to shift the TOTEM data and related uncertainties.

\subsection{Questions and objections raised concerning the analysis and the interpretation}

The authors of the PDG review of High Energy Soft QCD and Diffraction~\cite{PDG} claim that analyzing the whole ensemble of TeV-range elastic $pp$ and $p\bar{p}$ low $|t|$ data including the TOTEM measurements at LHC, a reasonable description can be obtained using a C-even amplitude (pomeron) only, that is, without an odderon, in contradiction with the conclusion by TOTEM. This statement does not hold once one start to examine the exact predictions. For example, the model of the authors~\cite{Durham} fails to describe both the TOTEM $\rho$ and $\sigma_{tot}$ measurements in $pp$ at 7, 8 and 13 TeV ($\sim$ 3.4$\sigma$ difference) and especially the elastic $d\sigma/dt$ in $p\bar{p}$ for the dip and bump region at 1.96 TeV ($\sim$ 4.3$\sigma$ difference). A good description of the LHC $pp$ data without the odderon, leads inevitably to a significantly worse description of the Tevatron $p\bar{p}$ data and vice versa.

\begin{figure}
\begin{center}
\centerline{\includegraphics[width=0.735\linewidth]{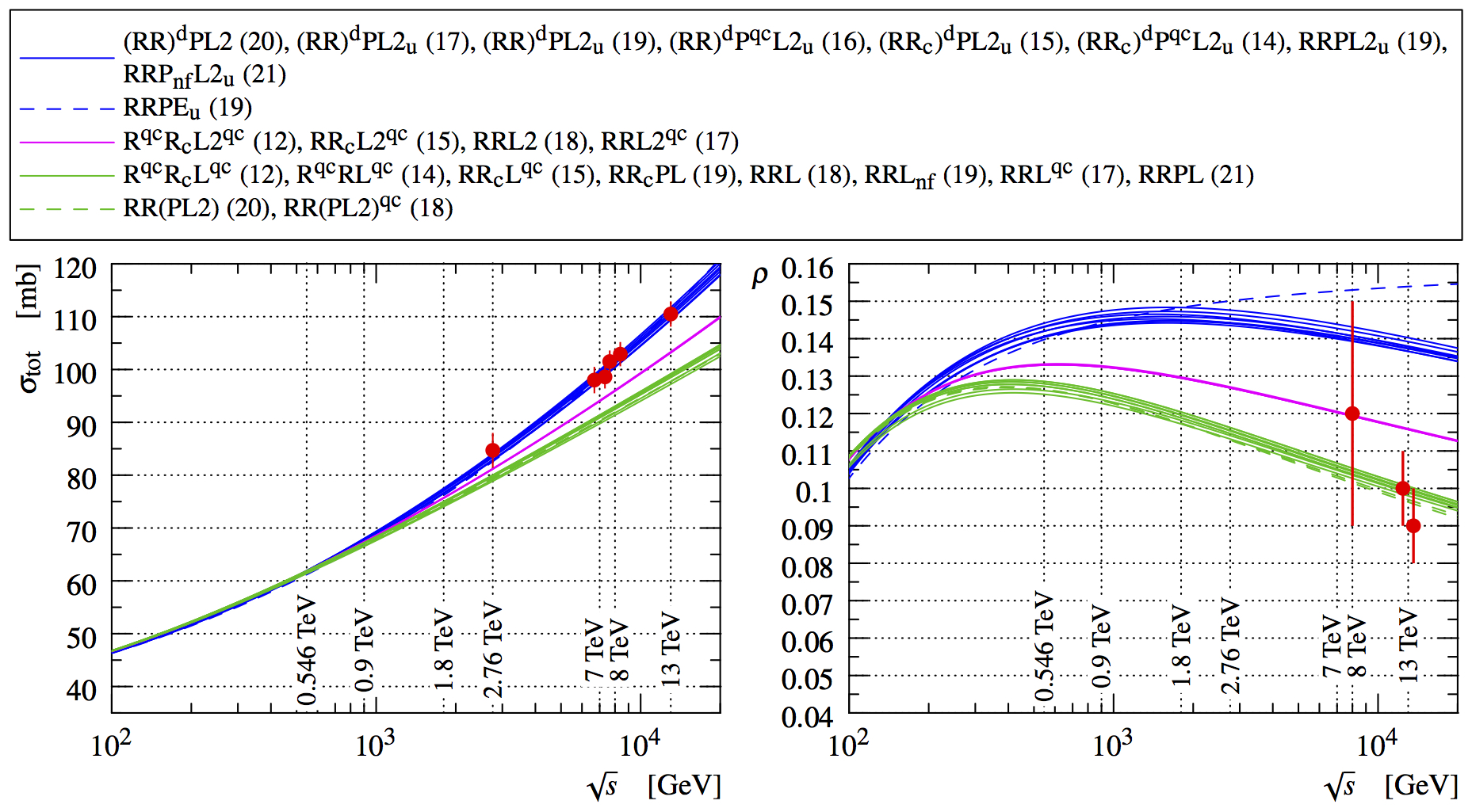}}
\caption{Predictions of the $pp$ total cross section ($\sigma_{tot}$) and $\rho$ parameter as function of $\sqrt{s}$ by each COMPETE~\cite{Compete} model (see legend for model) with the TOTEM measurements marked in red.}
\label{fig:compete_sigmatot_rho}
\end{center}
\end{figure}

\begin{figure}
\begin{minipage}{0.50\linewidth}
\centerline{\includegraphics[width=0.885\linewidth]{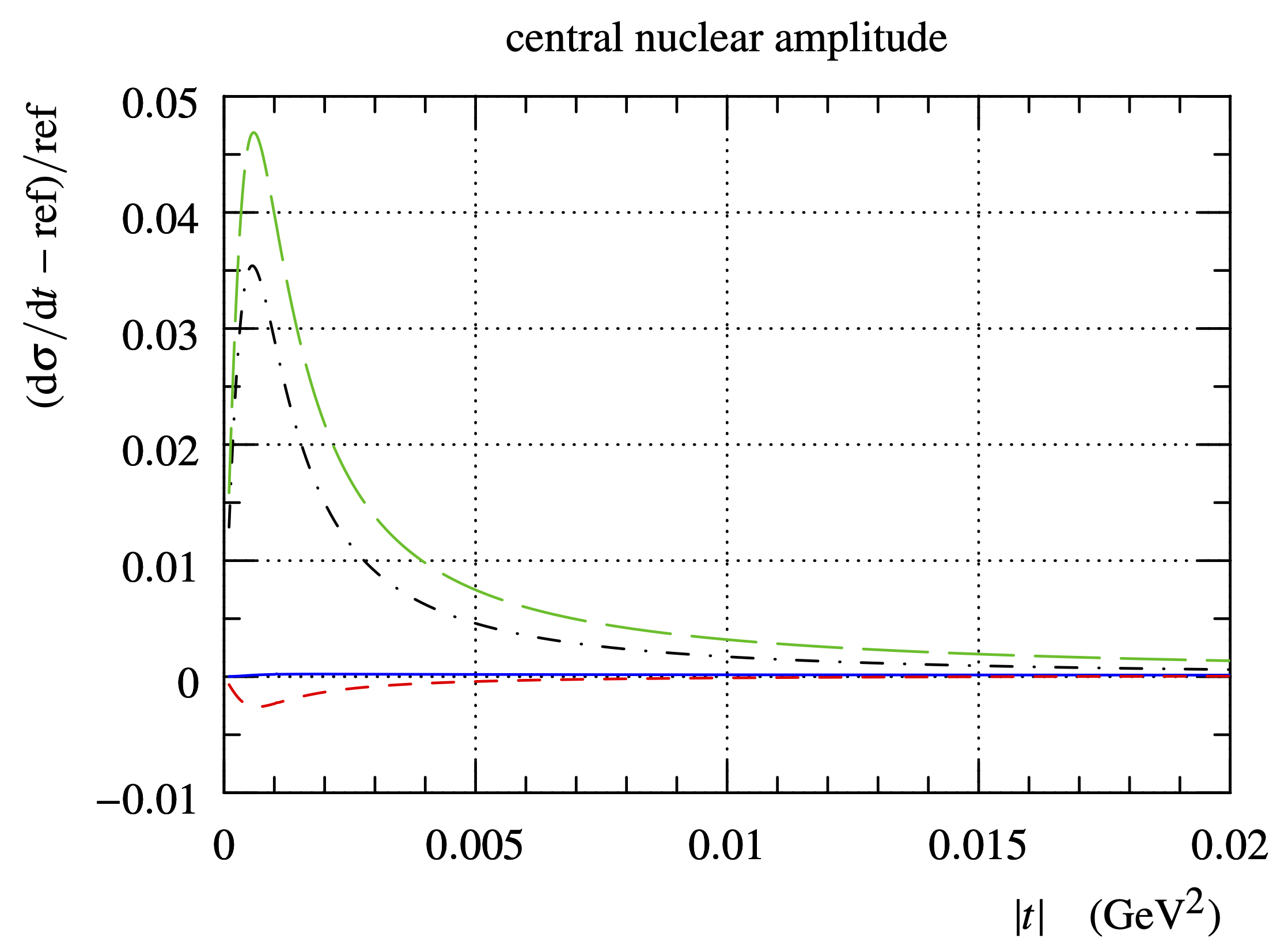}}
\end{minipage}
\hfill
\begin{minipage}{0.50\linewidth}
\centerline{\includegraphics[width=0.885\linewidth]{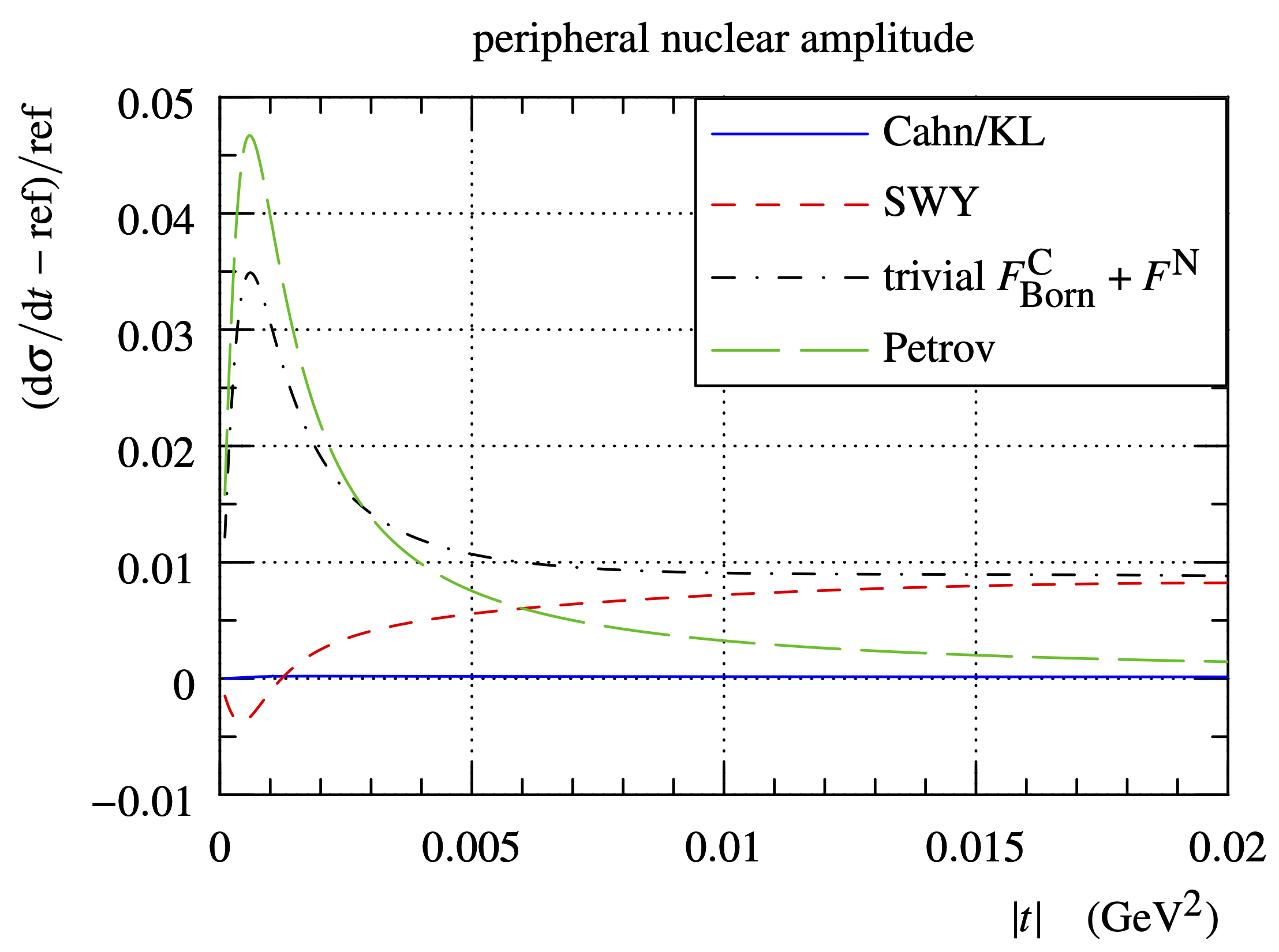}}
\end{minipage}
\hfill
\caption[]{The relative difference of the differential elastic cross section in the CNI region between various CNI formulae (see text) and the numerical calculation of the Coulomb and nuclear eikonals to all orders of $\alpha$ (denoted "ref") ~\cite{Kaspar} for central (left) and peripheral (right) nuclear amplitudes. The labelling refects the impact parameter behaviour: central nuclear amplitudes yield profile functions peaking at smaller impact parameter value than peripheral amplitudes.}
\label{fig:Kaspar_diff}
\end{figure}

In the PDG review, the statement of the authors is backed up by a similar attempt by Donnachie-Landshoff~\cite{Donnachie_Landshoff}, which claim to describe the elastic $d\sigma/dt$ data at small $|t|$ from 13.76 GeV to 13 TeV without the odderon. Donnachie-Landshoff obtain a $\rho$ = 0.14 in $pp$ at $\sqrt{s}$ = 13 TeV, when using the TOTEM 8 TeV CNI data~\cite{TOTEM-rho-8TeV} in addition to the TOTEM 13 TeV CNI data~\cite{TOTEM-rho-13TeV}, whereas when using only the TOTEM 13 TeV CNI data they obtain a $\rho$ = 0.10. This is not possible, if experimental uncertainties are treated correctly, since the TOTEM 13 TeV CNI data is about a factor three more precise than the TOTEM 8 TeV CNI data when the normalisation uncertainty is not taken into account. It is likely that the normalisation uncertainty that is common to all data points has not been treated correctly as a separate term $A$ in the $\chi^2$ as in Eq.~\ref{eq:chi2} in the fits by Donnachie-Landshoff. Otherwise one cannot explain the large weight the TOTEM 8 TeV CNI data obtains in the Donnachie-Landshoff fits. The normalisation uncertainty is the dominating uncertainty in the TOTEM CNI data except at the smallest $|t|$ values and  smaller in the TOTEM 8 TeV CNI data than in the TOTEM 13 TeV CNI data, 4.2 \% compared to 5.5 \%. Note also that in Ref.~\cite{Donnachie_Landshoff}, a trivial sum of Coulomb and nuclear elastic amplitudes is used, ignoring completely CNI effects on the amplitude, leading to relative deviations in the total elastic amplitude of several percent, see Fig.~\ref{fig:Kaspar_diff}.   

The PDG review also states that the model RR(PL2)$^{qc}$ of COMPETE (dashed green line in Fig.~\ref{fig:compete_sigmatot_rho}) is consistent with the TOTEM 13 TeV $\rho$ and $\sigma_{tot}$ within 1$\sigma$~\cite{Cudell_Selyugin}, in contradiction with the statement that all COMPETE models are incompatible with the TOTEM $\rho$ and $\sigma_{tot}$ measurements. This agreement with the RR(PL2)$^{qc}$ model is obtained by modifying the normalization of the TOTEM 13 TeV elastic $d\sigma/dt$ by $\sim$2$\sigma$ (when including the Coulomb normalization that was not taken into account in Ref.~\cite{Cudell_Selyugin}). Since the normalization of the TOTEM 13 TeV CNI data is obtained from two completely independent data sets and methods (optical theorem and Coulomb amplitude) that agree very well, it is unlikely that it is off by $\sim$2$\sigma$. The standard approach in physics is not to modify the data but instead adjust the model to describe the data and not vice versa. Without modifying the normalization of the TOTEM 13 CNI data, the original version of the RR(PL2)$^{qc}$ model~\cite{Compete} fails to describe the $\sigma_{tot}$ in pp at $\sqrt{s}$ = 2.76, 7, 8 and 13 TeV ($\sim$5.4$\sigma$ difference).

Regarding the determination of $\rho$, it important to stress that most of the sensitivity to $\rho$ is contained in only a few data bins in the CNI region, between those at very low $|t|$ with a significantly larger Coulomb than CNI contribution and the large majority of data bins at higher $|t|$, where the hadronic amplitude dominates. Experience from TOTEM has shown that the fits should be done in several steps in separate $|t|$ ranges, first to fix the other parameters (hadronic amplitude and Coulomb normalisation) before the $\rho$ to avoid any bias in the $\rho$ determination from data bins with very little or without any sensitivity to $\rho$, see e.g. section 6.3 in Ref.~\cite{TOTEM-rho-13TeV}. In Refs.~\cite{Donnachie_Landshoff, Cudell_Selyugin} it is not stated whether the fits have been performed in several steps to avoid bias in the $\rho$ determination from data bins with minimal sensitivity to $\rho$ or whether they have been performed in a single step.   

In addition, the CNI formulae of Cahn~\cite{Cahn} and Kundr{\' a}t-Locaji{\v c}ek (KL)~\cite{KL} used for the $\rho$ determination at 13 TeV have been claimed to contain flaws including inexact approximation of the Coulomb amplitude and too early truncation of the power series of the electromagnetic coupling $\alpha$~\cite{Petrov}. A numerical calculation of the Coulomb and nuclear eikonals to all orders of $\alpha$~\cite{Kaspar} verified that the CNI formulae of Cahn and KL reproduce the numerical estimate for the phase and the $d\sigma/dt$ at a precision significantly below the current experimental one, as shown by Fig.~\ref{fig:Kaspar_diff}. Hence any approximations done by Cahn and KL do not have any detrimental effect on the $\rho$ determination. Instead, the CNI formula of Ref.~\cite{Petrov} and the sum of Coulomb and nuclear amplitudes~\cite{Godizov} were found to deviate from the numerical estimate by several percent. The SWY formula~\cite{SWY} reproduces the numerical estimate for central nuclear amplitudes but not for peripheral ones, see Fig.~\ref{fig:Kaspar_diff}. Also the effect of not including excited proton states in the eikonal have been estimated to be negligible compared to the current experimental precision~\cite{Bethe}. In conclusion, the formulae used for the 13 TeV $\rho$ determination provide more than adequate models for the CNI effects.     

\section{The combination of the $pp$ and $p\bar{p}$ comparison and $\rho$ and $\sigma_{tot}$ measurements}
The significances of the measurements are combined using the Stouffer's method~\cite{Stouffer} in the order of sensitivity, starting from the $pp$ and $p\bar{p}$ comparison, adding the 13 TeV $\rho$ measurement and then finally if needed the $\sigma_{tot}$ measurements using the freedom provided by Stouffer's method to use only a subset of the significances (e.g. $\rho$ and the $pp$ and $p\bar{p}$ comparison) for testing the exclusion of a model. The $\chi^2$ for
the $\sigma_{tot}$ measurements at 2.76, 7, 8 and 13 TeV is computed with respect to the model predictions without odderon exchange~\cite{Compete, Durham, Block_Halzen} including also model uncertainties when specified. Same was done separately for the 13 TeV $\rho$ measurement. Unlike the COMPETE~\cite{Compete} and Block-Halzen~\cite{Block_Halzen} models, the Durham model~\cite{Durham} provides the predicted $d\sigma_{el}/dt$ without odderon exchange contribution. Therefore a direct comparison of the predicted Durham $d\sigma_{el}/dt$ at 1.96 TeV with the D0 $p\bar{p}$ $d\sigma_{el}/dt$ that gives a significance of 4.3$\sigma$ is used for the combined significance instead of the $pp$ and $p\bar{p}$ comparison. The 1.96 TeV $d\sigma_{el}/dt$ of the model is chosen since it is most sensitive to odderon exchange after the model has been tuned to the LHC elastic $pp$ data.

The 13 TeV $\rho$ measurement provides a 4.6 and 3.9$\sigma$ significance for the COMPETE "blue band" (see Fig.~\ref{fig:compete_sigmatot_rho}) and the Block-Halzen models~\cite{Block_Halzen}, respectively. The comparison of $\rho$ and $\sigma_{tot}$ measurements with the predictions of the Durham~\cite{Durham}, the COMPETE "magenta band" and "green band" (see Fig.~\ref{fig:compete_sigmatot_rho}) models give significances of 3.4, 4.0 and 4.6$\sigma$, respectively. Combining them with the significance of the $pp$ and $p\bar{p}$ comparison (or for Durham the one with D0) give combined significances ranging
from 5.2 to 5.7$\sigma$ for odderon exchange for all examined models~\cite{Compete, Durham, Block_Halzen}.

\subsection{Questions and objections raised about the combination}
The Stouffer's method~\cite{Stouffer} combines significances following $z_{\rm comb} = \sum^{k}_{i=1} z_i/\sqrt{k}$, where $z_i$ is the individual significances and $k$ the number of significances to be combined. The method is valid for independent measurements, whose significances obey the normal distribution. This is true for the odderon significances obtained from the $pp$ and $p\bar{p}$ comparison in the dip-bump region and the $pp$ $\rho$ and $\sigma_{tot}$ measurements at very low $|t|$, since they are based on results from completely separate $|t|$ regions and TOTEM data sets. When the 13 TeV  $\rho$ and $\sigma_{tot}$ measurements are both used for the combined significance, values determined from independent TOTEM data sets are used. 

It has also been questioned whether the $pp$ and $p\bar{p}$ comparison and the $\rho$ and $\sigma_{tot}$ measurements can be combined, since the former is a data to data comparison and the latter a data to model comparison. However, since the only way to produce a significant difference between the $pp$ and $p\bar{p}$ $d\sigma_{el}/dt$ at TeV energy scale is through odderon exchange, a model without odderon exchange would produce a $p\bar{p}$ $d\sigma_{el}/dt$ at 1.96 TeV similar to the extrapolated $pp$ $d\sigma_{el}/dt$ if the model still has to describe the $pp$ $d\sigma_{el}/dt$'s measured at LHC. This is illustrated by the Durham model without odderon contribution that fails to describe the D0 $p\bar{p}$ $d\sigma_{el}/dt$ at 1.96 TeV (at a 4.3$\sigma$ significance). Also the failure of the models to describe simultaneously both the $\rho$ and $\sigma_{tot}$ measurements in $pp$ points to a difference in elastic $pp$ and $p\bar{p}$ scattering and therefore to be quantitatively assessing the same thing, the existence of odderon exchange in elastic scattering, as the $pp$ and $p\bar{p}$ comparison. 

\section{Conclusions}
Issues and objections raised regarding the D0-TOTEM comparison of the elastic $d\sigma/dt$ of $pp$ and $p\bar{p}$, the TOTEM $\rho$ and $\sigma_{tot}$ measurements in $pp$ as well as their combination and odderon interpretation have been adequately addressed. Both provide evidence of odderon exchange in elastic scattering and their combination constitute the first experimental observation of the odderon, acknowledged as convincing evidence of the existence of the odderon after a quest of almost 50 years~\cite{Leader}.

\nocite{*}
\bibliographystyle{auto_generated}
\bibliography{Odderon_lowx_Osterberg/Osterberg}

%% file: Petrov_article/Petrov.tex
\vspace*{1.2cm}

\thispagestyle{empty}
\begin{center}
{\LARGE \bf Odderon: Lost or/and Found?}

\par\vspace*{7mm}\par

{

\bigskip

\large \bf Vladimir Petrov\footnote{Speaker}  and Nikolay Tkachenko }

\bigskip

{\large \bf  E-Mail: Vladimir.Petrov@ihep.ru}

\bigskip

{Logunov Institute for High Energy Physics, NRC "Kurchatov Institute", Protvino, RF}

\bigskip

{\it Presented at the Low-$x$ Workshop, Elba Island, Italy, September 27--October 1 2021}

\vspace*{15mm}

\end{center}
\vspace*{1mm}

\begin{abstract}
    
    This is a quick survey of theoretical and experimental efforts to understand and identify the Odderon.
\end{abstract}
 \part[Odderon: Lost or/and Found?\\ \phantom{x}\hspace{4ex}\it{Vladimir Petrov and Nikolay Tkachenko}]{}
 
 \section{Introduction}
 In a long and rich history of the studies on high energy hadron interactions two kinds of strong ("nuclear") forces are constantly featured: C-even and C-odd ones. 
 C-even forces represented by the Pomeron and f-Reggeon bear a universal character, i.e. 
they are like the forces of gravity acting as the universal attraction of all hadrons to each other. On the contrary, C-odd forces, which in high-energy physics are associated with $ \rho $-, $ \omega $-, etc. Reggeons resemble electromagnetic interactions, when charges of the same sign (particle-particle) repel, and charges of different signs (particle-antiparticle) are attracted to each other. 

Until the beginning of the 70s, the belief reigned that with the increase in collision energy, the main C-even agent, the Pomeron, plays an increasingly dominant role, while the C-odd forces become less and less significant ("die out") and finally can be neglected.

Such a paradigm was challenged in the works of B. Nicolescu et al.\cite{luk}, , in which a new notion was introduced, later dubbed "Odderon", which, being a part of the amplitude subleading (w.r.t.the Pomeron) in the imaginary part of the scattering amplitude, becomes the leading one in the real part. 
It worth noticing that the very term" Odderon" looks akin to the name of some new Reggeon but according to \cite{luk} this was but a specific contribution to the C-odd part of the scattering amplitude. No Reggeon (or a particle) was associated with this "Odderon"\footnote{Later the Odderon option as formulated in \cite{luk} was dubbed (after serious modfifications) the "Maximal Odderon" in contrast to other Odderon incarnations (e.g. as a C-odd Reggeon). As we will not concern these models , we will  use the term "Odderon" in the sense of "Maximal Odderon" as well.}.

Below we will try to trace almost a half-of-century history of the Odderon concept and attempts and efforts to find its experimental manifestations.
 
\section{Odderon: Predictions and Nature}
Which observables are potentially suitable for identifying the possible existence of the Odderon?  Below we briefly describe a few options used.

\subsection{Difference of the proton-proton  and anti proton-proton total cross-sections. }

\begin{wrapfigure}[36]{l}{110mm}
\vspace{-6.1mm}
\includegraphics[width=110mm]{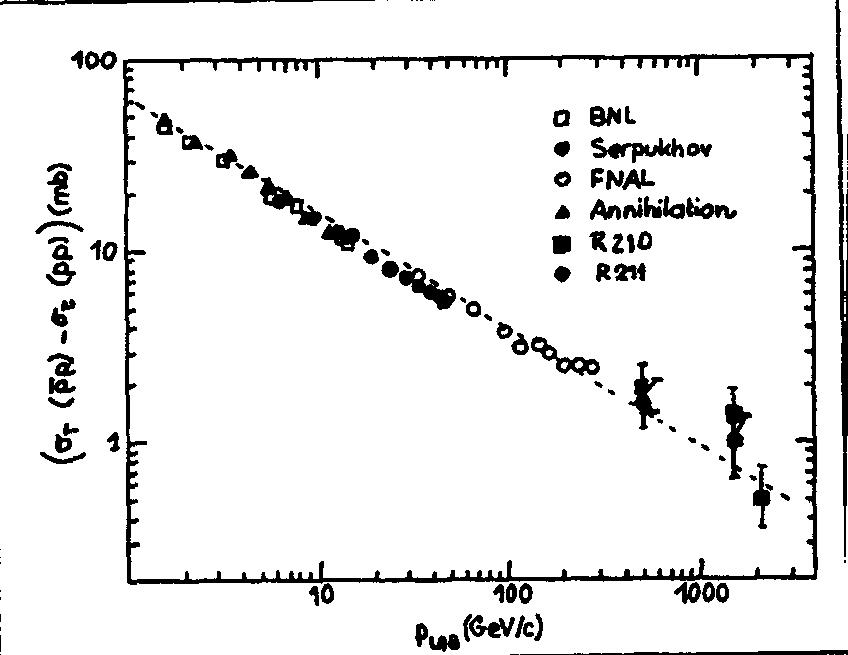}
\vspace{-7.6mm}
\caption{The energy evolution of the difference $ \Delta\sigma = \sigma_{tot}^{\bar{p} p} - \sigma_{tot}^{pp}$ at $ \sqrt{s} \leqslant 60\: GeV $.}
\label{p1}
\end{wrapfigure}
\begin{wrapfigure}[24]{l}{110mm}
\includegraphics[width=110mm]{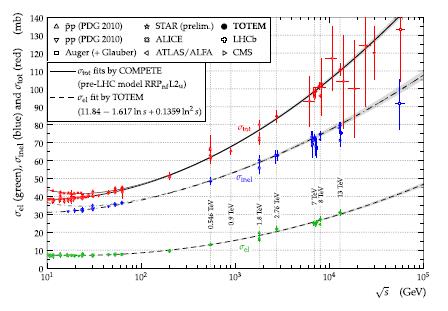}
\caption{The energy evolution of the cross-sections in  $ \bar{p} p $ and $ pp $ collisions.}
\label{p2}
\end{wrapfigure}

The simplest one is the difference between the total cross-sections of $ \bar{p} p $ and $ p p $ interactions 
\begin{center}
$ \Delta\sigma = \sigma_{tot}^{\bar{p} p} - \sigma_{tot}^{pp}.$
\end{center}
because it is exactly a C-odd quantity.

The first paper in Ref.\cite{luk} predicted that "at high energies($ \sqrt{s} $)" 
\begin{center}
$ \mid \Delta\sigma \mid \sim \ln s  $
\end{center}
i.e. grows indefinitely with energy. 

Pre-ISR data showed that $ \Delta\sigma $ is positive and decreases with energy growth.
The ISR data ($ \sqrt{s}= 20\div 63\: GeV $) confirmed this trend and gave the last opportunity to compare $ \sigma_{tot}^{\bar{p} p} $ and $ \sigma_{tot}^{pp} $ at the same energy. The minimum value of the difference as measured at the ISR \cite {amb} was
 \begin{center}
$ \Delta\sigma (52.8\mbox{ GeV}) = 1.49\pm 0.35\mbox{ mb}$
\end{center}
Fig.1 \cite{Ow}  shows the early result for $\Delta\sigma$ for laboratory energies
$$E_{\mbox{\mbox{lab}}}\approx p_{\mbox{\small{lab}}}c \leqslant 2000\mbox{ GeV}$$
(cms energy up to $ 60\mbox{ GeV}$). 

However, in the mentioned first article on the Odderon it was argued that $\Delta\sigma $ should drop till
 $\sqrt{s} \approx 24$ GeV
where it should disappear and then (after it would turn negative) would begin to indefinitely grow in absolute value achieving $  - \:\mathcal{O}( 10 mb) $ at $ p_{\mbox{lab}}c = 10^{4}\mbox{ GeV}~ (\sqrt{s} \approx
150\mbox{ GeV}$). This would be a clear evidence in favour of the Odderon as formulated in \cite{luk}  but the ISR measurements, as we see,  ruled out such an option.

Meanwhile $\Delta\sigma  $ well collaborated with the prediction of the Regge pole scheme
\begin{center}
$ \Delta\sigma \sim s^{\alpha_{-}(0)-1} $
\end{center}
where $ \alpha_{-}(0) $ is the intercept of the "secondary" Reggeons ($ \rho, \omega $ etc). As generically $ \alpha_{-}(0)\approx 0.5 $ we see that Fig.1 seemed to confirm the asymptotic disappearing of $ \Delta\sigma  $.

This , however, did not discourage the Odderon proponents who argued that the crossover of $ \sigma_{\mbox{tot}}^{\bar{p} p}$ and
$\sigma_{\mbox{tot}}^{pp}$ had a chance to show up  at higher energies.

\begin{wrapfigure}[18]{l}{110mm}
\vspace{-0.1mm}
\includegraphics[width=110mm]{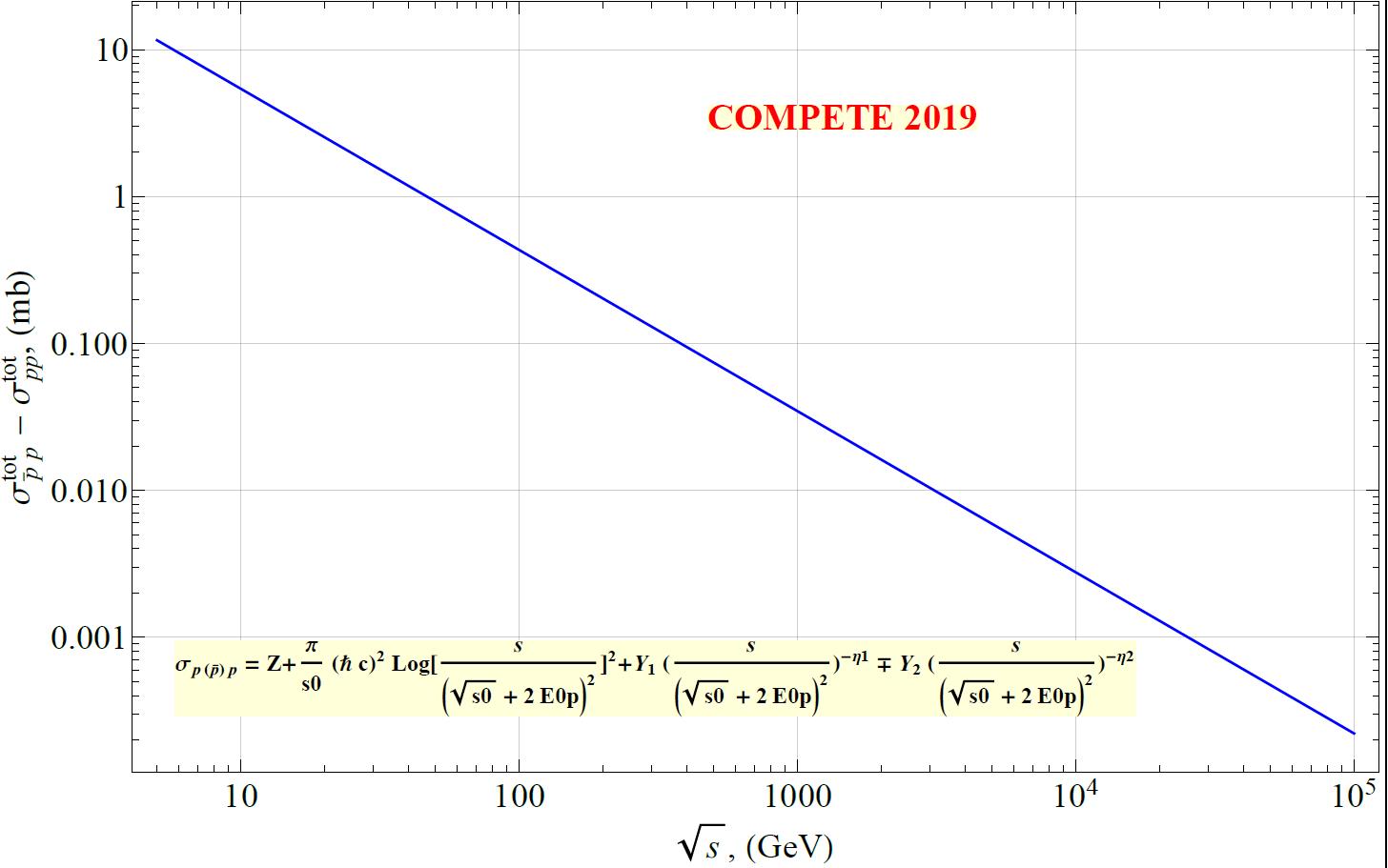}
\vspace{-4.6mm}
\caption{The energy evolution of $ \Delta\sigma $ as given by the COMPETE parametrization.}
\label{p3}
\end{wrapfigure}

Postponing the $ Sp\bar{p}S $ results for a bit later, let us  come to the highest energies achieved by now, i.e. 2 TeV for $ \bar{p} p $ and 13 TeV for $ pp $. A straightforward comparison between the two channels is still impossible because of the absence of the relevant data at the same energy, so we take the COMPETE parametrization \cite{ed} which describes the data on $ \sigma_{tot}^{\bar{p}p} $ and $ \sigma_{tot}^{pp} $ very well. This is pictured in Fig.2 \cite{Ant2} .

\hfill \break

The COMPETE predicts for $ \Delta\sigma $ the stable decrease  as is seen in Fig.3.

So, it seems that the difference in total cross sections is not the best place to look for manifestations of the Odderon\footnote{The enthusiasts of the "Maximal Odderon" still insist that the cross-over will occur though such claims already do not look very convincing} . However, this could only mean that the Odderon does not couple significantly to the imaginary part of the \textit{forward} scattering amplitude while a possibility of a noticeable coupling  to the real part of the forward scattering amplitude is quite conceivable.

\subsection{Early sounding of the Odderon via ReF(s,0)/ImF(s,0).}
So, in addition to the difference between the cross sections , the quantity
\begin{center}
$  \rho = \mbox{Re}F(s,0)/\mbox{Im}F(s,0).$
\end{center} 
($F(s,t)$ stands for the elastic scattering amplitude) seemed to be  a suitable observable quite accessible at the $ Sp\bar{p}S $ collider. Although there were no corresponding $ pp $ data, the Odderon contribution could manifest itself in $ \rho^{\bar{p}p} $ through the dispersion relations, as a sort of "echo" from the u-channel. 
During the functioning of the $ Sp\bar{p}S $, two dedicated measurements were made in the UA4 and UA4/2 experiments. 
The result obtained in UA4 \cite{ua4/1}  became a sensation: instead of the expected value of about $ 0.10\div 0.15 $, it turned out that
\begin{center}
$\rho^{\bar{p}p}~(\mbox{UA4}) = 0.24 \pm 0.04 !$
\end{center}
The number caused a flow of publications, often containing the most fantastic scenarios, but the enthusiasts of the "Maximal Odderon" felt themselves to be the main beneficiaries \cite{nic2} . It was the notorious "maximality" that seemed to be the reason for such a large value of $ \rho $.

Six years passed in discussions, conferences , talks and articles till a new sensation
broke out. A "remeasurement" undertaken by the UA4 collaboration (under the nickname UA4/2) produced the following result \cite{ua4/2} :
\begin{center}
$  \rho^{\bar{p}p}~ (\mbox{UA4}/2) = 0.135 \pm 0.015 .$
\end{center} 
Concerning the 1987 result it was said:
"The previous result
$ \rho= 0.24 + 0.04 $ obtained with a poor beam optics, a factor eleven less statistics and much less control of systematic effects should be considered as superseded."\cite{ua4/2}.

So the "Maximal Odderon" was again not lucky: after several years of triumph, disappointment came. 
But ahead there  was a new take-off, although one had to wait a long time, almost a quarter of a century. 
\subsection{A resurrection of the "Maximal Odderon" or...?}

In December 2017, one of us (V.P.) had a long discussion with S. Giani and J. Ka\v{s}par about their just obtained result on extracting the value of $ \rho $ from the data of the TOTEM collaboration on the differential cross section for elastic proton-proton scattering in the region of Coulomb-nuclear interference at 13 TeV.

The point was that this value  ($ \rho = 0.1 $) essentially coincided with the value of $ \rho $ obtained earlier in the theoretical
article by B. Nicolescu and E. Martynov, in which an attempt was made to describe a large amount of data in the framework of a highly modified version of the "Maximum Odderon" model. 

On this basis, the conclusion was made: a new particle was discovered, the "Odderon"consisting of 3 gluons! 
The news attracted attention of the public media. For instance, a few months later an article appeared in  "The Newsweek" under the title:
"What's an Odderon and Did CERN Just Revealed it Exists?" The more professional CERN Courier placed in March 2018 an article " Oddball Antics in pp Collisions" and finally in Match 2021 "Odderon Discovered" with a statement:
"The TOTEM collaboration at the LHC, in collaboration with the DØ collaboration at the former Tevatron collider at Fermilab, have announced the discovery of the Odderon – an elusive three-gluon state predicted almost 50 years ago." 

Without a doubt, the discovery of a new particle is a great event and an outstanding achievement for any experiment.

Let us now look at two publications concerning these findings. 

The first one appeared in 2019 \cite{TOT1} and was devoted to "probing
the existence of a colourless C-odd three-gluon compound state" on the basis of the retrieval of the parameter $ \rho $ from the data on elastic proton-proton scattering in the region of Coulomb-nuclear interference at 13 TeV. The second one will be commented in the next subsection.  

As was mentioned above, the conclusion about the discovery  of the "C-odd three-gluon compound state" was made because of coincidence of the measured value of $ \rho $ with the approximately the same value appearing in the model of "Maximal Odderon" \cite{nic3}. 

What is interesting, in the model suggested in \cite{nic3} the  authors do not deal with "gluons" at all ( because their arguments do not use QCD) and define the Odderon as follows:

"The Odderon is defined as a singularity in the complex $ j $-plane,
located at $ j=1 $ when $ t=0 $ and which contributes to the odd-
under-crossing amplitude $ F_{-} $". 

However, the crossing-odd amplitude (negative signature)with  $ j=1 $ cannot have singularity at $ t=0 $ because this is the \textit{physical} p-wave partial amplitude in $ \bar{p}p $ channel. Otherwise the axiomatic bounds (assuming non-zero mass gaps in any channel)would be violated while the authors use precisely these bounds to justify their "maximum" choice  for the C-odd amplitude. Otherwise, we would be forced to assume that there is no color confinement. This in turn would naturally give rise to an infrared singularity at $ t=0 $. I do not believe that authors of \cite{nic3} meant such a radical scenario. Although, in this case, gluons would appear, indeed, but in an amount significantly exceeding 3. 

In contrast, the C-even (positive signature) amplitude $ F_{+}(1,t) $  may well have a singularity at $ t=0 $ , since for it the value $ j=1 $ is not physical.

A detailed criticism of the "Maximal Odderon model" in both conceptual and descriptive aspects can be found in Ref.\cite{ptr1}.

There is one more aspect of this topic that I would like to touch upon.  
In the article \cite{TOT1} some phenomenological model for the strong interaction amplitude was used for description of the data and hence, for retrieving parameters, e.g.  $ \rho $. However, this model does not exhibit the Odderon singularity as does the strong interaction amplitude described in \cite{nic3}.
So, the coincidence of the values of $ \rho $ seems accidental, not related with the presence or absence of the Odderon singularity as assumed in \cite{TOT1}.

Our conclusion from this reasoning is that no specific value of the parameter $ \rho $ can be considered as an evidence of presence or absence of the Odderon.

Being the $ \rho $-parameter of the "forward origin" this is in line with the above conclusion about another forward observable $ \Delta\sigma $ and implies that the Odderon, if exists, should be probed at non-zero transferred momenta.

Its decoupling at $ t=0 $ is evidently related with absence of massless states in the 
p-wave partial amplitude $ F_{-}(1,t) $ in the $ \bar{p}p $-channel, i.e. actually with confinement. 

\pagebreak

\subsection{A step aside: the Odderon at nonzero t. An old friend? }

In no way all above said means that the Odderon does not exist or is unobservable. We argued only about forward observables. If $ t\neq 0 $ the only way to search for it is a comparison of differential cross-sections in $ pp $ and $ \bar{p}p $-channels.

\begin{wrapfigure}[25]{l}{90mm}
\vspace{-0.1mm}
\includegraphics[width=90mm]{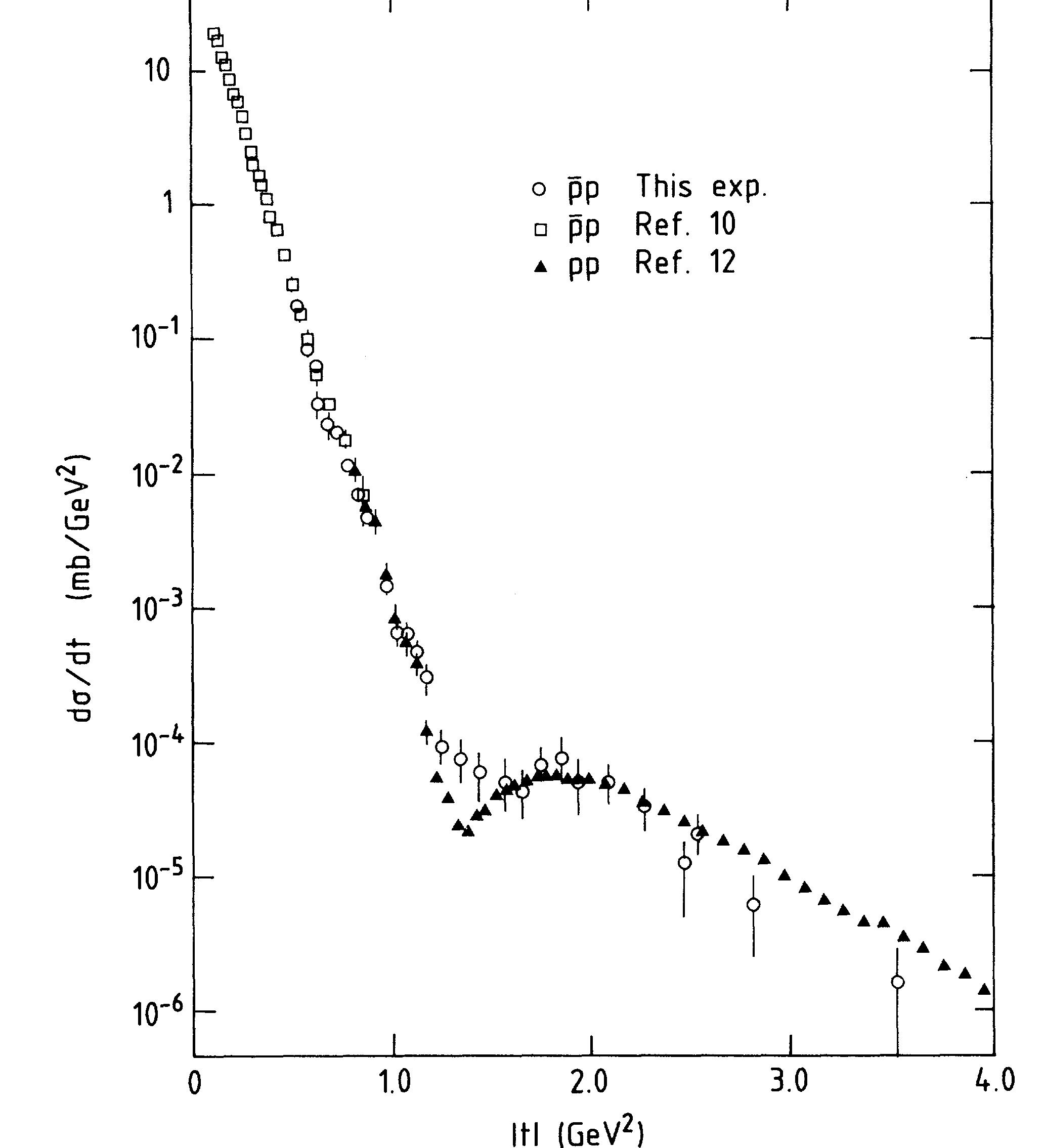}
\vspace{-4.6mm}
\caption{The comparison of ($ pp $) and  $ \bar{p}p $ cross-sections.}
\label{p4}
\end{wrapfigure}

And here we cannot help but recall the good old ISR. We find that as early as in 1985 a dedicated measurements were made \cite{isrod}  to compare  the differential cross-section of elastic $ pp $ and $ \bar{p}p $ scattering at the same energy ($ \sqrt{s}= 53 $ GeV).
Fig.4  \cite{isrod} shows that the two cross-sections almost coincide except the vicinity of the dip ($ pp $) and shoulder ( $ \bar{p}p $).
Fig.4. The comparison of ($ pp $) and  $ \bar{p}p $ cross-sections.

Fig.5 shows the ratio of the two cross-sections which differs from 1 only at $ t $ in the vicinity of dip/shoulder.

It is clear that the cause of the difference is a C-odd force. But which one? Is it the manifestation of the well known secondary Reggeons which are responsible for a non zero  
$ \Delta\sigma $  at low energies ? If we try to blame them for the said difference, we will see that their contribution is only a small part of the visible effect.

It was understood by the authors of Ref.\cite{isrod} as they mentioned:

"{\it When we compare the available models to these data we find that none of them describes the data adequately.}"

That was true that no "adequate description" was provided that time but, nonetheless, the result did not remain unnoticed. In Ref.\cite{gaur} it was even called "the new great ISR discovery" with an "intriguing question":
"Is it the maximal odderon growth ?"
As we already mentioned the "Maximal Odderon", unfortunately,  was not acceptable on conceptual grounds. 

Howbeit, we have to admit that there was some new C-odd interaction agent observed in the experiment \cite{isrod}  which was not of pure quark origin as $ \rho, \omega$ etc. In other words that, in this blurry meaning, the Odderon was  discovered already 36 years ago.

But even if we admit this we do not know if this effect survives at high energies or dies off?
Meanwhile, the TOTEM Collaboration made measurements of the $ pp $ elastic scattering at 2.76 TeV. The closest results in energy was the DO (FNAL) measurement of the $\bar{p}p$ elastic scattering at 1.96 TeV. For lack of anything better, it was decided to compare the cross sections, albeit not at the same, but at relatively close energies.The comparison has shown that the effect persists, although less pronounced.

 Soon afterwards an attempt was  made\cite{odde}  with help of a specially designed extrapolation technique of the "data transfer" to provide the comparison at the same energy (1.96 TeV). The result appeared qualitatively the same with minor quantitative differences (Fig.6).
 
 Some qualitative estimate of the energy dependence can be made if to consider the ratio

 \begin{center}
 $d\sigma^{\bar{p}p}/d\sigma^{pp} = f(\sqrt{s}, \tau= \mid t\mid/\mid
 t_{\mbox{dip}}(s)\mid). $
 \end{center}

\begin{wrapfigure}[39]{l}{90mm}
\vspace{-0.1mm}
\includegraphics[width=90mm]{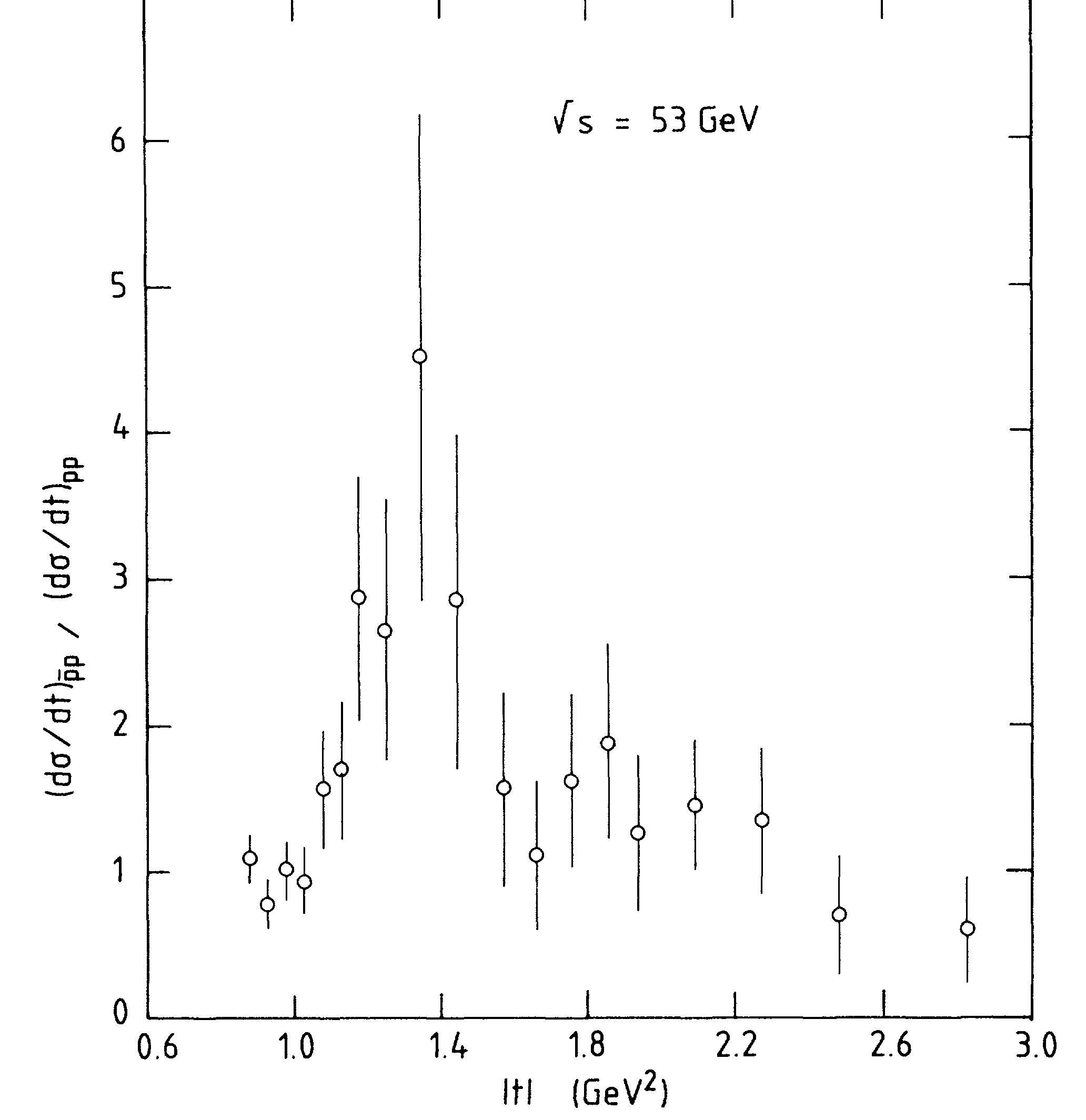}
\vspace{-6.6mm}
\caption{The ratio $d\sigma^{\bar{p}p}/d\sigma^{pp}$ at 53 GeV \cite{isrod}.}
\label{p5}
\vspace{1.1mm}
\includegraphics[width=90mm]{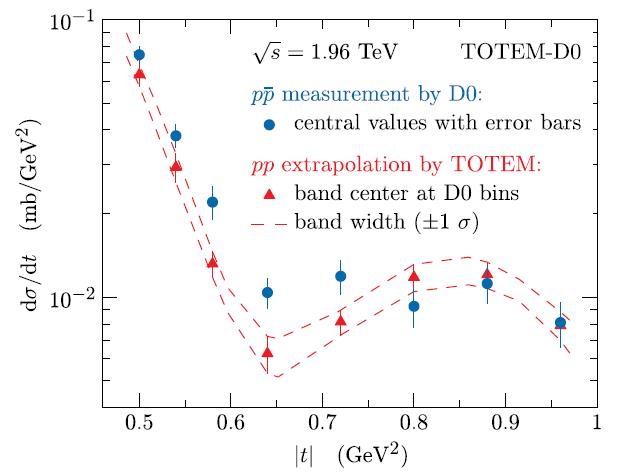}
\vspace{-6.6mm}
\caption{The $\bar{p}p$ and (extrapolated) $pp$ data at $\sqrt{s}=
1.96 TeV$ \cite{odde}.}
\label{p6}
\end{wrapfigure}

 The function $ f( \sqrt{s} , \tau) $ seems to be 1 almost at all $ \tau $ except a "bell" in the vicinity of $ \tau =1 $.
 
 Then we can obtain for the height of the bell that \[f(53\mbox{ GeV},1)-1
 = 3.5\pm 1.7 \]
 while
 \[f(1960\mbox{ GeV},1) -1 = 0.67\pm 1\sigma (?) .\]
 Unfortunately, the values of the extrapolated $ pp $ data at 1.96 TeV are still kept secret, so we could not estimate the errors better and make a picture like Fig.5.
 
 What can we conclude from this story?
 
1. The discovery of the Odderon as a new C-odd force superior to the "old" C-odd forces from the secondary quark Reggeons was successfully confirmed in the energy interval $ 53 \div 2000 $ GeV.

2. 
The Odderon effect in the sense described above weakens with energy albeit very slow.

It remains to understand the Odderon nature in terms of, say, $ j $-plane singularity and to clarify its particle content.
We cannot agree that , as done in Ref.\cite{odde}, with a reference to the paper \cite{nicla}, that there was a "colorless C-odd gluonic compound" observed because the present data cannot inform us about quark-gluon content of the exchange. Only a direct detection of a state associated with this exchange with definition of its mass, width and spin-parity can be qualified as such an evidence. This can be compared with the discovery of pion occured only 12 years after publication of the Yukawa paper. Unfortunately, we can not  say that the paper \cite{nicla} with its 
erroneous theoretical content and bad description quality ($  p-value=8.5\cdot 10^{-71}$) \footnote{One can find the corresponding criticism in Ref.\cite{ptr1}. }  can be likened to the Yukawa paper.

At the same time, we would like to pay due tribute to the commendable tenacity of the main and pioneering proponent of the Odderon, Basarab Nicolescu, who devoted many years to enthusiastic promotion of this idea. 
 
Comments: Presented at the Low-$x$ Workshop, Elba Island, Italy, September 27--October 1 2021.
 
\section*{Acknowledgements}

V.P. thanks Christophe Royon for invitation to give this talk at a very interesting Workshop.

\nocite{*}
\bibliographystyle{auto_generated}
\bibliography{Petrov_article/Petrov}

%% file: proceedings_elba2021/proceedings_elba2021/Ribeiro.tex
\vspace*{1.2cm}

\thispagestyle{empty}
\begin{center}
{\LARGE \bf CMS results on photon-induced processes}

\par\vspace*{7mm}\par

{

\bigskip

\large \bf Beatriz Ribeiro Lopes \\ on behalf of the CMS Collaboration}

\bigskip

{\large \bf  E-Mail: beatriz.ribeiro.lopes@cern.ch}

\bigskip

{CMS group, DESY (Deutsches Elekronen-Synchrothron) Hamburg, Germany }

\bigskip

{\it Presented at the Low-$x$ Workshop, Elba Island, Italy, September 27--October 1 2021}

\vspace*{15mm}

\end{center}
\vspace*{1mm}

\begin{abstract}

Photon-induced processes can be measured at the Large Hadron Collider (LHC) at CERN, mainly in the form of photon-induced central exclusive production (CEP). In practice, measuring photon-induced CEP means using the LHC as a photon-photon or photon-proton collider, which offers a rich additional physics programme, that is complementary to the standard LHC programme. 
At the Compact Muon Solenoid (CMS), CEP are measured by imposing cuts on the detector activity in each event (taking advantage of the so-called rapidity gap), and in case of proton-proton collisions, also by tagging the outgoing intact protons using the Precision Proton Spectrometer (PPS).
In this conference talk, an overview of the most recent results on this topic by the CMS collaboration is discussed.

\end{abstract}
  \part[CMS results on photon-induced processes\\ \phantom{x}\hspace{4ex}\it{Beatriz Ribeiro Lopes on behalf of the CMS Collaboration}]{}
\section{Introduction}

Photon-induced processes at the LHC can be measured as CEP processes. The CEP of any system X (where X can be, among many others, $ee$, $\mu\mu$, $\gamma\gamma$, $WW$, $ZZ$, $Z\gamma$, $t\bar{t}$) occurs when X is produced at a hadron collider by photon or gluon exchange and the interacting protons (or ions) are not disrupted but leave the collision intact, and stay in the beam pipe at very small angles. The photon-exchange case is the topic of this talk.


The most distinctive characteristic of this kind of process is that there are no proton (or ion) remnants: since the interacting particles remain intact, the only particles that can be detected around the interaction point are the decay products of X. The intact protons can be measured separately, using dedicated forward detectors.

\begin{figure}
\begin{center}
\epsfig{figure=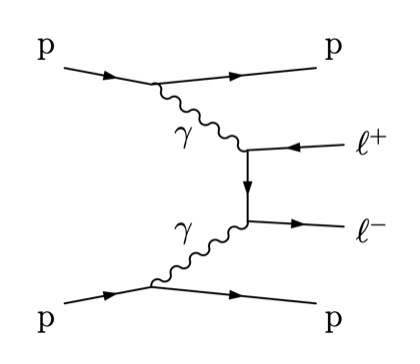,height=0.26\textwidth}
\caption{Diagram showing the central exclusive production of a lepton pair, a pure QED process.}
\label{dilepton}
\end{center}
\end{figure}

\begin{figure}
\begin{center}
\epsfig{figure=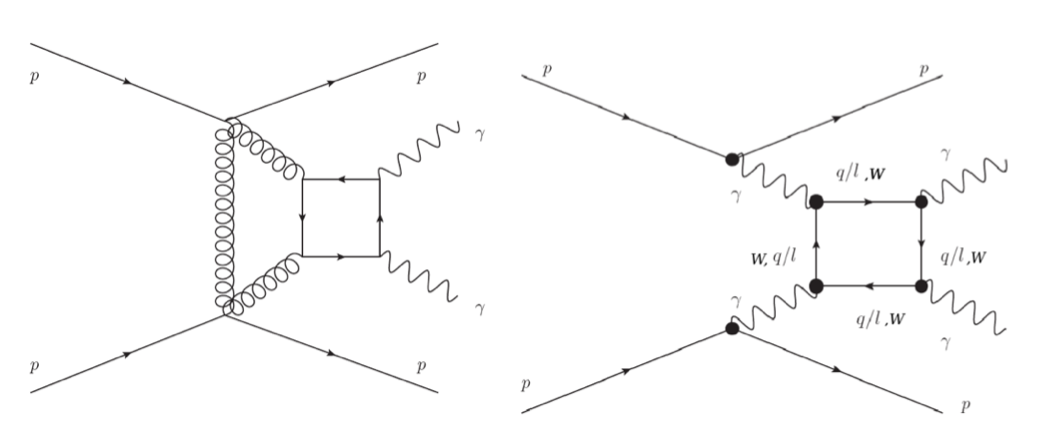,height=0.30\textwidth}
\caption{Central exclusive production of a photon pair. Left: gluon-induced process (QCD), right: photon-induced process, or light-by-light scatteringn(QED), from \cite{royon}.}
\label{feynmanLbyL}
\end{center}
\end{figure}

Not all CEP processes are photon-induced, as already mentioned. In fact, while processes such the exclusive production of dileptons ($X=ee$ or $\mu\mu$, see diagram in figure \ref{dilepton}) are purely quantum electrodynamics (QED) processes, other processes like the exclusive production of a photon pair ($X=\gamma\gamma$) have both a gluon-induced (QCD) and a photon-induced (QED) component, as seen in figure \ref{feynmanLbyL}. The photon-induced component of this process is often referred to as light-by-light scattering.

\begin{figure}
\begin{center}
\epsfig{figure=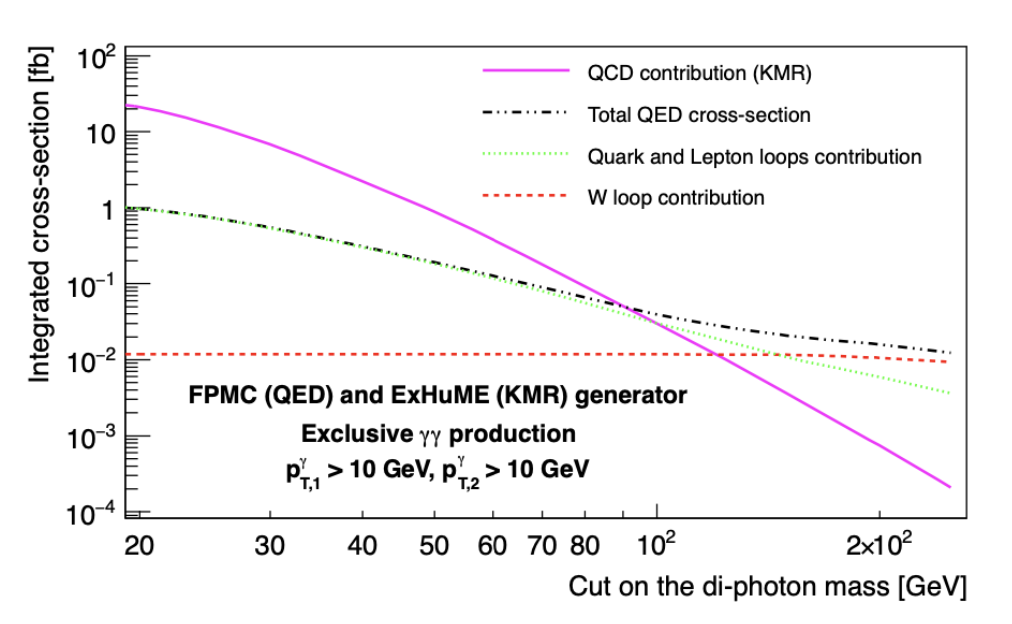,height=0.42\textwidth}
\caption{Contribution of different processes to the production cross-section of exclusive $\gamma\gamma$, as a function of the invariant mass of the two photons, $m_{\gamma\gamma}$, from \cite{royon}.}
\label{fig:xsec_vs_mass}
\end{center}
\end{figure}

The CEP of diphoton is dominated by the QCD contribution at low invariant mass of the diphoton system $m_{\gamma\gamma}$, while the photon-induced component dominates at high masses, starting at a few hundred GeV. This can be seen in figure \ref{fig:xsec_vs_mass}, taken from \cite{royon}, where the production cross-section of exclusive $\gamma\gamma$ is shown as a function of $m_{\gamma\gamma}$. The figure corresponds to the calculation done specifically for exclusive $\gamma\gamma$, but the same is qualitatively true for other CEP processes such as $WW$, $ZZ$, $Z\gamma$ and $t\bar{t}$.

\begin{figure}
\begin{center}
\epsfig{figure=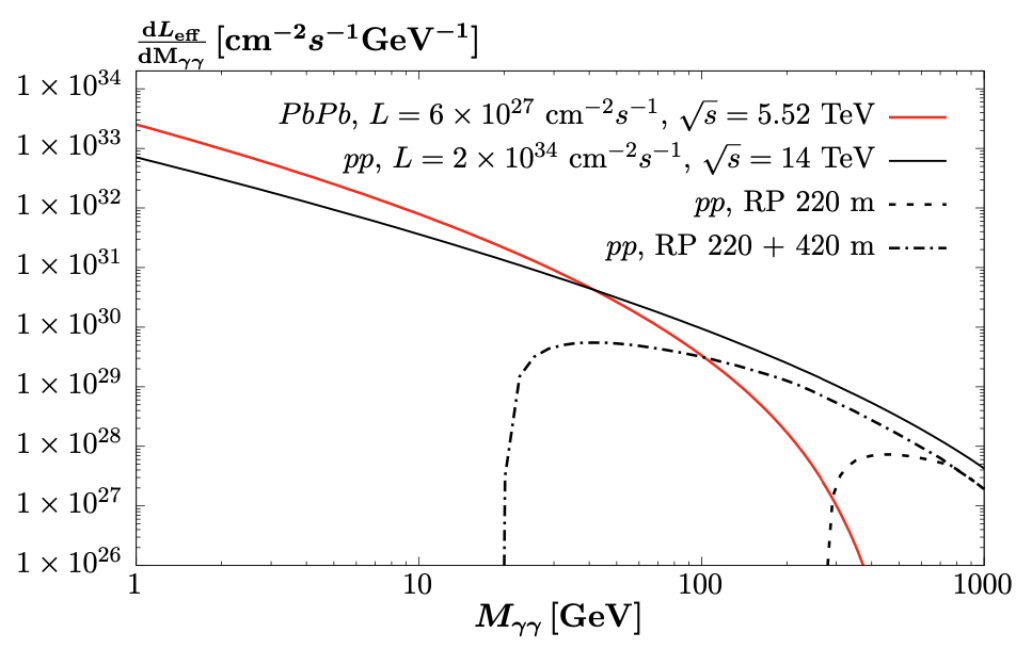,height=0.42\textwidth}
\caption{Effective luminosity accessible at the LHC experiments for CEP of diphotons, as a function of the invariant mass of the two photon system. The red solid line shows what is reachable with PbPb collisions. The black solid line shows the luminosity with pp collisions, without tagged protons, while the 2 different black dashed lines show the luminosity for pp collisions with tagged protons, with two different configurations of the proton detectors. The one with the caption "RP 220m" corresponds to what is achievable, for example, with the Precision Proton Spectrometer (PPS) at CMS. Figure from \cite{bruce}.}
\label{fig:pp_vs_PbPb}
\end{center}
\end{figure}

To study photon-induced CEP experimentally at the LHC, in proton-proton interactions, we only have access to the mass range where this contribution is dominant, i.e., the high-mass region. However, if we consider lead-lead interactions instead, then the cross-section is enhanced by a factor $Z^4$ (where $Z$ is the atomic number of the colliding particles), and consequently we gain access to the lower mass region as well. The accessible effective luminosities for CEP of diphotons at the LHC experiments for pp and PbPb are compared in figure \ref{fig:pp_vs_PbPb}, from \cite{bruce}.

In general, photon-induced processes are a promising way to look for new physics, since they are sensitive to anomalous couplings between the SM particles such as the gauge bosons and the top quark.
A unique feature of these processes is an excellent mass resolution, irrespective of the decay mode of the central system, since the energy loss of the outgoing protons is directly related to the invariant mass of central system. In other words, if we are able to measure the outgoing protons, we have an independent and high resolution handle on the mass of system. This high mass resolution also opens the possibility for precision tests of the SM couplings.

Furthermore, by matching protons to the central system, most backgrounds that would normally be irreducible can be eliminated, and a high signal-to-background ratio is achievable.

The physics programme at CMS that aims to measure CEP can be divided in three categories, according to the invariant mass of the system X which is exclusively produced: the low mass region, with $m_X$ up to a few GeV, is accessible only with heavy ion collisions; the intermediate mass region, with $m_X$ up to a few hundred GeV, is accessible with proton-proton collisions without tagging intact protons, and the high mass region, $m_X$ starting at around 400 GeV, is accessible with proton-proton collisions combined with tagged protons.

\subsection*{Low mass region}

\begin{figure}
\begin{center}
\epsfig{figure=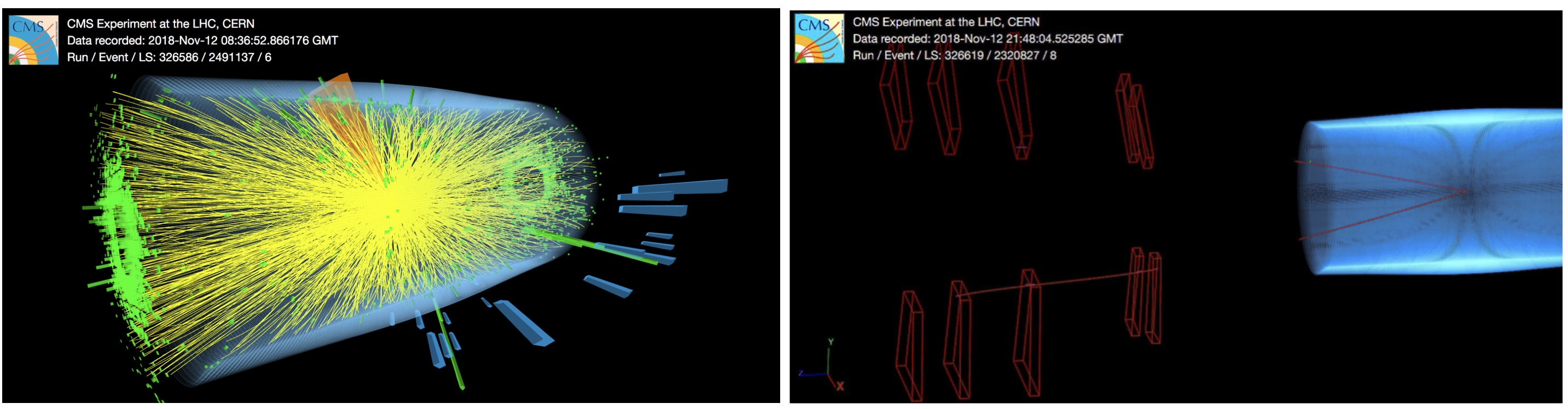,height=0.26\textwidth}
\caption{Left: CMS event display of a typical PbPb collision, where each yellow track represents a charged particle track in the CMS tracker and each green area represents an energy deposit in the calorimeters. Right: CMS event display of $\gamma\gamma\rightarrow\mu\mu$ candidate event, where the two red tracks represent two muons. Event displays taken from \cite{eventdisplayPbPb}.}
\label{PbPbeventdisplay}
\end{center}
\end{figure}

In typical PbPb collisions, where the impact parameter $b$ is smaller than twice the atomic radius ($b<2R_A$), hundreds of particles are produced and events are very "crowded" (see left of figure \ref{PbPbeventdisplay}). However, in the case where $b>2R_A$, normally called ultra-peripheral collisions (UPCs), the Pb ions can interact via photon exchange and remain intact. This means that the only particles observed in the final state are the ones produced via CEP or the respective decay products, resulting in a very distinctive signature with low number of tracks (see right side of figure \ref{PbPbeventdisplay}). This distinctive signature is used to measure CEP in the low mass region, without the need for tagging the outgoing ions.

In the results section, two results are discussed in this mass region: the exclusive dimuon production \cite{dimuonPbPb} and the light-by-light scattering \cite{LbyLPbPb}.

\subsection*{Intermediate mass region}

In pp collisions, processes like the exclusive production of dimuon ($\gamma\gamma\rightarrow \mu\mu$) can be measured even without resorting to proton tagging, since the muons are the only expected tracks in the event, and a requirement for the presence of rapidity gaps can be used as selection criterion.

There is no recent public result in this category, but some analyses of this type were done using data from the Run 1 of the LHC, for example \cite{wwRun1}.

\subsection*{High mass region}

\begin{figure}
\begin{center}
\epsfig{figure=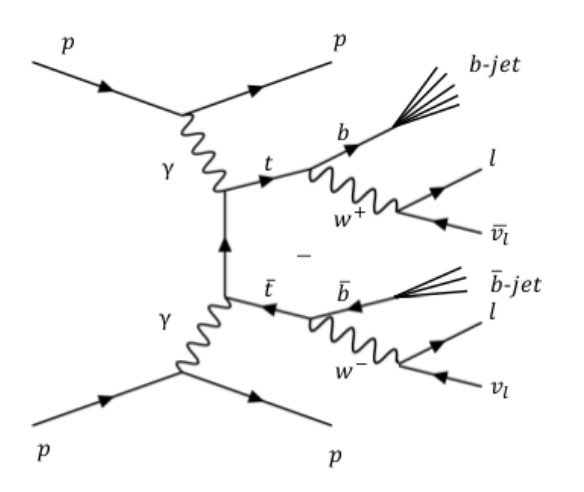,height=0.42\textwidth}
\caption{Diagram illustrating the exclusive production of a pair of top quarks, with both top quarks decaying leptonically.}
\label{ttbar}
\end{center}
\end{figure}

Some photon-induced processes, in particular the production of more massive particles, result in several particles in the final state, and are difficult to tag using only the requirement of no extra activity in the central detector. A good example is the exclusive production of a top quark pair, $\gamma\gamma\rightarrow t\bar{t}$, illustrated in figure \ref{ttbar} for the case where both top quarks decay leptonically. In this case, we have at least two jets in the final state, as well as two leptons and missing energy from the neutrinos. This is a more complex signature and very similar to the inelastic QCD $t\bar{t}$ production at the LHC, which has a cross-section that is larger by several orders of magnitude. In order to tag this process, one needs an additional signature.

This signature can be obtained by tagging the outgoing intact protons, which will only be present in CEP and not in the inelastic $t\bar{t}$ production. Tagging protons requires the development of dedicated detectors, and is extremely challenging because these protons travel very close (few mm) to the beam, where the radiation environment will damage significantly any detector. In CMS, the detector that was developed for this purpose is the CMS-TOTEM Precision Proton Spectrometer (PPS), a detector consisting of movable structures called roman pots (RPs), placed at ~200 m from the interaction point. These can contain 3D-pixel or silicon strip detectors, that are able to track the protons as they leave the collision and pass through the RPs. The acceptance of PPS covers protons that lost ~2-20\% of their momentum.
This range of momentum loss can be translated into an acceptance in terms of the mass of the system X, using equation \ref{eq:mass}:

\begin{equation}
    m_X=\sqrt{s\xi_1\xi_2}
    \label{eq:mass}
\end{equation}

\begin{equation}
    y_X=\frac{1}{2}\log(\xi_1/\xi_2)
    \label{eq:y}
\end{equation}

where $\xi_i$ is the fraction of momentum loss of each interacting proton, computed as $\xi_i=\frac{p_f-p_i}{p_i}$, for initial proton momentum $p_i$ and final momentum $p_f$. This results in a good acceptance at high masses (starting ~400~GeV). This detector was installed during the most part of the Run 2 of the LHC (2016-2018), and there is more than 100~fb$^{-1}$ of data available.

In this talk, two results are presented concerning lead-lead (PbPb) collisions and two concerning proton-proton (pp) collisions. In PbPb collisions, we present the recent works on exclusive dimuon production \cite{dimuonPbPb} and exclusive diphoton production, often called light-by-light scattering \cite{LbyLPbPb}. In pp collisions, we discuss the recent results on (semi)exclusive dilepton production \cite{semiexclusivedilepton} and light-by-light scattering \cite{LbyLpp}. Limits on anomalous couplings and on pseudoscalar axion-like particles are also discussed.

\section{Results}

\subsection{Results in PbPb collisions}

\subsubsection*{Exclusive dimuon production}

The analysis in \cite{dimuonPbPb} aims to study the exclusive production of muon pairs in UPCs, using PbPb collision data collected in 2018 during the LHC Run 2, corresponding to an integrated luminosity of 1.5 nb$^{-1}$, at $\sqrt{s_{NN}}=5.02$~TeV. The two lead nuclei produce each a photon flux, and the average transverse momentum $<p_T>$ of the exclusively produced muon pair depends on the overlap between these photon fluxes, and thus could depend on the impact parameter $b$.

In \cite{QEDcalculation}, the authors perform a calculation based on QED, which predicts that the dimuon $<p_T>$ should increase as $b$ decreases. The main goal of the analysis is to test this calculation and prove the dependence experimentally.

Instead of measuring the dimuon $<p_T>$ directly, another quantity, which has better experimental resolution, is used, the acoplanarity $\alpha$:

\begin{equation}
    \alpha=1-|\Delta\phi_{\mu\mu}|/\pi 
    \label{eq:acoplanarity}
\end{equation}

where $\Delta\phi_{\mu\mu}$ is the distance in azimuthal angle between the two muons. Larger dimuon $<p_T>$ results in a larger acoplanarity.

The impact parameter $b$ cannot be measured directly, so a good experimental handle on this parameter is needed. We can take advantage of the fact that the smaller the $b$, the larger the overlap between the photon fluxes of the two nuclei, and the higher the probability for the excitation of one or both ions via photon absorption into giant dipole resonances or higher excited states. The giant dipole resonances typically decay with the emission of one neutron, while higher excited states may emit two or more neutrons. These forward neutrons are emitted at very low rapidity, and can be detected at CMS thanks to the zero-degree calorimeters (ZDC), part of the CMS forward calorimetry system.

\begin{figure}
\begin{center}
\epsfig{figure=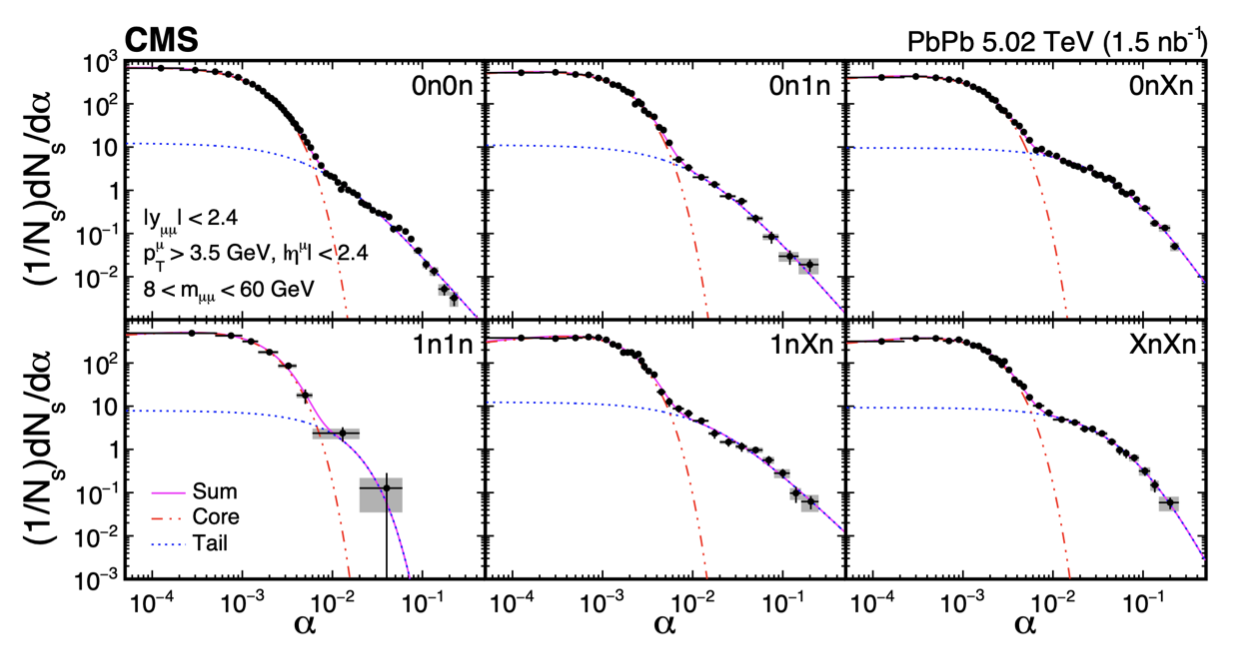,height=0.47\textwidth}
\caption{For 6 classes of neutron multiplicity, acoplanarity distributions for $\gamma\gamma\rightarrow\mu\mu$ in ultraperipheral PbPb collisions at $\sqrt{s_{NN}}=5.02$~TeV. The distributions are normalized to unit integral over the measured range. The dot-dot-dashed line denotes the fit to the core contribution, while the dotted line denotes the fit to the tale. The sum of the two is indicated with a pink solid line. The statistical uncertainty is shown as vertical black lines on the points, while the systematic uncertainty is depicted by the gray boxes. Figure from \cite{dimuonPbPb}.}
\label{dimuon6plots}
\end{center}
\end{figure}

\begin{figure}
\begin{center}
\epsfig{figure=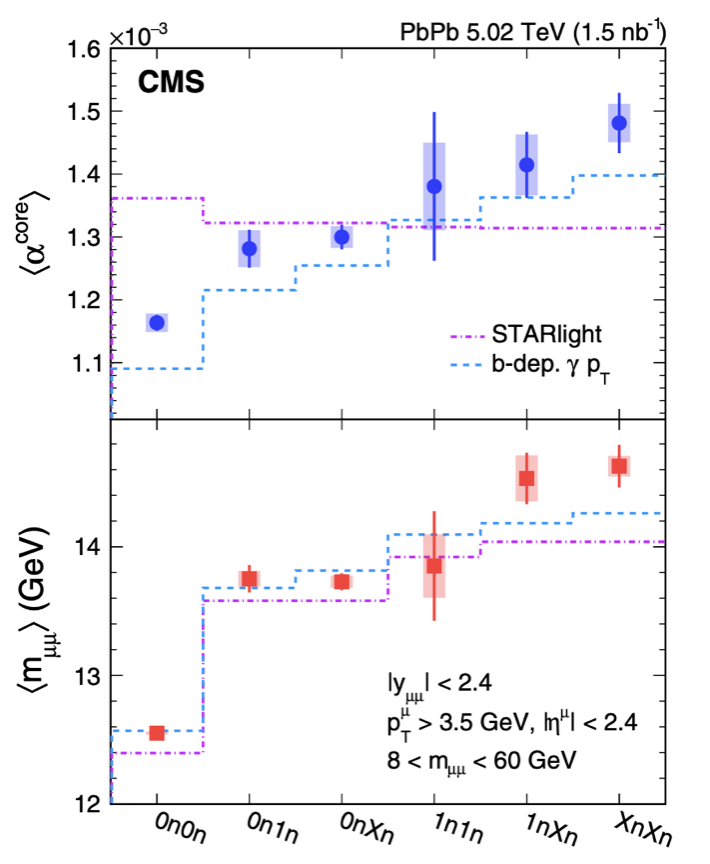,height=0.6\textwidth}
\caption{Top: Average acoplanarity at the core of the distribution as a function of the forward neutron multiplicity. Bottom: Average dimuon invariant mass as a function of the forward neutron multiplicity. Both distribution are shown for $\gamma\gamma\rightarrow\mu\mu$ in ultraperipheral PbPb collisions at $\sqrt{s_{NN}}=5.02$~TeV. The experimental data (dark blue/red) are compared with a QED calculation (light blue) and with the predictions from the MC generator STARlight (purple). Figure from \cite{dimuonPbPb}.}
\label{dimuonsummary}
\end{center}
\end{figure}

The results are summarized in figure \ref{dimuon6plots}. The average acoplanarity at the core of the distributions $<\alpha^{\text{core}}>$ is taken from each fit (red lines in the figure), and plotted against the neutron multiplicity, as can be seen in the top panel of figure \ref{dimuonsummary}. A clear dependence is observed in the data (dark blue), in qualitative agreement with the calculation (light blue). The bottom panel shows a similar distribution, but this time for the average dimuon invariant mass. Again the data (red) are in qualitative agreement with the calculation.
In purple are the predictions from the STARlight \cite{starlight} MC generator. It is clear that the MC does not fully account for the observed dependence between the acoplanarity (or dimuon average $p_T$) and the impact parameter $b$. The authors conclude by calling for theoretical effort in the direction of improving the simulation of these interactions.

\subsubsection*{Light-by-light scattering}

The CMS analysis providing evidence for light-by-light scattering is described in \cite{LbyLPbPb}. It is based on data collected during the several LHC Run 2 PbPb campaigns at $\sqrt{s}=5.02$~TeV, corresponding to a total integrated luminosity of 390~$\mu\text{b}^{-1}$. The signal in this analysis is characterised by two back-to-back photons in the final state and no extra activity. The main background arises from CEP of electron pairs, which are misidentified as photon pairs, and the QCD (gluon-induced) production of photon pairs.

\begin{figure}
\begin{center}
\epsfig{figure=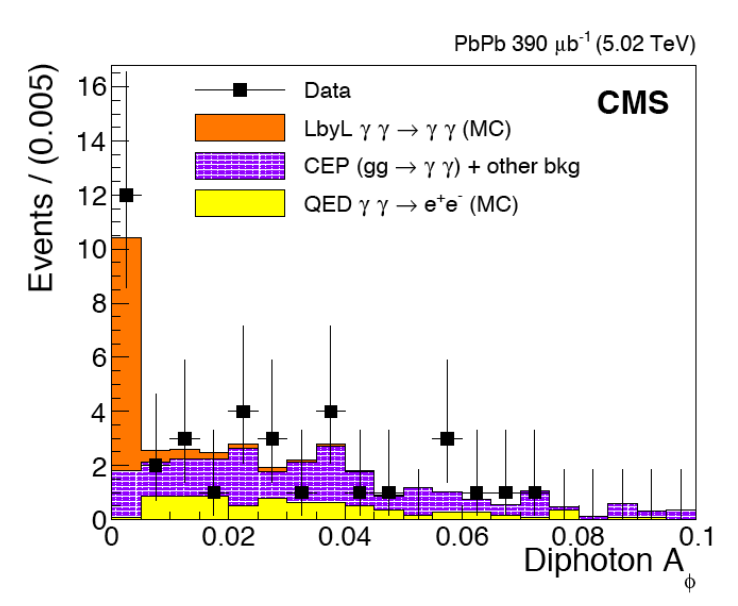,height=0.42\textwidth}
\caption{Distribution of the diphoton acoplanarity, for data (black points) superimposed with the prediction from MC simulation for the signal (orange) and the two main backgrounds (purple and yellow). Figure from \cite{LbyLPbPb}.}
\label{plotLbyL_pbpb}
\end{center}
\end{figure}

An event selection is applied requiring two photons with $E_T > 2$~GeV, pseudorapidity $|\eta_\gamma| < 2.4$, diphoton invariant mass $m_\gamma\gamma > 5$~GeV, diphoton transverse momentum $p_T(\gamma\gamma)<1$~GeV, and diphoton acoplanarity $\alpha=1-|\Delta\phi_{\gamma\gamma}|/\pi$ below 0.01 (first two bins of the distribution in figure \ref{plotLbyL_pbpb}). The number of observed events after selection is 14, with 9.0$\pm$0.9 signal expected and 4.0$\pm$1.2 background, corresponding to a significance of 3.7$\sigma$, above the 3$\sigma$ threshold normally required to claim evidence for this process.

The fiducial cross-section is measured to be

$$\sigma_{\text{fid}}(\gamma\gamma\rightarrow\gamma\gamma)=120\pm46~\text{(stat)}\pm12~\text{(theo)}~\text{nb}$$

consistent with the SM prediction of 116$\pm$12~nb \cite{LbyLPbPb_theory}.

\begin{figure}
\begin{center}
\epsfig{figure=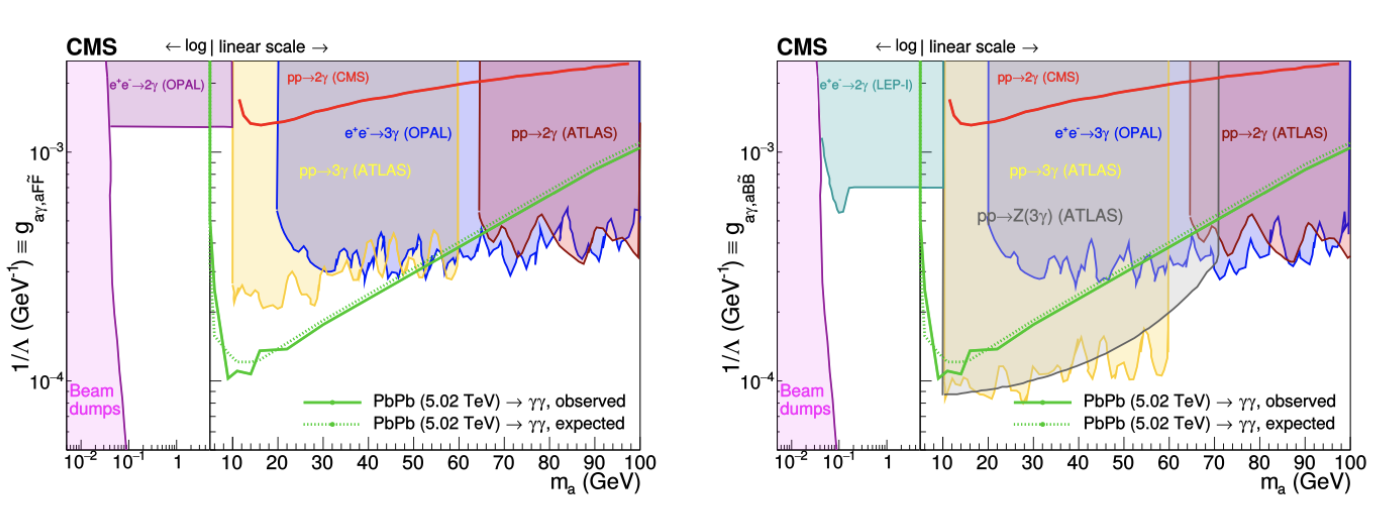,height=0.35\textwidth}
\caption{Limits on ALPs production set by CMS in \cite{LbyLPbPb}, compared to previous works, left: assuming ALPs coupling to photons only, right: assuming also the hyper charge coupling. Figure from \cite{LbyLPbPb}.}
\label{alps_limits}
\end{center}
\end{figure}

New spin-even particles like pseudo scalar axion-like particles (ALPs) can contribute to the light-by-light scattering continuum or to new diphoton resonances. This work sets limits on ALPs production in the range 5-90~GeV. These were the best limits at date of publication over the mass range 5-50~GeV (5-10~GeV) for ALPs coupling to electromagnetic (electroweak) current. The limits are shown compared to previous results in figure \ref{alps_limits}, on the left in the case of electromagnetic coupling only, and on the right in case of electroweak coupling. There is an equivalent analysis that was recently published by the ATLAS collaboration, which sets competitive limits on the production of these particles \cite{atlas_LbyL}.

\subsection{Results in pp collisions with tagged protons}

\subsubsection*{Semi-exclusive dilepton production}

\begin{figure}
\begin{center}
\epsfig{figure=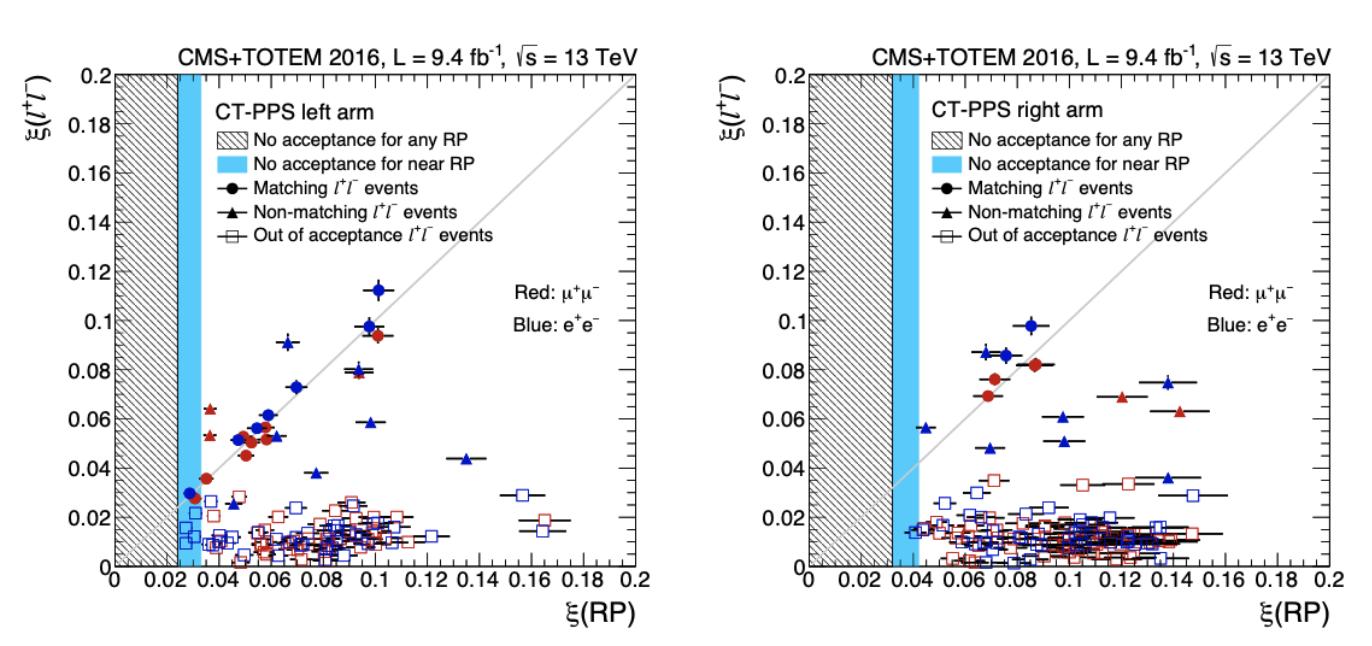,height=0.4\textwidth}
\caption{Distribution of $\xi$ of the lepton pair, computed from lepton kinematics, versus the $\xi$ of the protons detected in each PPS arm (left: left arm, right: right arm). The circle points correspond to events matching the diagonal (signal-like), within the uncertainties, while triangles are events that do not match (background like). The non-filled squares correspond to events outside of the PPS acceptance. Blue points are $e^+e^-$ candidates, while red points are $\mu^+\mu^-$ candidates. From \cite{semiexclusivedilepton}.}
\label{xiplot}
\end{center}
\end{figure}

The exclusive and semi-exclusive production of lepton pairs is dominated by photon interaction. We call semi-exclusive the case where one of the protons remains intact but the other is dissociated in the collision. The analysis in \cite{semiexclusivedilepton} aims to tag one (or two) intact protons with PPS and combine them with a lepton pair in the central CMS apparatus, using data collected in 2016 by CMS and TOTEM at $\sqrt{s}=13$~TeV, corresponding to an integrated luminosity of 9.4~fb$^{-1}$. The analysis aims mainly at single-tagged (one proton) events, since the double-arm acceptance of PPS starts at an invariant mass of the lepton pair of around 400~GeV, where a low number of events is expected. A $e^+e^-$ / $\mu^+\mu^-$ selection in the central system is combined with a proton(s) requirement in PPS, and the resulting events are shown in figure \ref{xiplot}. This plot shows the $\xi$ of the dilepton pair, computed from lepton kinematics, and the $\xi$ of the protons detected in each PPS arm (left: left arm, right: right arm). The events along the diagonal represent the signal-like events. 12 (8) single-tagged events are observed in the  $\mu^+\mu^-$ ($e^+e^-$) channel, with 1.49 (2.36) expected background, corresponding to a significance of 4.3 (2.6) $\sigma$. The combined significance between the two channels exceeds 5 $\sigma$. This is the first observation of proton-tagged CEP of a lepton pair, and is a very important result to validate the PPS functioning (alignment, optics, etc.).

\subsubsection*{Light-by-light scattering}

Light-by-light scattering has been observed by both CMS and ATLAS at low invariant diphoton mass (up to a few GeV). The analysis in \cite{LbyLpp} is the first study of light-by-light scattering at high mass ($m_{\gamma\gamma}>350$~GeV) at a hadron collider. This process is sensitive to anomalous $\gamma\gamma\gamma\gamma$ couplings, in the context of an effective dimension-8 extension of the SM , which can be written as:

\begin{equation}
    L_8^{\gamma\gamma\gamma\gamma}=\zeta_1 F_{\mu\nu}F^{\mu\nu}F_{\rho\sigma}F^{\rho\sigma}+\zeta_2 F_{\mu\nu}F^{\mu\rho}F_{\rho\sigma}F^{\rho\nu}
\end{equation}

\begin{figure}
\begin{center}
\epsfig{figure=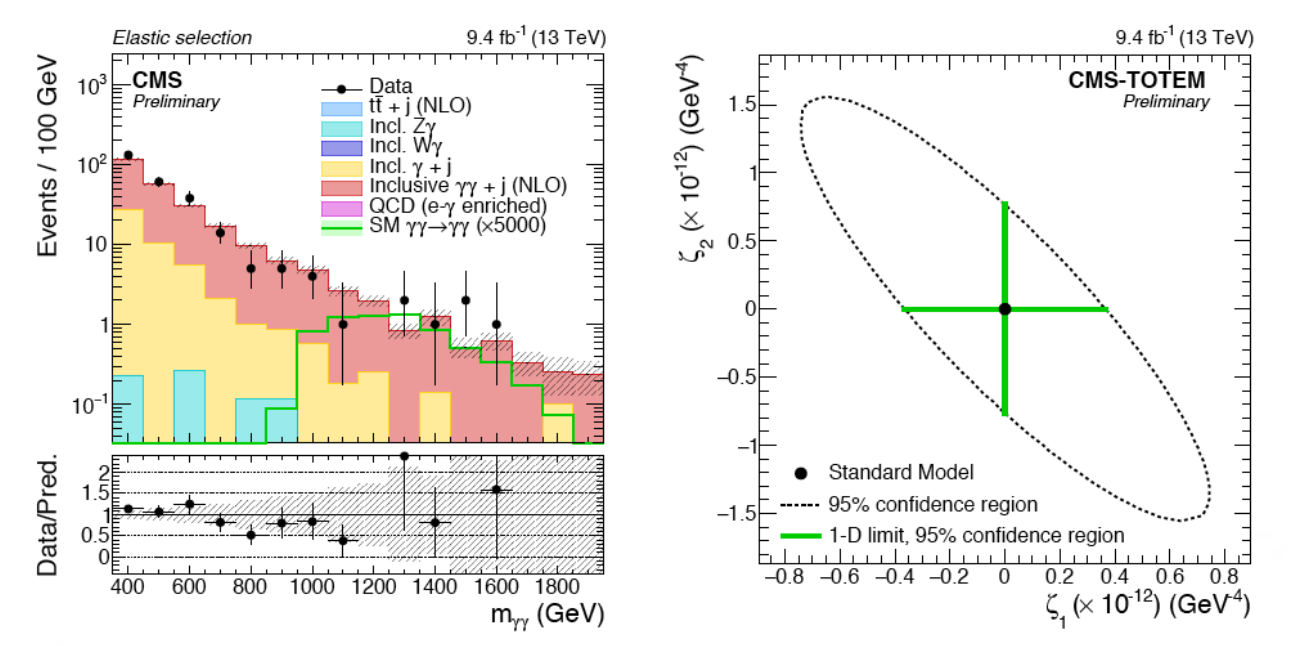,height=0.4\textwidth}
\caption{Left: Distribution of the invariant mass of the photon pair, after central selection but before proton matching. The signal expectation, depicted by the green line, is multiplied by 5000 so it can be visible on the plot. Right: limits on the coupling parameters $\zeta_1$ and $\zeta_2$, compared to the SM prediction, and respective 95\% confidence region. From \cite{LbyLpp}.}
\label{plots_diphoton}
\end{center}
\end{figure}

The data that are used were collected by CMS and TOTEM during 2016 at $\sqrt{s}=13$~TeV, corresponding to an integrated luminosity of 9.4~fb$^{-1}$. The selection consists on two photons at CMS matched with two outgoing protons in PPS. 
Events are selected that have two photons, with photon $p_T > 75$~GeV, $m_{\gamma\gamma}>350$~GeV (compatible with PPS double-arm acceptance) and $1-|\Delta\phi_{\gamma\gamma}|< 0.005$. The events observed and expected, after this selection, are shown in figure \ref{plots_diphoton} (left-hand side). Then, a proton matching requirement is applied, but 0 events are observed with matching protons.

This work sets limits on quartic gauge couplings using the coupling parameters $\zeta_1$ and $\zeta_2$, as seen on the right-hand side of figure \ref{plots_diphoton}. The limits are:

$$|\zeta_1|<3.7\times10^{-13}~\text{GeV}^{-4} \quad(\zeta_2=0)$$
$$|\zeta_2|<7.7\times10^{-13}~\text{GeV}^{-4} \quad(\zeta_1=0)$$

\section{Conclusion}

In this conference talk, the most recent results on photon-induced exclusive production of $e^+e^-$, $\mu^+\mu^-$ and $\gamma\gamma$ in pp and PbPb collisions were presented. Some of these results are able to set competitive limits on anomalous couplings and ALPs production. The results shown include data with an integrated luminosity up to 9.4 fb$^{-1}$, however there are more than 100 fb$^{-1}$ of data currently being analysed, and many results will be published soon.
In the future, with more data available and the improved PPS setup, we will gain additional sensitivity, and be able to measure a wider variety of processes, as well as perform precision measurements.

\section*{Acknowledgements}

The author thanks the organising committee of the Low-x 2021 conference for making this very interesting event possible. Many thanks also to the author's supervisor, Abideh Jafari, for all the guidance and for the encouragement to present at this conference. Finally, thanks to everyone at the DESY CMS-Top group, as well as to Cristian Baldenegro, Michele Gallinaro, Matteo Pisano, Michael Pitt, Enrico Robutti, Pedro Silva and Silvano Tosi for the insightful comments and suggestions.

\nocite{*}
\bibliographystyle{auto_generated}
\bibliography{proceedings_elba2021/proceedings_elba2021/Ribeiro}

%% file: proceed_lowx2021/royon.tex
\vspace*{1.2cm}

\thispagestyle{empty}
\begin{center}
{\LARGE \bf Anomalous coupling studies with intact protons at the LHC}

\par\vspace*{7mm}\par

{

\bigskip

\large \bf Christophe Royon}

\bigskip

{\large \bf  E-Mail: christophe.royon@cern.ch}

\bigskip

{Department of Physics and Astronomy, The University of Kansas, Lawrence KS 66047, USA}

\bigskip

{\it Presented at the Low-$x$ Workshop, Elba Island, Italy, September 27--October 1 2021}

\vspace*{15mm}

\end{center}
\vspace*{1mm}

\begin{abstract}

Measuring intact protons in the final state at the LHC at high luminosity allows increasing the reach on quartic $\gamma \gamma \gamma \gamma$, $\gamma \gamma WW$ and $\gamma \gamma \gamma Z$  anomalous  couplings  and on the search for Axion-Like Particles by about two to three orders of magnitude with respect to standard methods at the LHC.
\end{abstract}
\part[Anomalous coupling studies with intact protons at the LHC\\ \phantom{x}\hspace{4ex}\it{Christophe Royon}]{}

 \section{Photon induced processes at the LHC}

\begin{figure}[h]
\centering
\includegraphics[width=0.25\textwidth]{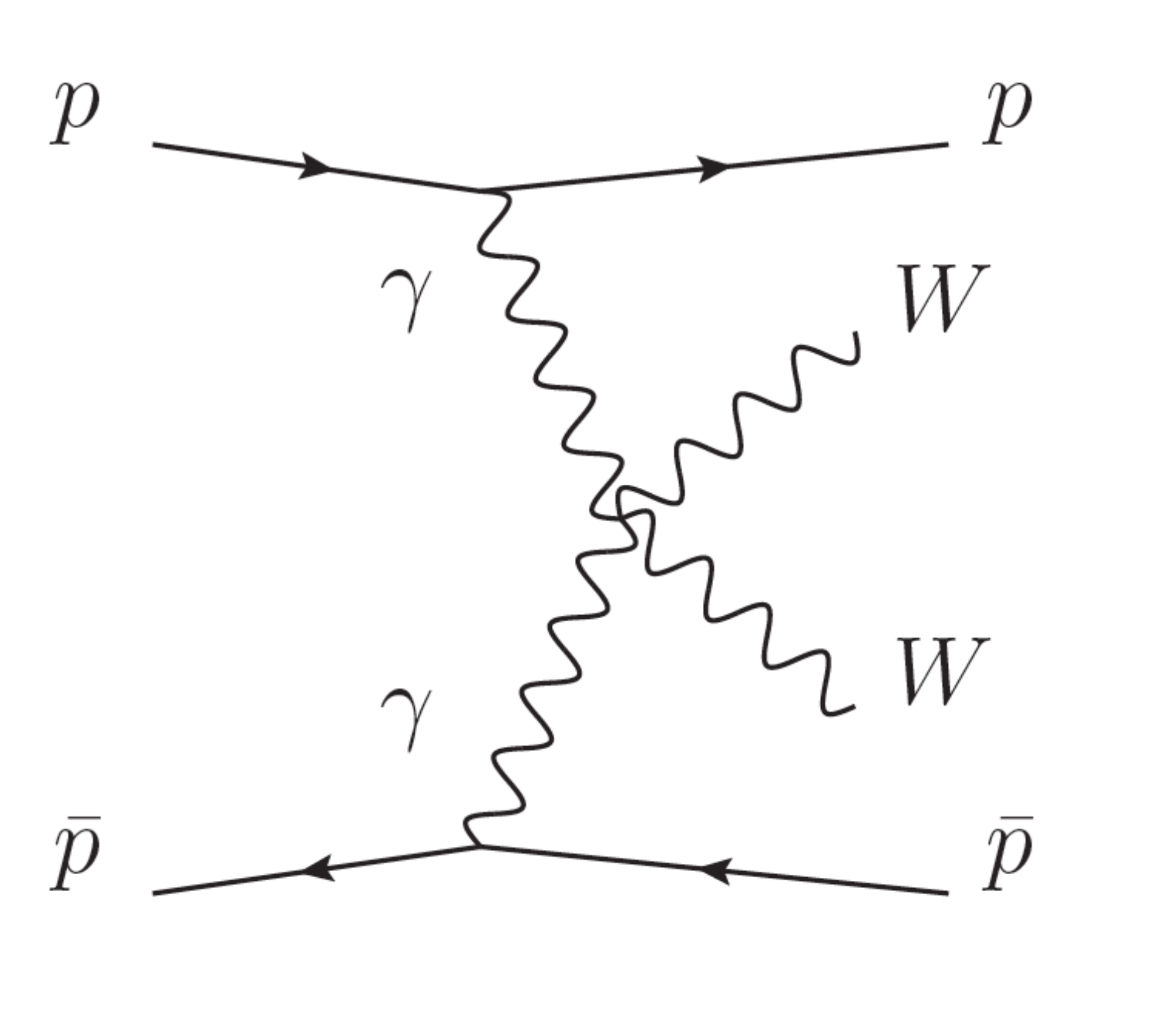}
\includegraphics[width=0.25\textwidth]{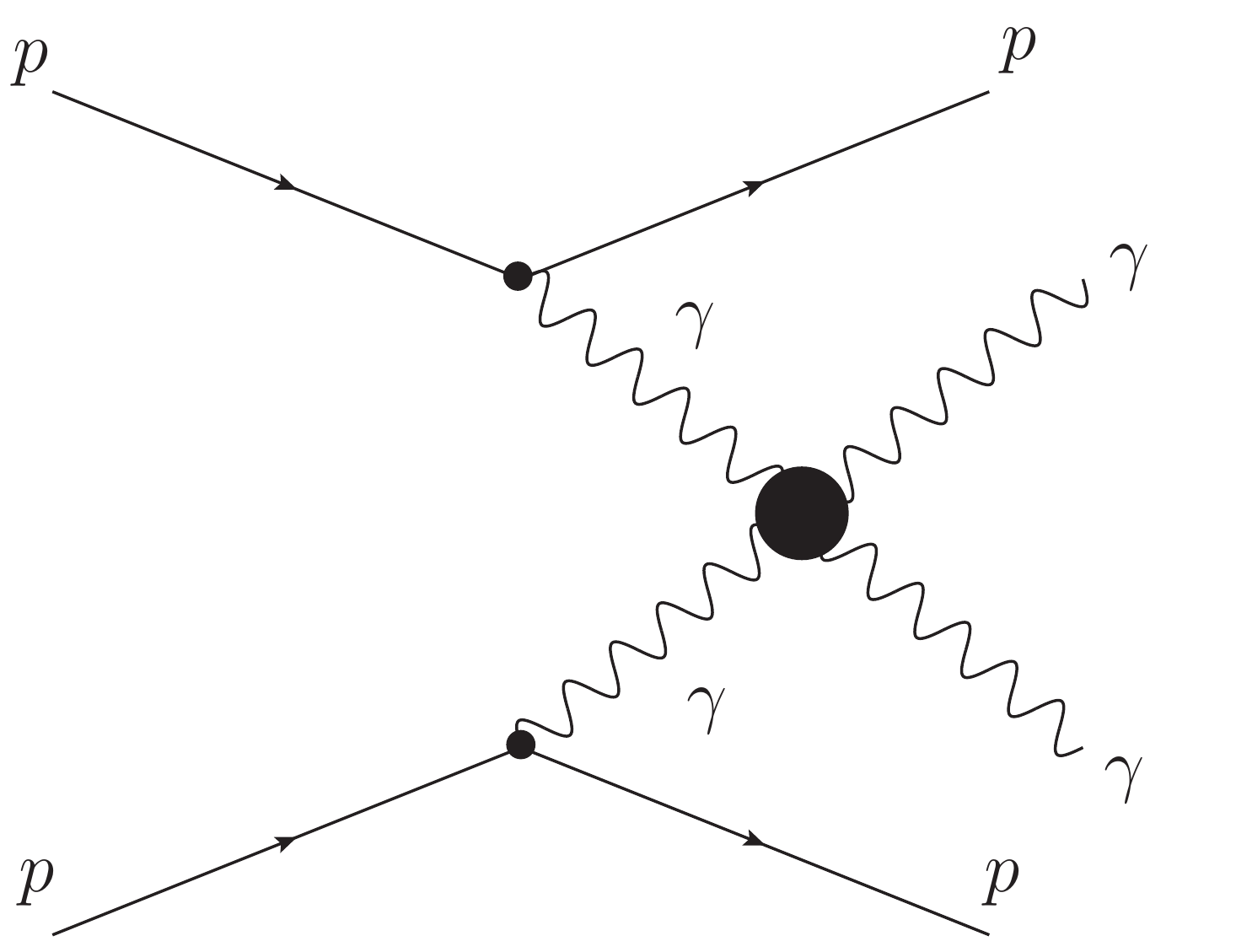}
\caption{Example of $WW$ and $\gamma \gamma$ exclusive production by photon exchanges.}
\label{fig0}
\end{figure}

We consider special events at the LHC that correspond to photon-induced processes where quasi-real photons are emitted by the incoming interacting protons as shown in Fig.~\ref{fig0} (we will see in the following why photon exchanges dominate at high energy).  Protons can be  intact after interactions and can be detected and measured in special detectors called roman pots. 
These events are especially clean since all particles in the final state (including the intact protons) are measured. We can produce exclusively pairs of photons and $W$ bosons in addition of the two intact protons in the final state  (see Fig.~\ref{fig0}). In the same way, one can look for the photon induced production of $ZZ$, $\gamma Z$, $t \bar{t}$, etc. 

As an example we will consider the production of two $\gamma$'s in the central ATLAS or CMS detectors, and of two intact protons. Both the ATLAS and CMS-TOTEM collaborations installed roman pots detectors at about 220 meters from the interaction point that can measure intact protons at high luminosity at the LHC, the so-called ATLAS Forward Proton (AFP) and CMS-TOTEM Precision Proton Spectrometer (PPS)~\cite{AFPPPS,AFPPPS2}. At high luminosity (standard runs with $\beta^* \sim 0.5$ at the LHC), the acceptance in mass of the two photons or the two $W$ bosons (see Fig.~\ref{fig0})  with intact protons tagged  in the roman pot detectors typically covers the domain 400-2300 GeV. We can thus get sensitivity to beyond standard model physics since we can produce high mass objects.

Quartic photon couplings $\zeta_1$ can be modified via loops of new particles or new resonances that couple to photons~\cite{pap2_1,pap2_2}. In the case of loops of new heavy particles, we get
\begin{eqnarray}
\zeta_1 = \alpha_{em}^2 Q^4 m^{-4} N c_{1,s} \nonumber
\end{eqnarray}
where $\zeta_1$ is proportional to the 4th power of the charge and inversely proportional to the 4th power of the mass of the charged particle, and on its spin, $c_{1,s}$. This
leads to $\zeta_1$ of the order of 10$^{-14}$ -10$^{-13}$ GeV$^{-4}$ depending on beyond standard model theories (extra-dimensions, composite Higgs...).
 $\zeta_1$ can also be modified by neutral particles
at tree level (extensions of the SM including scalar,
pseudo-scalar, and spin-2 resonances that couple to the photon). In that case
\begin{eqnarray}
\zeta_1 =
(f_s m)^{-2} d_{1,s} \nonumber
\end{eqnarray} 
where $f_s$ is the $\gamma \gamma X$ 
coupling of the new particle to the
photon, and $d_{1,s}$ depends on the spin of the particle. For instance, 2 TeV
dilatons lead to $\zeta_1 \sim$ 10$^{-13}$ GeV$^{-4}$. All these couplings were implemented in the FPMC generator~\cite{FPMC} that will be used in the following for all predictions. 

\section{Diphoton exclusive production: SM and BSM contributions}

\begin{figure}[h]
\centering
\includegraphics[width=0.8\textwidth]{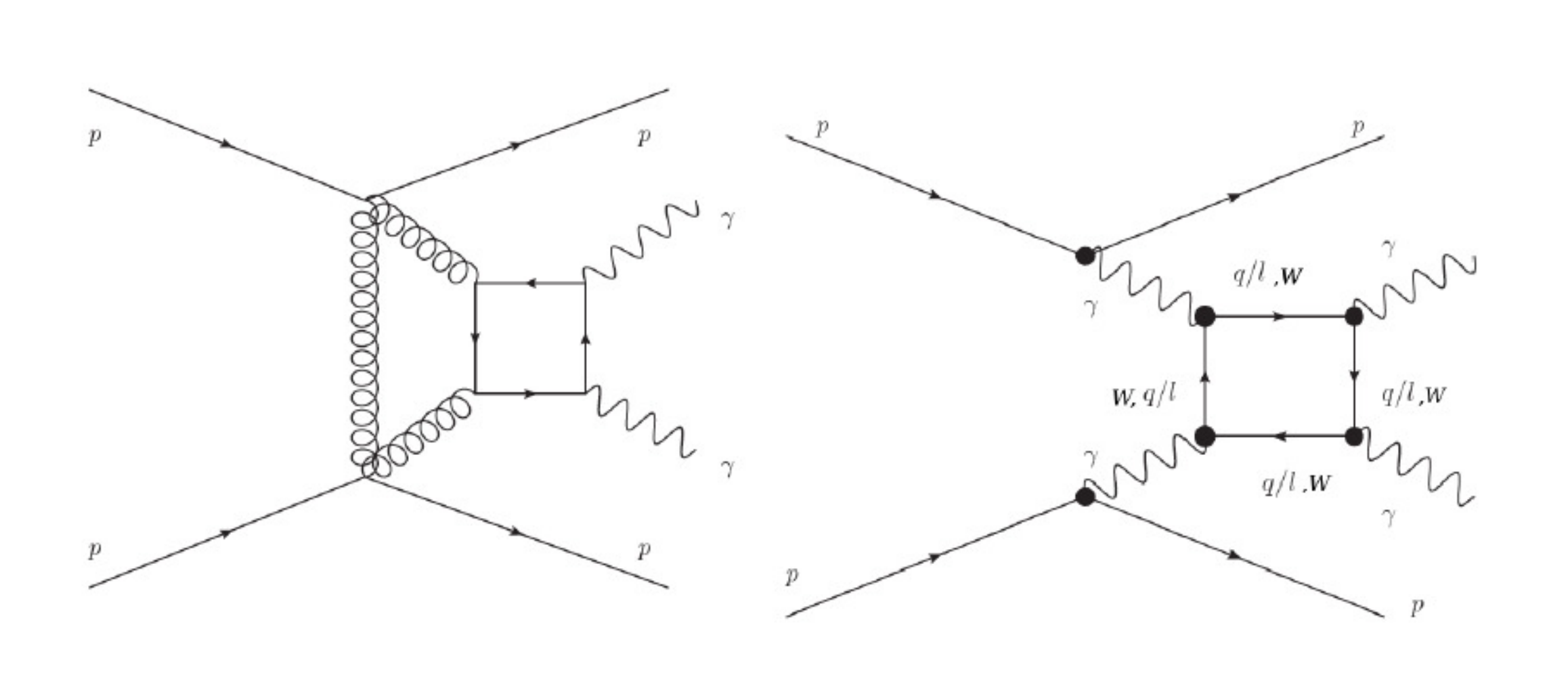}
\caption{Exlusive production of diphoton vis QCD processes (left) and QED photon exchanges (right).}
\label{fig1}
\end{figure}

\begin{figure}[h]
\centering
\includegraphics[width=0.6\textwidth]{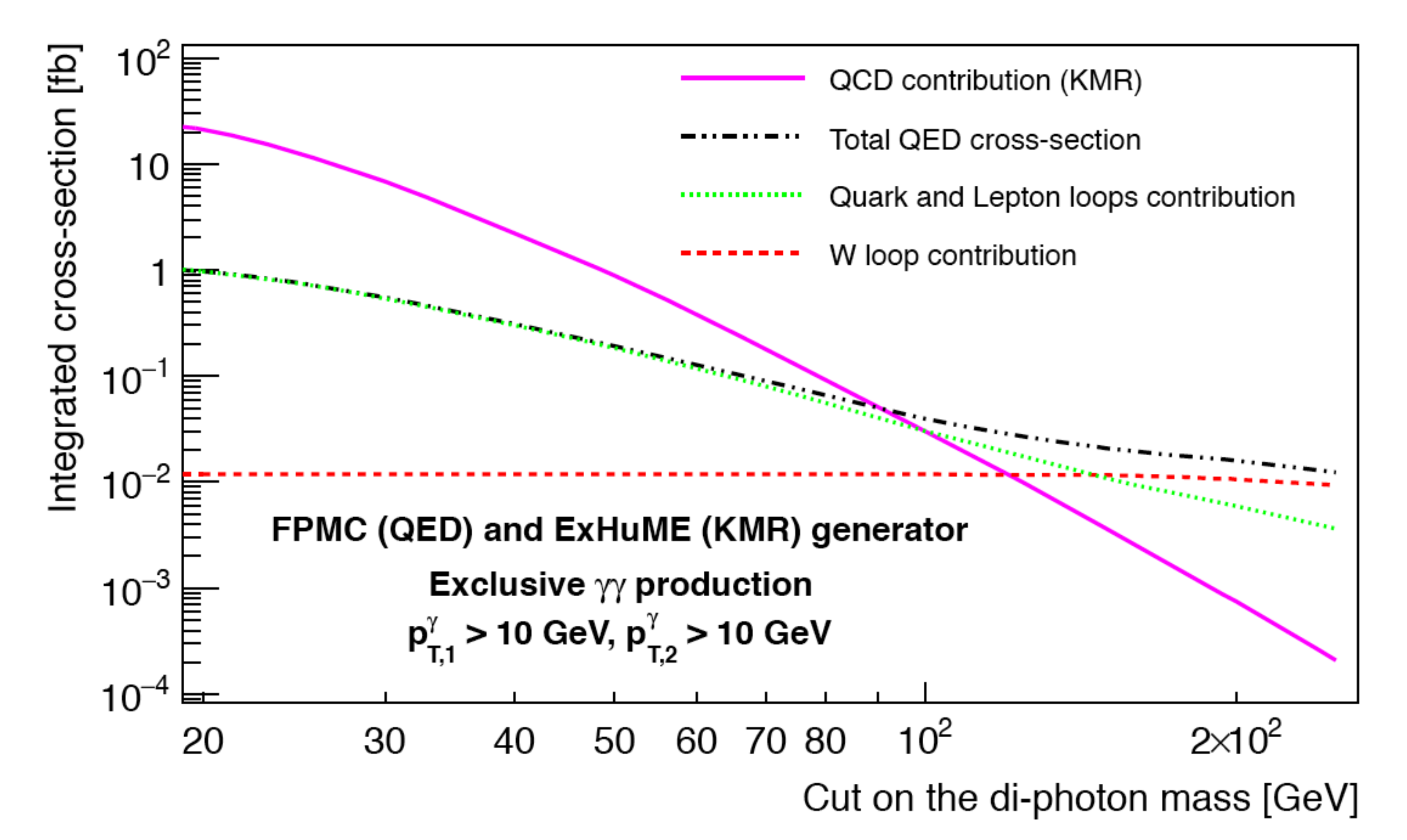}
\caption{Cross section of exclusive diphoton production above a given diphoton mass given in abscissa for QCD (full line) and QED (dashed dotted line) processes (see text).}
\label{fig2}
\end{figure}

In this section, we will concentrate on diphoton exclusive production  and the conclusions can be generalized to exclusive $WW$, $ZZ$, $\gamma Z$, and $t \bar{t}$ production via photon exchanges. We will start by examining the standard model (SM) production of exclusive diphotons as shown in Fig.~\ref{fig1}. Diphotons can be produced exclusively either via QCD (Fig.~\ref{fig1}, left) or QED processes (Fig.~\ref{fig1}, right). The cross sections for a diphoton mass above the value in abscissa are shown in Fig.~\ref{fig2}. In purple full line, we display the QCD contribution and in black dashed dotted line the sum of the three QED photon-induced contributions (in green dotted lines, the quarks and leptons loop contribution, and in red dashed line the $W$ loop contribution)~\cite{pap1_1,pap1_2,pap1_3,pap1_4}. We note that above a diphoton mass of 200 GeV, the QCD contribution becomes negligible. Recalling the fact that the acceptance of the roman pot detectors starts at about 400 GeV for standard running at the LHC, it is clear that observing two photons in ATLAS/CMS and two tagged protons means
a photon-induced process.

\begin{figure}[h]
\centering
\includegraphics[width=0.65\textwidth]{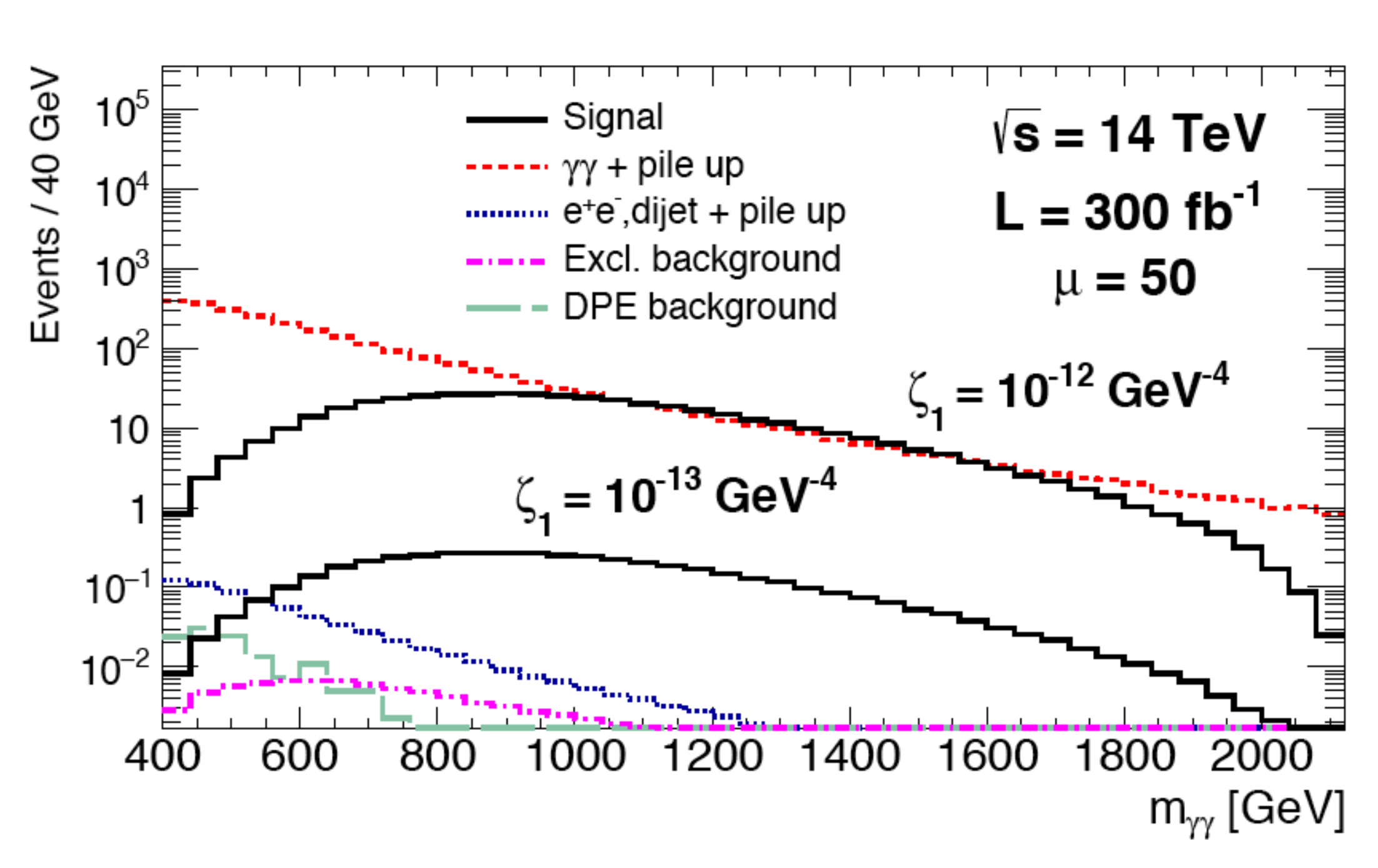}
\caption{Number of events as a function of the diphoton mass for signal and background for exclusive $\gamma \gamma$ production for 300 fb${-1}$.}
\label{fig3}
\end{figure}

Let us new give some details about the exclusive diphoton production analysis for a luminosity of 300 fb$^{-1}$ at the LHC. The number of events is shown in Fig~\ref{fig3}. The number of signal events is shown as a black line for two values of anomalous couplings. We also notice that the number of SM exclusive diphotons (red dashed dotted line) or exclusive dileptons with leptons misidentified as photons (blue dotted line) is quite low and can be neglected. The only background that matters is shown in red dashed lines, and corresponds to the non-exclusive standard diphoton production (with protons destroyed in the final state) superimposed with intact protons originating from secondary interactions called pile up. These events are due to the fact that we have up to 50 interactions per bunch crossing at the LHC at standard luminosities and diphoton productions can be easily superimposed with soft interactions producing intact protons. This is basically the only background that we have to consider.

\begin{figure}[h]
\centering
\includegraphics[width=0.95\textwidth]{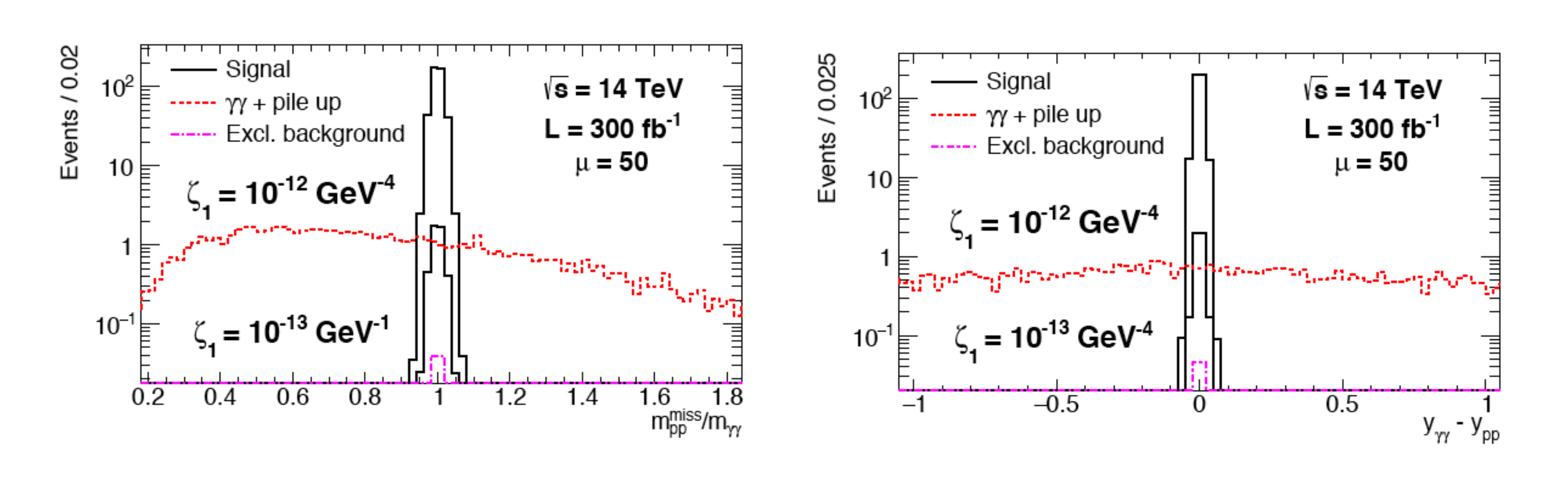}
\caption{Mass ratio and rapidity difference between the $pp$ and $\gamma \gamma$ information for signal (in full line) and pile up background (dashed line).}
\label{fig4}
\end{figure}

Measuring intact protons is crucial in order to suppress the pile up background. The method is quite simple. Since, for signal, we detect and measure all particles in the final state (namely the two photons, and the two intact protons), we can match the kinematical information as measured by the two photons with the one using the two protons, namely the rapidity and mass defined as
\begin{eqnarray}
M_{pp} &=& \sqrt{\xi_1 \xi_2 s} = M_{\gamma \gamma} \nonumber \\
y_{pp} &=& \frac{1}{2} \log \left( \frac{\xi_1}{\xi_2} \right) = y_{\gamma \gamma} \nonumber
\end{eqnarray}
where $\xi_1$ and $\xi_2$ are the proton fractional momentum loss. The results are shown in Fig.~\ref{fig4}, left, for the mass ratio and in Fig.~\ref{fig4}, right for the rapidity difference between the $pp$ and $\gamma \gamma$ information for signal in black full line and for pile up background in red dashed lines. It is clear that this variable can reject most of the pile up background and we obtain indeed less than 0.1 event of background for 300 fb$^{-1}$. The sensitivity on quartic photon anomalous coupling is thus up to a few
$10^{-15}$ GeV$^{-4}$, which is better by more than two orders of magnitude with respect to ``standard" methods
at the LHC~\cite{pap2_1,pap2_2}. 
Let us note that exclusivity cuts using proton tagging are crucial  to suppress backgrounds since,
without matching mass and rapidity requirements, the 
background would be about 80.2 events for 300 fb$^{-1}$. Running roman pot detectors at high luminosity at the LHC both in ATLAS and CMS-TOTEM at high luminosity was indeed motivated by the gain that we obtain on the reach on anomalous couplings~\cite{pap1_1,pap1_2,pap1_3,pap1_4}. This is now becoming a reality and both CMS-TOTEM and ATLAS reported recently some observation of QED exclusive dilepton production~\cite{dilepton1, dilepton2} and CMS-TOTEM the first limits on quartic photon anomalous couplings with about 9.4 fb$^{-1}$ of data~\cite{cmsdiphoton}. The analysis with the total accumulated luminosity (about 110 fb$^{-1}$) is in progress.

\begin{figure}[h]
\centering
\includegraphics[width=0.7\textwidth]{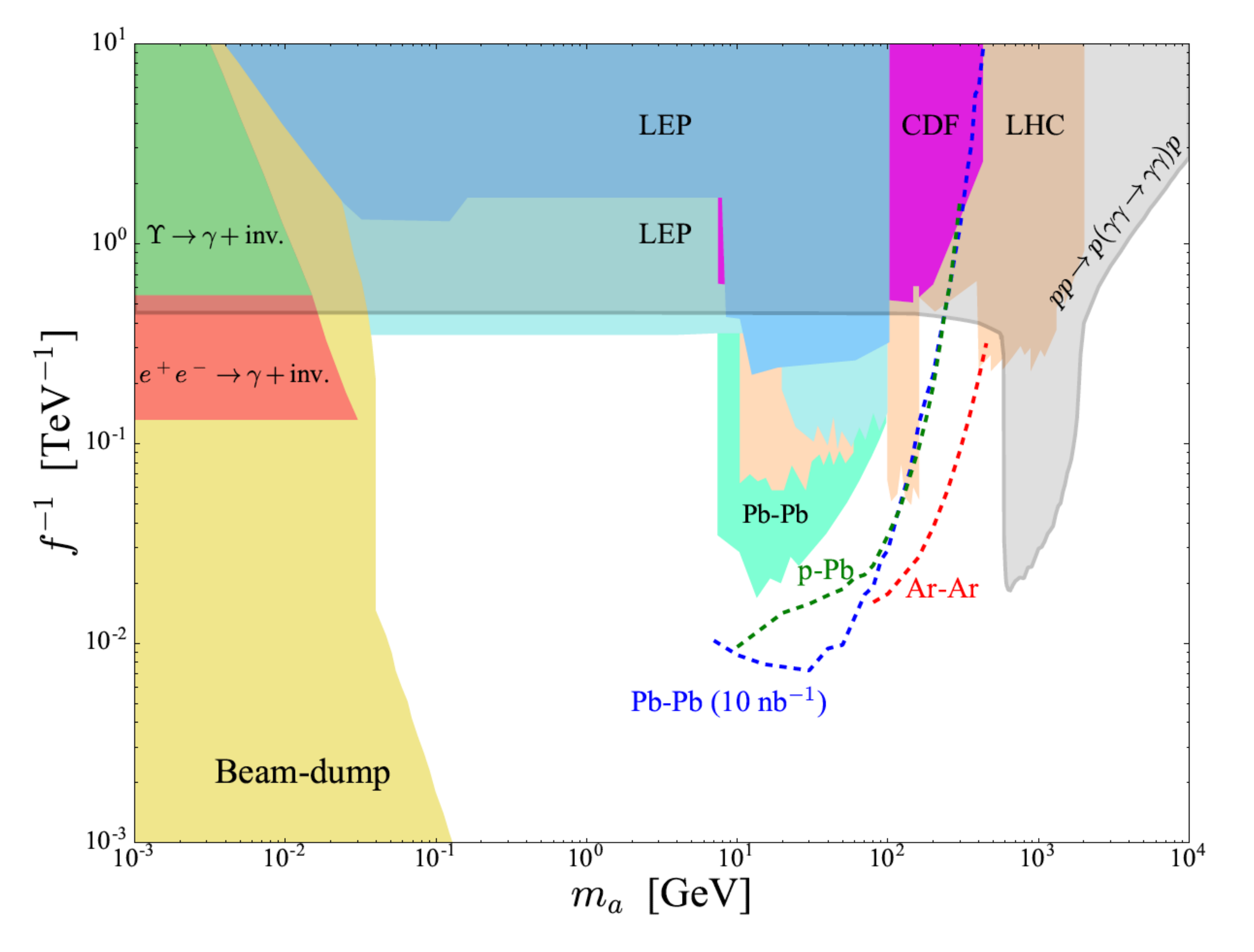}
\caption{Coupling vs ALP mass sensitivity plot. The reach using the measurement of two intact protons and the two photons for photon-induced processes is shown as a grey area for $pp$ collisions, and we also indicate the reach using heavy ion runs at the LHC covering the intermediate mass region.}
\label{fig5}
\end{figure}

This method can be applied directly to the search for axion-like particles (ALP) as an example. ALPs can be produced as a resonance via photon induced processes, and we can detect them using the method described above if they decay into two photons as an example. The sensitivity plot (coupling versus mass of the ALP) is shown in Fig.~\ref{fig5} for $pp$ interactions with 300 fb$^{-1}$ of data as a grey region at high ALP masses~\cite{alp1,alp2}. We gain about two orders of magnitude on sensitivity to ALP masses of the order of 1 TeV with respect to standard LHC methods and we reach a new domain at high mass that cannot be reached without tagging the protons. In addition, we also show for reference the complementarity with $Pb Pb$ runnings that cover the region  at lower masses (typically  ALP masses in the range 10-500 GeV) since the cross section is enhanced by a factor of $Z^4$ for $Pb Pb$ running~\cite{alp1,alp2}. In this case, we do not detect the intact or dissociate heavy ions in roman pot detectors but we use the rapidity gap method since the amount of pile up in heavy ion runs at the LHC is negligible.

\section{Anomalous production of $Z \gamma$ and $WW$ vis photon-induced processes}

\begin{figure}[h]
\centering
\includegraphics[width=0.5\textwidth]{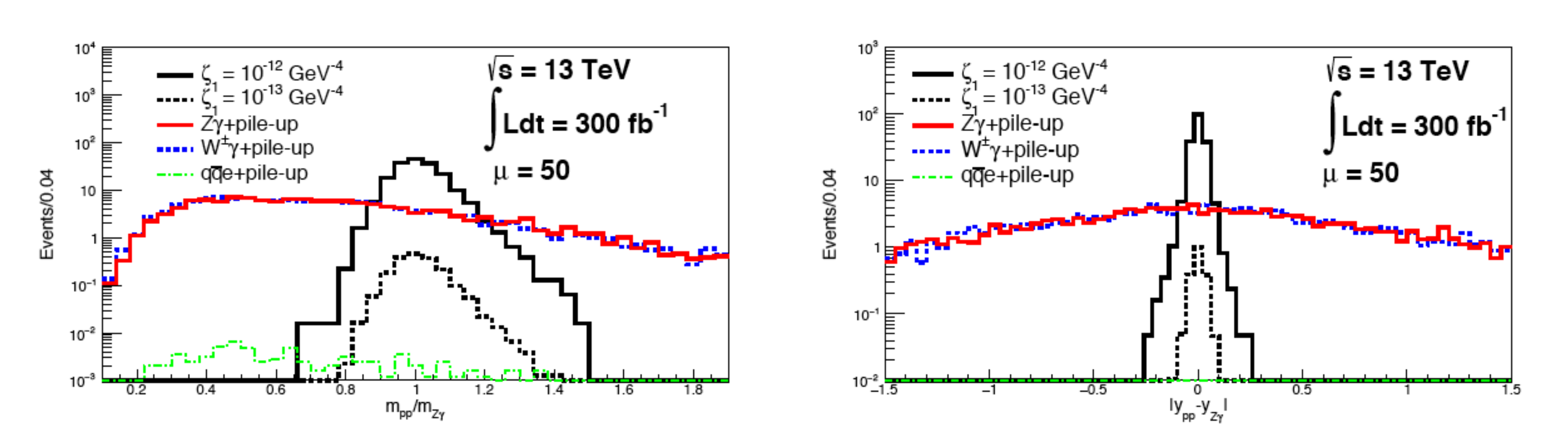}
\caption{Missing mass ratio and rapidity difference between the $Z \gamma$ and the di-proton system in the case when the $Z$ boson decays into two jets.}
\label{fig6b}
\end{figure}

Our previous study can be extended to other exclusive productions via photon exchanges and we will discuss briefly the production of $Z \gamma$ and $WW$ events. Exactly the same method of matching the mass and rapidity measurements of the $Z \gamma$ system with the tagged proton information can be used. The new aspect of this study is that we can consider both leptonic and hadronic decays of the $Z$ boson. Of course the resolution on mass and rapidity matching is worse since the jet resolution is worse than for leptons as illustrated in Fig.~\ref{fig6b}, but it leads to unprecedented sensitivities to $\gamma \gamma \gamma Z$ anomalous couplings, up to 10$^{-13}$, better by three orders of magnitude~\cite{gammaz} than the sensitivities without tagging the protons at the LHC (the usual method being to look for the three photon decay of the $Z$ boson).

\begin{figure}[h]
\centering
\includegraphics[width=0.5\textwidth]{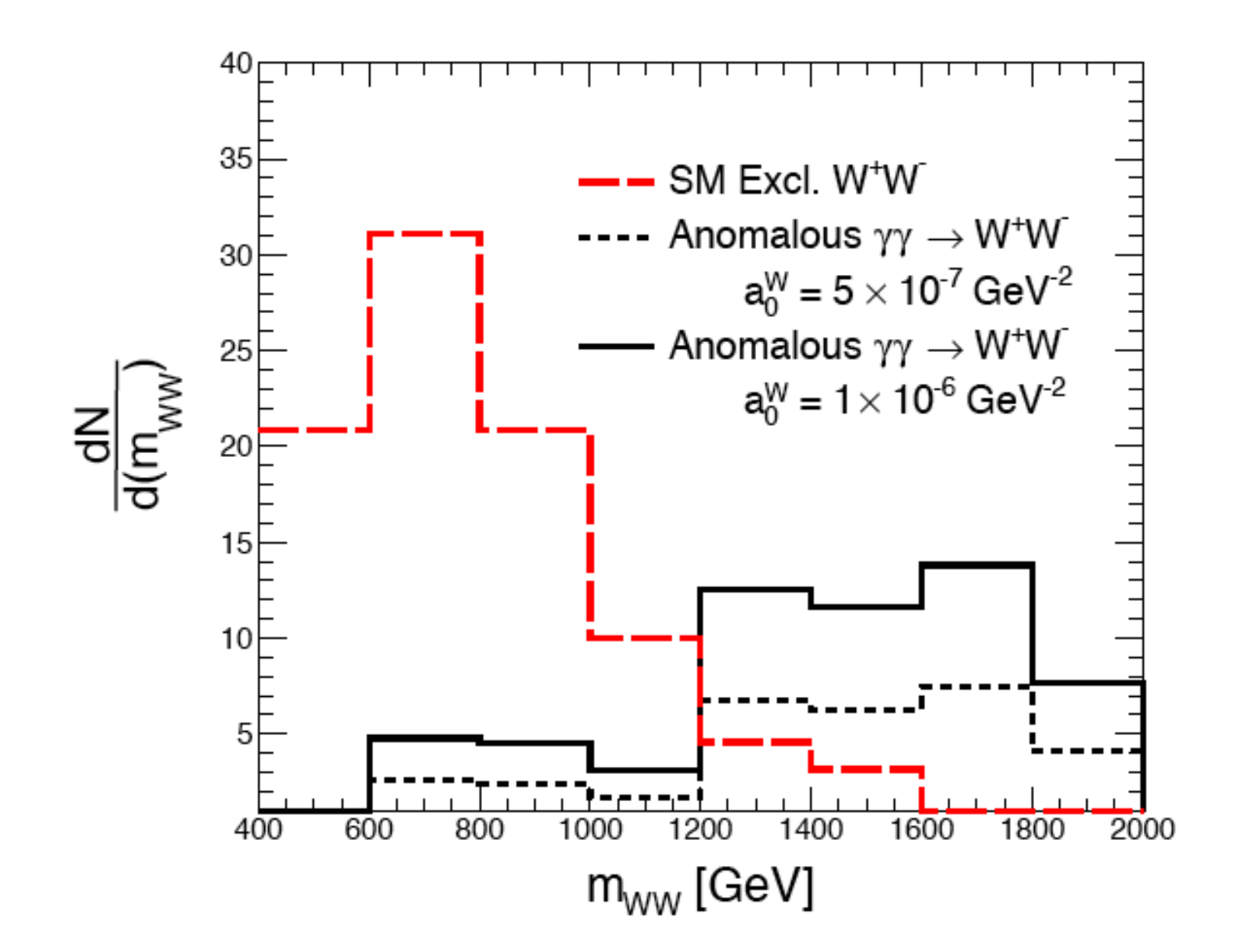}
\caption{$WW$ mass distribution for exclusive $WW$ production (SM is red dashed line and anomalous couplings in full ball line).}
\label{fig6}
\end{figure}

The same study can be used to observe the SM exclusive production of $WW$ bosons via photon exchanges and also to increase our sensitivity to quartic $\gamma \gamma WW$ anomalous couplings. Recent studies~\cite{ww} showed that the strategy is somewhat different for SM and BSM studies.  To measure the SM exclusive $WW$ production (the cross section is of the order of 95.6 fb at the LHC), the best sensitivity originates from the leptonic decays of the $W$s where we can obtain about 50 events with 2 events of background for 300 fb$^{-1}$. The non-zero background originates from the fact that the neutrinos originating from the leptonic decay of the $W$ bosons cannot obviously be measured and this is why the mass and rapidity matching does not work so nicely. Fast timing detectors are needed to suppress further the background in this case. The strategy to look for $\gamma \gamma WW$ quartic anomalous couplings is slightly different since the anomalous coupling events appear at high $WW$ mass as shown in Fig.~\ref{fig6}. The best sensitivity to quartic $\gamma \gamma WW$ couplings appear by looking at the hadronic decay of the $W$ bosons even if the dijet background is quite high. The sensitivity with 300 fb$^{-1}$ is of the order of 3.7 10$^{-7}$ GeV$^{-2}$, better by almost three orders of magnitude that the present LHC sensitivity. This can be further improved by using more advanced jet variables such as subjettiness in order to reject further the dijet background.

\section{Conclusion}
In this report we considered the exclusive production of $\gamma \gamma$, $WW$ and $Z \gamma$ via photon induced processes, considering the LHC as a $\gamma \gamma$ collider. Tagging the protons in dedicated ATLAS-AFP or CMS-TOTEM-PPS roman pot detector as well as the $\gamma \gamma$, $WW$, $Z \gamma$ in the main ATLAS or CMS detector ensures that we have a photon-induced process since 
gluon exchanges are suppressed at high mass in the acceptance of the roman pot detectors.
Matching the kinematical information of the central system with the tagged protons ensures that we have a background-free experiment  and any
observed event is a signal. This leads to better sensitivities to quartic anomalous coupling by two or three order of magnitude with respect to the standard methods at the LHC depending on the process.

\nocite{*}
\bibliographystyle{auto_generated.bst} 
\bibliography{proceed_lowx2021/royon.bib}


%% file: santimaria_proceedings_elba2021/santimaria_proceedings_elba2021.tex
\vspace*{1.2cm}

\thispagestyle{empty}
\begin{center}
{\LARGE \bf The LHCspin project}

\par\vspace*{7mm}\par

{
\bigskip
\large \bf 
M. Santimaria\textsuperscript{1$\star$},
V. Carassiti\textsuperscript{2},
G. Ciullo\textsuperscript{2,3},
P. Di Nezza\textsuperscript{1},
P. Lenisa\textsuperscript{2,3},
S. Mariani\textsuperscript{4,5},
L. L. Pappalardo\textsuperscript{2,3} and
E. Steffens\textsuperscript{6}
}

\bigskip

\begin{center}
{\bf 1} INFN Laboratori Nazionali di Frascati, Frascati, Italy
\\
{\bf 2} INFN Ferrara, Italy
\\
{\bf 3} Department of Physics, University of Ferrara, Italy
\\
{\bf 4} INFN Firenze, Italy
\\
{\bf 5} Department of Physics, University of Firenze, Italy
\\
{\bf 6} Physics Dept., FAU Erlangen-Nurnberg, Erlangen, Germany
\\
* marco.santimaria@lnf.infn.it
\end{center}

\bigskip

{\it Presented at the Low-$x$ Workshop, Elba Island, Italy, September 27--October 1 2021}

\end{center}
\vspace*{1mm}

\begin{abstract}

The goal of LHCspin is to develop, in the next few years, innovative solutions and cutting-edge technologies to access spin physics in polarised fixed-target collisions at high energy, exploring the unique kinematic regime offered by LHC and exploiting new final states by means of the LHCb detector.
The forward geometry of the LHCb spectrometer is perfectly suited for the reconstruction of particles produced in fixed-target collisions. 
This configuration, with centre of mass energies
ranging from $\sqrt{s_{\rm{NN}}}=115~\rm{GeV}$ in $p-p$ interactions to
$\sqrt{s_{\rm{NN}}}=72~\rm{GeV}$ in heavy ion collisions,
allows to cover a wide backward rapidity region, 
including the poorly explored high$-x$ regime.
With the instrumentation of the proposed target system, LHCb will become the first experiment simultaneously collecting unpolarised beam-beam collisions at $\sqrt{s_{\rm{NN}}}=14~\rm{TeV}$ and both unpolarised and polarised beam-target collisions.
The status of the project is presented along with a selection of physics opportunities.
\end{abstract}
 \part[The LHCspin project\\ \phantom{x}\hspace{4ex}\it{M. Santimaria et al}]{}
\section{Introduction}
\label{sec:intro_santi}
The LHC delivers proton and lead beams with an energy of $7~\rm{TeV}$ and $2.76~\rm{TeV}$ per nucleon, respectively, with world's highest intensity. A short run with xenon ions was also performed in 2017, while an oxygen beam is currently foreseen for the Run 3~\cite{Citron:2018lsq_santi}.
Fixed-target proton-gas collisions occur at a centre of mass energy per nucleon of up to $115~\rm{GeV}$.
This corresponds to a centre of mass rapidity shift of
$y-y_{\rm{CM}} \approx \rm{arcsinh}(\sqrt{E_{\rm{N}} / 2M_{\rm{N}}})=4.8$, so that the LHCb acceptance $(2<\eta<5)$ covers the backward and central rapidities in the centre of mass frame.
Such a coverage offers an unprecedented opportunity to investigate partons carrying a large fraction of the target nucleon momentum, i.e.~large Bjorken$-x$ values, corresponding to large and negative Feynman$-x$ values\footnote{$x_{\rm{F}} \approx x_1 - x_2$ where $x_1$ and $x_2$ are the Bjorken$-x$ values of the beam and target nucleon, respectively.}.

\begin{figure}[ht]
\centering
\includegraphics[width=0.9\textwidth]{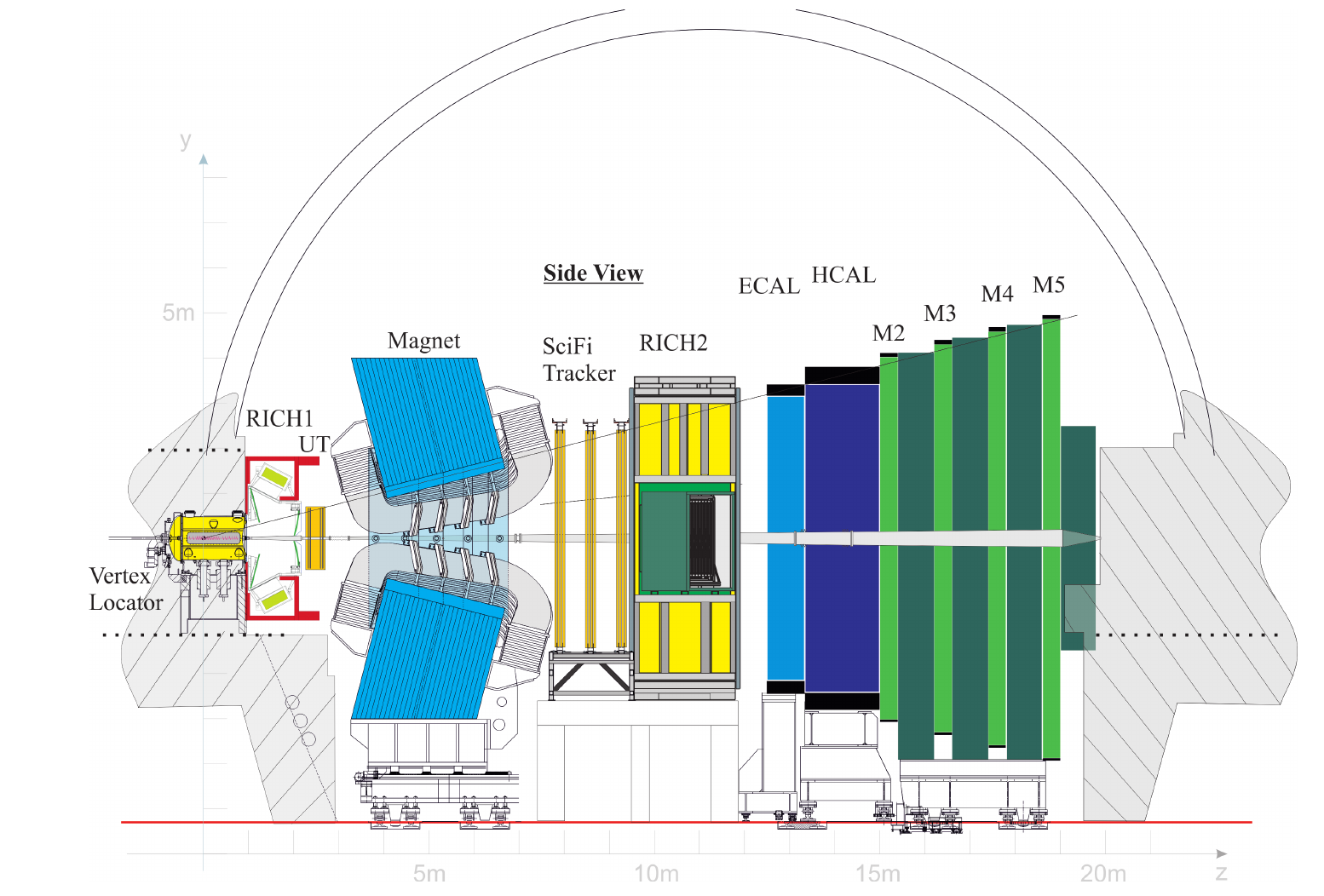}
\caption{The Run 3 LHCb detector.}
\label{fig:lhcb}
\end{figure}

The LHCb detector~\cite{Alves:2008zz} is a general-purpose forward spectrometer specialised in detecting hadrons containing $c$ and $b$ quarks, and the only LHC detector able to collect data in both collider and fixed-target mode.
It is fully instrumented in the $2<\eta<5$ region with a vertex locator (VELO), a tracking system, two Cherenkov detectors, electromagnetic and hadronic calorimeters and a muon detector.
Fig.~\ref{fig:lhcb} shows a scheme of the upgraded LHCb detector which is currently being installed for the Run 3, starting in 2022.

\begin{figure}[ht]
\centering
\includegraphics[width=0.99\textwidth]{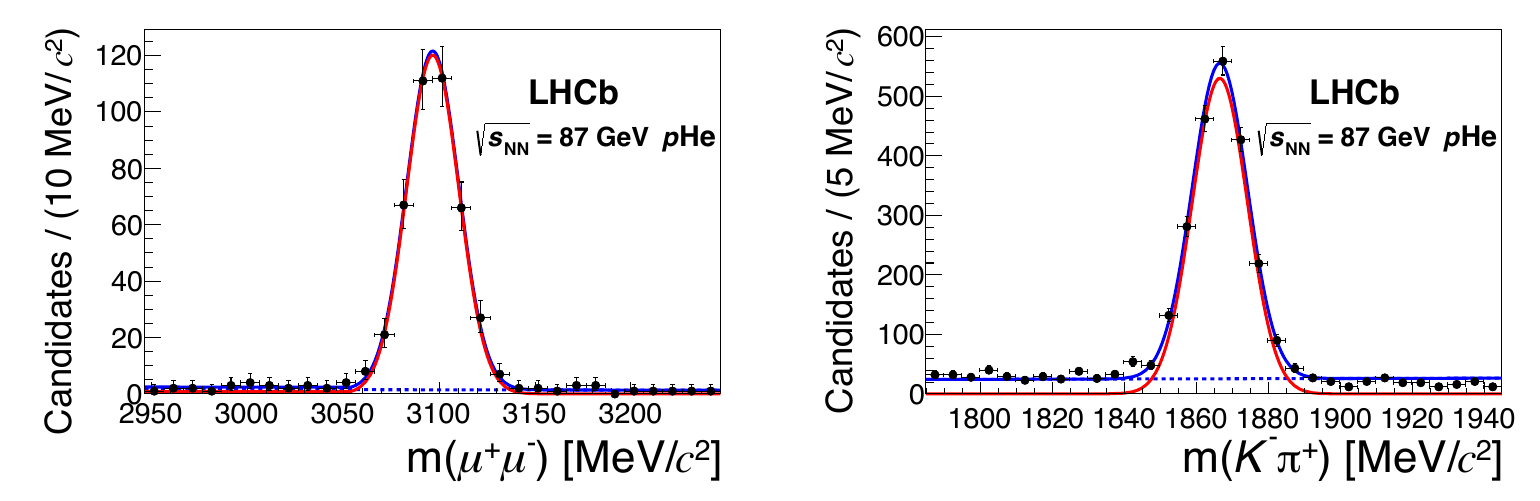}
\caption{$J/\psi\to\mu^+\mu^-$ (left) and $D^0\to K^-\pi^+$ (right) SMOG samples from~\cite{LHCb:2018ygc}.}
\label{fig:smog_phe}
\end{figure}

The fixed-target physics program at LHCb is active since the installation of the SMOG (System for Measuring the Overlap with Gas) device~\cite{Aaij:2014ida}, enabling the injection of noble gases in the beam pipe section crossing the VELO detector at a pressure of $\mathcal{O}(10^{-7})~\rm{mbar}$.
Precise measurements of charm~\cite{LHCb:2018ygc} and antiproton~\cite{LHCb:2018jry} production have been published based on $p-\rm{Ar}$ and $p-\rm{He}$ fixed-target collisions. Fig.~\ref{fig:smog_phe} shows the high-quality and low-background samples collected in just one week of dedicated SMOG runs during Run 2.

\begin{figure}[ht]
\centering
\includegraphics[width=0.49\textwidth]{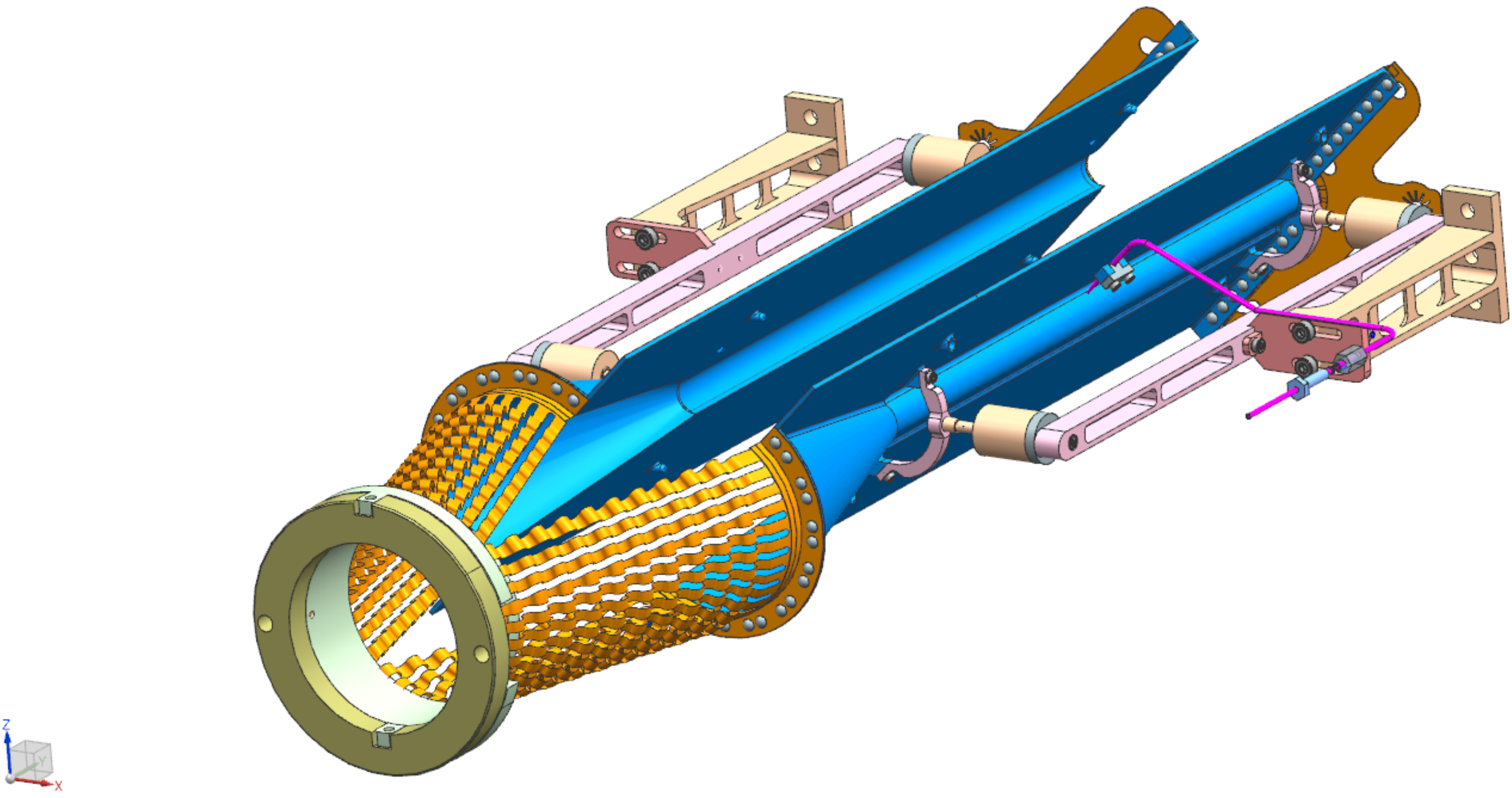}
\includegraphics[width=0.49\textwidth]{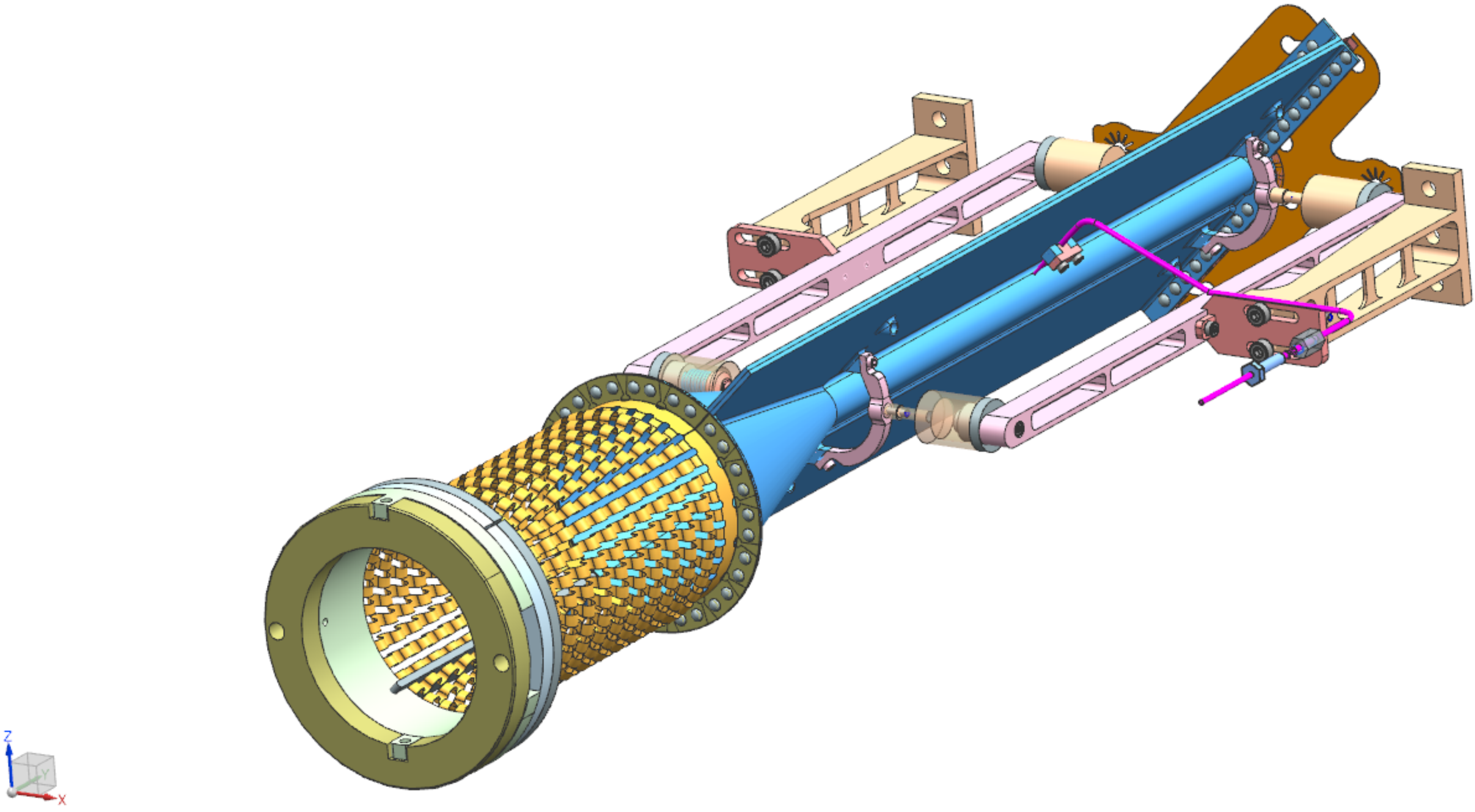}
\caption{The SMOG2 storage cell in the open (left) and closed (right) configuration.}
\label{fig:smog2}
\end{figure}

\begin{figure}[h]
\centering
\includegraphics[width=0.52\textwidth]{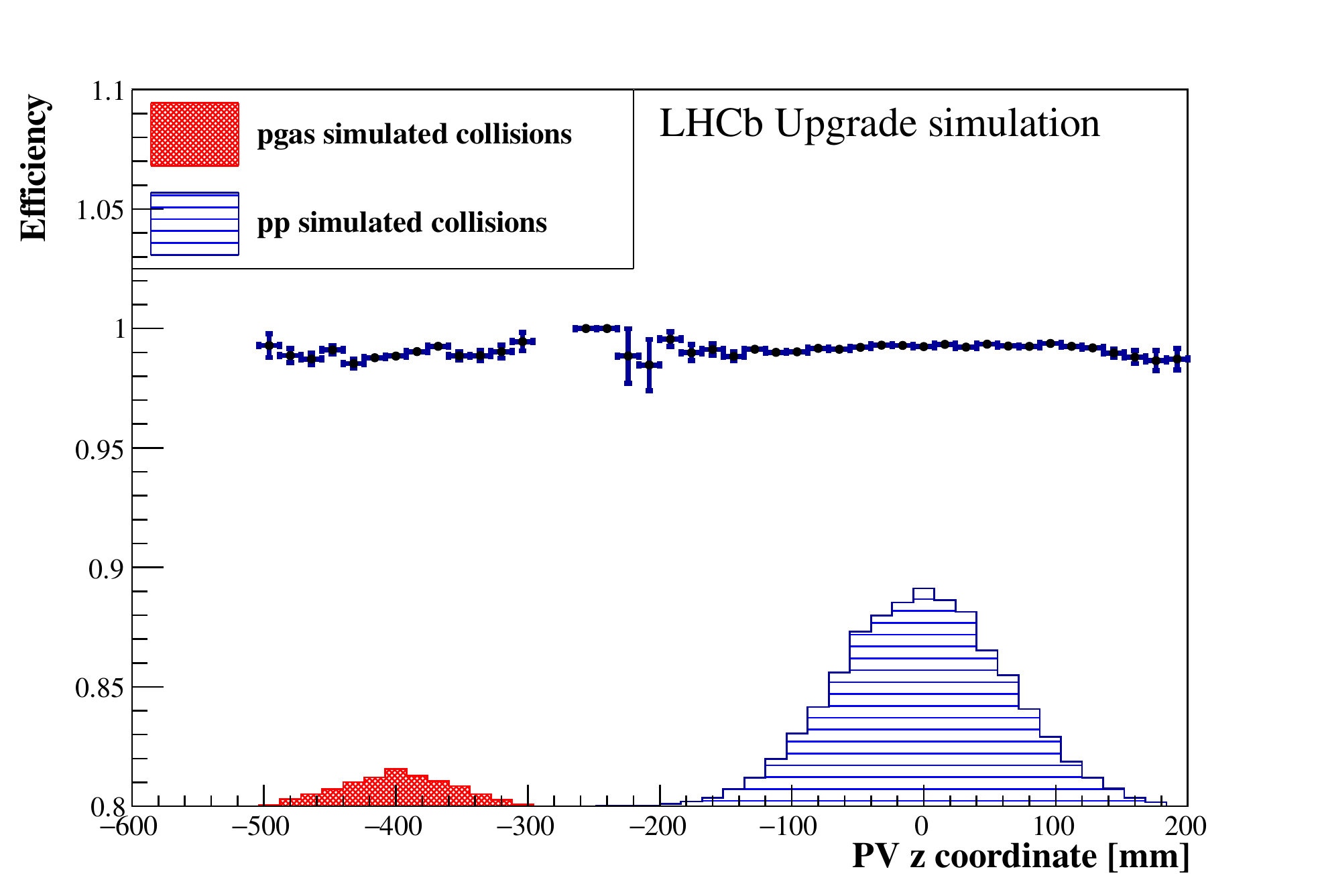}
\hfill
\includegraphics[width=0.47\textwidth]{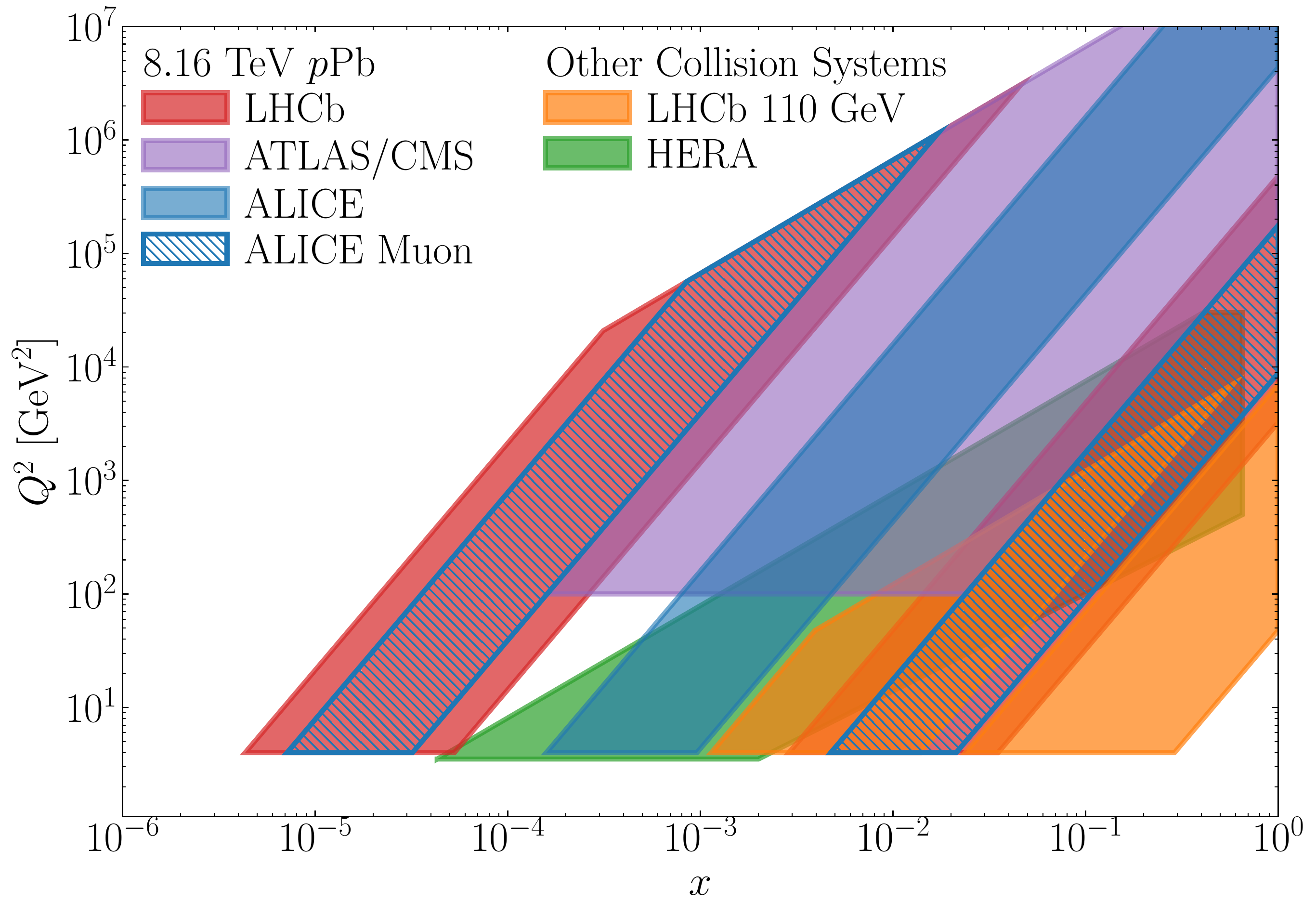}
\caption{Left: VELO track reconstruction efficiency for beam-gas (red) and beam-beam (blue) primary vertices (PV)~\cite{LHCB-FIGURE-2019-007}. Right: kinematic coverage of LHCspin (orange) and other existing facilities.}
\label{fig:kin}
\end{figure}

With the SMOG2 upgrade~\cite{LHCbCollaboration:2673690}, an openable gas storage cell, shown in Fig.~\ref{fig:smog2}, has been installed in 2020 in front of the VELO. 
The cell boosts the target areal density by a factor of $8$ to $35$ depending on the injected gas species. In addition, SMOG2 data will be collected in the upcoming Run 3 with a novel reconstruction software allowing simultaneous data-taking of beam-gas and beam-beam collisions. Very high tracking efficiency is expected in the beam-gas interaction region, despite its upstream position with respect to the VELO. Furthermore, beam-gas and beam-beam vertices are well detached along the $z$ coordinate, as shown in Fig.~\ref{fig:kin}.
SMOG2 will offer a rich physics program for the Run 3 and, at the same time, allows to investigate the dynamics of the beam-target system, setting the basis for future developments.
\\
The LHCspin project~\cite{Aidala:2019pit} aims at extending the SMOG2 physics program in Run 4 (expected to start in 2028) and, with the installation of a polarised gas target, to bring spin physics at LHC by exploiting the well suited LHCb detector. 
A selection of physics opportunities accessible at LHCspin is presented in Sec.~\ref{sec:phys}, while the experimental setup is discussed in Sec.~\ref{sec:det}.

\section{Physics case}
\label{sec:phys}
The physics case of LHCspin covers three main areas: exploration of the wide physics potential offered by unpolarised gas targets, investigation of the nucleon spin and heavy ion collisions.

\subsection{Measurements with unpolarised gases}
\label{ssec:pdfs}
\begin{figure}[ht]
\centering
\includegraphics[width=0.40\textwidth]{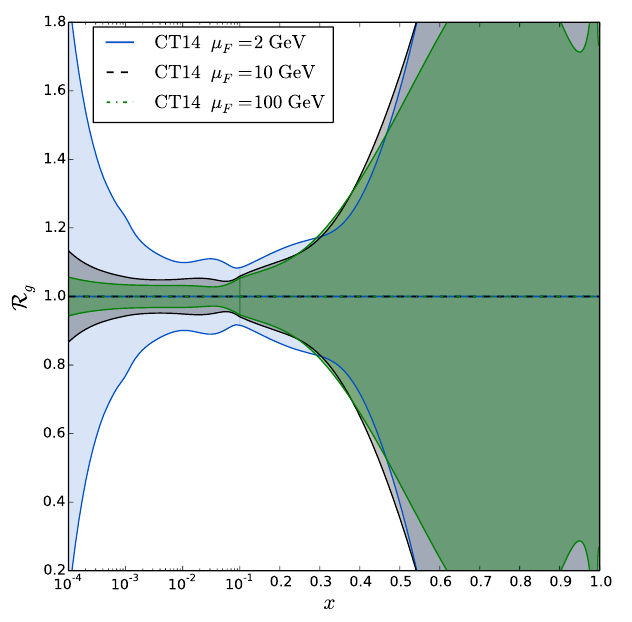}
\hfill
\includegraphics[width=0.50\textwidth]{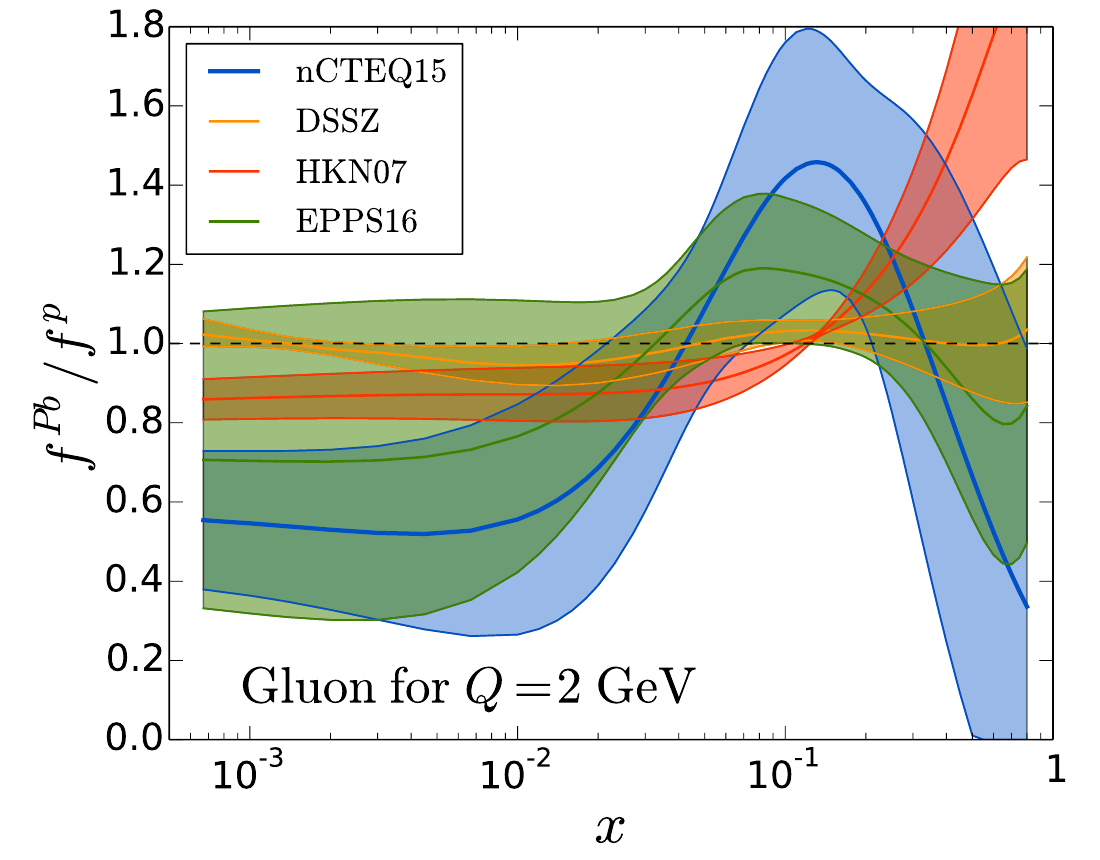}
\caption{Left: Relative uncertainty on the CT$14$ gluon PDF in a proton~\cite{Dulat:2015mca_santi}. Right: relative uncertainties on a set of gluon nuclear PDFs in a lead nucleus~\cite{Aidala:2019pit}.}
\label{fig:gpdf}
\end{figure}

Similarly to SMOG2, LHCspin will allow the injection of several species of unpolarised gases: $\rm{H}_2$, $\rm{D}_2$, He, $\rm{N}_2$, $\rm{O}_2$, Ne, Ar, Kr and Xe. The impact of the gas on the LHC beam lifetime is negligible: the luminosity loss due to collisions on hydrogen in the cell has a characteristic time of around $2000$ days, whereas typical runs last for $10-20$ hours.
\\
Injecting unpolarised gases yields excellent opportunities to investigate parton distribution functions (PDFs) in both nucleons and nuclei in the large-$x$ and intermediate $Q^2$ regime (Fig.~\ref{fig:kin}, right), which are especially affected by lack of experimental data and impact several fields of physics from QCD to astrophysics.
For example, the large acceptance and high reconstruction efficiency of LHCb on heavy flavour states enables the study of gluon PDFs, which represent fundamental inputs for theoretical predictions~\cite{Hadjidakis:2018ifr} and are currently affected by large uncertainties, as shown in the example of Fig.~\ref{fig:gpdf} (left).
In addition, the structure of heavy nuclei is known to depart from that obtained by the simple sum of free protons and neutrons: within the unique acceptance of LHCb, a large amount of data can be collected to shed light on the intriguing anti-shadowing effect expected at $x\sim 0.1$~\cite{Eskola:2016oht}, as shown in Fig.~\ref{fig:gpdf} (right).

With the large amount of data to be collected with LHCspin, several measurements impacting astrophysics and cosmic ray physics become possible.
For example, heavy-flavour hadroproduction directly impacts the knowledge of prompt muonic neutrino flux~\cite{Garzelli:2016xmx}, which is especially affected by PDF uncertainties, while large samples of proton collisions on helium, oxygen and nitrogen provide valuable inputs to improve the understanding of the compositions of ultra-high energy cosmic rays. Moreover, the possibility of injecting an oxygen beam opens new and exciting prospects for antiproton measurements~\cite{Brewer:2021kiv_santi}. 

\subsection{Spin physics}
\begin{figure}[ht]
\centering
\includegraphics[width=0.99\textwidth]{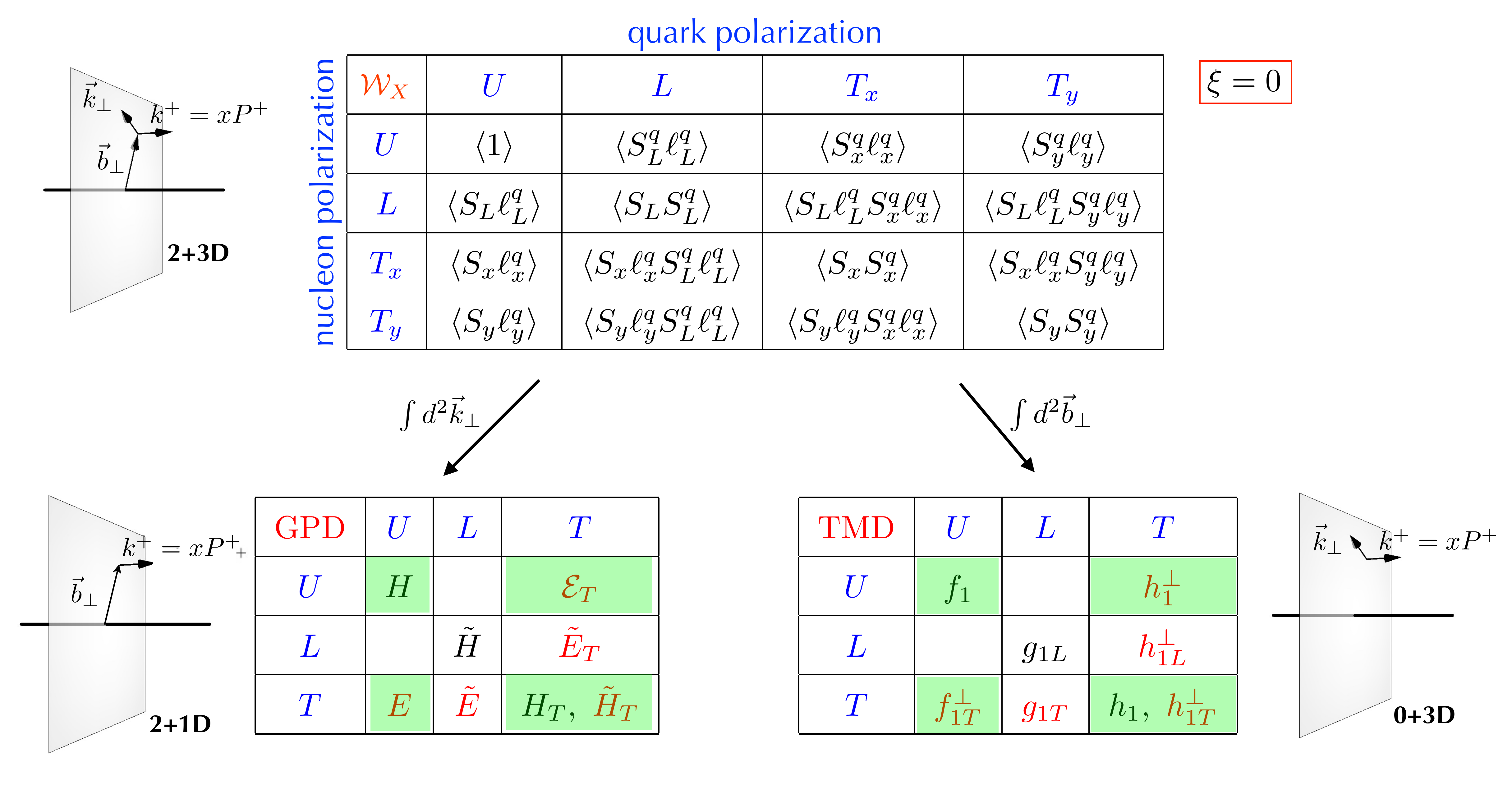}
\caption{Wigner distributions (top) and leading-twist GPDs and TMDs for different combinations of quark$\times$nucleon polarisation states (bottom). Distributions marked in red vanish for no orbital angular momentum contribution to the nucleon spin, while the quantities highlighted in green can be accessed at LHCspin~\cite{pasquini}.}
\label{fig:wigner}
\end{figure}

Beside the colliner PDFs mentioned in Sec.~\ref{ssec:pdfs}, polarised quark and gluon distributions can be probed at LHCspin by means of proton collisions on polarised hydrogen and deuterium.
Fig.~\ref{fig:wigner} shows the 5D Wigner distributions~\cite{Bhattacharya:2017bvs} which, upon integration on the transverse momentum, lead to the observable generalised parton distributions (GPDs), while transverse momentum dependent distributions (TMDs) are obtained when integrating over the transverse coordinate. There are several leading-twist distributions that can be probed with unpolarised and transversely polarised quarks and nucleons, giving independent information on the spin structure of the nucleon.

\begin{figure}[ht]
\centering
\includegraphics[width=0.56\textwidth]{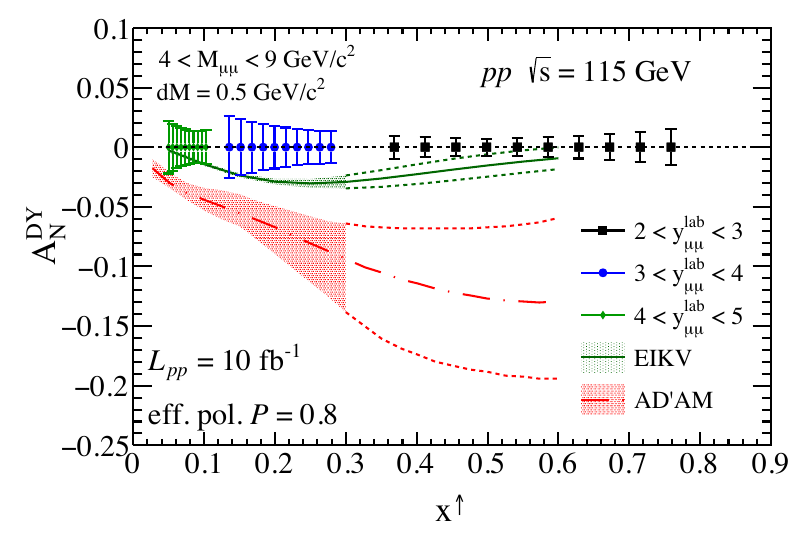}
\hfill
\includegraphics[width=0.42\textwidth]{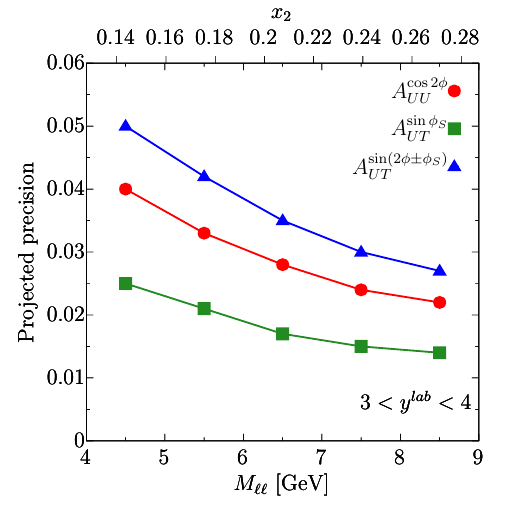}
\caption{Left: Measurements of $A_N$ in DY events as a function of $x$ compared to two theoretical predictions~\cite{Aidala:2019pit}. Right: projected precision for some azimuthal asymmetry amplitudes with DY data as a function of the dilepton invariant mass~\cite{Hadjidakis:2018ifr}.}
\label{fig:dy}
\end{figure}

To access the transverse motion of partons within a polarised nucleon, transverse single spin asymmetries (TSSAs) can be measured. For example
\begin{equation}
    A_N=\frac{1}{\it{P}}\frac{\sigma^{\uparrow}-\sigma^{\downarrow}}{\sigma^{\uparrow}+\sigma^{\downarrow}} \sim \frac{f^q_1(x_1,k^2_{T1}) \otimes f^{\perp\overline{q}}_{1T}(x_2,k^2_{T2})}{f^q_1(x_1,k^2_{T1}) \otimes f^{\overline{q}}_1(x_2,k^2_{T2})}
\end{equation}
in the polarised Drell-Yan (DY) channel probes the product of $f_1$ (unpolarised TMD) and $f^{\perp}_{1T}$ (Sivers function) in quarks and antiquarks in the low $(x_1)$ and high $(x_2)$ $x$ regimes. Projections for the uncertainty of such measurements are shown in Fig.~\ref{fig:dy} (left) based on an integrated luminosity of $10~\rm{fb}^{-1}$. Being T-odd, it is theoretically established that the Sivers function changes sign in polarised DY with respect to semi-inclusive deep inelastic scattering~\cite{Collins:2002kn}. This fundamental QCD prediction can be verified by exploiting the large sample of DY data expected at LHCspin. In addition, isospin effects can be investigated by comparing $p-\rm{H}$ and $p-\rm{D}$ collisions.
\\
Several TMDs can be probed by evaluating the azimuthal asymmetries of the produced dilepton pair: projected precisions for three such asymmetries are shown in Fig.~\ref{fig:dy} (left).

\begin{figure}[ht]
\centering
\includegraphics[width=0.49\textwidth]{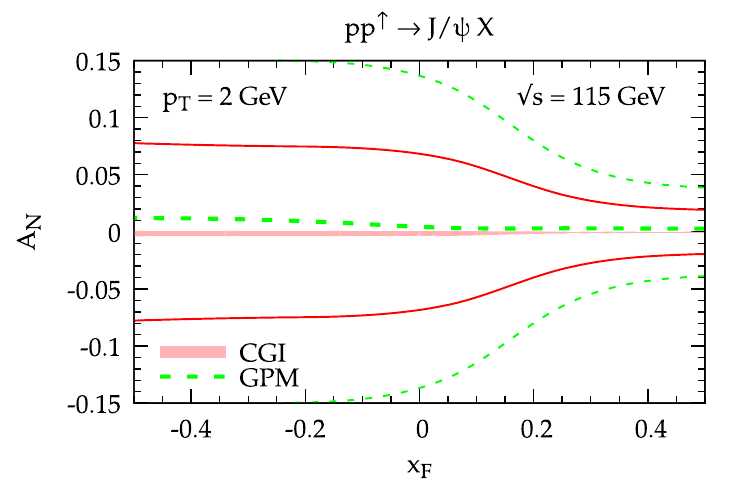}
\hfill
\includegraphics[width=0.49\textwidth]{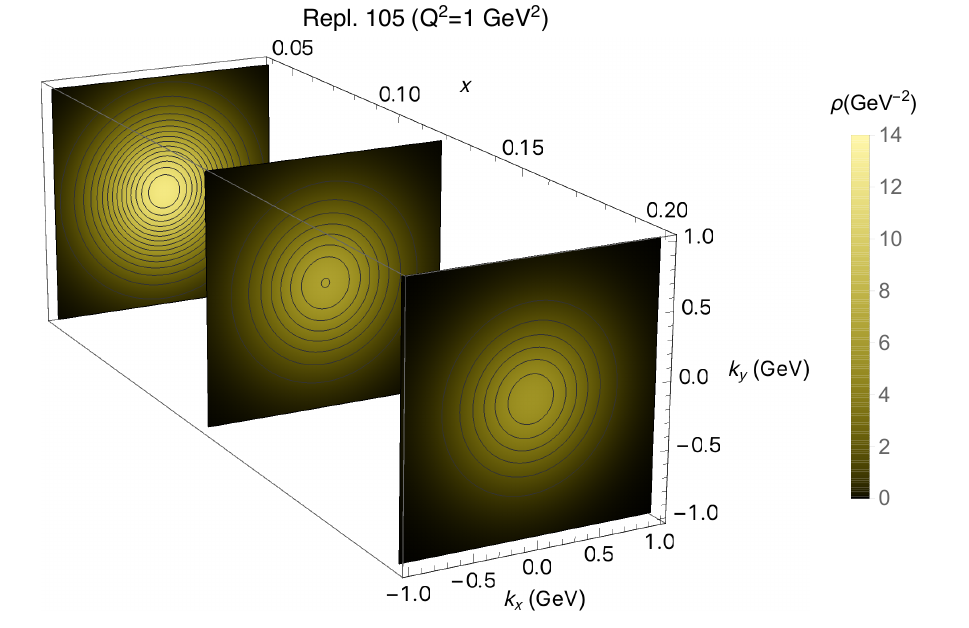}
\caption{Left: theoretical predictions for $A_N$ in inclusive $J/\psi$ production~\cite{DAlesio:2018rnv}. Right: up quark densities in momentum space~\cite{Bacchetta:2017gcc}.}
\label{fig:tmds}
\end{figure}

Heavy flavour states will be the strength point of LHCspin. Being mainly produced via gluon fusion at LHC, quarkonia and open heavy flavour states will allow to probe the unknown gluon Sivers function via inclusive production of $J/\psi$, $D^0$ but also with several unique states like $\eta_c$, $\chi_c$, $\chi_b$ or $J/\psi J/\psi$.
Fig.~\ref{fig:tmds} (left) shows two predictions for $A_N$ on $J/\psi$ events: $5-10\%$ asymmetries are expected in the $x_F<0$ region, where the LHCspin sensitivity is the highest.
Heavy flavour states can be exploited as well to probe the gluon-induced asymmetries $h^{\perp g}_{1}$ (Boer-Mulders) and $f_1^g$ (always present at the denominator of $A_N$), which are both experimentally unconstrained. 

\begin{figure}[ht]
\centering
\includegraphics[width=0.8\textwidth]{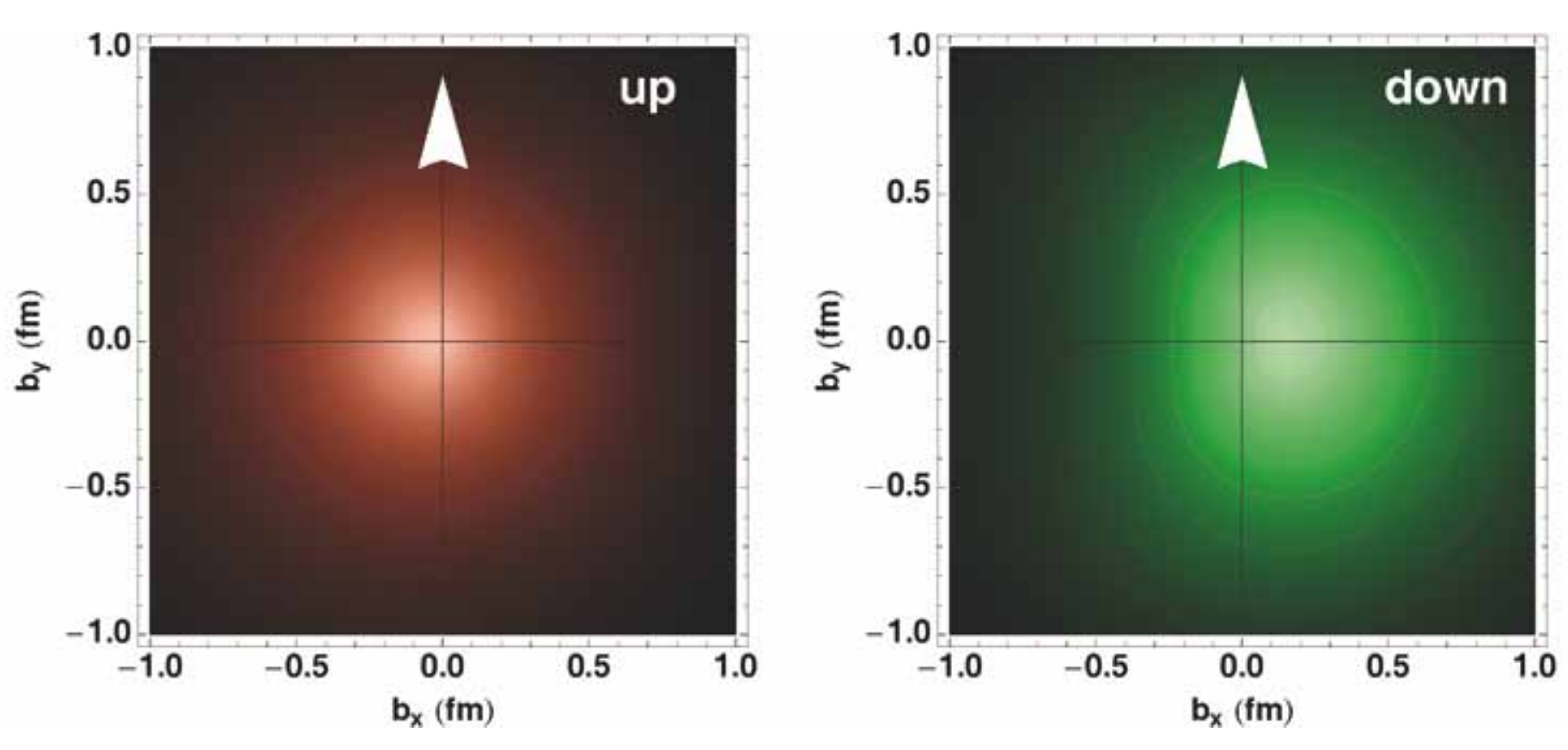}
\caption{Distortion of the up and down quark distributions in the impact parameter space when spin is taken into account~\cite{saggiatore}.}
\label{fig:pic}
\end{figure}

While TMDs provide a ``tomography'' of the nucleon in momentum space (Fig.~\ref{fig:tmds}, right), a 3D picture in the spatial coordinates can be built by measuring GPDs, as shown in Fig.~\ref{fig:pic}.
Correlating position and momentum, GPDs also quantify the parton orbital angular momentum, whose contribution to the total nucleon spin can be inferred, for example, via the Ji sum rule~\cite{Ji:1996ek}.
GPDs can be experimentally probed with exclusive dilepton and quarkonia productions in ultra-peripheral collisions, which are dominated by the electromagnetic interaction.
For example, TSSAs can be exploited to access the $E_g$ function, which has never been measured and represents a key element of the proton spin puzzle.
It is also attractive to measure the elusive
transversity PDF, whose knowledge is currently limited to valence quarks at the leading order~\cite{Radici:2018iag}, as well as its integral, the tensor charge, which is of direct interest in constraining physics beyond the Standard Model~\cite{Courtoy:2015haa}. 

\subsection{Heavy ion collisions}
\begin{figure}[h]
\centering
\includegraphics[width=0.47\textwidth]{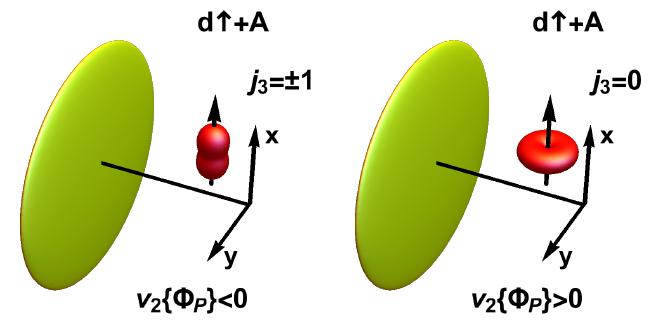}
\hspace{1cm}
\includegraphics[width=0.42\textwidth]{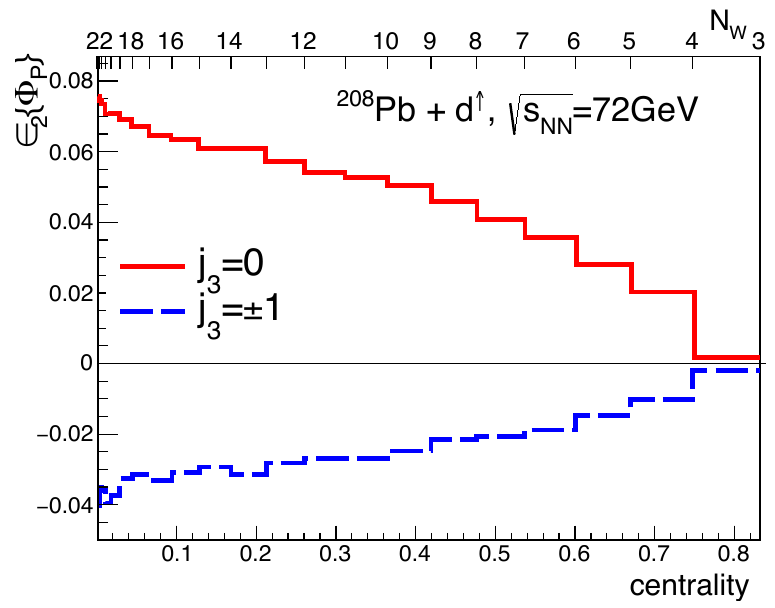}
\caption{Left: sketch of a ultra-relativistic collision of a lead nucleus against a transversely polarised deuteron in two different angular momentum projections. Right: ellipticity with respect to the polarisation axis as a function of the collision centrality with LHCspin kinematics~\cite{Broniowski:2019kjo}.}
\label{fig:deuteron}
\end{figure}

Thermal heavy-flavour production is negligible at the typical temperature of few hundreds MeV of the system created in heavy-ion collisions. Quarkonia states ($c\overline{c}$, $b\overline{b}$) are instead produced on shorter timescales, and their energy change while traversing the medium represents a powerful way to investigate Quark-Gluon Plasma (QGP) properties. LHCb capabilities allow to both cover the aforementioned charmonia and bottomonia studies and to extend them to beauty baryons as well as to exotic probes.
The QGP phase diagram exploration at LHCspin can be performed with a rapidity scan at a centre of mass energy of $\sqrt{s_{\rm{NN}}}=72~\rm{GeV}$ which is in-between those accessed at RHIC and SPS. 
In addition, flow measurements will greatly benefit from the excellent identification performance of LHCb on charged and neutral light hadrons.
\\
The dynamics of small systems is an interesting topic joining heavy-ion collisions and spin physics.
In the spin $1$ deuteron nucleus, the nucleon matter distribution is prolate for $j_3=\pm1$ and oblate for $j_3=0$, where $j_3$ is the projection of the spin along the polarisation axis.
In ultra-relativistic lead ion collisions on transversely polarised deuteron, the deformation of the target deuteron can influence the orientation of the fireball in the transverse plane, quantified by the ellipticity, as shown in Fig.~\ref{fig:deuteron}.
The measurement proposed in~\cite{Broniowski:2019kjo} can easily be performed at LHCspin on minimum bias events thanks to the high-intensity LHC beam.

\section{Experimental setup}
\label{sec:det}
\begin{figure}[ht]
\centering
\includegraphics[width=0.9\textwidth]{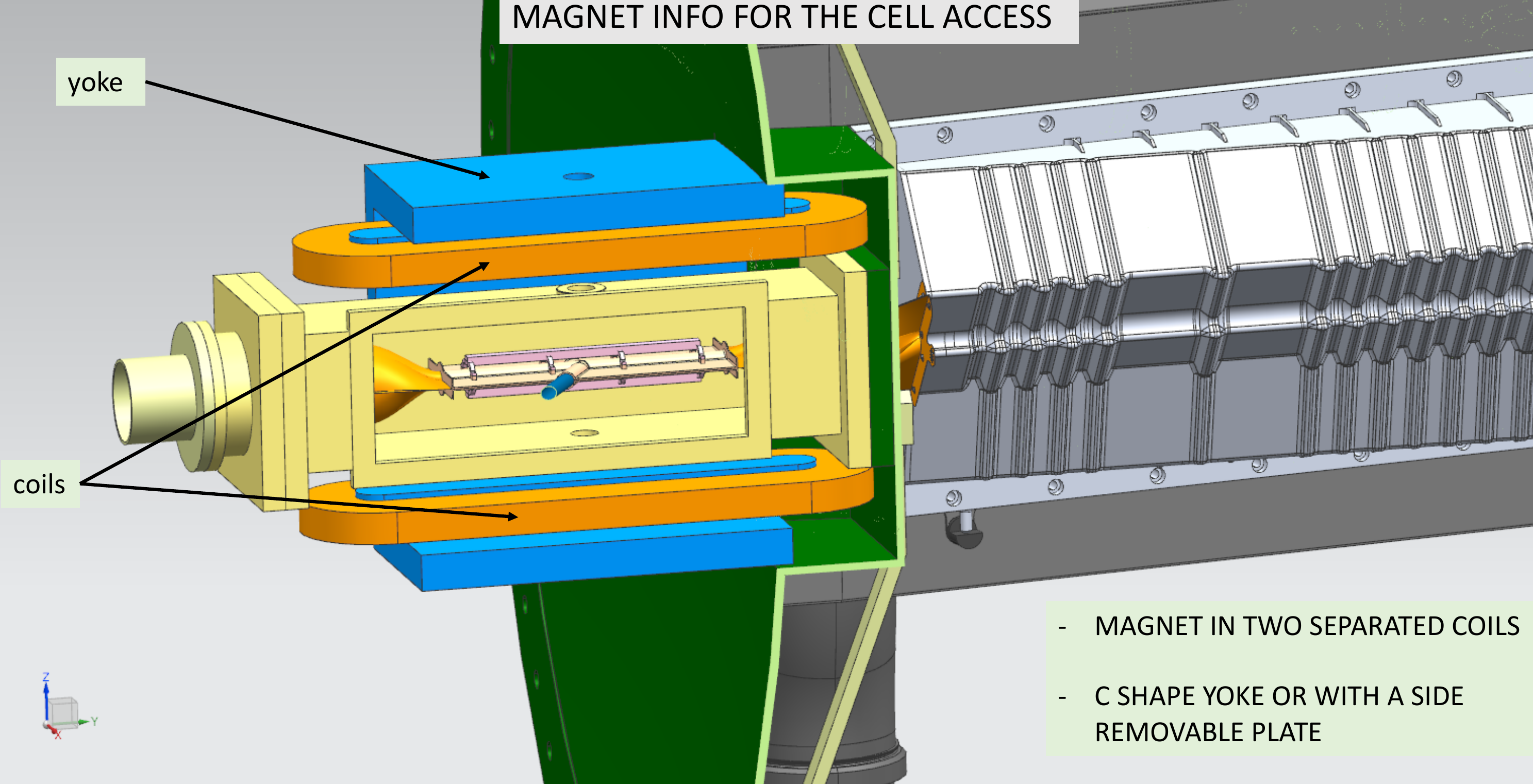}
\caption{A drawing of the PGT with the magnet coils (orange) and the iron return yoke (blue) enclosing the storage cell. The VELO vessel and detector box are shown in green and grey, respectively.}
\label{fig:rd}
\end{figure}

The LHCspin experimental setup is in R\&D phase and calls for the development of a new generation polarised target. The starting point for this ambitious task is the setup of the polarised target system employed at the HERMES experiment~\cite{Airapetian:2004yf} and comprises three main components: an Atomic Beam Source (ABS), a Polarised Gas Target (PGT) and a diagnostic system.
The ABS consists of a dissociator with a cooled nozzle, a Stern-Gerlach apparatus to focus the wanted hyperfine states, and adiabatic RF-transitions for setting and switching the target polarisation between states of opposite sign.
The ABS injects a beam of polarised hydrogen or deuterium into the PGT, which is located in the LHC primary vacuum.
The PGT hosts a T-shaped openable storage cell, sharing the SMOG2 geometry, and a compact superconductive dipole magnet, as shown in Fig.~\ref{fig:rd}. The magnet generates a $300~\rm{mT}$ static transverse field with a homogeneity of 10\%, which is found to be suitable to avoid beam-induced depolarisation~\cite{Steffens:2019kgb}.
Studies for the inner coating of the storage cell are currently ongoing, with the aim of producing a surface that minimises the molecular recombination rate as well as the secondary electron yield.

\begin{figure}[ht]
\centering
\includegraphics[width=0.49\textwidth]{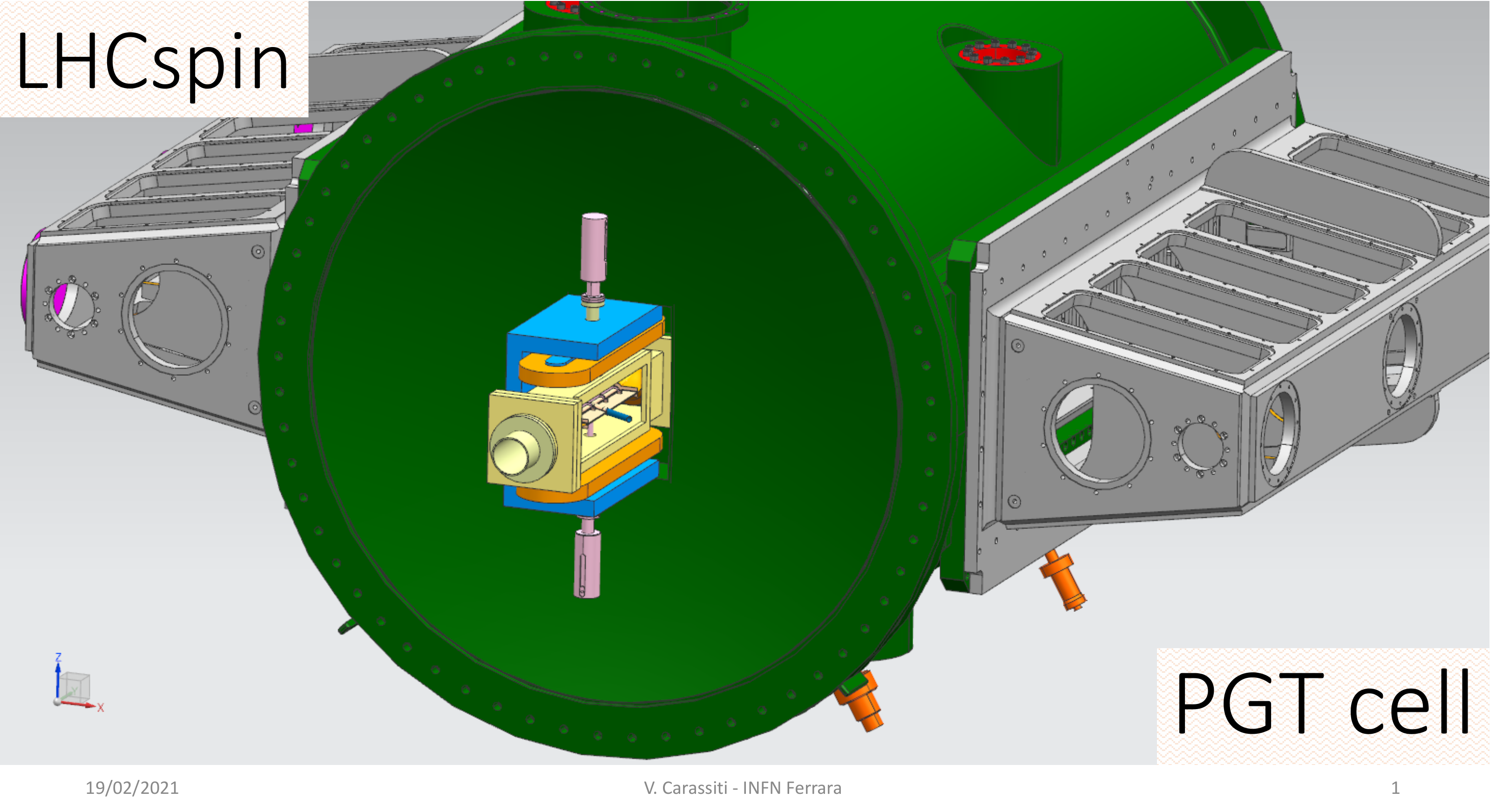}
\hfill
\includegraphics[width=0.45\textwidth]{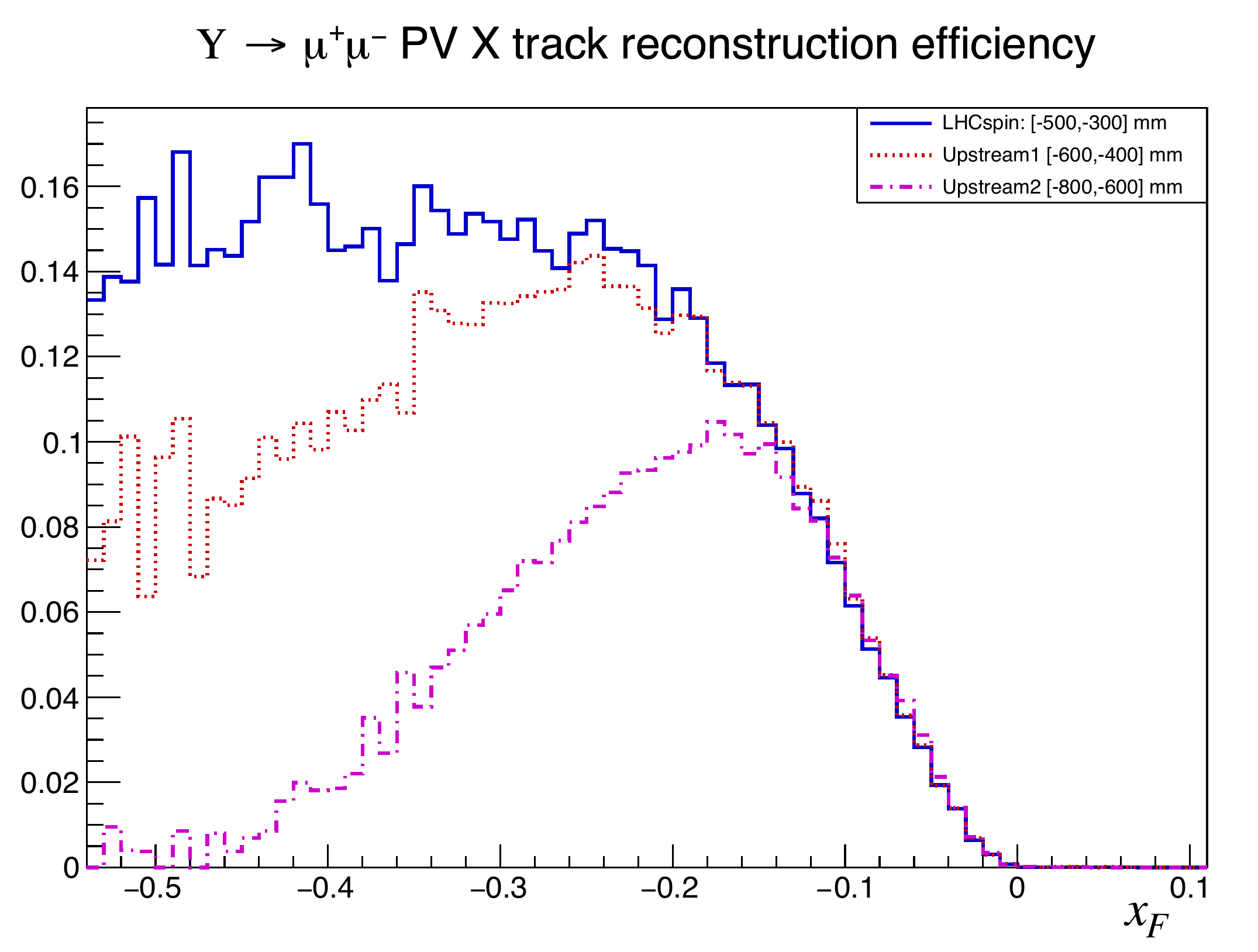}
\caption{Left: The PGT and the VELO vessel (green). Right: simulated reconstruction efficiency for $\Upsilon\to\mu^+\mu^-$ events with three different cell positions, the blue line corresponding to the SMOG2 location.}
\label{fig:pgtvelo}
\end{figure}

In Fig.~\ref{fig:pgtvelo} (left), the PGT is shown in the same location of the SMOG2 cell, a configuration that offers a large kinematic acceptance and does not require additional detectors in LHCb. Fig.~\ref{fig:pgtvelo} (right) shows the efficiency to reconstruct a primary vertex and both tracks in simulated \mbox{$\Upsilon\to\mu^+\mu^-$} events as a function of $x_F$ under three locations of a $20~\rm{cm}$-long storage cell.
The simulation is performed within the GAUSS framework~\cite{Clemencic:2011zza} with upgrade LHCb conditions. New algorithms are currently being developed for the Run 3 fixed-target reconstruction and are expected to sensibly improve the current performance as well as to enable to record LHCspin data in parallel with $p-p$ collisions~\cite{LHCB-FIGURE-2019-007}.

The diagnostic system continuously analyses gas samples drawn from the PGT and comprises a target gas analyser to detect the molecular fraction, and thus the degree of dissociation, and a Breit-Rabi polarimeter to measure the relative population of the injected hyperfine states.
\\
An instantaneous luminosity of $\mathcal{O}(10^{32})~\rm{cm}^{-2}\rm{s}^{-1}$ is foreseen for fixed-target $p-\rm{H}$ collisions in Run 4, with a further factor $3-5$ increase for the high-luminosity LHC phase from Run 5 (2032). 

\section{Conclusions}
The fixed-target physics program at LHC has been greatly enhanced with the recent installation of the SMOG2 gas storage cell at LHCb.
LHCspin is the natural evolution of SMOG2 and aims at installing a polarised gas target to bring spin physics at LHC for the first time, opening a whole new range of exploration. With strong interest and support from the international theoretical community, LHCspin is a unique opportunity to advance our knowledge on several unexplored QCD areas, complementing both existing facilities and the future Electron-Ion Collider~\cite{Accardi:2012qut}.

\paragraph{Funding information}
This project has received funding from the European Union’s Horizon 2020 research and innovation programme under grant agreement STRONG – 2020 - No 824093.

\nocite{*}
\bibliographystyle{auto_generated}
\bibliography{santimaria_proceedings_elba2021/santimaria_proceedings_elba2021}

%% file: proceedings_elba2021_SalimCerci/proceedings_elba2021_SalimCerci/proceedings_elba2021/SalimCerci.tex
\vspace*{1.2cm}

\thispagestyle{empty}
\begin{center}
{\LARGE \bf Search for BFKL signatures in CMS}

\par\vspace*{7mm}\par

{

\bigskip

\large \bf Salim Cerci\footnote{On behalf of the CMS Collaboration}}

\bigskip

{\large \bf  E-Mail: Salim.Cerci@cern.ch}

\bigskip

{Department of Physics, Faculty of Science and Letters, Adiyaman University, 02040 Turkey}

\bigskip

{\it Presented at the Low-$x$ Workshop, Elba Island, Italy, September 27--October 1 2021}

\vspace*{15mm}

\end{center}
\vspace*{1mm}

\begin{abstract}

Results by the CMS Collaboration on the measurements of dijet production processes are presented. Such processes are expected to be  sensitive to Balitsky-Fadin-Kuraev-Lipatov (BFKL) evolution effects. In particular, the measurements of  azimuthal angle correlations between two jets separated by a large rapidity interval at different centre-of-mass energies are discussed. The measurements are corrected for detector effects and compared with the predictions of various Monte Carlo event generators, Dokshitzer– Gribov–Lipatov–Altarelli–Parisi (DGLAP) and BFKL-based calculations. 
\end{abstract}
  \part[Search for BFKL signatures in CMS\\ \phantom{x}\hspace{4ex}\it{Salim Cerci on behalf of the CMS Collaboration}]{}

 \section{Introduction}
 
Hadronic jet measurements are important probes for investigating the low-$x$ structure of the proton, where $x$ represents the fractional momentum carried by the incoming partons. In the quantum chromodynamics (QCD) prediction, two partons are produced with a back-to-back topology in azimuthal plane. Hence, two jets show a strong correlation in their azimuthal angle which is a sensitive probe for better understanding of the QCD radiation in hard processes in high energy particle collisions. In the Bjorken limit which can be accessible at the large centre-of-mass energies,  the scaling variable $x \sim p_T/ \sqrt s$ is kept fixed near unity whereas the transverse momentum $p_T$ becomes approximately equals to  the square of the four-momentum $Q^2$. Since $x$ is not small, $Q^2$ tends to infinity. Thus, the calculations can be resummed within the collinear factorization framework with the Dokshitzer–Gribov–Lipatov– Altarelli–Parisi (DGLAP) formalism, where parton emissions are strongly ordered in transverse momentum. 
In another kinematic regime: $x \rightarrow finite$, $p_T >> \Lambda_{QCD}$ and $\sqrt s \rightarrow \infty$, the large rapidity separation  between the scattered partons occur, thus the DGLAP dynamics fails at low $x$.  Such processes can be described by the Balitsky–Fadin–Kuraev–Lipatov (BFKL) evolution equations.  
 In the following the measurements searching for the  BFKL evolution equation effects are presented. The measurements are performed with data collected in proton-proton collisions by the CMS experiment~\cite{cms}.
 
\section{Azimuthal decorrelation of jets at $\sqrt s = 7$~TeV}

\begin{figure}[hh]
\begin{center}
\includegraphics[height=0.43\textwidth]{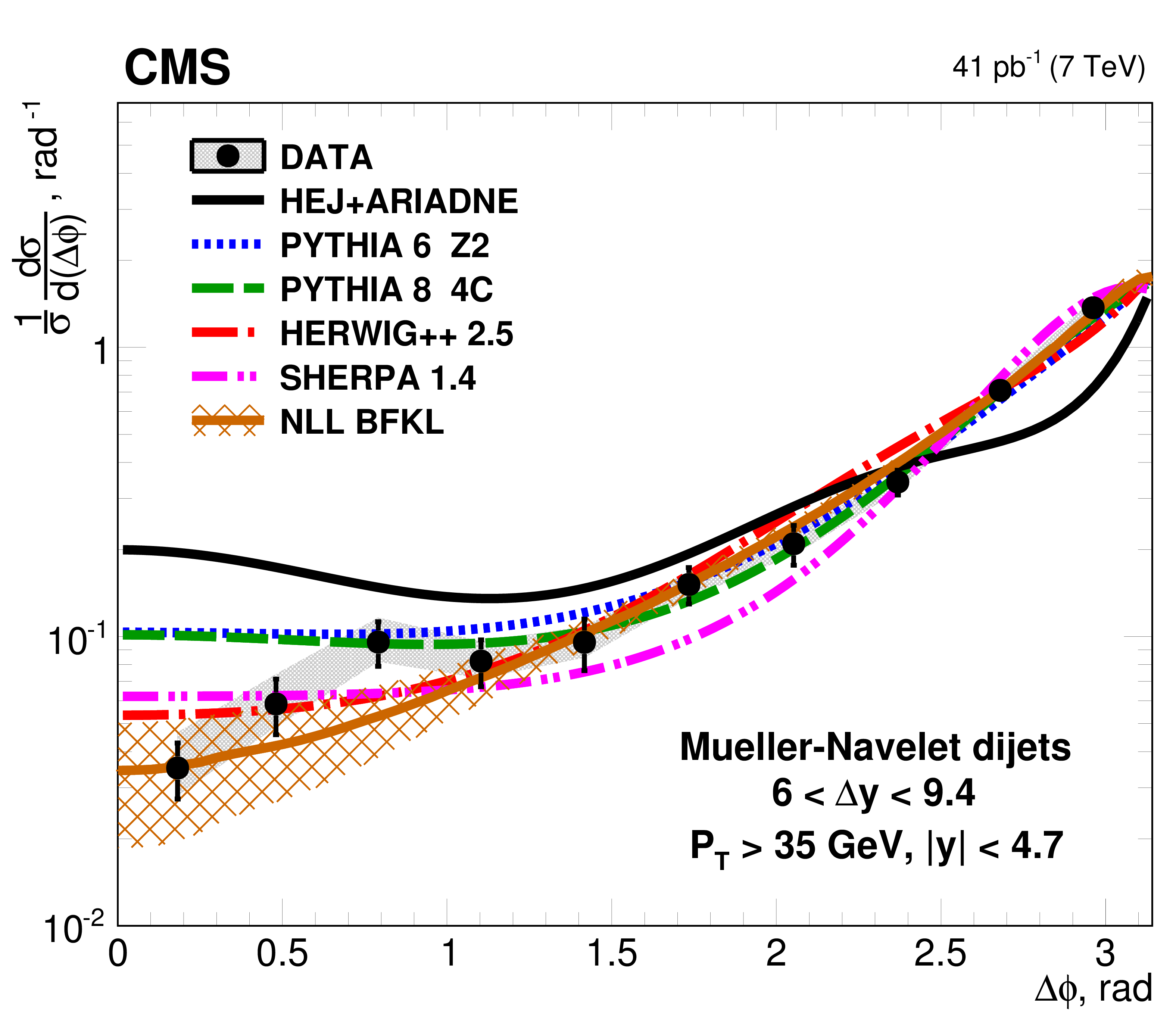}
\includegraphics[height=0.43\textwidth]{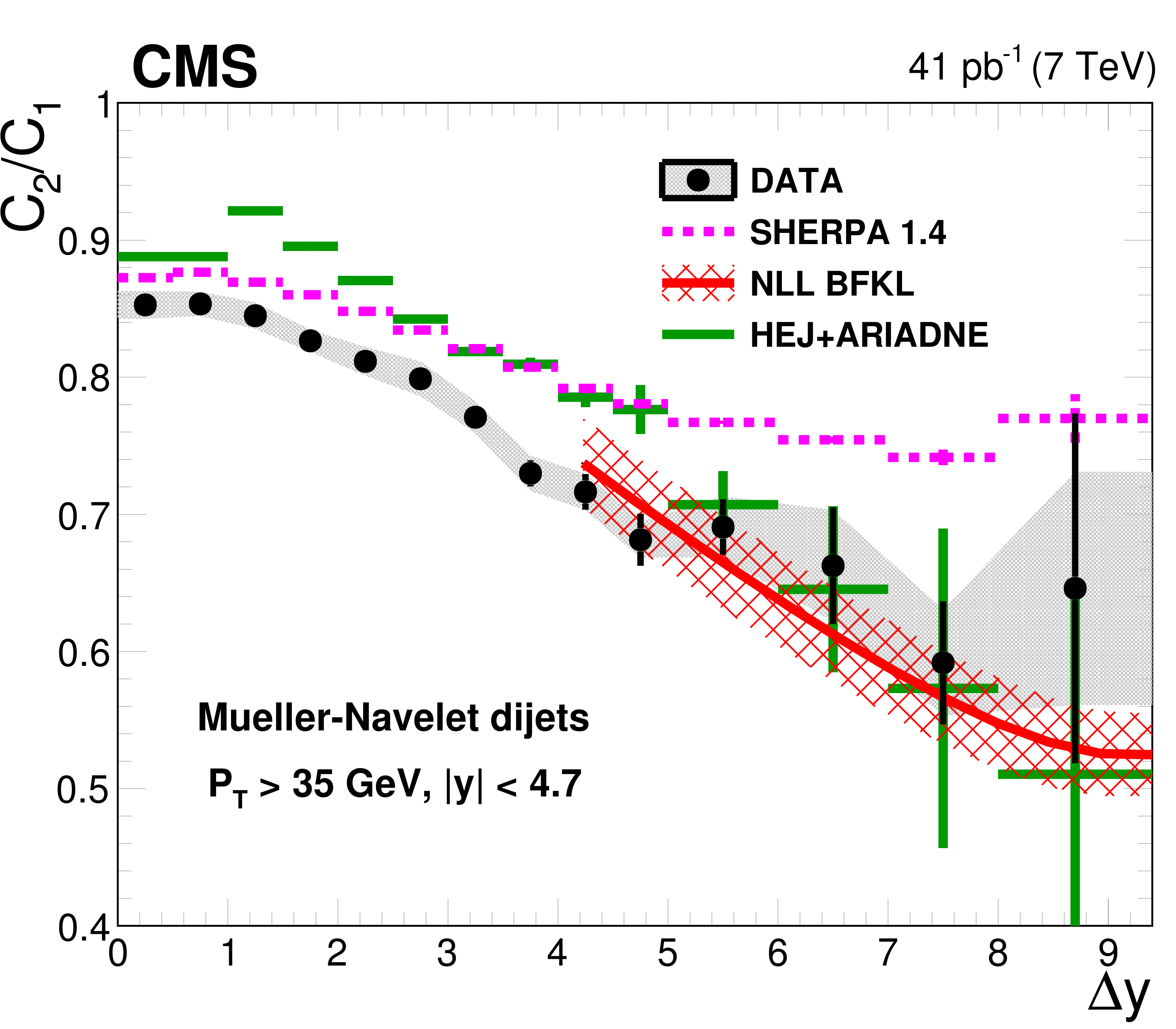}
\caption{Left: The azimuthal-angle difference distribution measured for Mueller-Navelet jets in the rapidity interval $6.0 < \Delta y < 9.4$. Right: Comparison of the measured ratio $C_2/C_1$ as a function of rapidity difference $\Delta y$ to SHERPA, HEJ+ARIADNE and analytical NLL BFKL calculations at the parton level~\cite{fsq-12-002}.}
\label{fig:1_scerci}
\end{center}
\end{figure}

The CMS Collaboration reported a measurement of azimuthal angle decorrelation between the most forward and the most backward jets (so-called Mueller-Navelet jets) in proton-proton collisions at $\sqrt s = 7$ TeV~\cite{fsq-12-002}. In the analysis, jets with transverse momentum, $p_T$ > 35 GeV and absolute pseudorapidity, |y| < 4.7 are considered. The normalised cross sections are compared with various Monte Carlo generators and analytical predictions based on the DGLAP and BFKL parton evolution equations.
In Fig.~\ref{fig:1_scerci}, the azimuthal angle decorrelation of dijets and ratio of its average cosines ($C_{n} = <\mathrm{cos}(n(\pi - \phi_{dijet}))>$) are shown as a function of rapidity separation between the jets, $\Delta y$, reaching up to $\Delta y = 9.4$ for the first time.
 
\section{Dijets with large rapidity separation at $\sqrt s = 2.76$~TeV}

 A measurement of inclusive and Mueller-Navelet dijet differential cross sections as a function of 
rapidity separation between the jets, $\Delta y$ in pp collisions at $\sqrt s = 2.76$~TeV 
 is presented~\cite{fsq-13-004}. The present study extends the results of 7 TeV measurement~\cite{fsq-12-002} by measuring cross section ratios. The same event selection and jet definition are applied hence a direct comparison of the results is allowed. 
 
 The inclusive dijet production cross
section, $\sigma^{incl}$, is defined as the cross section for events with at least one pair of jets with $p_T > p_{Tmin} = 35$~GeV where $p_{Tmin}$ represents the transverse momentum threshold. The "exclusive" dijet production cross section, $\sigma^{excl}$, corresponds to dijet events if only two jets with $p_{Tmin} > 35$~GeV. The Mueller-Navelet cross section, $\sigma^{MN}$, denotes the dijet events with the most forward and most backward jets with $p_T > 35$~GeV. Finally, the cross section of events with no extra jets above $p_{Tveto} = 20$~GeV is represented as $\sigma^{excl}_{veto}$. 

\begin{figure}[hhh]
\begin{center}
\includegraphics[height=0.45\textwidth]{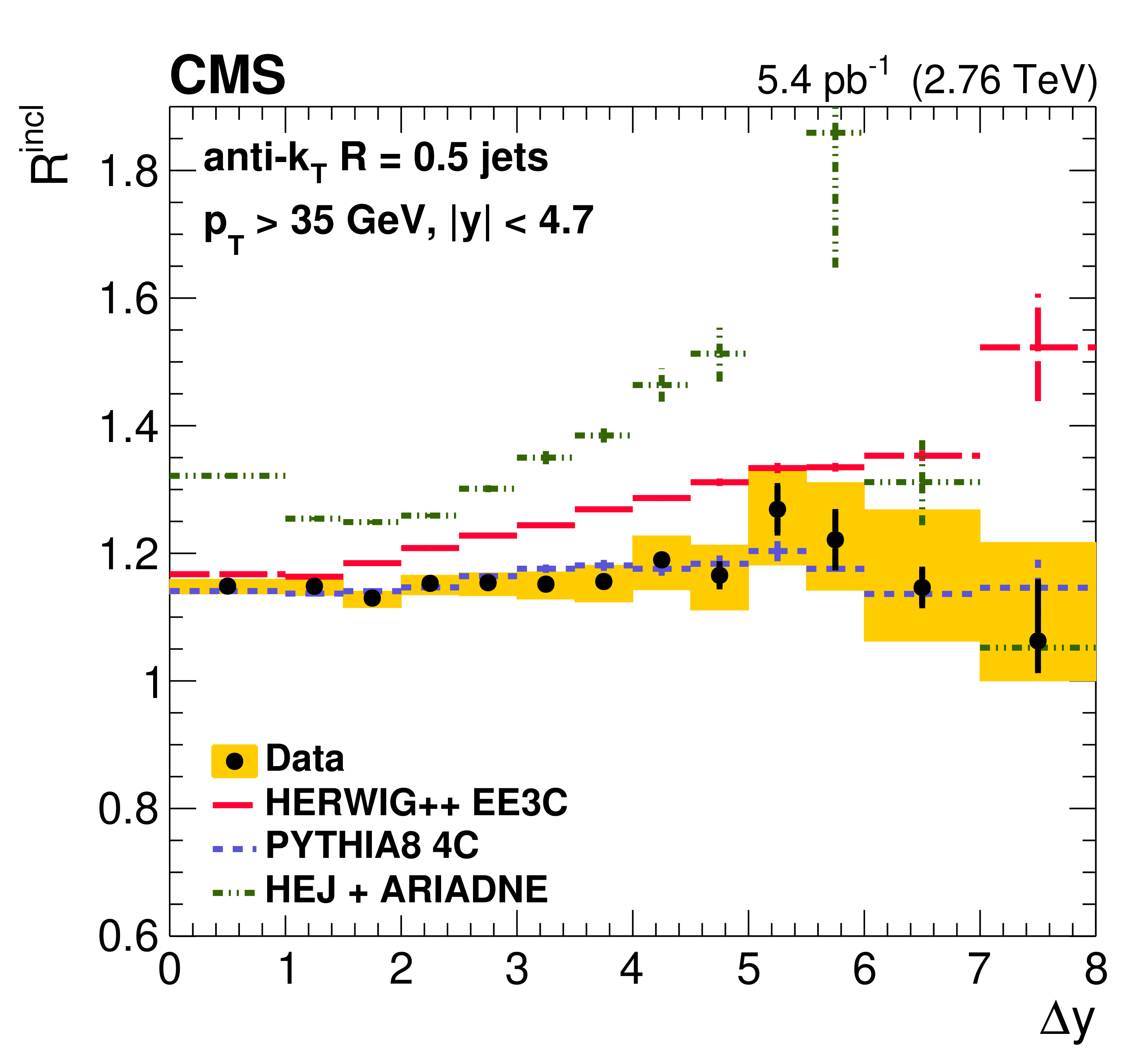}
\includegraphics[height=0.45\textwidth]{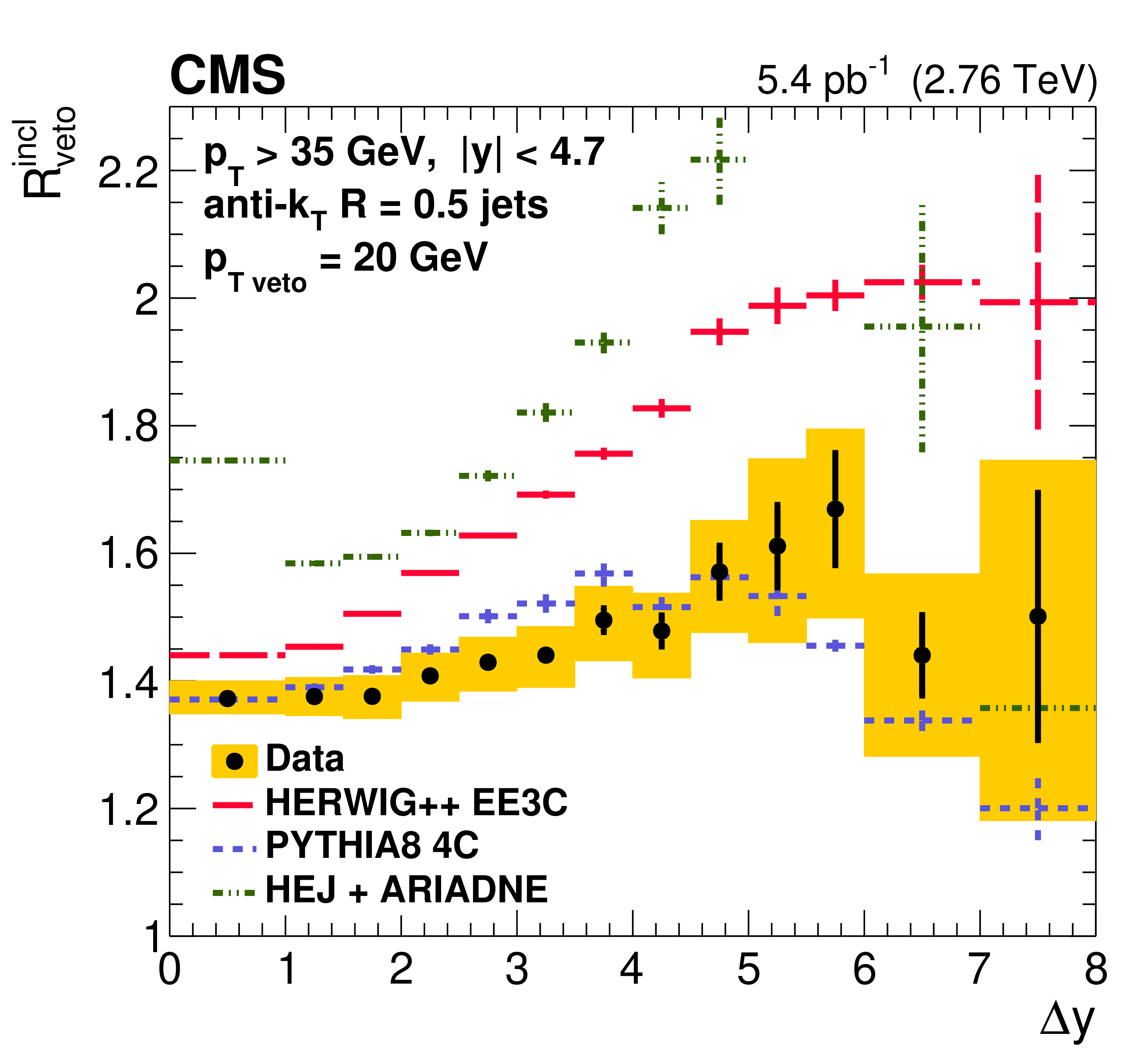}
\caption{The ratio of differential inclusive dijet cross section to "exclusive" (left) and to "exclusive" with veto dijet production (right). Vertical error bars represent the statistical uncertainties, whereas the systematic uncertainties are indicated as shaded bands~\cite{fsq-13-004}.}
\label{fig:2_scerci}
\end{center}
\end{figure}

The ratios of the inclusive to the “exclusive” dijet production cross sections, $R^{incl}$, and to  "exclusive" with veto dijet production, $R^{incl}_{veto}$, are shown in Fig.~\ref{fig:2_scerci}. The results are compared with the predictions of PYTHIA8 (tune 4C)~\cite{p8}, HERWIG++ (tune EE3C)~\cite{herwig} and HEJ+ARIADNE~\cite{hej} event generators. PYTHIA8 shows an agreement with the data, whereas HEJ+ARIADNE and HERWIG++ significantly overestimate the ratio. In the case of ratio $R^{incl}_{veto}$, PYTHIA8 gives the best description of the data, however it still fails to model the shape of the $\Delta y$ dependence.

\begin{figure}[hh]
\begin{center}
\includegraphics[height=0.45\textwidth]{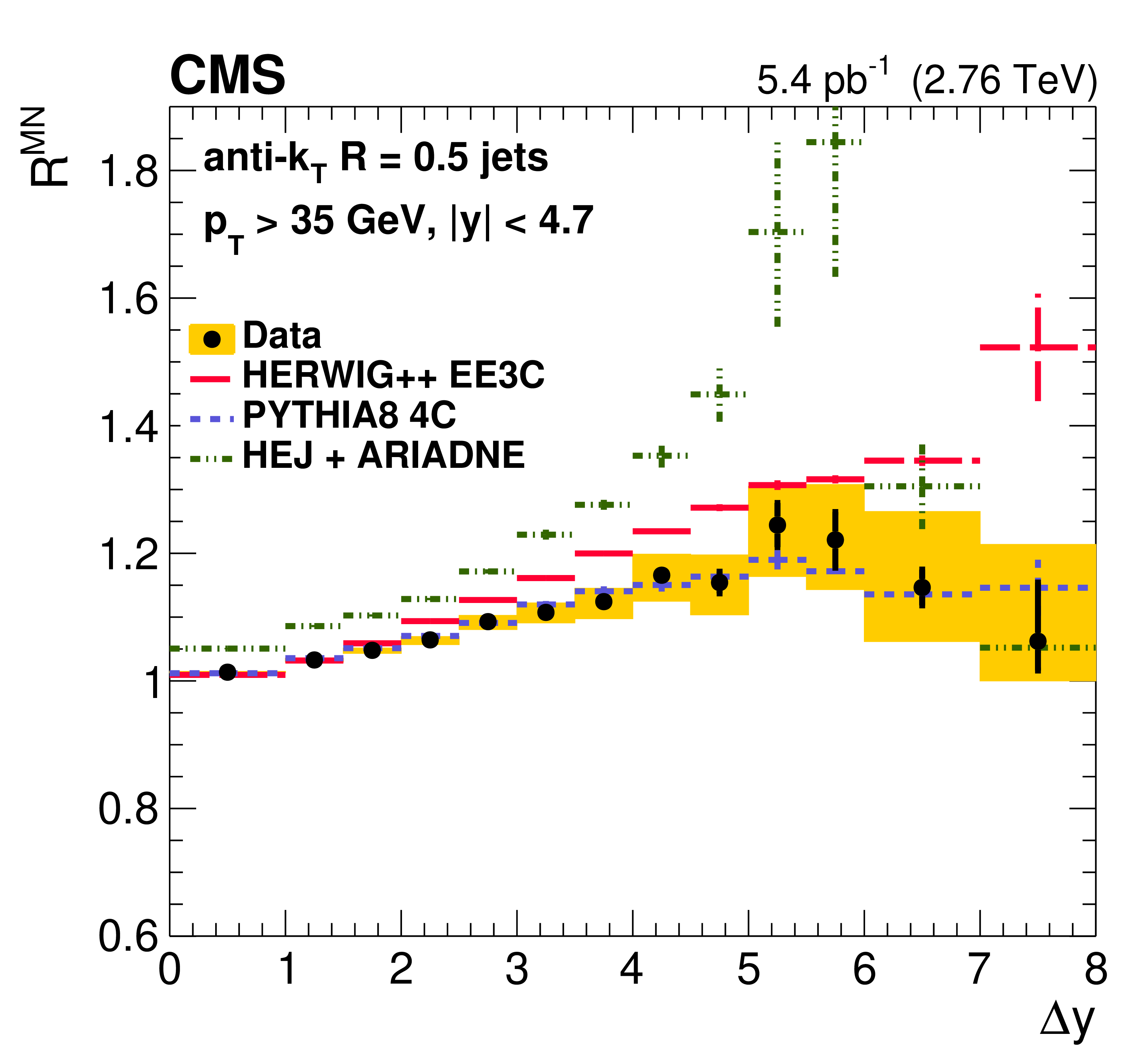}
\includegraphics[height=0.45\textwidth]{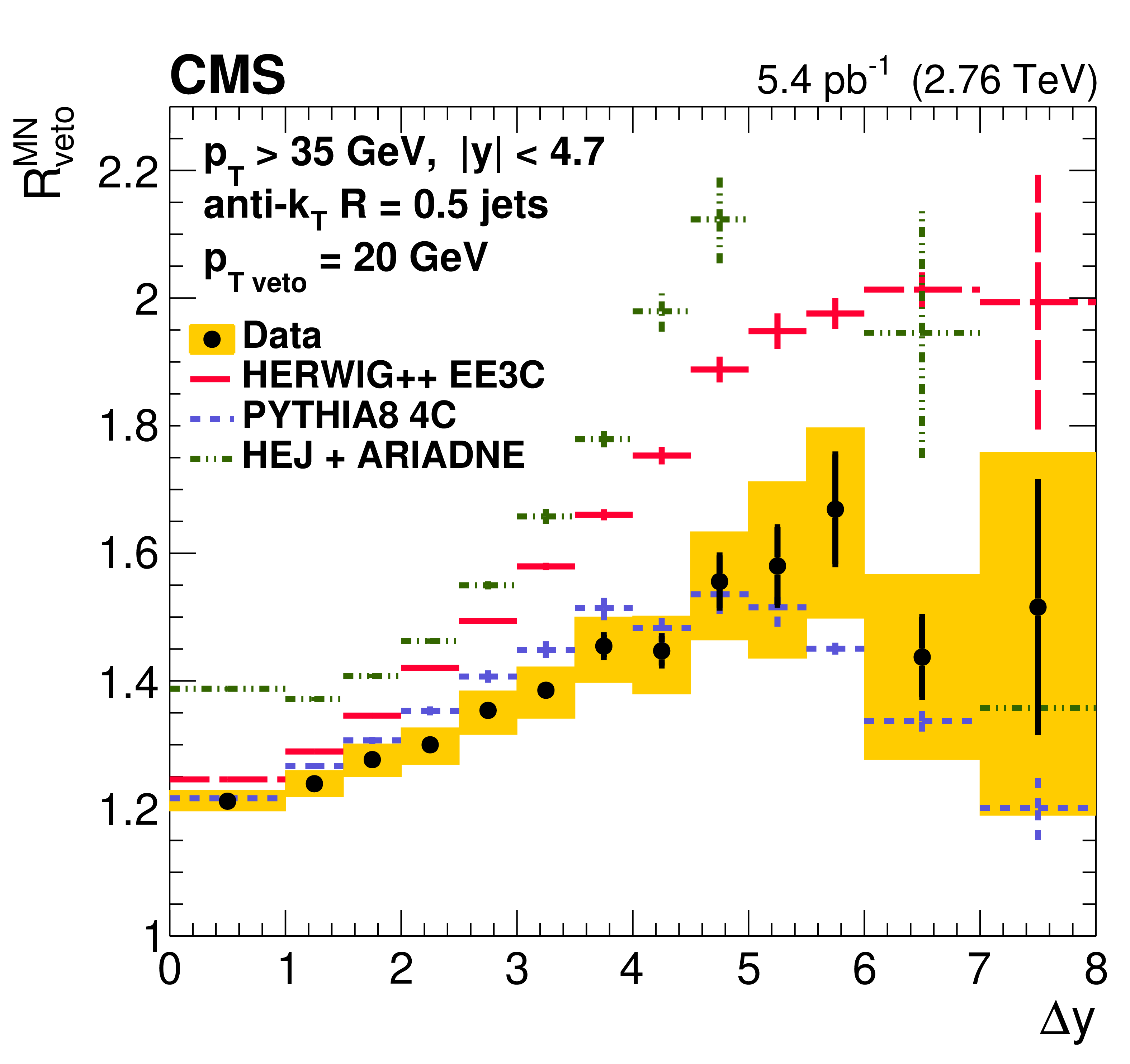}
\caption{The ratio of Mueller-Navelet dijet cross section to "exclusive" (left) and to "exclusive" with veto dijet production (right). Vertical error bars represent the statistical uncertainties, whereas the systematic uncertainties are indicated as shaded bands~\cite{fsq-13-004}.}
\label{fig:3_cserci}
\end{center}
\end{figure}

The ratios of the Mueller-Navelet dijet production cross section to  “exclusive”, $R^{MN}$, and to  “exclusive” with veto dijet production, $R^{incl}_{veto}$, are shown in Fig.~\ref{fig:3_cserci}. Generally, only the leading order DGLAP based event generator PYTHIA8 describes the ratios $R^{incl}$ and $R^{MN}$.

\begin{figure} [hhh]
\begin{center}
\includegraphics[height=0.45\textwidth]{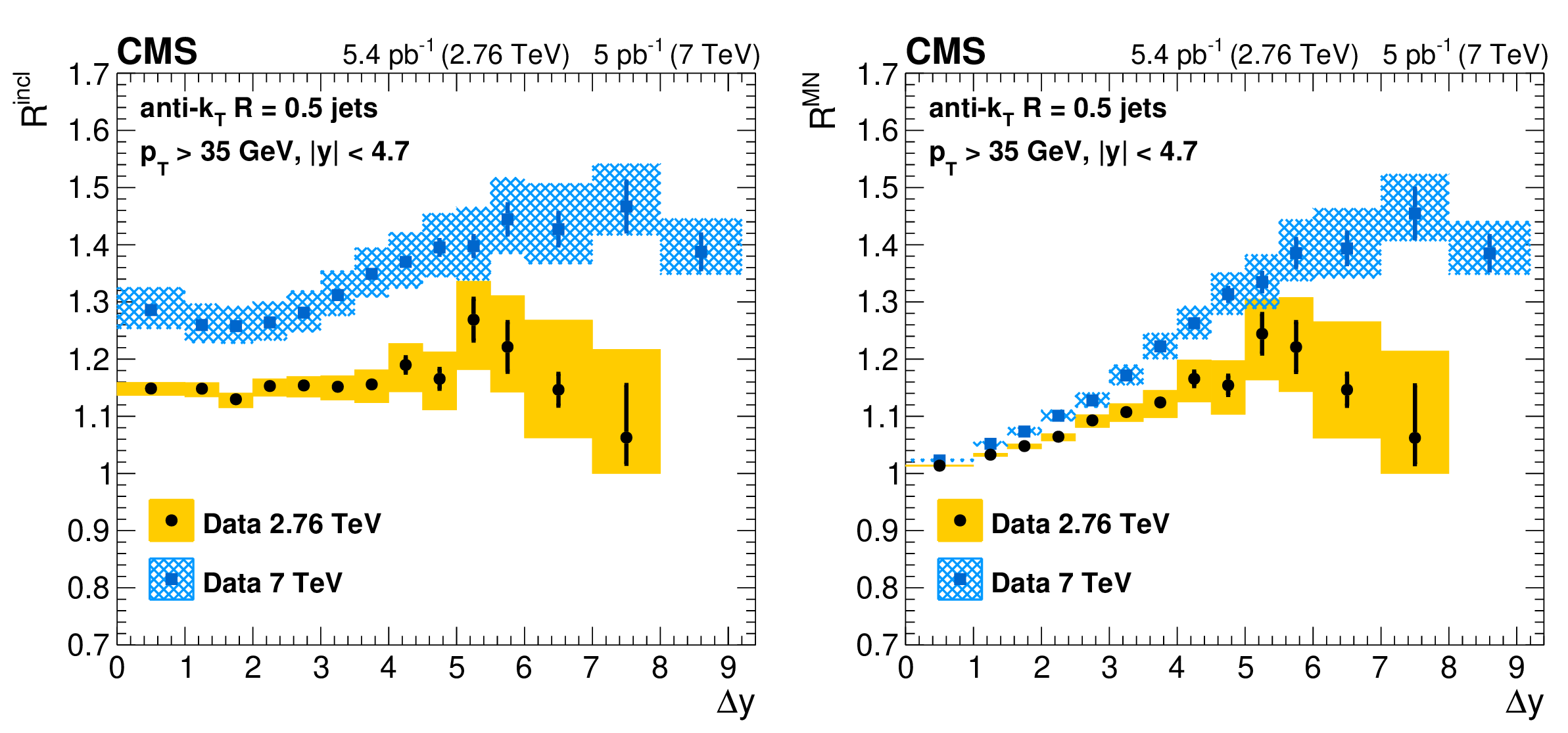}
\caption{Comparison of ratios of the cross sections for inclusive (left) and MN (right) at  the collision energies of  7 TeV and 2.76 TeV~\cite{fsq-13-004}.}
\label{fig:4_scerci}
\end{center}
\end{figure}

 The results of the ratios of $R^{incl}$ and $R^{MN}$ are compared with the previous measurement performed at $\sqrt s = 7$~TeV~\cite{fsq-12-002} are shown in Fig.~\ref{fig:4_scerci}. A strong rise with $\Delta y$ is observed at high energies. This can be explained as a reflection of both the increasing available phase space and BFKL dynamics. Due to the increasing phase space volume for hard parton radiation, the ratios show a strong rise with the increasing $\Delta y$. As a result of the kinematic limitations on the events with more than two jets with $p_T > 35$~GeV, the ratios decrease at vert large $\Delta y$. BFKL calculations at the next-to-leading-logarithm level are needed to compare the results obtained with this analysis. 

\section{Summary}

Recent dijet production studies at different centre-of-mass energies performed by the CMS Collaboration are presented.  The measurements are compared to various Monte Carlo event generators as well as the predictions based on BFKL calculations. The results on azimuthal angle decorrelations in dijet events, where the two jets are separated by a large rapidity interval, are consistent with the predictions based on BFKL calculations.

\nocite{*}
\bibliographystyle{auto_generated}
\bibliography{proceedings_elba2021_SalimCerci/proceedings_elba2021_SalimCerci/proceedings_elba2021/SalimCerci}


%% file: chachamis_proceedings_elba2021/chachamis_proceedings_elba2021.tex

\vspace*{1.2cm}

\thispagestyle{empty}
\begin{center}
{\LARGE \bf Two-particle Correlations in multi-Regge Kinematics }

\par\vspace*{7mm}\par

{

\bigskip

\large \bf N. Bethencourt de Le\'on$^1$, G. Chachamis$^2$, A. Sabio Vera$^{1,3}$}

\bigskip

{\large \bf  E-Mail: chachamis@gmail.com}

\bigskip

{$^1$ Instituto de F{\'\i}sica Te{\'o}rica UAM/CSIC, Nicol{\'a}s Cabrera 15, E-28049 Madrid, Spain.\\
$^2$ Laborat{\' o}rio de Instrumenta\c{c}{\~ a}o e F{\' \i}sica Experimental de Part{\' \i}culas (LIP),\\
Av. Prof. Gama Pinto, 2, P-1649-003 Lisboa, Portugal.\\
$^3$ Theoretical Physics Department, Universidad Aut{\' o}noma de Madrid, E-28049 Madrid, Spain.\\ }

\bigskip

{\it Presented at the Low-$x$ Workshop, Elba Island, Italy, September 27--October 1 2021}

\vspace*{15mm}

\end{center}
\vspace*{1mm}

\begin{abstract}.
Multi-jet production at the LHC is an important  process to study. In particular, events with final state kinematic configurations where we have two jets widely separated in rapidity with similar $p_T$ and lots of mini-jets or jets populating the space in between are relevant for the high energy limit of QCD.
Keeping the jet multiplicity fixed, the study of these events is a good ground to test different models of multi-particle production in hadron-hadron collisions. We report on a comparison between the predictions of the old multiperipheral Chew-Pignotti model and those of BFKL for the single jet rapidity distributions and for jet-jet rapidity correlations.
\end{abstract}
 \part[Two-particle Correlations in multi-Regge Kinematics\\ \phantom{x}\hspace{4ex}\it{ G. Chachamis et al}]{}
 \section{Introduction}

An important question in Quantum Chromodynamics (QCD) is the structure of the high energy limit dynamics of multi-particle production at hadron and hadron-lepton colliders. Multi-particle events, and in particular, multi-jet
events, are difficult to describe when one wants to go beyond the leading order (LO) approximation in fixed order perturbation theory. Even in the early days of hadron colliders (Intersecting Storage Rings, ISR at CERN) when 
the colliding energy was of the order of few tens of GeV and QCD was not yet the established theory of the strong interaction, multi-particle production was one of the key problems to tackle in order to probe the underlying dynamics
even on heuristic terms.  The experimental data amassed in the last 50-60 years, starting from those early attempts,  
have generally reinforced the view that the final state particles (pions, kaons, protons, etc) that end up on the 
detector seem to belong in correlated ``clusters''~\cite{Dremin:1977wc} that are emitted in the hard scattering part
of the process or on the partonic level. 

Our modern day picture of a generic partonic cross section
of two incoming partons that interact and produce two or more outgoing partons which in turn,
through a parton shower, radiate new quarks and gluons and form jets does not contradict 
the heuristic notion of  clusters.
The study of differential distributions and particle-particle correlations between final states had given
important insights about the strong interaction no matter whether
 the  final states were considered before hadronizations effects (final state partons/jets) or after (hadrons in the detector calorimeters).
Especially for jets, correlations summarize important properties without being too sensitive
to unresolved soft particles in the jet~\cite{Tannenbaum:2005by} (see also the introduction of Ref.~\cite{SanchisLozano:2008te}). It is thus very interesting to see whether the rapidity distributions of final state hadrons are any similar to the rapidity distributions of jets.

Quite a few different approaches exist that are useful to probe multi-particle hadroproduction beyond fixed order. These are  resummation frameworks that resum different leading contributions to amplitudes (e.g. DGLAP~\cite{Gribov:1972ri,Gribov:1972rt,Lipatov:1974qm,Altarelli:1977zs,Dokshitzer:1977sg}, BFKL~\cite{Kuraev:1977fs,Kuraev:1976ge,Fadin:1975cb,Lipatov:1976zz,Lipatov:1985uk}, CCFM~\cite{Ciafaloni:1987ur,Catani:1989yc}, Linked Dipole Chain~\cite{Gustafson:1986db,Gustafson:1987rq,Andersson:1995ju}, Lund Model~\cite{Andersson:1998tv}, see also~\cite{LHCForwardPhysicsWorkingGroup:2016ote}). 

The run 2 of the LHC at 13 TeV has provided lots of data which should be analyzed in detail. Here, we want to focus on a final state configurations which could be quite important to answer the question of what
is the range of applicability of different multi-particle production models. These configurations correspond to events with several final state jets where the two outermost in rapidity ones are also well separated in rapidity and can be  tagged requiring that their transverse momenta are similar and generally large. These corresponds to a subset of the so-called Mueller-Navelet jets~\cite{Mueller:1986ey} and if additionally we require a very similar $p_T$ for the outermost jets (however, not within overlapping ranges) we can enforce that their influence on 
the rapidity distributions of the jets in between will be symmetric.

Our aim here
is to report on our recent work~\cite{deLeon:2020myv} where we studied some of the characteristics that can be attributed to these processes. As the most striking features in multiperipheral models can be found in the rapidity space --the origin of these features can be traced at the decoupling of the transverse degrees of freedom from the longitudinal ones--, we will present single rapidity differential distributions and two-jet rapidity-rapidity correlations. 

Assuming configurations in a hadron-hadron collider with fixed final state jet multiplicity $N$,
the single and double rapidity distributions for a given jet and a pair of jets are given respectively by
$\rho_{1}(y_i)$ and $\rho_{2}\left(y_{i}, y_{j}\right)$, where $i$ and $j$ are the positions of the jets
once they are ordered in rapidity with $1 \le i,j \le N$. We can formally define for the 
single and double differential normalized distributions
\begin{equation}
\rho_{1}(y_i)=\frac{1}{\sigma} \int d^{2} p_{\perp} \frac{d^{3} \sigma}{d y_i d^{2} p_{\perp i}}
\end{equation}
 and
\begin{equation}
\rho_{2}\left(y_{i}, y_{j}\right)=\frac{1}{\sigma} \int d^{2} p_{\perp i} d^{2} p_{\perp j} \frac{d^{6} \sigma}{d y_{i} d^{2} p_{\perp i} d y_{j} d^{2} p_{\perp j}}\,.
\end{equation}
After integrating over their transverse components we will have
\begin{equation}
\rho_{1}(y_i)=\frac{1}{\sigma} \frac{d \sigma}{d y_i}
\end{equation}
 and
\begin{equation}
\rho_{2}\left(y_{i}, y_{j}\right)=\frac{1}{\sigma}  \frac{d^{2} \sigma}{d y_{i}  d y_{j} }\,.
\end{equation}
The two-particle rapidity-rapidity correlation function is then given by
\begin{eqnarray}
C_{2}\left(y_{i},  y_{j}\right)&=&\frac{1}{\sigma} \frac{d^{2} \sigma}{d y_{i} d y_{j} }-\frac{1}{\sigma^{2}} \frac{d \sigma}{d y_{i} } \frac{d \sigma}{d y_{j}}\\
&=& \rho_{2}\left(y_{i}, y_{j}\right) - \rho_{1}(y_i)  \rho_{1}(y_j)
\end{eqnarray}
In practice however, the correlation function is usually computed with the following expression which is less sensitive
to experimental errors:
\begin{equation}
R_{2}\left(y_{1}, y_{2}\right)=\frac{C_{2}\left(y_{1}, y_{2}\right)}{\rho_{1}\left(y_{1}\right) \rho_{1}\left(y_{2}\right)}=\frac{\rho_{2}\left(y_{1}, y_{2}\right)}{\rho_{1}\left(y_{1}\right) \rho_{1}\left(y_{2}\right)}-1\,.
\label{correlation_function}
\end{equation}

In Section 2, we calculate the analytic expressions for the double and single rapidity distributions within an
old multiperipheral model, namely the Chew-Pignotti model~\cite{Chew:1968fe}, whereas, In Section 3,
we explain how we compute the same quantities for the gluon Green's function of a collinear BFKL model by
using Monte Carlo techniques. We conclude in Section 4. 
\section{The Chew-Pignotti Model}

As we mentioned in the introduction, we will work with a Chew-Pignotti type of multiperipheral model~\cite{Chew:1968fe} following the analysis by DeTar in Ref.~\cite{Detar:1971qw}. The simplified features of the model
 will give us the opportunity to produce analytic results that could in principle
 be compared to the experimental data. For a more detailed review on multiperipheral models and the cluster concept in multiparticle production at hadron colliders, see Ref.~\cite{Dremin:1977wc} and references therein. 
The key point is that in these types of models, the transverse coordinates decouple from the longitudinal degrees
of freedom (rapidity) and that is what allows to obtain analytic expressions for the rapidity distributions.
 
 While the multiperipheral models were devised to describe multiple particle production, we will use
 the Chew-Pignotti here for multiple jet production. We will assume that the rapidity separation of
 the the bounding jets is Y and also that the jets in between have a fixed multiplicity $N$ such that in total
 we will have $N+2$ final state jets in each event. The outermost in rapidity jets (the most forward/backward jets)  are having rapidities $\pm \frac{Y}{2}$. It should be clear that we choose to work with limits $y_0=-\frac{Y}{2}$ and $y_{N+1}=\frac{Y}{2}$ because we want to cast our analytic expressions for the distributions in a symmetric
 way with respect to the forward and backward rapidity direction. Naturally, one could also work with limits $y_0=0$ and $y_{N+1}=Y$. Actually for the results from the BFKL approach in section 3,  we choose to present our plots in
the range $\left[0, Y \right]$ as we want to be closer to an experimental analysis setup.

The cross section for the production of  $N+2$ final state jets is given by
\begin{eqnarray}
\sigma_{N+2} &=& \alpha^{N+2} \int_{0}^{Y} \prod_{i=1}^{N+1} dz_i \delta \left(Y-
\sum_{s=1}^{N+1} z_s \right) \nonumber\\
&=&  \alpha^{N+2}
\int_{-\frac{Y}{2}}^{\frac{Y}{2}} dy_N \int_{-\frac{Y}{2}}^{y_N} dy_{N-1} \cdots \int_{-\frac{Y}{2}}^{y_3} dy_2 
\int_{-\frac{Y}{2}}^{y_2} dy_1\, = \,  \alpha^2 
\frac{\left(\alpha Y\right)^N}{N!} \, ,
\label{sigma_N2}
\end{eqnarray}
which leads to a total cross section $\sigma_{\rm total} = 
\sum_{N=0}^\infty \sigma_{N+2} = \alpha^2 e^{\alpha Y}$ and the rise with $Y$ would need to be tamed
by introducing unitarity corrections in transverse space. A rapidity 
$y_l$, with $l=0, \dots, N+1$ is assigned to each of the final-state jets. At $y_0=-\frac{Y}{2}$ and $y_{N+1}=\frac{Y}{2}$ we have the positions of the outermost jets with the jet vertex reduced to a simple factor equal to $\alpha$, the strong coupling constant. The in-between jets will have rapidities $y_l = -\frac{Y}{2}+\sum_{j=1}^l z_j$,
where $l=1, \dots N$.

We are mainly interested in the description of the differential distributions for events with fixed final state multiplicity, firstly on a qualitative level. At this point we need to note the following:
the final jet multiplicity for a given final state is uniquely defined, it actually depends on the lower $p_T$ cutoff
we set for a mini-jet to qualify as a jet as well as on the chosen jet radius $R$ (in the rapidity-azimuthal angle plane) for the jet clustering algorithm. Nevertheless, we believe that our analysis in the following is valid
 once a well defined mechanism for deciding the multiplicity of a final state is established. 
This is not a trivial statement at is it implies that if a final state has initially been assigned  multiplicity $N_1+2$ complies with the $N_1+2$ differential distributions, then, if a different set of parameters is chosen for the jet clustering algorithm which results in a different number of final state jets giving a shift
in multiplicity from $N_1$ to $N_2$,  then that final state will also comply with the $N_2+2$ differential distributions. 

\begin{figure}
\begin{center}
\begin{flushleft}
\hspace{1.cm}\includegraphics[width=7cm]{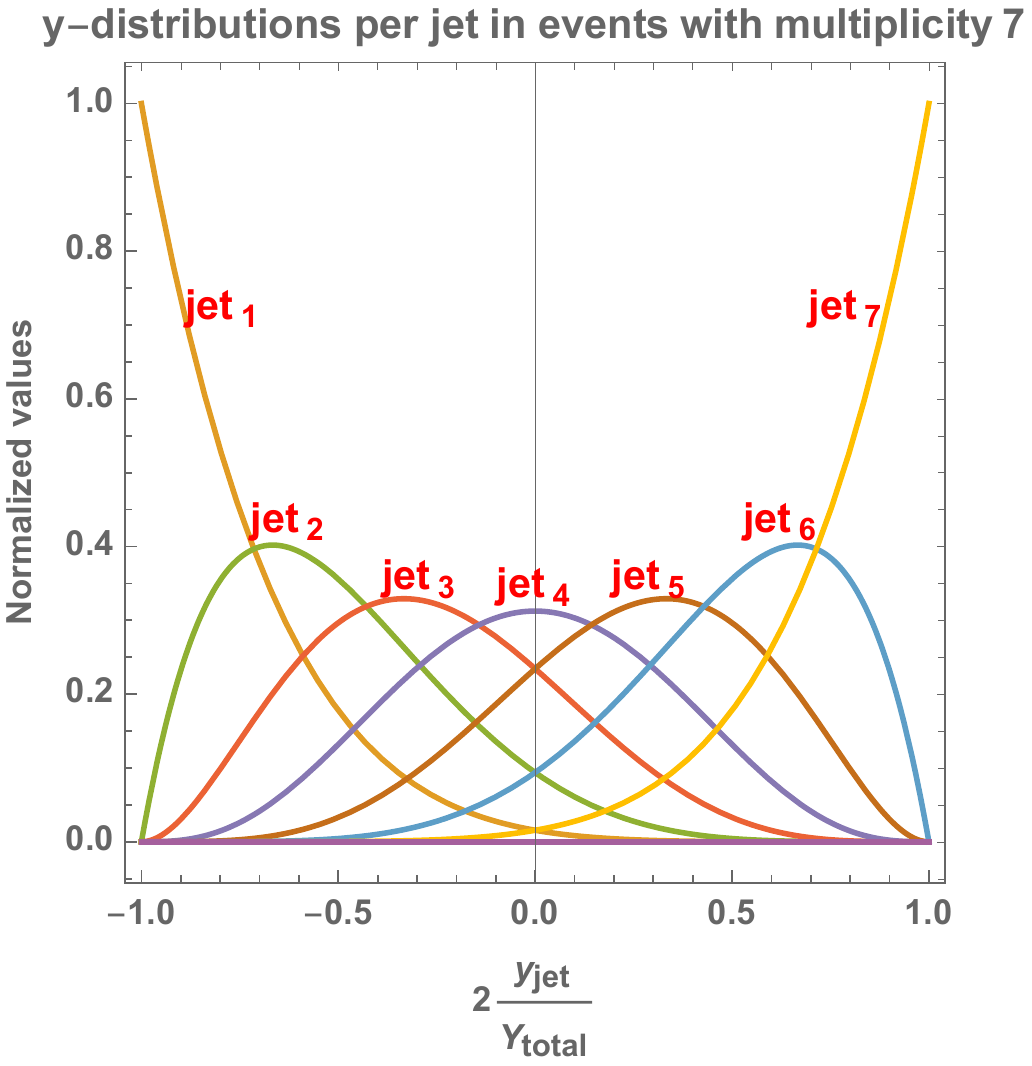}
\end{flushleft}
\vspace{-7.6cm}
\begin{center}
\hspace{8cm}\includegraphics[width=7.cm]{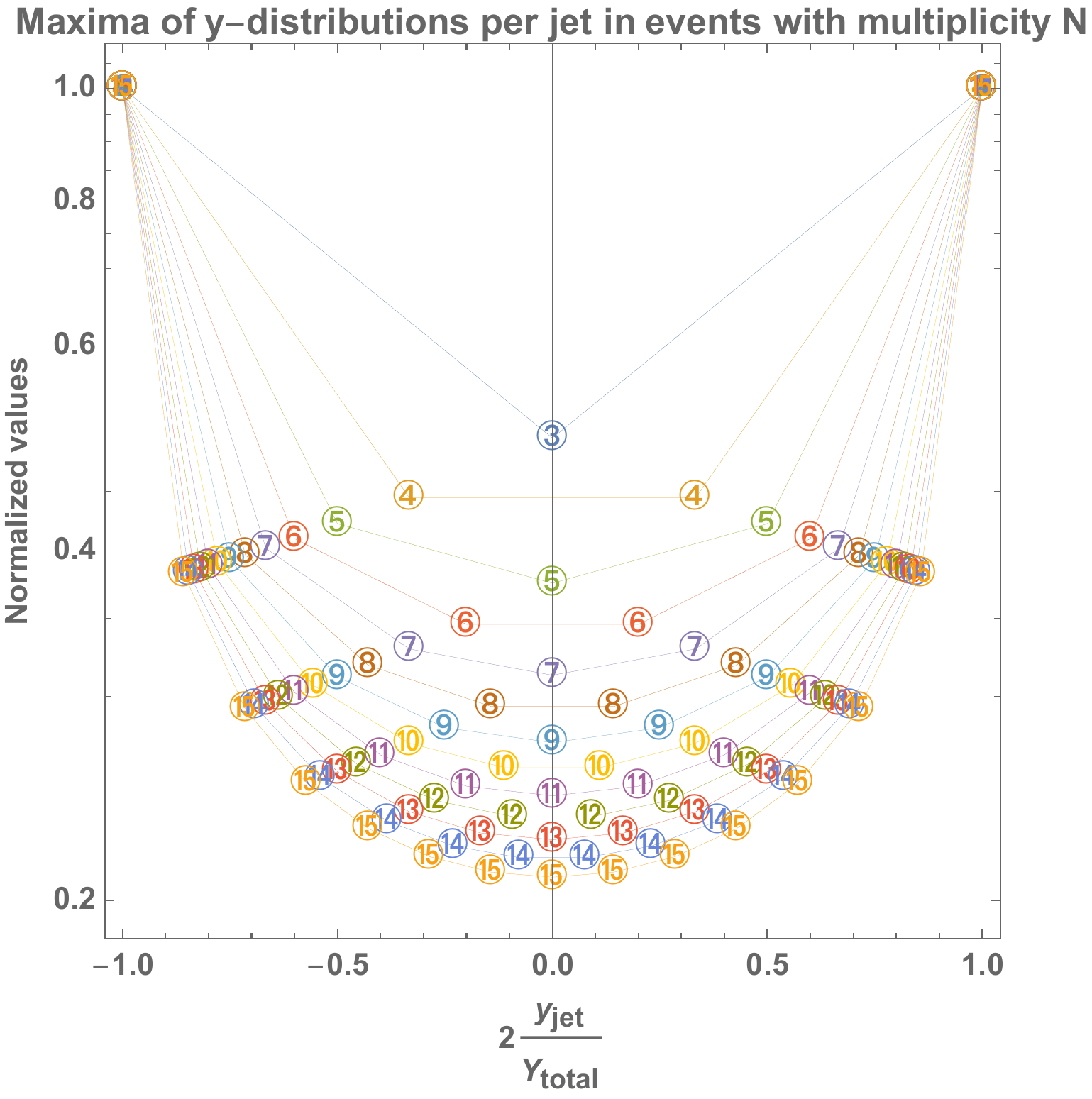}
\end{center}
\vspace{-.5cm}
\caption{Rapidity distributions for each of the jets in a final state with seven mini-jets (left).  Their maxima are indicated by the symbol \textcircled{\tiny 7} (right). The positions of the $y$-distribution maxima in configurations with multiplicity $N$ are marked by \textcircled{\tiny N} (right).}
\label{Multiplicity7yjets}
\end{center}
\end{figure}

The contribution for jet $l$ in a $N+2$ final state event, will have the following differential distribution in rapidity 
\begin{eqnarray}
\frac{d \sigma_{N+2}^{(l)}}{d y_l} &=&  \alpha^{N+2} \int_0^Y 
  \prod_{i=1}^{N+1} dz_i \delta \left(Y-\sum_{s=1}^{N+1} z_s \right) 
\delta \left(y_l +\frac{Y}{2}- \sum_{j=1}^l z_j\right) \nonumber\\
&=& \alpha^{N+2} \int_{y_l}^\frac{Y}{2} dy_N \int_{y_l}^{y_N} dy_{N-1} \cdots \int_{y_l}^{y_{l+2}} dy_{l+1}
\int_{-\frac{Y}{2}}^{y_l} dy_{l-1}  \cdots \int_{-\frac{Y}{2}}^{y_3} dy_2 
\int_{-\frac{Y}{2}}^{y_2} dy_1\nonumber\\
&=&  \alpha^{N+2} \frac{\left(\frac{Y}{2}-y_l \right)^{N-l}}{(N-l)!} \frac{\left(y_l+\frac{Y}{2}\right)^{l-1}}{(l-1)!}  \, ,
\label{dsdy}
\end{eqnarray}
as derived from Eq.~(\ref{sigma_N2}).
For very large multiplicities this results to an asymptotic Poisson distribution as one can verify, {\it e.g.}, in the region $y \simeq - \frac{Y}{2}$ with $y = \left(\frac{\lambda}{N}-\frac{1}{2}\right) Y$ where 
\begin{eqnarray}
\lim_{N \to \infty}{N-1 \choose l-1}   
\left(1-\frac{\lambda}{N}\right)^{N-l} 
\left(\frac{\lambda}{N}\right)^{l-1}
&=& e^{- \lambda} \frac{\lambda^{l-1}}{(l-1)!} \, .
\end{eqnarray}
Taking the limits $l \to 1$ and $y_l \to - \frac{Y}{2}$ in 
Eq.~(\ref{dsdy}) we derive a normalized universal distribution for each $N$ when plotted versus $2y/Y$. The case for  N=7+2 is plotted in Fig.~\ref{Multiplicity7yjets} (left) which is very characteristic for multiperipheral models. We remind the reader that the notation jet$_{i=1,2, \dots, N}$ is introduced for jets with ordered rapidities $y_1 < y_2 < \dots < y_N$.
All these seven normalized $y$-distributions (one for each of the seven jets) spans an area of $\frac{2}{N}$. Their maxima are found  at $y=\frac{2l-N-1}{2(N-1)} Y$ with a value that is
\begin{eqnarray}
{N-1 \choose l-1} \frac{(l-1)^{l-1}}{(N-1)^{N-1}} \left(N- l\right)^{N-l} \, .
\end{eqnarray}
In Fig.~\ref{Multiplicity7yjets} (right), we show the maxima for multiplicities  $N$, where $\left(3+2\right) \le \left(N+2\right) \le \left(15+2\right)$.

In a similar manner, we get expressions for the double differential rapidity distributions for jet pairs, {\it i.e.}
\begin{eqnarray}
\frac{d^2 \sigma_{N+2}^{(l,m)}}{d y_l d y_m} &=&  \alpha^{N+2} 
\int_{0}^{Y}
  \prod_{i=1}^{N+1} dz_i \delta \left(Y-\sum_{s=1}^{N+1} z_s \right) 
\delta \left(y_l +\frac{Y}{2}- \sum_{j=1}^l z_j\right)
\delta \left(y_m +\frac{Y}{2}- \sum_{k=1}^m z_k\right) \nonumber\\
&=&  \alpha^{N+2} \frac{\left(\frac{Y}{2}-y_l \right)^{N-l}}{(N-l)!}
\frac{(y_l-y_m)^{l-m-1}}{(l-m-1)!} 
\frac{\left(y_m+\frac{Y}{2}\right)^{m-1}}{(m-1)!}  \, .
\label{d2sdydy}
\end{eqnarray}

To calculate the correlation between the rapidities of jet $l$ and jet $m$ we use Eq.~(\ref{correlation_function}),
more precisely 
\begin{eqnarray}
{\cal R}_{N+2} \left(x_l,x_m\right) = \sigma_{N+2} 
\frac{ \frac{ d^2 \sigma_{N+2}^{(l,m)}}{d y_l d y_m} }{\frac{d \sigma_{N+2}^{(l)}}{d y_l} \frac{d \sigma_{N+2}^{(m)}}{d y_m}}-1 
 =  
\frac{2^N}{N!}\frac{(N-m)!(l-1)!}{(l-m-1)!}   
\frac{(x_l-x_m)^{l-m-1}}{\left(1+x_l\right)^{l-1}\left(1-x_m \right)^{N-m}}
 -1
\, ,
\label{R}
\end{eqnarray}
where $Y > y_l > y_m > 0$, $l>m$ and $x_J = 2 y_J / Y$.

\begin{figure}
\begin{subfigure}{.5\textwidth}
  \centering
  \includegraphics[width=1\linewidth]{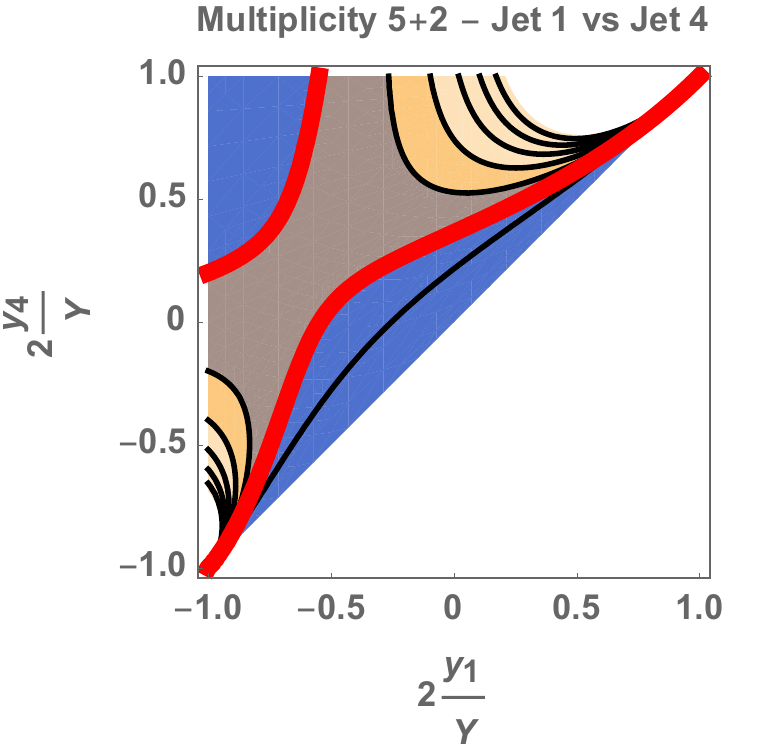}
  \label{fig:sfig1}
\end{subfigure}%
\begin{subfigure}{.5\textwidth}
  \centering
  \includegraphics[width=1\linewidth]{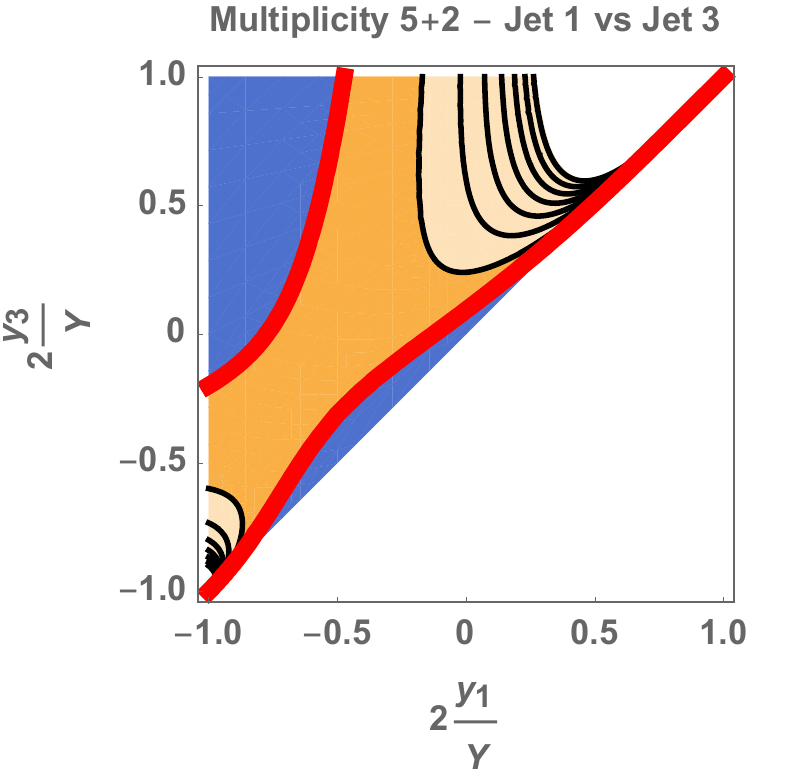}
  \label{fig:sfig2}
\end{subfigure}
\caption{Left:  ${\cal R}_{5+2} \left(x_4,x_1\right) = \sigma_{5+2} 
\frac{ \frac{ d^2 \sigma_{5+2}^{(4,1)}}{d y_4 d y_1} }{\frac{d \sigma_{5+2}^{(4)}}{d y_4} \frac{d \sigma_{5+2}^{(1)}}{d y_1}}-1$. 
Right:  ${\cal R}_{5+2} \left(x_3,x_1\right) = \sigma_{5+2} 
\frac{ \frac{ d^2 \sigma_{5+2}^{(3,1)}}{d y_3 d y_1} }{\frac{d \sigma_{5+2}^{(3)}}{d y_3} \frac{d \sigma_{5+2}^{(1)}}{d y_1}}-1$. }
\label{R7-1425}
\end{figure}

\begin{figure}
\begin{subfigure}{.5\textwidth}
  \centering
  \includegraphics[width=1\linewidth]{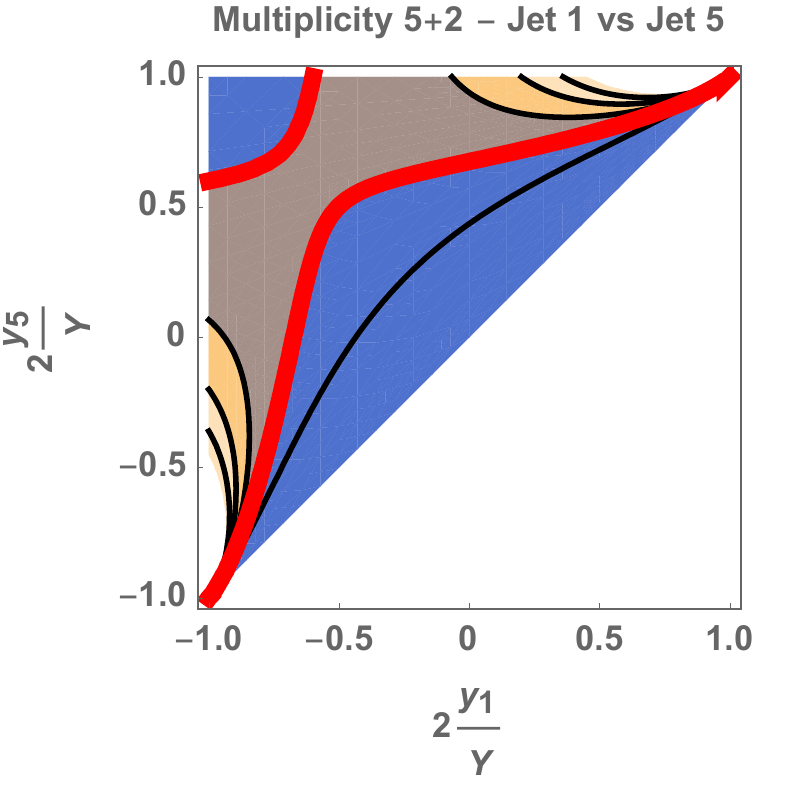}
  \label{fig:sfig1}
\end{subfigure}%
\begin{subfigure}{.5\textwidth}
  \centering
  \includegraphics[width=1\linewidth]{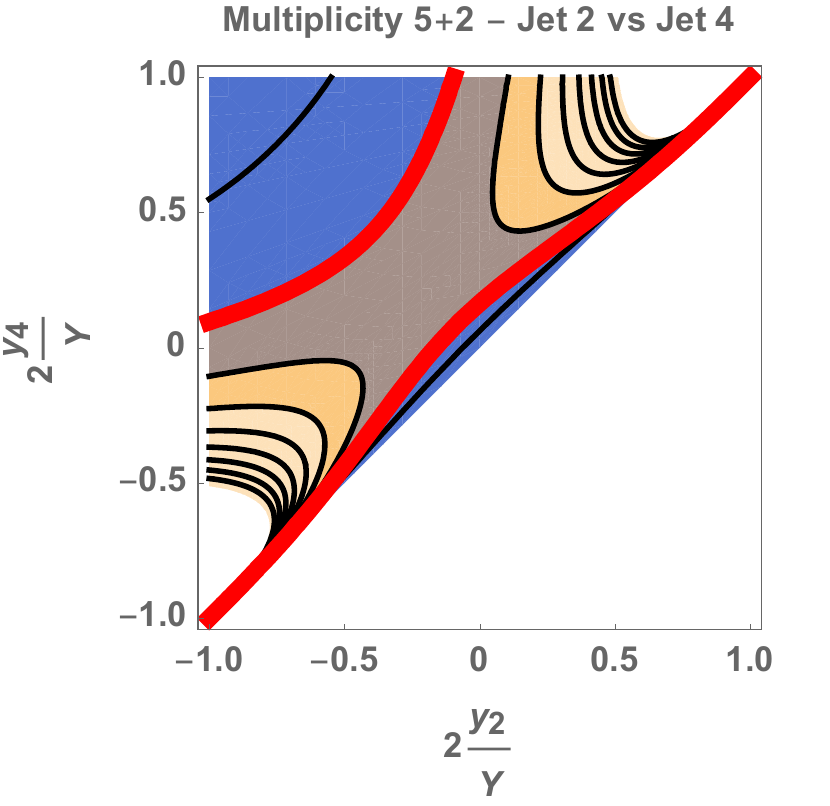}
  \label{fig:sfig2}
\end{subfigure}
\caption{Left:  ${\cal R}_{5+2} \left(x_5,x_1\right) = \sigma_{5+2} 
\frac{ \frac{ d^2 \sigma_{5+2}^{(5,1)}}{d y_5 d y_1} }{\frac{d \sigma_{5+2}^{(5)}}{d y_5} \frac{d \sigma_{5+2}^{(1)}}{d y_1}}-1$. 
Right:  ${\cal R}_{5+2} \left(x_4,x_2\right) = \sigma_{5+2} 
\frac{ \frac{ d^2 \sigma_{5+2}^{(4,2)}}{d y_4 d y_2} }{\frac{d \sigma_{5+2}^{(4)}}{d y_4} \frac{d \sigma_{5+2}^{(2)}}{d y_2}}-1$. }
\label{R7-1524}
\end{figure}

We plot the correlation functions using Eq.~(\ref{R}) in Figs.~\ref{R7-1425} and ~\ref{R7-1524}. We choose
multiplicity $5+2$ and we show the correlation between the rapidities of jet 1 and jet 4 (Fig.~\ref{R7-1425}, left),
jet 1 and jet 3 (Fig.~\ref{R7-1425}, right), jet 1 and jet 5 (Fig.~\ref{R7-1524}, left), jet 2 and jet 4 (Fig.~\ref{R7-1524}, right). The red lines in each of them underline the contour for which ${\cal R}=0$ while the white regions correspond to sectors of very rapid growth of ${\cal R}$. 
\begin{figure}
\begin{subfigure}{.5\textwidth}
  \centering
  \includegraphics[width=.7\linewidth]{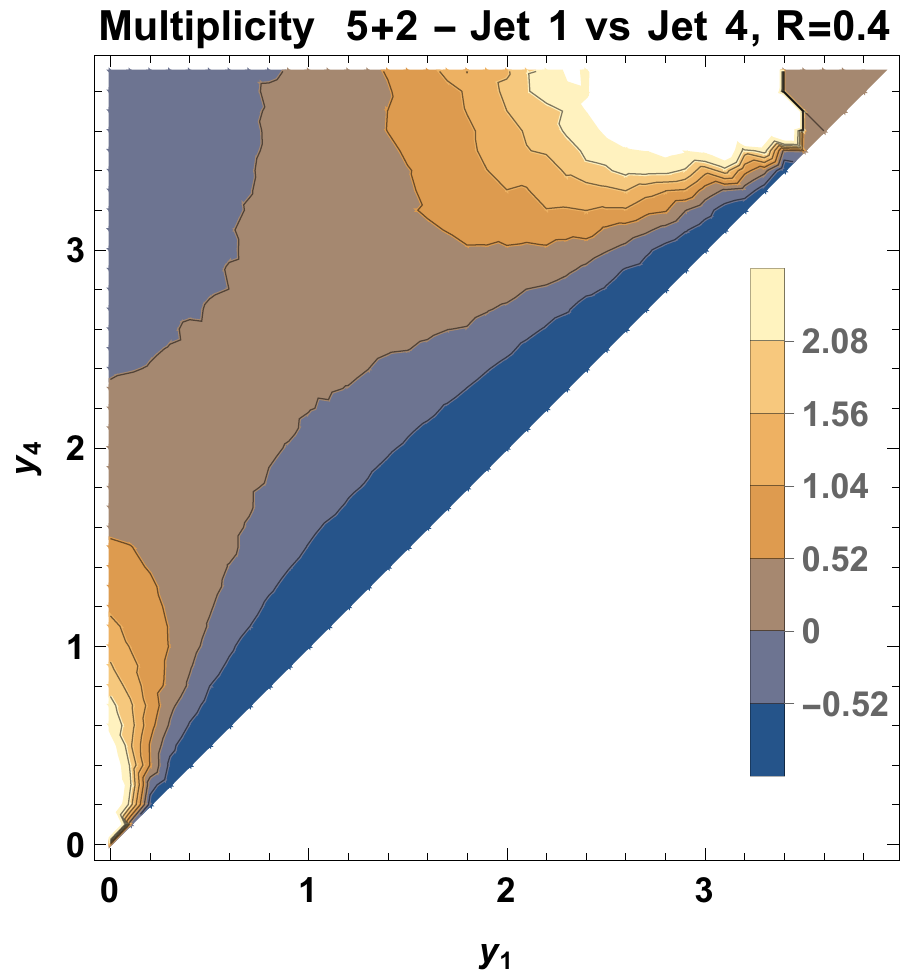}
  \caption{}
  \label{fig:sfig1}
\end{subfigure}%
\begin{subfigure}{.5\textwidth}
  \centering
  \includegraphics[width=.7\linewidth]{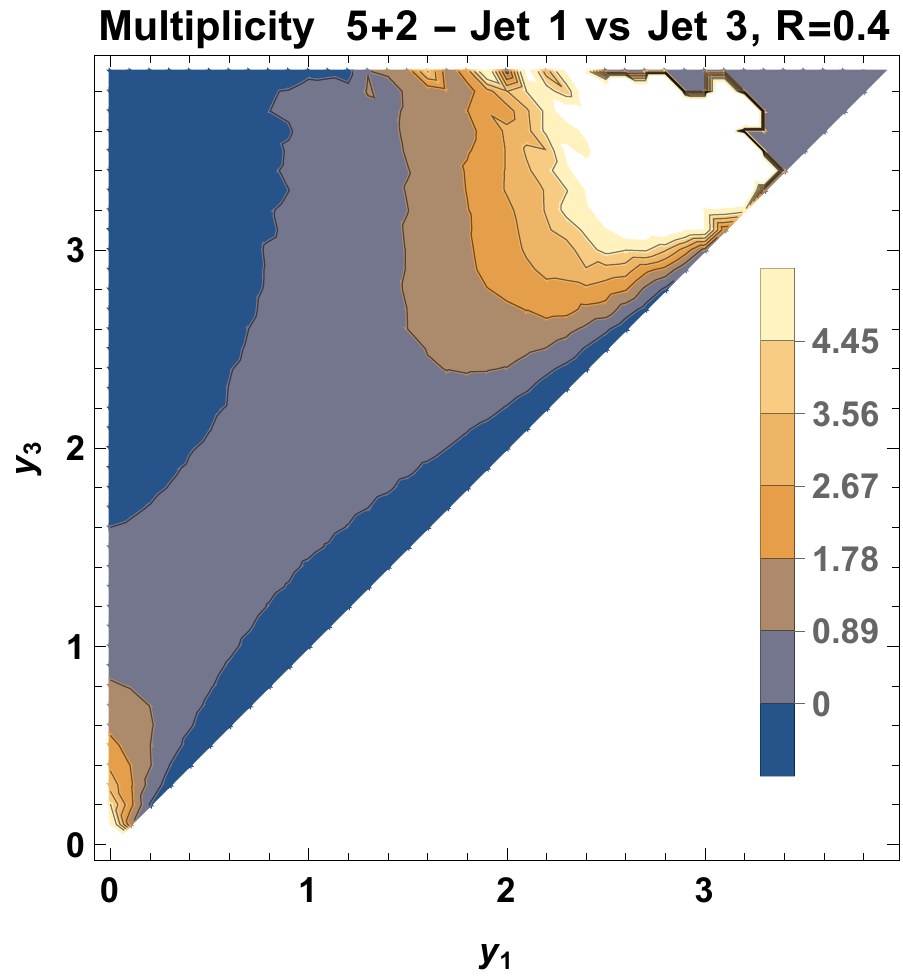}
  \caption{}
  \label{fig:sfig2}
\end{subfigure}
\\
\begin{subfigure}{.5\textwidth}
  \centering
  \includegraphics[width=.7\linewidth]{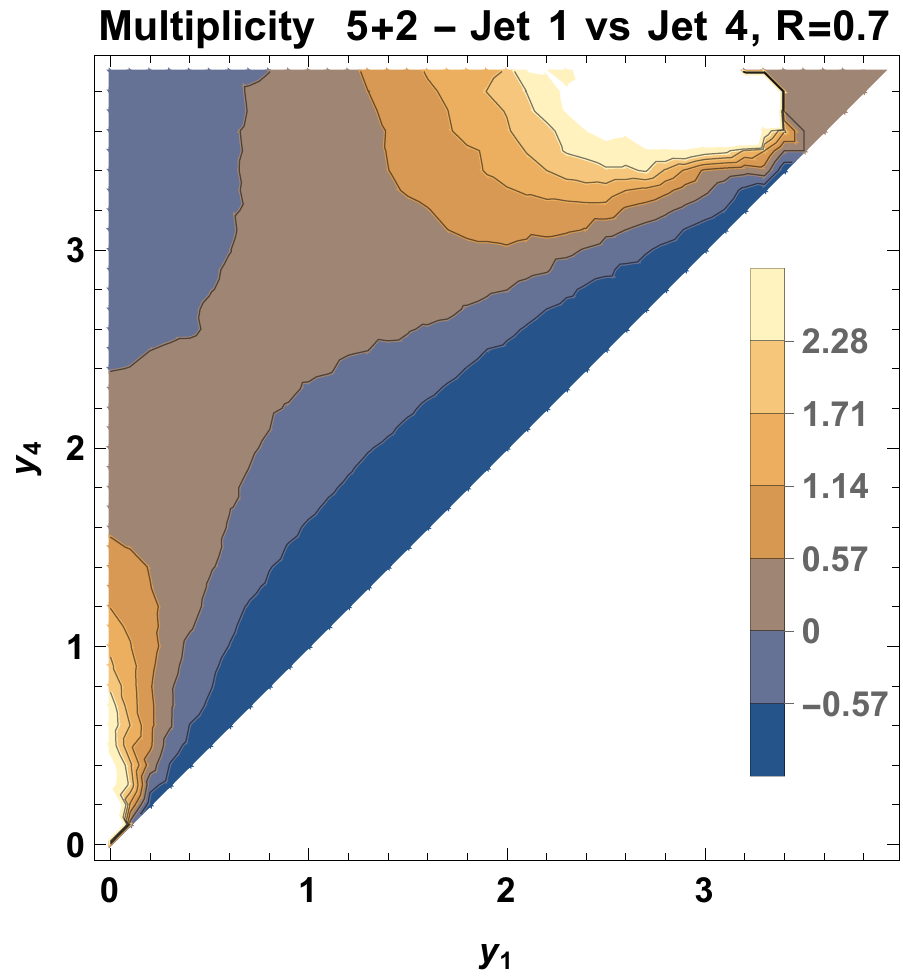}
  \caption{}
  \label{fig:sfig1}
\end{subfigure}%
\begin{subfigure}{.5\textwidth}
  \centering
  \includegraphics[width=.7\linewidth]{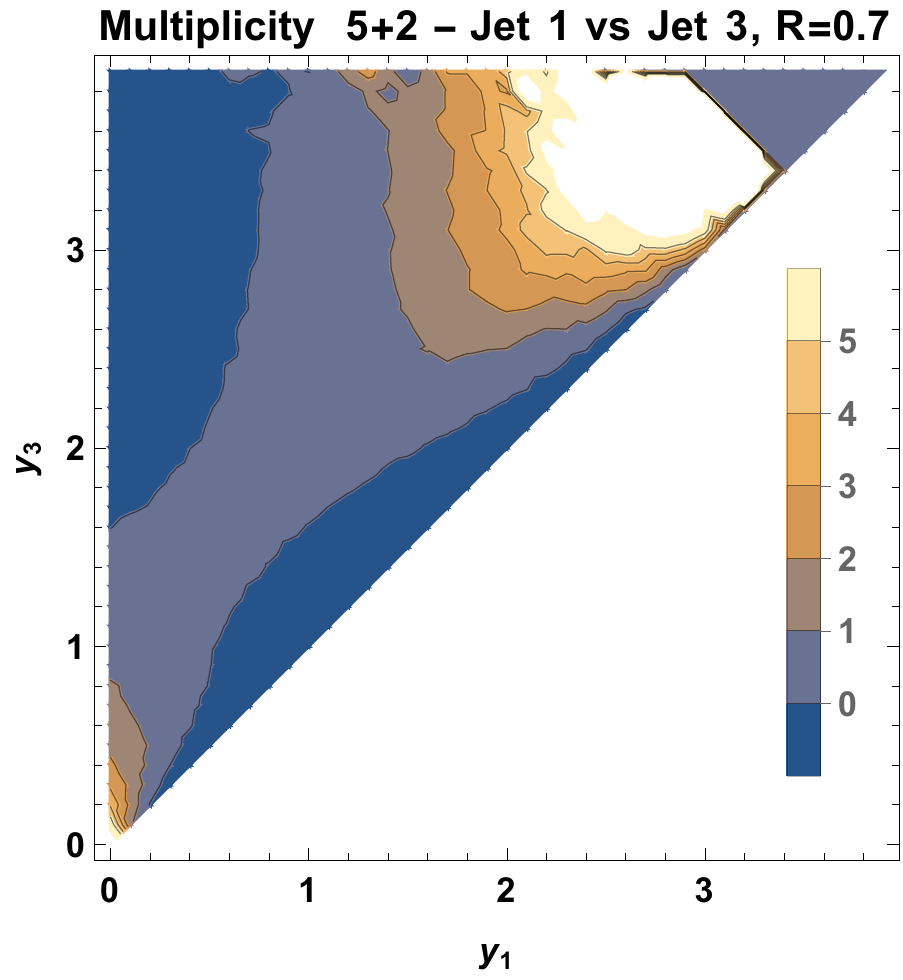}
  \caption{}
  \label{fig:sfig2}
\end{subfigure}
\caption{Top: The correlation functions of Fig.~\ref{R7-1425} with the collinear BFKL model and for jet radius $R = 0.4.$
Bottom: The same but for jet radius $R = 0.7$.}
\label{fig:coll1}
\end{figure}

\begin{figure}
\begin{subfigure}{.5\textwidth}
  \centering
  \includegraphics[width=.7\linewidth]{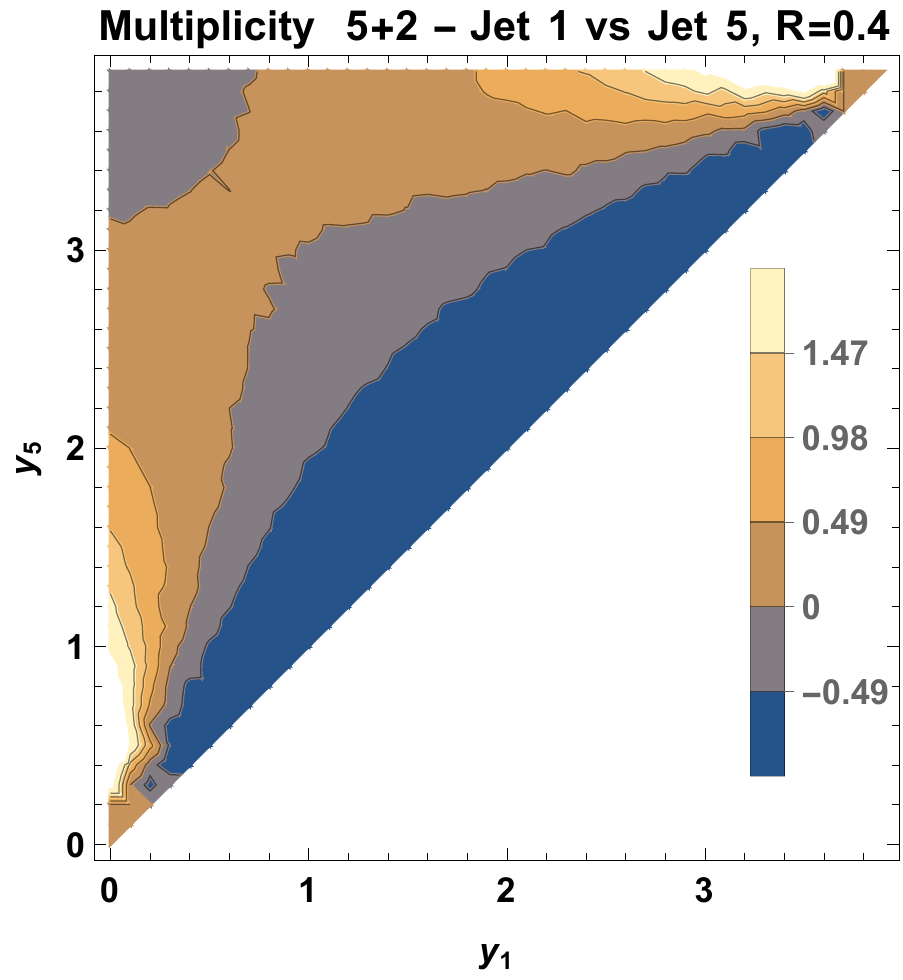}
  \caption{}
  \label{fig:sfig1}
\end{subfigure}%
\begin{subfigure}{.5\textwidth}
  \centering
  \includegraphics[width=.7\linewidth]{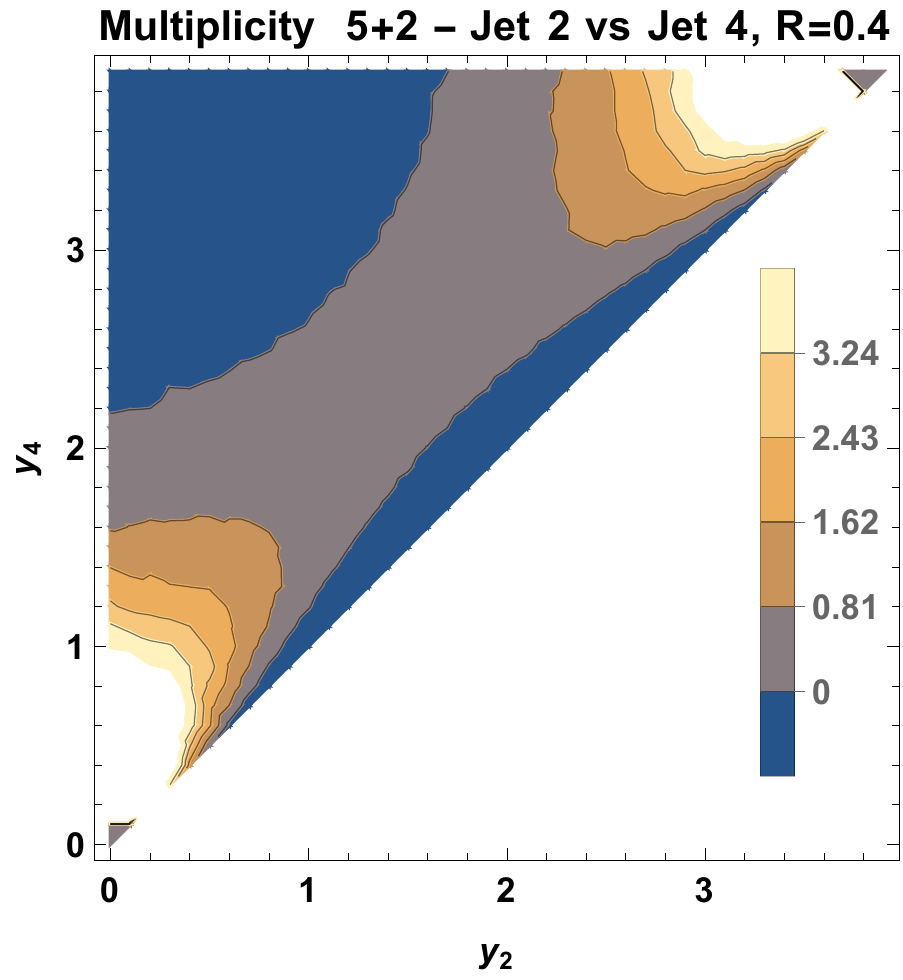}
  \caption{}
  \label{fig:sfig2}
\end{subfigure}
\\
\begin{subfigure}{.5\textwidth}
  \centering
  \includegraphics[width=.7\linewidth]{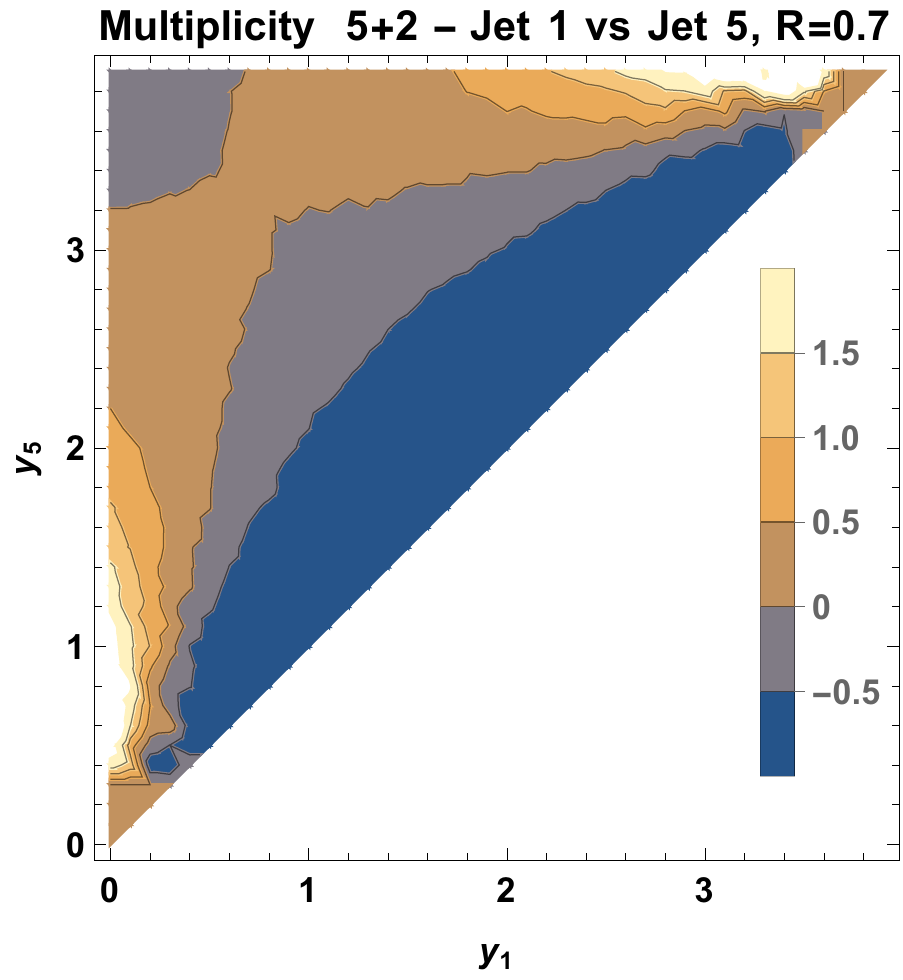}
  \caption{}
  \label{fig:sfig1}
\end{subfigure}%
\begin{subfigure}{.5\textwidth}
  \centering
  \includegraphics[width=.7\linewidth]{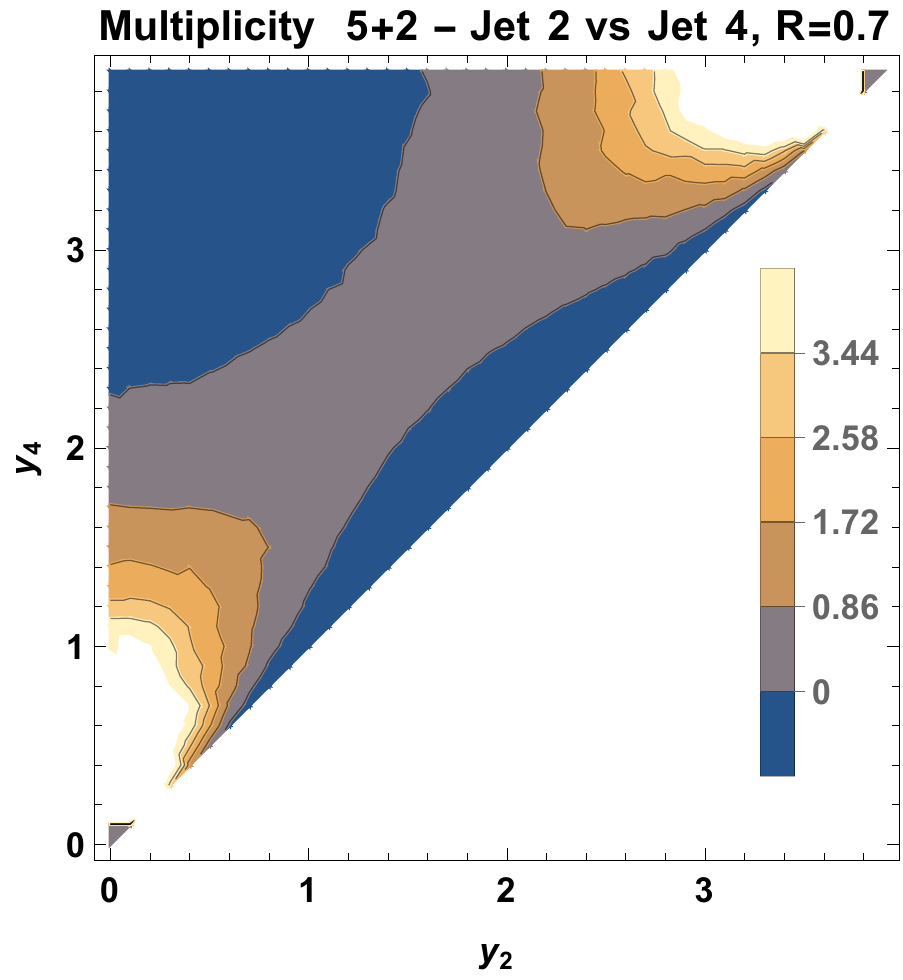}
  \caption{}
  \label{fig:sfig2}
\end{subfigure}
\caption{Top: The correlation functions of Fig.~\ref{R7-1524} with the collinear BFKL model and for jet radius $R = 0.4.$
Bottom: The same but for jet radius $R = 0.7$.}
\label{fig:coll2}
\end{figure}

Our analysis here is quite simple as no relevant dynamics from the  transverse coordinates of the phase space is introduced. Nevertheless, it should still be enough to capture the gross features of single rapidity distributions and double rapidity correlations which should be generally independent of the selected $p_T$ of the outermost
in rapidity jets. In the next section, we will investigate whether these gross
features are any similar to predictions derived from the BFKL formalism using a simple collinear BFKL model
implemented as a special case in our Monte Carlo code {\tt BFKLex}~\cite{Chachamis:2011rw,Chachamis:2011nz,Chachamis:2012fk,Chachamis:2012qw,Chachamis:2015zzp,Chachamis:2015ico,deLeon:2021ecb}.
 
\section{Correlations in BFKL with BFKLex}
 
 In this section we evaluate the correlation functions for the rapidities of jets within the BFKL formalism. We 
 work with the gluon Green's function, that is at partonic level, where an emitted gluon is named a jet if its
 $p_T$ is above some cutoff and we leave for a future work the 
 more complicated and detailed study of the full BFKL dynamics at hadronic level.
 
At a collider with colliding energy $s$, at the leading logarithmic approximation,
where large logarithmic terms of the form ${\bar \alpha}_s^n \ln^n{s}$ are resummed 
(using ${\bar \alpha}_s= {\bar \alpha}_s N_c  / \pi$), 
the differential partonic cross section for the production of two well separated in
rapidity jets and transverse momenta $\vec{p}_{i=1,2}$ is given by
 \begin{eqnarray}
\frac{d \hat{\sigma}}{d^2 \vec{p}_1 d^2 \vec{p}_2} &=& 
\frac{\pi^2 {\bar \alpha}_s^2}{2} \frac{f(\vec{p}_1^{~2}, \vec{p}_2^{~2},Y)}{\vec{p}_1^{~2} \vec{p}_2^{~2}} \,.
\end{eqnarray}
where we take the longitudinal momentum fractions of the colliding partons to be $x_{i=1,2}$ 
and the rapidity difference between the two jets $Y \sim \ln{x_1 x_2 s / \sqrt{\vec{p}_1^{~2} \vec{p}_2^{~2}}}$ .

Within the BFKL resummation framework, one can show that the gluon Green's function $f$ follows, in a collinear approximation, the integro-differential equation
\begin{eqnarray}
\frac{\partial f (K^2,Q^2,Y) }{\partial Y} &=& \delta (K^2-Q^2) \nonumber \\
&&\hspace{-1cm}+ \, 
 {\bar \alpha}_s \int_0^\infty d q^2 \left(\frac{\theta (K-q)}{K^2}+\frac{\theta (q-K)}{q^2} + 4 (\ln{2}-1) \delta(q^2-K^2)\right)
f (q^2,Q^2,Y) \, ,
\end{eqnarray}
the solution of which can be cast in an iterative form:
\begin{eqnarray}
f (K^2,Q^2,Y) &=& e^{4(\ln{2}-1){\bar \alpha}_s Y} \Bigg\{\delta (K^2-Q^2)    \nonumber\\
&&\hspace{-1cm}+ \,    \sum_{N=1}^\infty 
 \frac{({\bar \alpha}_s Y)^N}{N!} \left[ \prod_{L=1}^N
\int_0^\infty d x_L \left(\frac{\theta(x_{L-1}-x_L)}{x_{L-1}} + \frac{\theta(x_L-x_{L-1})}{x_{L}}\right) \right]
\delta (x_N-Q^2) \Bigg\} \,.
\label{FCSumMC}
\end{eqnarray}
Using 
$\delta(K^2-Q^2)=\int \frac{d \gamma}{2 \pi i Q^2} \left(\frac{K^2}{Q^2}\right)^{\gamma-1}$ 
and 
\begin{eqnarray}
&&\int_0^\infty d x_N 
\left(
\frac{\theta(x_{N-1}-x_N)}{x_{N-1}} 
+ \frac{\theta(x_N-x_{N-1})}{x_{N}}\right) 
\left(\frac{x_N}{K^2}\right)^{\gamma-1} = \left(\frac{1}{\gamma}+\frac{1}{1-\gamma}\right) \left(\frac{x_{N-1}}{K^2}\right)^{\gamma-1},
\end{eqnarray}
which is valid for $0<\gamma<1$, we can write
\begin{eqnarray}
\left[ \prod_{L=1}^N \int_0^\infty d x_L 
\left(\frac{\theta(x_{L-1}-x_L)}{x_{L-1}} 
+ \frac{\theta(x_L - x_{L-1})}{x_{L}}\right) \right]
\left(\frac{x_N}{K^2}\right)^{\gamma-1}  
= \left(\frac{1}{\gamma}+\frac{1}{1-\gamma}\right)^N
\end{eqnarray}
so that we then have
\begin{eqnarray}
f (K^2,Q^2,Y) &=&  \frac{e^{4(\ln{2}-1) {\bar \alpha}_s Y}}{Q^2} \int \frac{d \gamma}{2 \pi i} \left(\frac{K^2}{Q^2}\right)^{\gamma-1}
 \sum_{N=0}^\infty \frac{\left({\bar \alpha}_s Y\right)^N}{N!}
\left(\frac{1}{\gamma}+\frac{1}{1-\gamma}\right)^N  \nonumber\\
&=&  \int \frac{d \gamma}{2 \pi i Q^2} \left(\frac{K^2}{Q^2}\right)^{\gamma-1} e^{{\bar \alpha}_s Y \chi\left( \gamma\right)} \, ,
\label{FCMM}
\end{eqnarray}
with $\chi(\gamma) = 4 (\ln{2}-1)+ \gamma^{-1}+ (1-\gamma)^{-1}$. 

From a simple inspection of Eq.~\ref{FCSumMC} one can infer that for the class of jet-jet rapidity correlations we studied in the Section 2, the predictions from this collinear BFKL model
should be very similar to those in the Chew-Pignotti simple model. 
Actually, in Eq.~(\ref{FCSumMC}) every $N$ increase by one unit
accounts for the emission of a new final state gluon.  
After making use of Eq.~(\ref{sigma_N2}), we can write
\begin{eqnarray}
f (K^2,Q^2,Y) &=&  \sum_{N=0}^\infty  {\bar \alpha}_s^{N} 
\int_{0}^{Y} \prod_{i=1}^{N+1} dz_i \delta \left(Y-
\sum_{s=1}^{N+1} z_s \right) \xi^{(N)} (K^2,Q^2) 
\end{eqnarray}
where
\begin{eqnarray}
\xi^{(N)} (K^2,Q^2) &=& \int \frac{d \gamma}{2 \pi i Q^2} \left(\frac{K^2}{Q^2}\right)^{\gamma-1}  \chi^N\left( \gamma\right) \, .
\end{eqnarray}
Working in a manner that follows the logic we used to obtain Eqs.~(\ref{dsdy}) and 
(\ref{d2sdydy}), {\it i.e.} we get here
\begin{eqnarray}
\frac{d f_{N}^{(l)} (K^2,Q^2,Y, y_l) }{d y_l} &=&  {\bar \alpha}_s^{N} \frac{\left(\frac{Y}{2}-y_l \right)^{N-l}}{(N-l)!} \frac{\left(y_l+\frac{Y}{2}\right)^{l-1}}{(l-1)!} \xi^{(N)} (K^2,Q^2)  \, ,  \\
 \frac{d f_{N}^{(l,m)} (K^2,Q^2,Y, y_l,y_m) }{d y_l d y_m} &=& 
 {\bar \alpha}_s^{N} \frac{\left(\frac{Y}{2}-y_l \right)^{N-l}}{(N-l)!}
\frac{(y_l-y_m)^{l-m-1}}{(l-m-1)!} 
\frac{\left(y_m+\frac{Y}{2}\right)^{m-1}}{(m-1)!} \xi^{(N)} (K^2,Q^2)   \, .
\end{eqnarray}
Obviously,  the 
$\xi^{(N)}$ factor cancels out  for normalized quantities
 and thus we end up with the same expressions as for 
the Chew-Pignotti model. 
It is important to remember at this point that the full BFKL formalism 
carries non-trivial dependences in rapidity, transverse momenta and azimuthal angles 
which need to be studied in detail in future works. 
Nevertheless, our findings here in rapidity space suggest that the 
full BFKL predictions might not be totally different from
the old multiperipheral model approach. 

To connect with future works, we have implemented the BFKL collinear model within our {\tt BFKLex} Monte Carlo code, setting the transverse momenta of the most forward and backward jets to be 30 and 35 GeV 
respectively and their difference in rapidity $Y=4$. We also set the multiplicity of the emitted gluons (jets) to
be $N=5+2$. We use the anti-kt jet clustering algorithm in its implementation
 in {\tt fastjet}~\cite{Cacciari:2011ma}. 
We present results from the collinear model  in Figs.~\ref{fig:coll1} and~\ref{fig:coll2}.
The jet radius was chosen to take two values, $R = 0.4$ and $R = 0.7$. In Fig.~\ref{fig:coll1} we plot
the same correlation functions as in Fig.~\ref{R7-1425} whereas in Fig.~\ref{fig:coll2}
the corresponding ones for Fig.~\ref{R7-1524}. It is not surprising, that the collinear model results are 
very similar to the Chew-Pignotti ones and obviously the actual jet radius R does not
affect significantly the plots. Let us also note that in Figs.~\ref{fig:coll1} and~\ref{fig:coll2}
we kept the rapidity range from 0 to $Y$  to make the association to experimental data setups easier. Such
can be found for example in relevant dijet experimental analyses for 7 TeV data from both ATLAS and CMS~\cite{Aad:2011jz,Chatrchyan:2012pb,Aad:2014pua,Khachatryan:2016udy}.

\section{Conclusions}
The 13 TeV data from the run 2 of the LHC at low luminosity are suitable for
various studies of multi-jet physics. Following our recent work in Ref.~\cite{deLeon:2020myv}, 
we want to suggest the investigation of a particular subset of Mueller-Navelet jet events where the outermost jets
are very similar in $p_T$ and the jet multiplicity is kept fixed.
We believe that their experimental study is interesting as it might be possible to  identify features of different 
multi-particle production models such as those predicted by the BFKL formalism. 
We have presented predictions for single and double differential distributions in jet rapidity as well as
the jet-jet correlation functions
from an old multiperipheral model, namely the Chew-Pignotti model, using analytic expressions we obtained
after performing an analysis based on the decoupling of the longitudinal and transverse coordinates. 
We have also presented results for the jet-jet correlation
functions from a collinear BFKL model implemented in our Monte Carlo code {\tt BFKLex}.
In the future, we plan to perform a more complete study of these observables in high energy QCD
including the full dependence on the transverse coordinates
and moving from a partonic level analysis of the BFKL gluon Green's function 
to the hadronic level with PDFs included and suitable phenomenological kinematic cuts.

Comments: Presented at the Low-$x$ Workshop, Elba Island, Italy, September 27--October 1 2021.

\section*{Acknowledgements}
We would like to thank the organizers of the Low-x Workshop for their excellent work. 
This work has been supported by the Spanish Research Agency (Agencia Estatal de Investigaci{\'o}n) through the grant IFT Centro de Excelencia Severo Ochoa SEV-2016-0597 and the Spanish Government grant FPA2016-78022-P.  It has also received funding from the European Union's Horizon 2020 research
and innovation programme under grant agreement No. 824093. The work of GC was supported by the Funda\c{c}{\~ a}o para a Ci{\^ e}ncia e a Tecnologia (Portugal) under project CERN/FIS-PAR/0024/2019 and contract 'Investigador auxiliar FCT - Individual Call/03216/2017'.

\nocite{*}
\bibliographystyle{auto_generated}
\bibliography{chachamis_proceedings_elba2021/chachamis_proceedings_elba2021}

%% file: colferai/proceedings/Colferai.tex

\vspace*{1.2cm}
\thispagestyle{empty}

\begin{center}
{\LARGE \bf Is BFKL factorization valid \\[2mm]
    \hspace{4em} for Mueller-Tang jets?}

\par\vspace*{7mm}\par

{\bigskip \large \bf Dimitri Colferai}
\bigskip
{\large \bf  E-Mail: colferai@fi.infn.it}

\bigskip

{Department of Physics, University of Florence and INFN Florence}

\bigskip

{\it Presented at the Low-$x$ Workshop, Elba Island, Italy, September 27--October 1 2021}

\vspace*{15mm}
\end{center}
\vspace*{1mm}

\begin{abstract}

  The perturbative QCD description of high-energy hadroproduction of two hard
  jets separated by a large rapidity gap void of emission (also called
  Mueller-Tang jets) is based on a factorization formula of BFKL type that
  represents exchanges of colour-singlet objects among the external
  particles. This formula resums to all perturbative orders a certain class of
  Feynman diagrams that are supposed to dominate the cross-section in the Regge
  limit. However, the explicit calculations at next-to-leading logarithmic order
  questions the validity of such factorization when an IR safe jet algorithm is
  used to reconstruct jets. We show the origin of such violation of
  factorization, and quantify its impact for LHC phenomenology.  In this
  connection, we estimate the impact of other contributions to the cross-section
  that are not included in the Mueller-Tang factorization formula --- colour
  non-singlet exchanges --- that, at low rapidity separation, compete with the
  singlet ones.
\end{abstract}
  \part[Is BFKL factorization valid for Mueller-Tang jets?\\ \phantom{x}\hspace{4ex}\it{Dimitri Colferai}]{}
\section{Introduction}

\begin{figure}[ht]
\begin{center}
  \includegraphics[width=0.35\linewidth]{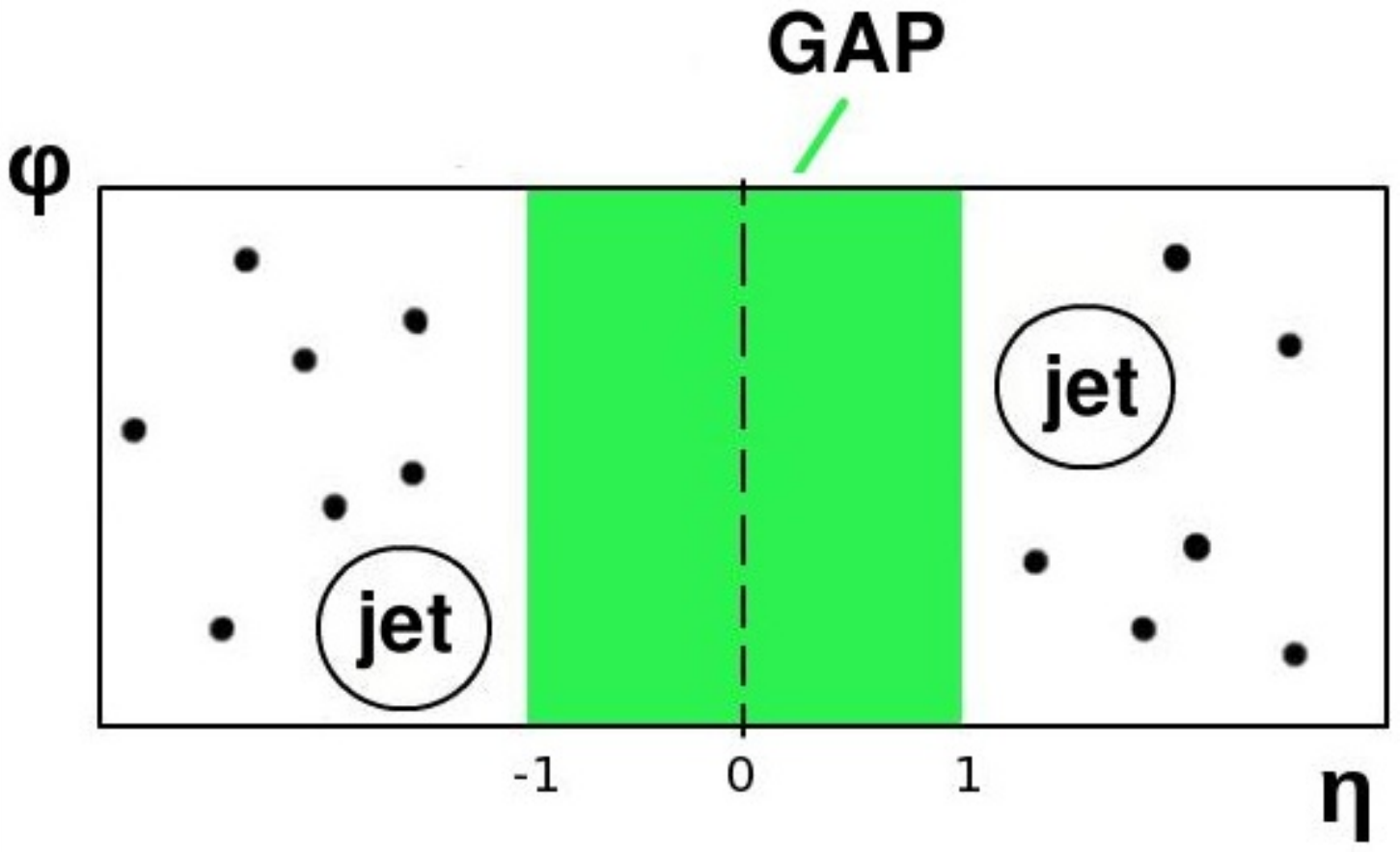}
  \hspace{0.1\linewidth}
  \includegraphics[width=0.35\linewidth]{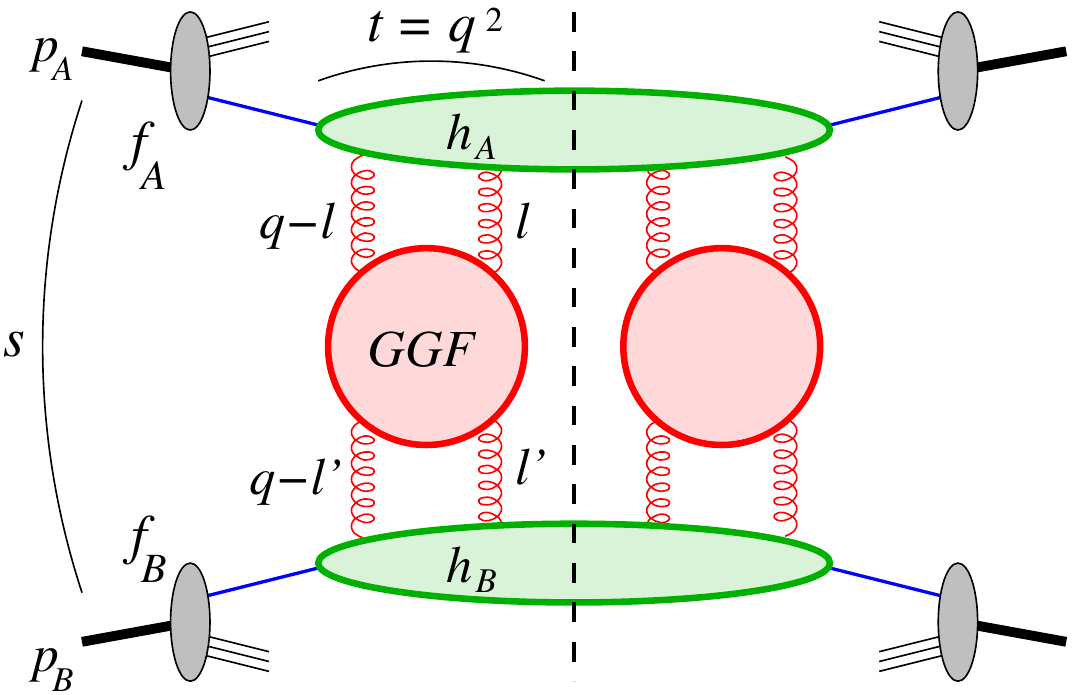}\\
  (a)\hspace{0.45\linewidth}(b)
  \caption{(a) Sketch of particle detection of a jet-gap-jet event in the
    azimuth-rapidity plane. (b) Diagrammatic representation or the factorization
    formula for MT jets.}
  \label{f:jgj}
\end{center}
\end{figure}

Mueller-Tang (MT) jets~\cite{MuTa92} are important for studying perturbative
high-energy QCD and the Pomeron at hadron colliders. They are characterized by
final states with at least 2 jets with comparable hard transverse momenta
($\ktcolf_{J1}\sim\ktcolf_{J2} \gg \Lqcd$), well separated in rapidity
$Y\equiv y_{J1}-y_{J2}$, and absence of emission in a given interval of
pseudo-rapidity $\Delta\eta\lesssim Y$ in the central region (the so-called
gap). For this reason, they are also called ``jet-gap-jet'' events, and a
typical final state is depicted in fig.~\ref{f:jgj}a.

The presence of the gap suggests that these events mainly occur when the momentum
exchange between the forward and backward systems is due to a colour-singlet
virtual state: a non-singlet exchange would be characterized most of the times
by final state radiation deposited in the central region.

A large rapidity interval is possible because at LHC the
center-of-mass (CM) energy is much larger than the jet transverse energy.  In
this case, the coefficients of the perturbative series are enhanced by powers of
$Y\simeq \log(s/\ktcolf_J^2)$, and an all-order resummation of the leading terms
$\sim(\alpS Y)^n$ is needed for a proper determination of the amplitude.

\subsection{Cross section in leading logarithmic approximation}

At lowest perturbative order, a colour-singlet exchange in the $t$-channel is
due to two gluons in colour-singlet combination.  At higher orders, as just
mentioned, the partonic elastic amplitude is affected by powers of
$Y\simeq \log(s/\ktcolf_J^2)$ due to gluon ladder-like diagrams. Such contributions
can be resummed into the so-called BFKL gluon Green function (GGF)~\cite{BFKL}.
It is interesting to observe that such LL loop diagrams are both UV and IR
finite.

By squaring the partonic amplitude, the LL partonic cross-section is then given
by the product of 2 GGFs, which embody the energy-dependence, and two impact
factors (IFs), that couple the gluons to the external particles. In the LL
approximation the IFs are just a trivial product of coupling constants and
colour factors.

Finally, the cross section for MT jets in the LL approximation can be expressed by the
factorization formula (see fig.~\ref{f:jgj}b)
\begin{align}
  \frac{\dif\sigma^{(LL)}}{\dif J_1 \dif J_2} \simeq \int
  \dif (x_1, x_2, \lt_1, \lt'_1, \lt_2, \lt'_2)\; &
  f_A(x_1) \Phi_A(x_1,\lt_1,\lt_2;J_1) G(x_1 x_2 s,\lt_1,\lt_2) \nonumber\\
  &\times G(x_1 x_2 s,\lt'_1,\lt'_2)
  \Phi_B(x_2,\lt'_1,\lt'_2;J_2) f_B(x_2) \;. \label{ff}
\end{align}
Here $J=(y_J,\ktcolf_J)$ represents the set of jet variables,
the GGFs $G$ describe universal gluon dynamics, the IFs $\Phi_i$
describe the coupling of the reggeized gluons or pomerons to the external
particles, and the PDFs $f_i$ describe the partonic content of hadrons.

\subsection{First phenomenological analyses of jet-gap-jet events}

\begin{figure}[ht]
\begin{center}
\includegraphics[width=0.5\linewidth]{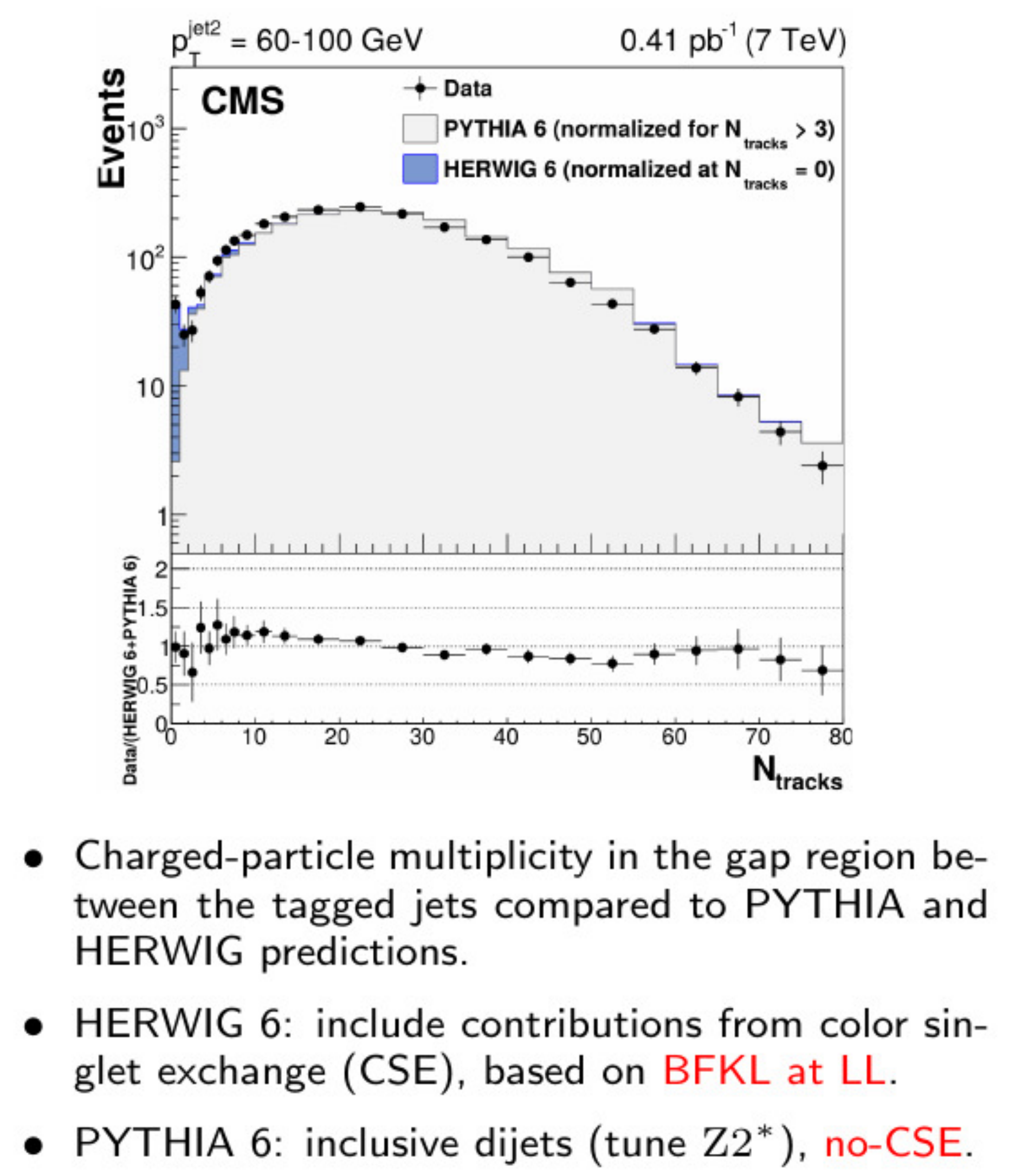}
\caption{CMS measurements of multiplicity of charged particles in the gap
  region, and comparison with Pythia and Herwig predictions.}
\label{f:multiplicity}
\end{center}
\end{figure}

The importance of considering such BFKL contributions to the cross section has
been emphasized since the first analysis by CMS. The plot in
fig.~\ref{f:multiplicity} shows the number of events as a function of the
multiplicity of charged particles in the gap region. We see that both Herwig and
Pythia are able to describe the data if one or more particles are observed
between the jets, but only Herwig agrees in the first bin without observed
particles, and this happens because Herwig includes the contribution of
colour-singlet exchange from BFKL at LL

\begin{figure}[hb]
\begin{center}
\includegraphics[width=0.9\linewidth]{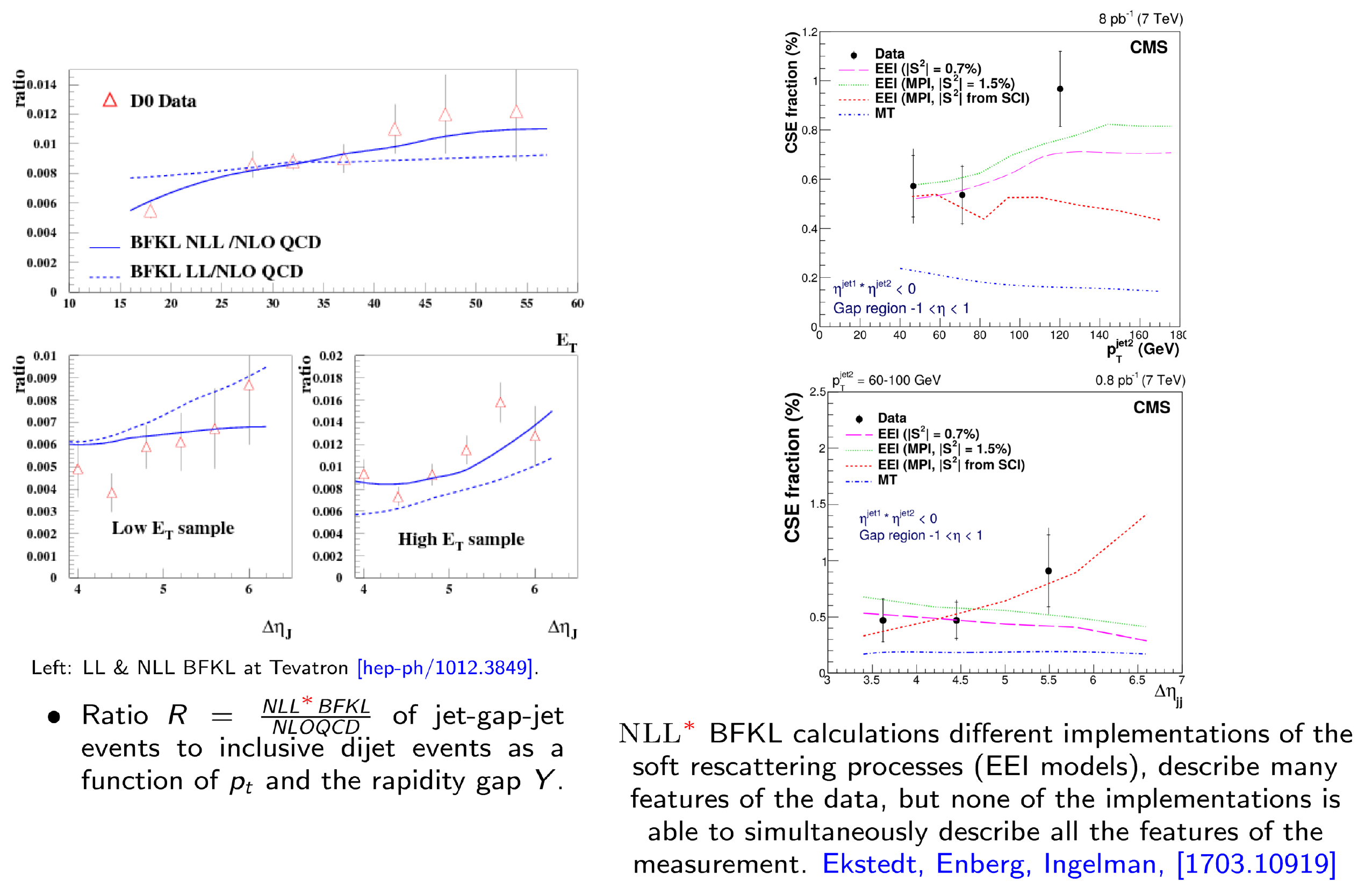}\\
\caption{Comparison of various differential measurements by D0 for jet-gap-jet
  events, and comparison with various theoretical models implementing the BFKL
  GGF in NLL approximation.}
\label{f:d0}
\end{center}
\end{figure}

If one looks at differential distributions of JGJ events, like distributions in
$p_\perp$ or in rapidity distance $Y$, however, the situation is not so
nice. Here LL predictions are unable to describe data (see
fig.~\ref{f:d0}a). Even if one improves the BFKL GGF by adding next-to-leading
logarithmic (NLL) contributions~\cite{NLLFL,NLLCC,KMR10,EEI17}, none of the
implementations is able to simultaneously describe all the features of the
measurements (see fig.~\ref{f:d0}b).

\section{Impact factor in next-to-leading logarithmic approximation}

\subsection{Structure of the calculation and final result}

It appears thus compelling to provide a full NLL description of MT jets. The
idea is to generalize the BFKL factorization formula for MT jets to the NLL
approximation.


The NLL BKFL GGF is known in the non-forward case~\cite{NLLnf}, but due to its
complexity, only the forward version~\cite{NLLFL,NLLCC}, has been used in order
to estimate the contribution of NL logarithmic terms to the cross
section~\cite{KMR10}. However, this is not relevant for our study of the impact
factors.

\begin{figure}[ht]
\begin{center}
\includegraphics[angle=-90,width=0.5\linewidth]{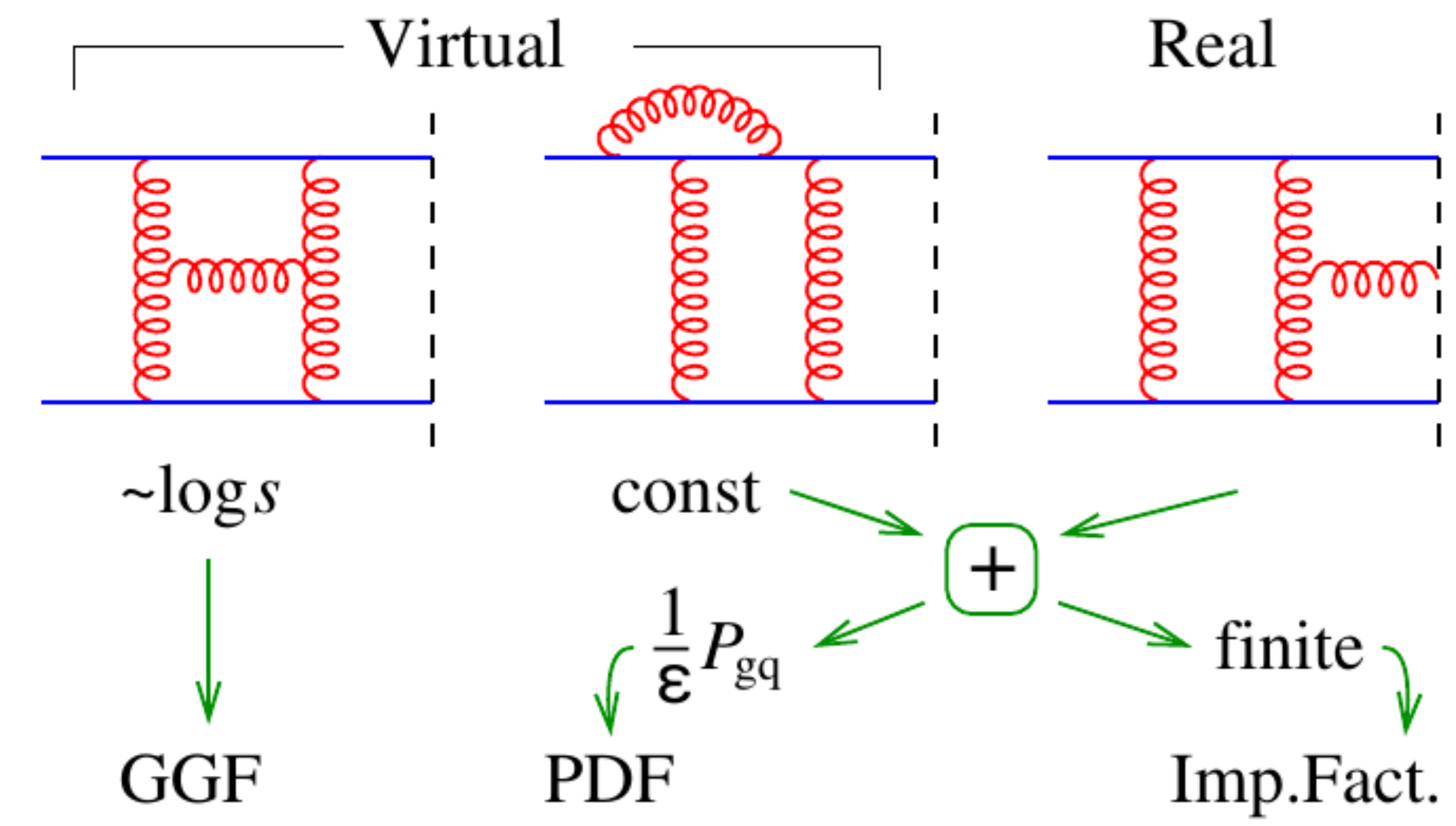}
\caption{Schematics of the decomposition of the (real + virtual) NLO calculation
  for the determination of the NL impact factors.}
\label{f:mtdec}
\end{center}
\end{figure}

The determination of the NL IF can be done with a NLO calculation, which is
affected, of course, by IR (soft and collinear) divergencies.  Actually, the
very existence of NL IF is not a trivial statement. By summing virtual and real
contributions at first perturbative order, one has to prove that
\begin{itemize}
\item the $\log(s)$ terms from virtual corrections reproduce the BFKL kernel
  (and this we already know);
\item the const term of the virtual corrections, which are IR divergent and
  constitutes the virtual part of the IF, when combined with real emission
  terms, must provide a finite remainder, after subtraction of the collinear
  singularities (proportional to the Altarelli-Parisi splitting functions) to be
  absorbed in the PDF;
\end{itemize}
Such finite remainder defines the next-to-leading impact factor. A sketch of
this decomposition is depicted in fig.~\ref{f:mtdec}

\begin{figure}
  \begin{center}
    \includegraphics[width=0.27\linewidth]{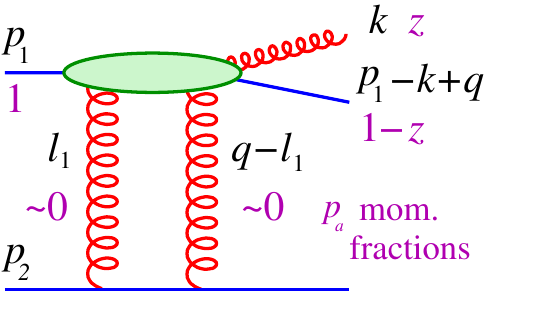}%
    \raisebox{5ex}{
      $\times$
      \begin{minipage}[c]{0.10\linewidth}
      h.c.\\
      $(\lt_1\leftrightarrow\lt_2)$
    \end{minipage}}
    \caption{Kinematics of the calculation of the NL impact factor. Black
      symbols denote 4-vectors, while purple ones denote longitudinal momentum
      fraction.}
    \label{f:kinematics}  
  \end{center}
\end{figure}

The calculation of NL IF for MT jets was performed~\cite{HMMS14a,HMMS14b} using
Lipatov's effective action, and has been confirmed by our independent
calculation~\cite{CoDeRo21}. This is the structure of the result in the case
of incoming quark:
\begin{align*}
  \Phi(\lt_1,\lt_2,\qt)&=\frac{\alpS^3}{2\pi(N_c^2-1)}\int_0^1\dif z\int\dif\ktcolf\;
  S_J(\ktcolf,\qt,z)\, C_F\frac{1+(1-z)^2}{z}\\
  & \times \left\{C_F^2\frac{z^2\qt^2}{\ktcolf^2(\ktcolf-z\qt)^2}
  +C_F C_A\, f_1(\lt_{1,2},\ktcolf,\qt,z) + C_A^2\, f_2(\lt_{1,2},\ktcolf,\qt)
  \right\}
\end{align*}
It is important to understand the kinematics of the process (see
fig.~\ref{f:kinematics}): after the ``upper'' incoming quark interacts with the
two gluons in colour-singlet, a quark and a gluon can be found in the forward
hemisphere of the final state; the ``lower'' parton $p_2$ remains intact and is
just slightly deflected in the backward hemisphere.

Let me denote with $k$ the outgoing gluon momentum, with $\ktcolf$ its transverse
momentum and with $z$ its longitudinal momentum fraction with respect to the
parent quark; $q$ is the overall $t$-channel transferred momentum.  $\ktcolf$ and
$z$ are integration variables. Virtual contributions are contained as
delta-function contributions at $z = 0$ and $\ktcolf = 0$.

We can see the quark-to-gluon splitting function $P_{gq}$ as overall factor, and
then three terms with different colour structures. The integration in the phase
space of the gluon and quark final state has to be restricted by an IR-safe jet
algorithm $S_J$, such as the $k_\perp$-algorithm.

\subsection{Violation of BFKL factorization}

In the diffractive process we are considering, one quark moves in the backward
direction and is identified with the backward jet. The other two partons, whose
distance in azimuth-rapidity is denoted by
$\Delta\Omega = \sqrt{\Delta\phi^2+\Delta y^2}$, are emitted in the forward
hemisphere, so as to produce at least one jet.  There are 3 possibilities:
\begin{itemize}
\item $\Omega < R$ corresponding to a composite jet;
\item $\Omega > R$ where the {\em gluon is the jet} and the quark is outside the
  jet cone;
\item $\Omega > R$ where the {\em quark is the jet} and the gluon is outside the
  jet cone.
\end{itemize}
In the last configuration there is a problem due to the $\dif z/z$ integration
of the $C_A^2\, f_2$ term. In fact, when the quark is the jet, the gluon can
become soft and its phase-space integration is essentially unconstrained.

The limit $z\to 0$ at fixed $\ktcolf$ corresponds to find the gluon in the central
(and backward) region, where the emission probability of the gluon turns out to
be flat in rapidity, and formally the $z$ (or $y_g$) integration diverges.

If we believe the above transition probability to be reliable at least in the
forward hemisphere ($y_g>0$), the longitudinal integration yields a
$\log(s)$. But a $\log(s)$ in the IF is not acceptable, being against the spirit
of BFKL factorization where all the energy-dependence is embodied in the GGFs.

All this looks strange, because one would have argued that gluon emission in the
central region should be dynamical suppressed, due to the singlet exchange in
the $t$-channel. We will solve this puzzle later on. For the moment, we discuss
a proposal to cope with this fact in practice.

In order to avoid this problem, the authors of~\cite{HMMS14a,HMMS14b} impose an
upper limit $M_{X,\max}$ on the invariant mass of the forward diffractive
system. In that case, the $z$ variable is bounded from below and the $z$-integral
is finite.  However, a crucial question then arises: do we really need to impose
a cut on the diffractive mass?  Actually, {\em can} we impose such a constraint?

If it were possible, then one could avoid $\log(s)$ terms in the IF, though at the
price of introducing logarithms of the diffractive mass: $\log(M^2_{X,\max}/\ktcolf_J^2)$.
However, from the experimental point of view, in order to impose such constraint
one should be able to measure the spectator partons, i.e., the proton remnants
(or the intact proton in case of diffractive events).
Since this is not possible with the present experimental detectors at hadron
colliders, other solutions must be found. 

\begin{figure}[ht]
\begin{center}
  \includegraphics[width=0.6\linewidth]{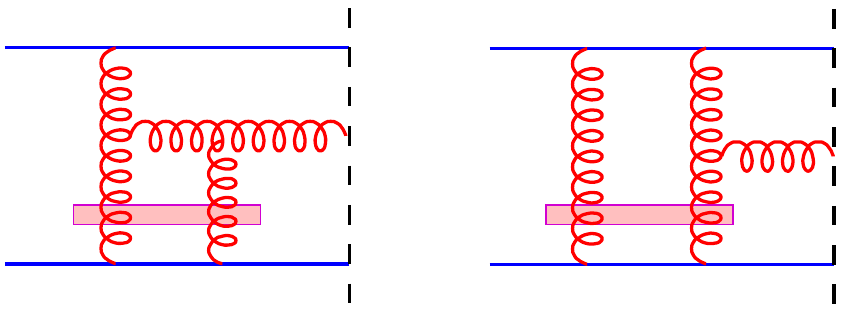}
\caption{Examples of diagrams that involve a non-singlet emission ``above'' the
  emitted gluon, thus producing a $\log(s)$ term in the impact factor. The pink
  rectangles denote colour-singlet projection.}
  \label{f:ec}
\end{center}
\end{figure}

In order to find the origin of such logarithmic contributions in the IF, let us
consider a pair of diagrams contributing to the $C_A^2\,f_2$ term and drawn in
fig.~\ref{f:ec}. It is clear that, if the two $t$-channel (vertical) gluons
emitted by the lower quark are in a colour-singlet state, by colour conservation
the (one or two) upper gluons cannot be in such a state, since a (coloured)
gluon is emitted in the final state. Therefore we cannot claim that this diagram
involves a colour-singlet exchange between the upper and lower system --- the
final state gluon being in the central region.

The option of defining MT jets by selecting those diagrams that involve only
colour-singlet exchanges is not viable, in particular this would end up in a
non-gauge-invariant procedure. We therefore claim that this problem cannot be
avoided and that MT jets are not describable by the naive factorization formula
originally proposed.

At this point, it remains to estimate the size of such violation and possibly to
resum another set of diagrams, if the violation is sizeable.  Given the fact
that we cannot measure particles (partons, hadrons) below some energy threshold
$E_{\mathrm{th}}\sim 200\MeVns$ , we can at most require no activity above that
threshold within the rapidity gap.  This prescription is IR safe, because it is
inclusive for gluon energies $E_g < E_{\mathrm{th}}$ (our analysis here proceeds
at partonic level). Since such soft gluons can have arbitrary values of rapidity
between the two jets, we can easily estimate that the logarithmic contribution
to the impact factor is of the order
\begin{equation}
  \Phi_{log} \sim C_A^2 \frac{E_{\mathrm{th}}^2}{\ktcolf_J^2}\log\frac{s}{\ktcolf_J^2}
\end{equation}
Note that this term is regular for $E_{\mathrm{th}}\to 0$, actually it vanishes, at
variance with the $\ord{C_F^2}$ term in the IF which diverges in the same limit.
When evaluated with the values of energies and momenta of typical processes
analysed at LHC, this term turns out to be small, of order $1\%$ or less with
respect to other terms~\cite{CoDeRo21}.

\begin{figure}[ht]
\begin{center}
  \includegraphics[width=0.5\linewidth]{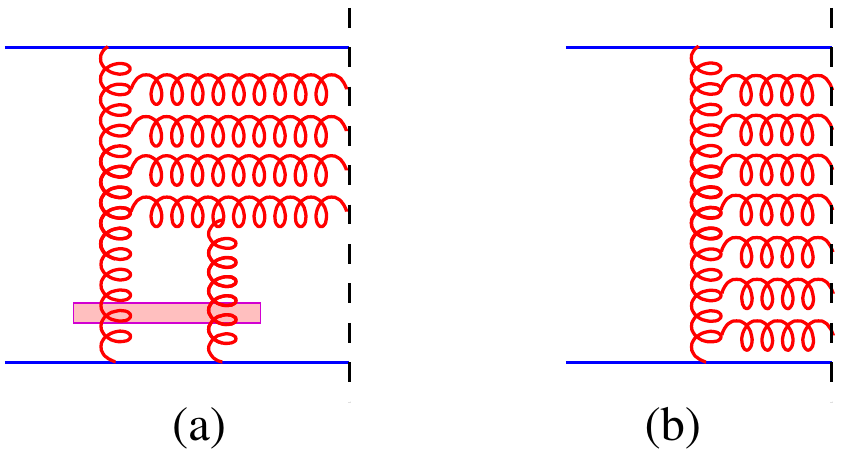}
\caption{Examples of diagrams resumming gluon emission from non-singlet gluon
  exchange. (a) Corrections to the IF; the pink rectangles denote colour-singlet
  projection. (b) Mueller-Navelet-like contributions outside the BFKL
  factorization formula~\eqref{ff}.}
  \label{f:resum}
\end{center}
\end{figure}

Although not really needed from a quantitative point of view at the moment, one
could envisage to resum such logs in the same BFKL spirit, i.e. by considering
diagrams where an arbitrary number of soft (below threshold) gluons are emitted
in the gap (without being detected). This enlarge considerably the number of
diagrams to be taken into account. Some of them, like those in
fig.~\ref{f:resum}a, could be incorporated in the IF, which however acquire a
dependence on the jet rapidity as well as on the gap extension and threshold.

Other diagrams, like those of fig.~\ref{f:resum}b, contribute in a completely
different way, outside the structure of the MT factorization formula. Actually,
they are just diagrams of a Mueller-Navelet (two jet inclusive) process, with
the restriction that the energy of all gluons emitted in the gap region is below
threshold. For threshold energies $E_{\mathrm{th}}\ll |\ktcolf|$, this contribution
can be easily estimated in LL approximation, at least as far as the
$E_{\mathrm{th}}$-dependence is concerned: for soft emissions, virtual and real
corrections cancel each other. Since virtual corrections are always fully taken
into account, while imposing a void gap constrains only real emission, what
remains is essentially equal to the virtual corrections with momenta above
$E_{\mathrm{th}}$. In the LL approximation, virtual corrections are provided by
the exponentiation of the intercept of the reggeized gluon $\omega(\ktcolf)$ with
its internal momentum integrated below $E_{\mathrm{th}}$, resulting in
\begin{align*}
  \omega_{\mathrm{th}}(\ktcolf) &\simeq -\frac{\alpS N_c}{\pi}\log\frac{|\ktcolf|}{E_{\mathrm{th}}} \\
  \frac{\dif\sigma_{\mathrm{oct}}}{\dif t} &\simeq \frac{\dif\sigma_0}{\dif t}
  \exp\left(-\frac{\alpS N_c}{\pi}\log\frac{\ktcolf^2}{E_{\mathrm{th}}^2} Y\right)
\end{align*}
to be compared with the MT asymptotic cross section
\begin{align*}
  \frac{\dif\sigma_{\mathrm{sing}}}{\dif t} \simeq \frac{\dif\sigma_0}{\dif t}
  \frac{(\alpS C_F \pi)^2}{2}\frac{
    \exp\left(\frac{\alpS N_c}{\pi}8\log2 Y\right)}{[\frac72\alpS N_C\zeta(3)Y]^3} \;,
\end{align*}
$\dif\sigma_0/\dif t$ being the lowest order (one-gluon exchange) cross section.

\begin{figure}[ht]
\begin{center}
  \includegraphics[width=0.5\linewidth]{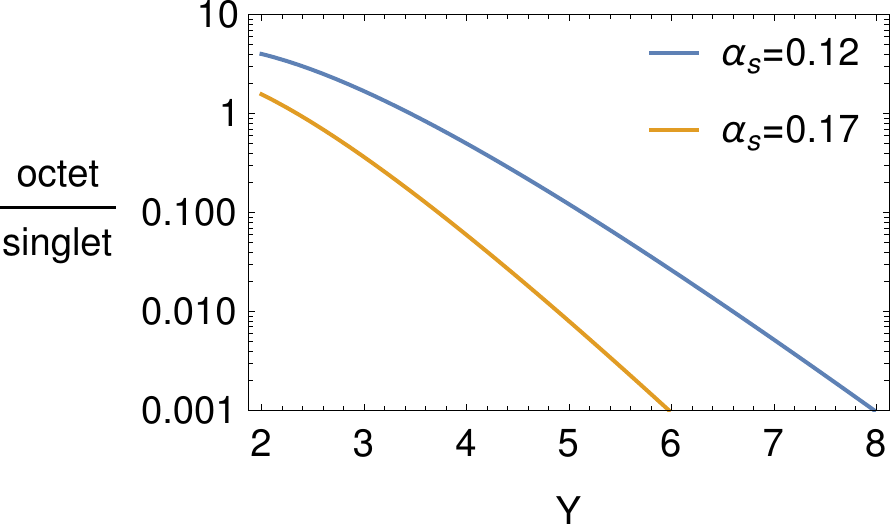}
  \caption{Ratio of differential cross sections in rapidity $Y$ of non-singlet
    (octet) exchanges (fig.~\ref{f:resum}b) emitting gluon having energies below
    threshold and singlet-exchange (fig.~\ref{f:jgj}b). Two values of $\alpS$ are
    considered.}
  \label{f:singVSoct}
\end{center}
\end{figure}

Such comparison was already made by Mueller and Tang in their original
paper~\cite{MuTa92}, but for very large values of $Y\simeq 12$ and other
parameters not corresponding to LHC kinematics, where the non-singlet exchanges
is strongly suppressed with respect to the singlet ones. Repeating such
comparison with realistic LHC parameters ($\ktcolf= 30 \GeVns$, $E_{\mathrm{th}}=0.2\GeVns$) 
we find (fig.~\ref{f:singVSoct}) that for $Y\sim 3$ the two contributions are of
the same order, and at $Y\sim 4$ the non-singlet one is still important, about
10\% of the singlet one.

\section{Conclusions}

To summarize, we have demonstrated that, for jet-gap-jet observables, there is
violation of the standard BFKL factorization at NLL level, since the IFs present
logarithmically enhanced energy-dependent contributions. However such terms are
rather small, below 1\% for current measurements of Mueller-Tang jets at LHC,
and their resummation looks not compelling.

Nevertheless, colour non-singlet contributions are expected to be non-negligible
at LHC, in particular for small values of the rapidity distance $Y$ between
jets. Mueller-Navelet contribution below threshold should in this case be
included, unless NLL corrections to the latter provide a further suppression so
as to render them irrelevant. But this requires further studies.

\vspace{2ex}

Comments: Presented at the Low-$x$ Workshop, Elba Island, Italy, September 27--October 1 2021.

\section*{Acknowledgements}

I thank Krzysztof Kutak and Leszek Motyka for useful discussions on this
subject.\\[1mm]
\noindent
\includegraphics[width=2.2em,angle=90]{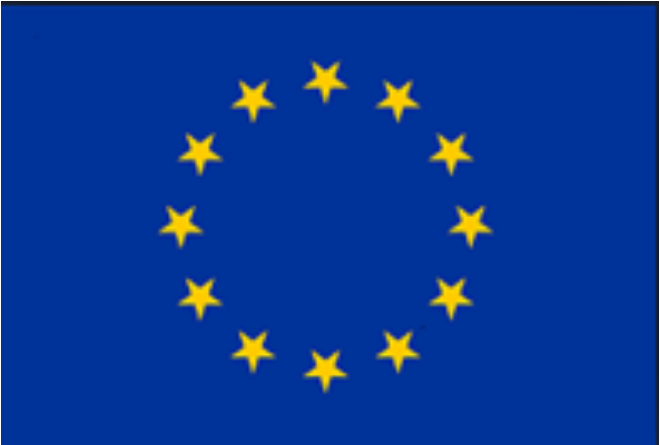}~
\begin{minipage}[b]{0.9\linewidth}
  This project has received funding from the European Union’s Horizon 2020 research and innovation programme under grant agreement No 824093.
\end{minipage}

\nocite{*}
\bibliographystyle{auto_generated}
\bibliography{colferai/proceedings/Colferai}

%% file: Boettcher_Lowx2021_proceedings/proceedings/Boettcher_Lowx2021_proceedings.tex
\vspace*{1.2cm}

\thispagestyle{empty}
\begin{center}
{\LARGE \bf Light hadron and photon production in $p{\rm Pb}$ collisions at LHCb}

\par\vspace*{7mm}\par

{

\bigskip

\large \bf Thomas Boettcher}

on behalf of the LHCb Collaboration

\bigskip

{\large \bf  E-Mail: boettcts@ucmail.uc.edu}

\bigskip

{Department of Physics, University of Cincinnati}

\bigskip

{\it Presented at the Low-$x$ Workshop, Elba Island, Italy, September 27--October 1 2021}

\vspace*{15mm}

\end{center}
\vspace*{1mm}

\begin{abstract}
  
Light hadron and photon production in $p{\rm Pb}$ collisions are
sensitive to nuclear effects on the initial state of the colliding
nucleus. Because of the LHCb detector's forward acceptance,
measurements of hadron and photon production at LHCb provide
information about the partonic structure of the nucleus for $x$ as low
as $10^{-6}$. This report presents measurements of hadron production
and discusses how these measurements constrain nuclear parton
distribution functions at low $x$. Prospects for future measurements
of light hadron and photon production are also discussed.

\end{abstract}
  \part[Light hadron and photon production in $p{\rm Pb}$ collisions at LHCb\\ \phantom{x}\hspace{4ex}\it{Thomas Boettcher on behalf of the LHCb Collaboration}]{}
\section{Introduction}

\begin{figure}[h]
  \centering
  \includegraphics[width=0.8\textwidth]{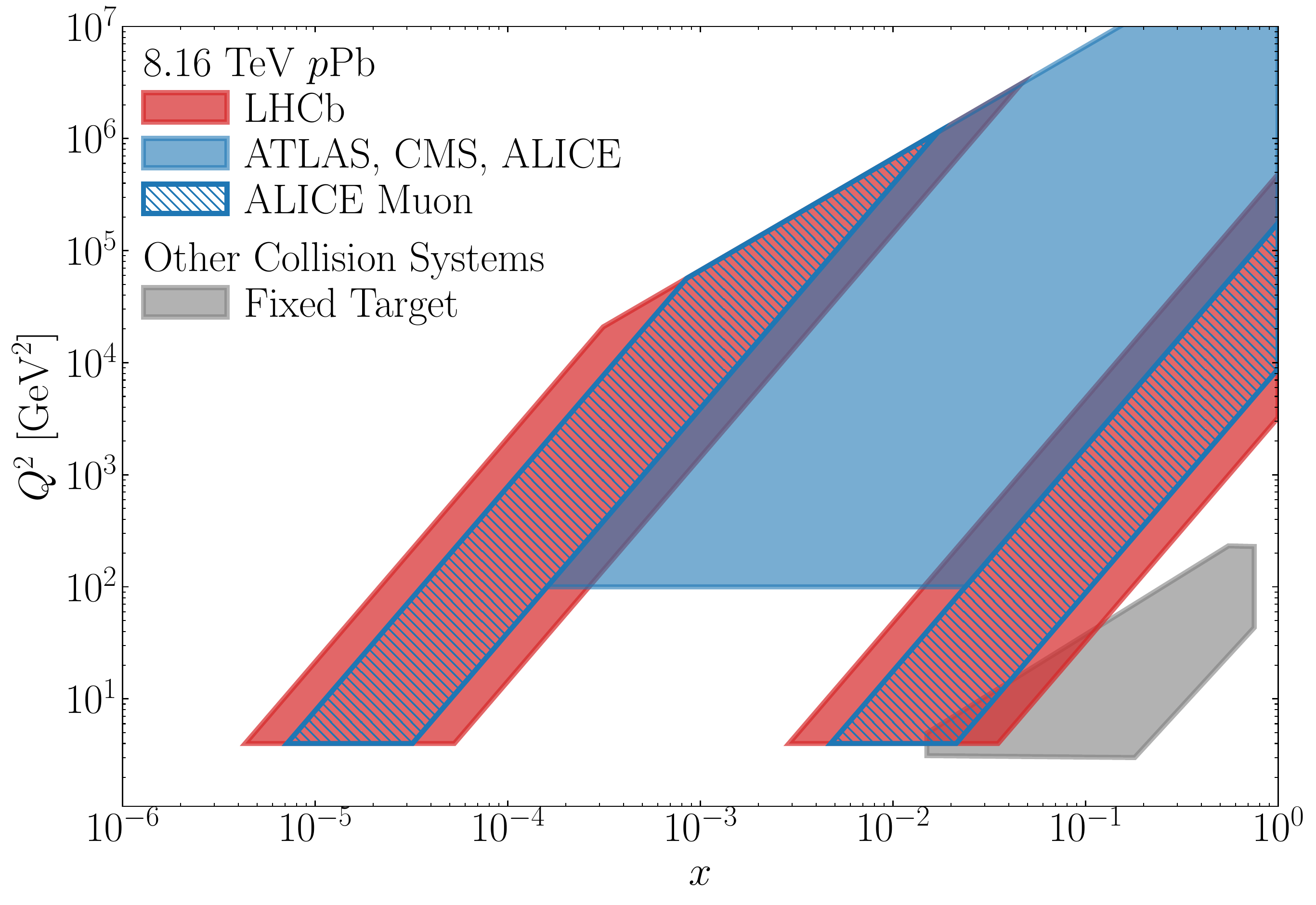}
  \caption{LHCb coverage in the $x$-$Q^2$ plane for $p{\rm Pb}$
    collisions at $\sqrt{s_{\rm NN}}=8.16\,{\rm TeV}$.}
  \label{fig:coverage}
\end{figure}

The LHCb detector is a general purpose detector at forward rapidity at
the LHC~\cite{LHCb:2014set}. Because of its forward acceptance, LHCb
is also able to study the structure of colliding hadrons in a
kinematic regime complementary to that probed by central region
detectors. LHCb's kinematic coverage is shown in
Fig.~\ref{fig:coverage}. Proton-lead collisions at LHCb probe
interactions between low- and high-$x$ partons. Hadron and photon
production in proton-lead collisions at LHCb are sensitive to nuclear
partons with momentum fractions $x$ as small as $10^{-6}$. As a
result, LHCb data can constrain nuclear parton distribution functions
(nPDFs) in the relatively unexplored low-$x$
regime~\cite{Helenius:2014qla}. Data in this regime can also constrain
models of parton saturation, such as the color glass condensate (CGC)
effective field theory~\cite{Lappi:2013zma}. Furthermore, by
alternating the directions of the proton and lead beams, LHCb is able
to collect data at large backward rapidities. The LHCb forward and
backward data cover a large range in $x$ and can be used to constrain
models of particle production in nuclear collisions and to study the
onset of low-$x$ nuclear effects.

\section{$D^0$ meson production}

\begin{figure}[h]
  \centering
  \includegraphics[width=0.48\textwidth]{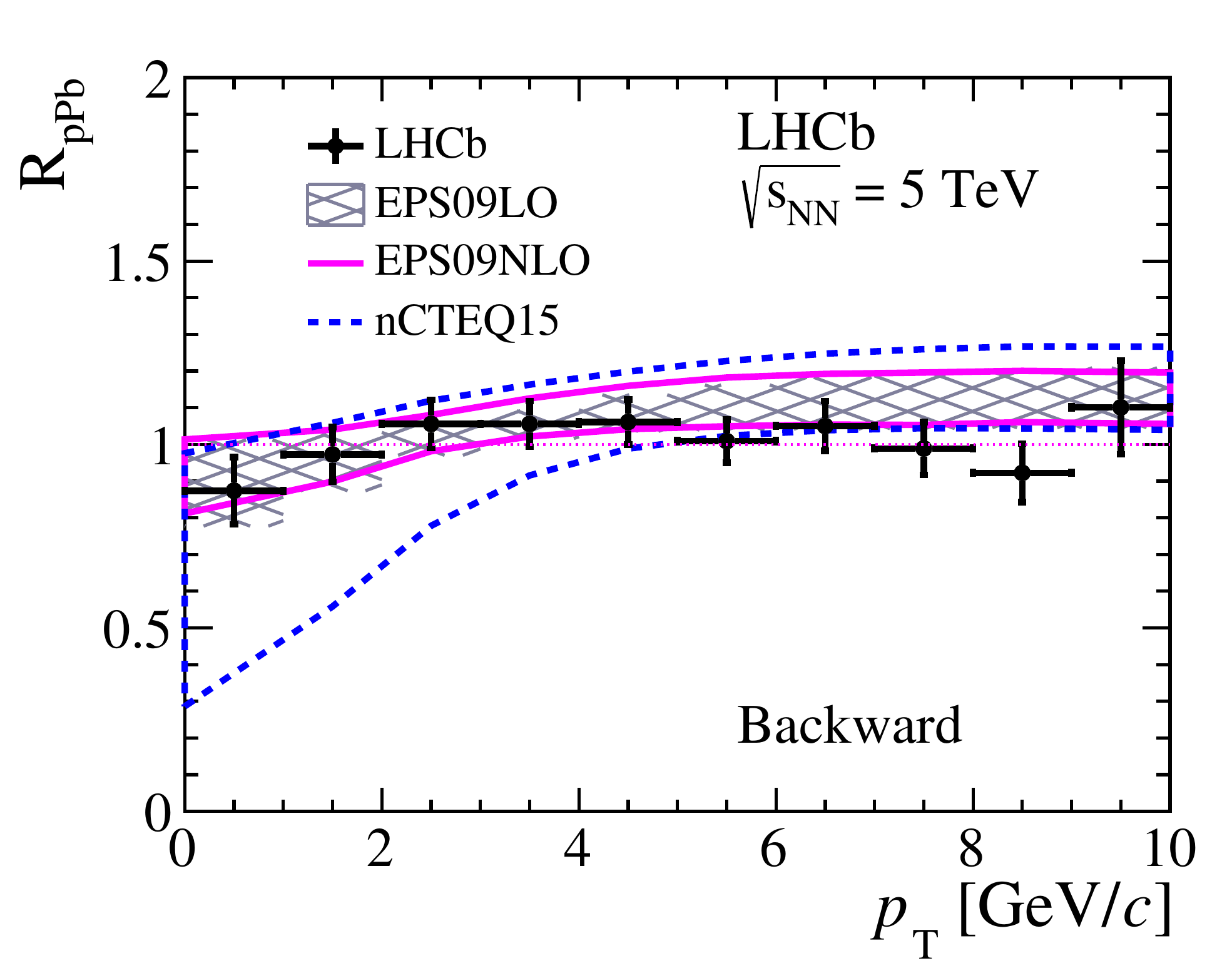}
  \includegraphics[width=0.48\textwidth]{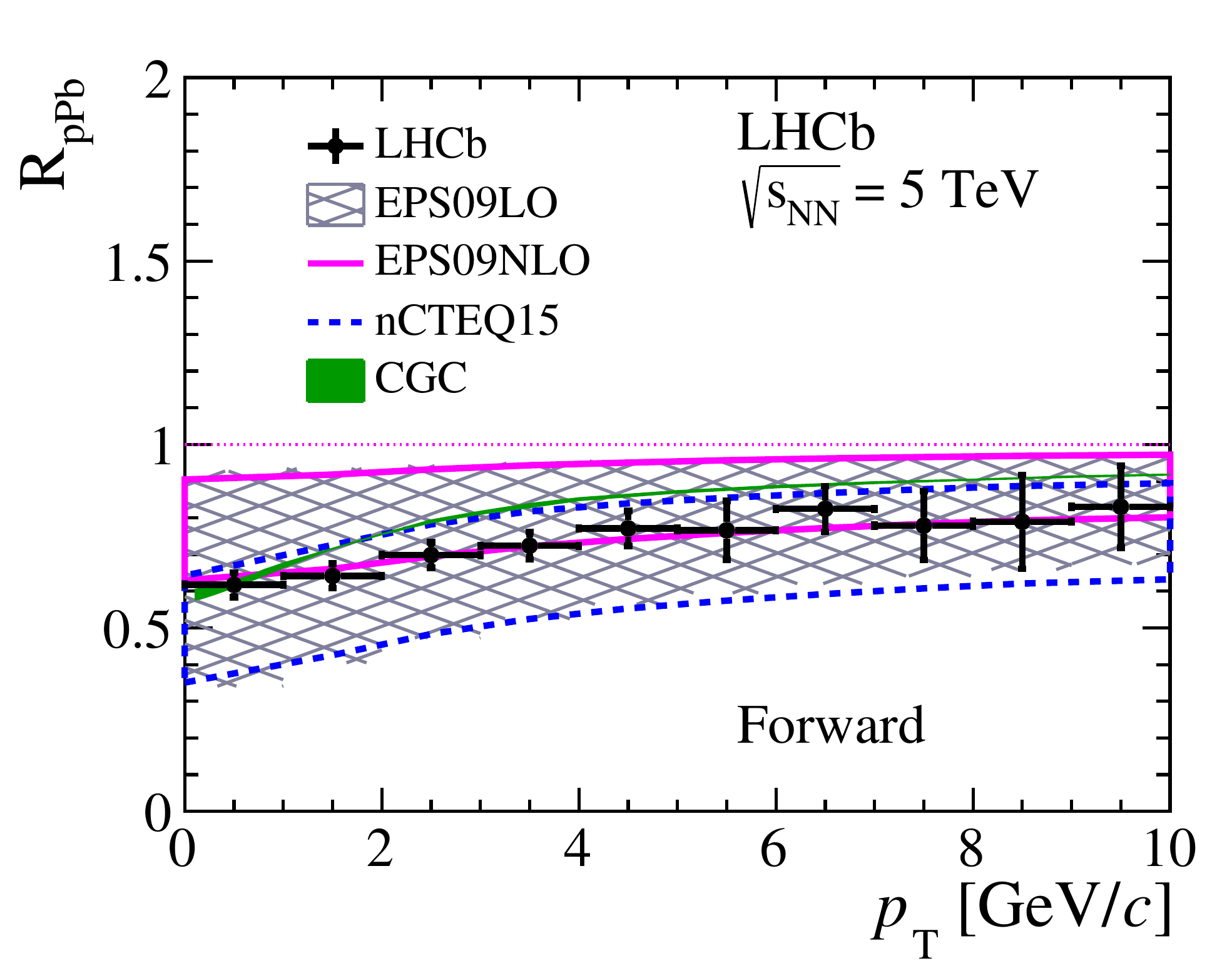}
  \caption{LHCb measurement of the $D^0$ nuclear modification factor
    in $p{\rm Pb}$ collisions at $\sqrt{s_{\rm NN}}=5.02\,{\rm
      TeV}$~\cite{LHCb:2017yua} at (left) forward and (right) backward
    rapidity. The measurement is performed for center of mass rapidity
    $2.5<|y_{\rm CM}|<4.0$ for $p_{\rm T}<6\,{\rm GeV}/c$ and
    $2.5<|y_{\rm CM}|<3.5$ for \mbox{$6<p_{\rm T}<10\,{\rm GeV}/c$}.}
  \label{fig:dprod}
\end{figure}

LHCb has measured $D^0$ meson production in $p{\rm Pb}$ collisions at
nucleon-nucleon center-of-mass energy $\sqrt{s_{\rm NN}}=5.02\,{\rm
  TeV}$~\cite{LHCb:2017yua}. The measured nuclear modification factor
is shown in Fig.~\ref{fig:dprod}. The measurement was performed in
both the proton-going (forward) and lead-going (backward) directions
and is the first measurement of its kind at zero transverse
momentum. As a result, the LHCb $D^0$ production measurement is
particularly sensitive to nPDFs at low $x$ and low $Q^2$ and is
potentially sensitive to saturation effects. Results are compared to
perturbative QCD (pQCD) calculations using the
EPS09~\cite{Eskola:2009uj} and nCTEQ15~\cite{Kovarik:2015cma} nPDF
sets. The LHCb results agree with the nPDF predictions in both the
forward and backward directions. Additionally, the forward result is
also compared to a CGC calculation, showing good
agreement~\cite{Ducloue:2015gfa}. The forward measurement in
particular is much more precise than the nPDF prediction, suggesting
that this measurement can help constrain nPDFs at low $x$.

The impact of LHCb's measurement of $D^0$ production on nPDFs has been
studied in Refs.~\cite{Kusina:2017gkz} and \cite{Eskola:2019bgf}. Both
analyses incorporate LHCb data using a Hessian reweighting
technique~\cite{Eskola:2019dui}. Incorporating the LHCb $D^0$
production data results in much smaller gluon nPDF uncertainties,
especially at low $x$. These analyses demonstrate LHCb's unique
ability to study nuclear structure at low $x$.

\section{Charged hadron production}

\begin{figure}[h]
  \centering
  \includegraphics[width=\textwidth]{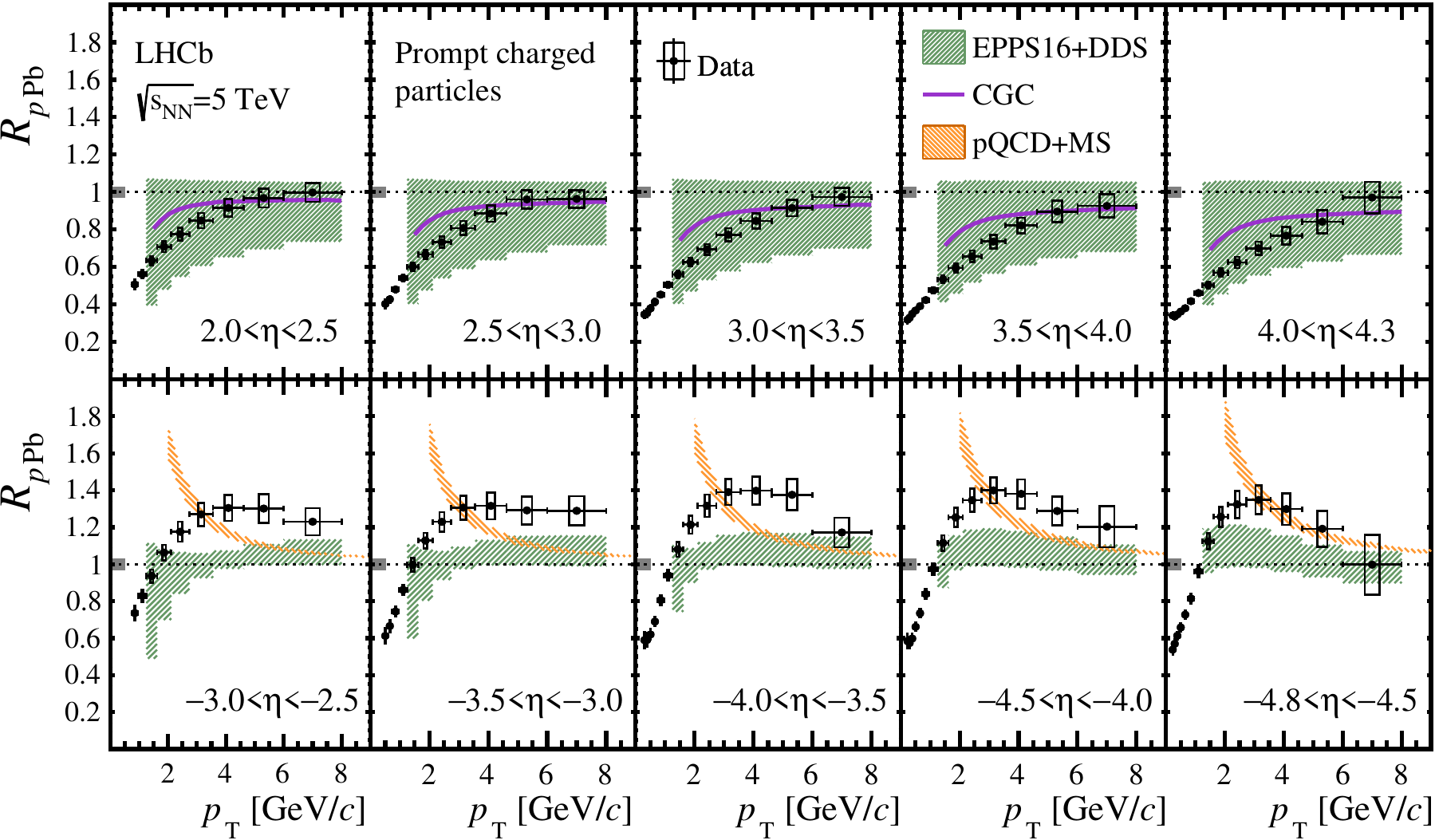}
  \caption{LHCb measurement of the charged-particle nuclear
    modification factor in $p{\rm Pb}$ collisions at $\sqrt{s_{\rm
        NN}}=5.02\,{\rm TeV}$~\cite{LHCb:2021vww}.}
  \label{fig:chg}
\end{figure}

Measurements of $D^0$ production are not sensitive to nPDFs at
momentum scales $Q$ below the $D^0$ mass. Measurements of inclusive
charged particle production can be used to access lower scales. LHCb
has measured the nuclear modification factor of charged particles in
$p{\rm Pb}$ collisions at $\sqrt{s_{\rm NN}}=5.02\,{\rm
  TeV}$~\cite{LHCb:2021vww}. Results are shown in
Fig.~\ref{fig:chg}. The measurement is performed as a function of both
$p_{\rm T}$ and center-of-mass pseudorapidity $\eta$. Total
uncertainties of less than $5\%$ are achieved in most bins.

Results are compared to next-to-leading order (NLO) pQCD calculations
using the EPPS16 nPDF set~\cite{Eskola:2016oht} and DSS14
fragmentation functions~\cite{deFlorian:2014xna}. In addition, the
forward data are compared to CGC calculations~\cite{Lappi:2013zma}. At
forward $\eta$, the data agrees with the pQCD prediction and is much
more precise than the theoretical calculation. The data also shows
greater suppression at low $p_{\rm T}$ than predicted by the CGC
calculation. At backward $\eta$, the data shows a much larger
enhancement than predicted by the nPDF calculation. The backward data
are also compared to a pQCD calculation that includes fully coherent
energy loss in the nucleus~\cite{Kang:2013ufa}. This model
successfully describes a similar enhancement observed in $p{\rm Au}$
collisions at $\sqrt{s_{\rm NN}}=200\,{\rm GeV}$ by the PHENIX
collaboration~\cite{PHENIX:2019gix}, but fails to describe the LHCb
data. The failure of these calculations to describe the LHCb data
suggests contributions from other nuclear effects, such as the Cronin
effect, radial flow, or final state
recombination~\cite{Hwa:2004zd}. Measurements of the nuclear
modification factors of identified particles are needed to determine
the origin of the enhancement.

\begin{figure}[h]
  \centering
  \includegraphics[width=\textwidth]{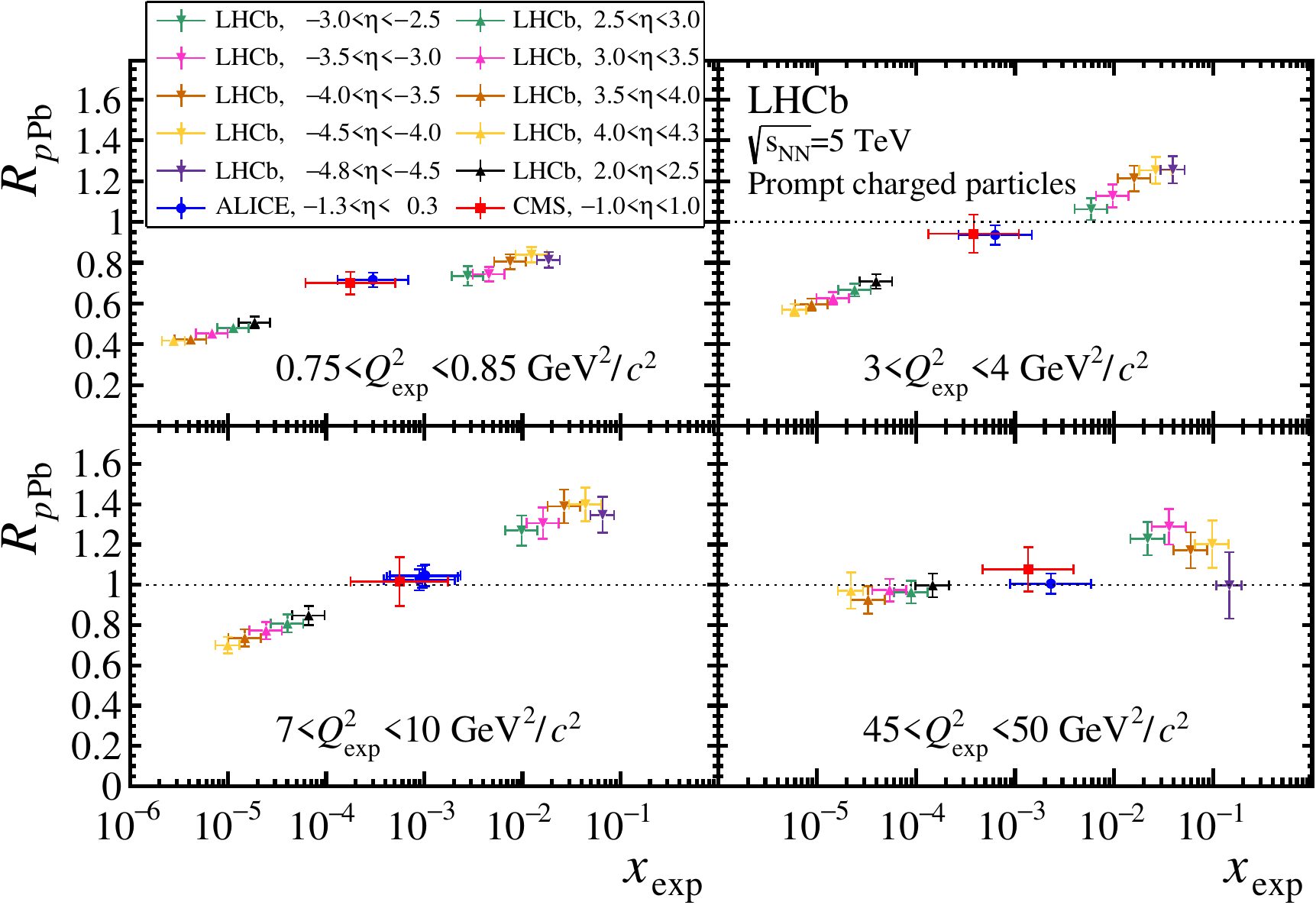}
  \caption{Comparison of LHCb~\cite{LHCb:2021vww},
    ALICE~\cite{ALICE:2018vuu}, and CMS~\cite{CMS:2016xef}
    charged-particle nuclear modification factors as a function of
    $x_{\rm exp}$ for various values of $Q^2_{\rm exp}$.}
  \label{fig:chgcomp}
\end{figure}

The LHCb charged particle production data are compared to measurements
from ALICE~\cite{ALICE:2018vuu} and CMS~\cite{CMS:2016xef} using
approximations of $x$ and $Q^2$ given by
\begin{align}
  Q^2_{\rm exp}&=m^2+p_{\rm T}^2,\\
  x_{\rm exp}&=\frac{Q_{\rm exp}}{\sqrt{s_{\rm NN}}}e^{-\eta_{\rm CM}},
\end{align}
where $m=256~{\rm MeV}/c^2$ is the average mass of charged particles
in $p{\rm Pb}$ collisions generated by EPOS-LHC. The comparison is
shown in Fig.~\ref{fig:chgcomp}. The results show continuous evolution
in $x_{\rm exp}$ at various values of $Q^2_{\rm exp}$ across multiple
experiments.

\section{Neutral pion and direct photon production}

Inclusive charged particles and $\pi^0$s share similar production
processes and probe similar kinematic regimes. Additionally, $\pi^0$s
are reconstructed using their decays to two photons, so many of the
systematic uncertainties will be independent from systematic
uncertainties on charged particle production. As a result, a
measurement of $\pi^0$ production at LHCb would provide a
complementary probe of nPDFs at low $x$.

Identified particle measurements could also help explain the origin of
the charged-particle excess observed by LHCb in $p{\rm Pb}$ collisions
at $\sqrt{s_{\rm NN}}=5.02\,{\rm TeV}$. Radial flow, for example,
would produce a larger charged particle enhancement than $\pi^0$
enhancement due to the small $\pi^0$
mass~\cite{Ayala:2006bc}. Alternatively, enhanced nuclear parton
densities would produce similar charged particle and $\pi^0$
enhancements~\cite{Helenius:2014qla}.

\begin{figure}[h]
  \centering
  \includegraphics[width=0.8\textwidth]{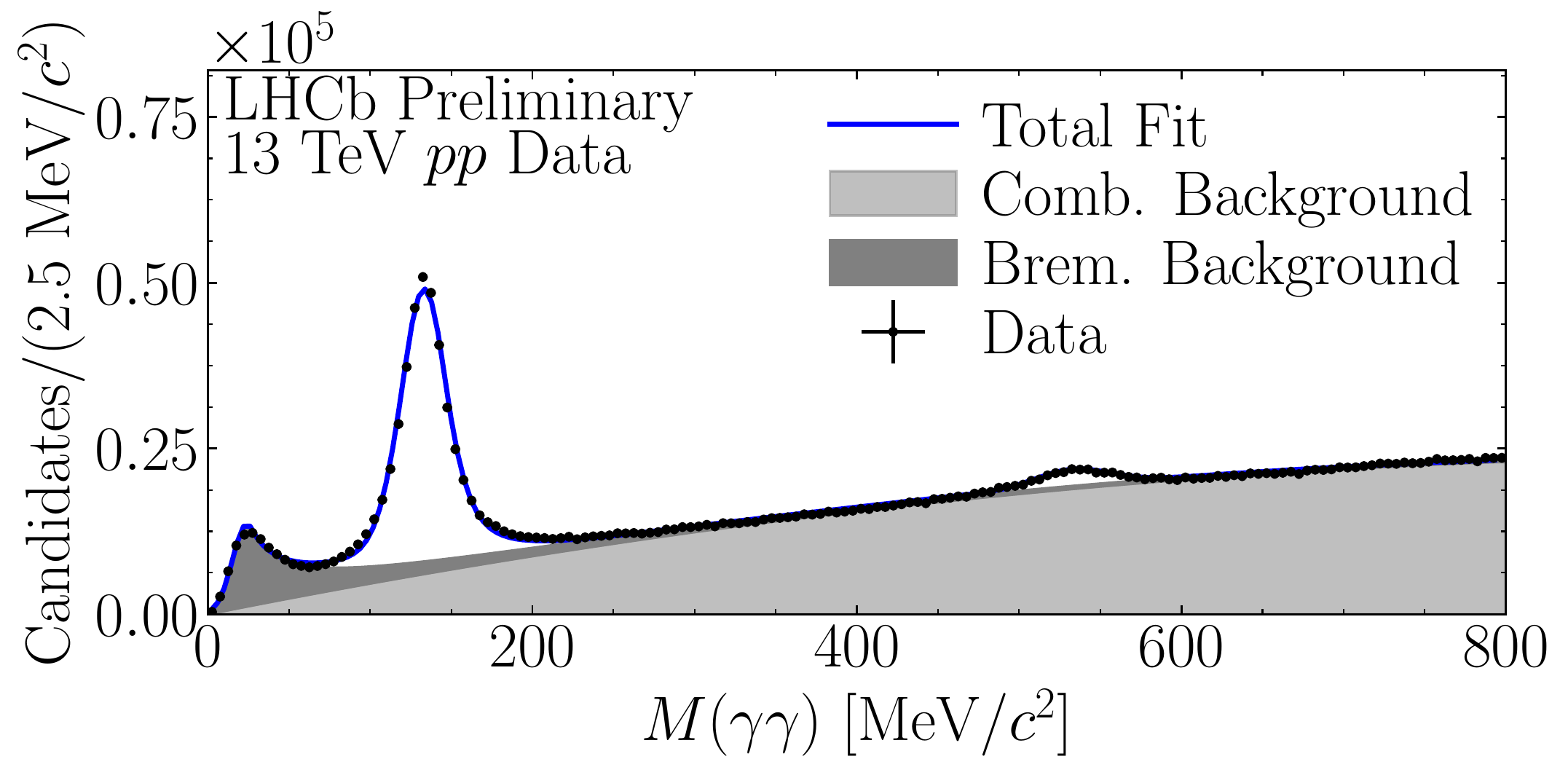}
  \caption{Fit of the diphoton mass spectrum measured at LHCb in $pp$
    collisions at $\sqrt{s}=13\,{\rm TeV}$. The total fit is given by
    the solid blue line. The combinatorial background is shown in
    light gray and the bremsstrahlung background is shown in dark
    gray.}
  \label{fig:fit}
\end{figure}

A measurement of the $\pi^0$ nuclear modification factor in $p{\rm
  Pb}$ collisions at $\sqrt{s_{\rm NN}}=8.16\,{\rm TeV}$ is in
progress at LHCb. Photon pairs from high-$p_{\rm T}$ forward $\pi^0$
decays are often reconstructed as single calorimeter clusters. To
avoid these so-called ``merged $\pi^0$s'', $\pi^0$s are reconstructed
using photons that convert to electron-positron pairs in the detector
material. Converted photons are combined with photons reconstructed in
the ECAL to produce $\pi^0$ candidates. Yields are extracted using
fits to the diphoton mass spectrum. An example fit is shown in
Fig.~\ref{fig:fit}. Backgrounds include combinatorial background, as
well as bremsstrahlung background consisting of converted photons
combined with the bremsstrahlung radiation from one of the conversion
electrons.

\begin{figure}[h]
  \centering
  \includegraphics[width=\textwidth]{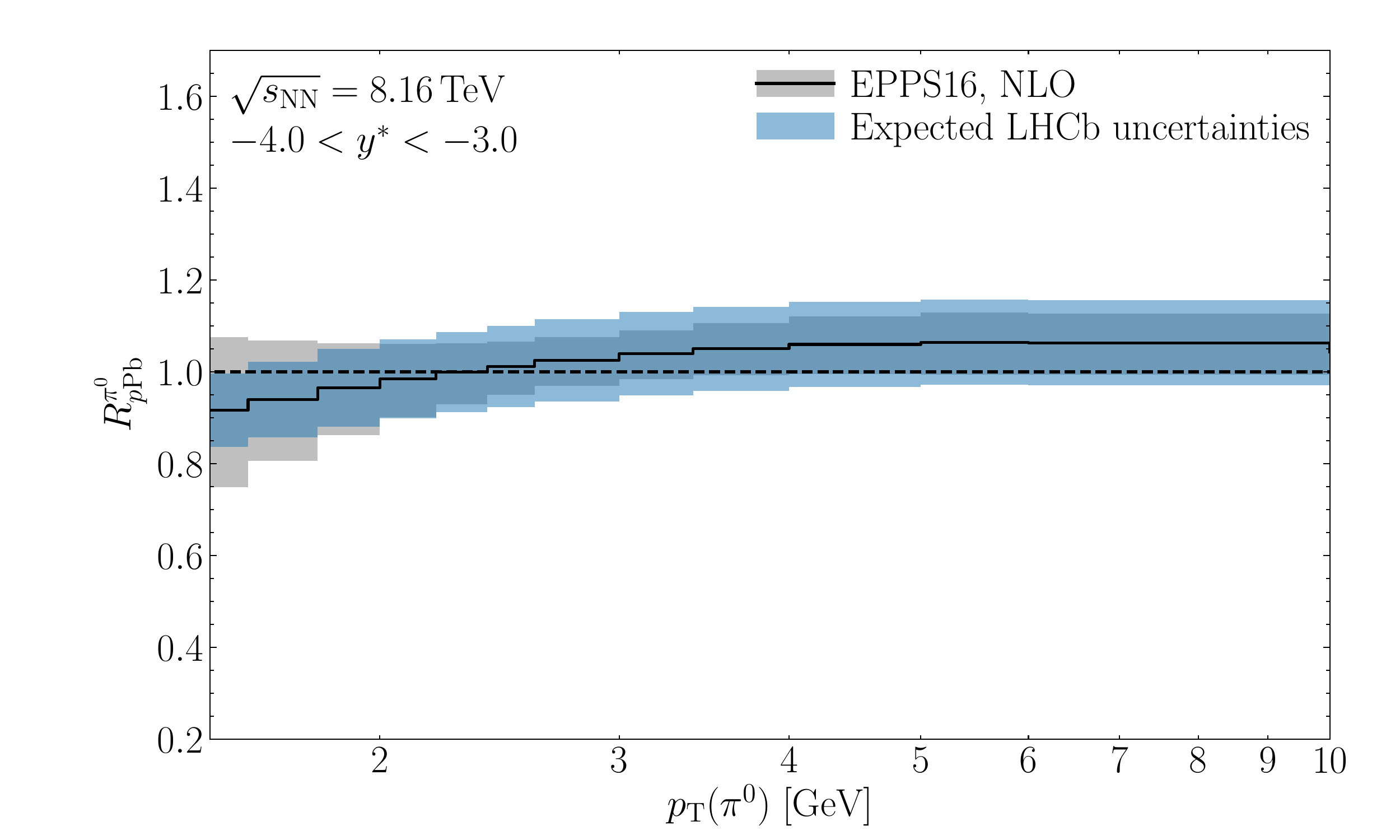}

  \includegraphics[width=\textwidth]{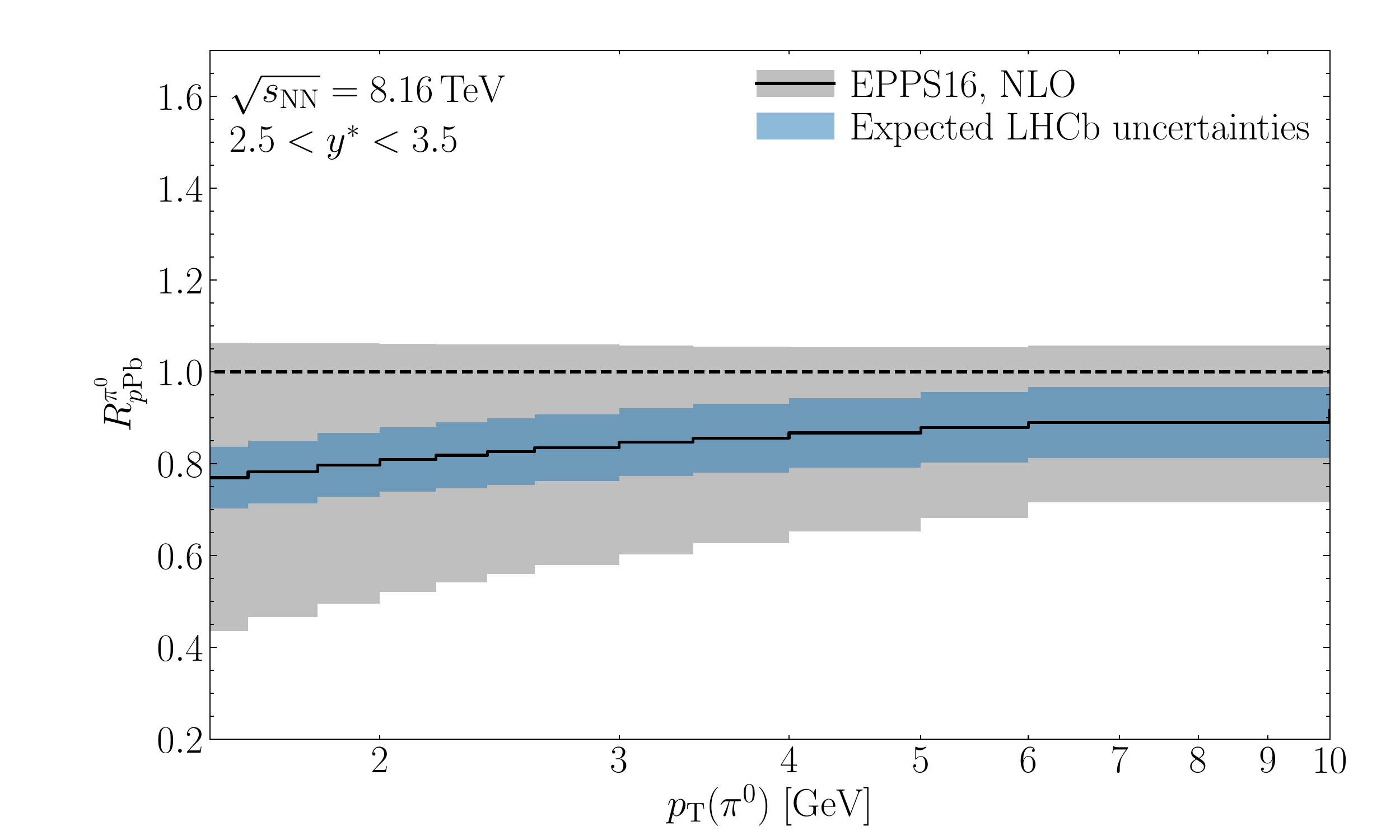}

  \caption{Next-to-leading-order pQCD calculations of the $\pi^0$
    nuclear modification factor~\cite{Helenius:2014qla} and expected
    LHCb uncertainties at (top) backward and (bottom) forward rapidity
    and $\sqrt{s_{\rm NN}}=8.16\,{\rm TeV}$.}
  \label{fig:rpafake}
\end{figure}

Fig.~\ref{fig:rpafake} shows NLO pQCD predictions for the $\pi^0$
nuclear modification factor~\cite{Helenius:2014qla}, as well as
expected LHCb uncertainties. These uncertainties are dominated by the
photon detection efficiency determination. The LHCb measurement is
expected to be much more precise than the nPDF calculation in the
forward region and could provide strong constraints at low $x$ in
future global nPDF fits.

Measuring the $\pi^0$ $p_{\rm T}$ spectrum is also necessary for
direct photon searches. Direct photons are photons that are not
produced in meson decays. They may be produced promptly in hard QCD
processes or via fragmentation in the parton shower. Direct photon
production at forward rapidity is an ideal probe of gluon
saturation. Furthermore, if a collision produces quark gluon plasma
(QGP), the QGP will radiate low-$p_{\rm T}$ thermal photons. Because
most photons produced in hadron collisions come from $\pi^0$ decays,
precise knowledge of the $\pi^0$ $p_{\rm T}$ spectrum is needed in
order to subtract backgrounds from decay photons. As a result, a
measurement of $\pi^0$ production at LHCb will provide the basis for
future direct photon searches.

\section{Conclusions}

The LHCb detector's unique angular acceptance allows it to study
nuclear effects in unexplored kinematic regimes. LHCb charm production
data has already been used to constrain nPDFs, especially at
low-$x$. Measurements of light hadron and photon production can
provide additional constraints at low $Q^2$, potentially probing the
effects of parton saturation. Additionally, LHCb sees a large
enhancement of charged-particle production in $p{\rm Pb}$ collisions
at backward rapidity that is not adequately explained by available
theoretical calculations. A measurement of $\pi^0$ production at LHCb
is underway and could clarify the origin of this enhancement.

Comments: Presented at the Low-$x$ Workshop, Elba Island, Italy, September 27--October 1 2021. 
 
\section*{Acknowledgements}

The author is supported by the U.S. National Science Foundation.
\nocite{*}
\bibliographystyle{auto_generated}
\bibliography{Boettcher_Lowx2021_proceedings/proceedings/Boettcher_Lowx2021_proceedings}

%% file: proceeding_Lowx_2021_Celiberto/proceeding_Lowx_2021_Celiberto.tex
\vspace*{1.2cm}

\thispagestyle{empty}
\begin{center}
{\LARGE \bf Phenomenology of the
hadronic structure at small-$x$}

\par\vspace*{7mm}\par

{

\bigskip

\large \bf Francesco Giovanni Celiberto}

\bigskip

{\large \bf  E-Mail: fceliberto@ectstar.eu}

\bigskip

{
European Centre for Theoretical Studies in Nuclear Physics and Related Areas (ECT*),
\\
I-38123 Villazzano, Trento, Italy
\vskip .18cm
Fondazione Bruno Kessler (FBK)
I-38123 Povo, Trento, Italy
\vskip .18cm
INFN-TIFPA Trento Institute of Fundamental Physics and Applications
\\
I-38123 Povo, Trento, Italy
}

\bigskip

{\it Presented at the Low-$x$ Workshop, Elba Island, Italy, September 27--October 1 2021}

\vspace*{15mm}

\end{center}
\vspace*{1mm}

\begin{abstract}

We present exploratory studies of the proton structure via two distinct kinds of gluon densities: the transverse-momentum dependent functions, whose evolution is determined by the CSS equation, and the unintegrated gluon distribution, whose definition is given in the BFKL formalism.
Our analyses are relevant for the exploration of the intrinsic motion of gluons inside protons in the small-$x$ regime, and for a comprehensive tomographic imaging of the proton at new-generation colliding facilities.
\end{abstract}
 \part[Phenomenology of the hadronic structure at small-$x$\\ \phantom{x}\hspace{4ex}\it{Francesco Giovanni Celiberto}]{} 

\vspace{-0.20cm}
\section{Introduction}
\label{sec:intro_celib}
\vspace{-0.20cm}

Unraveling the inner structure of hadrons through a multi-dimensional study of their constituents represents a frontier research of phenomenological studies at new-generation colliding facilities.
The well-established collinear factorization that relies on a one-dimensional description of the proton content via collinear parton distribution functions (PDFs) has collected a long chain of achievements in describing data at hadronic and lepton-hadronic accelerators.

There are however vital questions on the dynamics of strong interactions which still do not have an answer.
Unveiling the origin of spin and mass of the nucleons requires a stretch of our vision from the collinear description to a tomographic viewpoint in three dimensions, naturally afforded by the \emph{transverse-momentum-dependent} (TMD) factorization.

In the small-$x$ regime a purely TMD-based approach may be not adequate, since large contributions proportional to $\ln (1/x)$ enter the perturbative series with a power that grows with the order, and need to be accounted for via an all-order resummation procedure.
The most powerful tool to resum those large logarithms is the Balitsky--Fadin--Kuraev--Lipatov (BFKL) formalism~\cite{Fadin:1975cb,Kuraev:1976ge,Kuraev:1977fs,Balitsky:1978ic} in the leading approximation (LL$x$), which means inclusion of all terms proportional to $\alpha_s^n \ln (1/x)^n$, and in the next-to-leading approximation (NLL$x$), including all terms proportional to $\alpha_s^{n+1} \ln (1/x)^n$.

In this paper we report progresses on the study of the proton structure at small-$x$ via two distinct kinds of gluon distributions.
In Section~\ref{sec:TMDs} we present the main features of a quite recent calculation of all (un)polarized gluon TMD distributions at leading twist, whose definition genuinely embodies BFKL effects.
In Section~\ref{sec:UGD} we provide with an evidence that helicity amplitudes for the exclusive leptoproduction of $\rho$-mesons act as discriminators for the existing models of the BFKL \emph{unintegrated gluon distribution} (UGD).

We come out with the message that forthcoming analyses at new-generation colliders will help to shed light on the hadronic structure at small-$x$ and to explore the interplay between different formalisms, such as the TMD factorization and the BFKL resummation.

\vspace{-0.20cm}
\section{Small-$x$ improved transverse-momentum dependent gluon distributions}
\label{sec:TMDs}
\vspace{-0.20cm}

The complete list of unpolarized and polarized gluon TMDs at leading twist (twist-2) was given for the first time in Ref.~\cite{Mulders:2000sh}. In Tab.~\ref{tab:gluon_TMDs} we present the eight twist-2 gluon TMDs for a spin-1/2 target, using the nomenclature convention as in Refs.~\cite{Meissner:2007rx,Lorce:2013pza}.
The two functions on the diagonal in Tab.~\ref{tab:gluon_TMDs} respectively represent the density of unpolarized gluons inside an unpolarized nucleon, $f_1^g$, and of circularly polarized gluons inside a longitudinally polarized nucleon, $g_1^g$.
In the collinear regime they correspond to the well-known unpolarized and helicity gluon PDFs.

TMD distributions receive contributions from the resummation of transverse-momentum logarithms which enter perturbative calculations. Much is know about this resummation~\cite{Bozzi:2003jy,Catani:2010pd,Echevarria:2015uaa}, but very little is known about the genuinely non-perturbative TMD content.
The distribution of linearly polarized gluons in an unpolarized hadron, $h_1^{\perp g}$, is particularly relevant, since it gives rise to spin effects in collisions of unpolarized hadrons~\cite{Boer:2010zf,Sun:2011iw,Boer:2011kf,Pisano:2013cya,Dunnen:2014eta,Lansberg:2017tlc}, whose size is expected to increase at small-$x$ values.
The Sivers function, $f_{1T}^{\perp g}$, gives us information about the density of unpolarized gluons in a transversely polarized nucleon, and plays a key role in the description of transverse-spin asymmetries that can be studies in collisions with polarized-proton beams.
Notably, in Ref.~\cite{Boussarie:2019vmk} it was argued that in the forward limit the Sivers function can be accessed in unpolarized electron-nucleon collisions via its connection with the QCD Odderon.

At variance with collinear distributions, TMDs are process-dependent via the \emph{gauge links} (or \emph{Wilson lines})~\cite{Brodsky:2002cx,Collins:2002kn,Ji:2002aa}.
Quark TMDs depend on the $[+]$ and $[-]$ staple links, which set the direction of future- and past-pointing Wilson lines, respectively. 
Gluon TMDs exhibit a more involved gauge-link dependence, since they are sensitive on combinations of two or more staple links. This brings to a more diversified kind of \emph{modified universality}.

Two major gluon gauge links appear: the $f\text{-type}$ and the $d\text{-type}$ ones. They are also known in the context of small-$x$ studies as Weisz\"acker--Williams and dipole structures, respectively.
The antisymmetric $f_{abc}$ QCD color structure appears in the $f$-type $T$-odd gluon-TMD correlator, whereas the symmetric $d_{abc}$ structure characterizes the $d$-type $T$-odd one. This fact leads to a dependence of $f$-type gluon TMDs on the $[\pm,\pm]$ gauge-link combinations, while $d$-type gluon TMDs are characterized by the $[\pm,\mp]$ ones.
Much more complicate, box-loop gauge links emerge in processes where multiple color exchanges connect both initial and final state states~\cite{Bomhof:2006dp}. This leads to a violation of the TMD factorization~\cite{Rogers:2010dm} (see also Ref.~\cite{Rogers:2013zha}).

{
\renewcommand{\arraystretch}{1.7}

 \begin{table}
\centering
 \hspace{1cm} gluon pol. \\ \vspace{0.1cm}
 \rotatebox{90}{\hspace{-1cm} nucleon pol.} \hspace{0.1cm}
 \begin{tabular}[c]{|m{0.5cm}|c|c|c|}
 \hline
 & $U$ & circular & linear \\
 \hline
 $U$ & $f_{1}^{g}$ & & \textcolor{blue}{$h_{1}^{\perp g}$} \\
 \hline	
 $L$ & & $g_{1}^{g}$ & \textcolor{red}{$h_{1L}^{\perp g}$} \\
 \hline	
 $T$ & \textcolor{red}{$f_{1T}^{\perp g}$} & \textcolor{blue}{$g_{1T}^{g}$} & \textcolor{red}{$h_{1}^{g}$}, \textcolor{red}{$h_{1T}^{\perp g}$} \\
 \hline
  \end{tabular}
 \caption{A table of leading-twist gluon TMDs for spin-$1/2$ targets. 
 $U$, $L$, $T$ stand for unpolarized, longitudinally polarized and transversely polarized hadrons, whereas
 $U$, `circular', `linear' depict unpolarized, circularly polarized and linearly polarized gluons, respectively. 
 $T$-even (odd) functions are given in blue (red). 
 Black functions are $T$-even and survive the integration over the gluon transverse momentum.}
 \label{tab:gluon_TMDs}
 \end{table}

}

From the experimental point of view, the gluon-TMD sector is a largely unexplored field. 
First attempts at phenomenological analyses of the unpolarized gluon function have been presented in Refs.~\cite{Lansberg:2017dzg,Gutierrez-Reyes:2019rug,Scarpa:2019fol}. Phenomenological studies of the gluon Sivers TMD can be found in Refs.~\cite{Adolph:2017pgv, DAlesio:2017rzj,DAlesio:2018rnv,DAlesio:2019qpk}.
Therefore, exploratory analyses of gluon TMDs via simple and flexible models are needed. Pioneering studies along this direction were carried out in the so-called \emph{spectator-model} framework~\cite{Lu:2016vqu,Mulders:2000sh,Pereira-Resina-Rodrigues:2001eda}.
Formerly employed in the description of quark TMD distributions~\cite{Bacchetta:2008af,Bacchetta:2010si,Gamberg:2005ip,Gamberg:2007wm,Jakob:1997wg,Meissner:2007rx}, it is based on the assumption that the struck hadron with mass $\cal M$ and momentum $\cal P$ emits a gluon with longitudinal fraction $x$, momentum $p$, and transverse momentum $\boldsymbol{p}_T$, and what remains is treated as an effective on-shell particle having ${\cal M}_X$ and spin-1/2.
Within this model taken at tree level all the leading-twist TMDs in Table~\ref{tab:gluon_TMDs} can be calculated. Spectator-model gluon $T$-even densities were recently calculated in Ref.~\cite{Bacchetta:2020vty} and presented in Refs.~\cite{Celiberto:2021zww,Bacchetta:2021oht}.
In those works the nucleon-gluon-spectator vertex was modeled as follows
\begin{equation}
 \label{eq:form_factor}
 {\cal G}^{\, \mu} = \left( \tau_1(p^2) \, \gamma^{\, \mu} + \tau_2(p^2) \, \frac{i}{2{\cal M}} \sigma^{\, \mu\nu}p_\nu \right) \,,
\end{equation}
the $\tau_1$ and $\tau_2$ functions being dipolar form factors in $\boldsymbol{p}_T^2$. A dipolar profile for the couplings is useful to fade gluon-propagator divergences, quench large-$\boldsymbol{p}_T$ effects which are beyond the reach of a pure TMD description, and remove logarithmic singularities coming from $\boldsymbol{p}_T$-integrated distributions.
Furthermore, the spectator mass was allowed to take a continuous range of values weighed by a spectral function ${\cal S}_{\rm } ({\cal M}_X)$, which provides the necessary flexibility to reproduce both the small- and the moderate-$x$ shape of gluon collinear PDFs.
The analytic expression of the spectral function contains seven parameters and reads
\begin{equation}
\label{eq:rhoX}
 {\cal S}_{\rm } ({\cal M}_X) = \mu^{2a} \left( \frac{A}{B + \mu^{2b}} + \frac{C}{\pi \sigma} e^{-\frac{({\cal M}_X - D)^2}{\sigma^2}} \right) \,.
\end{equation}
Model parameters were fixed by performing a simultaneous fit of our unpolarized and helicity TMDs, $f_1^g$ and $g_1^g$, to the corresponding collinear PDFs from {\tt NNPDF}~\cite{Ball:2017otu,Nocera:2014gqa} at the initial scale $Q_0 = 1.64$ GeV. The statistical uncertainty of the fit was obtained via the widely known bootstrap method.
We refer to Ref.~\cite{Bacchetta:2020vty} for details on the fitting procedure and quality.
We stress that since our tree-level approximation does not take into account the gauge link, our model $T$-even TMDs are process-independent.
Preliminary results for spectator-model $T$-odd TMDs at twist-2 and their dependence on the gauge link can be found in Refs.~\cite{Bacchetta:2021lvw,Bacchetta:2021twk,Bacchetta:2022esb}.

Pursuing the goal of shedding light on the the full 3D dynamics of gluons inside the proton, we consider the following densities which describe the 2D $\boldsymbol{p}_T$-distribution of gluons for different combinations of their polarization and the proton spin. For an unpolarized proton, we identify the unpolarized density 
\begin{equation}
 x \rho (x, p_x, p_y) = x f_1^g (x, \boldsymbol{p}_T^2) 
\label{eq:rho_unpol}
\end{equation}
as the probability density of finding unpolarized gluons at given $x$ and $\boldsymbol{p}_T$, while the Boer--Mulder distribution 
\begin{equation}
 x \rho^{\leftrightarrow} (x, p_x, p_y) = \frac{1}{2} \bigg[ x f_1^g (x, \boldsymbol{p}_T^2) + \frac{p_x^2 - p_y^2}{2 M^2} \, x h_1^{\perp g} (x, \boldsymbol{p}_T^2) \bigg]
\label{eq:rho_T}
\end{equation}
represents the probability density of finding linearly-polarized gluons in the transverse plane at $x$ and $\boldsymbol{p}_T$.

\begin{figure}[tb]
 \centering
 \includegraphics[scale=0.22,clip]{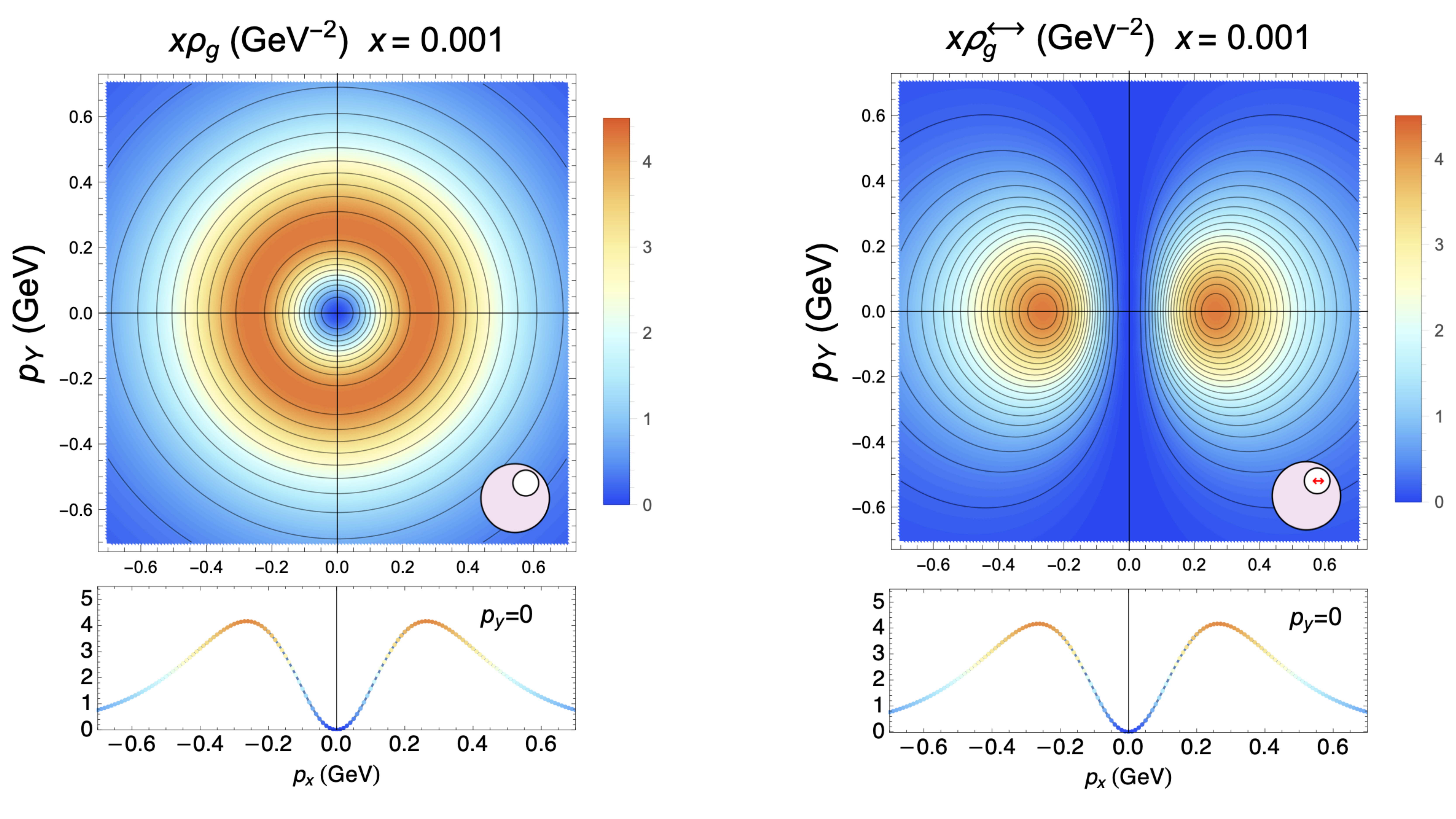}

 \caption{
 3D tomographic imaging of the proton  unpolarized (left) and Boer--Mulders (right) gluon TMD densities as functions of the gluon transverse momentum, for $x = 10^{-3}$ and at the initial energy scale, $Q_0 = 1.64$ GeV. 1D ancillary panels below main contour plots show the density at $p_y = 0$.
 Figures from~\cite{Bacchetta:2020vty}.
 }
 \label{fig:gluon_TMDs}
\end{figure}

Contour plots in Fig.~\ref{fig:gluon_TMDs} show $\boldsymbol{p}_T$-shape of the $\rho$-distributions in Eqs.~\eqref{eq:rho_unpol} and~\eqref{eq:rho_T}, respectively, obtained at $Q_0 = 1.64$ GeV and $x=10^{-3}$ for an unpolarized proton virtually moving towards the reader. The color code quantifies the size of the oscillation of each distribution along the $p_x$ and $p_y$ directions. To better catch these oscillations, ancillary 1D plots representing the corresponding density at $p_y = 0$ are shown below each contour plot. As expected, the density of Eq.~\eqref{eq:rho_unpol} exhibits a cylindrical symmetry around the direction of motion of the proton pointing towards the reader. Since the nucleon is unpolarized but the gluons are linearly polarized along the $p_x$ direction, the Boer--Mulders $\rho$-density in Eq.~\eqref{eq:rho_T} shows a dipolar structure. The departure from the cylindrical symmetry is emphasized at small-$x$, because the Boer--Mulders function is particularly large. 
From the analytic point of view, one has that the ratio between $f_1^g$ and $h_1^{\perp g}$ TMDs goes to a constant in the asymptotic small-$x$ limit, $x \to 0^+$.
This is in line with the prediction coming from the linear BFKL evolution, namely that at low-$x$ the "number" of unpolarized gluons is equal to the number of linearly-polarized ones, up to higher-twist effects (see, \emph{e.g.}, Refs.~\cite{Dominguez:2011br,Marquet:2016cgx,Taels:2017shj,Marquet:2017xwy,Petreska:2018cbf}). 
Thus, a connection point between our model gluon TMDs and the high-energy dynamics has been discovered.

\vspace{-0.20cm}
\section{Unintegrated gluon distribution}
\label{sec:UGD}
\vspace{-0.20cm}

The BFKL approach~\cite{Fadin:1975cb,Kuraev:1977fs,Balitsky:1978ic} affords us a factorized formula for scattering amplitudes (and thence, in inclusive reactions, for cross sections) given as a convolution of the universal BFKL Green's function and two process-dependent impact factors describing the transition from each initial-state particle to the corresponding detected object.

The connection between the UGD and gluon TMDs is still largely uncharted. From a formal perspective, the TMD formalism relies on parton correlators and thus on Wilson lines, whereas the BFKL approach ``speaks" the language of Reggeized gluons. From a phenomenological viewpoint, TMD factorization is expected to hold at low transverse momenta, whereas the BFKL resummation requires large transverse-momentum emissions.
A first connection between the UGD and the unpolarized and linearly polarized gluon TMDs, $f^g_1$ and $h^{\perp g}_1$,
was investigated in Refs.~\cite{Dominguez:2011wm,Hentschinski:2021lsh,Nefedov:2021vvy}.

The first class of processes that serves as probe channels of the BFKL dynamics is represented by the inclusive \emph{semi-hard} emission~\cite{Gribov:1983ivg} of two particles with high transverse momenta and well separated in rapidity (see Refs.~\cite{Celiberto:2017ius,Celiberto:2020wpk} for an overview of recent applications). Here the established factorization is \emph{hybrid}. Indeed the pure high-energy factorization is supplemented by collinear densities which enter expressions of impact factors. 

In the last thirty years several phenomenological studies have been proposed for different semi-hard final states. 
An incomplete list includes: the inclusive hadroproduction of two jets featuring large transverse momenta and well separated in rapidity (Mueller--Navelet channel~\cite{Mueller:1986ey}), for which several phenomenological studies have appeared so far~(see, \emph{e.g.},~Refs.~\cite{Colferai:2010wu,Caporale:2012ih,Ducloue:2013hia,Ducloue:2013bva,Caporale:2013uva,Caporale:2014gpa,Colferai:2015zfa,Caporale:2015uva,Ducloue:2015jba,Celiberto:2015yba,Celiberto:2015mpa,Celiberto:2016ygs,Celiberto:2016vva,Caporale:2018qnm}), the inclusive detection of two rapidity-separated light-flavored bound states~\cite{Celiberto:2016hae,Celiberto:2016zgb,Celiberto:2017ptm,Celiberto:2017uae,Celiberto:2017ydk}, three- and four-jet hadroproduction~\cite{Caporale:2015vya,Caporale:2015int,Caporale:2016soq,Caporale:2016vxt,Caporale:2016xku,Celiberto:2016vhn,Caporale:2016djm,Caporale:2016lnh,Caporale:2016zkc}, $J/\Psi$-plus-jet~\cite{Boussarie:2017oae,Boussarie:2017xdy}, hadron-plus-jet~\cite{Bolognino:2019cac,Bolognino:2019yqj,Bolognino:2019cac}, Higgs-plus-jet~\cite{Celiberto:2020tmb,Celiberto:2021fjf,Celiberto:2021tky,Celiberto:2021txb,Celiberto:2020rxb,Celiberto:2021xpm}, heavy-light dijet system~\cite{Bolognino:2021mrc,Bolognino:2021hxx}, heavy-flavor~\cite{Celiberto:2017nyx,Bolognino:2019yls,Bolognino:2019ccd,Celiberto:2021dzy,Celiberto:2021fdp,Bolognino:2022wgl}, and forward Drell–Yan dilepton production with a possible backward-jet detection~~\cite{Golec-Biernat:2018kem}. 
This permitted us to define BFKL-sensitive observables as well as to disengage the BFKL dynamics from collinear contaminations~\cite{Celiberto:2015yba,Celiberto:2015mpa,Celiberto:2020wpk}.

\begin{figure}[t]
\centering
\includegraphics[width=0.47\textwidth]{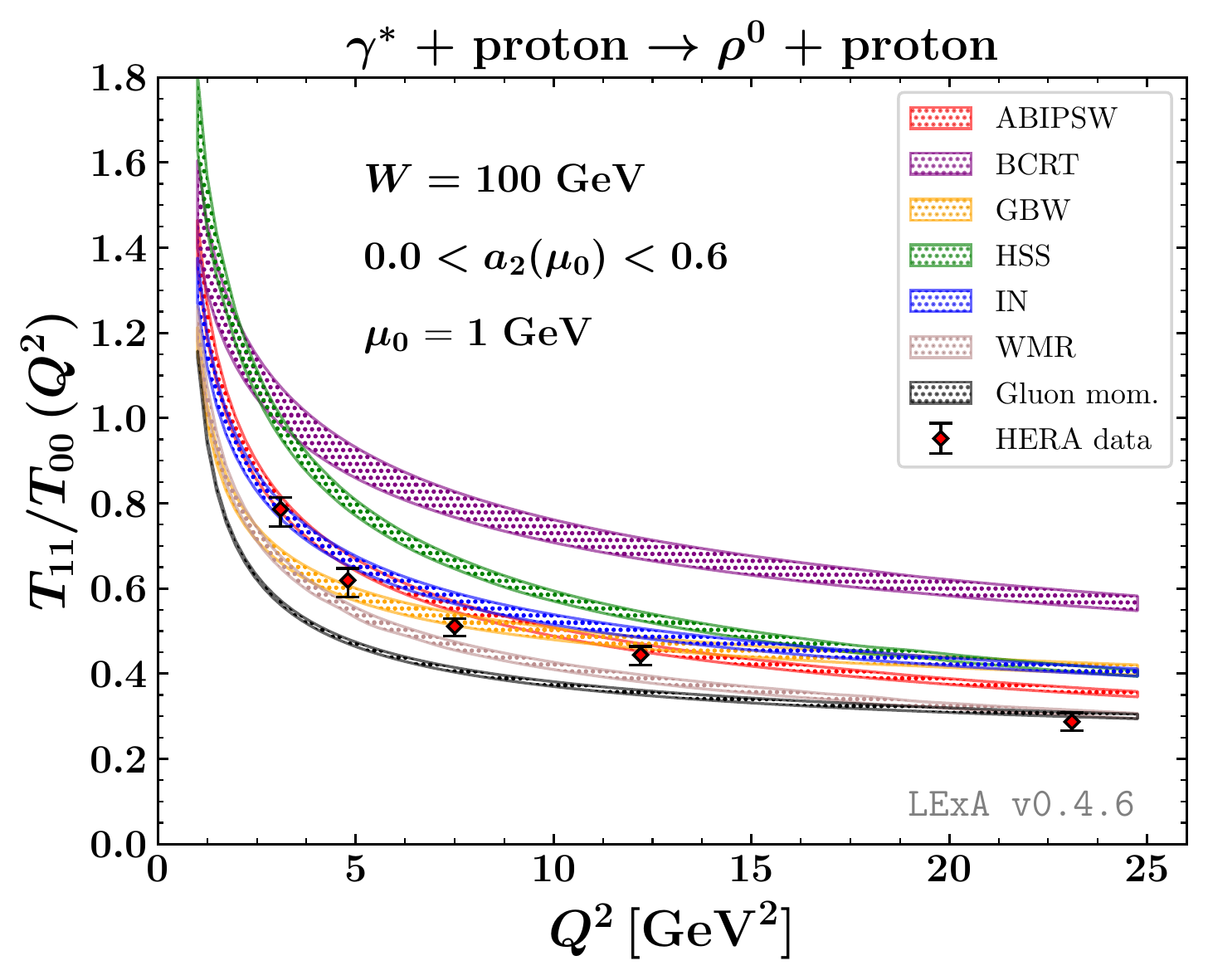} \hspace{0.5cm}
\includegraphics[width=0.47\textwidth]{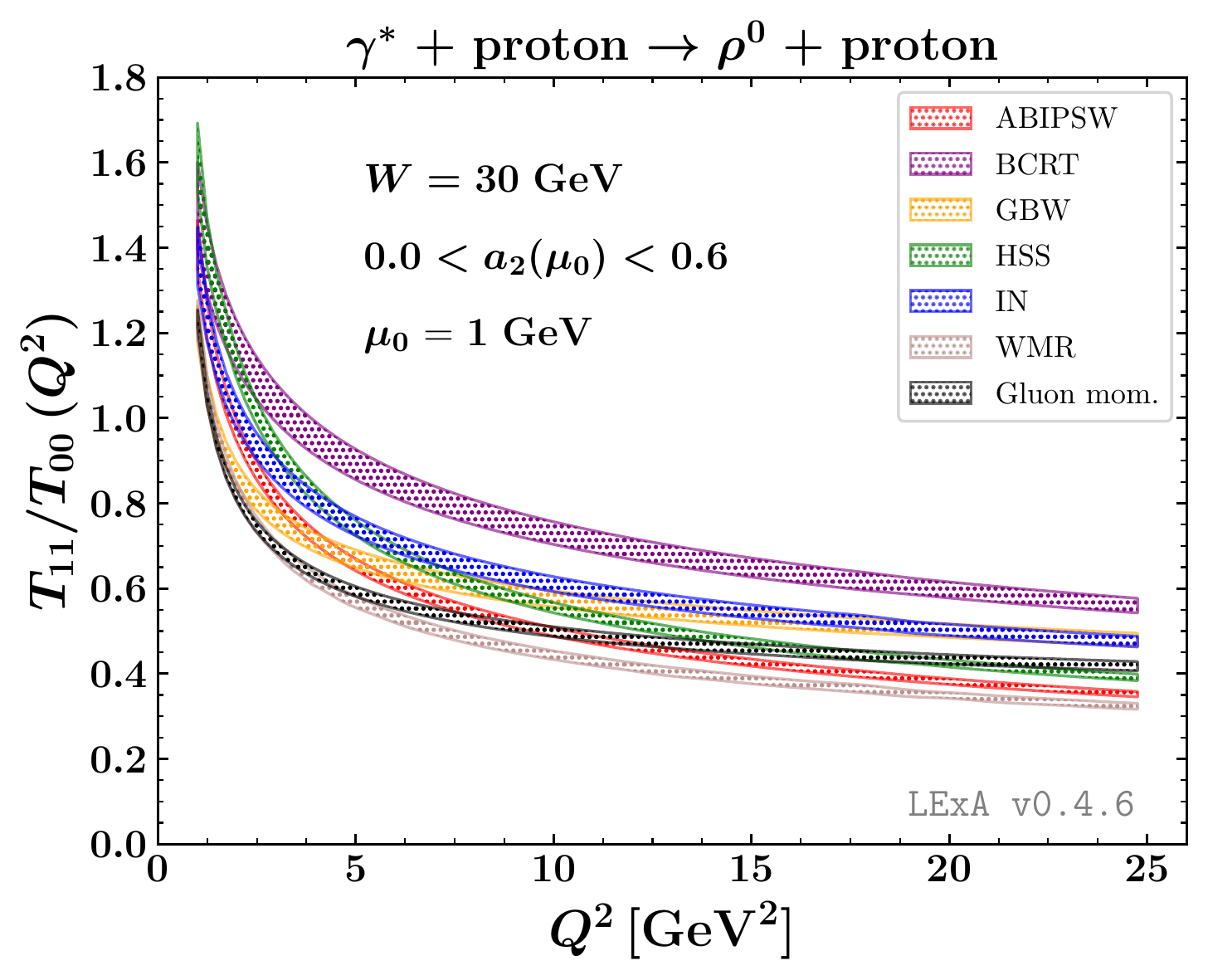}
\caption{$Q^2$-dependence of the polarized $T_{11}/T_{00}$ ratio, for all the considered UGD models, at $100$ GeV HERA (left) and at $30$ GeV EIC (right). 
Uncertainty bands represent the effect of varying $a_2(\mu_0 = 1\,$\rm GeV$)$ between $0.0$ and $0.6$.
Numerical results were obtained through the \emph{Leptonic-Exclusive-Amplitudes} ({\tt LExA}) super-module of the {\tt JETHAD} interface~\cite{Celiberto:2020wpk}. Figures from~~\cite{Bolognino:2021niq}.}
\label{fig:rho}
\end{figure}

The second kind of high-energy probe channels is represented by single emissions of forward particles. Here we access the proton content through the UGD, whose evolution at small-$x$ is controlled by the BFKL Green's function. Being a non-perturbative object, the UGD in not well known and several phenomenological models for it have been built so far. The UGD has been probed via the inclusive deep inelastic scattering~\cite{Hentschinski:2012kr,Hentschinski:2013id}, the exclusive electro- or photo-production of vector mesons at HERA~\cite{Anikin:2009bf,Anikin:2011sa,Besse:2013muy,Bolognino:2018rhb,Bolognino:2018mlw,Bolognino:2019bko,Bolognino:2019pba,Celiberto:2019slj,Bautista:2016xnp,Garcia:2019tne,Hentschinski:2020yfm} and the EIC~\cite{Bolognino:2021niq,Bolognino:2021gjm,Bolognino:2022uty}, the single inclusive heavy-quark emission at the LHC~\cite{Chachamis:2015ona}, and the forward Drell--Yan production at LHCb~\cite{Motyka:2014lya,Brzeminski:2016lwh,Motyka:2016lta,Celiberto:2018muu}.

We consider the exclusive $\rho$-meson production in lepton-proton collisions via the subprocess
\begin{equation}
\label{eq:subprocess}
 \gamma^*_{\lambda_i} (Q^2) \, p \; \to \; \rho_{\lambda_f} p \;,
\end{equation}
where a photon with virtuality $Q^2$ and polarization $\lambda_i$ interacts with the proton and a $\rho$-meson with polarization $\lambda_f$ is produced. The two polarization states $\lambda_{i,f}$ can assume values $0$ (longitudinal) and $\pm 1$ (transverse).
Since a strict semi-hard scale ordering holds, $W^2 \gg Q^2 \gg \Lambda^2_{\rm QCD}$ ($W$ is the subprocess center-of-mass energy), one enters the small-$x$ regime given by $x = Q^2/W^2$. The BFKL approach provides us with a high-energy factorized formula for polarized amplitudes
\begin{equation}
\label{eq:ampltude}
 {\cal T}_{\lambda_i \lambda_f}(W^2, Q^2) = \frac{i W^2}{(2 \pi)^2} \int \frac{\drv^2 \boldsymbol{p}_T}{(\boldsymbol{p}_T^2)^2} \; \Phi^{\gamma^*_{\lambda_i} \to \rho_{\lambda_f}}(\boldsymbol{p}_T^2, Q^2) \, f_g^{\rm BFKL} (x, \boldsymbol{p}_T^2, Q^2) \;,
\end{equation}
with $\Phi^{\gamma^*_{\lambda_i} \to \rho_{\lambda_f}}(q^2, Q^2)$ being the impact factor that describes the $\gamma^* \to \rho$ transition and encodes collinear distribution amplitudes (DAs, for further details see Section~2 of Ref.~\cite{Bolognino:2021niq}), and $f_g^{\rm BFKL}$ is the BFKL UGD. We consider in our study the seven UGD models given in Section~3 of Ref.~\cite{Bolognino:2021niq}.

In Fig.~\ref{fig:rho} we show the $Q^2$-dependence of ${\cal T}_{11} / {\cal T}_{00}$. We compare our predictions with HERA data~\cite{Aaron:2009xp} at $W = 100$ GeV (left panel), and we present new results for the EIC at the reference energy of $W = 30$ GeV (right panel). We use the twist-2 (twist-3) DAs for the longitudinal (transverse) case, and we gauge the impact of the collinear evolution of the DAs via a variation of the non-perturbative parameter $a_2(\mu_0 = 1\,$\rm GeV$)$ in the range 0.0 to 0.6.

We note that our predictions are spread over a large range and none of the UGD models is in agreement with HERA data over the whole $Q^2$-window, the ABIPSW, IN and GBW ones better catching the intermediate-$Q^2$ range. Results at EIC energies show a reduction of the distance between models, together with a hierarchy inversion for some regions of $Q^2$. This provides us with a clear evidence that the ${\cal T}_{11} / {\cal T}_{00}$ helicity ratio act as a discriminator for the UGD.

\vspace{-0.20cm}
\section{Future perspectives}
\label{sec:conclusions}
\vspace{-0.20cm}

We reported progresses on the study of the proton structure at small-$x$ via two distinct kinds of gluon distributions: the (un)polarized gluon TMD functions and the BFKL UGD.
All the presented results are relevant to explore the proton content at small-$x$, where the intrinsic motion of the constituent gluons plays an important role in the description of observables sensitive to different combinations of the hadron and the parton polarization states.
Here, a key ingredient to get a consistent description of the proton structure is interplay between genuine TMD and high-energy effects.
We believe that a path towards the first extraction of the small-$x$ improved gluon distributions from a global fit on data coming from new-generation colliding facility, such the EIC~\cite{Accardi:2012qut,AbdulKhalek:2021gbh}, the HL-LHC~\cite{Chapon:2020heu}, the FPF~\cite{Anchordoqui:2021ghd}, and NICA-SPD~\cite{Arbuzov:2020cqg} has been traced.

\vspace{-0.20cm}
\section*{Acknowledgements}
\vspace{-0.20cm}

We thank Alessandro Bacchetta, Andr\`ee Dafne Bolognino, Dmitry Yu. Ivanov, Alessandro Papa, Marco Radici, Wolfgang Sch\"afer, Antoni Szczurek, and Pieter Taels for collaboration.


\vspace{-0.20cm}
\nocite{*}
\bibliographystyle{auto_generated}
\bibliography{proceeding_Lowx_2021_Celiberto/proceeding_Lowx_2021_Celiberto.bib}


%% file: proceedings_elba2021_DenizSunarCerci/proceedings_elba2021_DenizSunarCerci/SunarCerci.tex
\vspace*{1.2cm}

\thispagestyle{empty}
\begin{center}
{\LARGE \bf Jet cross section measurements in CMS}

\par\vspace*{7mm}\par

{

\bigskip

\large \bf Deniz Sunar Cerci\footnote{On behalf of the CMS Collaboration}}

\bigskip

{\large \bf  E-Mail: deniz.sunar.cerci@cern.ch}

\bigskip

{Department of Physics, Faculty of Science and Letters, Adiyaman University, 02040 Turkey}

\bigskip

{\it Presented at the Low-$x$ Workshop, Elba Island, Italy, September 27--October 1 2021}

\vspace*{15mm}

\end{center}
\vspace*{1mm}

\begin{abstract}

Recent results are presented of differential jet cross section measurements using proton--proton collision data recorded with the CMS experiment at the Large Hadron Collider (LHC).
The measurements range from inclusive jets to multijet final states. The impact of these jet measurements on parton density functions as well as the strong coupling $\alpha_S$ is also reported. The measurements are corrected for detector effects and compared with predictions in perturbative QCD.
    \end{abstract}
  \part[Jet cross section measurements in CMS\\ \phantom{x}\hspace{4ex}\it{Deniz Sunar Cerci on behalf of the CMS Collaboration}]{}
 \section{Introduction}
In high energy proton-proton (pp) collisions jets, i.e.,  collimated spray of particles, are abundantly produced. Inclusive jet production in pp collisions is a useful tool to test perturbative Quantum Chromodynamics (QCD) predictions. In addition, this provides important constraints on the description of the proton structure, expressed by the parton distribution functions (PDFs) and the value of the strong coupling constant $\alpha_S$. 

The CMS Collaboration has performed many measurements of inclusive jet production and multi-jets production at different centre-of-mass energies. In the following the most recent results are presented.
 
\section{Results}
\subsection{Inclusive jet measurements}
Measurements of inclusive jet production in proton-proton collisions have been performed with the data collected by the CMS experiment~\cite{cms} at different centre-of-mass energies, i.e. 7 TeV~\cite{cms2}, 8 TeV~\cite{cms3} and 13 TeV~\cite{cms4}. 

\begin{figure}
\begin{center}
\epsfig{figure=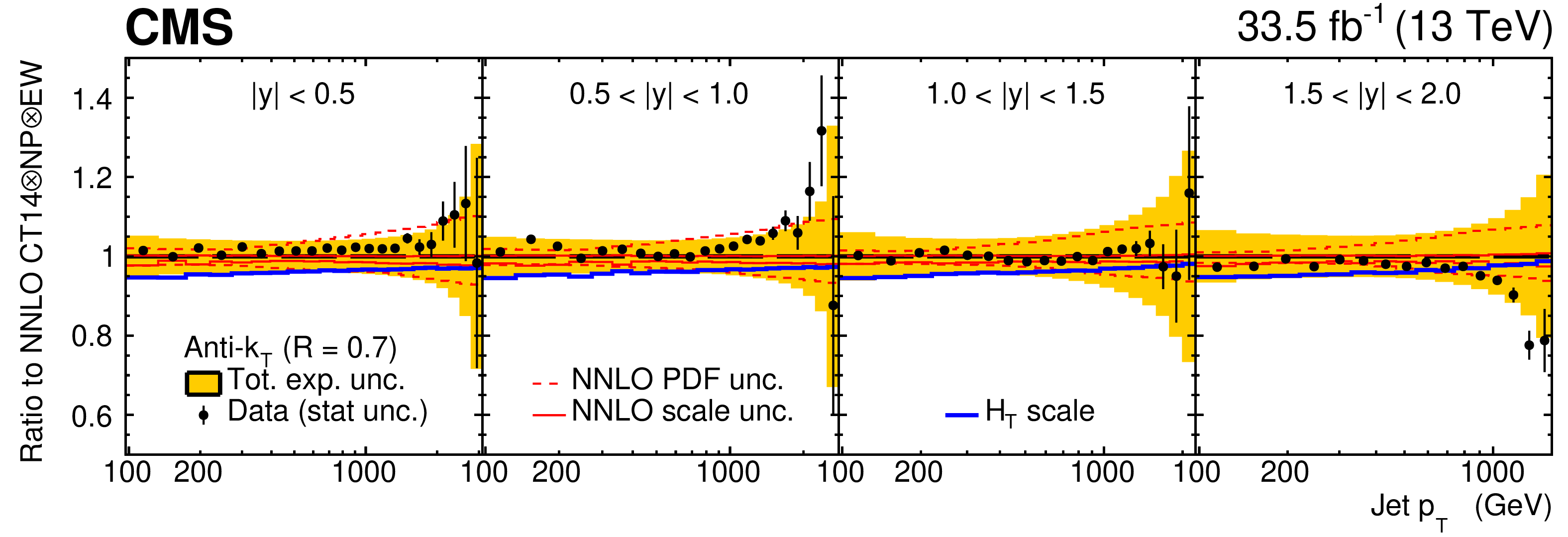,height=0.35\textwidth}
\epsfig{figure=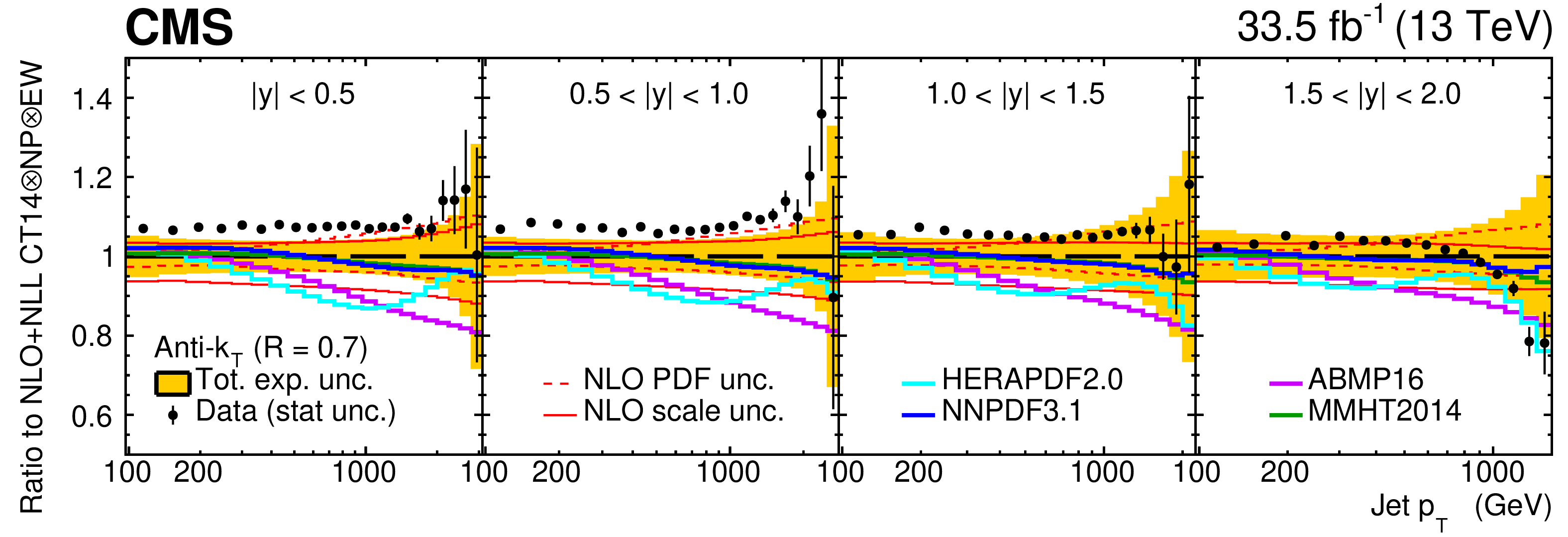,height=0.35\textwidth}
\caption{The double-differential inclusive jet cross sections as a function of jet $p_T$ measured in intervals of $|y|$ shown with jet distance parameter $R = 0.7$. The data are divided by NNLO (upper panel) and NLO+NLL predictions (lower panel) ~\cite{cms5}.}
\label{fig:1}
\end{center}
\end{figure}

The CMS Collaboration has recently published an inclusive jet cross section measurement with the data collected in 2016, corresponding to an integrated luminosity of up to 36.3 fb$^{-1}$~\cite{cms5}. The double differential inclusive jet cross sections are measured as a function of jet transverse momentum $p_T$ and rapidity $y$. The jets clustered with the anti-$k_T$ jet algorithm are used with two jet distance parameters, $R = 0.4$ and 0.7. A comprehensive QCD analysis at next-to-next-to-leading order (NNLO) is performed to study the PDFs of the proton as well as to extract the strong coupling constant. The inclusive jet cross sections as functions of the jet $p_T$ and $|y|$ for R = 0.7 is shown in Fig.~\ref{fig:1}. The measured jet cross sections are compared with  fixed order NNLO QCD predictions using CT14PDF. In the measurement, a wide range of the jet $p_T$ from 97 GeV up to 3.1 TeV is covered. The prediction using parton $H_T$  as renormalisation and factorisation scale results in a softer $p_T$ spectrum than in case of set to jet $p_T$. The NLO+NLL calculations predict harder $p_T$ spectrum than the NNLO calculations. The data are well described by all predictions within the experimental and theory uncertainties. 

\begin{figure}
\begin{center}
\epsfig{figure=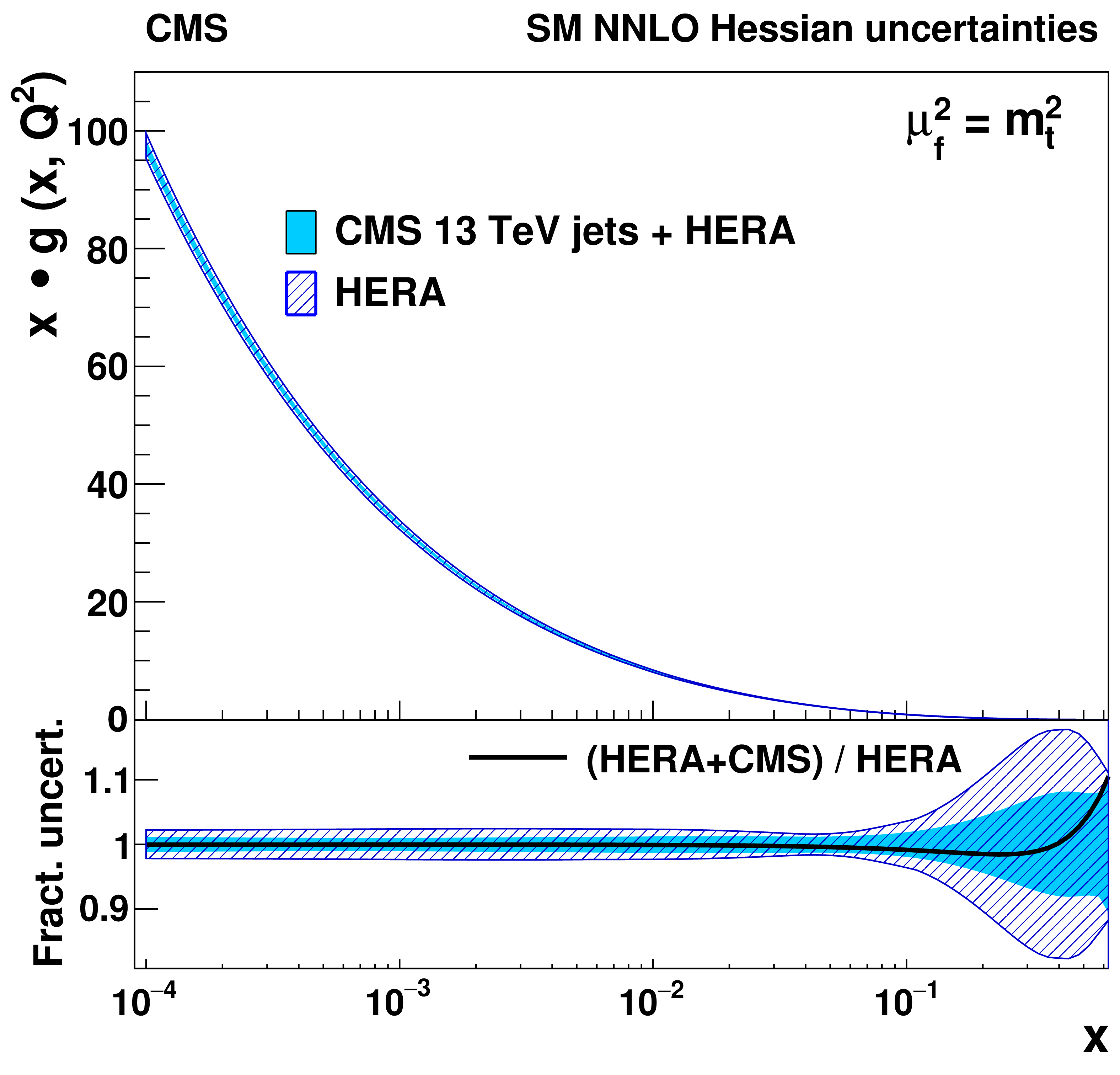,height=0.45\textwidth}
\epsfig{figure=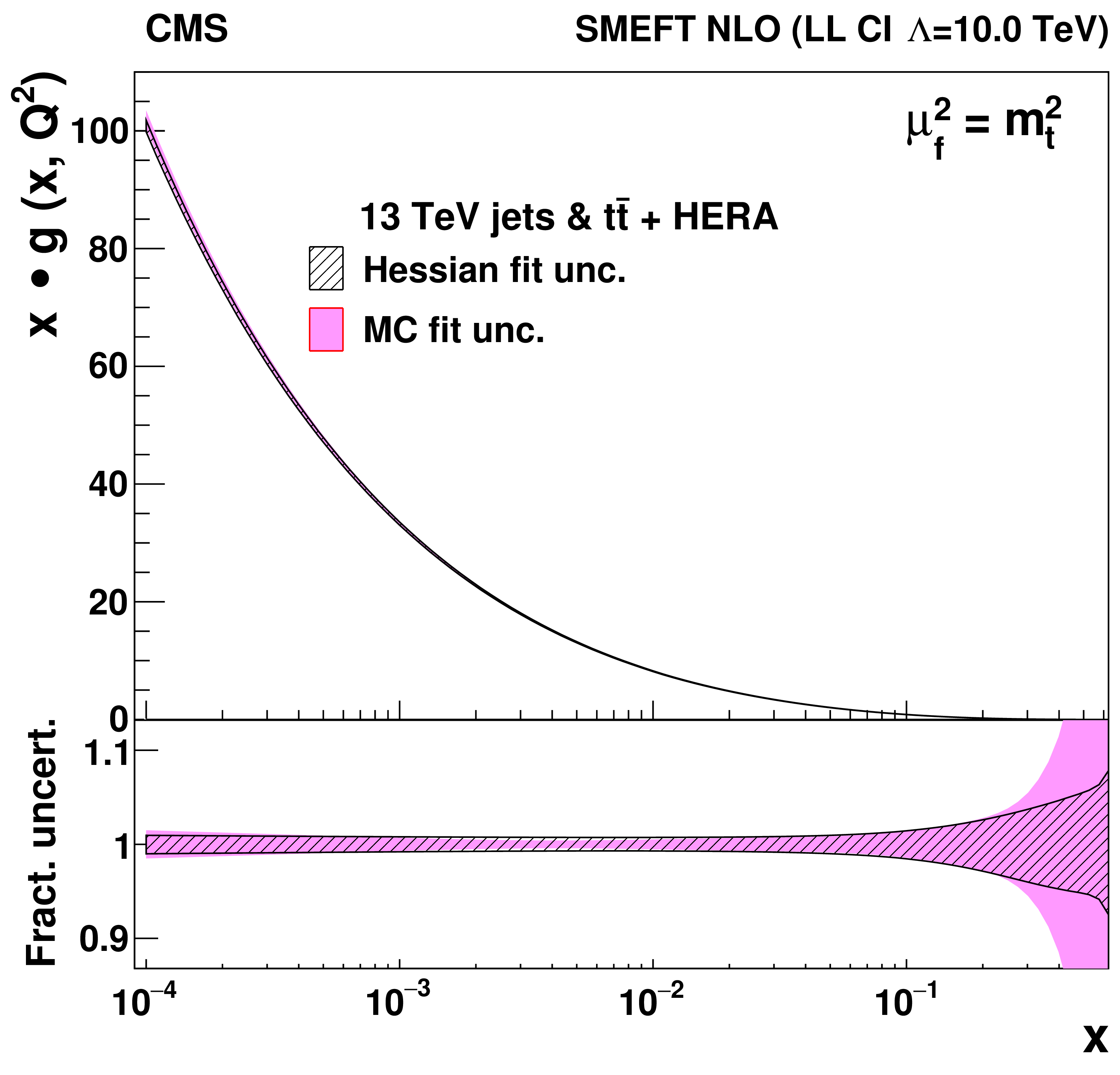,height=0.45\textwidth}
\caption{The gluon distributions shown as a function of $x$ at the scale $\mu = {m_t}^{2} $ resulting from the NNLO fit using HERA DIS and the CMS 13 TeV jets data (left) and from the SMEFT fit with the left-handed CI model with $\Lambda = 10$~TeV (right)~\cite{cms5}.}
\label{fig:2}
\end{center}
\end{figure}

A comprehensive QCD analysis is performed to investigate the sensitivity of the presented measurement on the proton PDFs and  $\alpha_S$. Due to the small out-of-cone radiation effects, the jet cross section for $R = 0.7$ is used. The results obtained with both CMS data and HERA DIS data to the fit on the gluon PDF is shown on Fig.~\ref{fig:2} (left). Significant improvement on  the accuracy of the PDFs is observed  by using the present measurement in the QCD analysis. For the first time, the value of the strong coupling constant at the Z boson mass is extracted in a QCD analysis at NNLO using these data and results in $\alpha_S = (m_Z) = 0.1170 \pm 0.0019$. Furthermore, the model of contact interactions (CI) is used for investigation of the effect of beyond standard model particle exchanges between the quarks. In the context of the effective field theory (EFT)-improved SM (SMEFT) fit, the CI Wilson coefficient $c_1$ is taken as a free parameter. The obtained result from the SMEFT fit with the left-handed CI model with $\Lambda = 10$~TeV is shown in Fig.~\ref{fig:2} (right).

\begin{figure} [hhh]
\begin{center}

\epsfig{figure=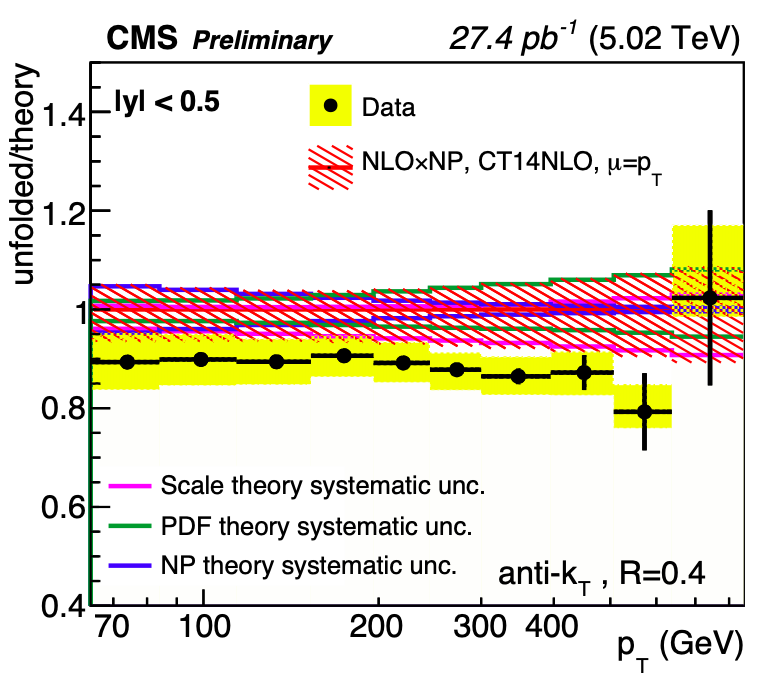,height=0.29\textwidth}
\epsfig{figure=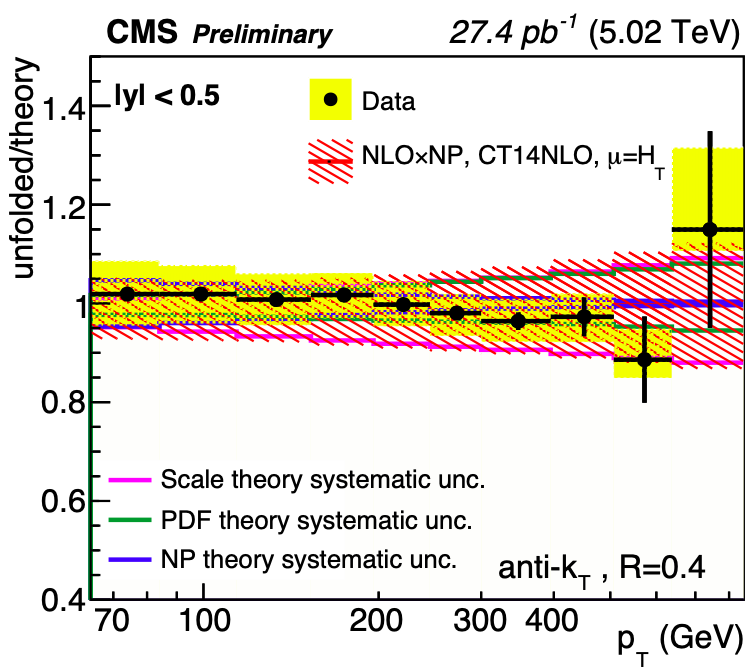,height=0.29\textwidth}
\epsfig{figure=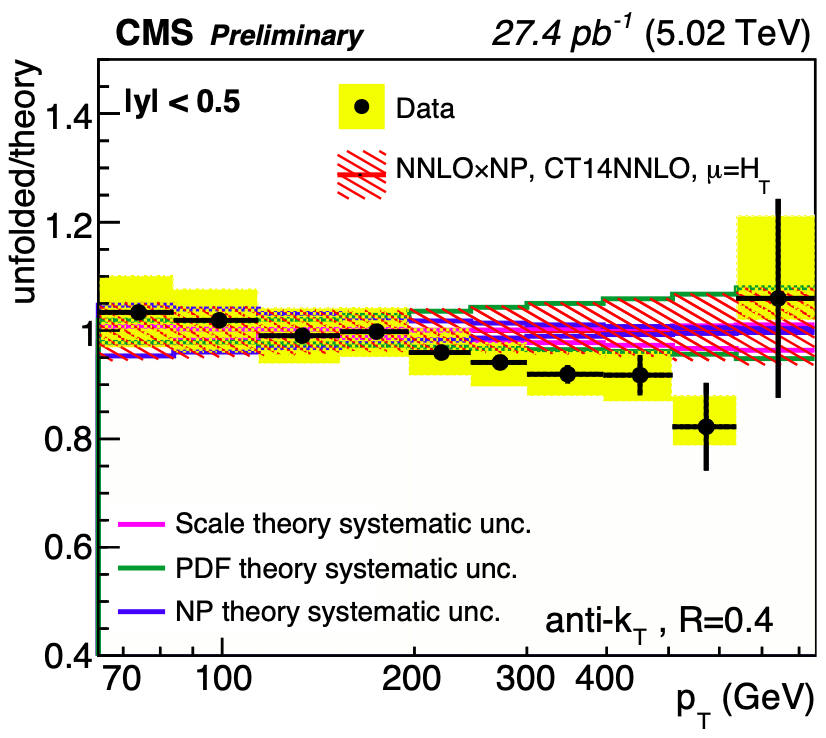,height=0.29\textwidth}
\caption{Ratios of the unfolded inclusive jet cross section to the NLO theoretical prediction, using the CT14nlo PDF set, with $\mu_R = \mu_F = p_T$ (left) and with $\mu_R = \mu_F = H_T$ (middle). Ratio of the unfolded inclusive jet cross section to the NNLO theoretical prediction, using the CT14nlo PDF set, with $\mu_R = \mu_F = H_T$ (right)~\cite{smp-21-009}.}
\label{fig:3}
\end{center}
\end{figure}

  A measurement of the differential inclusive jet production cross section is performed by the CMS Collaboration~\cite{smp-21-009}. The measurement is based on pp collisions at $\sqrt s = 5$ TeV corresponding to a total integrated luminosity of 27.4 pb$^{-1}$.  The present measurement provides a valuable reference data for the analysis of heavy ion collisions probing quark-gluon plasma (QGP). The reconstruction of jets with the anti-$k_T$ algorithm using $R = 0.4$ is carried out within in the kinematic range of $|y| < 2$ and $0.06 < p_T < 1 $~TeV. The unfolded jet cross section is compared with pQCD predictions, calculated at both NLO and NNLO with jet $p_T$ and $H_T$ parton scale choices. The predictions are corrected for nonperturbative (NP) and electroweak (EW) effects. The comparison of the measurement to NLO and NNLO predictions with jet $p_T$ and parton $H_T$ scale is shown in Fig.~\ref{fig:3}.  

\subsection{Multijet production}

\begin{figure}[hh]
\begin{center}
\epsfig{figure=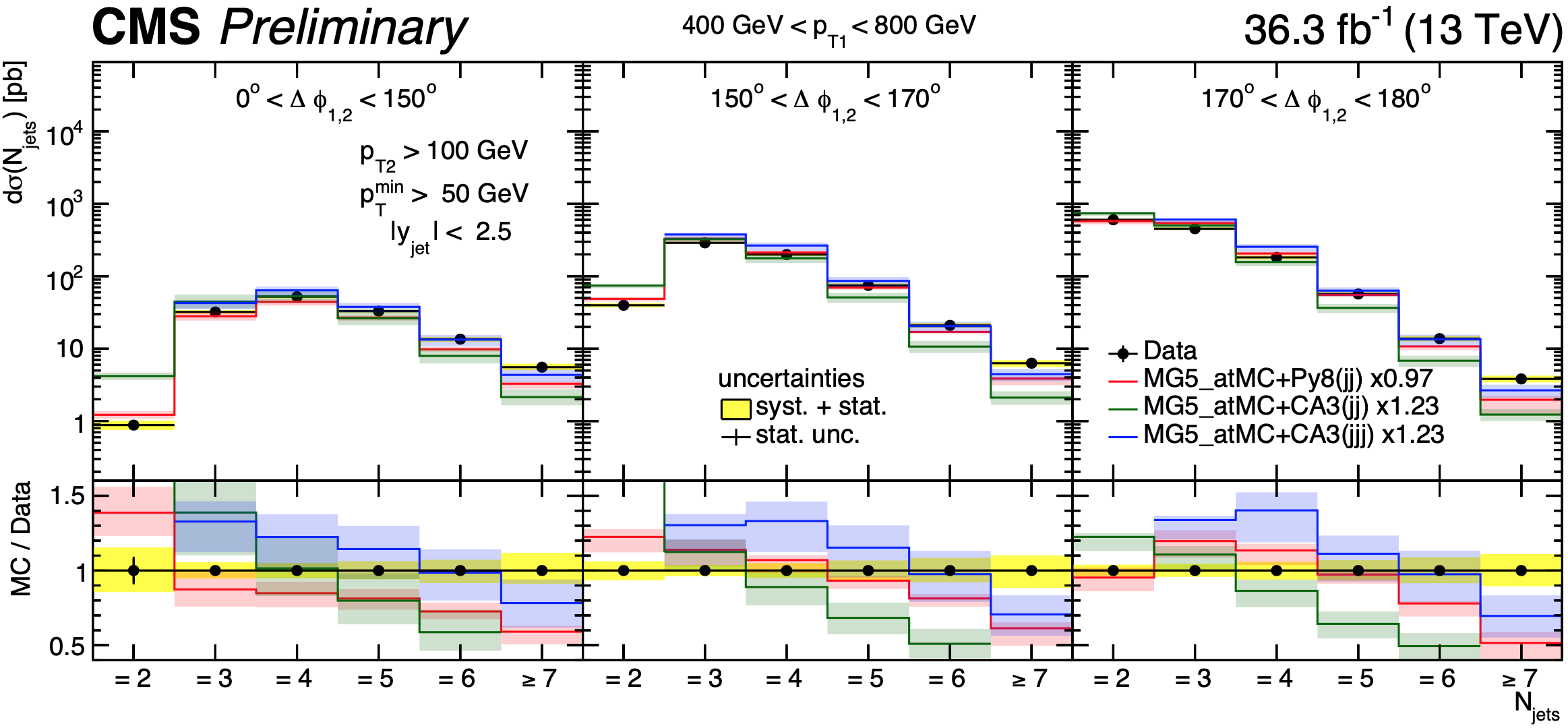,height=0.45\textwidth}
\caption{Comparison of the differential cross section of two leading jets as a function of the exclusive jet multiplicity (inclusive for 7 jets) in bins of $p_{T1}$ and $\Delta \phi_{1,2}$~\cite{smp-21-006}.}
\label{fig:4}
\end{center}
\end{figure}

The differential cross-section of the four jets leading in $p_T$ as a function of their transverse momentum is measured with the data recorded with the CMS detector in pp collisions at $\sqrt s = 13$~TeV~\cite{smp-21-006}. The same analysis strategy as the inclusive jet measurement at 13 TeV is followed except the jets clustered with $R= 0.4$ are used. The events which have at least two jets with the leading jet of $p_{T1} > 200$ GeV and subleading jet of $p_{T2} > 100$ GeV are considered. All jets must satisfy the range of $|y| < 2.5$. The multiplicity of additional jets with  $p_T > 50$~GeV is measured in bins of the 
azimuthal separation between leading and subleading jets ($\Delta \phi_{1,2}$) and  transverse momenta of the leading jet ($p_{T1}$).  Comparisons of data to NLO dijet predictions MG5 AMC+PY8 (jj) and MG5 AMC+CA3 (jj) as well as the NLO three-jet prediction of MG5 AMC+CA3 (jjj) are shown in Fig.~\ref{fig:4}. Reasonable agreement is observed with the normalization of MG5 AMC+PY8 (jj) NLO calculation even for three jets. The measurement is larger than the predictions particularly in the low $p_{T1}$ region.  

\section{Summary}

The CMS Collaboration has performed extensive jet studies in proton-proton collisions at different centre-of-mass energies. The most recent measurements of inclusive jet and multijet production are presented. The measurements are compared to various Monte Carlo event generators as well as the fixed order NLO, NLO+NLL and NNLO predictions. The QCD analysis is also performed at next-to-leading order. The PDFs, the values of the strong coupling constant and of the pole mass of the top quark are extracted. 

\nocite{*}
\bibliographystyle{auto_generated}
\bibliography{proceedings_elba2021_DenizSunarCerci/proceedings_elba2021_DenizSunarCerci/SunarCerci}

%% file: draft_Weisong_proceedingslowx2021/Weisong.tex
\vspace*{1.2cm}

\thispagestyle{empty}
\begin{center}
{\LARGE \bf Charmonia photo-production in ultra-peripheral and peripheral PbPb collisions with LHCb}

\par\vspace*{7mm}\par

{

\bigskip

\large \bf Weisong Duan on behalf of the LHCb Collaboration}

\bigskip

{\large \bf  E-Mail: weisong.duan@cern.ch}

\bigskip

{South China Normal University, Guangzhou, China}

\bigskip

{\it Presented at the Low-$x$ Workshop, Elba Island, Italy, September 27--October 1 2021}

\vspace*{15mm}

\end{center}
\vspace*{1mm}

\begin{abstract}

The LHCb recorded ∼ 210 $\mu b^{-1}$ integrated luminosity of PbPb collisions at $\sqrt{s_{NN}}$ = 5.02 TeV in 2018. With an increase of the luminosity by a factor of 20 compared to the previous 2015 PbPb dataset, precise measurements on photo-produced charmonia in ultra-peripheral collisions are now possible. Moreover, the great momentum resolution of the detector allows photo-produced $J/\psi$ in collisions with a nuclear overlap to be studied. This new type of probe is sensitive to the geometry of the collisions but also to the electromagnetic field of the Pb nuclei. In this contribution, we present the latest results on $J/\psi$ photo-production measured by LHCb in peripheral and ultra-peripheral PbPb collisions.
\end{abstract}
  \part[Charmonia photo-production in ultra-peripheral and peripheral PbPb collisions with LHCb\\ \phantom{x}\hspace{4ex}\it{Weisong Duan on behalf of the LHCb collaboration}]{}

\section{Introduction}
The LHCb detector is a single-arm forward spectrometer fully instrumented in the pseudorapidity range 2 < $\eta$ < 5~\cite{Collaboration_2008}. It has a high precision tracking system, which provides excellent vertex and momentum resolution, and full particle identification. Compared to the ALICE, CMS and ATLAS, LHCb covers the forward rapidity region, providing better access to the gluon distribution at small $x$.

The photonuclear production of vector mesons such as $J/\psi$ is sensitive to the gluon parton distribution function in the nucleus at small Bjorken-$x$, which is estimated by $x \approx (m_{J/\psi}\cdot e^{-y})/\sqrt{s_{NN}}$, where $m_{J/\psi}$ is mass of $J/\psi$ and $y$ is its rapidity. Coherent photoproduction of $J/\psi$ meson provides a way to study the nuclear shadowing effects at small Bjorken-$x$ ranging from $10^{-5}$ to $10^{-2}$ at LHC energies.
\section{Study of coherent $J/\psi$ production in ultra-peripheral lead-lead collisions at $\sqrt{s_{NN}}$ = 5 TeV}

The ultra-peripheral collisions, UPCs, are $\rm{PbPb \to Pb + Pb + X}$ in which two ions interact via their cloud of virtual photons. If the photon couples coherently to the nucleus as a whole, it is called coherent production. If the photon couples with one nucleon leading to the breakup of the target nucleus, it is called incoherent production.

In UPCs, coherent $J/\psi$ meson production can be described by the interaction between photons and gluons, according to the Regge theory~\cite{PhysRevC.86.014905, PhysRevC.93.055206}, gluons are considered as a single object with vacuum quantum numbers, which is called pomeron ($\enspace \rm P \kern -1em I$ $\enspace$). The cross-section for photoproduction gives constraints on the gluon parton distribution functions. This process has low multiplicities and very low transverse momentum $p_{T}$.

The $J/\psi$ mesons are reconstructed through the $J/\psi \to \mu^{+}\mu^{-}$ decay channel, using 2015 Pb-Pb data samples, corresponding to an integrated luminosity of 10 $\mu b^{-1}$. An ultra-peripheral electromagnetic interaction could occur simultaneously with the hadron collision. The HeRSCheL detector~\cite{Akiba_2018} is used to reject backgrounds from hadronic interactions.

\begin{figure}[htbp]
\begin{center}
\includegraphics[height=0.35\textwidth]{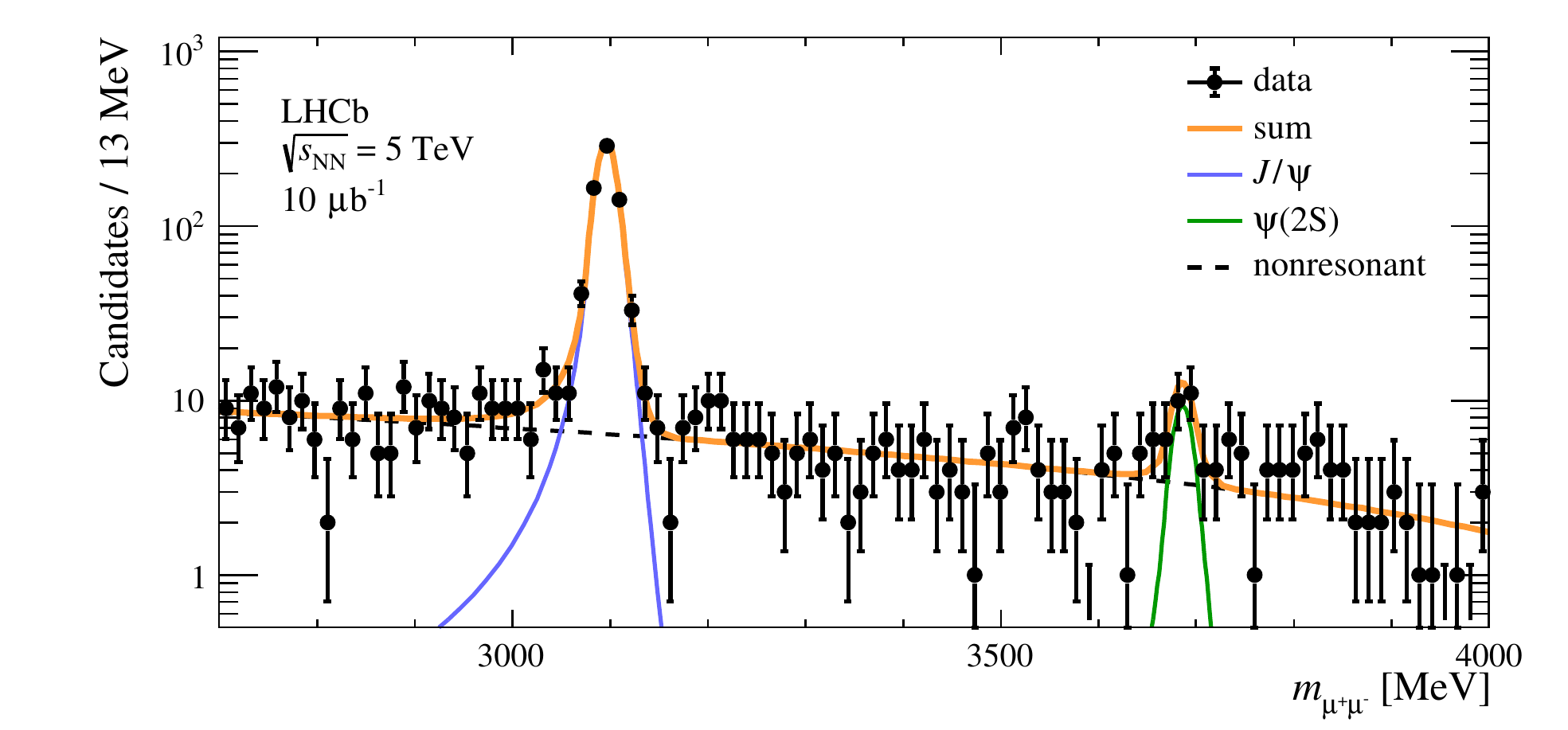}
\caption{The di-muon invariant mass distribution in the rapidity range between 2.0 and 4.5. The solid blue line corresponds to the $J/\psi$ meson. The solid green line represents the $\psi(2S)$ meson. The dashed black line represents non-resonant background.}
\label{massfit}
\end{center}
\end{figure}

The number of candidates are obtained by fitting the di-muon spectrum as shown in Fig.~\ref{massfit}. The $J/\psi$ and $\psi(2S)$ mass spectrum are modeled by a double-sided Crystal ball function, and the non-resonant background is modeled by exponential function multiplied by an first-order polynomial function. Thus we determine the number of $J/\psi$ candidates within the $J/\psi$ mass window 3040 MeV/$c^{2}$ $\sim$ 3165 MeV/$c^{2}$ and the number of $\psi(2S)$ candidates within the $\psi(2S)$ mass window 3608 MeV/$c^{2}$ $\sim$ 3763 MeV/$c^{2}$.

\begin{figure}[htbp]
\begin{center}
\includegraphics[height=0.35\textwidth]{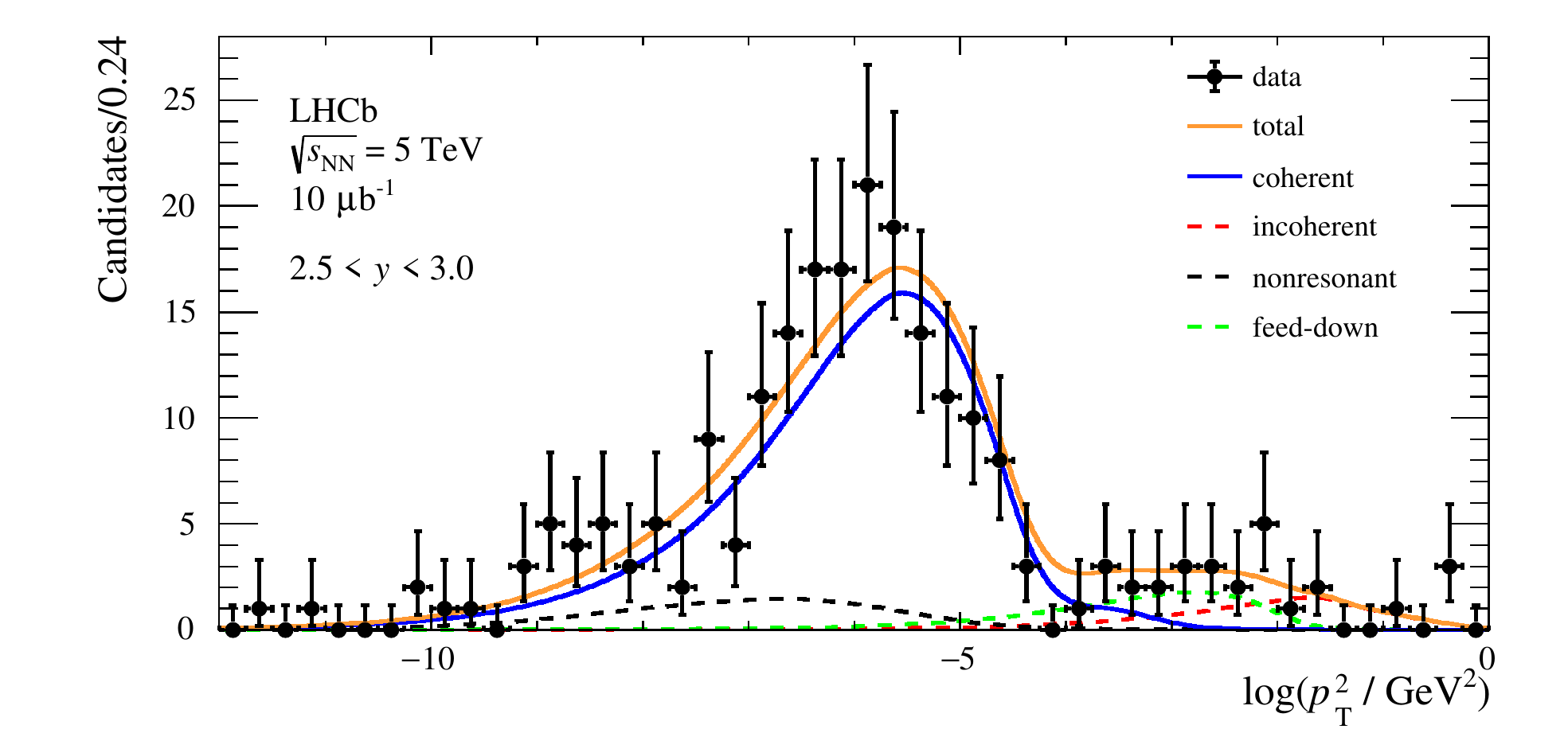}
\caption{The ${\rm log} (p^{2}_{T})$ distribution of $J/\psi$ candidates in the interval 2.5 < $y$ < 3.0, where the unit of $p_{T}$ is GeV/c. The solid blue line represents coherent $J/\psi$ distribution. The dashed red line represents the incoherent distribution. The dashed green line represents feed down from $\psi(2S)$. The dashed black line represents non-resonant background.}
\label{logpt2fit}
\end{center}
\end{figure}

To determine the coherent $J/\psi$ production, a fit to the ${\rm log} (p^{2}_{T})$ is performed to extract the coherent $J/\psi$ mesons within the $J/\psi$ mass window. The ${\rm log} (p^{2}_{T})$ distribution of the $J/\psi$ mesons is shown in Fig.~\ref{logpt2fit}. In short, the number of inclusive $J/\psi$ mesons is obtained by the invariant mass fit, and the number of coherent $J/\psi$ mesons is obtained by the ${\rm log} (p^{2}_{T})$ fit.

\begin{figure}
\begin{center}
\includegraphics[height=0.45\textwidth]{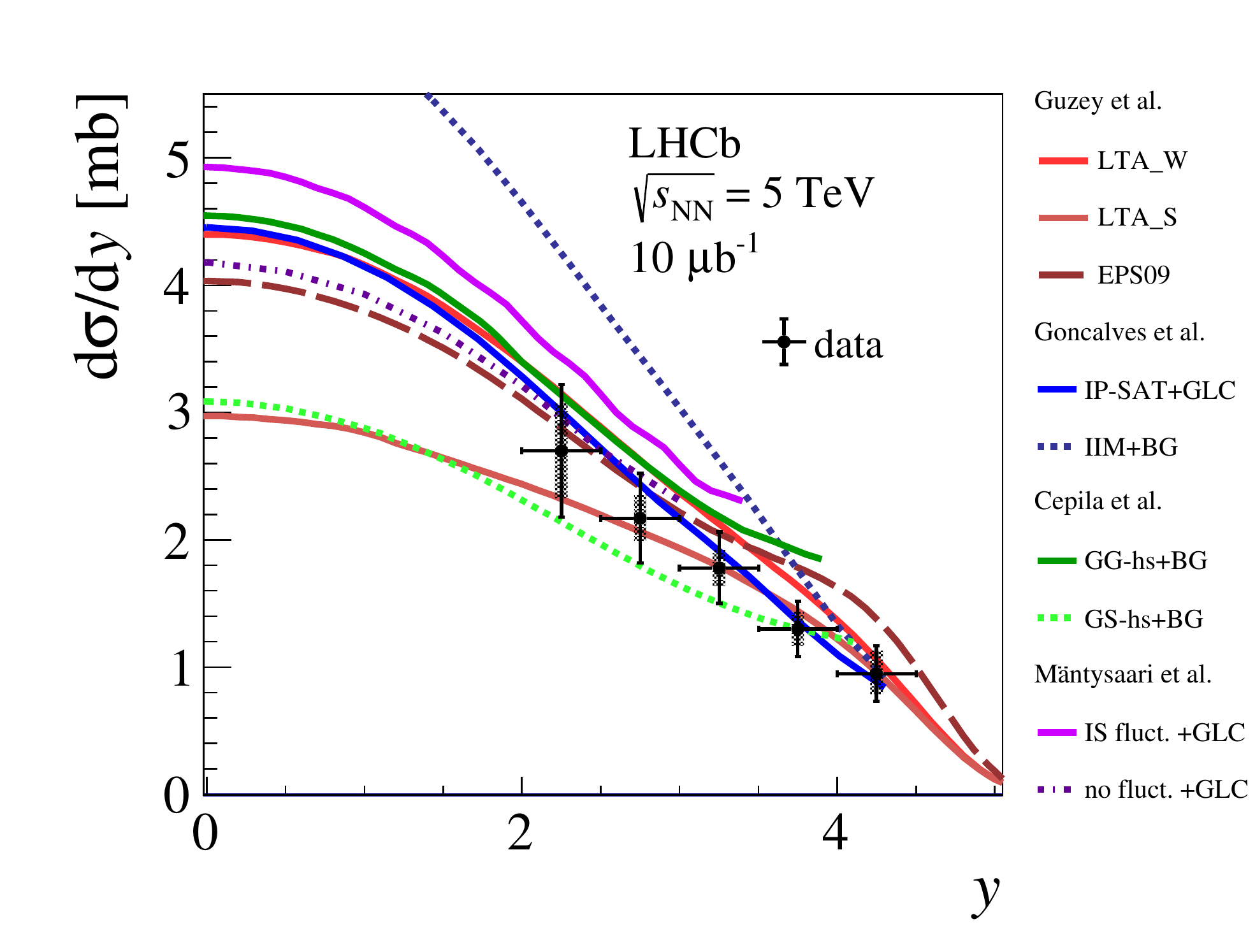}
\caption{Differential cross-section of coherent $J/\psi$ production as a function of rapidity, with comparisons to phenomenological models.}
\label{csresults}
\end{center}
\end{figure}

The results of the differential cross section are calculated in five rapidity bins as shown in Fig.~\ref{csresults}.  We compared the results between experimental results and theoretical predictions~\cite{PhysRevC.93.055206, PhysRevC.84.011902, PhysRevC.97.024901, PhysRevD.96.094027, MANTYSAARI2017832}. The coherent $J/\psi$ cross-section production is given by:

\begin{equation}
    \frac{\mathrm{d}\sigma_{coh,J/\psi}}{\mathrm{d}y} = \frac{N_{coh,J/\psi}}{\varepsilon_{t}\cdot \mathcal{L}\cdot\Delta y \cdot \mathcal{B}(J/\psi \to \mu^{+}\mu^{-})}
\end{equation}

where $\varepsilon_{t}$ is the total efficiency, $\mathcal{L}$ is an integrated luminosity of the Pb-Pb data sample, and $\mathcal{B} = (5.961 ± 0.033)\%$ is the $J/\psi \to \mu^{+}\mu^{-}$ branching ratio.

\begin{figure}
\begin{center}
\includegraphics[height=0.45\textwidth]{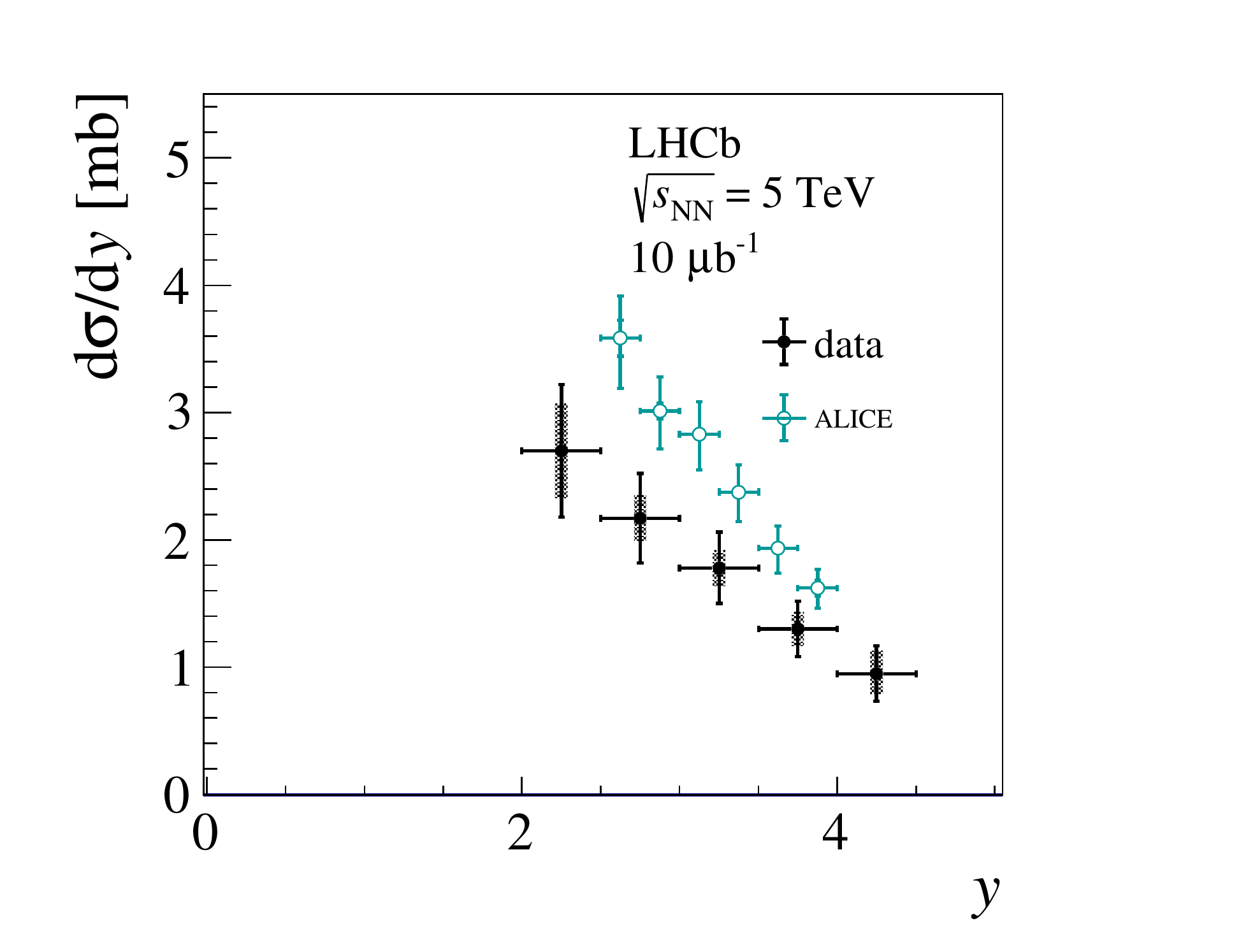}
\caption{Differential cross-section of coherent $J/\psi$ production as a function of rapidity, with comparison to ALICE measurements.}
\label{cscompare}
\end{center}
\end{figure}

\begin{figure}
\begin{center}
\includegraphics[height=0.45\textwidth]{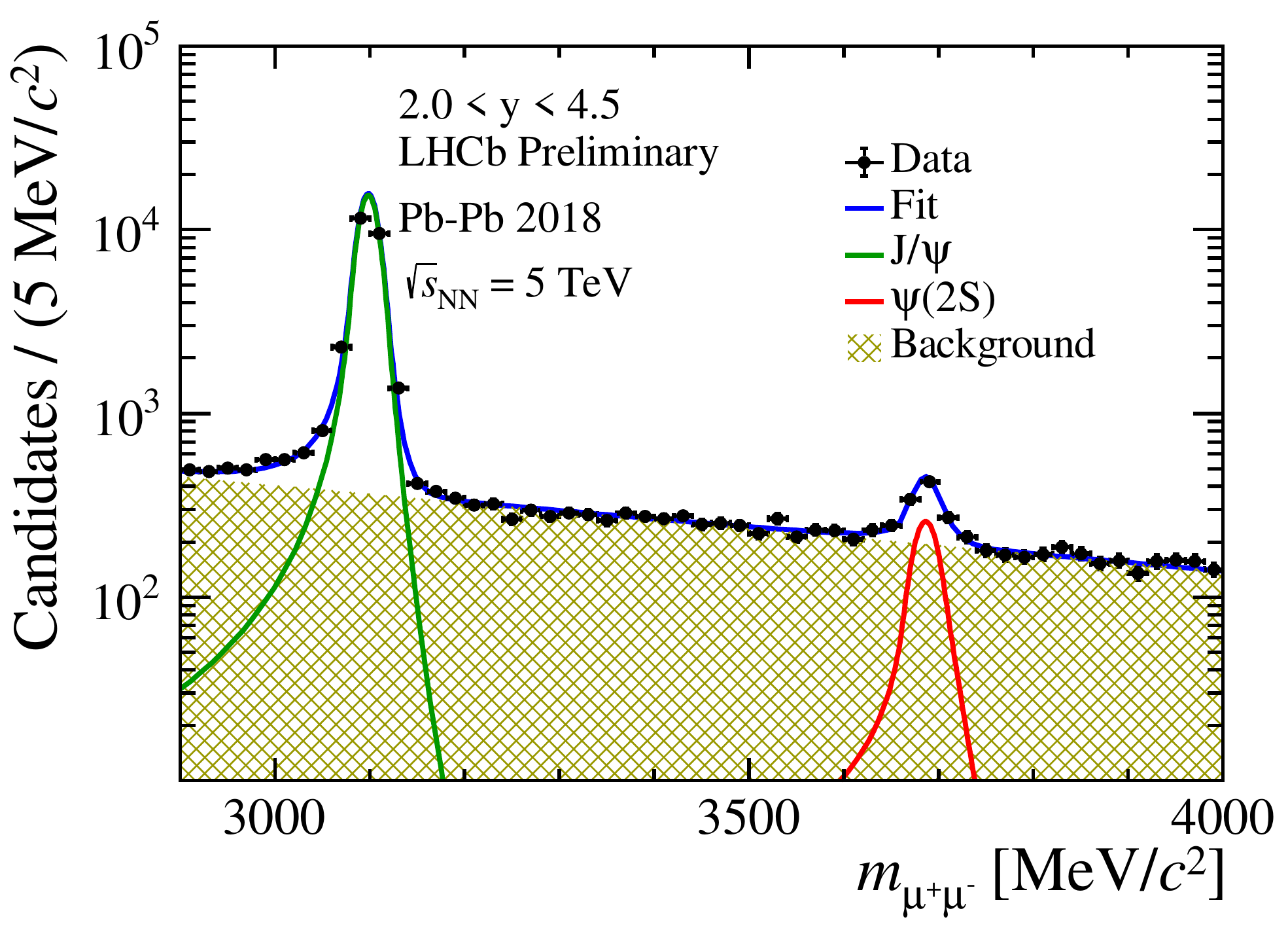}
\caption{The invariant mass distribution of $J/\psi$ and $\psi(2S)$ candidates. The solid green line represents $J/\psi$ signal, the solid red line represents $\psi(2S)$ signal, and the yellow regions represents non-resonant background.}
\label{2018massfit}
\end{center}
\end{figure}

Similarly, ALICE measured the coherent $J/\psi$ production cross-section~\cite{2019134926}, so we also compared the coherent $J/\psi$ production cross-section between ALICE and LHCb, as shown in Fig.~\ref{cscompare}. The LHCb result is slightly lower than the ALICE measurement by around 1.3 $\sigma$. Measurements of the coherent $J/\psi$ and $\psi(2S)$ are currently underway using 2018 Pb-Pb data, corresponding to an integrated luminosity of 210 $\mu b^{-1}$. Fig.~\ref{2018massfit} shows the di-muon invariant mass distribution in the range between 2.7 and 4.0 GeV. The final results are expected in the near future.

\section{Study of $J/\psi$ photo-production in lead-lead peripheral collisions at $\sqrt{s_{NN}}$ = 5 TeV}

The second result in this contribution is photo-production of $J/\psi$ at low $p_{T}$, studied in  peripheral Pb-Pb collisions at $\sqrt{s_{NN}}$ = 5 TeV, using the data sample collected by LHCb in 2018, with an integrated luminosity of about 210 $\mu b^{-1}$~\cite{lhcbcollaboration2021study}. 

\begin{figure}[htbp]
\begin{center}
\includegraphics[height=0.33\textwidth]{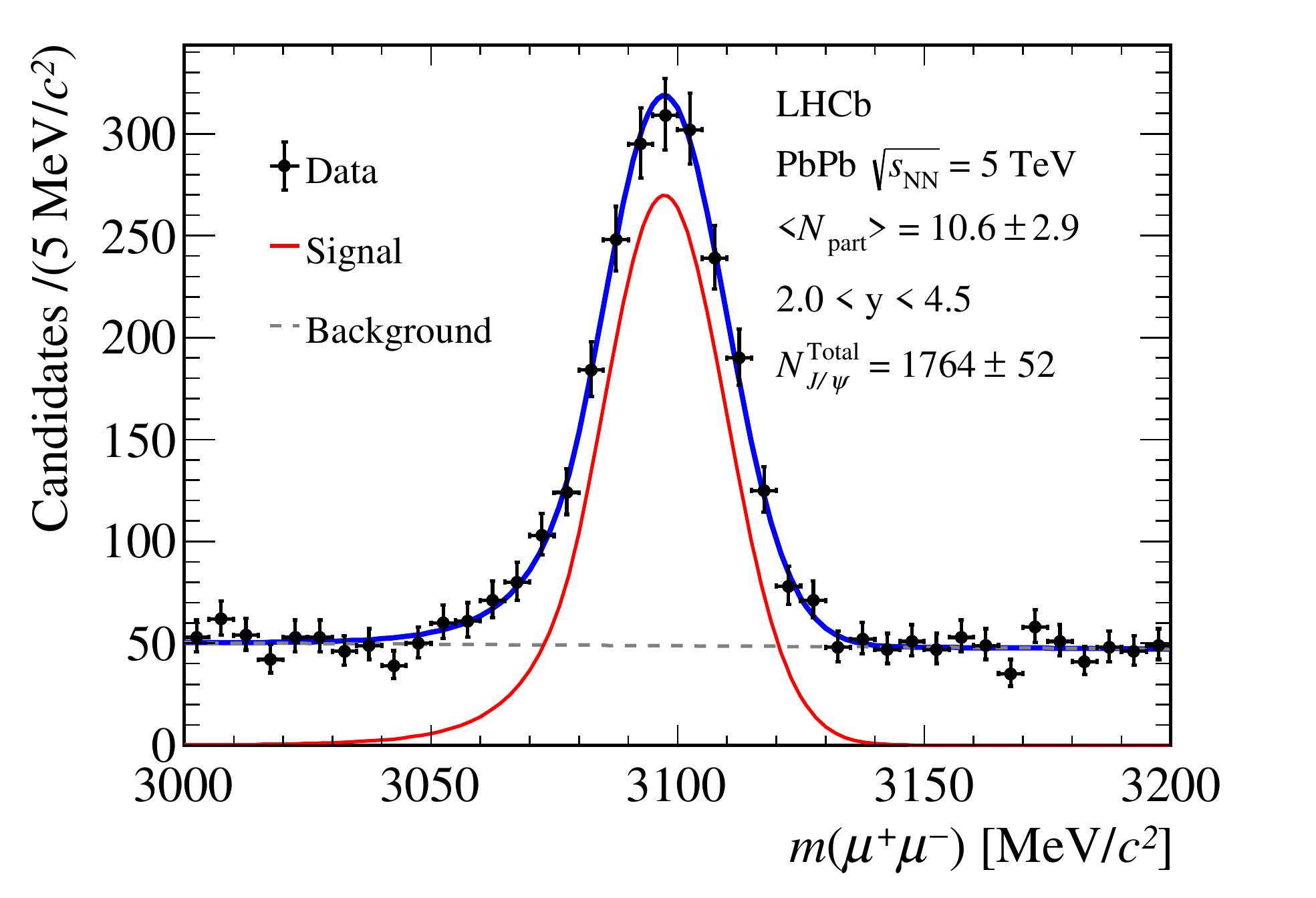}
\includegraphics[height=0.33\textwidth]{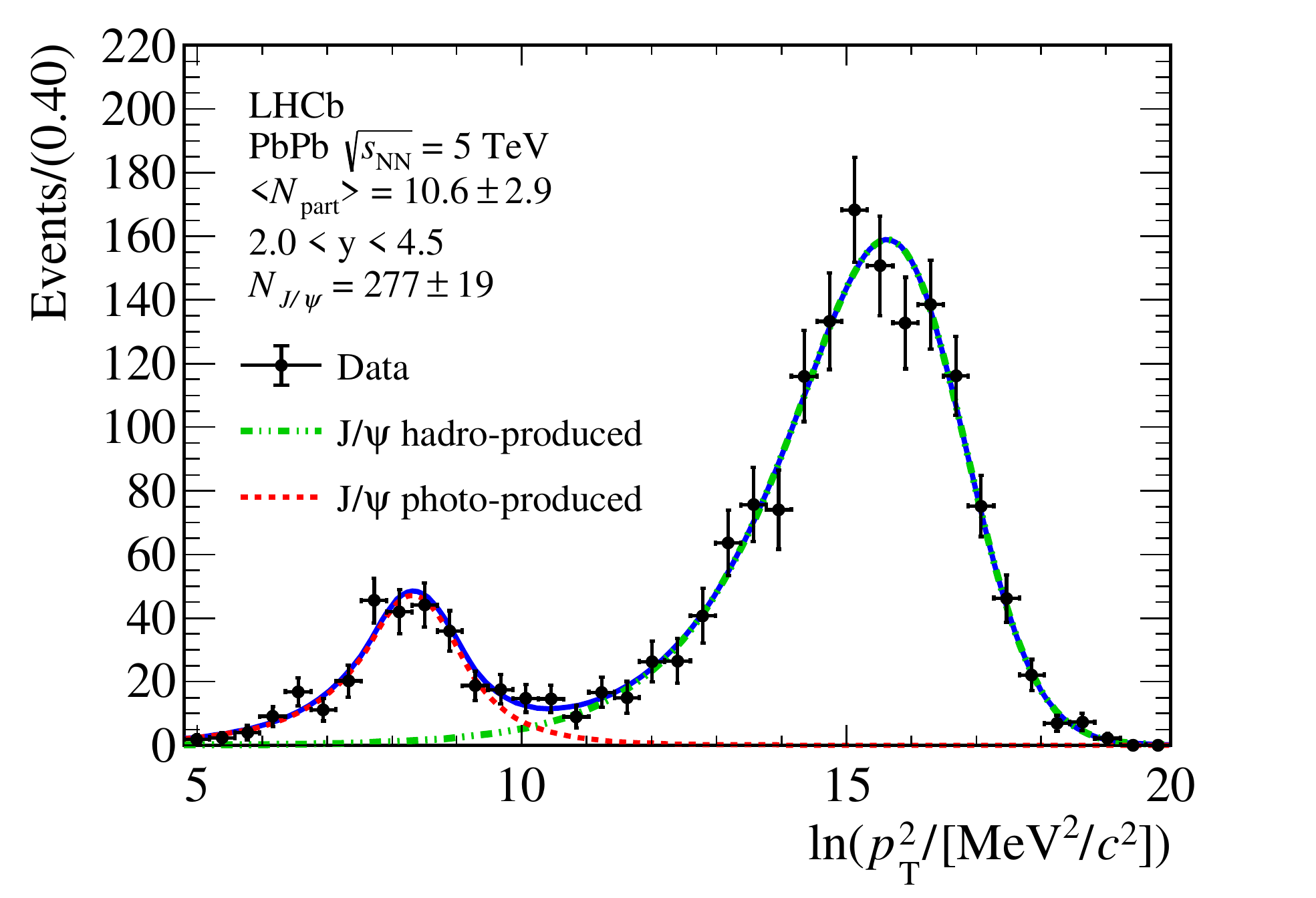}
\caption{The left plot is the invariant mass distribution of $J/\psi$ candidates in $p_{T}$ < 15.0 GeV/c and 2 < $y$ <4.5, with $\left<N_{part}\right>$ = 10.6 $\pm$ 2.9. The right plot is the ${\rm ln} (p^{2}_{T})$ distribution of $J/\psi$ candidates after background subtraction for the same kinematic interval.}
\label{mass_and_pt}
\end{center}
\end{figure}

The $J/\psi$ candidates are selected through the $J/\psi \to \mu^{+}\mu^{-}$ decay channel. The di-muon invariant mass spectrum of the selected candidates in the range between 3.0 and 3.2 GeV is shown in the left of Fig.~\ref{mass_and_pt}, for the $J/\psi$ mesons with $\rm{p_{T}}$ <15.0 GeV/c and the number of participants $\left<N_{part}\right>$ = 10.6 $\pm$ 2.9 in full LHCb rapidity coverage 2 < $y$ <4.5. 
 
The inclusive $J/\psi$ candidates consists of photo-produced and hadronically produced $J/\psi$ mesons, which are separated by an unbinned  maximum likelihood fit to the ${\rm log} (p^{2}_{T})$ distribution, as shown in the right of Fig.~\ref{mass_and_pt}. In this figure, the transverse momentum of photo-produced $J/\psi$ yields (red dotted line) are visible in the range between 0 and 250 MeV/c.

\begin{figure}[htbp]
\begin{center}
\includegraphics[height=0.33\textwidth]{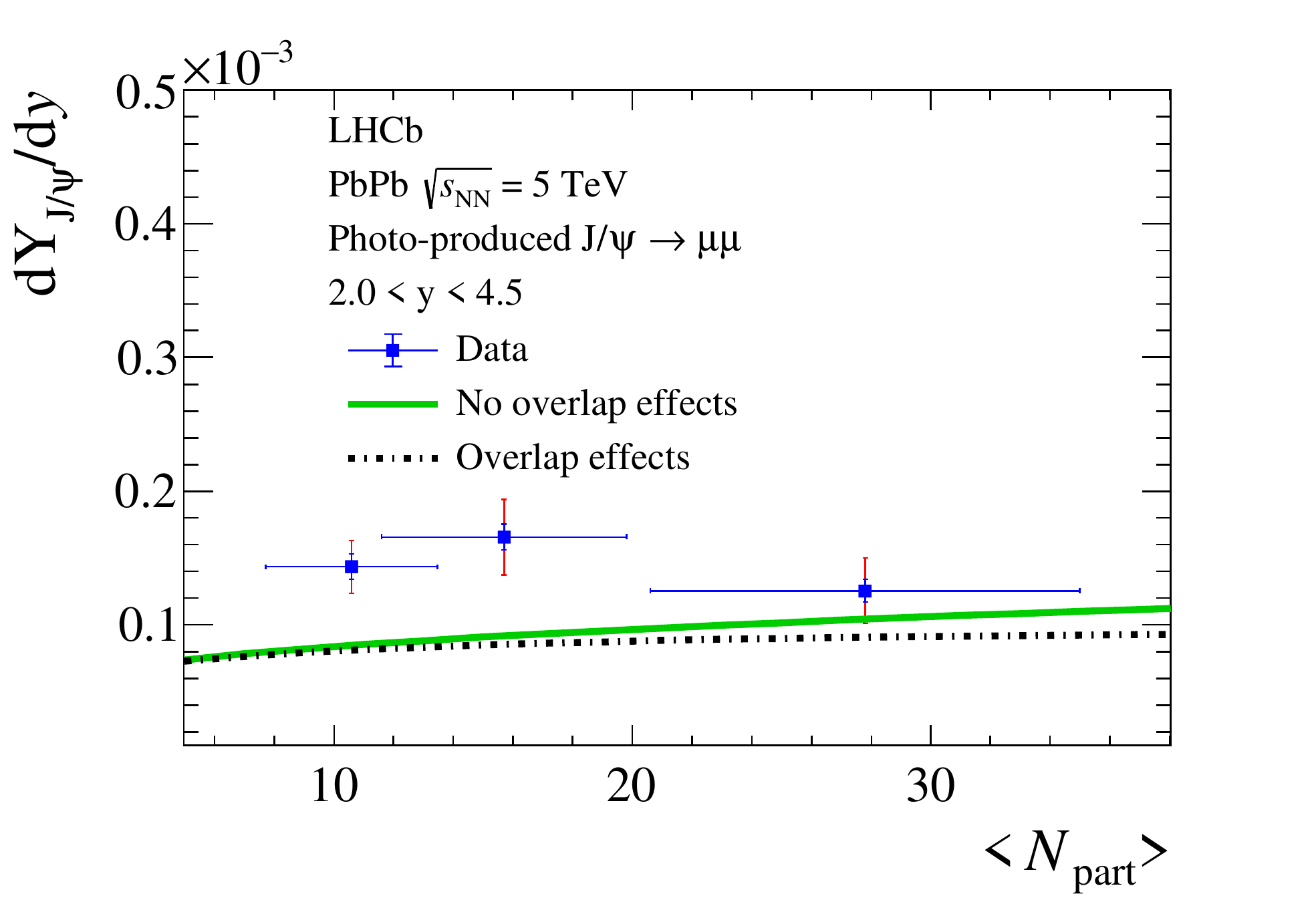}
\includegraphics[height=0.33\textwidth]{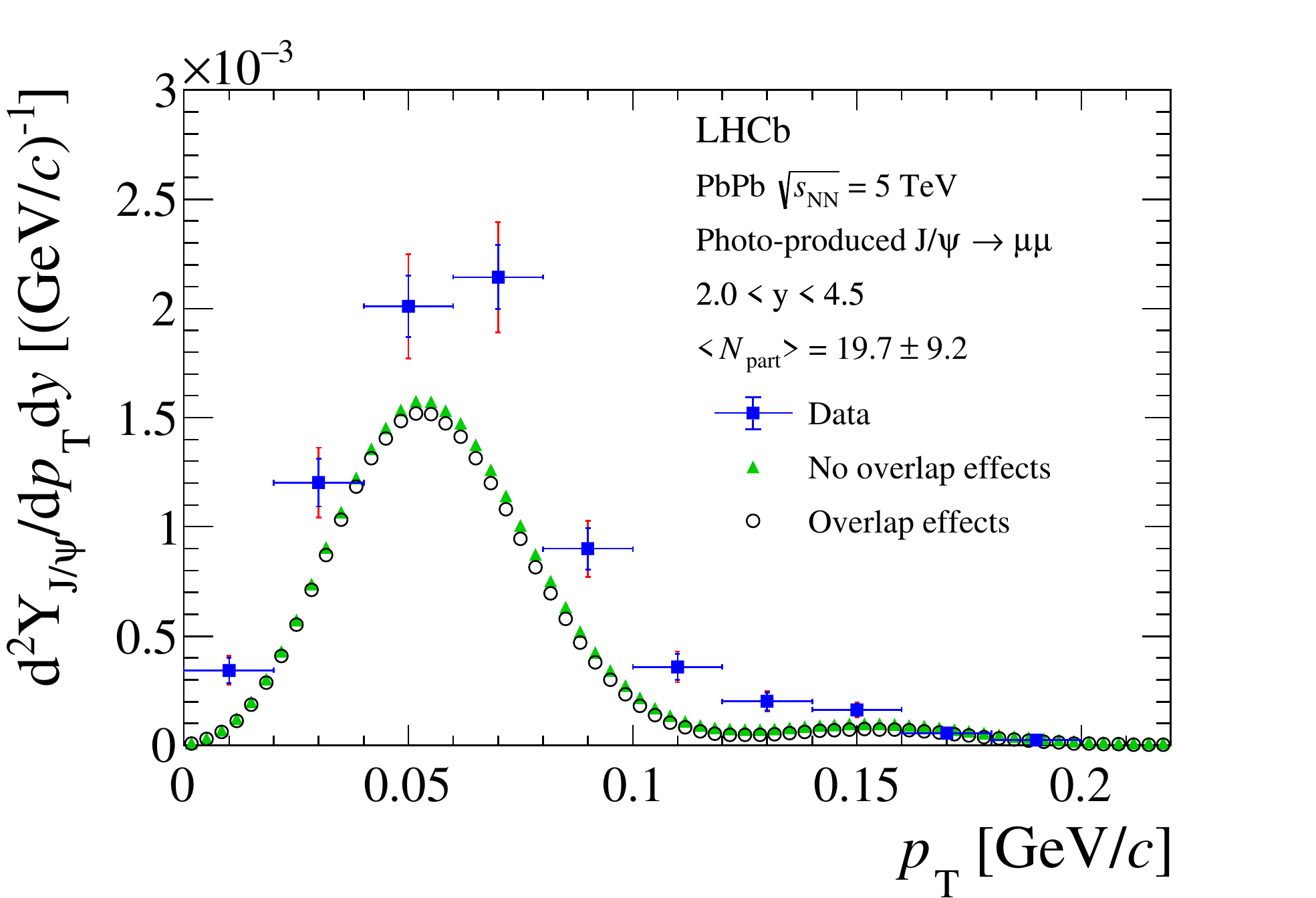}
\caption{Illustration of invariant yields of photo-produced $J/\psi$ mesons as a function of $N_{part}$ and $p_{T}$. Note that the blue error bars represent the statistical uncertainty and the red error bars the total uncertainty. Theoretical model predictions~\cite{PhysRevC.97.044910, PhysRevC.99.061901} are shown in open circles in black and green.}
\label{compare}
\end{center}
\end{figure}

Fig.~\ref{compare} shows the photo-produced $J/\psi$ meson yields as a function of $p_{T}$(right), and $\left<N_{part}\right>$ (left). The mean $p_{T}$ of the coherent $J/\psi$ is estimated to be $\left<p_{T}\right>$ = 64.9 $\pm$ 2.4 MeV/c. Theoretical predictions~\cite{PhysRevC.97.044910, PhysRevC.99.061901} are drawn in open circles, and are qualitatively in agreement with the experimental results in shape.

\section{Conclusions}

LHCb detector's unique geometry acceptance allows us to study the nuclear shadowing in small $x$ region through the UPC collisions. The $J/\psi$ photoproduction in UPC using PbPb collisions at 5TeV is measured by LHCb, and is compared to results by ALICE. The photoproduced $J/\psi$ production in peripheral PbPb collisions is also measured with high precision at very low $p_{T}$, and is compared to theoretical calculations. These results demonstrate the capabilities of the LHCb detector in studying nuclear effects. More results from the large 2018 PbPb dataset are expected in the future.

\nocite{*}
\bibliographystyle{auto_generated}
\bibliography{draft_Weisong_proceedingslowx2021/Weisong}

%% file: giugli_francesco/proceedings_elba2021/giugli_francesco.tex
\vspace*{1.2cm}

\thispagestyle{empty}
\begin{center}
{\LARGE \bf Precision measurements of jet production at the ATLAS experiment}

\par\vspace*{7mm}\par

{

\bigskip

\large \bf Francesco Giuli on behalf of the ATLAS Collaboration\footnote{Copyright 2021 CERN for the benefit of the ATLAS Collaboration. Reproduction of this article or parts of it is allowed as specified in the CC-BY-4.0 license.}}

\bigskip

{\large \bf  E-Mail: francesco.giuli@cern.ch}

\bigskip

{Department of Physics, University of Rome ``Tor Vergata'' and INFN, Section of Rome 2,\\
Via della Ricerca Scientifica 1, 00133, Rome, Italy}

\bigskip

{\it Presented at the Low-$x$ Workshop, Elba Island, Italy, September 27--October 1 2021}

\vspace*{15mm}

\end{center}
\vspace*{1mm}

\begin{abstract}

Measurements of jet production are sensitive to the strong coupling constant, high order perturbative calculations and parton distribution functions. In this talk we present the most recent ATLAS measurements using data from $pp$ collisions at a centre-of-mass energy of $\sqrt{s}$ = 13~TeV. We present measurements of variables probing the properties of the multijet energy flow and of the Lund Plane using charged particles. We will also present new measurements sensitive to the strong coupling constant. For jet fragmentation, we present a measurement of the fragmentation properties of $b$-quark initiated jets, studied using charged $B$ mesons. All results are corrected for detector effects and compared to several Monte Carlo predictions with different parton shower and hadronisation models.
\end{abstract}
  \part[Precision measurements of jet production at the ATLAS experiment\\ \phantom{x}\hspace{4ex}\it{Francesco Giuli on behalf of the ATLAS Collaboration}]{}
 \section{Introduction}
In this proceeding, a review of the most recent ATLAS measurements on jet production is reported. Four different analyses are presented: the first one refers to a measurement of soft-drop jet observables~\cite{ATLAS:2019mgf}, the second one to a measurement of hadronic event shapes in high-$p_{\mathrm{T}}$ multijet final states~\cite{ATLAS:2020vup}, the third one to a measurement of the Lund jet plane using charged particles~\cite{ATLAS:2020bbn} and the last one to a measurement of $b$-quark fragmentation properties in jets using the decay $B^{\pm}\rightarrow J/\psi K^{\pm}$~\cite{ATLAS:2021agf}. All these analyses use $pp$ collisions data collected with the ATLAS detector~\cite{ATLAS:2008xda} at $\sqrt{s}$ = 13~TeV at the Large Hadron Collider (LHC).

\section{Measurement of soft-drop jet observables}
\begin{figure}[t!]
\begin{center}
\includegraphics[width=0.443\textwidth]{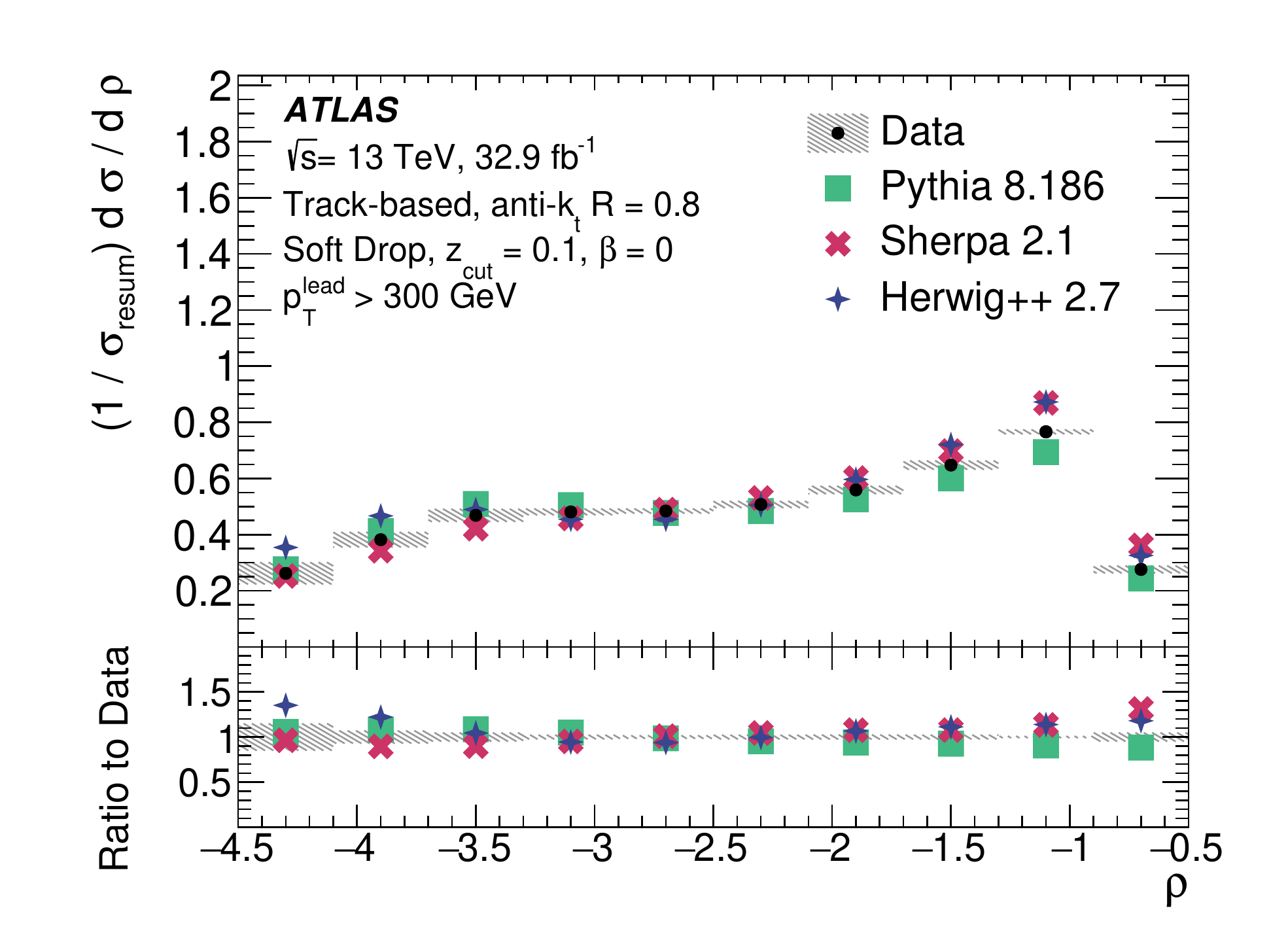}
\includegraphics[width=0.443\textwidth]{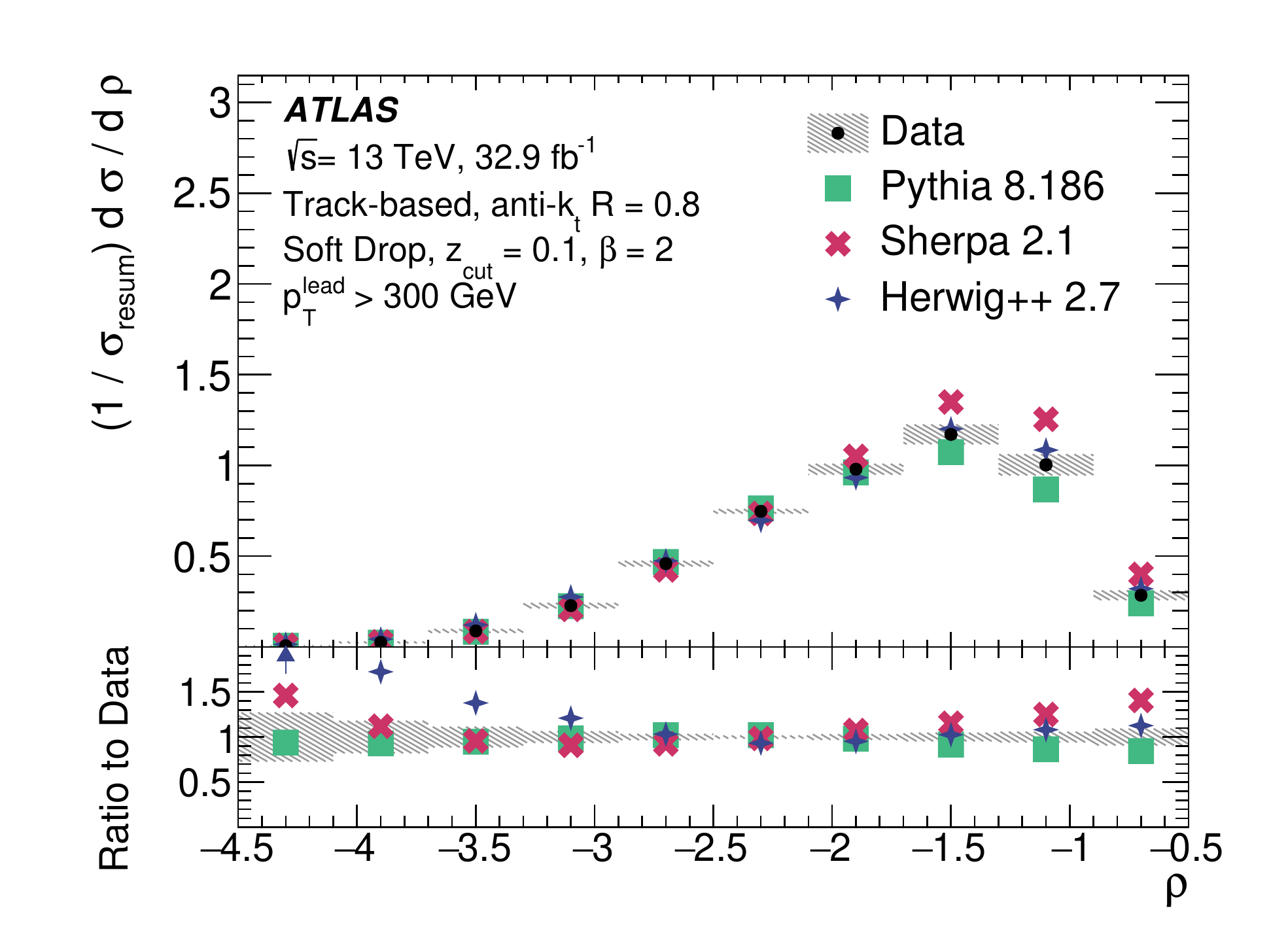}
\end{center}
\caption{Comparison of the unfolded distributions with MC predictions. The uncertainty bands include all sources of systematic uncertainties. Top left: $\rho$, $\beta=0$. Top right: $\rho$, $\beta=0$. These plots are taken from Ref.~\cite{ATLAS:2019mgf}.} 
\label{fig:softdrop}
\end{figure}

\begin{figure}[t!]
\begin{center}
\includegraphics[width=0.443\textwidth]{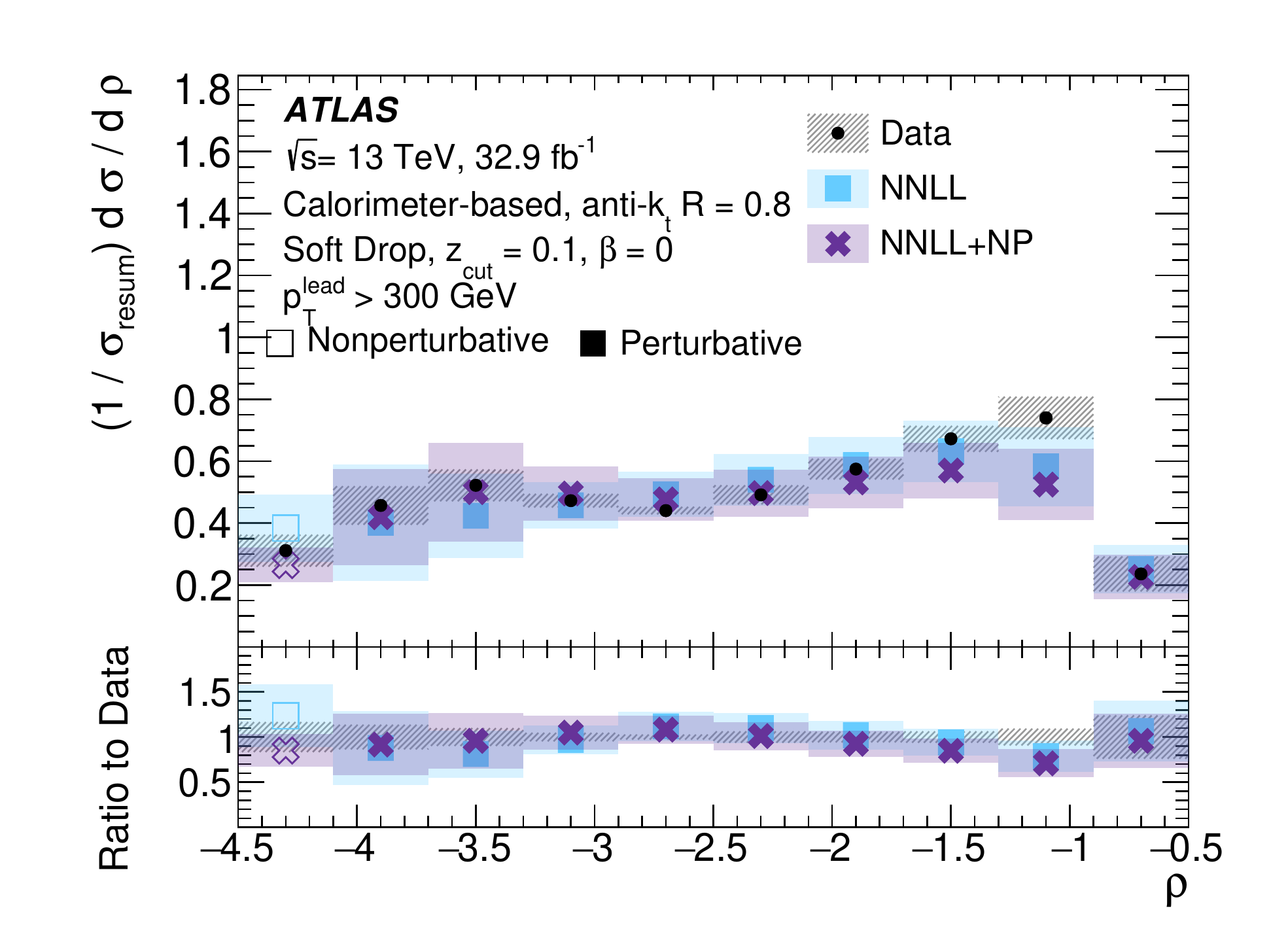}
\includegraphics[width=0.443\textwidth]{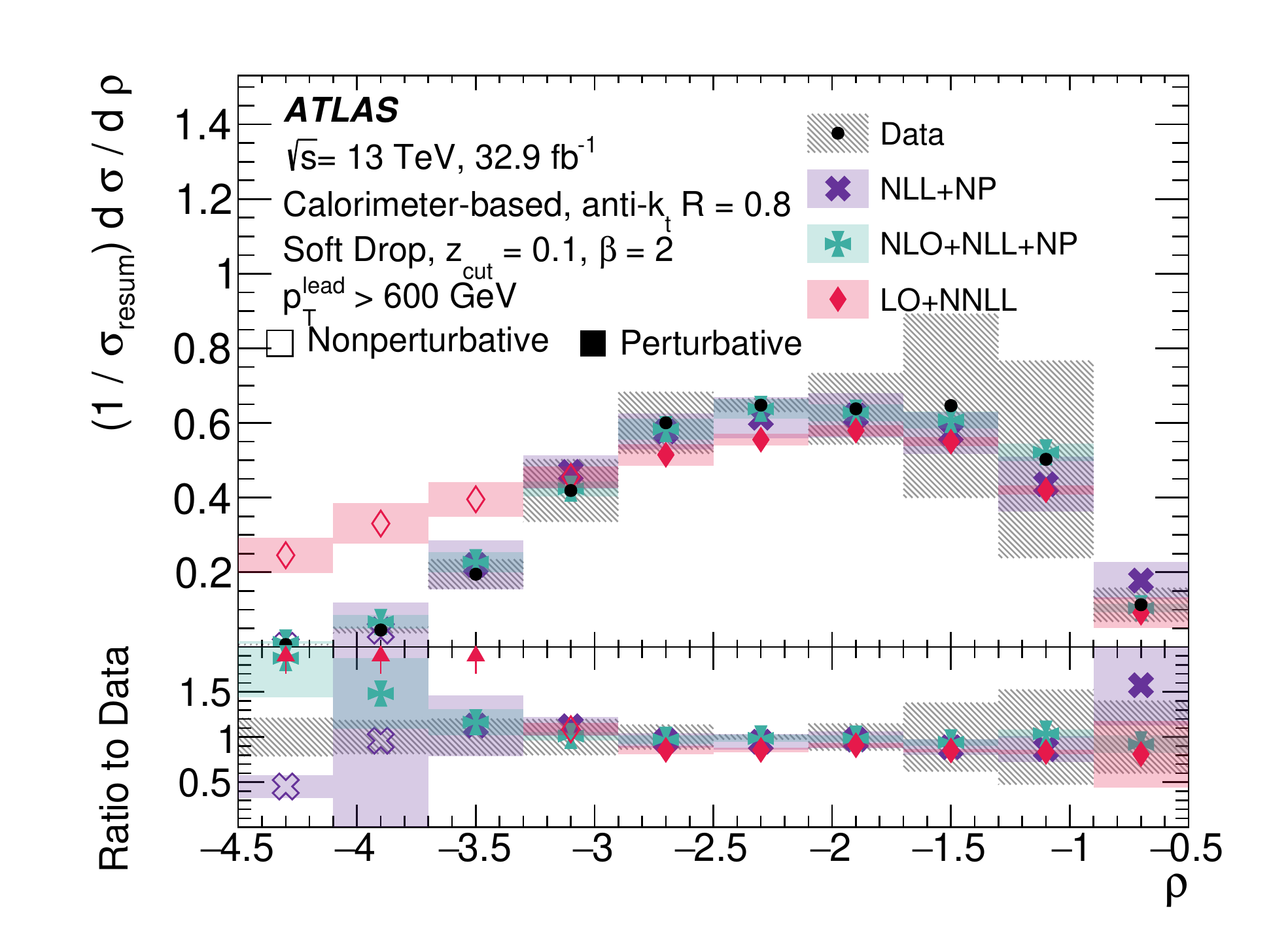}
\end{center}
\caption{Comparison of the unfolded $\rho$ distribution with different theory predictions. The open marker style indicates that non-perturbative effects on the calculation are expected to be large. ``NP'' indicates that non-perturbative corrections have been applied. The uncertainty bands include all sources of systematic uncertainties. Left: $\beta=0$. Right: $\beta=2$. These plots are taken from Ref.~\cite{ATLAS:2019mgf}.} 
\label{fig:softdrop_LL}
\end{figure}

Jet substructure quantities are measured using jets groomed with the \textit{soft-drop} grooming procedure~\cite{Larkoski:2014wba} in dijet events from data corresponding to an integrated luminosity of 32.9 fb$^{-1}$. This algorithm proceeds as follows. After a jet is clustered using any algorithm, its constituents are reclustered using the Cambrigde-Aachen (C/A) algorithm~\cite{Dokshitzer:1997in,Wobisch:1998wt}, which iteratively clusters the closest constituents in azimuth and rapidity. Then, the last step of the C/A clustering algorithm is undone, breaking the jet $j$ into two subjets, namely $j_{1}$ and $j_{2}$, which are used to evaluate the soft-drop condition:
\begin{equation}
\dfrac{\min(p_{\mathrm{T},j_{1}},p_{\mathrm{T},j_{2}})}{p_{\mathrm{T},j_{1}} + p_{\mathrm{T},j_{2}}} > \left(\dfrac{\Delta R_{12}}{R}\right)^{\beta},
\end{equation}
where $\Delta R_{12}$ is the distance between the two subjets, $R$ represents the jet radius and $p_{\mathrm{T},j_{i}}$ is the transverse momentum of the subjet $j_{i}$. The parameters $\beta$ and $z_{\mathrm{cut}}$ are algorithm parameters which determine the sensitivity of the algorithm to soft and wide-angle radiation. If the two subjets fail the soft-drop condition, the subjet characterised by the lower $p_{\mathrm{T}}$ is removed, and the other subjet is relabelled as $j$ and the procedure is iterated. When the soft-drop condition is satisfied, the algorithm is stopped, and the resulting jet is the soft-dropped jet.\\
This analysis presents two closely related substructure observables, which are calculated from jets after they have been groomed with the soft-drop algorithm: 
\begin{itemize}
\item the dimensionless version of the jet mass, $\rho=\log(m^{2}/p_{\mathrm{T}}^{2})$;
\item the opening angle between the two subjets that pass the soft-drop condition, $r_{g}$.
\end{itemize}
The unfolded data are compared to Monte Carlo (MC) events generated at leading order (LO) with \texttt{PYTHIA8.186}~\cite{Sjostrand:2006za,Sjostrand:2007gs}, \texttt{SHERPA2.1}~\cite{Gleisberg:2008ta,Sherpa:2019gpd} and \texttt{HERWIG++ 2.7}~\cite{Bahr:2008pv,Corcella:2000bw}, as reported in Figure~\ref{fig:softdrop}.
Several trends are visible in these results. For $\rho$, the MC predictions are mostly accurate within 10\% except for the lowest relative masses, which are dominated by non-perturbative physical effects. This becomes more visible for larger values of $\beta$, where more soft radiation is included within the jet, increasing the size of the non-perturbative effects. In addition, in the high-relative-mass region, where the effects of the fixed-order (FO) calculation are relevant, some differences between MC generators are seen. A similar trend may be seen for $r_{g}$, where the small-angle region (i.e. where non-perturbative effects are largest) shows more pronounced differences between MC generators.\\
Several calculations have been performed to predict the $\rho$ distributions, and unfolded data are compared with these predictions (more details on the predictions can be found in Ref.~\cite{ATLAS:2019mgf}), as shown in Figure~\ref{fig:softdrop_LL}. The LO+next-to-next-to-leading-logarithm (NNLL) and next-to-leading-order+next-to-leading-logarithm (NLO+NLL) calculations are able to model the data in the resummation region ($-3\lesssim\rho\lesssim -1$), with the NLO+NLL calculation providing an accurate description of the data for high values of $\rho$. We can also observe how, in the region where the FO effects are dominant, the LO+NNLL and NNLL calculations do not model data well. This behaviour is expected, since the calculations do not include terms beyond LO at Matrix-Element (ME) level.

\section{Measurement of hadronic event shapes}
Event shapes~\cite{Banfi:2004nk,Banfi:2010xy} are a class of observables that describe the dynamics of energy flows in multijet final states, they are usually defined to be infrared and collinear safe and they are sensitive to different aspects of the theoretical description of strong-interaction processes. For example, hard, wide-angle radiation is studied by investigating the tails of these distributions, while other regions of the event-shape distributions provide information about anisotropic, back-to-back configurations, which are sensitive to the details of the resummation of soft logarithms in the theoretical predictions.\\
The dataset used in this analysis comprises the 2015-2018 data taking period, corresponding to an integrated luminosity of 139 fb$^{-1}$. In this paper, several event-shape variable are presented. For each event, the thrust axis $\hat{n}_{\mathrm{T}}$ is defined as the direction with respect to the jet momentum $p_{\mathrm{T}}$ is maximised~\cite{Brandt:1964sa,Farhi:1977sg}. The transverse thrust $T_{\perp}$ and its minor component $T_{\mathrm{m}}$ can be expressed as:
\begin{equation}
T_{\perp} = \dfrac{\sum_{i}|\vec{p}_{\mathrm{T},i}\cdot\hat{n}_{\mathrm{T}}|}{\sum_{i}|\vec{p}_{\mathrm{T},i}|}; \;\;\;\;\; T_{\mathrm{m}} = \dfrac{\sum_{i}|\vec{p}_{\mathrm{T},i}\times\hat{n}_{\mathrm{T}}|}{\sum_{i}|\vec{p}_{\mathrm{T},i}|},
\end{equation}
where the index $i$ runs over all jets in the event. These two quantities are useful to define $\tau_{\perp}=1-T_{\perp}$.\\
Several MC samples were used for this analysis, and they were produced using \texttt{PYTHIA8.235}~\cite{Sjostrand:2014zea}, \texttt{SHERPA2.1}~\cite{Gleisberg:2008ta,Sherpa:2019gpd}, \texttt{HERWIG7.1.3}~\cite{Bellm:2017bvx} and \texttt{MadGraph5{\_}aMC@NLO 2.3.3}~\cite{Alwall:2014hca_giugli}, together with \texttt{PYTHIA8.212}~\cite{Sjostrand:2014zea} (hereafter referred to as \texttt{MG5{\_}aMC}). Unfolded data are compared to the above-mentioned MC predictions in various bins of the jet multiplicity, $n^{\mathrm{jet}}$ ($=$ 2, 3, 4, 5 and $\geq$ 6), and the scalar sum of transverse momenta of the two leading jets, $H_{\mathrm{T}2}=p_{\mathrm{T1}}+p_{\mathrm{T2}}$ (1~TeV $<H_{\mathrm{T}2}<$ 1.5~TeV, 1.5~TeV $<H_{\mathrm{T}2}<$ 2.0~TeV and $H_{\mathrm{T}2}>$ 2.0~TeV).\\
\begin{figure}[t!]
\begin{center}
\includegraphics[width=0.8\textwidth]{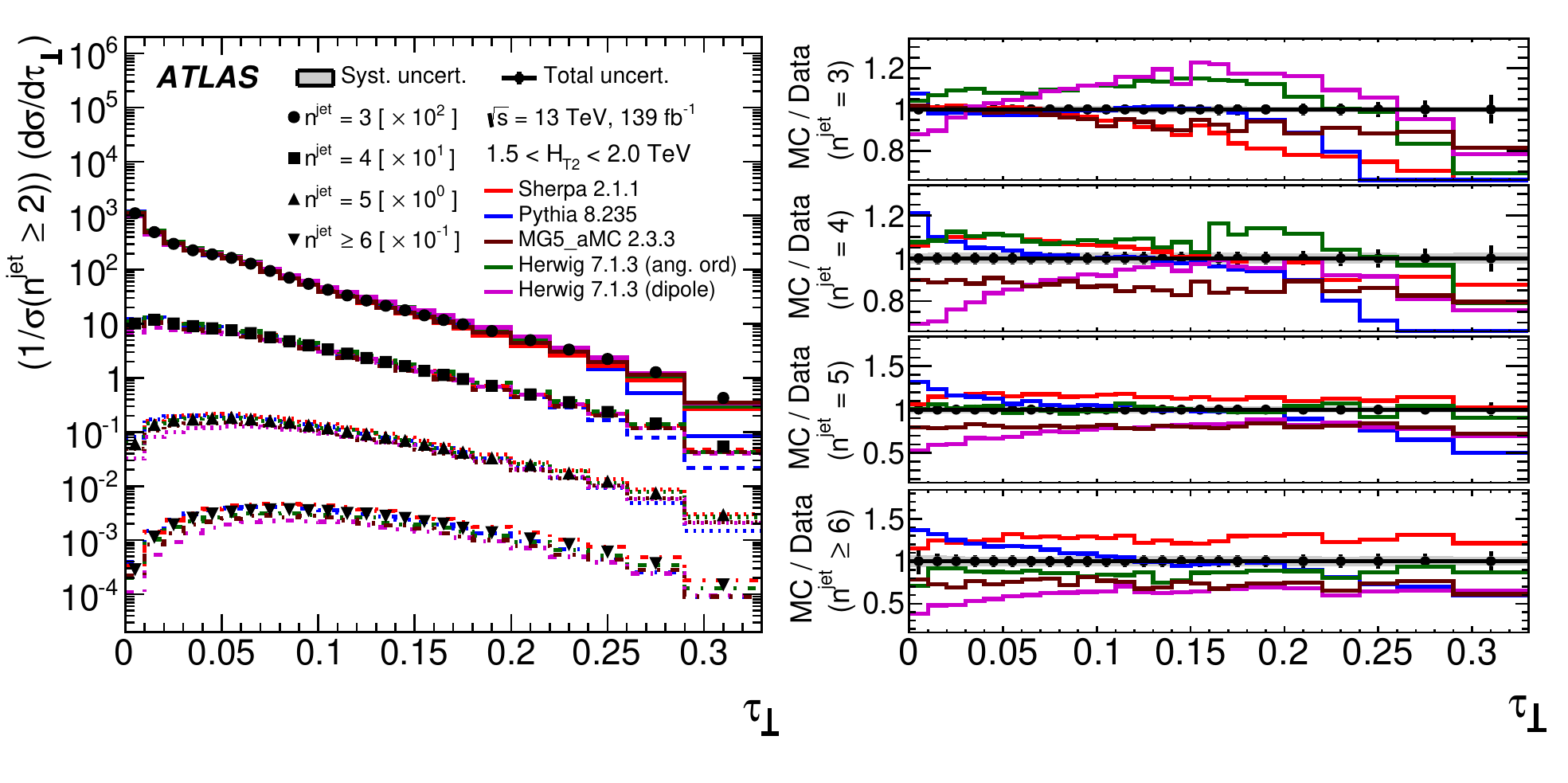}
\includegraphics[width=0.8\textwidth]{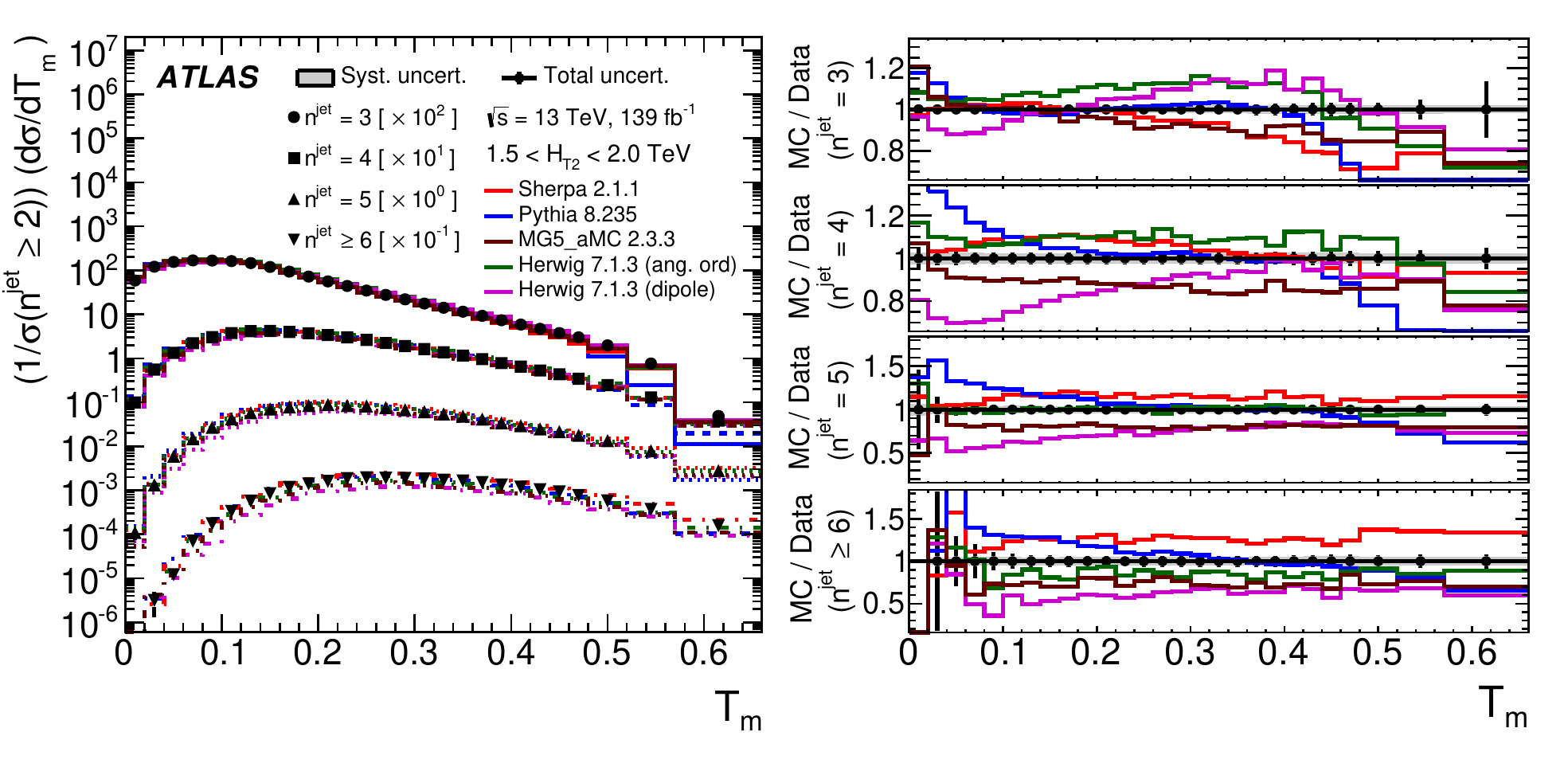}
\end{center}
\caption{Comparison between data and MC simulation for different jet multiplicities in the 1.5~TeV $<H_{\mathrm{T}2}<$ 2.0~TeV bin. The right panels show the ratios between the MC and the data distributions. The error bars show the total uncertainty (statistical and systematic added in quadrature) and the grey bands in the right panels show the systematic uncertainty. Top: normalised cross section as a function of $\tau_{\perp}$. Bottom: normalised cross section as a function of $T_{\mathrm{m}}$. These plots are taken from Ref.~\cite{ATLAS:2020vup}.} 
\label{fig:eventshape}
\end{figure}
The normalised cross section as a function of $\tau_{\perp}$ and $T_{\mathrm{m}}$ is shown in Figure~\ref{fig:eventshape}. The MC simulations tend to underestimate the data in the intermediate region of $\tau_{\perp}$ for low jet multiplicities, while the measurements are underestimated by all MC predictions at high $\tau_{\perp}$ values. The shape of the distributions tends to agree with data for larger $n^{\mathrm{jet}}$. \texttt{HERWIG7} prediction based on dipole shower highly underestimates the ATLAS data at low values of $\tau_{\perp}$, whereas the measurements are overestimated by \texttt{PYTHIA8} in such region. Very similar conclusions can be drawn looking at the normalised cross section as a function of $T_{\mathrm{m}}$. \texttt{Sherpa} simulations predict fewer isotropic events than in data, while the \texttt{MG5{\_}aMC} predictions are closer to the measurements. As regards the $H_{\mathrm{T}2}$-dependence of the depicted results, it has been found that there are more isotropic events at low energies, with increasing alignment of jets with the thrust jet axis for higher energy scales.\\
In summary, none of the MC predictions provide a good description of the ATLAS measurements in all the regions of the phase space. \texttt{HERWIG7} and \texttt{MG5{\_}aMC} computations are closer to data (but with significant discrepancies), remarking the limited ability of parton shower (PS) models to simulate hard and wide angle radiation and further emphasising that the addition of $2\rightarrow3$ processes in the ME allows to improve the description of measured data.

\section{Measurement of the Lund jet plane}
\begin{figure}[t!]
\begin{center}
\includegraphics[width=0.43\textwidth]{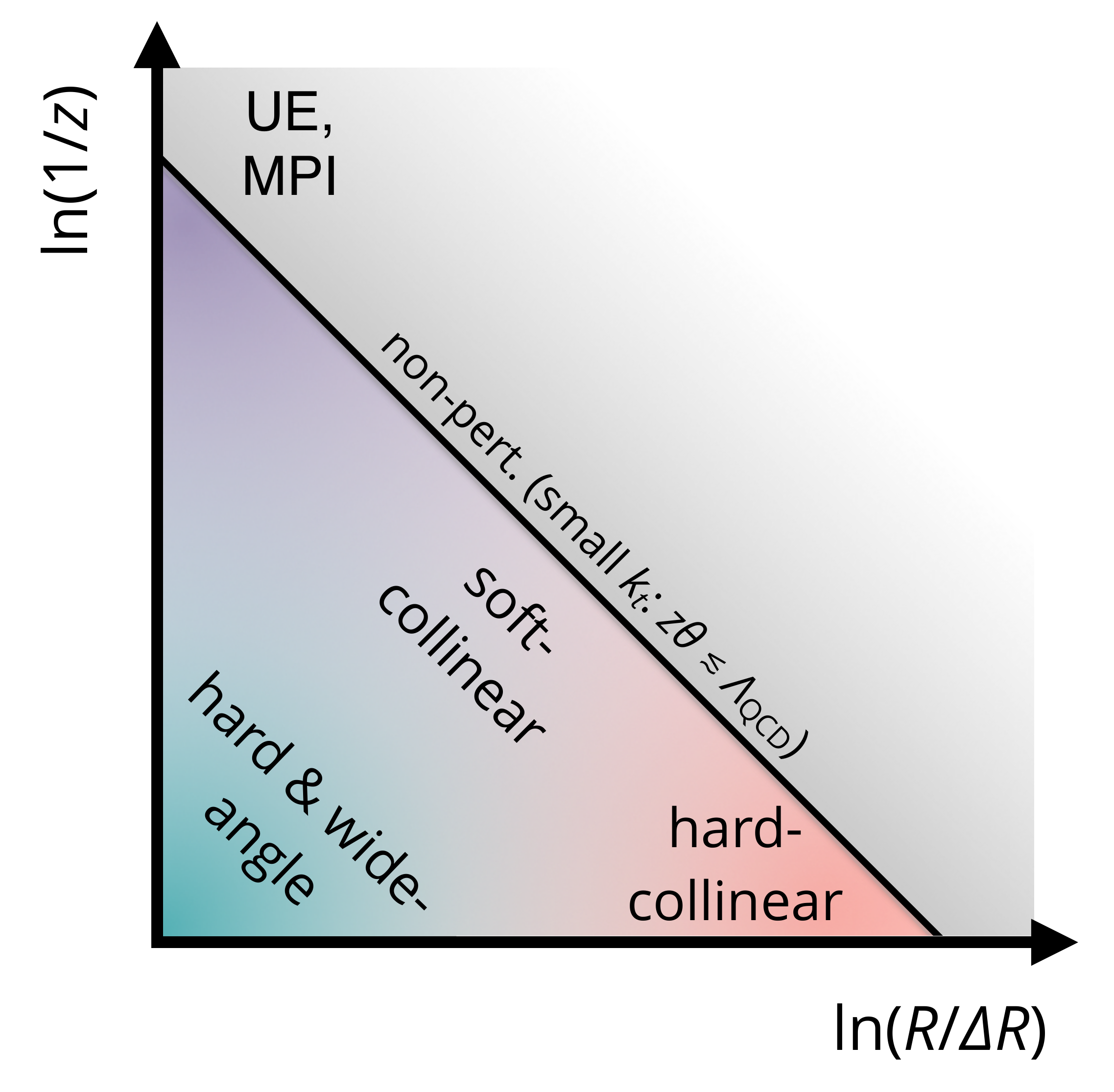}
\includegraphics[width=0.43\textwidth]{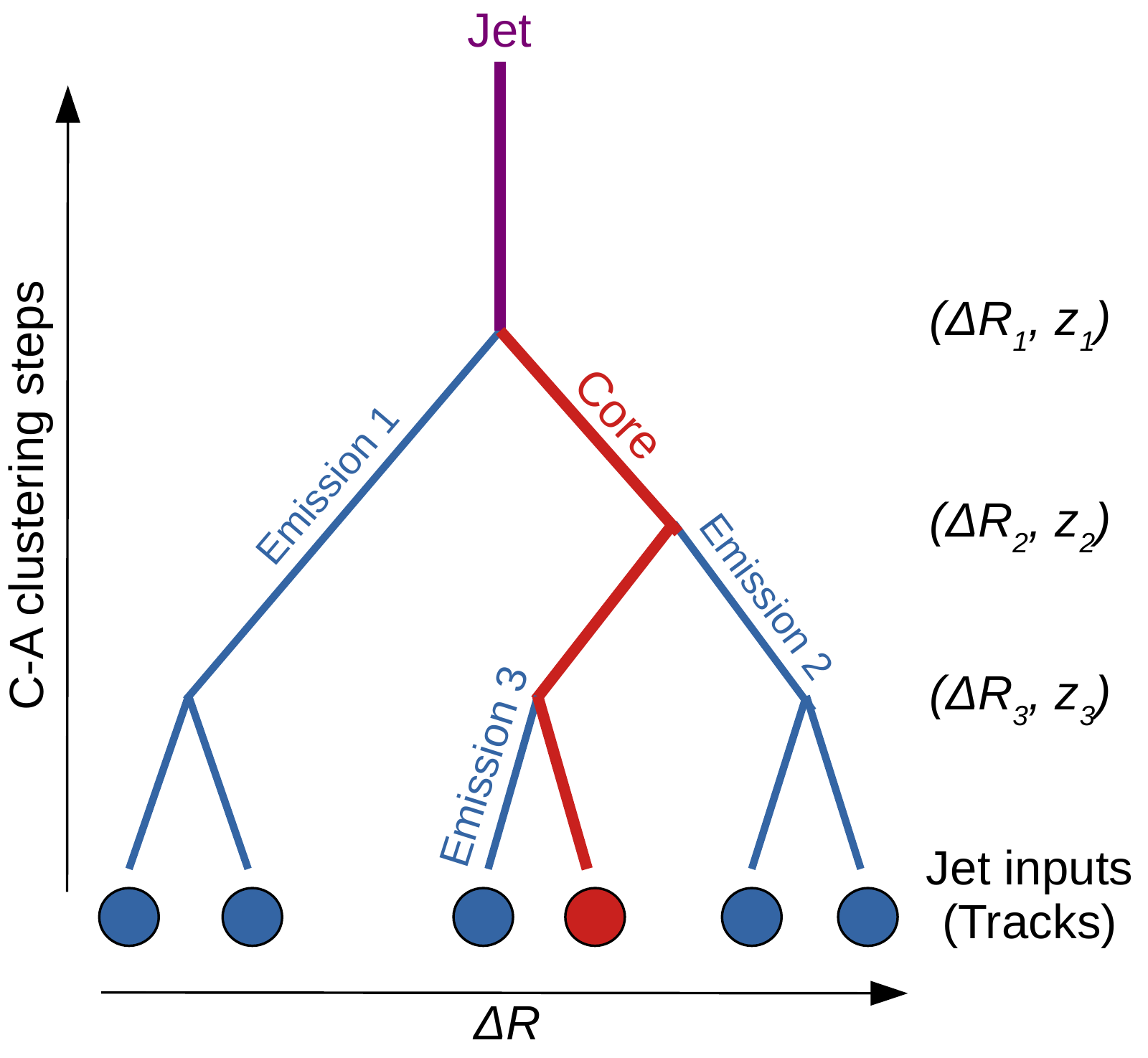}
\end{center}
\caption{Schematic representation of the LJP. The left-hand side plots is taken from Ref.~\cite{ATLAS:2020bbn}.} 
\label{fig:LJP}
\end{figure}
\begin{figure}[t!]
\begin{center}
\includegraphics[width=0.443\textwidth]{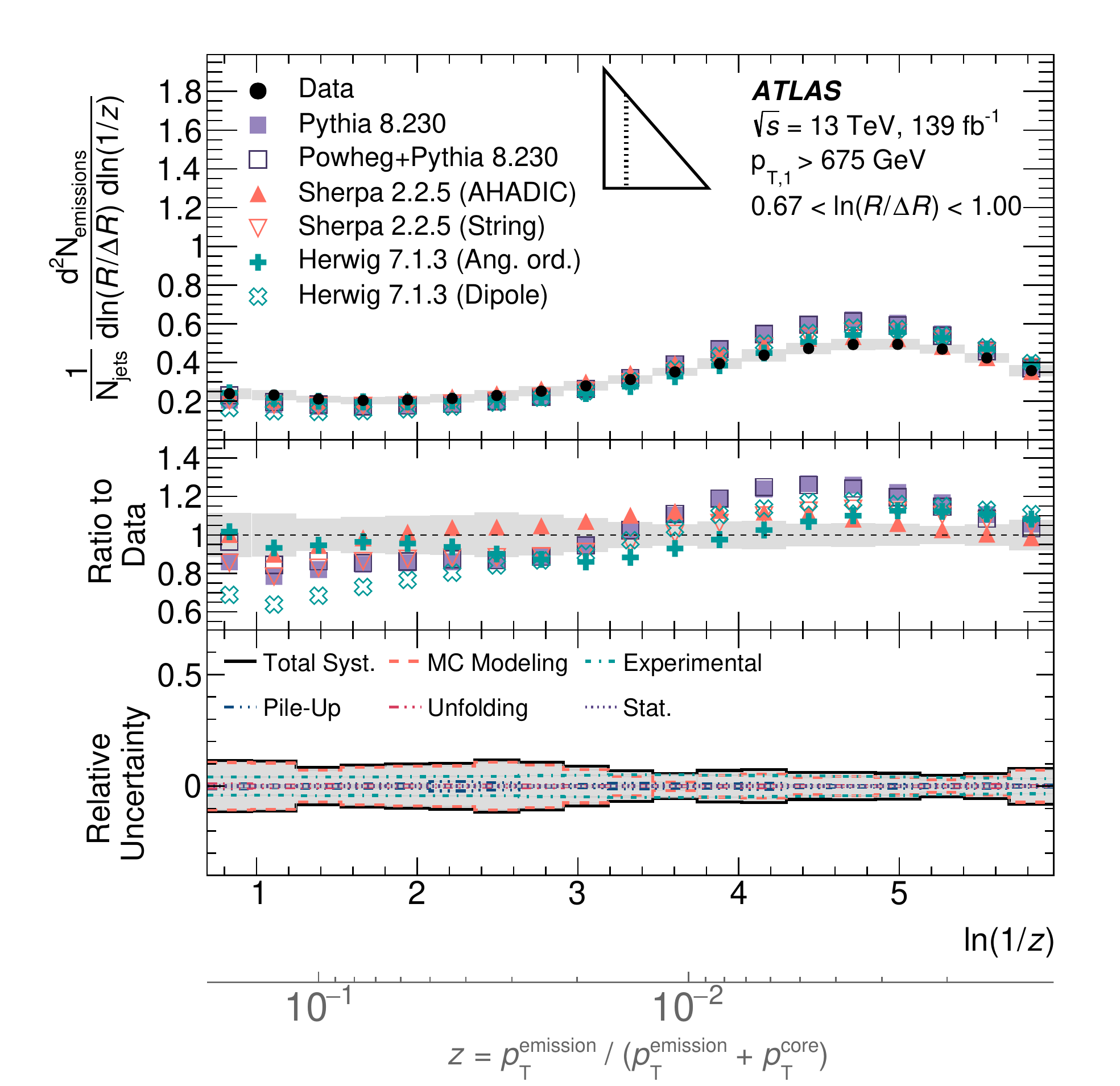}
\includegraphics[width=0.443\textwidth]{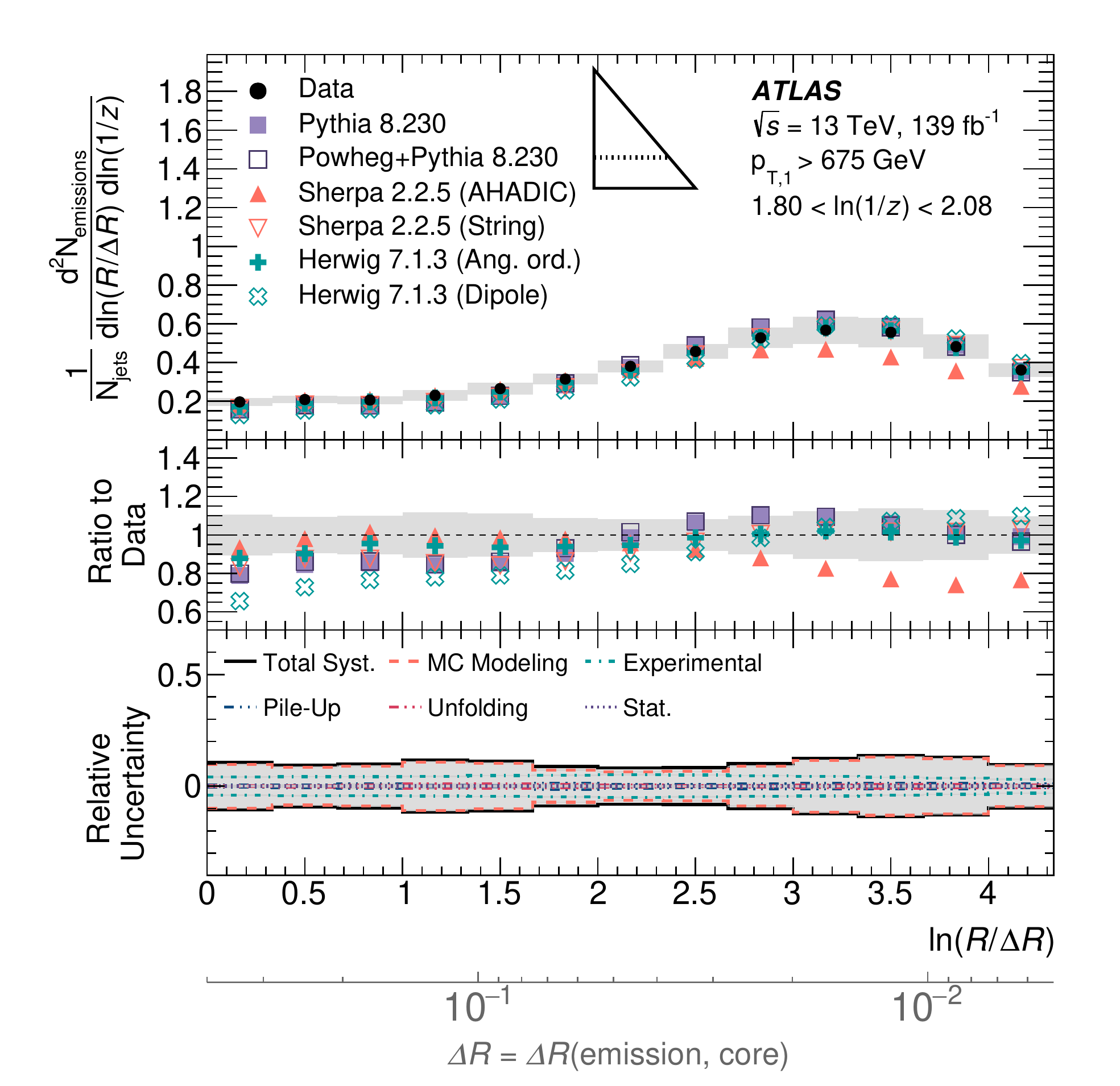}
\end{center}
\caption{Representative horizontal and vertical slices through the LJP. Unfolded data are compared with particle-level simulation from several MC generators. The uncertainty band includes all sources of systematic and statistical uncertainty. The inset triangle illustrates which slice of the plane is depicted Left: 0.67 $<\ln(R/\Delta R)<$ 1.0. Right: 1.80 $<\ln(1/z)<$ 2.08. These plots are taken from Ref.~\cite{ATLAS:2020bbn}.} 
\label{fig:LJP_data}
\end{figure}

The Lund plane is a powerful representation for providing insight into jet substructure. A recent proposal~\cite{Dreyer:2018nbf} describes a method to construct an observable analog of the Lund plane using jets, which captures the salient features of this representation. Jets are formed using clustering algorithms that sequentially combine pairs of proto-jets starting from the initial set of constituents~\cite{Salam:2010nqg}. In this proposal, a jet's constituents are reclustered using the C/A algorithm~\cite{Dokshitzer:1997in,Wobisch:1998wt}. Then, the C/A history is reversed, and each jet is declustered, starting from the hardest proto-jet. The Lund plane can be approximated by using the harder (softer) proto-jet to present the core (emission) in the original theoretical depiction. For each proto-jet pair, at each step in the C/A declustering sequence, an entry is made in the primary Lund Jet plane (LJP) through the observables $\ln(1/z)$ and $\ln(R/\Delta R)$, with
\begin{equation}
z=\dfrac{p_{\mathrm{T}}^{\mathrm{emission}}}{p_{\mathrm{T}}^{\mathrm{emission}}+p_{\mathrm{T}}^{\mathrm{core}}} \;\; \mathrm{and} \;\; \Delta R^{2} = (y_{\mathrm{emission}} - y_{\mathrm{core}})^{2} + (\phi_{\mathrm{emission}} - \phi_{\mathrm{core}})^{2}.
\end{equation}
A schematic representation of the LJP can be found in Figure~\ref{fig:LJP}.\\
This measurement is conducted using the full Run 2 statistics, for an integrated luminosity of 139 fb$^{-1}$. To perform the data unfolding, several samples of dijet events were simulated. \texttt{PYTHIA8.186}  \cite{Sjostrand:2006za,Sjostrand:2007gs} was used for simulating the nominal sample. Additional samples were simulated using NLO MEs from \texttt{POWHEG}~\cite{Nason:2004rx_giugli,Frixione:2007vw_giugli,Alioli:2010xa,Alioli:2010xd_giugli} and \texttt{Sherpa2.2.5}~\cite{Sherpa:2019gpd} or \texttt{HERWIG 7.1.3}~\cite{Bellm:2017bvx}.\\
The data from two seleceted slices of the LJP, together with the breakdown of the major systematic uncertainties, are shown in Figure~\ref{fig:LJP_data}. ATLAS data and several MC predictions are compared. It is visible how the \texttt{Herwig7.1.3} angle-ordered prediction provides the best description across most of the plane, while any prediction describes the data accurately in all the regions. The differences in the hadronization algorithms implemented in \texttt{Sherpa2.2.5} are particularly visible at the transition between perturbtive and non-perturbative regions of the plane. The \texttt{POWHEG+PYTHIA} and \texttt{PYTHIA} predictions only differ significantly for hard and wide-angle perturbative emissions, where ME corrections are relevant.

\section{Measurement of $b$-quark fragmentation properties}
\begin{figure}[H]
\begin{center}
\includegraphics[width=0.443\textwidth]{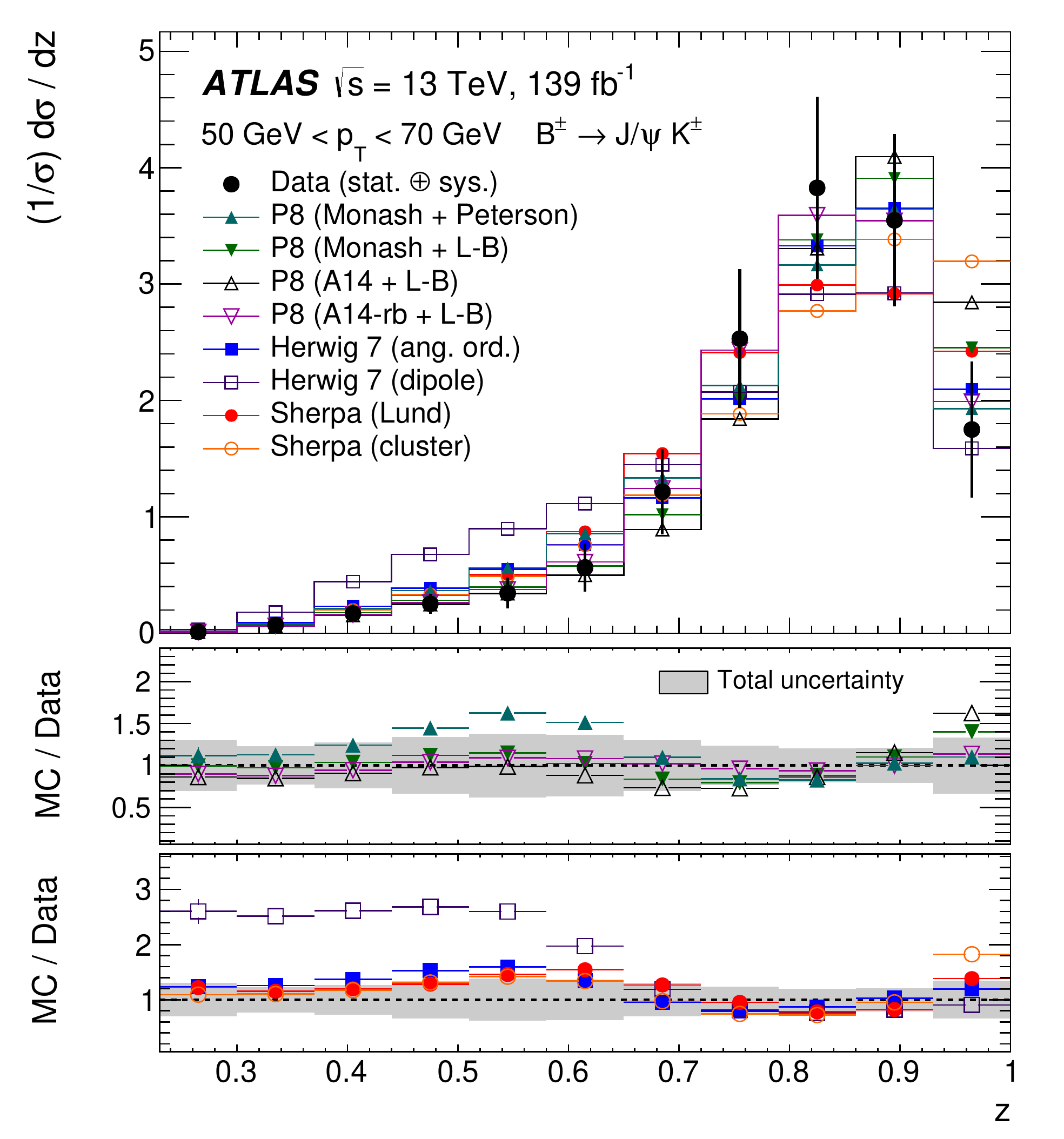}
\includegraphics[width=0.443\textwidth]{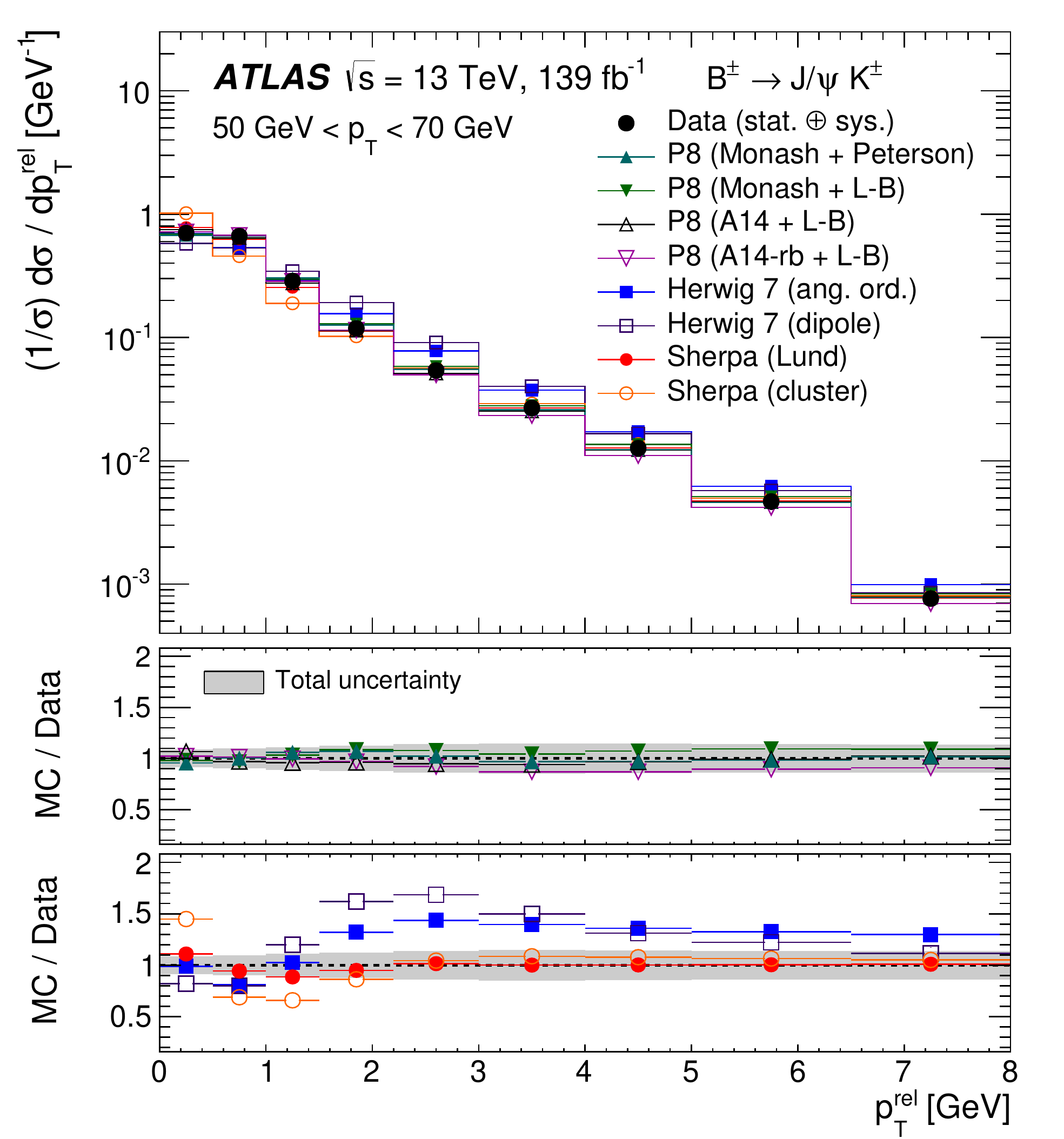}
\includegraphics[width=0.443\textwidth]{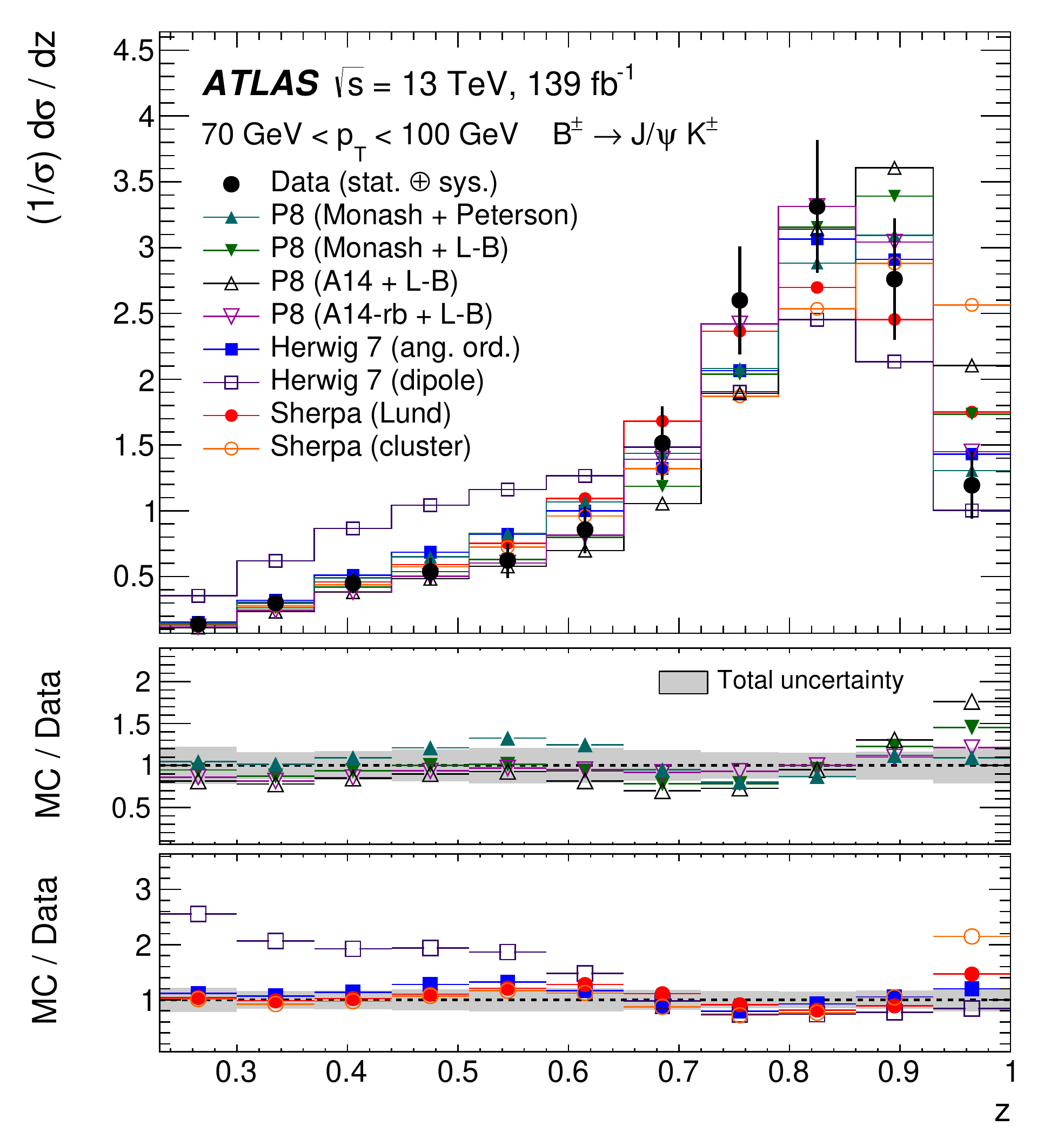}
\includegraphics[width=0.443\textwidth]{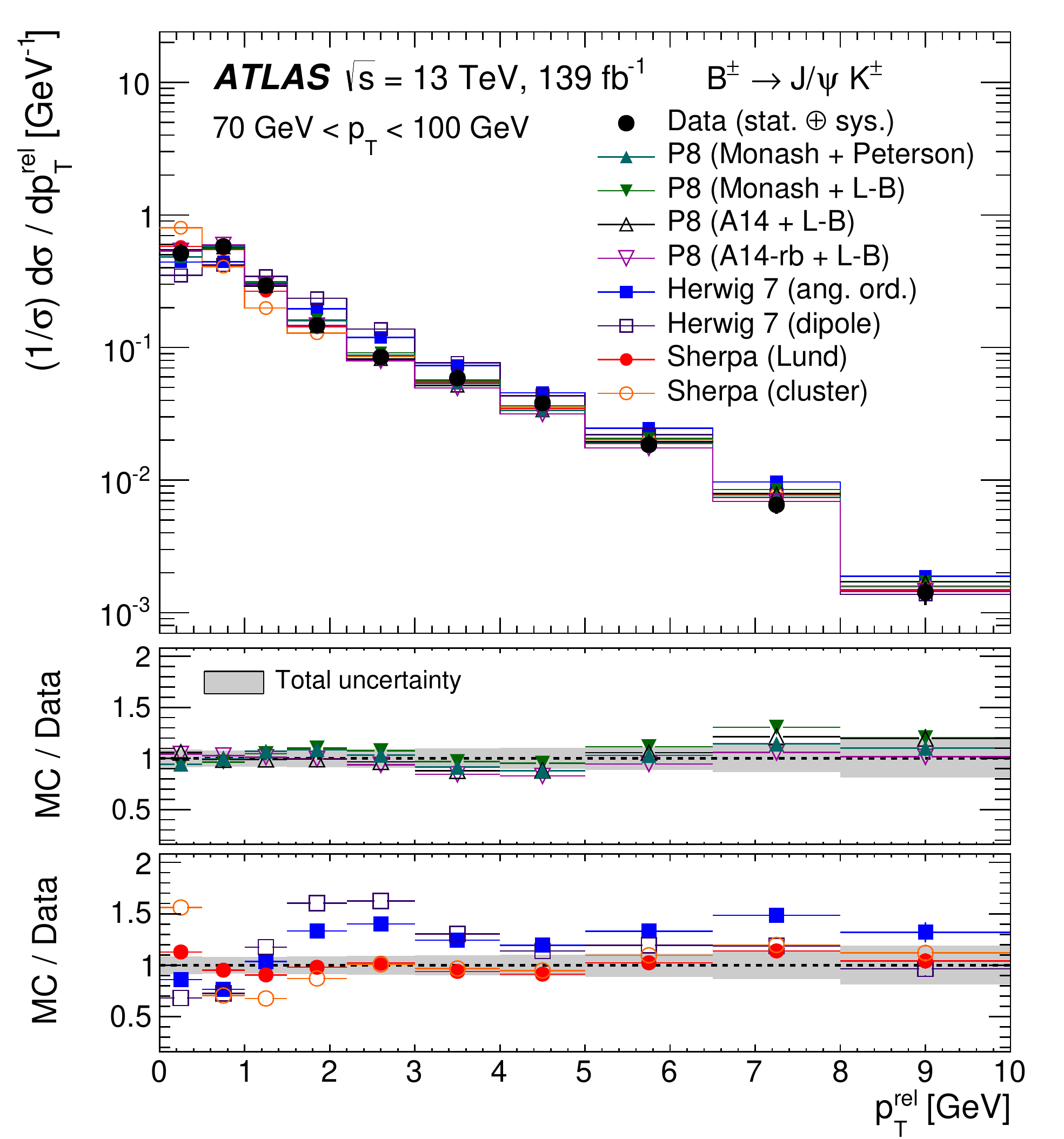}
\includegraphics[width=0.443\textwidth]{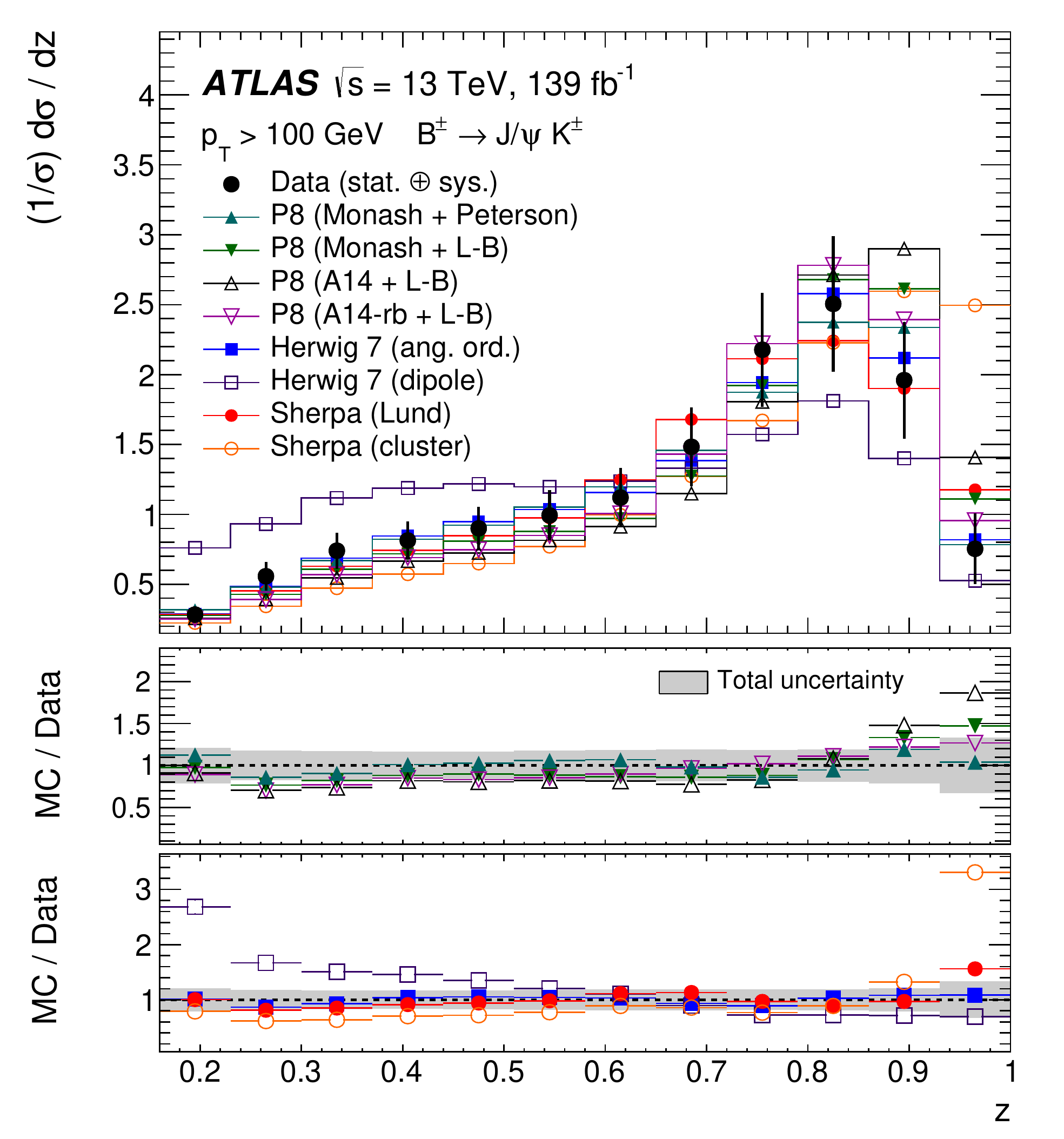}
\includegraphics[width=0.443\textwidth]{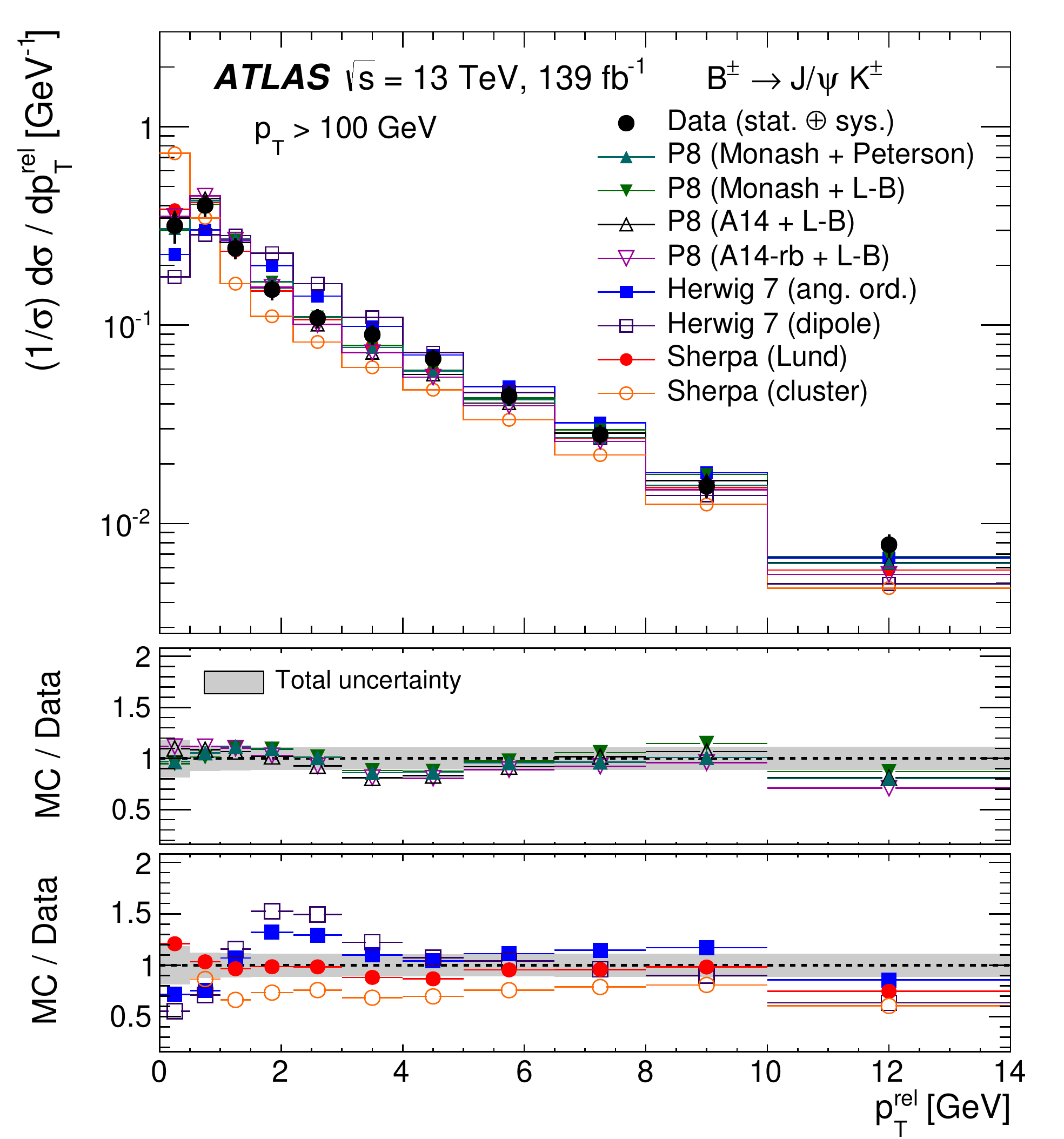}
\end{center}
\caption{Distributions of the longitudinal profile $z$ and the transverse profile $p_{\mathrm{T}}^{\mathrm{rel}}$ together with different predictions from \texttt{PYTHIA8}, \texttt{SHERPA} and \texttt{HERWIG 7}. The vertical error bars represent the total experimental uncertainties. Top: 50 $<p_{\mathrm{T}}^{\mathrm{jet}}<$ 70~GeV. Middle: 70 $<p_{\mathrm{T}}^{\mathrm{jet}}<$ 100~GeV. Bottom: $p_{\mathrm{T}}^{\mathrm{jet}}>$ 100~GeV. These plots are taken from Ref.~\cite{ATLAS:2021agf}.} 
\label{fig:bfrag}
\end{figure}

The fragmentation of heavy quarks is a crucial aspect of Quantum ChromoDynamics (QCD). Detailed studies and precision measurements of the heavy-quark fragmentation properties allow a deeper understanding of QCD. The MC predictions used at the LHC are tuned to describe the measurements in $e^{+}e^{-}$ collisions at relatively low $\sqrt{s}$. Therefore, new measurements of $b$-quark fragmentation can be used to improve MC simulations at LHC energy scales.\\
This analysis presents a measurement of $b$-quark fragmentation into $B^{\pm}$ mesons and it uses the full Run 2 data set, corresponding to an integrated luminosity of 139 fb$^{-1}$. The $B^{\pm}$ mesons are then recostructed via the $B^{\pm}\rightarrow J/\psi K^{\pm}\rightarrow\mu^{+}\mu^{-}K^{\pm}$ decay chain. After the matching between the jet and the reconstructed $B$ meson, two variables of interest are built as follows:
\begin{equation}
z=\dfrac{\vec{p}_{B}\cdot\vec{p}_{j}}{|\vec{p}_{j}|^{2}} \;\; \mathrm{and} \;\; p_{\mathrm{T}}^{\mathrm{rel}}=\dfrac{\vec{p}_{B}\times\vec{p}_{j}}{|\vec{p}_{j}|},
\end{equation}
where $\vec{p}_{B}$ is the three-momentum of the $B$ hadron and $\vec{p}_{j}$ is the jet three-momentum. The measurement is performed in three different intervals of the jet transverse momentum, namely: 50 $<p_{\mathrm{T}}^{\mathrm{jet}}<$ 70~GeV, 70 $<p_{\mathrm{T}}^{\mathrm{jet}}<$ 100~GeV and $p_{\mathrm{T}}^{\mathrm{jet}}>$ 100~GeV.\\
Several different models of multijet production are used. These samples have been generated using \texttt{SHERPA2.2.5}~\cite{Sherpa:2019gpd}, \texttt{PYTHIA8.240}~\cite{Sjostrand:2006za,Sjostrand:2007gs} and \texttt{HERWIG7.2.1}~\cite{Bellm:2017bvx}, with substantial differences in the ME calculations, as well as PS algorithms and hadronisation models. The decays of the $B$ mesons were modelled using \texttt{EVTGEN1.6.0} code~\cite{Lange:2001uf} for all the above-mentioned samples.\\
These predictions are then comprared with the particle-level results, as shown in Figure~\ref{fig:bfrag}, where the longitudinal ($z$) and transverse  ($p_{\mathrm{T}}^{\mathrm{rel}}$) profiles for each $p_{\mathrm{T}}$ bin are reported. The results show important differences between the low and high $p_{\mathrm{T}}$ bins. Particularly, the lower tails of the $z$ distributions contain a larger fraction of the high-$p_{\mathrm{T}}$ data due to the larger probability of having gluon splitting, $g\rightarrow b\bar{b}$ when considering high values of the jet $p_{\mathrm{T}}$.\\
The \texttt{SHERPA} predictions give a reasonable description of the $z$ distributions in the low and medium $p_{\mathrm{T}}$ bins, although they differ from data for very high values of $z$. They also show large discrepancies for low values of $p_{\mathrm{T}}^{\mathrm{rel}}$, which increase when moving towards higher bins of the jet $p_{\mathrm{T}}$. All the various \texttt{PYTHIA8} samples provide a good description of the $z$ and $p_{\mathrm{T}}^{\mathrm{rel}}$ distributions, being compatible with data within the systematic uncertainties across the different jet-$p_{\mathrm{T}}$ bins. The results for the longitudinal profile show reasonable agreement with the \texttt{HERWIG7} prediction with the angle-ordered PS, while large discrepancies are observed with the dipole parton shower. Due to the larger gluon splitting fractions, the \texttt{HERWIG7} sample with the dipole PS significantly overestimates the data in the tails of the $p_{\mathrm{T}}^{\mathrm{rel}}$ distributions at low $p_{\mathrm{T}}$, while the differences are smaller with increasing $p_{\mathrm{T}}$. The Herwig angle-ordered PS gives a better description of the $p_{\mathrm{T}}^{\mathrm{rel}}$ distributions, although non-negligible discrepancies are also observed.

\section{Conclusion}
Measurements of variables probing the properties of the multijet energy flow and of the Lund Plane using charged particles, as well as a measurement of the fragmentation properties of $b$-quark initiated jets, have been presented in this proceeding. Ref.~\cite{ATLAS:2019mgf} demonstrates differences between the soft-drop jet substructure observables in their sensitivity to the quark and gluon composition of the sample, which are most pronounced for the least amount of grooming. In Ref~\cite{ATLAS:2020vup} the discrepancies between event shapes data and all the investigated MC show that further refinement of the current MC predictions is needed to describe the data in some regions, particularly at high jet multiplicities. Ref.~\cite{ATLAS:2020bbn} illustrates the ability of the Lund jet plane to isolate various physical effects, and will provide useful input to both perturbative and nonperturbative model development and tuning. Finally, Ref.~\cite{ATLAS:2021agf} provides key measurements with which to better understand the fragmentation functions of heavy quarks. As has been shown, significant differences among different MC models are observed, and also between the models and the data. Some of the discrepancies are understood to arise from poor modelling of the $g\rightarrow b\bar{b}$ splittings, to which the present analysis has substantial sensitivity. Including the present measurements in a future tune of the MC predictions may help to improve the description and reduce the theoretical uncertainties of processes where heavy-flavour quarks are present in the final state, such as top quark pair production or Higgs boson decays into heavy quark pairs.

\nocite{*}
\bibliographystyle{auto_generated}
\bibliography{giugli_francesco/proceedings_elba2021/giugli_francesco}

%% file: Lowxsubmission/Lowxsubmission/klein.tex
\vspace*{1.2cm}

\thispagestyle{empty}
\begin{center}
{\LARGE \bf $\rho$ photoproduction in ALICE}

\par\vspace*{7mm}\par

{\bigskip \large \bf Spencer R. Klein on behalf of the ALICE Collaboration}

\bigskip

{\large \bf  E-Mail: srklein@lbl.gov}

\bigskip

{Nuclear Science Division, Lawrence Berkeley National Laboratory, Berkeley CA 94720 USA}

\bigskip

{\it Presented at the Low-$x$ Workshop, Elba Island, Italy, September 27--October 1 2021}

\vspace*{15mm}

\end{center}
\vspace*{1mm}

\begin{abstract}

The $\rho^0$ is copiously photoproduced in ultra-peripheral heavy-ion collisions at the LHC.   In this talk, I will present recent results on rho photoproduction with ALICE, including cross-section measurements from PbPb and XeXe collisions, including a discussion of the production of high-mass final states that decay to $\pi^+\pi^-,$ and of neutron production that accompanies $\rho$ photoproduction.  I will conclude by presenting some prospects for ALICE in Runs 3 and 4. 

\end{abstract}
 \part[$\rho$ photoproduction in ALICE\\ \phantom{x}\hspace{4ex}\it{Spencer R. Klein on behalf of the ALICE Collaboration}]{} 
 \section{Introduction}
Ultra-peripheral collisions (UPCs) at heavy-ion colliders are a prolific source of photonuclear interactions; they are the energy frontier for photon-mediated reactions \cite{Baltz:2007kq,Bertulani:2005ru,Klein:2020fmr,Contreras:2015dqa}.   In UPCs, the ions do not interact hadronically, so the product of the photon-mediated interaction is visible.  In a simple approximation, the impact parameter $b$ must be more than twice the nuclear radius, $R_A$. UPCs include both two-photon interactions and photoproduction.   UPCs at the Large Hadron Collider (LHC) represent the energy frontier for photon-mediated interactions.  In pp collisions, photon-proton center of mass energies up to about 3 TeV are accessible, while in pPb collisions, the maximum energy is about 700 GeV. The photons have a  small $p_T$, roughly $p_Z/\gamma$, where $\gamma$ is the ion Lorentz boost, so it is possible to use the $p_T$ distribution to probe the size of the nuclei.  Unfortunately, the photon $p_T$ is a conjugate variable to the impact parameter, so if restrictions are imposed on $b$ (such as $b>2R_A$), the mean $p_T$ will increase; this increase is not calculable without some assumptions \cite{Klein:2020jom}. 

In vector meson photoproduction, an incident photon fluctuates to a quark-antiquark dipole which then scatters elastically from a target nucleus, emerging as a real vector meson.   During the elastic scattering, the vector meson retains the same quantum numbers (including helicity) as the incident photon.

The $\rho^0$ is of special interest as the most copiously photoproduced vector meson \cite{STAR:2002caw}.  It is the lightest vector meson, corresponding to the largest dipole, so is the most subject to nuclear effects.    In addition to $\rho$ photoproduction, the photon can fluctuate directly to a $\pi^+\pi^-$ pair, which then scatters, emerging as a real pion pair \cite{Bauer:1977iq}.    These two possibilities are indistinguishable, so interfere with each other, enhancing the spectrum below the $\rho^0$ mass, and suppressing it at higher masses. Higher-mass, excited $\rho$ states are also possible, and can lead to higher mass $\pi^+\pi^-$ pairs. 

One complication in PbPb UPCs is that the coupling constant $Z\alpha\approx 0.6$ is large, so for a collision with moderate impact parameters, it is very possible to exchange more than one photon, complicating the reaction \cite{Baltz:2002pp,Baur:2003ar}.   A second photon is likely to excite one of the nuclei, while a third photon may excite the other nucleus, leading to mutual Coulomb dissociation \cite{Baltz:1996as}.  The excitation may be collective, such as a Giant Dipole Resonance (GDR) or higher excitation, or an excitation of a single nucleon to a $\Delta$ or higher resonance, or, for higher energy photons, a more complex hadronic interaction.  Production of an additional vector meson is also possible.   Most of these reactions involve nuclear dissociation, leading to the emission of one or more neutrons, or, less frequently, one or more protons.   It is also possible to produce a vector meson and excite the nucleus via one-photon exchange leading to incoherent photoproduction.  

 In these reactions, the photons are emitted independently  \cite{Baur:2003ar}, connected only by a common impact parameter.  The impact parameter affects the photon flux and maximum energy.  And, since the photons are polarized with their electric field vectors parallel to the impact parameter vector, the photons share the same polarization.

These additional reaction products complicate the analysis of UPC photoproduction, since one can no longer focus exclusively on exclusive reactions.   In most cases, the additional reaction only leads to the production of neutrons, but sometimes $\pi^{\pm}$ may be created.      The cross-section for having multiple reactions may be computed in impact parameter ($b$) space.  For example, the cross-section to produce a $\rho$ with multiple Coulomb excitation is
\begin{equation}
\sigma = \int {\rm d}^2b P_{\rho} (b) P_{X1}(b) P_{X2}(b)
\label{eq:mult}
\end{equation}
where $P_{\rho} (b)$,  $P_{X1}(b)$, and  $P_{X2}(b)$ are the probabilities to produce a $\rho$, and excite the first and second nuclei respectively.   These probabilities are given by the product of the differential photon flux ${\rm d}^2N_\gamma/{\rm d}b^2$ and the $\gamma A$ cross-sections.   For nuclear excitations, the $\gamma A$ cross-sections must include a wide range of reactions, including collective nuclear excitations like the Giant Dipole Resonance (GDR), nucleon excitations, such as the $\Delta^+$ resonance, and partonic excitations from high-energy photons, spanning a wide photon energy range from about 10 MeV (in the target frame) up to the kinematic limit.   These cross-sections are usually determined using tabulations of data from multiple sources \cite{Baltz:1996as}.   When the cross-sections are large, it may be necessary to include a unitarity correction, since multiple photons may contribute to excitate a single target to a higher level.

For photons below a cutoff energy (when $k< \gamma \hbar c/b$), the photon flux has a $1/b^2$ dependence, so, the more photons that are exchanged, the smaller  $\langle b\rangle$  \cite{Baur:2003ar}.  So, one can use the number of exchanged photons to preferentially select different ranges of impact parameter. They also all share the same photon polarization, so polarization correlations are expected. 

\section{Detector and Data Analysis}

The data used here were collected using the ALICE detector, which comprises a large central detector and a forward muon spectrometer \cite{ALICE:2008ngc}.  For the analyses discussed here, the most important components are an inner silicon detector and large time projection chamber, in a 0.5 T solenoidal magnet. 

The events analyzed here were collected with a special trigger optimized for ultra-peripheral collisions \cite{ALICE:2020ugp,ALICE:2021jnv}.  It required two pairs of signals in the silicon detector, with azimuthal angular separation greater than 153 degrees.  Each signal pair, consisting of hits in different silicon layers, was consistent with one track.  The azimuthal angle requirement was to select pairs where the tracks were roughly back-to-back.  

The other trigger requirements provided vetos to reject events that contained additional particles.  The four veto detectors, and their pseudorapidity ($\eta$) coverage are: V0A ($2.8 < \eta < 5.1$), V0C ($-3.7 < \eta<-1.7$), ADA ($4.7 < \eta < 6.3$) and ADC ($-4.9 > \eta > -6.9$). 

Data from the zero degree calorimeters (ZDCs) were used in the analysis, but not in the trigger.

The analysis selected events with exactly two good oppositely charged tracks, consistent with a vertex in the interaction region.  The tracks were required to have specific energy loss (${\rm d}E/{\rm d}x$) in the TPC consistent with being $\pi^\pm$.

\begin{figure}
\begin{center}
\epsfig{figure=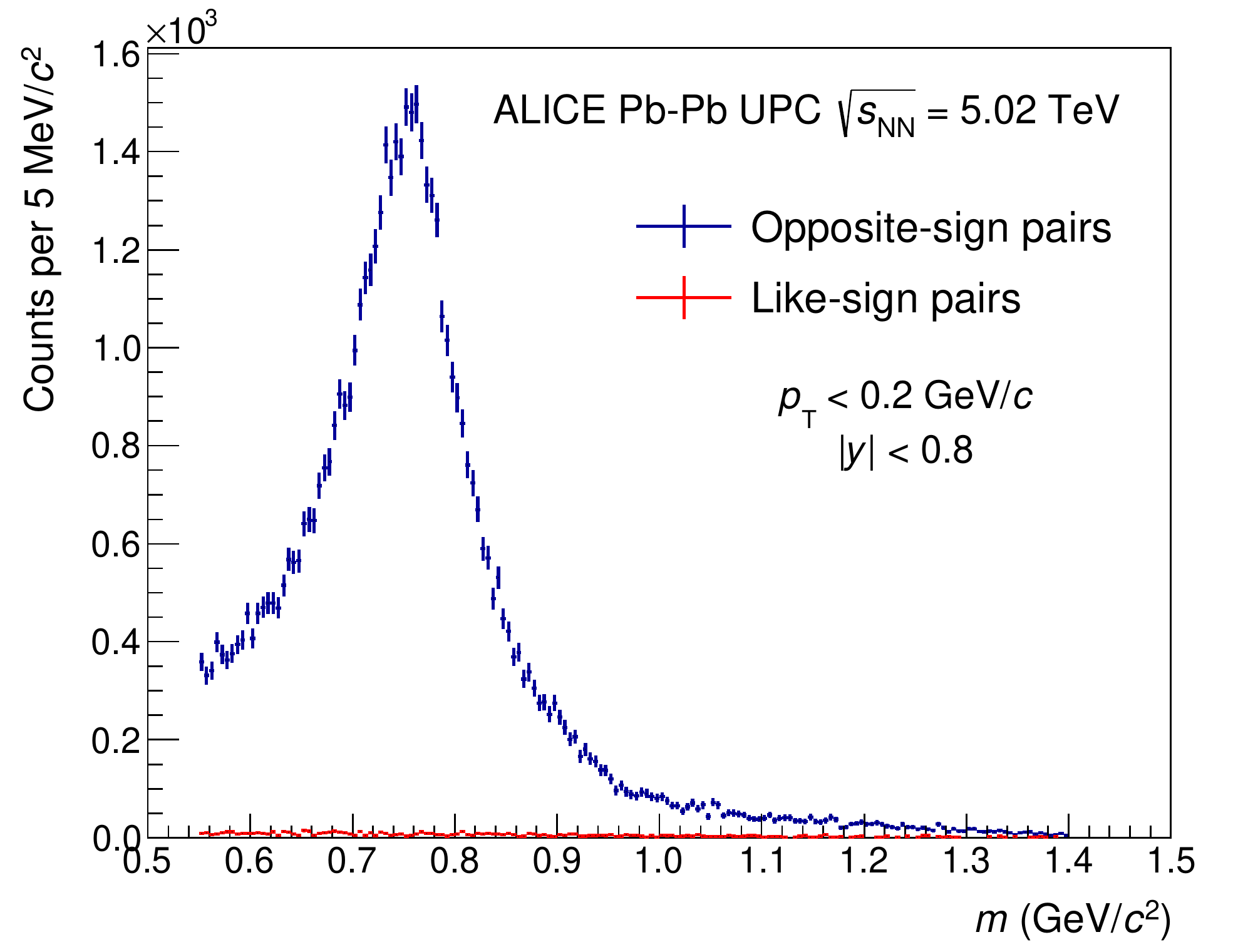,height=0.35\textwidth}
\epsfig{figure=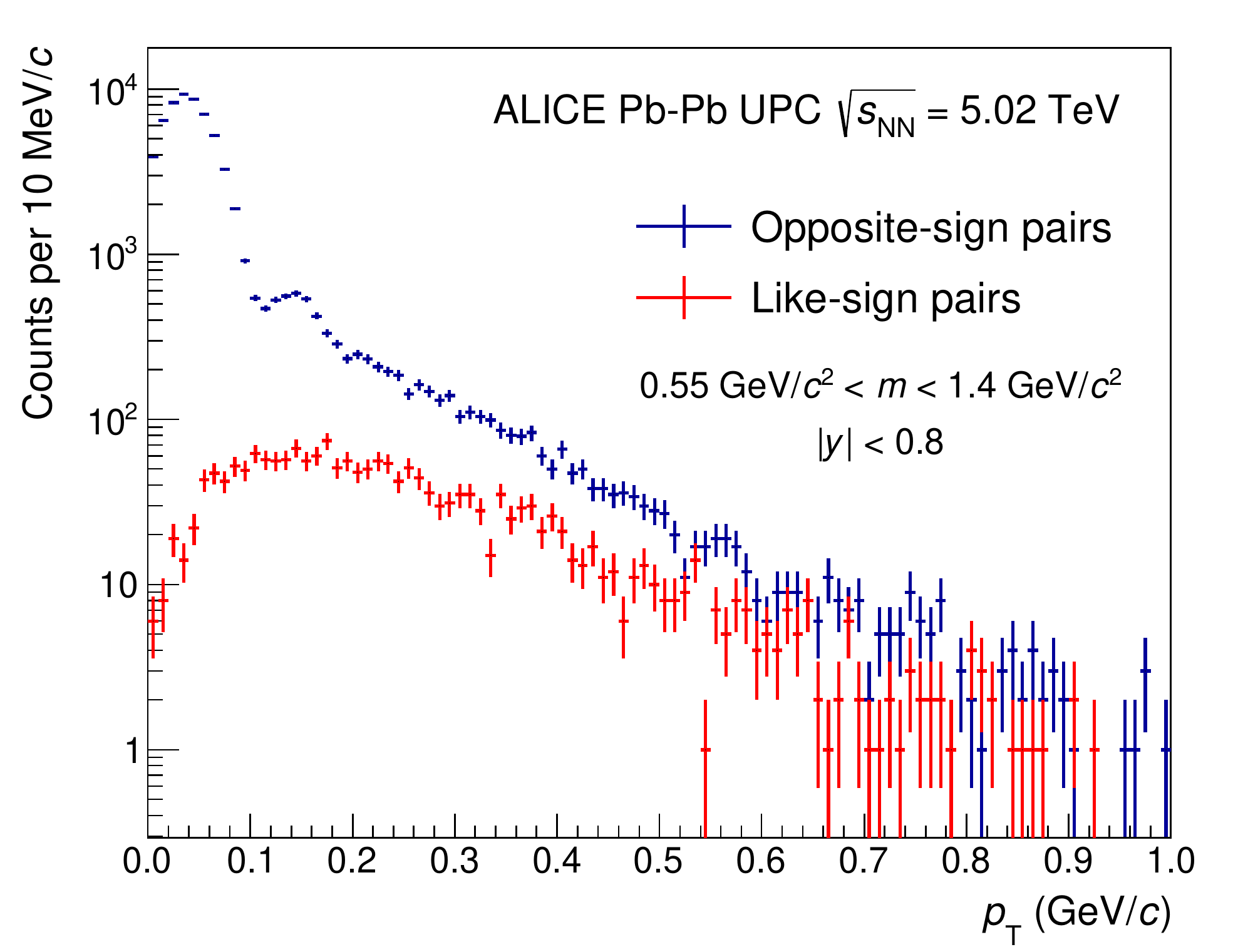,height=0.35\textwidth}
\caption{The mass (left) and $p_T$ spectra for selected pairs.  A cut of $p_T<200$ MeV/c is applied to the mass spectrum, to emphasize coherent production, while the $p_T$ spectrum includes a wide range of masses.  For $p_T < 200$ MeV/c, the like-sign background is orders of magnitude below the coherent production signal.  From \cite{ALICE:2020ugp}.}
\label{fig:first}
\end{center}
\end{figure}

\section{$\rho$ production results}

These cuts left a clean signal.  Figure \ref{fig:first} shows the dipion invariant mass and $p_T$ spectra.  The two peaks in the $p_T$ spectrum below  200 MeV/c, correspond to the first two diffractive maxima, clearly showing the diffractive nature of the production.  The like-sign pairs, which are a proxy for most backgrounds (notably grazing hadronic collisions), are far below the oppositely charged pairs, showing that there is little background in the events.  Two remaining backgrounds, from photoproduction of the $\omega$, followed by $\omega\rightarrow\pi^+\pi^-\pi^0$ and $\gamma\gamma\rightarrow\mu^+\mu^-$ are also small, and mostly concentrated at low $M_{\pi\pi}$.  The latter is a background since we cannot effectively distinguish $\mu^\pm$ and $\pi^\pm$. 

The $M_{\pi\pi}$ mass spectrum can be fit with a relativistic Breit-Wigner distribution for the $\rho^0$ plus a constant term for direct $\pi^+\pi^-$, and an interference term between the two.  $\gamma\gamma\rightarrow\mu^+\mu^-$ is included with a template based on STARlight \cite{Klein:2016yzr}, while $\omega\rightarrow\pi^+\pi^-\pi^0$ is removed by only fitting in the region $M_{\pi\pi}>0.62$ GeV.

\begin{figure}
\begin{center}
\epsfig{figure=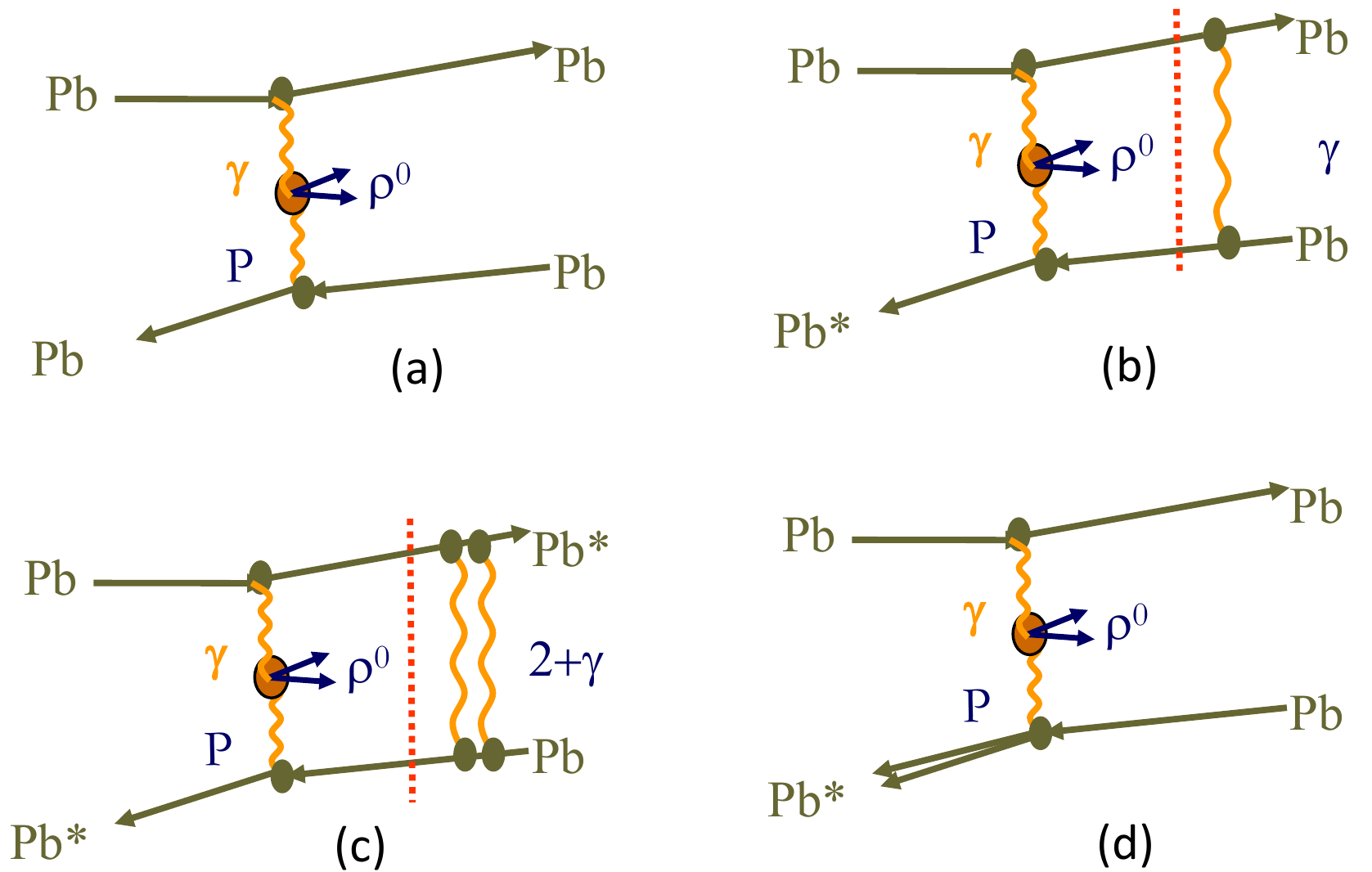,height=0.3\textwidth}
\epsfig{figure=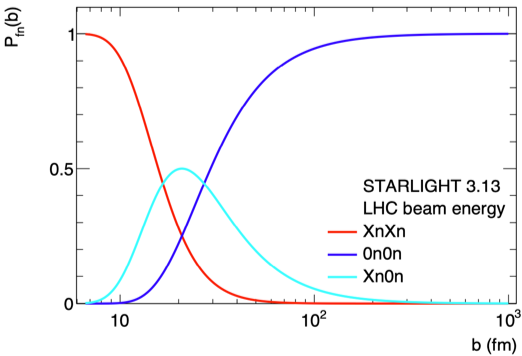,height=0.3\textwidth}
\caption{(left) Four diagrams that contribute to $\rho$ photoproduction: (a) coherent production, plus coherent production with an additional (b) one  or (c) two photons exchanged, and (d) incoherent photoproduction.  The products from reactions (b) and (d) are not completely distinguishable.   (Right) The impact parameter distributions for different nuclear excitations: no nuclear excitation (0n0n, diagram (a), single nuclear excitation (0nXn, diagram (b)) and mutual Coulomb excitation (XnXn, diagram (c)).  Nuclear excitation preferentially selects events with smaller impact parameters.  Incoherent photoproduction corresponds to single-photon exchange, so it has a similar impact parameter distribution as 0n0n.  The right panel is from Ref. \cite{Klein:2020fmr}.
}
\label{fig:diagrams}
\end{center}
\end{figure}

The produced $\rho$ and direct $\pi^+\pi^-$ may be accompanied by neutrons, which can occur when the two nuclei exchange additional photons, as in Figs. \ref{fig:diagrams}(b) and (c), or from 
an incoherent photoproduction reaction (Figs. \ref{fig:diagrams}(d)) involving a single photon.    The photons are expected to be emitted independently, sharing only a common impact parameter \cite{Gupta:1955zza,Baur:2003ar}.      

\begin{figure}
\begin{center}
\epsfig{figure=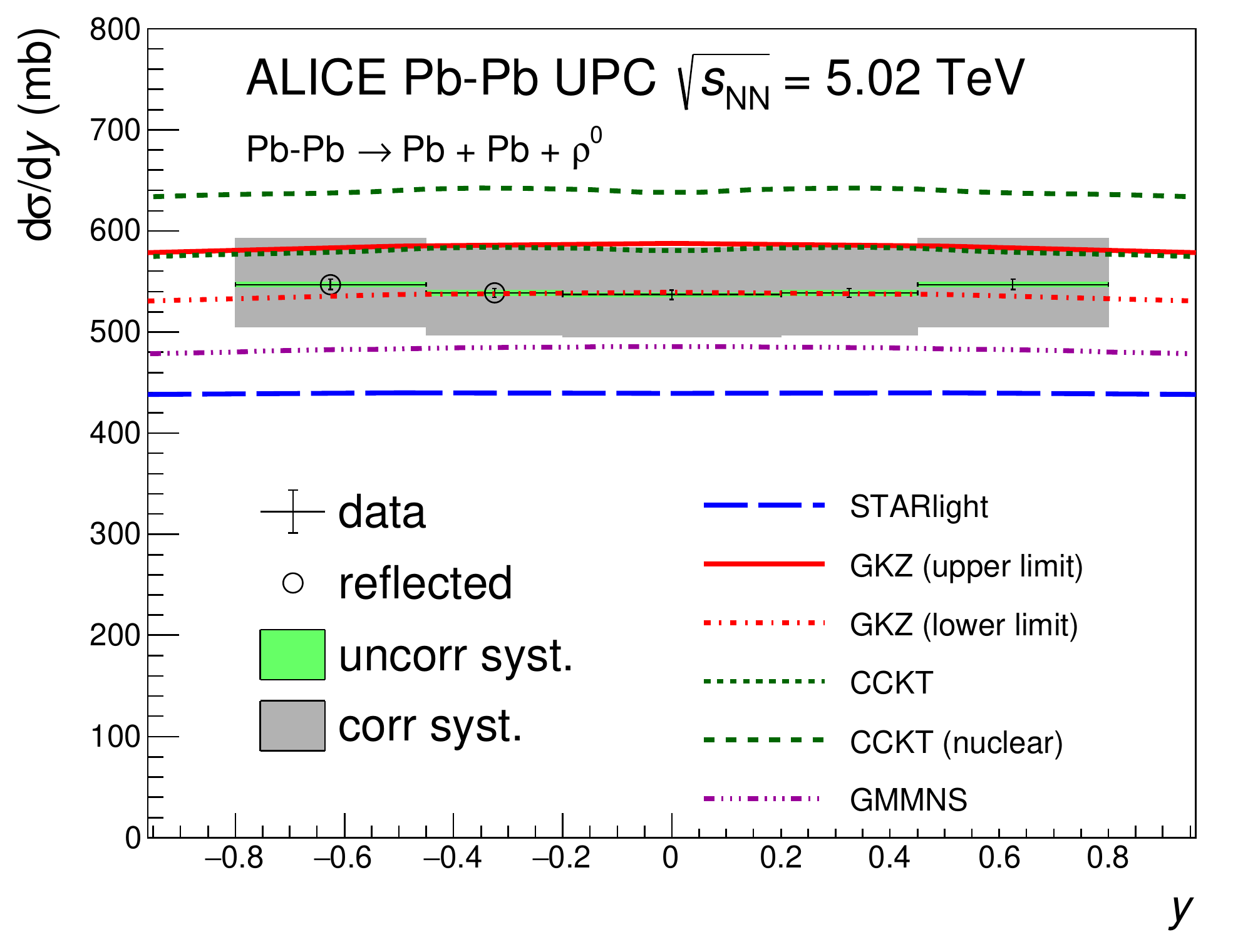,height=0.35\textwidth}
\epsfig{figure=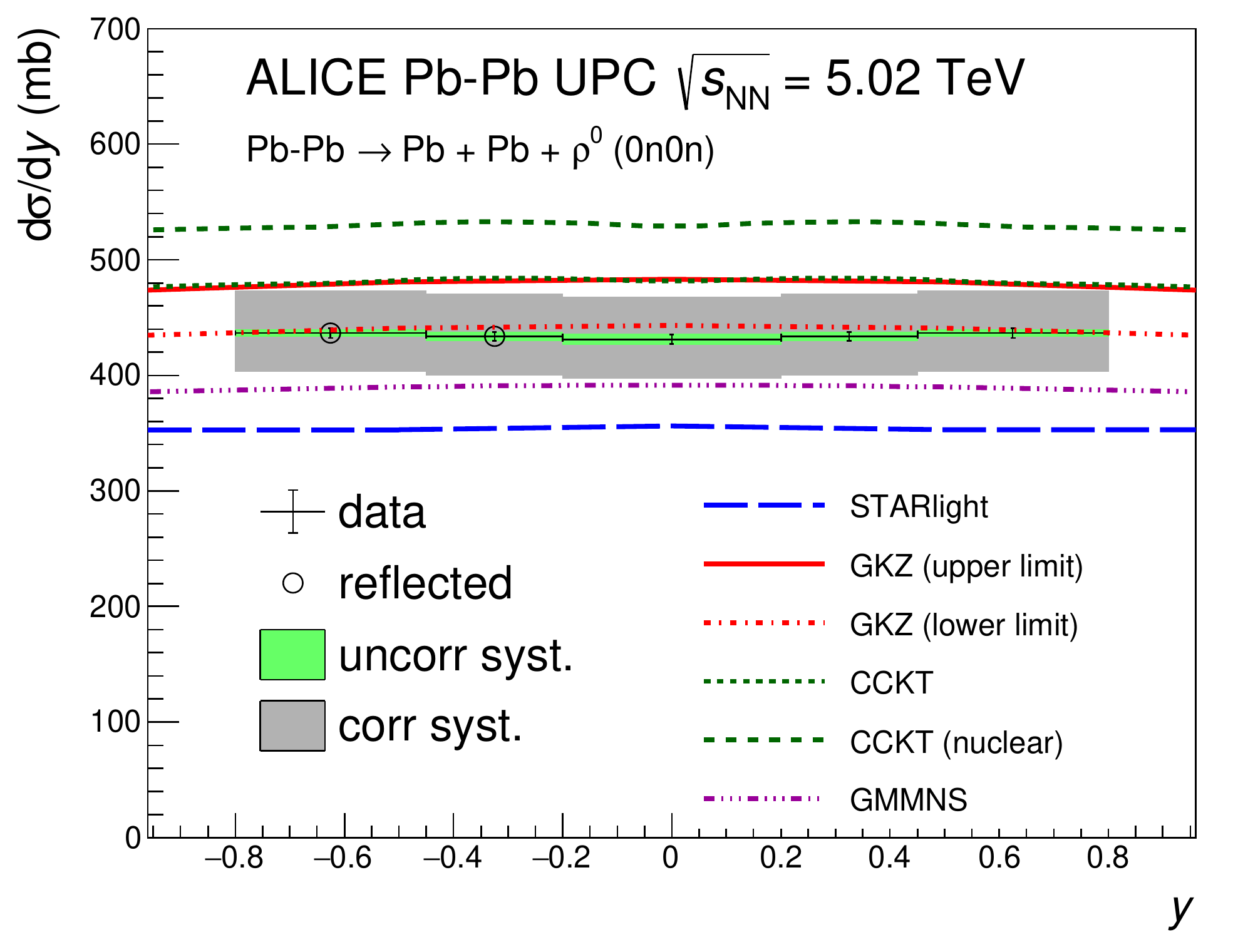,height=0.35\textwidth}
\epsfig{figure=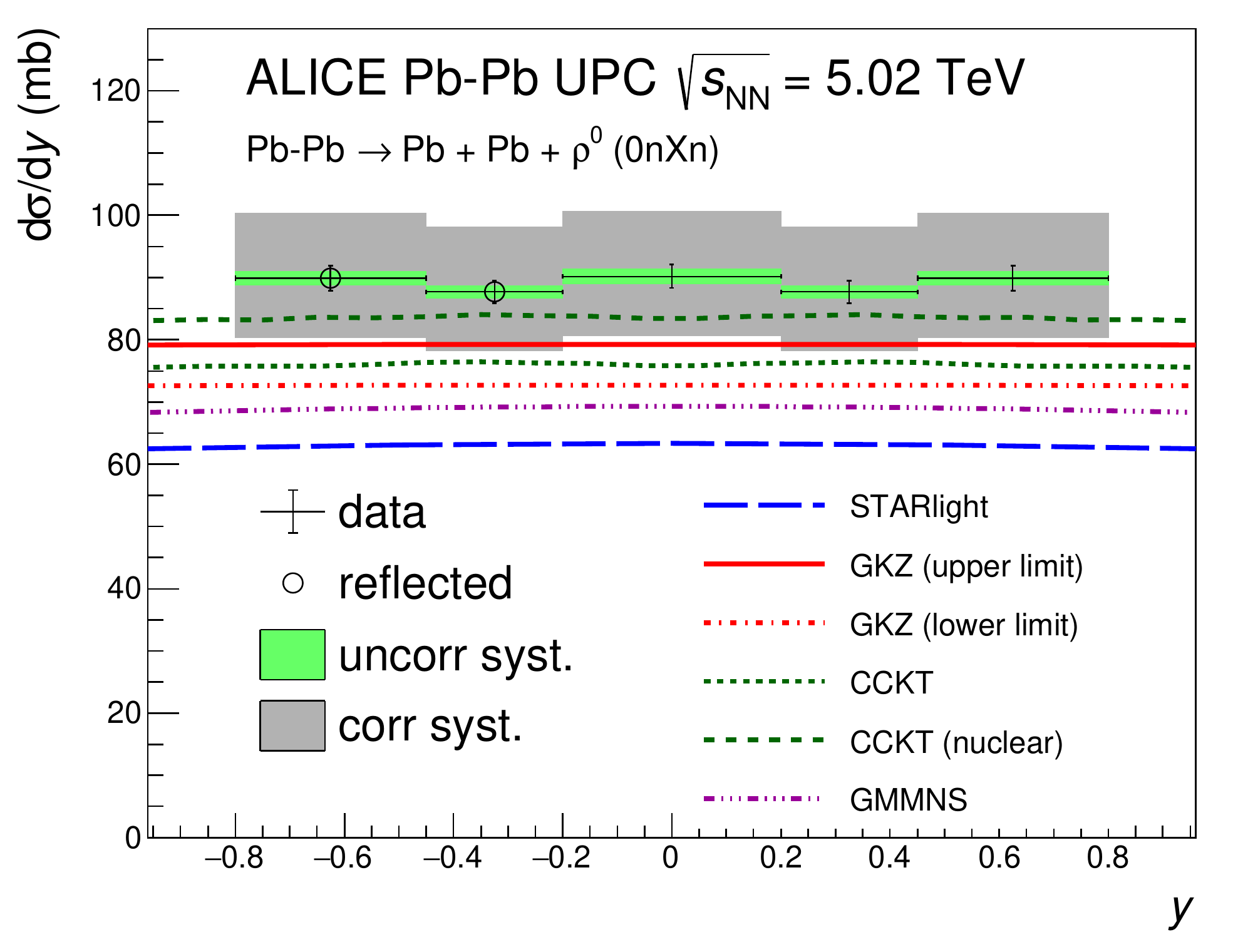,height=0.35\textwidth}
\epsfig{figure=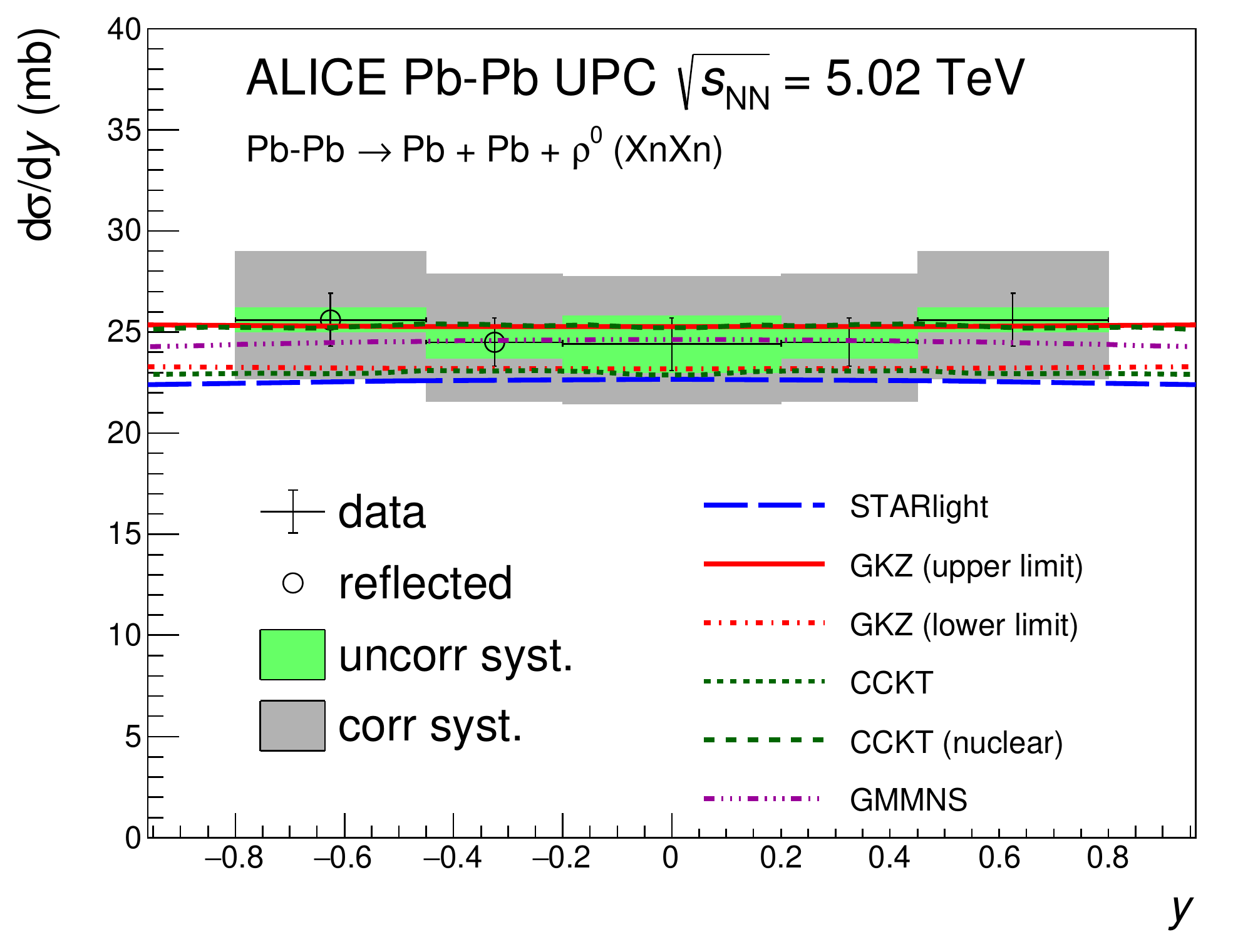,height=0.35\textwidth}
\caption{$d\sigma/dy$ for $\rho$ photoproduction for (top left) all events, and three different classes of neutron emission: (top right) no neutrons, (bottom left) neutrons in one ZDC only, and (bottom right) neutrons in both ZDCs.  Each panel is compared with several different theoretical calculations.  From
\cite{ALICE:2020ugp}.
}
\label{fig:dsdt}
\end{center}
\end{figure}

Figure \ref{fig:dsdt} shows the cross-sections for all events, and for the same events divided into three classes.   The top-left panel shows the total measured $\rho$ cross-section, compared to five models.  STARlight is based on parameterized HERA $\gamma {\rm p}$ data, with a Glauber-like eikonal formalism to handle nuclear targets \cite{Klein:1999qj}.   The GKZ predictions are based on a modified vector-meson-dominance model, using a Glauber-Gribov formalism for nuclear targets \cite{Guzey:2013jaa}.  The Glauber-Gribov approach allows for a dipole to interact multiple times as it traverses a target.  Each individual interaction can be inelastic, with the intermediate states (between interactions) allowed to include high-mass fluctuations. The CCKT predictions are based on a calculation of dipoles passing through a nuclear target, which is modeled in terms of gluon density as a function of impact parameter \cite{Cepila:2016uku}.  The gluon density includes gluonic hot-spots.  Finally, the GMMNS model is another dipole based calculation that includes an implementation of gluon saturation \cite{Goncalves:2017wgg}.   Most of the models do a reasonable job of matching the data, although STARlight is a bit on the low side, and the CCKT (nuclear) model is a bit high.

Per Eq. \ref{eq:mult}, the cross-sections for additional photon exchange may be easily calculated given a $\sigma_{\gamma p}$.  
The STARlight neutron calculation is done within the STARlight code \cite{Klein:2016yzr}, while the CCKT simulation used the {\bf $n_0^0n$} afterburner \cite{Broz:2019kpl}.  To the extent that these calculations are based on the same parameterized photoexcitation data, they should give the same relative cross-sections for $\rho^0$ production accompanied by neutron emission.   However, the relative cross-sections do not agree perfectly.  For example, in the two upper panels (total coherent $\rho$ cross-section and $\rho$ without neutron emission), 
the CCKT (nuclear) cross-section is well above the other calculations, while in the lower panels, where neutron emission is required, it is relatively lower.  A similar trend is evident for the CCKT curve.   It may be that {\bf $n_0^0n$} predicts lower excitation probabilities than STARlight. 

Experimentally, neutron emission, expected in most photonuclear reactions, is easy to detect using the ALICE ZDCs.  However, there is a complication.  Some of the photoexcitation occurs at high energies, and the reactions can lead to emission of one or more $\pi^\pm$ or heavier particles.  If these particles hit any of the detectors used as event vetoes (the ADA, ADC, V0A and V0C), causing a loss of $\rho$ signal.    The loss is substantial; it is $26\pm4\%$ for events with neutrons in one ZDC, and $43\pm5\%$ for events with neutrons in both ZDCs.   This loss is estimated using control triggers which do not include the veto, and appropriate corrections are applied.  
 
\section{$\rho$ photoproduction in XeXe collisions}

The $A$ dependence of $\rho$ photoproduction can provide an important clue about the presence of saturation or other high-density nuclear phenomena.     In 2017, the LHC collided xenon atoms, at $\sqrt{s_{NN}}=5.44$ TeV.  ALICE used the same UPC trigger as for lead-lead running and measured the cross-section, using similar methods \cite{ALICE:2021jnv}.  Figure \ref{fig:highmass} shows the $\rho$ photoproduction as a function of atomic number.   At mid-rapidity,
\begin{equation}
\frac{{\rm d}\sigma}{{\rm d}y} = 131.5 \pm 5.6 ({\rm stat.})^{+17.5}_{-16.9} ({\rm syst.})\ {\rm mb}.
\end{equation}
This is slightly below the STARlight predictions, slightly below the lower bound of the GMMNS prediction, and below the GKZ band.  However, none of these deviations are very significant. 

The cross-section scales as $A^\alpha$, with $\alpha=0.96\pm0.02$, dominated by systematic uncertainty.  This shows the presence of substantial nuclear effects.  Without nuclear effects, the coherent cross-section would scale as $A^{4/3}$.  This is the product of two scaling relations: the forward cross-section scales as $A^2$, while the $p_T$ range over which coherent production is possible scales as $A^{-2/3}$, leaving the $4/3$ exponent.   On the other hand, it is also considerably above the prediction of a black disk model, in which the cross-section scales as the frontal area of the nucleus, $A^{2/3}$.  

\section{A high-mass state}

\begin{figure}
\begin{center}
\epsfig{figure=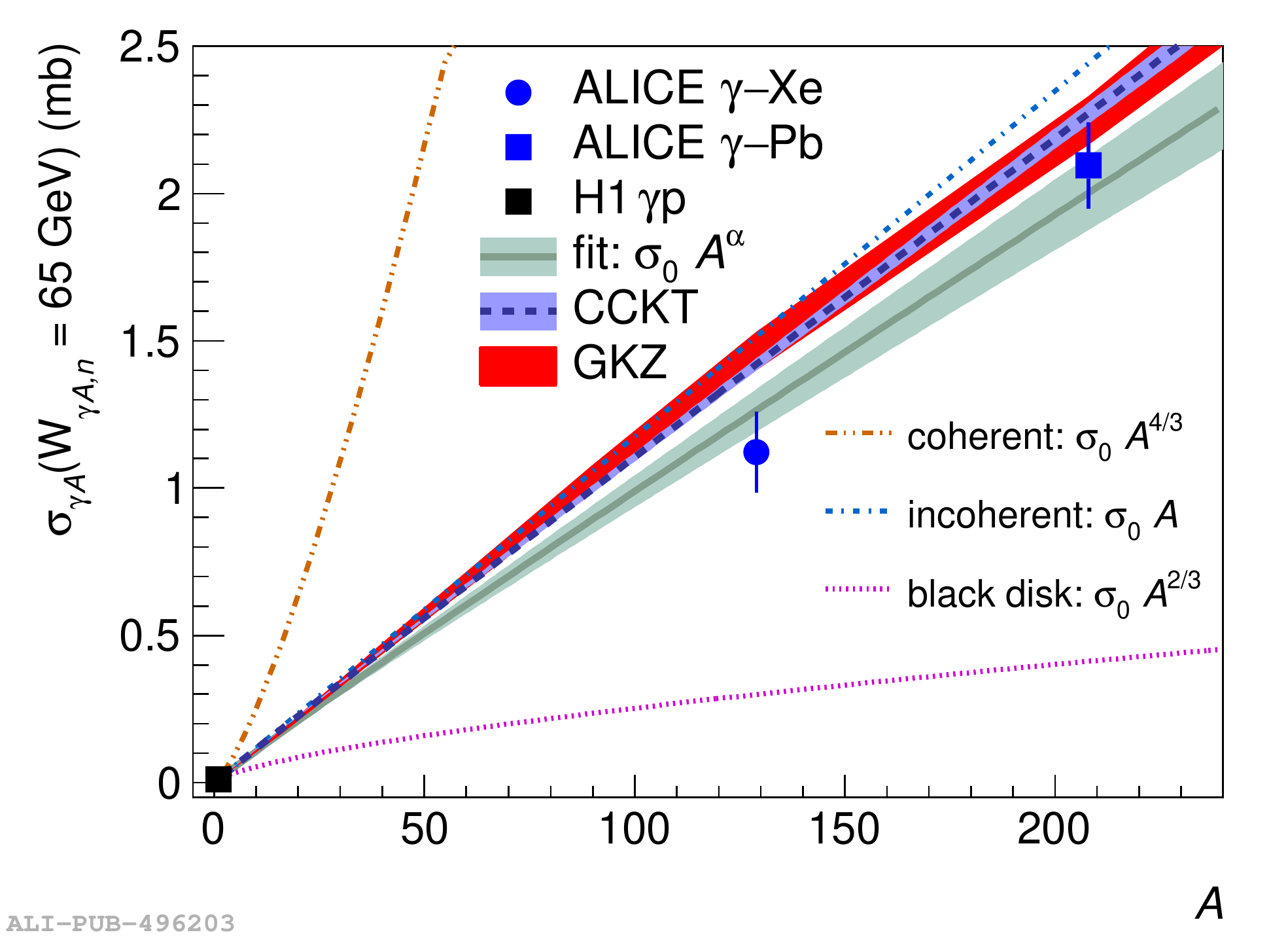,height=0.3\textwidth}
\epsfig{figure=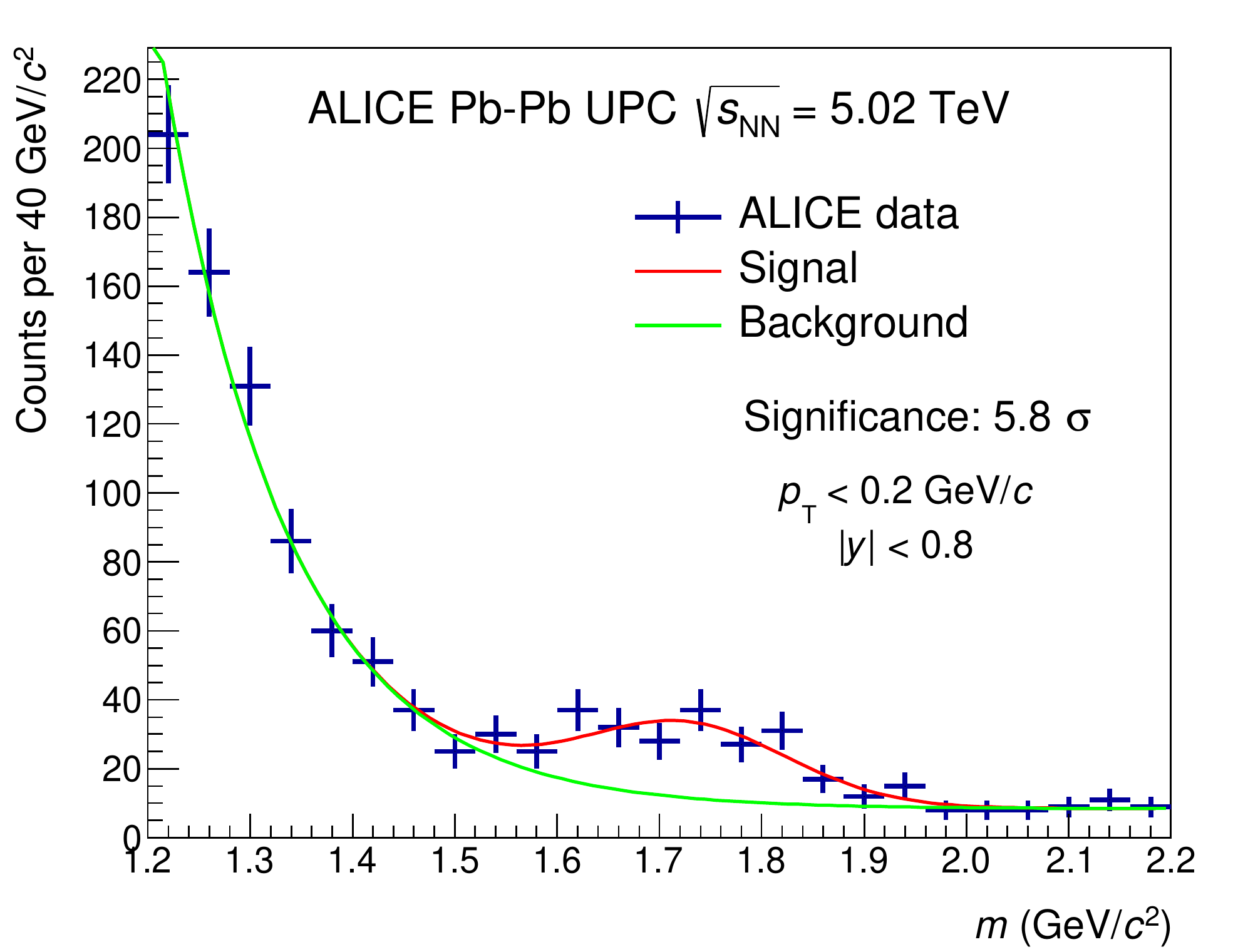,height=0.3\textwidth}
\caption{(left) $\gamma A\rightarrow \rho A$ cross-sections vs. atomic number $A$ for pp, XeXe and PbPb collisions, for 65 GeV photons. (right)$M_{\pi\pi}$ for exclusive production for events with pair $p_T<200$ MeV/c.   A broad resonance is visible over the high-mass tail of the $\rho^0$ plus direct $\pi^+\pi^-$.  From \cite{ALICE:2020ugp}.
}
\label{fig:highmass}
\end{center}
\end{figure}

A number of excited, higher-mass $\rho$ states can decay to $\pi^+\pi^-$.  Figure \ref{fig:highmass} (right) shows the $\pi^+\pi^-$ mass spectrum for events with $p_T < 200$ MeV/c in lead-lead collisions.  The expected tail of the $\rho^0$ is visible, with a broad resonance on top of it.   The spectrum is fit using an exponential for the  $\rho^0$ plus the direct $\pi^+\pi^-$ tail, plus a Gaussian for the resonance.  The null (no-resonance) hypothesis is rejected with 4.5 $\sigma$ significance.  The resonance best-fit parameters are mass $M=1725\pm 17$ MeV and width $\Gamma=143\pm21$ MeV.    The resonance is similar to that seen by STAR in gold-gold UPCs, with $M=1653\pm 10$ MeV and width $\Gamma=164\pm15$ MeV \cite{Klein:2016dtn}.   STAR pointed out that the peak might be compatible with photoproduction of the $\rho_3 (1690)$.  The ZEUS Collaboration also saw resonances in exclusive $\pi^+\pi^-$ electroproduction ($Q^2>2$ GeV$^2$), with masses of $1350 \pm 20  ^{+20}_{-30}$ MeV and $1780 \pm 20 ^{+15}_{-20}$ MeV \cite{ZEUS:2011tzw}.

\section{Future plans}

During LHC Run 3 and Run 4, ALICE will have many improvements which will improve charged particle reconstruction, raise ALICE's rate capability and remove trigger bottlenecks for UPC data collection.   The TPC endcaps have been replaced with GEM based readouts to allow continuous (rather than gated) TPC readout and a new ITS2 silicon tracker will use monolithic active pixel sensors to greatly improve vertexing, especially for open charm hadrons.   

For UPCs, the biggest improvement will be a new streaming readout which will do away with triggering.  All data will flow to the data acquisition system, where it can be studied with high-level event selection algorithms \cite{Antonioli:2013ppp}; for lead-lead running, all data will be saved.   This will give an enormous boost to UPC data collection, since triggering is usually the limiting factor for UPC studies.  During  Run 3 and Run 4, a total of 13 pb$^{-1}$ of lead-lead data should be collected.  This is equivalent to 5.5 billion $\rho^0\rightarrow\pi^+\pi^-$ within the ALICE acceptance, along with 210 million $\rho'\rightarrow\pi^+\pi^-\pi^+\pi^-$.  The $J/\psi$ sample should include 1.1 million $J/\psi\rightarrow\mu^+\mu^-$ in the central detector, a similar number of $e^+e^-$ plus about 600,000 $\mu^+\mu^-$ in the forward spectrometer \cite{Citron:2018lsq}.  For the $\psi'$, the rates are about 35,000 and 19,000 respectively. Photoproduction of $\Upsilon(1S)\rightarrow\mu^+\mu^-$ should also be visible, with 2,800 events expected in the central detector, and 880 in the forward muon spectrometer.   This should be enough for detailed studies of the spectroscopy of the light vector mesons, including of the substructure of the heavier mesons. It should also be possible to measure the production characteristics of heavy quarkonium, comparing the effect of shadowing on mesons of different masses.  The removal of the trigger bias and the improved vertex measurements will also facilitate the study of photoproduction of open charm.  

In addition to lead-lead running a short (0.5 nb$^{-1}$?) oxygen-oxygen run has been proposed for Run 3 \cite{ALICE:2021wim,Brewer:2021kiv}.  This offers two unique opportunities for UPCs. 

The first is to study incoherent photoproduction of the $\rho$ on oxygen targets.   Incoherent photoproduction is of great interest because, in the Good-Walker paradigm, it is related to event-by-event fluctuations in the nuclear configuration - the phase space that includes both the individual nucleon positions and, more importantly, the presence of gluonic `hot spot' fluctuations \cite{Klein:2019qfb}.  The $p_T$ spectrum for incoherent production is loosely tied to the length scale for these fluctuations, so it is desirable to measure over as wide a range in $p_T$ as possible.
  
\begin{figure}
\begin{center}
\epsfig{figure=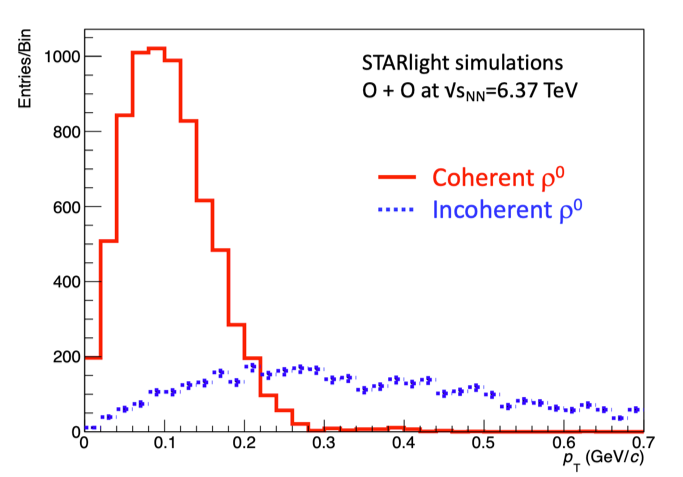,height=0.3\textwidth}
\caption{(left) Simulated $p_T$ spectrum for coherent and incoherent $\rho$ photoproduction in oxygen-oxygen collisions at $\sqrt{s_{NN}}=6.37$ TeV.   From \cite{ALICE:2021wim}.
}
\label{fig:oxygen}
\end{center}
\end{figure}

It is difficult to study incoherent photoproduction in lead-lead collisions because of the large background from coherent production.   In oxygen-oxygen running, the ratio of incoherent to coherent production is expected to be larger than in lead-lead.  As Fig. \ref{fig:oxygen} shows, the predicted coherent peak is still larger than the incoherent, but by less than in lead-lead collisions.  The other difference with lead-lead collisions is that oxygen is only charge eight, so most reactions only involve single photon exchange. Multi-photon exchange, as discussed above, is almost absent, so the presence of neutrons is a clear sign of nuclear breakup.  Although not all nuclear dissociation will produce neutrons, most will do so.   Photonic excitation is also possible, but oxygen is a very stable, doubly magic nucleus, with a lowest lying excited state at 6.05 MeV \cite{IAEA}, so nucleon emission is likely to be predominant. 

The second opportunity is to study the competition between photoproduction and double-diffractive interactions.  Both of these reactions can lead to $\pi^+\pi^-$ final states.  Photoproduction dominates in heavy-ion collisions, while double-diffractive interactions dominate in pp collisions.  With medium-charge nuclei, the amplitudes should be similar, so interference may be possible among final states with the same spin/parity.  

\section{Conclusions}

The $\rho^0$ is copiously photoproduced in ultra-peripheral collisions of heavy ions.  The cross-section for coherent $\rho$ photoproduction is quite well reproduced in models that use Glauber or dipole calculations to predict the cross-sections.   The cross-section scales with the atomic number $A$ as $A^{0.96\pm 0.02}$,  showing that nuclear effects substantially moderate the $A^{4/3}$ dependence expected for full coherence, without nuclear suppression.

The cross-section for coherent $\rho$ photoproduction accompanied by neutron emission is consistent with a model whereby the neutron production comes through the exchange of one or more additional photons, which are independent of the $\rho$ production, except for sharing a common impact parameter.  

We have also observed a heavy state, with a mass of 1650 MeV, decaying to $\pi^+\pi^-$.  The mass and cross-section may be consistent with that expected for the $\rho_3(1690)$. 

Looking ahead, ALICE expects a rich bounty of UPC results during LHC Run 3 and Run 4.   The new flow-through data acquisition system will eliminate the bottleneck formerly imposed by the requirements of a low-multiplicity UPC trigger.   Run 3 has been proposed to include a short oxygen-oxygen run, which should offer the opportunity to study incoherent $\rho$ photoproduction on an intermediate mass target. 


 
\section*{Acknowledgements}

This work is supported in part by the U.S. Department of Energy, Office of Science, Office of Nuclear Physics, under contract number DE- AC02-05CH11231.

\nocite{*}
\bibliographystyle{auto_generated}
\bibliography{Lowxsubmission/Lowxsubmission/klein}


%% file: LauraFabbri_Elba2021/LauraFabbri_Elba2021/LauraFabbri_Elba2021.tex
\vspace*{1.2cm}

\thispagestyle{empty}
\begin{center}
{\LARGE \bf Measurement of $W$ and $Z$ boson production in association with jets at ATLAS}

\par\vspace*{7mm}\par

{

\bigskip

\large \bf Laura Fabbri on behalf of the ATLAS Collaboration}
 \footnote {Copyright 2022 CERN for the benefit of the ATLAS Collaboration. CC-BY-4.0 license.}

\bigskip

{\large \bf  E-Mail: laura.fabbri@bo.infn.it}

\bigskip

{Dipartimento di Fisica e Astronomia {\it "Augusto Righi"}, Universit\`a di Bologna \\ e INFN Sezione di Bologna, Bologna, Italy }

\bigskip

{\it Presented at the Low-$x$ Workshop, Elba Island, Italy, September 27--October 1 2021}

\vspace*{15mm}

\end{center}
\vspace*{1mm}

\begin{abstract}

The electroweak sector of the Standard Model can be tested either via precision measurements of fundamental observables or via direct tests of its underlying gauge structure.  
The ATLAS collaboration has recently released a measurement of differential cross-sections for $Z$-boson produced in association to $b$-jets and a very recent result dedicated to $Z$-boson and jets with high transverse momentum. The measurements are performed using data of proton-proton collisions at the LHC collected at a centre-of-mass energy of 13 TeV and corresponding to a total integrated luminosity of 35.6 fb$^{-1}$ and 139 fb$^{-1}$ respectively.  
The results from these milestone analyses as well as their interpretation in the context of the Standard Model are presented in this proceeding.

\end{abstract}
 \part[Measurement of $W$ and $Z$ boson production in association with jets at ATLAS\\ \phantom{x}\hspace{4ex}\it{Laura Fabbri on behalf of the ATLAS Collaboration}]{}
 \section{Introduction}
From 2015 to 2018 the ATLAS experiment \cite{Atlas} collected nearly $140$ fb$^{-1}$ of "good for physics" data, most of which with a pile-up of up to 40 simultanous interactions per bunch-crossing. 
The physics program at the LHC is the most ambitious and successful plan at high energy physics. The huge dataset available and the well understood detector performance allow for precision measurements, access to rare processes, perform an extensive research program of weak interactions or physics at or above the TeV scale, study new states of matter. 
The present article focuses on the latest ATLAS results on Standard Model physics.

The Standard Model is an extremely predictive theory successfully verified by experiments for about fifty years. Since the discovery of the Higgs boson, ten years ago, the two main goals are to test and validate the model in a new energy regime, improving the accuracy of parameter measurements, and to search for new physics, both directly and indirectly, trying to access new physics effects in the collected events. 
One of the more promising signals for obtaining such results is the production of a vector boson in association with jets ($V$+jets).
Indeed, this signal has two advantages: a large production cross section and a clear experimental signature which can be precisely measured thanks to the easily identifiable decays of the $Z$ boson to charged leptonic final states. That makes the $Z$+jets a "standard candle" for testing Standard Model.
Furthermore, studying $Z$+jets is very important to improve our ability to select specific signals. In fact, such process constitute non-negligible background for studies of the Higgs bosons \cite{higgs1,higgs2} 
and in searches for new phenomena \cite{newP1,newP2,newP3}
, which often exploit the presence of high-$ p_\mathrm{T}$ jets to enrich a data sample with potential signal. The extrapolation of $W/Z$+jets backgrounds from control regions to signal regions and the modelling of the final discriminant distribution largely benefit from reliable predictions. Improving our prediction is mandatory for the success of the analysis.   

Additionally, this measurement constitutes a powerful test of perturbative quantum chromodynamics (pQCD) \cite{QCD1,QCD2} 
and, in the case of high-energy jets, it allows to probe the interplay of QCD with higher-order electroweak (EW) processes \cite{EW1,EW2,EW3,EW4}.  
Otherwise, looking at $Z +b$-jets, where the jets originate from $b$-quarks in proton-proton ($pp$) collisions, provides a test of the production mechanism of heavy-flavoured partons. Particularly, current predictions for $Z+b$-jets production are known at the next-to-leading-order (NLO) accuracy in pQCD, and they can be derived in either a 4-flavour number scheme (4FNS) or a 5-flavour number scheme (5FNS) \cite{FNS1,FNS2,FNS3,FNS4}
. In the 4FNS, $b$-quarks do not contribute to the parton distribution functions (PDFs) of the proton and, in QCD, they only appear in a massive final state due to gluon splitting ($g \rightarrow bb$). In the 5FNS, on the contrary, $b$-quark density is allowed in the initial state via a $b$-quark PDF, with the $b$-quark typically being treated as massless. The measurement of this cross-section is therefore a useful tool to constrain the $b$-quark PDF inside proton.  

\section{Differential cross-sections for $Z+b$-jets}

\begin{figure}[b]
\begin{center}
\epsfig{figure=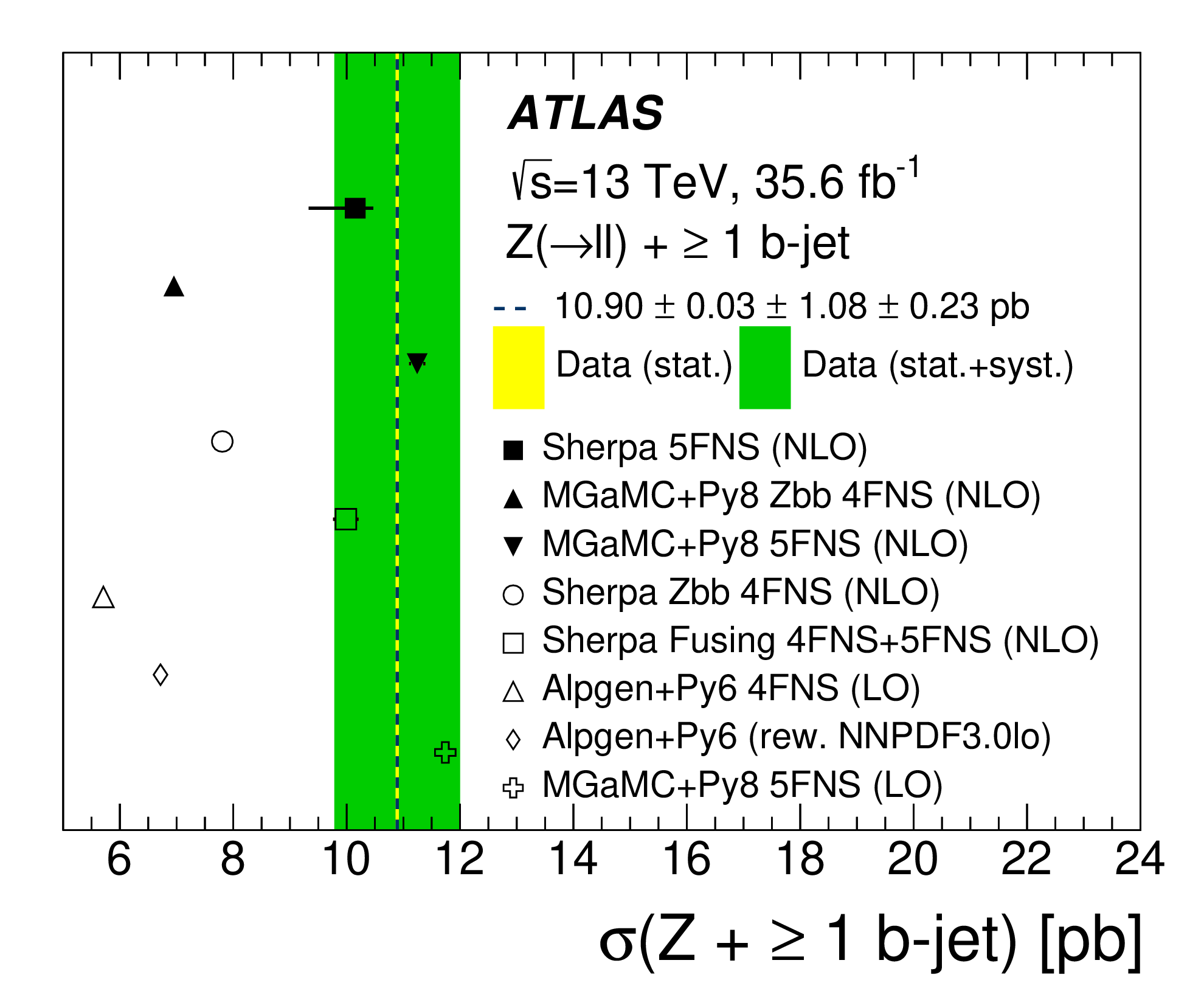,height=0.45\textwidth}
\caption{Measured cross-section for $Z$+≥ 1 $b$-jet. The data are compared with the predictions from Sherpa 5FNS (NLO), Alpgen + Py6 4 FNS (LO), Sherpa Fusing 4FNS+5FNS (NLO), Sherpa Zbb 4FNS (NLO), MGaMC + Py8 5FNS (LO), MGaMC + Py8 Zbb 4FNS (NLO) and MGaMC + Py8 5FNS (NLO). The yellow band corresponds to the statistical uncertainty of the data, and the green band to statistical and systematic uncertainties of the data, added in quadrature. The error bars on the Sherpa 5FNS (NLO) predictions correspond to the statistical and theoretical uncertainties added in quadrature. Only statistical uncertainties are shown for the other predictions. More details can be found in \cite{Z+bjets}.}
\label{inclusiveXS}
\end{center}
\end{figure}

To measure the $Z+b$-jets production cross-section a sample of  $35.6$ fb$^{-1}$ of $pp$ collision data collected by the ATLAS experiment at $\sqrt{s}$= 13 TeV in 2015 and 2016 has been used. 
Events are selected considering the decay of the $Z$-boson in muon or electron pairs passing specific kinematic criteria and containing at least one $b$-jet (jets passing the b-tagging algorithm at 70\% efficiency \cite{btag,btag2}). The background is dominated by two main contributions: $t\bar t$ process and $Z$ boson associated with light-jets or $c$-jets, misidentified as $b$-jets.
Integrated and differential cross sections as a function of several kinematic observables are compared with a variety of Monte Carlo generators. 
In general, the 5FNS calculations at NLO accuracy well predict inclusive cross sections, while the 4FNS LO calculations largely underestimate the data as reported in Figure \ref{inclusiveXS}. 

\begin{figure}
\centering
\begin{minipage}{.32\textwidth}
  \centering
	\epsfig{figure=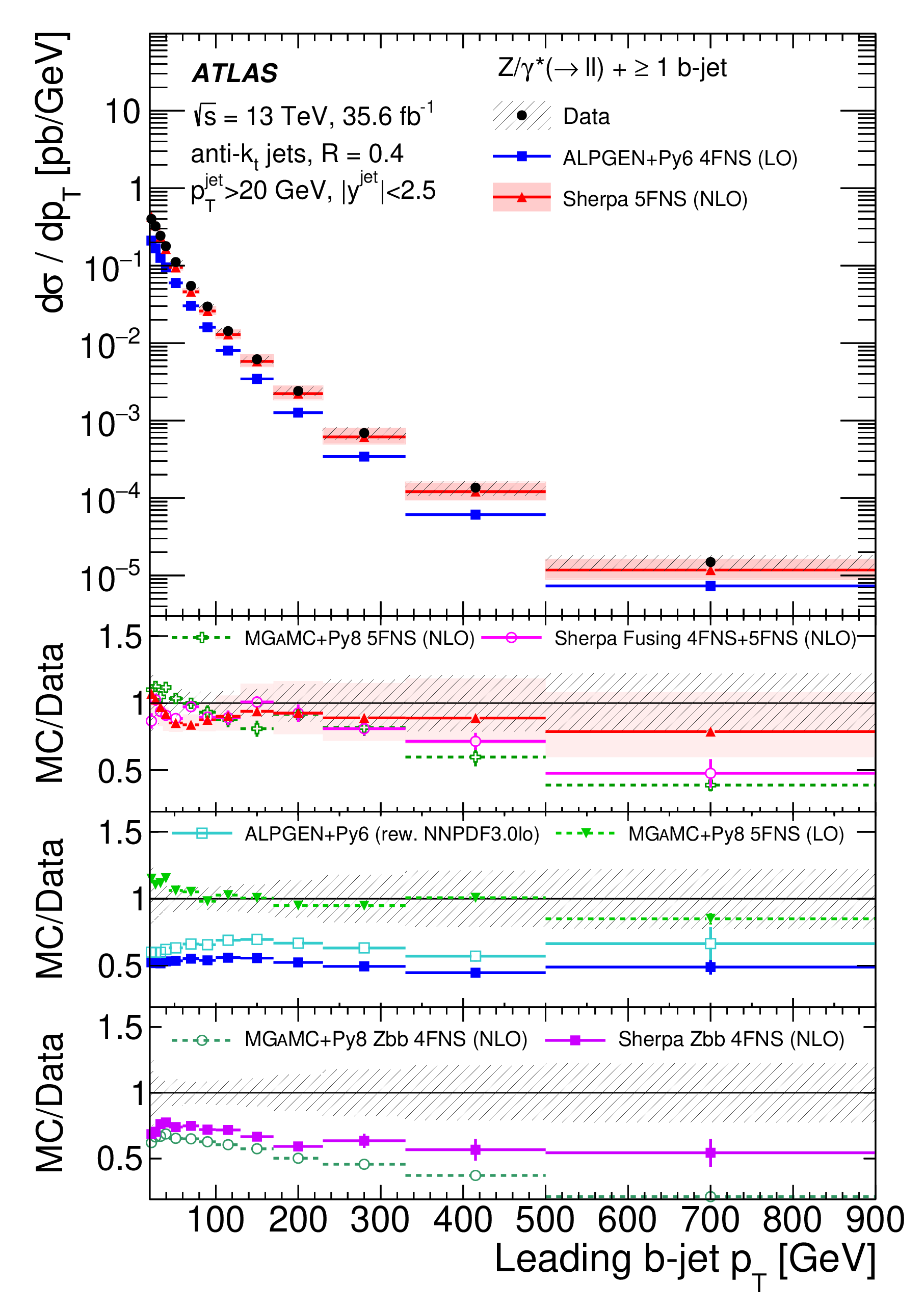,height=1.4\textwidth}
\end{minipage}
\begin{minipage}{.32\textwidth}
  \centering
	\epsfig{figure=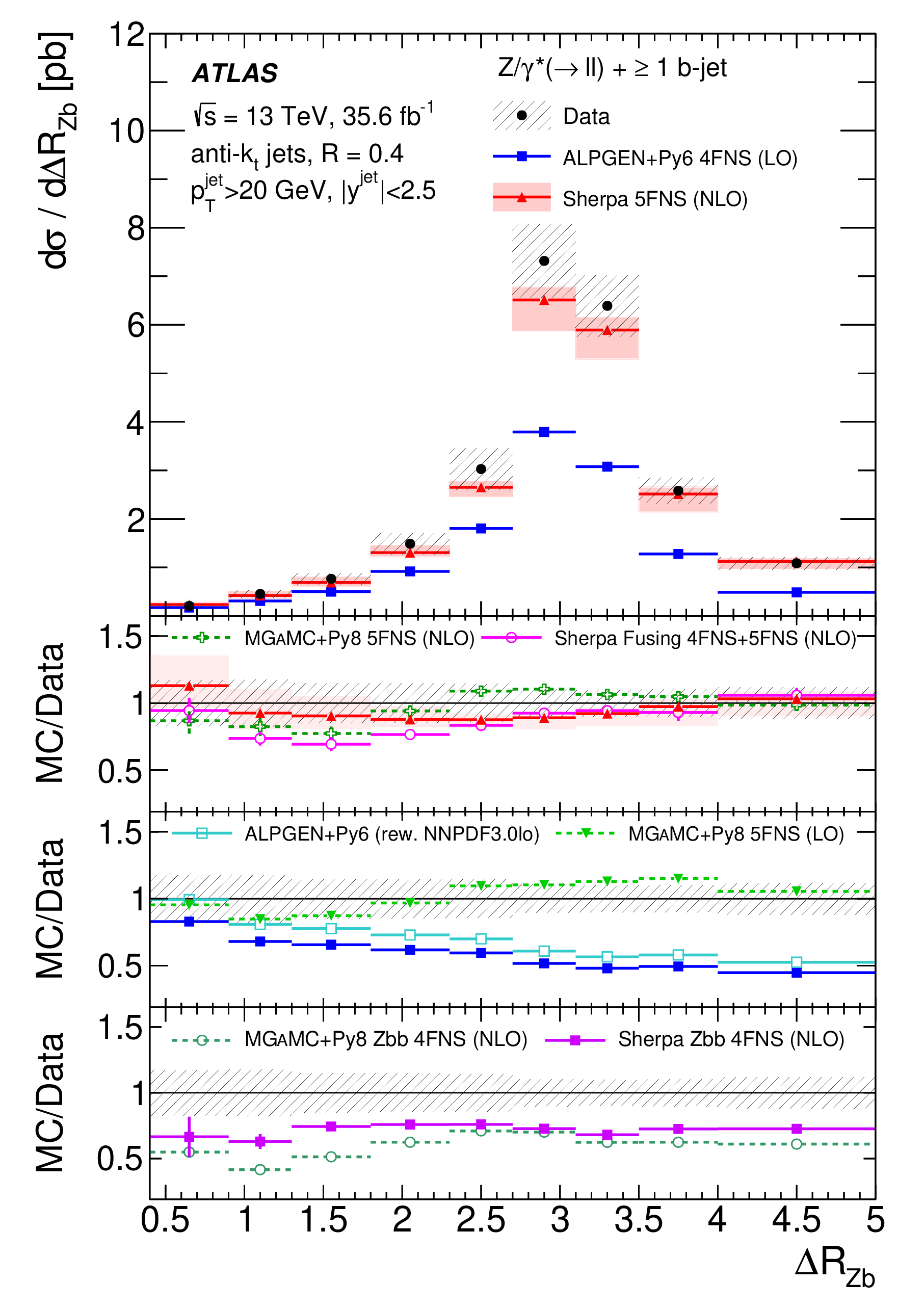, height=1.4\textwidth}
\end{minipage}
\begin{minipage}{.32\textwidth}
  \centering
	\epsfig{figure=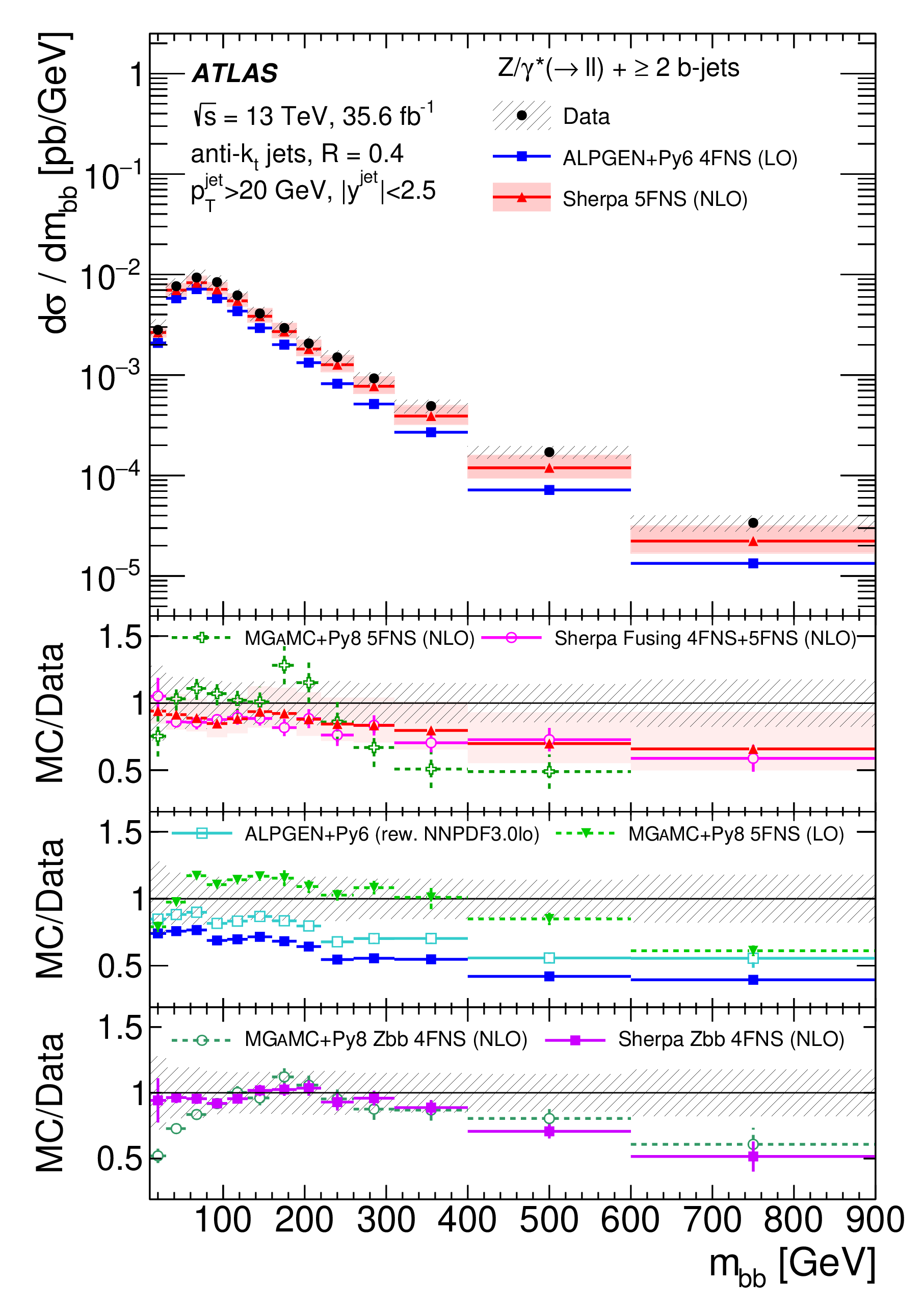, height=1.4\textwidth}
\end{minipage}

\caption{Measured cross-section as a function of leading $b$-jet $p_\mathrm{T}$ (left) and $\Delta R$ between the $Z$-boson candidate and the leading $b$-jet (center) in events with at least one $b$-jet. Measured cross-sections as a function of invariant mass of the two leading $b$-jets (right). The data are compared with different Monte Carlo generator predictions. The error bars correspond to the statistical uncertainty, and the hatched bands to the data statistical and systematic uncertainties added in quadrature. More details can be found in \cite{Z+bjets}. }	
\label{Z+b}
\end{figure}

Figure \ref{Z+b} shows the cross-sections distribution as a function of the $ p_\mathrm{T}$ of the leading $b$-jet (left) and the distance in the pseudo rapidity-azimuthal plane ($\Delta R = \sqrt{(\Delta \eta)^2 +(\Delta \phi)^2}$) between the Z-boson candidate and the leading $b$-jet (center) in events with at least one b-jet. The measured cross-section as a function of invariant mass of the two leading b-jets is presented in Figure \ref{Z+b} (right). 
Experimental data are compared with the predictions of different Monte Carlo generators. 
The best agreement with the data is provided by the NLO Sherpa2.2.1 5FNS predictions; the 5FNS LO calculation from MadGraph5\_aMC@NLO 
interfaced to Pythia8 (MG5aMC+Py8 in the following) better describes data with respect to the NLO calculation due to a larger number of partons in the matrix element. In general all the 4FNS predictions underestimate the cross section.
Figure \ref{Z+b} (center) shows that the 5FNS generators provide a good prediction also for this distribution that is more sensitive to the presence of additional QCD radiation, soft corrections and boosted particles decaying into $Z$-boson and $b$-quarks. 

The NLO 5FNS predictions from Sherpa2.2.1 are in good agreement with data in the low range of the di-$b$jets invariant mass distribution ($m_{bb}$) as reported in Figure \ref{Z+b} (right). None of the considered calculations provide a reasonable description of data for $m_{bb}$ >300 GeV.

\section{Z+jets at high $ p_\mathrm{T}$}
In the calculation of Z+jet production at NLO, real and virtual QCD and EW effects play a not negligible role, including topologies corresponding to dijet events, where a real Z boson is emitted from an incoming or outgoing quark leg \cite{EW1,EW2,EW3,EW4}
. These effects lead to enhancements in production that increase with the transverse momentum of the jets. 
To test predictions the measurement of the production cross-section of Z in association with high $ p_\mathrm{T}$ jets, where the additional contributions are more evident, was performed \cite{Z+jets}. 
Moreover, since QCD processes are sensitive to the kinematics between the Z boson and the closest jet, two topologies of events are identified: soft real emission of a Z boson from a jet ({\it “collinear”}), characterised by a small angular separation between the Z boson and the jet ($\Delta R(Z,j)<1.4$) and hard Z boson production ({\it“back-to-back”}) with $\Delta R(Z,j)>2$. 

\begin{figure}
\centering
\begin{minipage}{.48\textwidth}
  \centering
	\epsfig{figure=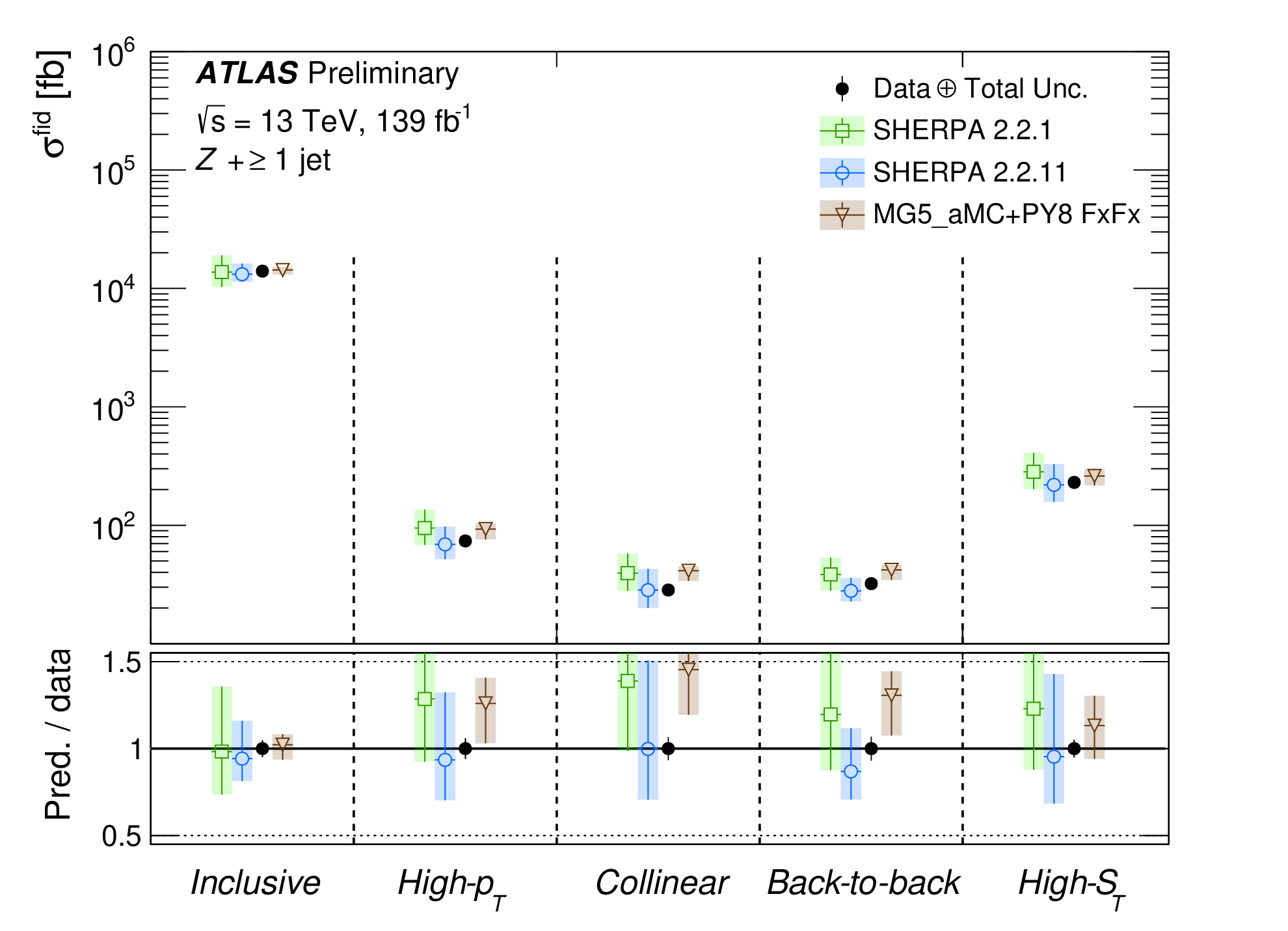,height=0.8\textwidth}
	\end{minipage}
\quad
\begin{minipage}{.48\textwidth}
  \centering
		\epsfig{figure=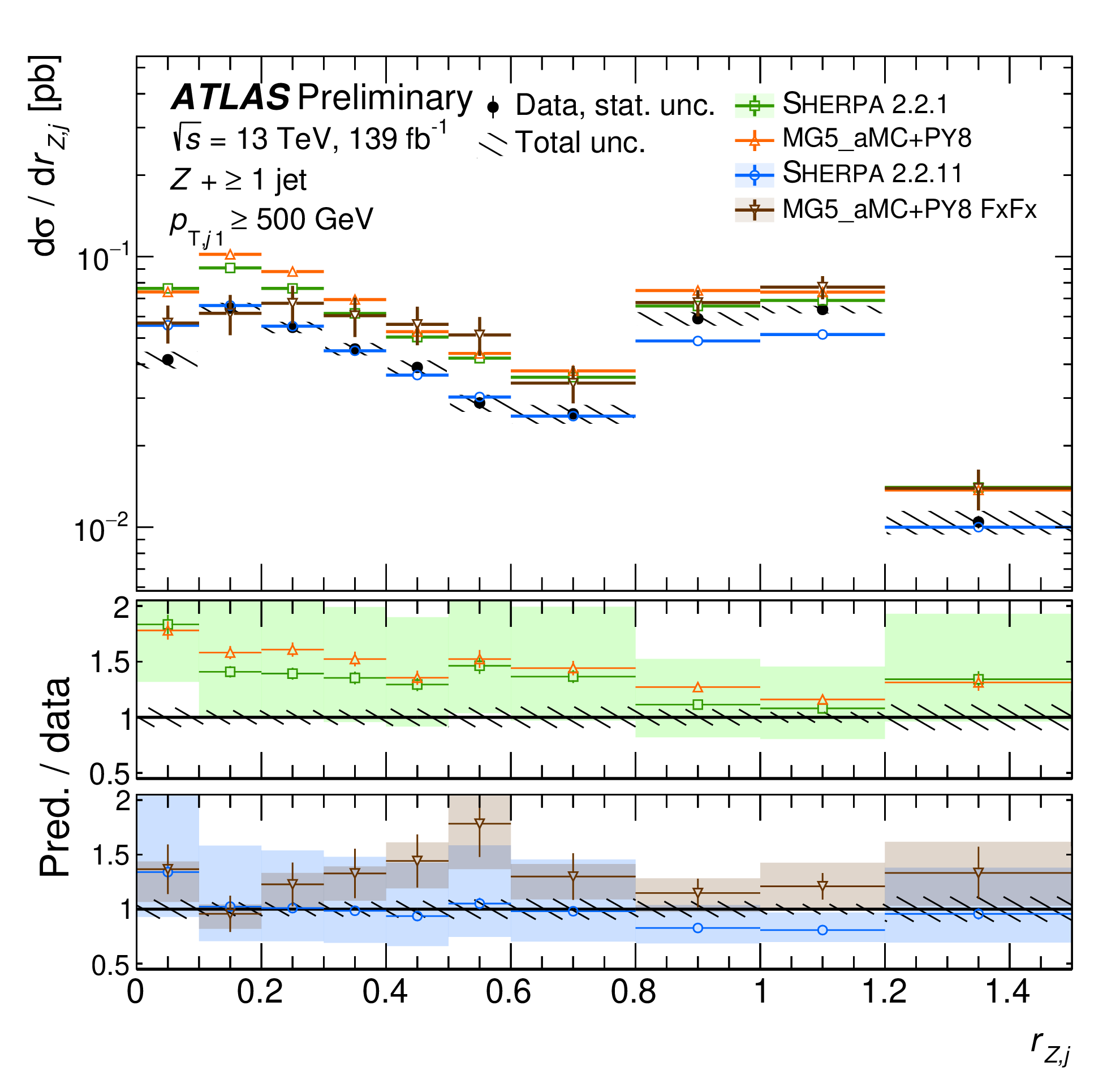,height=0.8\textwidth}
\end{minipage}
\caption{Summary of integrated cross-section results (left). Differential cross-sections for $Z$+jets at high $ p_\mathrm{T}$ as a function of the ratio of $Z$ and jet transverse momentum $r_{Z, j}$ (right).
The measured cross sections are shown with black points and the error bars represent the total uncertainty. 
Data are compared with several predictions. The uncertainties on predictions are given by the quadratic sum of the uncertainties from the variations of PDF, QCD scale and, for Sherpa v.2.2.11, virtual EW contributions. More details in \cite{Z+jets}.}
\label{Z+jet-fig}

\end{figure}

Measurements are performed in events containing a Z boson candidate reconstructed in the muon and electron decay channels and jets with $ p_\mathrm{T} >100$ GeV.
Only the QCD component of the Z+jets production is considered in the analysis, while the EW contribution is treated as background as well as $t \bar t$ and diboson, the main dominant background. 
Integrated and differential cross sections are measured in a fiducial phase space and compared with state-of-art Monte Carlo generator predictions. Predictions from Sherpa 2.2.1 and LO MG5\_ aMC+Py8 CKKWL overestimate the measured cross sections (see Figure \ref{Z+jet-fig} (left)). MG5\_ aMC+Py8 FxFx and Sherpa 2.2.11 provide a good description of data for the full set of observables considered in the analysis for both the collinear and the back-to-back topologies (see Figure \ref{Z+jet-fig} (right)). The improvement obtained by these two predictions with respect to the previous versions of the Monte Carlo generators can be explained with the higher number of parton in the matrix element. 
In addition Sherpa 2.2.11 contains NLO virtual EW corrections and treatment of unordered emissions in the PS.

\section{Conclusion}

The statistics collected by the ATLAS experiment during the LHC Run 2 allows extremely precise measurements that better probe MC generator performances. Perturbative QCD and quark PDF are tested in presence of b-jets. Data confirm that the 5FNS calculation at NLO better describes the inclusive and differential cross-sections.
All Sherpa predictions provide a good modelling of shape, while other predictions, more sensitive to gluon splitting, show various discrepancies. None of the generators correctly describe the region $m_{bb} > 300$ GeV region.
For the first time collinear and back-to-back $Z$ emission are disentangled and the kinematics between $Z$ and closest jet is studied. 
\\
\\
Comments: Presented at the Low-$x$ Workshop, Elba Island, Italy, September 27--October 1 2021.

%
%
%
\nocite{*}
\bibliographystyle{auto_generated}
\bibliography{LauraFabbri_Elba2021/LauraFabbri_Elba2021/LauraFabbri_Elba2021}

%% file: Precision_QCD_measurements_from_CMS/Precision_QCD_measurements_from_CMS/Precision_QCD_measurements_from_CMS.tex
\vspace*{1.2cm}

\thispagestyle{empty}
\begin{center}
{\LARGE \bf Precision QCD measurements from CMS}

\par\vspace*{7mm}\par

{

\bigskip

\large \bf Toni M\"{a}kel\"{a} on behalf of the CMS Collaboration}

\bigskip

{\large \bf  E-Mail: toni.maekelae@desy.de}

\bigskip

{Deutsches Elektronen Synchrotron DESY, Notkestraße 85, D-22607 Hamburg, Germany}

\bigskip

{\it Presented at the Low-$x$ Workshop, Elba Island, Italy, September 27--October 1 2021}

\vspace*{15mm}

\end{center}
\vspace*{1mm}

\begin{abstract}
~Jet production is an important probe of both quantum chromodynamics and new physics. Recent CMS measurements involving jet production at $5$ and $13\TeVns$ are presented, including the measurements of the $\PZ$ boson invisible width at $13\TeVns$, multijet production at $13\TeVns$, and inclusive jet production at $5$ and $13\TeVns$. The $13\TeVns$ inclusive jet data are used in a QCD analysis together with HERA inclusive deep inelastic scattering and CMS $13\TeVns$ triple-differential top quark-antiquark pair production cross sections. The parton distributions, the strong coupling constant and the top quark pole mass are extracted simultaneously. Further, a standard model effective field theory analysis is performed, in which the standard model is extended with 4-quark contact interactions, resulting in a first-ever simultaneous extraction of the contact interactions’ Wilson coefficient and the standard model parameters using LHC data.
\end{abstract}
  \part[Precision QCD measurements from CMS\\ \phantom{x}\hspace{4ex}\it{Toni M\"{a}kel\"{a} on behalf of the CMS Collaboration}]{}
\section{Introduction}

For a deeper understanding of quantum chromodynamics (QCD), it is essential to study the production of jets. Jets, and the objects produced in association with them, shed light on proton structure and can be used for extracting standard model (SM) parameters, such as the strong coupling and quark masses. Jets with high transverse momentum $\pt$ can also probe the scale of new physics and are utilized in indirect searches for physics beyond the standard model (BSM).

A selection of the relevant achievements of the CMS Collaboration is presented.
A precision measurement of the Z invisible width~\cite{CMS-PAS-SMP-18-014} is described in Section~\ref{sec:ZinvWidth} and inclusive jet production at $\sqrt{s}=5.02\TeVns$~\cite{CMS-PAS-SMP-21-009} in Section~\ref{sec:incJets5}.
Section~\ref{sec:multijet} discusses multijet production at 13$\TeVns$~\cite{CMS-PAS-SMP-21-006}. Finally, Section~\ref{sec:incJets13} details the measurement of inclusive jet production cross sections at $\sqrt{s}=13\TeVns$ along with a QCD analysis incorporating these data~\cite{CMS-PAS-SMP-20-011}. A detailed description of the CMS detector, together with a definition of the coordinate system and relevant kinematic variables is given in Ref.~\cite{CMS-JINST-3-S08004}. The contents reported here reflect those available at the time of the Low-$x$ 2021 Workshop. Since then, the work in Ref.~\cite{CMS-PAS-SMP-20-011} has been submitted to \textit{JHEP} and the preprint is available at~\cite{CMS:2021yzl}.

\section{Precision measurement of the Z invisible width at $\sqrt{s} = 13\TeVns$}
\label{sec:ZinvWidth}

A measurement~\cite{CMS-PAS-SMP-18-014} of the $\PZ$ boson invisible width is performed using data from $\Pp\Pp$ collisions at $\sqrt{s} = 13\TeVns$ recorded by the CMS experiment at the LHC, corresponding to an integrated luminosity of $36.3\,\mathrm{fb}^{-1}$. Jets are reconstructed with the anti-$\kt$ algorithm with the distance parameter $R=0.4$.

Besides being a search for dark matter using jets and missing transverse momentum $\pt^{\textrm{miss}}$, the precise measurement of the $\PZ$ boson invisible width constrains the number of neutrino species coupled to $\PZ$. The invisible width arises from the decays of the $\PZ$ boson to invisible final states, such as neutrinos, and is given by 

\begin{equation}
\Gamma(\PZ\rightarrow\nu\overline{\nu})
= \frac{ \sigma(\PZ + \jets) \mathcal{B}(\PZ\rightarrow\nu\overline{\nu}) }
            { \sigma(\PZ + \jets) \mathcal{B}(\PZ\rightarrow\ell\ell) }
    \Gamma(Z\rightarrow\ell\ell),
\label{ZinvisibleWidthEq}
\end{equation}
where $\Gamma(Z\rightarrow\ell\ell)$ is the decay width for visible leptonic decays. The $\sigma$ and $\mathcal{B}$ in Eq.~\eqref{ZinvisibleWidthEq} denote cross sections and branching ratios for the processes denoted in the brackets, respectively.

The extraction of the invisible width requires the $\PZ/\gamma^* \rightarrow \ell\ell$ process to be corrected to $\PZ \rightarrow \ell\ell$. The contribution of the $\gamma^* \rightarrow \ell\ell$ channel and its interference with $\PZ \rightarrow \ell\ell$ is simulated with \MGvATNLO 2.3~\cite{Alwall:2014hca, Alwall:2011uj}. The event selection criteria is given in detail in~\cite{CMS-PAS-SMP-18-014}. The signal processes of $\PZ$ decaying to neutrinos and the Drell-Yan process of $\PZ$ decaying to leptons as well as $\PW$ boson production are simulated at next-to-leading order (NLO) with \MGvATNLO. Parton shower and hadronisation are obtained by using \textsc{Pythia} 8.212~\cite{Sjostrand:2014zea} with the CUETP8M1 tune \cite{CMS:2015wcf}. Correction factors accounting for NNLO QCD and NLO electroweak effects are applied. To account for background processes, the generation of $\ttbar$ events is done with \MGvATNLO and normalized to the NNLO inclusive cross section with next-to-next-to leading logarithmic corrections. Single top processes are computed at leading order (LO) using \textsc{POWHEG}~\cite{Nason:2004rx, Frixione:2007vw, Alioli:2010xd} and normalized to NLO $t$-channel and $\PQt\PW$ production cross sections, whereas $s$-channel production is computed at NLO using \MGvATNLO. Multijet processes are considered at LO and computed with \textsc{Pythia}. All LO and NLO processes are generated using the NNPDF 3.0 PDFs~\cite{NNPDF:2014otw} at the corresponding order.

The dominant background contribution is due to $\PW+\jets$ events with a lepton outside of detector acceptance. It is estimated by using $\mu + \jets$ and $e + \jets$ as control samples and defining a transfer factor as the ratio of simulated event counts in the $\pt^{\textrm{miss}} + \jets$ signal and in the control samples, as a function of $\pt^{\textrm{miss}}$. This is used for correcting the $\mu + \jets$ and $e + \jets$ event yields, so that the expected contribution of $W + \jets$ to $\pt^{\textrm{miss}} + \jets$ can be obtained. 
The transfer factor is validated by simultaneous likelihood fits to channels involving different combinations of leptons and jets, extracting any differences in the the $\pt^{\textrm{miss}}$ distribution's shape and normalisation. The treatment of subdominant background sources is explained in Ref.~\cite{CMS-PAS-SMP-18-014}. Dominant systematic uncertainties arise from the identification of electrons and muons, the jet energy scale and pile-up~\cite{CMS-PAS-SMP-18-014}.

\begin{figure}[h]
\centering
\includegraphics[width=130mm,trim={0mm 0mm 3mm 0mm},clip]{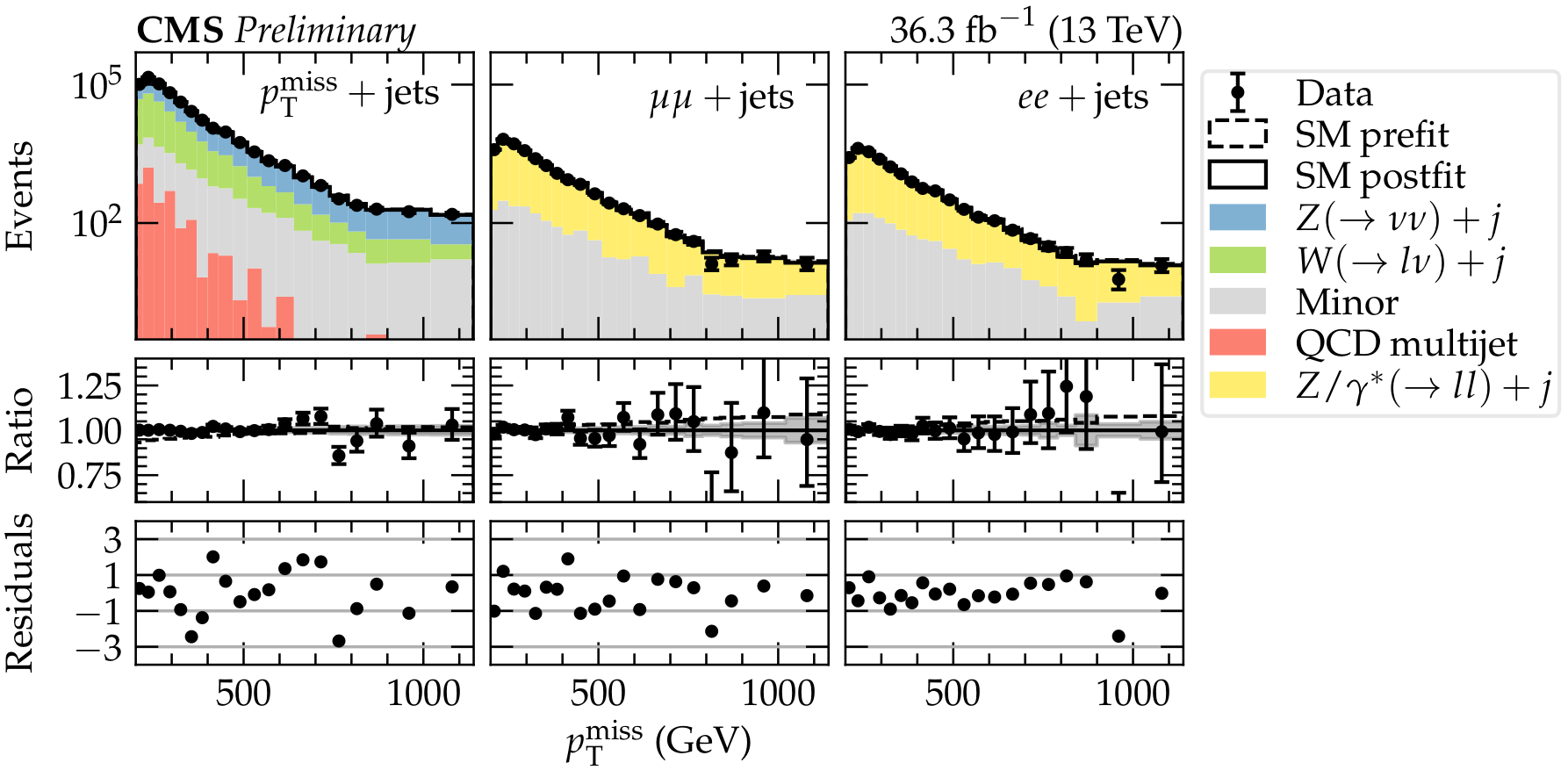}
\caption{Missing transverse momentum $\pt^{\mathrm{miss}}$ distributions. Selected charged leptons do not contribute to $\pt^{\mathrm{miss}}$. The plot also shows the ratios with respect to the SM postfit and the residuals as the difference between data and SM postfit.~\cite{CMS-PAS-SMP-18-014}}
\label{ZinvisibleWidthData}
\end{figure}

The $\PZ$ boson invisible width is extracted from a simultaneous likelihood fit to data for the 
$\pt^{\textrm{miss}} + \jets$, 
$\PZ/\gamma^* \rightarrow ee  + \jets$, 
$\PZ/\gamma*\rightarrow \mu\mu + \jets$, 
$\mu + \jets$, and 
$e + \jets$ 
channels. The transfer factor is included as a free unconstrained parameter in the fit, where it scales the $\PW + \jets$ process for $\pt^{\textrm{miss}} + \jets$ and $\ell + \jets$. The data are compared with the pre- and postfit in Figure~\ref{ZinvisibleWidthData}, which also indicates the contributions of different processes in each channel.

\begin{figure}[H]
\centering
\includegraphics[width=80mm]{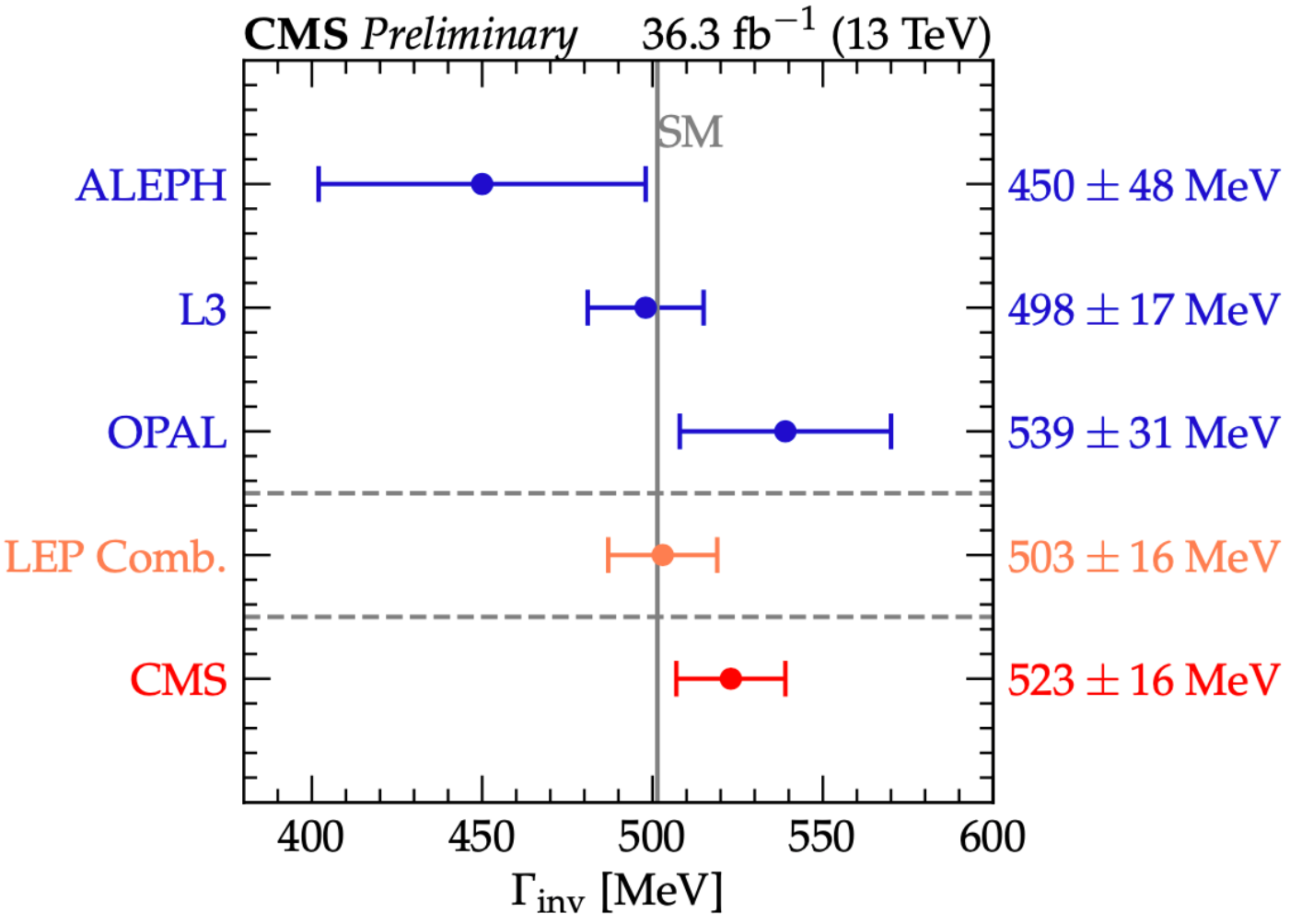}
\caption{Comparison of the $\PZ$ boson invisible width measurement with the direct measurements from the experiments at the LEP. The present measurement is in good agreement with the LEP combination~\cite{CMS-PAS-SMP-18-014}.}
\label{ZinvisibleWidthComparison}
\end{figure}

The resulting $\PZ$ boson invisible width is $523 \pm 3\,\mathrm{(stat)} \pm 16 \,\mathrm{(syst)}\MeVns$, making this the most precise direct measurement to date as well as the first one using hadron collider data. A comparison to the results of the LEP experiments is shown in Figure~\ref{ZinvisibleWidthComparison}.

\section{Measurement of the double-differential inclusive jet cross section at $5\TeVns$}
\label{sec:incJets5}

A measurement~\cite{CMS-PAS-SMP-21-009} of the inclusive jet production is performed using $\Pp\Pp$ collision data at $\sqrt{s}=5.02\TeVns$, corresponding to an integrated luminosity of $27.4\,\mathrm{pb}^{-1}$. The data was recorded by the CMS experiment during a special low-pileup run of the LHC, with 1.1 vertices per collision on average. The cross section is measured double-differentially as a function of jet transverse momentum $\pt$ and rapidity $y$, and the jets are reconstructed using the anti-$\kt$ algorithm with distance parameter $R=0.4$ in a phase space given by $64 \GeVns < \pt < 1 \TeVns$ and $|y| < 2.0$.

As the inclusive jet production is dominated by QCD and the background from electroweak processes is negligible, the data are compared to perturbative QCD predictions with a correction for nonperturbative effects~\cite{CMS-PAS-SMP-21-009}. The NLO prediction is obtained using NLOJet++~\cite{Nagy:2001fj, Nagy:2003tz} with the \textsc{FastNLO} framework~\cite{Britzger:2012bs}, and the NNLO prediction is determined with NNLOJET~\cite{Currie:2016bfm, Currie:2018xkj, Gehrmann:2018szu}.

\begin{figure}[H]
\includegraphics[width=50mm]{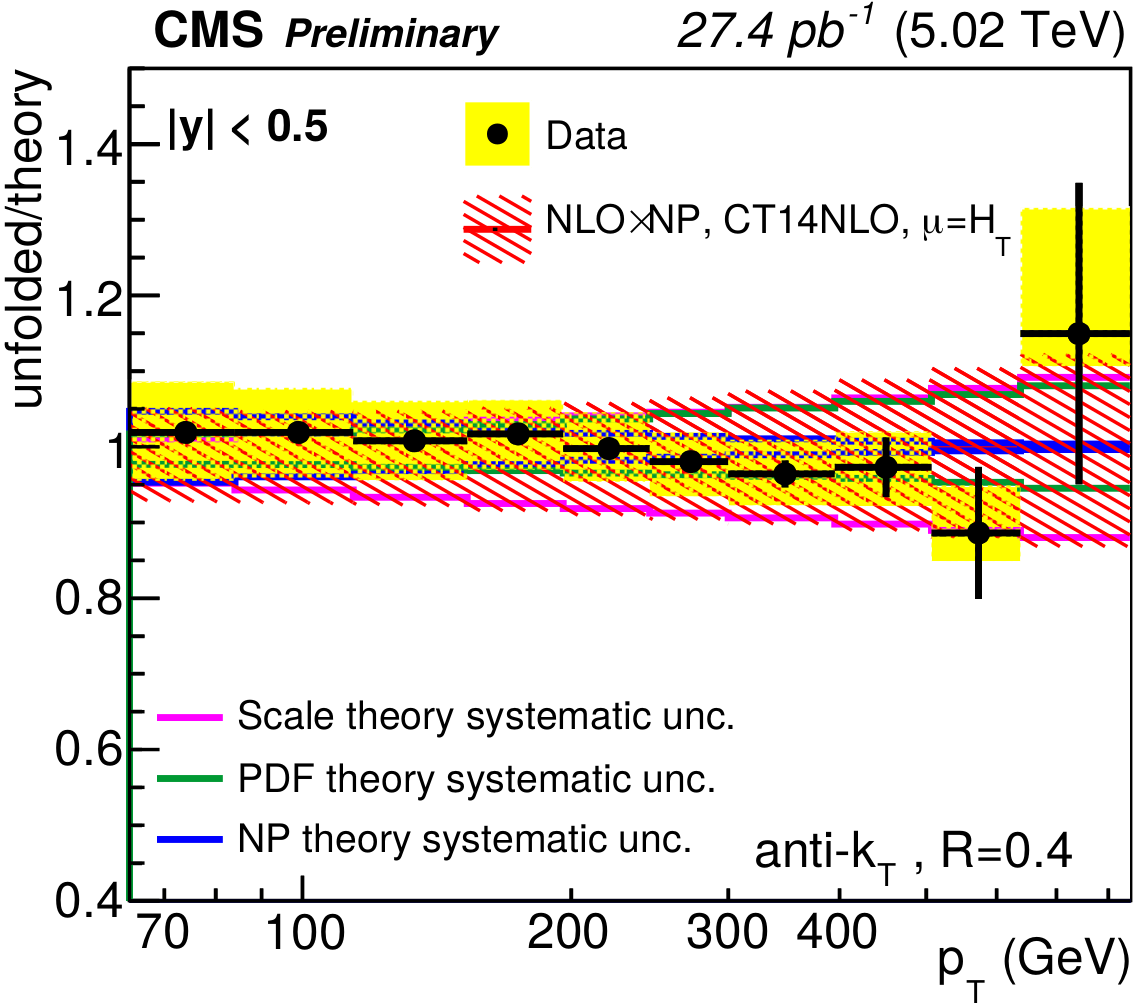}
\includegraphics[width=50mm]{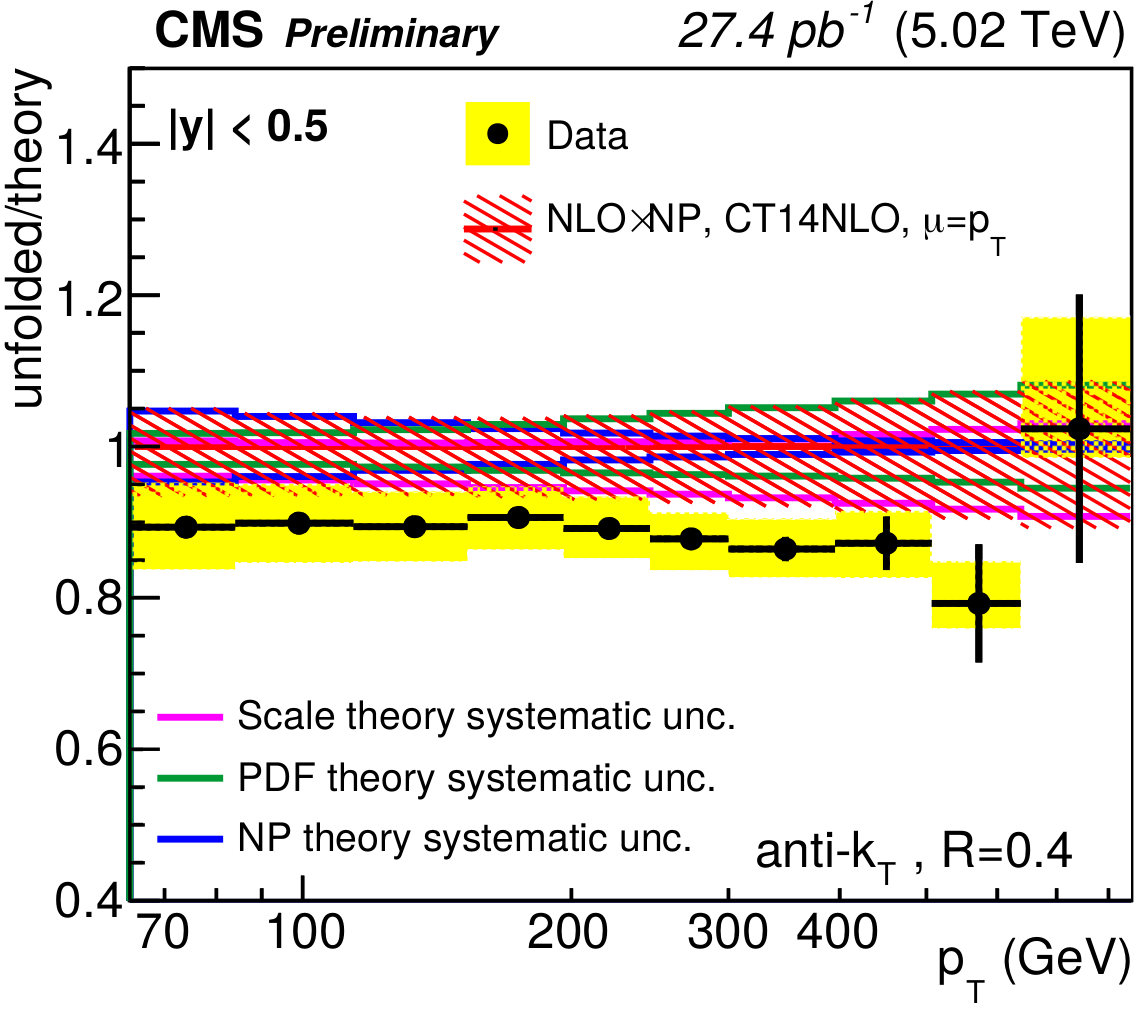}
\includegraphics[width=50mm]{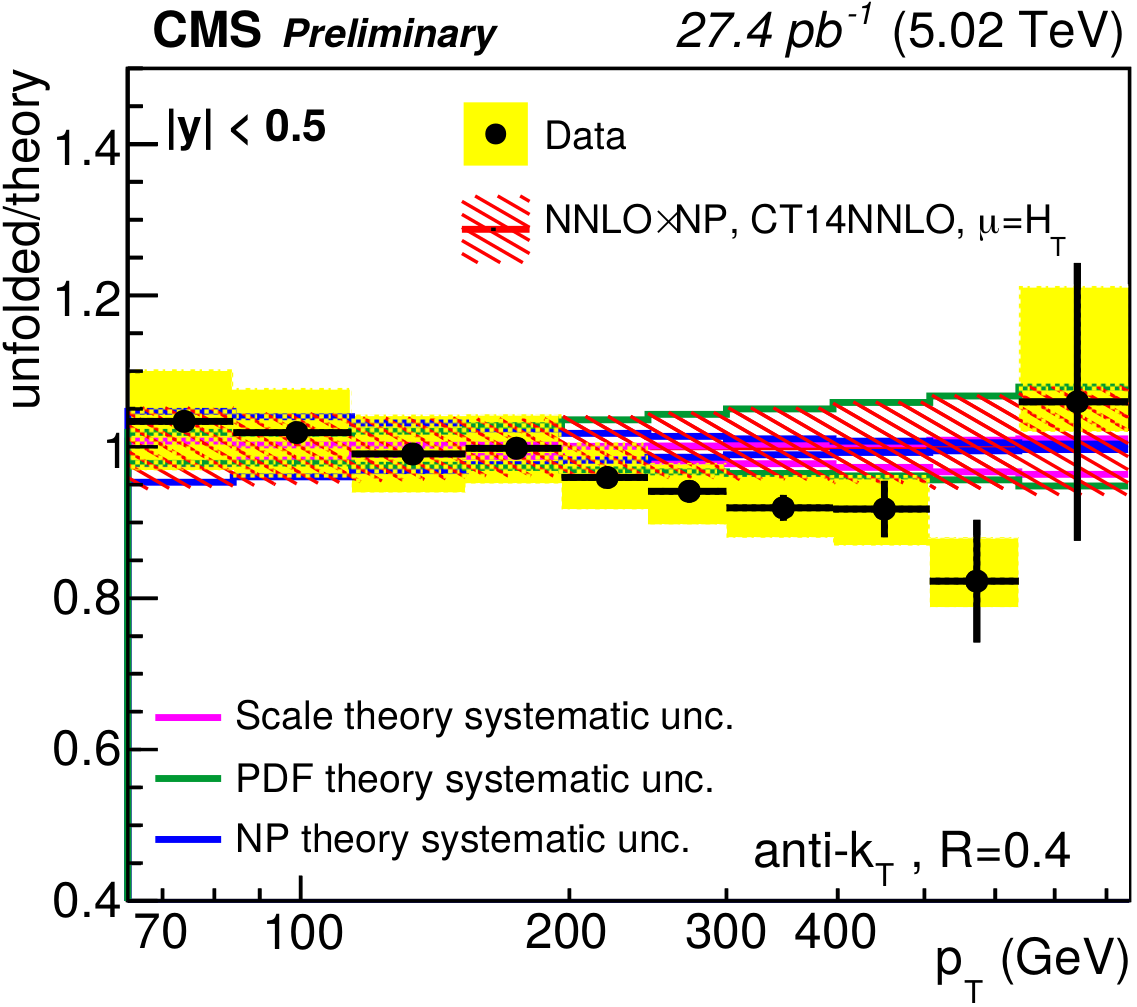}
\caption{A comparison of the 5\TeVns inclusive jet cross section data to theory predictions at NLO using the $\HT$ scale (left), at NLO using the $\pt$ scale (middle), and at NNLO using the $\HT$ scale~\cite{CMS-PAS-SMP-21-009}.}
\label{5TeVdataVsTheory}
\end{figure}

Figure~\ref{5TeVdataVsTheory} shows the ratio of the unfolded inclusive jet cross section to theoretical predictions computed with the CT14 PDF~\cite{Dulat:2015mca}. Using the NLO predictions with the factorisation and renormalisation scales set to $\mu_f = \mu_r = \pt$, the data is seen to be systematically below theory. However, $\HT$ scale gives a good description of the data at both NLO and NNLO. Furthermore, the NNLO prediction turns out less dependent on the choice of scale than NLO~\cite{CMS-PAS-SMP-21-009}.

\section{Cross section measurements of jet multiplicity and jet transverse momenta in multijet events at $13\TeVns$}
\label{sec:multijet}

Multijet production in $\Pp\Pp$ collisions is a probe of QCD at high $\pt$ and high jet multiplicities. A recent CMS measurement~\cite{CMS-PAS-SMP-21-006} of multijet production is performed at $13\TeVns$ corresponding to an integrated luminosity of $36.3\,\mathrm{fb}^{-1}$. The jets are reconstructed using the anti-$\kt$ algorithm with $R=0.4$. Events containing a leading jet with $p_{\mathrm{T}1} > 200\GeVns$ and a subleading jet with $p_{\mathrm{T}2} > 100\GeVns$ within $|y| < 2.5$ are selected. The leading and subleading jets form a dijet system, and the multiplicity of jets with $\pt > 50\GeVns$  is measured for different regions of the leading jet $\pt$ and in bins of the azimuthal angle $\Delta\phi_{1,2}$ between the jets in the dijet system. The differential cross section of four leading jets is measured as a function of their $\pt$.

The measurements are compared to perturbative QCD predictions interfaced with different models for hadronisation and parton showering. The NLO matrix elements (ME) of 2 jet and 3 jet production are computed using \MGvATNLO~\cite{Alwall:2014hca}. For hadronisation, it is interfaced with \textsc{Pythia 8}~\cite{Sjostrand:2014zea} using the CUETP8M1 tune~\cite{CMS:2015wcf} and the NNPDF 3.0 NLO PDF~\cite{NNPDF:2014otw}. Alternatively, CASCADE3~\cite{Baranov:2021uol} is used together with employing the \textsc{Herwig6} subtraction terms in MCatNLO and the NLO PB TMD set 2~\cite{BermudezMartinez:2018fsv} for transverse momentum dependent parton densities. All computations are normalised to the measured dijet cross section. The factorisation and renormalisation scales $\mu_f$, $\mu_r$ are set equal to  $1/2 \sum_i H_{\mathrm{T}i}$, with $H_{\mathrm{T}i}$ being the scalar transverse momenta and the sum taken over all produced partons. The scale uncertainty is obtained by varying $\mu_r$ and $\mu_f$ independently by a factor of 2 up and down, avoiding the cases with $\mu_r/\mu_f =4^{\pm 1}$.

\begin{figure}[h]
\centering
\includegraphics[width=110mm]{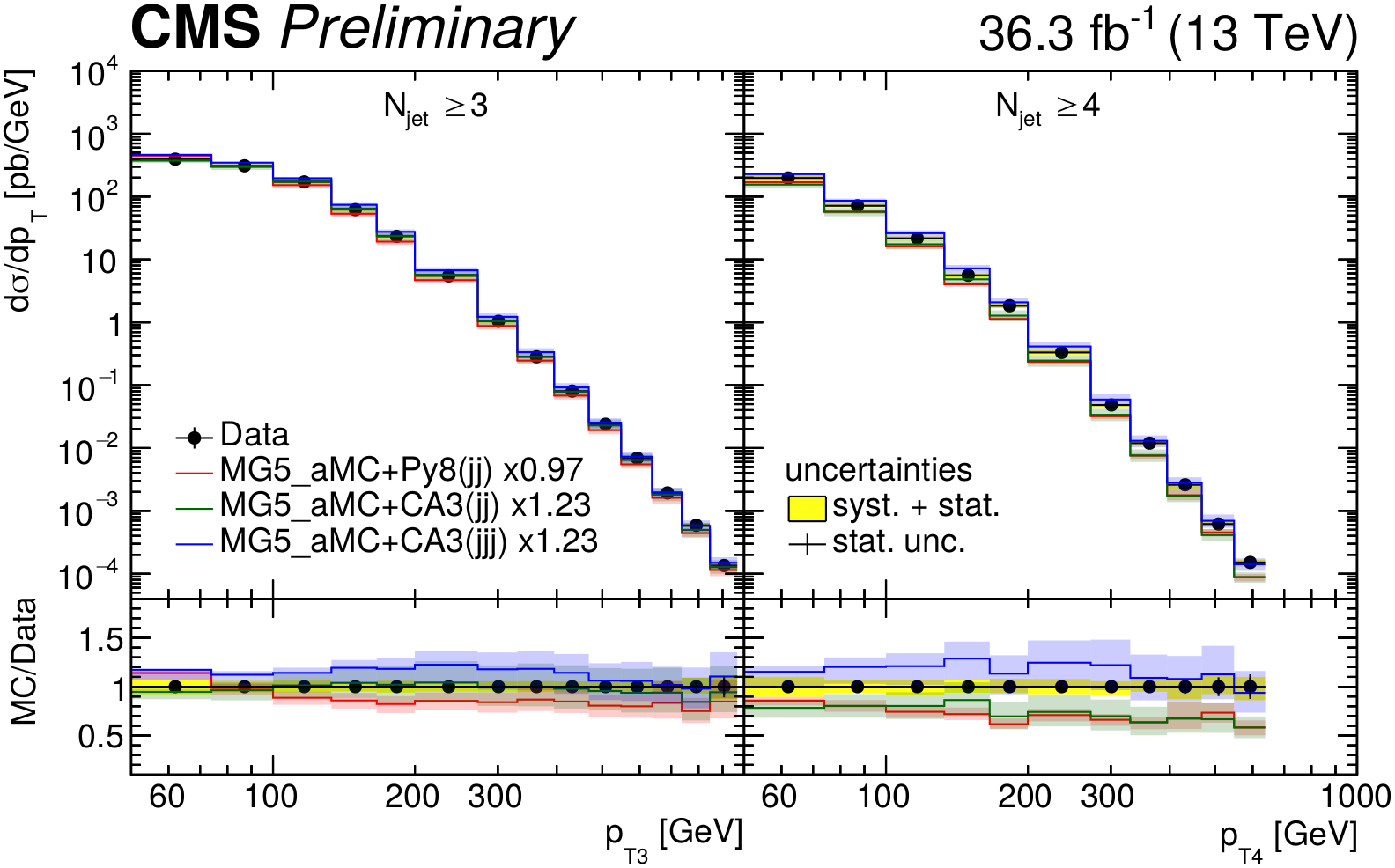}
\caption{A comparison of the measured $\pt$ distributions of the 3rd and 4th leading jets to NLO predictions. The label (jj) refers to 2 jet and (jjj) to 3 jet production in the ME~\cite{CMS-PAS-SMP-21-006}.}
\label{multijetFig}
\end{figure}

The NLO generator gives a good description of the experimental data, particularly when using the transverse momentum dependent PDFs~\cite{BermudezMartinez:2018fsv}. This is illustrated in Figure~\ref{multijetFig}, which portrays the $\pt$ distributions of the 3rd and 4th leading jets.

For the first time, jet multiplicity has been measured in bins of leading jet $\pt$ and the azimuthal angle between the two leading jets, with up to seven measurable jets. The results will be essential for comparisons of SM multijet production calculations, and are particularly beneficial for high jet multiplicity simulations with parton showers.

\section{Measurement and QCD analysis of double-differential inclusive jet cross sections at $13\TeVns$}
\label{sec:incJets13}

The $\Pp\Pp$ collison data at 13 TeV are used by the CMS Collaboration to measure the cross section of inclusive jet production~\cite{CMS-PAS-SMP-20-011}. The present results involve jets reconstructed using the anti-$\kt$ algorithm with the distance parameter $0.7$, for which the data correspond to an integrated luminosity of 33.5 $\mathrm{fb}^{-1}$.

\begin{figure}[h]
\includegraphics[width=77.5mm]{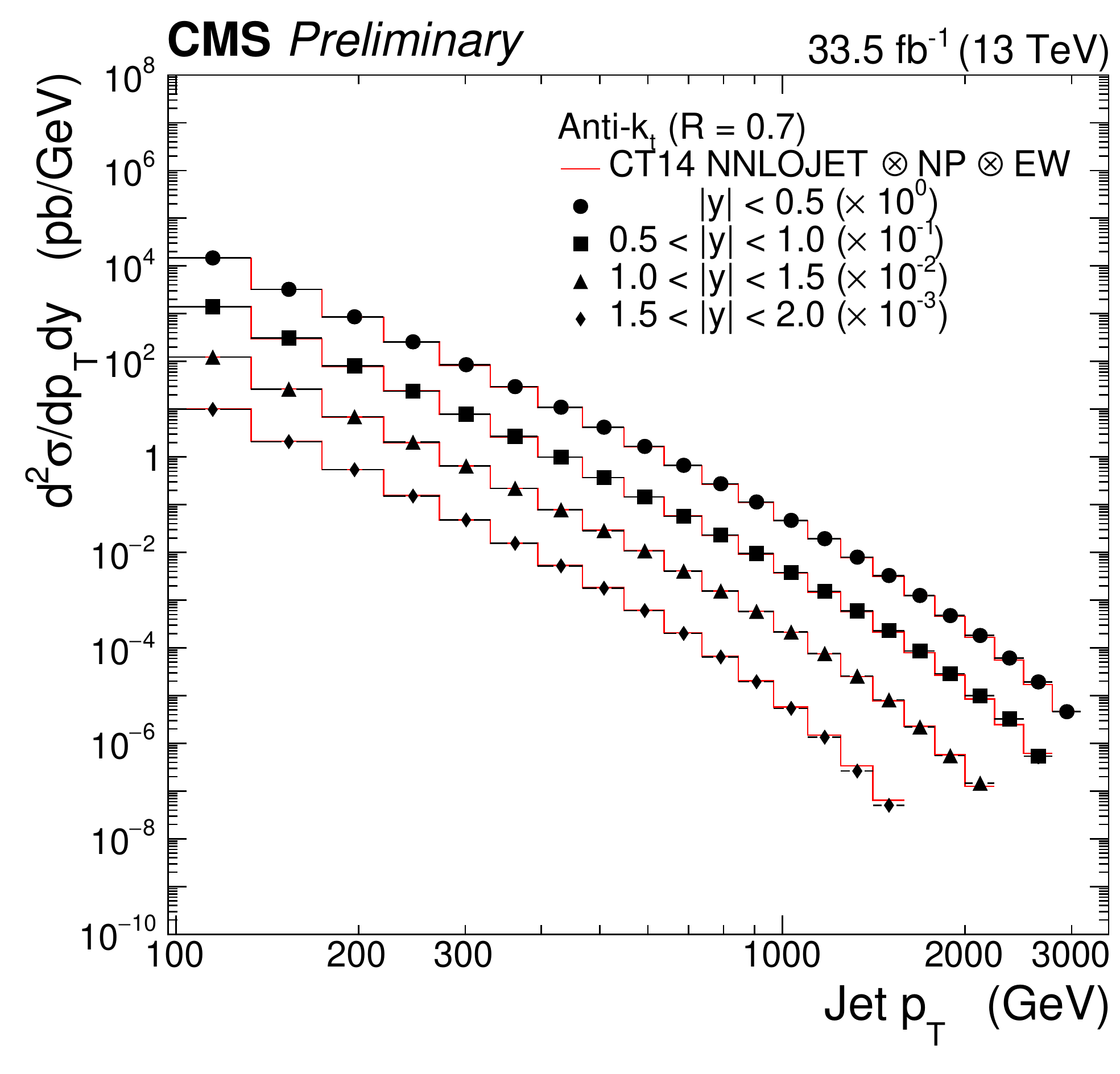}
\raisebox{4.3mm}{\includegraphics[width=71.2mm]{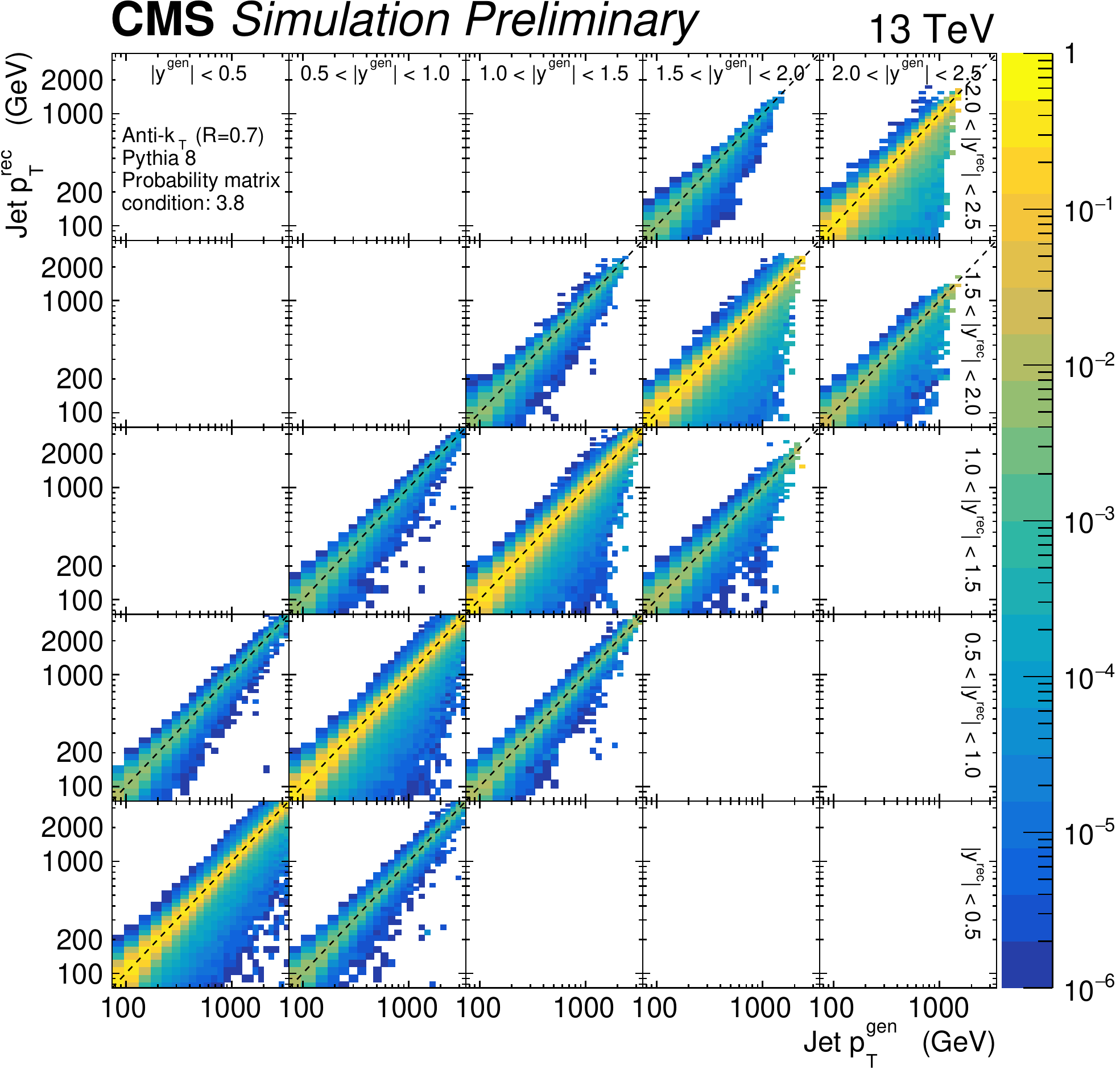}}
\caption{\textit{Left:} The inclusive jet cross sections with a comparison to NLO QCD predictions using the CT14 PDF~\cite{CMS-PAS-SMP-20-011}. \textit{Right:} The probability matrix. The horizontal and vertical axes correspond to particle and detector level jets, respectively.~\cite{CMS-PAS-SMP-20-011}}
\label{13TeV_cs_and_rm}
\end{figure}

The unfolding is performed two-dimensionally using least-square minimisation. Attention is paid to the smoothness of all bin-to-bin uncertainties, with tests of smoothness performed using Chebyshev polynomials. The data is shown in Figure~\ref{13TeV_cs_and_rm} along with the probability matrix, which is the response matrix normalised row-by-row.

\begin{figure}[H]
\centering
\includegraphics[width=150mm]{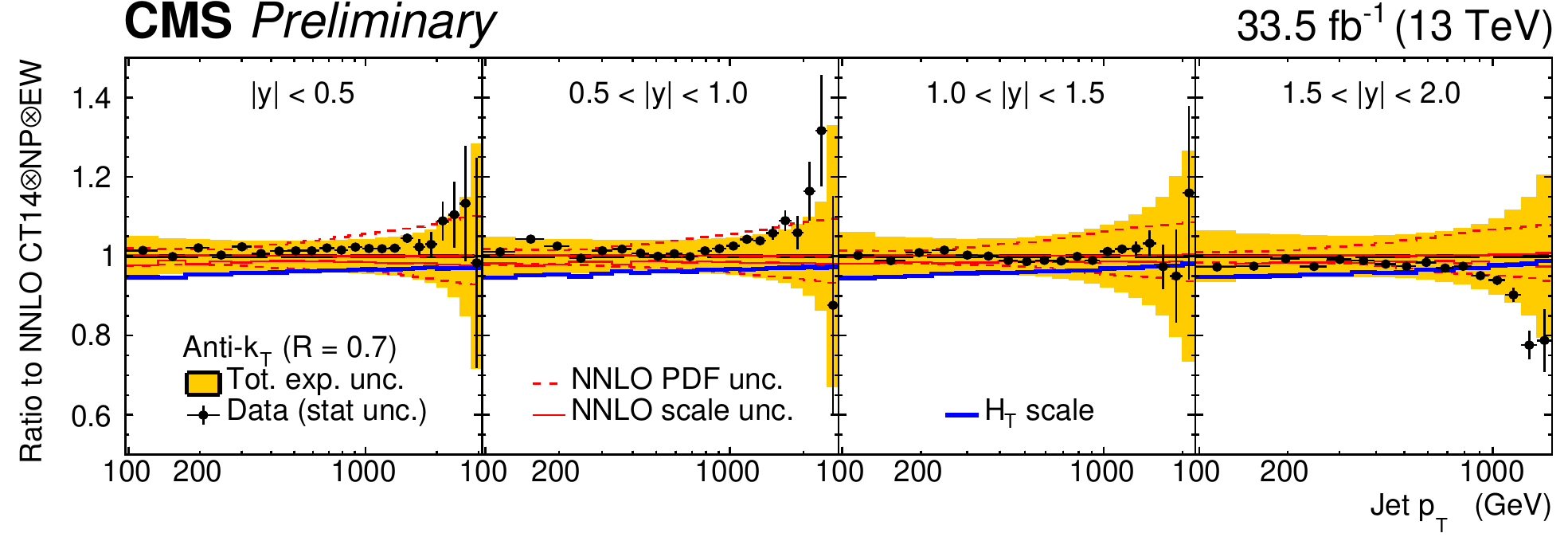}\\
\includegraphics[width=150mm]{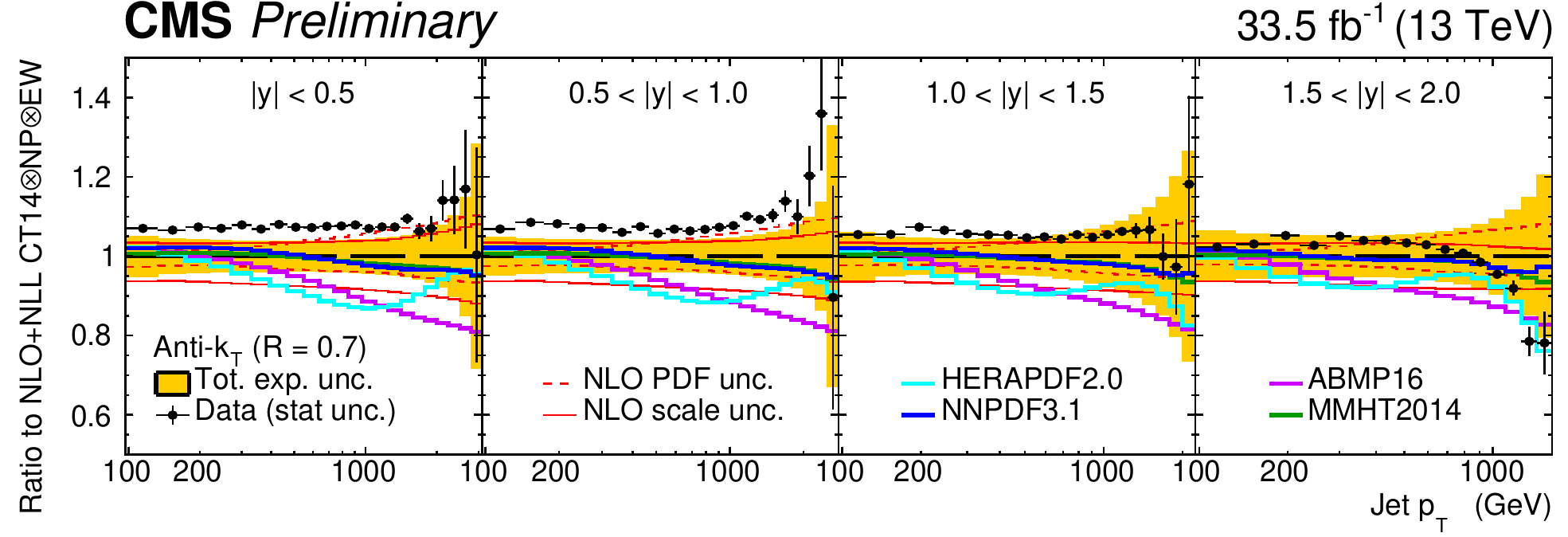}
\caption{Comparisons of the double differential cross section data to theoretical predictions using different PDFs and portraying the role of various uncertainties. In the bottom plot all histograms are divided by the NLO+NLL prediction and in the top plot by the NNLO prediction, obtained using the CT14 PDF~\cite{CMS-PAS-SMP-20-011}.}
\label{13TeV_data_vs_theory}
\end{figure}

The data are compared with fixed-order QCD predictions available at NLO and NNLO, obtained by using NLOJet++~\cite{Nagy:2001fj, Nagy:2003tz} and NNLOJET (rev5918)~\cite{Currie:2016bfm, Currie:2018xkj, Gehrmann:2018szu}. The NLO calculations are implemented in \textsc{FastNLO}~\cite{Britzger:2012bs}. The NLO cross-section is upgraded to NLO+NLL via correction factors obtained with the \textsc{NLL-Jet} calculation, provided by the authors of Ref.~\cite{Liu:2018ktv}, and the MEKS~\cite{Gao:2012he} code. Details of the electroweak and nonperturbative corrections are given in~\cite{CMS-PAS-SMP-20-011}. The comparison is performed using various global PDFs, and depicted in Figure~\ref{13TeV_data_vs_theory}. In particular, the scale uncertainty is observed to decrease noticeably at NNLO.

The sensitivity of the present measurement to the proton PDFs and $\alpS(m_{\PZ})$ is investigated in a comprehensive QCD analysis, where the double-differential inclusive jet production cross section is used together with the charged- and neutral-current deep inelastic scattering (DIS) cross sections of HERA~\cite{Abramowicz:2015mha}. In addition, the normalised triple-differential $\ttbar$ cross section~\cite{Sirunyan:2019zvx} from CMS is used. The scales  $\mu_f$ and $\mu_r$ are set to the four-momentum transfer $Q$ for the DIS data and to the individual jet $\pt$ for the inclusive jet production cross section. For $\ttbar$ production, they are set to half the sum of the transverse masses of the partons, as done in Ref.~\cite{Sirunyan:2019zvx}. The QCD analysis is performed in terms of the SM and standard model effective field theory (SMEFT) by using the \textsc{xFitter} QCD analysis framework~\cite{Bertone:2017tig, Alekhin:2014irh}, interfaced to \textsc{CIJET}~\cite{Gao:2012qpa, Gao:2013kp} for the SMEFT prediction. This allows for a simultaneous extraction of PDFs, $\alpS$, $m_{\PQt}^{\textrm{pole}}$ and the Wilson coefficient $c_1$ of 4-quark contact interactions (CI).

The CI are expected to appear as deviations from the SM jet cross section spectrum at low rapidity and high-$\pt$. However, SM predictions are based on PDFs which are derived assuming the validity of the SM at high jet $\pt$. Hence there is a possibility that BSM effects are absorbed in the PDF fit. To ensure that the search for CI is non-biased, the PDFs are fitted simultaneously when using a SMEFT prediction. The SM is extended by

\begin{equation}
\mathcal{L}_{\textrm{SMEFT}} = \mathcal{L}_{\textrm{SM}}
                             + \frac{4\pi}{2\Lambda^2}
                               \sum_{n} c_n O_n,
\label{SMEFT_Lagrangian}
\end{equation}
where $\Lambda$ is the scale of new physics, $c_n$ are Wilson coefficients and $O_n$ are dimension 6 operators for 4-quark CI corresponding to purely left-handed, vector-like or axial vector-like colour singlet exchanges. 

\begin{figure}[h]
\centering
\includegraphics[width=59.9mm]{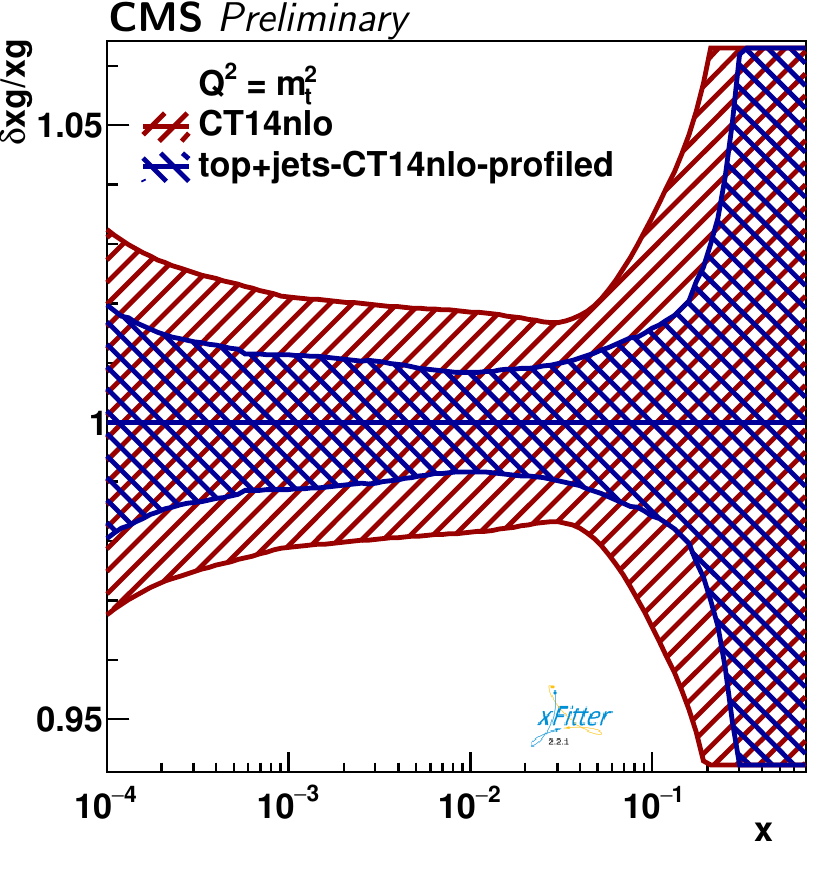}
\hspace*{3mm}
\raisebox{2.1mm}{\includegraphics[width=65mm]{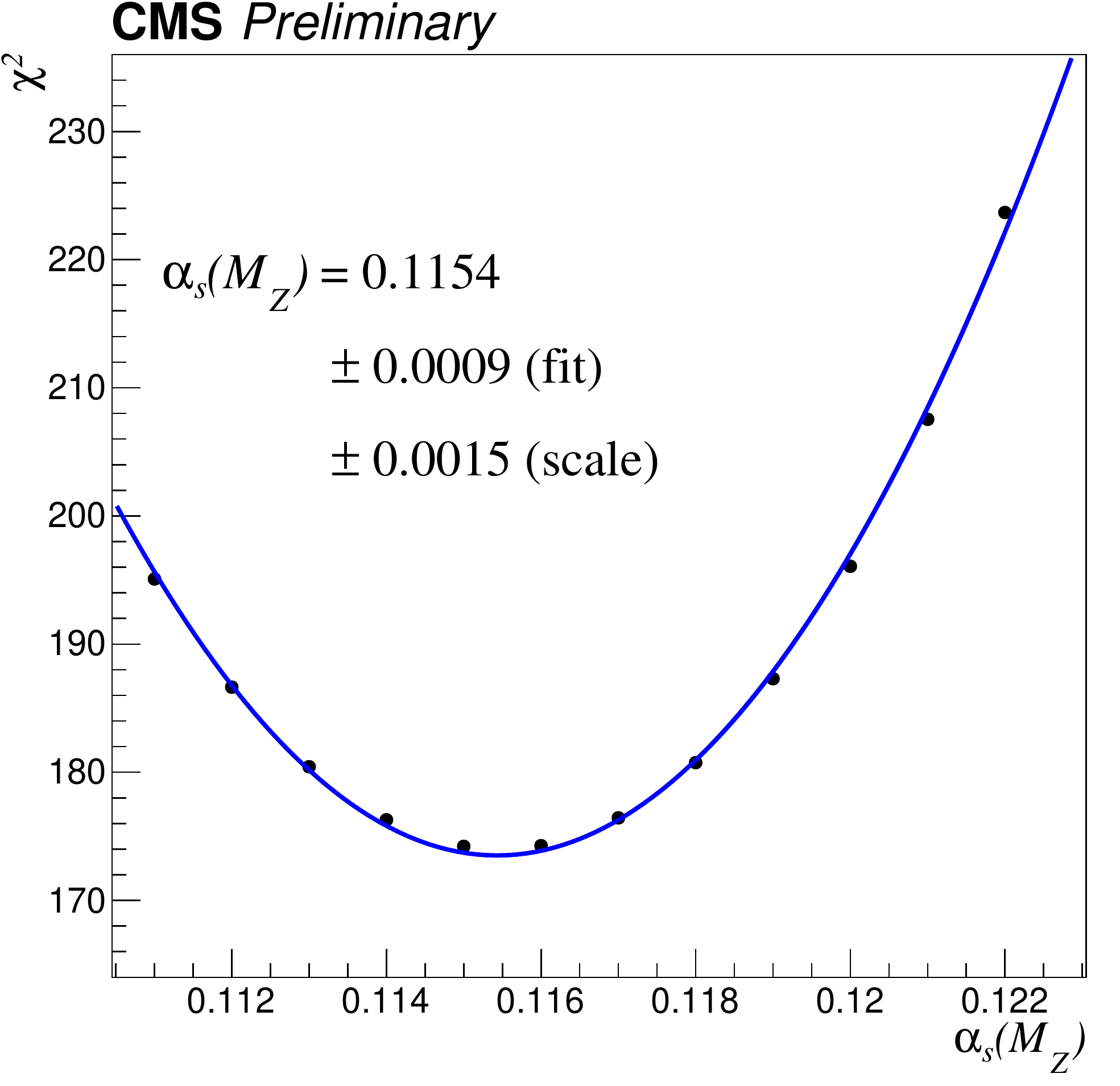}}
\caption{Profiling with CT14nlo at NLO, using the CMS inclusive jet and the triple-differential $\ttbar$ cross sections at $13\TeVns$. \textit{Left:} Relative uncertainty in the gluon PDF as a function of the momentum fraction $x$, at the scale $\mu_\text{f}=m_{\PQt}$. The CT14 uncertainty is shown in red and the profiling result in blue~\cite{CMS-PAS-SMP-20-011}. \textit{Right:} The $\chi^2$ scan for $\alpS(m_{\PZ})$ profiling with the CT14 PDF series~\cite{CMS-PAS-SMP-20-011}.}
\label{profilingPlots}
\end{figure}

The impact of the $13\TeVns$ data on a global PDF is assessed through a profiling procedure~\cite{Paukkunen:2014zia, Schmidt:2018hvu} performed using the CT14 PDF~\cite{Dulat:2015mca} at NLO and NNLO. The fractional uncertainty of the CT14 gluon PDF and the result of profiling with the CMS inclusive jet and $\ttbar$ data is shown in Figure~\ref{profilingPlots}, which also shows a scan for $\alpS$. The scan results in 
$\alpS(m_{\PZ}) = 0.1154 \pm 0.0009\,\mathrm{(fit)} \pm 0.0015\,\mathrm{(scale)}$, 
where the scale uncertainty is obtained by varying $\mu_r$ and $\mu_f$ independently by factors of 1/2 and 2, excluding the combinations with $\mu_r/\mu_f =4^{\pm 1}$. The gluon PDF precision is improved significantly by the CMS data, and the profiled top quark mass $m_{\PQt}=170.3 \pm 0.5\,\mathrm{(fit)} \pm 0.2\,\mathrm{(scale)}\GeVns$ is consistent with Ref.~\cite{Sirunyan:2019zvx}.

\begin{figure}[H]
\centering
\includegraphics[width=65mm]{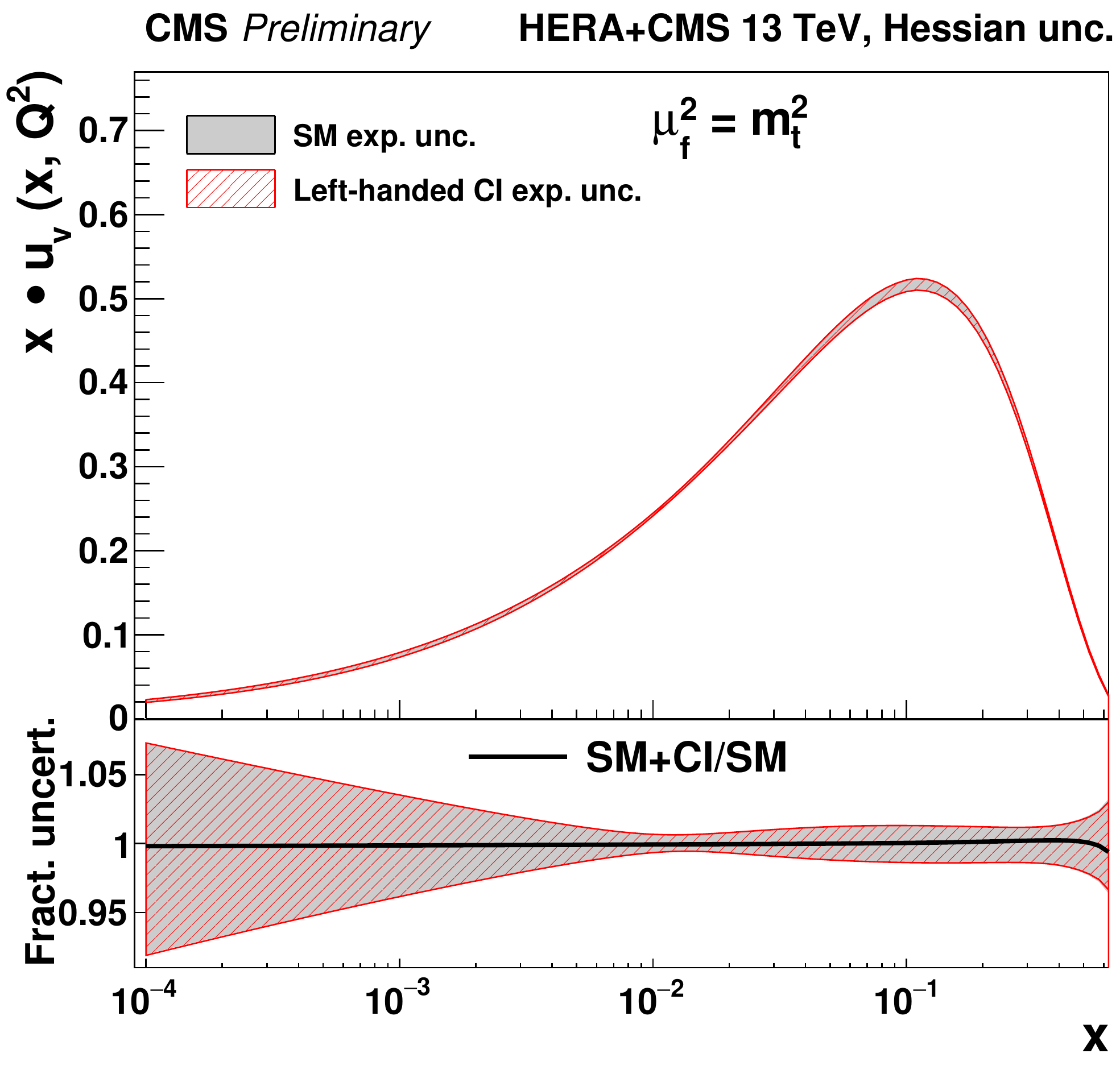}
\includegraphics[width=65mm]{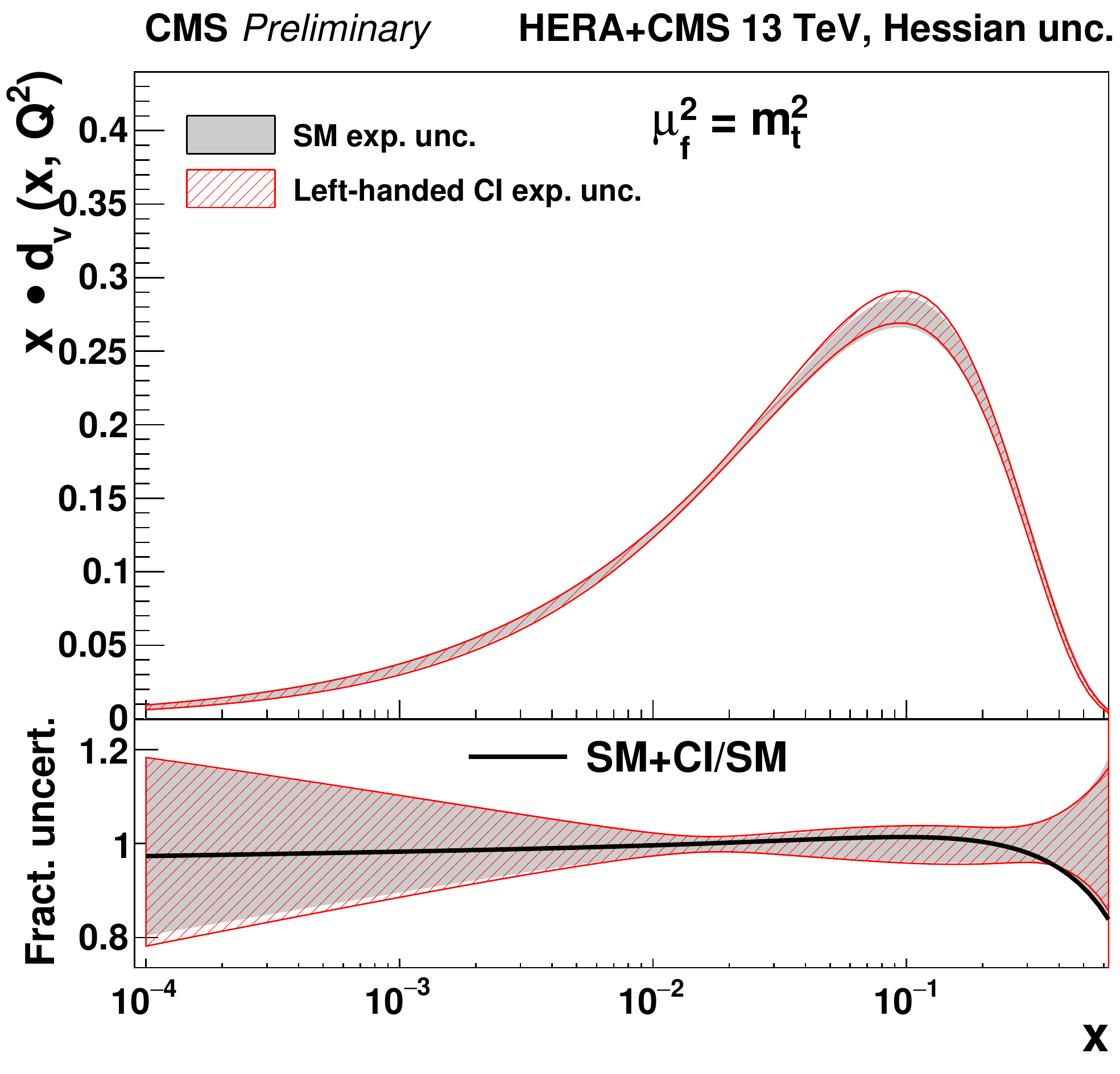}\\
\includegraphics[width=65mm]{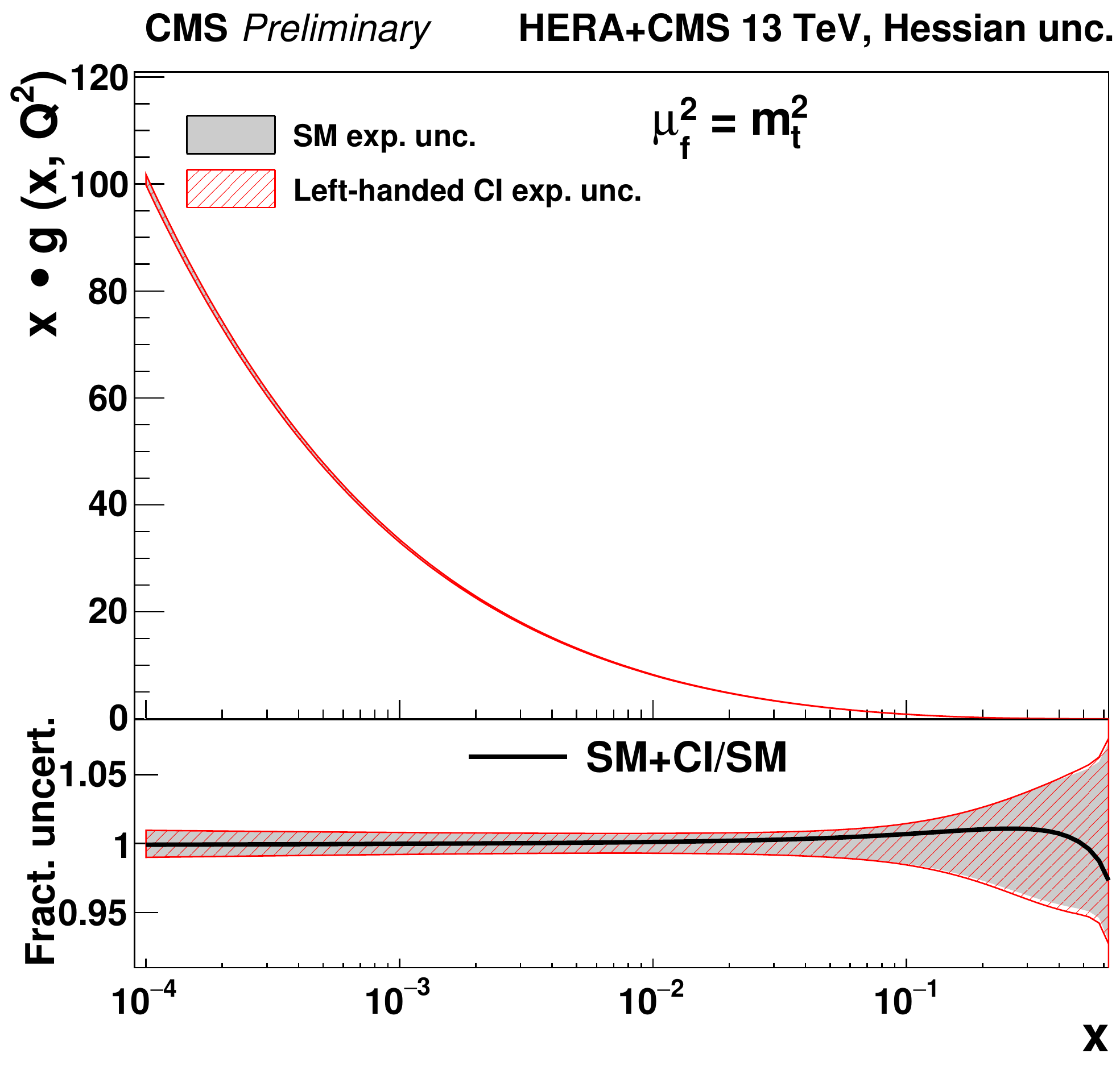}
\includegraphics[width=65mm]{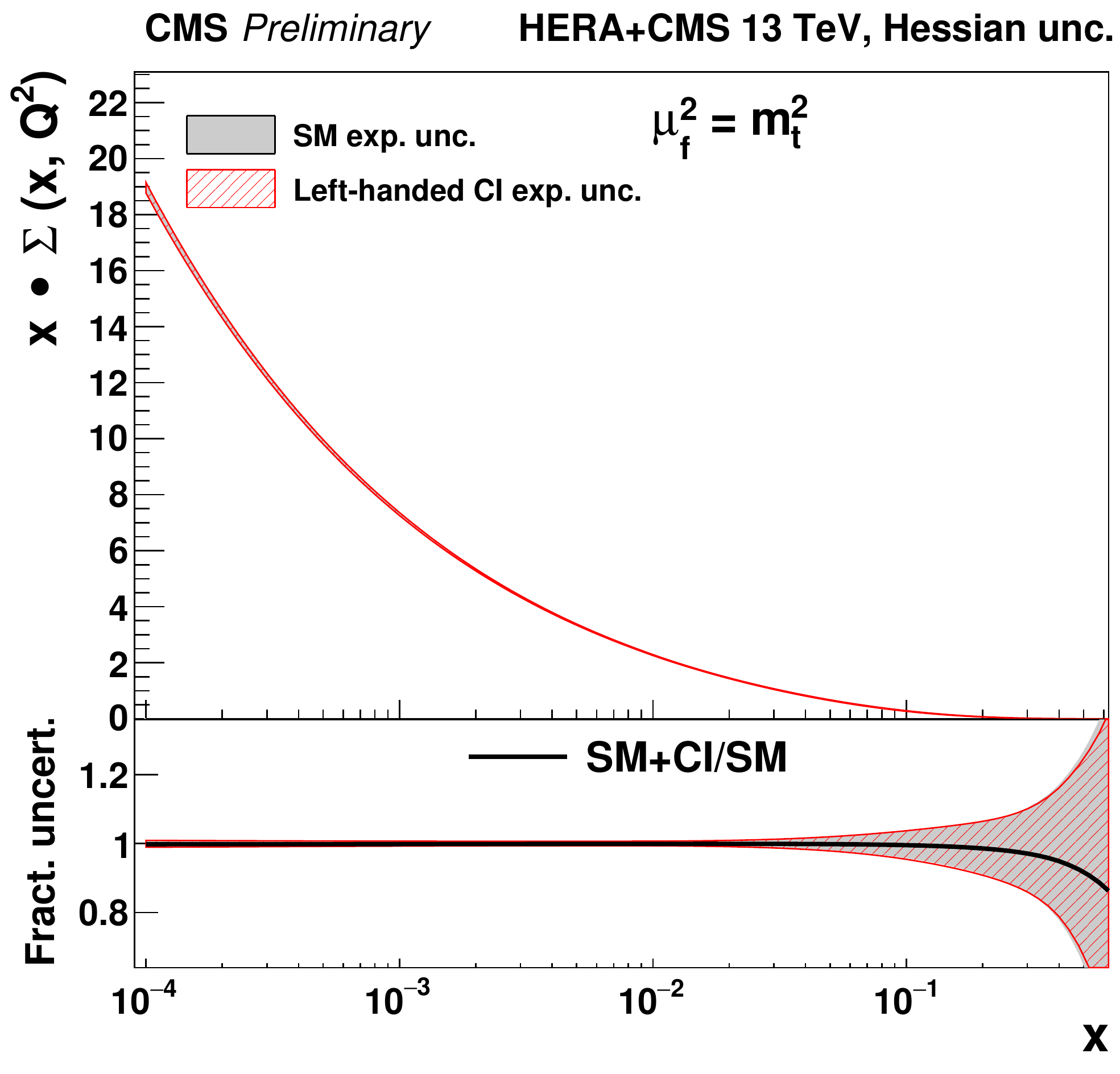}
\caption{The $\PQu$ valence~(upper left), $\PQd$ valence~(upper right), gluon~(lower left), and sea quark~(lower right) PDFs as functions of the momentum fraction $x$ at the scale $\mu_\text{f}=m_{\PQt}$. 
The red hashed band results from the SMEFT fit with the left-handed CI model using $\Lambda=10\TeVns$. It agrees with the gray band resulting from the SM fit; all differences are within the fit uncertainties~\cite{CMS-PAS-SMP-20-011}.}
\label{SM_vs_CI_PDFs}
\end{figure}

\begin{figure}[H]
\centering
\includegraphics[width=90mm]{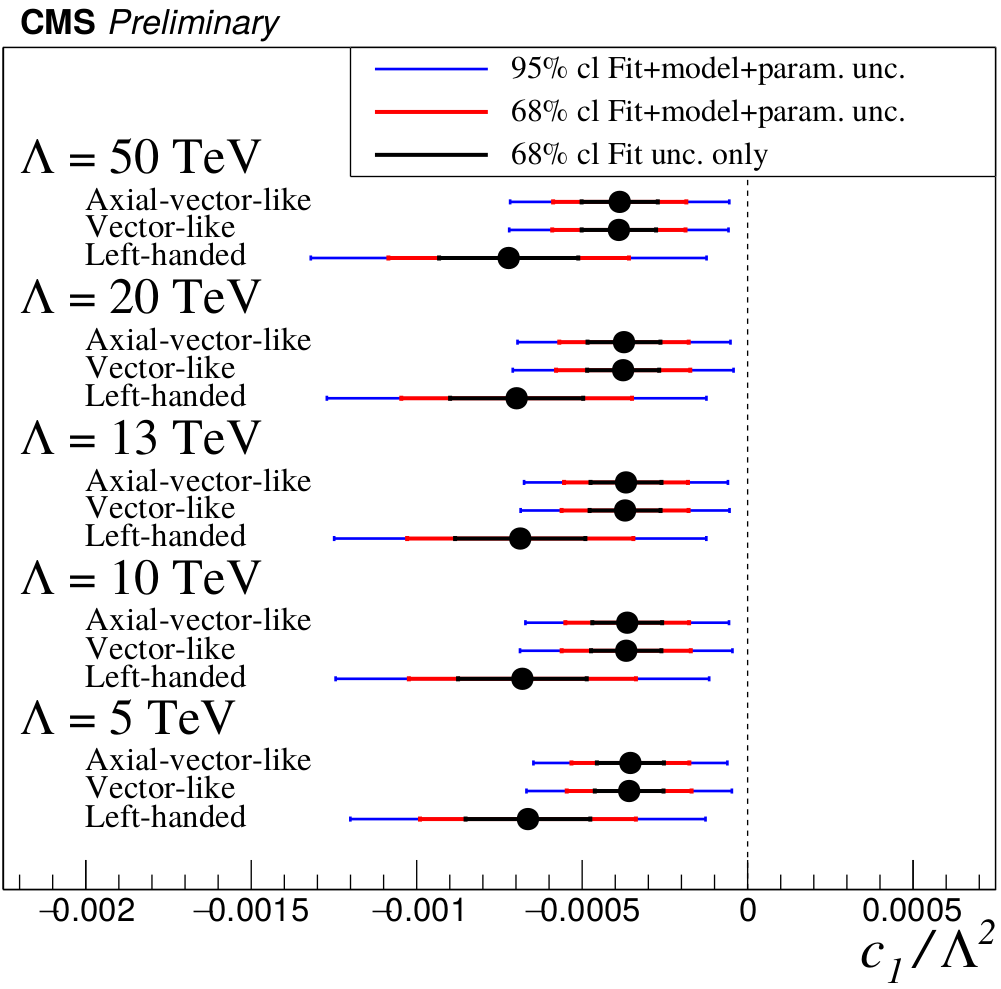}
\caption{The Wilson coefficients $c_1$ obtained in the NLO SMEFT analysis, divided by $\Lambda^2$. The black error bars show the fit uncertainty at 68\% CL. The red (blue) lines correspond to the total uncertainty at 68\% (95\%) CL~\cite{CMS-PAS-SMP-20-011}.}
\label{Wilson2perLambda}
\end{figure}

However, the profiling procedure does not allow for a simultaneous extraction of the PDFs and non-PDF parameters. Therefore, a full fit is performed using SM predictions and, alternatively, assuming a SM+CI model. Uncertainties are estimated similarly to the HERAPDF2.0 method~\cite{Abramowicz:2015mha}, accounting for the fit, parameterisation and model uncertainties. The model uncertainties arise from variations in the fixed non-PDF parameter values, including the QCD scales, and the parameterisation uncertainties from adding and removing new parameters in the PDF parameterisation, one at a time.
The SM fit results in 
$m_{\PQt}^{\textrm{pole}} = 170.4  \pm 0.6\,\mathrm{(fit)} \pm 0.3\,\mathrm{(model + par)}\GeVns$, 
compatible with the previous CMS result~\cite{Sirunyan:2019zvx}, and
$\alpS(m_{\PZ}) = 0.1187 \pm 0.0016\,\mathrm{(fit)} \pm 0.0030\,\mathrm{(model + par)}$,
compatible with the world average~\cite{10.1093/ptep/ptaa104}. 
The PDFs, $\alpS(m_{\PZ})$, and $m_{\PQt}^{\textrm{pole}}$ resulting from the SM fit and the SMEFT fits with all three CI models agree, and the PDFs are illustrated in Figure~\ref{SM_vs_CI_PDFs}. The SMEFT fits are sensitive to the ratio of the fitted $c_1$ to $\Lambda^2$, which expectedly remains constant as shown in Figure~\ref{Wilson2perLambda}. The negative $c_1$ imply a constructive interference with the SM gluon exchange, but are statistically compatible with zero. Thus, neither a deviation from the SM nor a risk of absorbing BSM effects in the SM PDF fit is observed.

Conventional searches for CI are performed by scanning for $\Lambda$ with $c_1$ fixed to $+1$ for destructive or $-1$ for constructive interference with the SM gluon exchange. The results of the present fit are translated into non-biased 95\% CL exclusion limits on $\Lambda$ with $c_1=-1$. These are $24\TeVns$ for left-handed, $32\TeVns$ for vector-like, and 31\TeVns for axial-vector-like CI. The most stringent comparable result is $22\TeVns$ for left-handed CI with constructive interference, obtained by the ATLAS Collaboration using $13\TeVns$ dijet cross sections data~\cite{ATLAS:2017eqx}.

\section{Summary}

Precision QCD measurements by the CMS Collaboration are reported, involving jet production in proton-proton collisions at $13\TeVns$ and $5\TeVns$. All results are in agreement with previous CMS results and world averages.

The first measurement of the $\PZ$ boson invisible width at a hadron collider is also the most precise to date. The study of multijet production at $13\TeVns$ is the first measurement of jet multiplicity with up to seven measurable jets and making use of transverse momentum dependent parton densities. Furthermore, measurements of inclusive jet production cross sections are presented at $5\TeVns$ and $13\TeVns$. The $5\TeVns$ results provide a valuable reference for probing quark-gluon plasma in lead--proton collisions at $5.02\TeVns$, whereas the $13\TeVns$ data are incorporated in an analysis following a non-biased strategy, resulting in the simultaneous extraction of PDFs, $\alpS(m_{\PZ})$, $m_{\PQt}^{\textrm{pole}}$ and the Wilson coefficient of 4-quark contact interactions for the first time using hadron collider data.
\nocite{*}
\bibliographystyle{auto_generated}
\bibliography{Precision_QCD_measurements_from_CMS/Precision_QCD_measurements_from_CMS/Precision_QCD_measurements_from_CMS}


%% file: proceedings-CSanchezGras/sanchez_cristina/Sanchez.tex
\vspace*{1.2cm}

\thispagestyle{empty}
\begin{center}
{\LARGE \bf Parton distribution functions and intrinsic charm at LHCb}

\par\vspace*{7mm}\par

{

\bigskip

\large \bf Cristina S\'{o}nchez Gras on behalf of the LHCb Collaboration}

\bigskip

{\large \bf  E-Mail: cristina.sanchez.gras@cern.ch}

\bigskip

{Nikhef National Institute for Subatomic Physics, Amsterdam, The Netherlands}

\bigskip

{\it Presented at the Low-$x$ Workshop, Elba Island, Italy, September 27--October 1 2021}

\vspace*{15mm}

\end{center}
\vspace*{1mm}

\begin{abstract}
	AAt LHCb, proton parton distribution functions (PDFs) can be studied in a unique phase space complementary to that accessible by ATLAS and CMS, corresponding to low and high values of Bjorken-$x$. The measurements of vector boson production in the forward region, with and without an associated jet, are presented. These measurements can be used to constrain the proton PDFs, and in particular, the production of a $Z$ boson in association with a $c$-jet can be studied to measure the intrinsic charm content of the proton.
\end{abstract}
  \part[Parton distribution functions and intrinsic charm at LHCb\\ \phantom{x}\hspace{4ex}\it{Cristina S\'{a}nchez Gras on behalf of the LHCb Collaboration}]{}
\section{Introduction}
Initially designed for the study of forward beauty and charm physics, the LHCb detector has the pseudorapidity ($\eta$) coverage $2 < \eta < 4.5$~\cite{lhcbdetector}. Its remarkable vertex reconstruction and particle identification performance, together with its high momentum resolution, have now established it as a general purpose detector. The forward coverage allows LHCb to reach \mbox{Bjorken-$x$} values (where $x$ is the fraction of momentum carried away by a parton) complementary to those accessible by other general purpose detectors, such as CMS~\cite{cms} and ATLAS~\cite{atlas_gras}. This allows to probe proton parton distribution functions (PDFs) at very low- and high-$x$ values.

On the low-$x$ side, central exclusive production (CEP) of $J/\psi$ and $\psi(2S)$ mesons in proton-proton ($pp$) collisions can probe the gluon PDF down to $x \sim 10^{-6}$~\cite{lhcbCEP}. For high values of $x > 0.1$, $Z$ boson production in association with charm jets can be used to determine the intrinsic charm content of the proton~\cite{lhcbIC}. The most recent LHCb results for these two cases are presented in this note.

\section{\boldmath Central exclusive production of $J/\psi$ and $\psi(2S)$ mesons in $pp$ collisions}
The diffractive process \mbox{$pp \rightarrow p + X + p$} in which two protons stay unscathed following their interaction is known as central exclusive production (CEP). The proton interaction takes place via the exchange of colourless objects. In the case of vector meson production, the exchange of a photon and a pomeron receives the name of photoproduction. The cross-section for photoproduction is proportional to the square of the gluon PDF (at leading order) in perturbative quantumchromodynamics (QCD). In $pp$ collisions with the LHCb coverage, the gluon PDF can then be proved at very low-$x$ values of $x \sim 10^{-6}$.

This process has a very distinctive signature: low final state multiplicity, as only the muons that follow the meson decay are present in the detector, and large rapidity gaps (regions of no activity) around the dimuon system. The latter can be spoiled when one of the protons dissociates after the interaction, making it an inelastic CEP process. The proton remnants in that case are produced outside the $2 < \eta < 5$ coverage and escape detection. To veto these processes, three (two) scintillator stations are in place around the beam pipe upstream (downstream) from the interaction point, conforming HeRSCheL. The HeRSCheL (High Rapidity Shower Counters for LHCb) detector consists on five 60 cm x 60 cm stations equipped with four scintillating pads each~\cite{herschel}. It increases the LHCb coverage to $1.5 < \eta < 10$ and $-10 < \eta < 5, \ -3.5 < \eta < -1.5$ in the forward and backward regions, respectively. Charged particles produced when a proton dissociates trigger detection by the scintillating pads. A $\chi^2$-like figure of merit built with the activity registered at the HeRSCheL stations is used to veto these events. 

\begin{figure}[hb]
	\begin{center}
		\begin{tabular}{c}
			\includegraphics[width=0.49\textwidth]{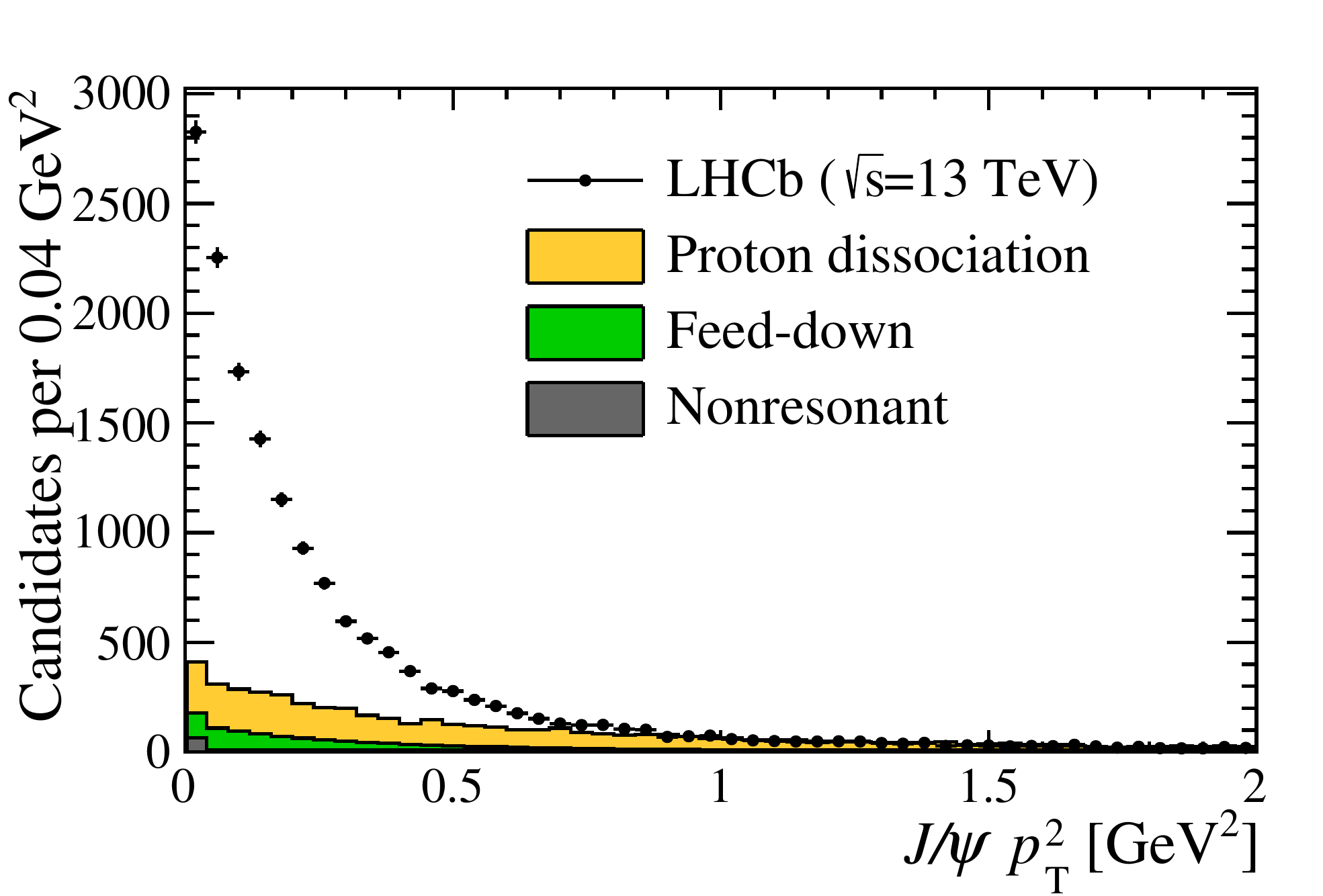}\hskip 0.01\textwidth
			\includegraphics[width=0.49\textwidth]{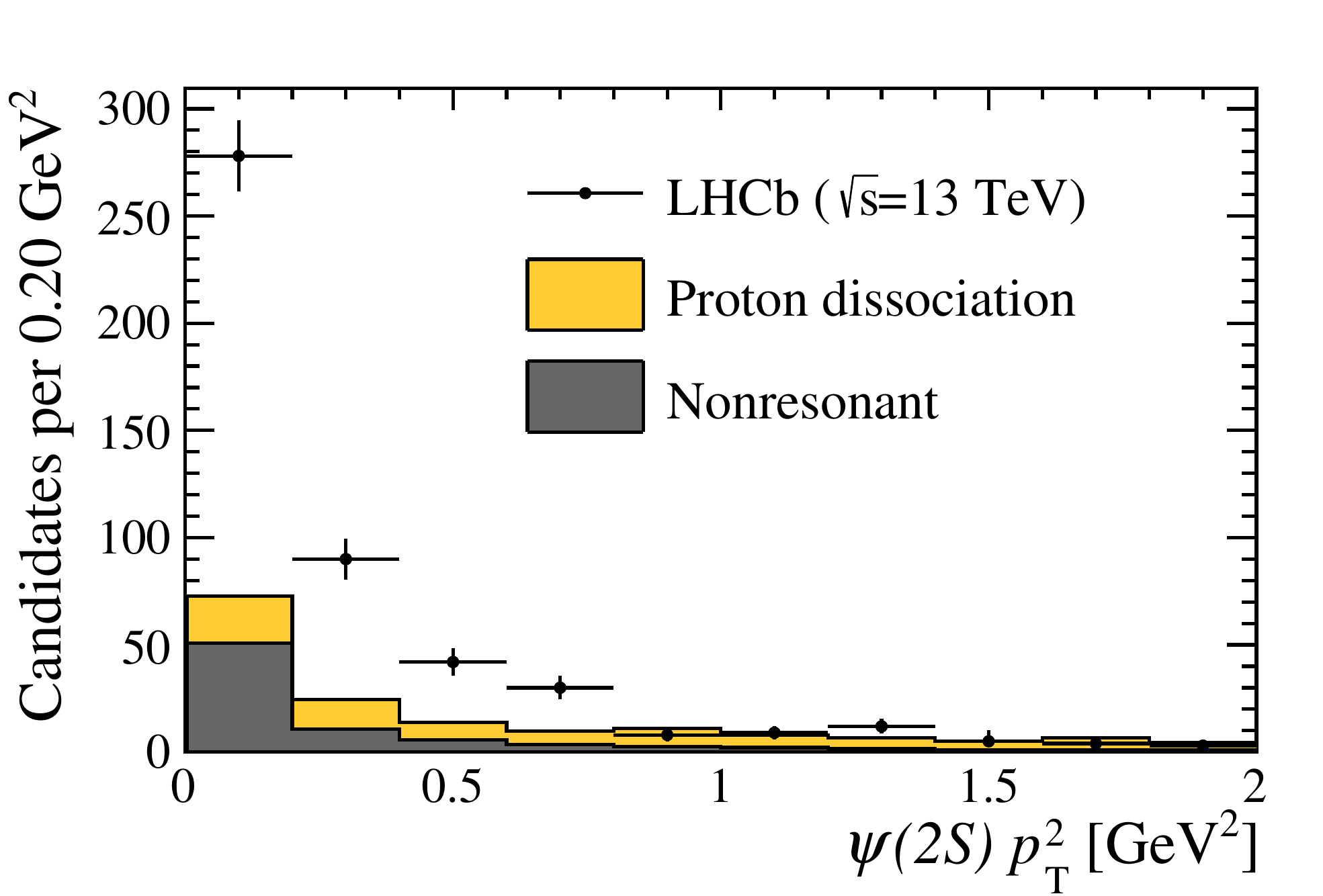}\\
		\end{tabular}
		\caption{Squared transverse momentum ($p_{\rm T}^2$) distribution for CEP $J/\psi \rightarrow \mu^+\mu^-$ (left) and \mbox{$\psi(2S) \rightarrow \mu^+\mu^-$} (right). The different backgrounds described in the text are indicated. }
		\label{fig:ptsqCEP}
	\end{center}
\end{figure}

CEP events are selected by requiring exactly two muon tracks and imposing a veto on the activity in HeRSCheL. The number of signal events is obtained by fitting the dimuon squared transverse momentum ($p_{\rm T}^2$) distribution. Following Regge theory, the cross-section for $J/\psi$ and $\psi(2S)$ CEP events follows \mbox{${\rm d \sigma}/{\rm d}y \sim \exp(-bp_{\rm T}^2)$}, with $b \sim 6$ (GeV$/c)^{-2}$. The background arising from the dissociation of one of the protons is parametrised in a sample with the HeRSCheL veto inverted. Aside from this, two more backgrounds need to be accounted for: nonresonant dimuon production and feed-down to $J/\psi$ from $\psi(2S)$ and $\chi_{c_{J}} \ (J=0,1,2)$. Nonresonant dimuon production takes place when both protons interact electromagnetically via photon-photon exchange. This background is measured by fitting the dimuon mass spectrum, and its $p_{\rm T}^2$ shape modelled from simulated samples. The $J/\psi$ feed-down background concerns $\chi_{c_{J}} \rightarrow J/\psi \gamma$ and $\psi(2S) \rightarrow J\psi X$ decays where only the $J/\psi$ is fully reconstructed. Its contribution is estimated combining data and simulation. The $J/\psi$ and $\psi(2S)$ $p_{\rm T}^2$ distributions are shown in Fig.~\ref{fig:ptsqCEP}. A fit to the background-subtracted shapes is used to determine the number of signal events.

\begin{figure}[t]
	\begin{center}
		\begin{tabular}{c}
			\includegraphics[width=0.49\textwidth]{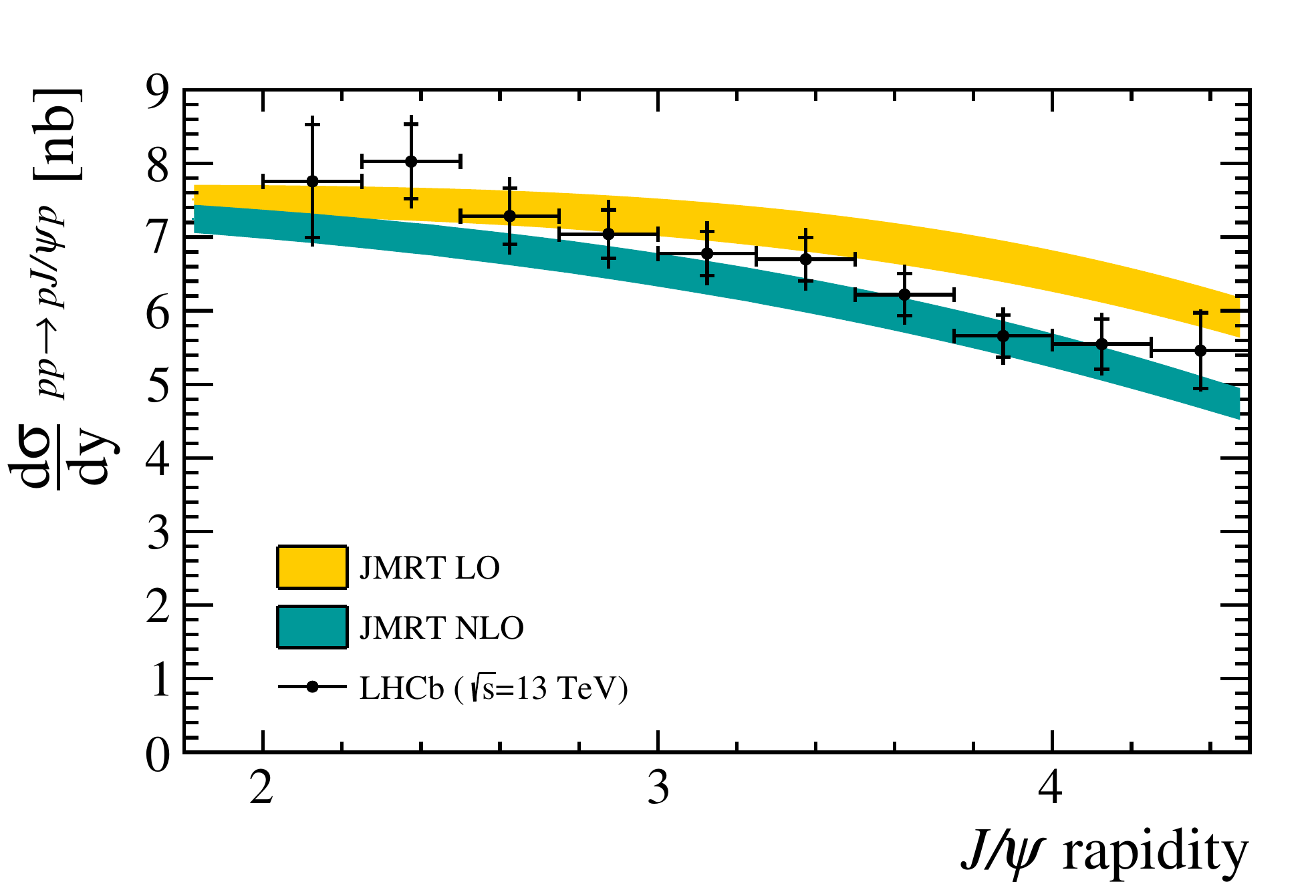}\hskip 0.01\textwidth
			\includegraphics[width=0.49\textwidth]{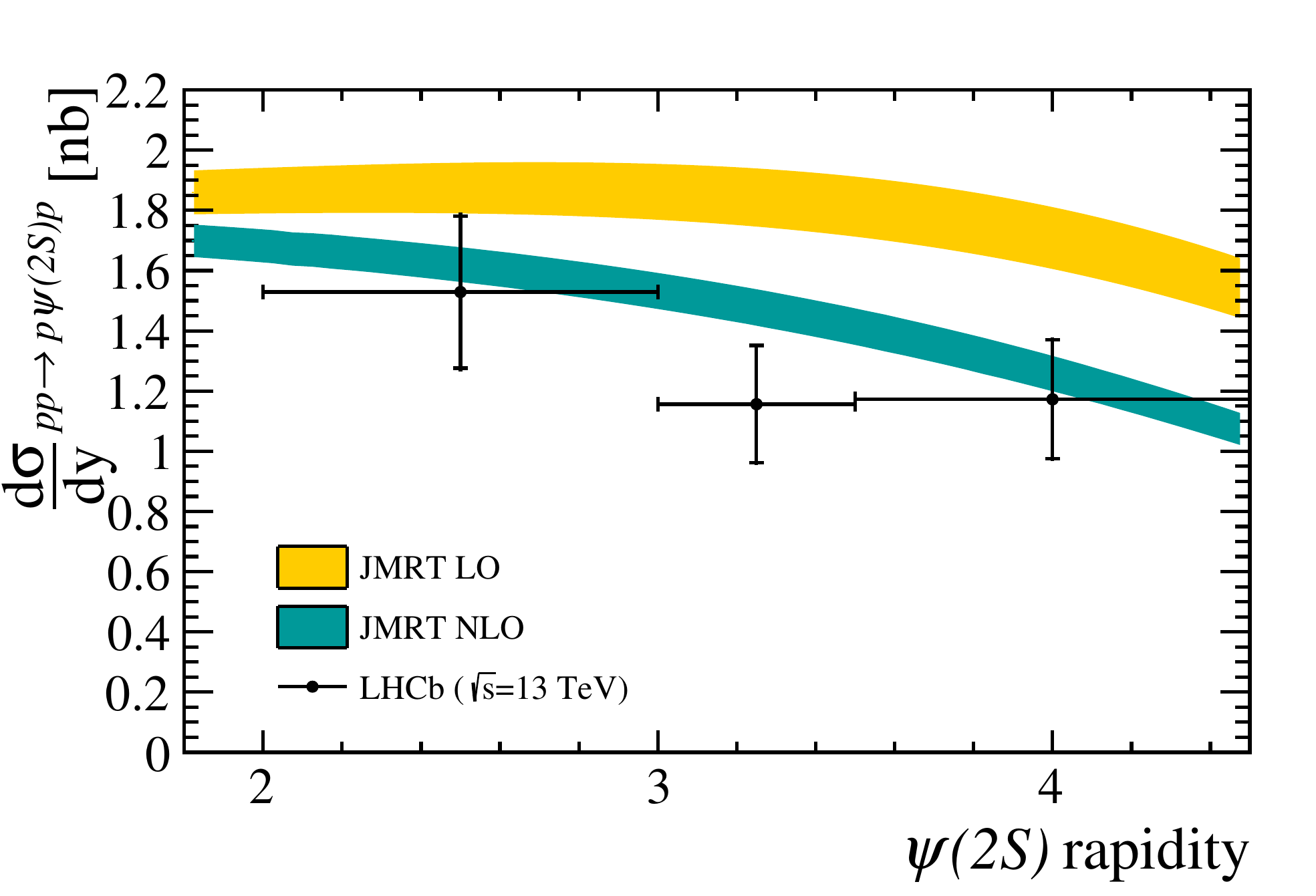}\\
		\end{tabular}
		\caption{Differential cross-section results for $J/\psi \rightarrow \mu^+\mu^-$ (left) and \mbox{$\psi(2S) \rightarrow \mu^+\mu^-$} (right) compared to LO and NLO JMRT theory descriptions~\cite{JMRTLO,JMRTNLO}. }
		\label{fig:resultCEP}
	\end{center}
	\begin{center}
		\begin{tabular}{c}
			\includegraphics[width=0.49\textwidth]{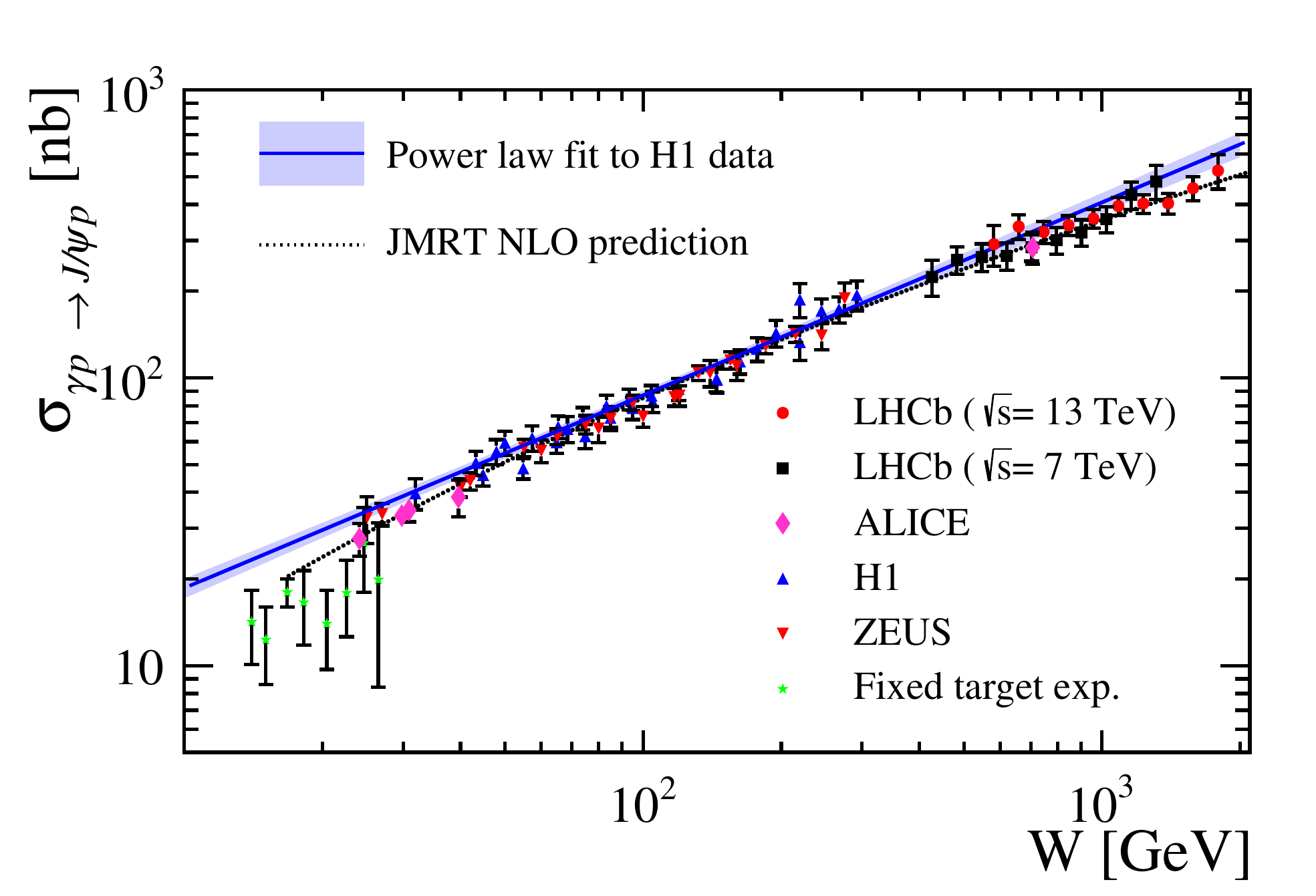}\hskip 0.01\textwidth
			\includegraphics[width=0.49\textwidth]{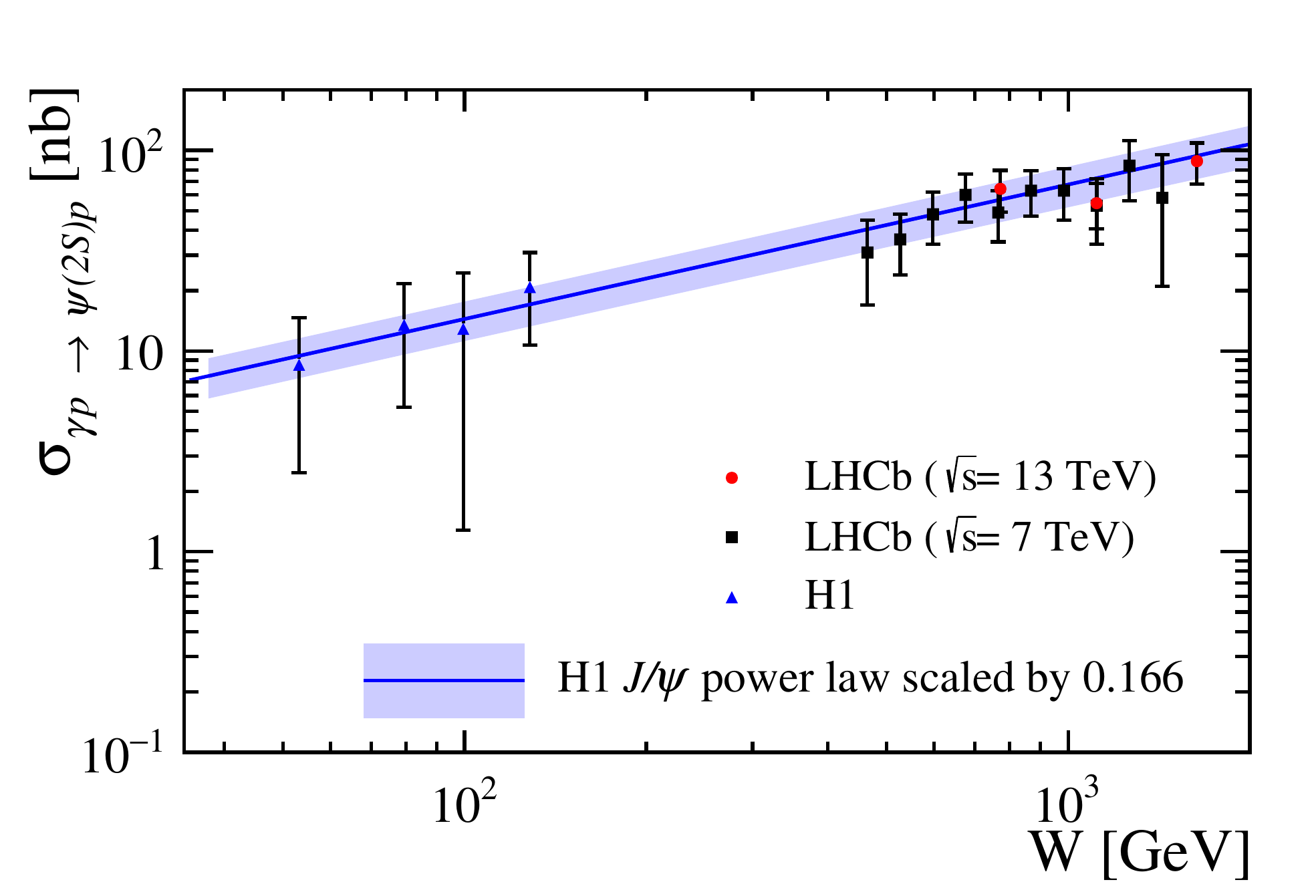}\\
		\end{tabular}
		\caption{Photoproduction cross-section results for $J/\psi$ (left) and $\psi(2S)$. The LHCb result at $\sqrt{s}=13 \ \mathrm{TeV}$ is shown together with the LHCb results at $\sqrt{s}=7 \ \mathrm{TeV}$~\cite{lhcb7tev} and those from the H1~\cite{H1param,H1psi2S} ZEUS~\cite{zeusCEP} and ALICE~\cite{aliceCEP} collaborations, and the results from the fixed target experiments E401~\cite{e401}, E516~\cite{e516} and E687~\cite{e687}.}
		\label{fig:photoprodCEP}
	\end{center}
\end{figure}

The result for the differential cross-section in rapidity bins is shown in Fig.~\ref{fig:resultCEP}, as well as the leading order (LO) and next-to-leading order (NLO) Jones-Martin-Ryskin-Teubner (JMRT) theory descriptions~\cite{JMRTLO,JMRTNLO}. In the $J/\psi$  case, the data is more in agreement with the NLO description, especially in the higher rapidity bins. For the $\psi(2S)$ the same trend is observed, but higher statistics are needed.

The measured cross-section per rapidity bins allow for the calculation of the photoproduction cross-section, $\sigma_{pp \rightarrow p\psi p}$, as:

\begin{equation}\label{eq:photoprod}
	\sigma_{pp \rightarrow p\psi p} = r(W_+)k_+ \frac{\mathrm{d}n}{\mathrm{d}k_+} \sigma_{\gamma p \rightarrow \psi p} (W_+) \ +  \ r(W_-)k_- \frac{\mathrm{d}n}{\mathrm{d}k_-} \sigma_{\gamma p \rightarrow \psi p} (W_-)\,,
\end{equation}
where $r(W_{\pm}$ is the gap survival factor, $k_{\pm} \equiv M_{\psi}/2e^{\pm |y|}$ is the photon energy, $\frac{\mathrm{d}n}{\mathrm{d}k_{\pm}}$ is the photon flux and $W_{\pm} = 2k_{\pm}\sqrt{s}$ is the photon-proton system invariant mass. The positive (negative) signs in Eq.~\ref{eq:photoprod} refer to the situation where the photon is emitted by the proton travelling parallel (antiparallel) to the LHCb beam axis. In LHCb, $W_+$ and $W_-$ contribute to the same rapidity bin and cannot be disentangled. However, given that only about a third of the data corresponds to $W_-$ and this low-energy contribution has been parametrised for the $J/\psi$ meson by the H1 collaboration~\cite{H1param}, their power-law parametrisation is used to fix it. This power-law is scaled by the ratio of the $\psi(2S)$ and $J/\psi$ cross-sections measured by H1~\cite{H1psi2S}. The estimated photoproduction cross-section is presented in Fig.~\ref{fig:photoprodCEP} and compared to the power-law H1 fit and their results~\cite{H1param,H1psi2S}, as well as to different results from other experiments. Also shown is the JMRT NLO theory description. In the case of the $J/\psi$, where more data is available, it is observed that the LHCb photoproduction cross-section values at $\sqrt{s} = 13 \ \mathrm{TeV}$ deviate from the power-law fit at higher rapidities and are in more agreement with the JMRT NLO description. More data is necessary to discern the behaviour of the $\psi(2S)$ photoproduction cross-section at high rapidities.

\section{\boldmath Intrinsic charm with $Z$ bosons produced in association with charm jets}
While the extrinsic charm ($c$) content of the proton (due to perturbative gluon radiation) has been widely established, several theory predictions suggest that the proton also contains charm intrinsically. This could take place in a sea-quark-like manner or as a valence-like $c$ quark, transforming the proton wave-function into $|uudc\bar{c}\rangle$, as predicted by Light Front QCD (LFQCD). Previous measurements have been performed at low-$Q^2$ \cite{measIC1,measIC2}. At such low energy, the theoretical treatment of hadronic nuclear effects is challenging, and it is difficult to understand the results as evidence or not of the proton intrinsic charm (IC) content. Nevertheless, global PDF analysis do not exclude it at the percent level \cite{percIC1,percIC2}.

\begin{figure}[ht]
	\begin{center}
		\includegraphics[width=0.49\textwidth]{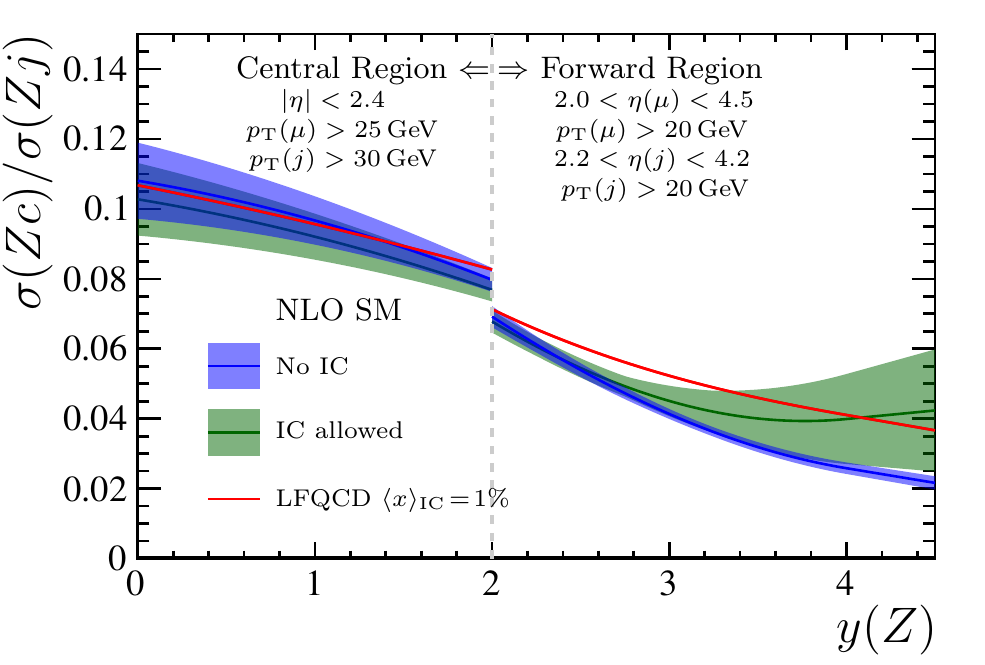}
		\caption{Theory predictions including and excluding intrinsic charm content in the proton for $\sigma(Zc)/\sigma(Zj)$~\cite{proposalIC}. The range $2 < y(Z) < 4.5$ corresponds to the LHCb forward region.}
		\label{fig:theoryIC}
	\end{center}
\end{figure}

A proposal was made to measure the ratio of $Z+c$ jets events to that of $Z+$jets, $\sigma(Zc)/\sigma(Zj)$ in the forward region~\cite{proposalIC}. Performing this measurement at high-$Q^2$ with $Z$ bosons at forward rapidities allows to access high Bjorken-$x$ values with $x>0.1$, where the hadronic and nuclear effects are negligible. Figure~\ref{fig:theoryIC} illustrates how the ratio $\sigma(Zc)/\sigma(Zj)$ at high $Z$ boson rapidities would allow to discriminate the intrinsic charm content of the proton.

\begin{figure}[ht]
	\begin{center}
		\begin{tabular}{c}
			\includegraphics[width=0.49\textwidth]{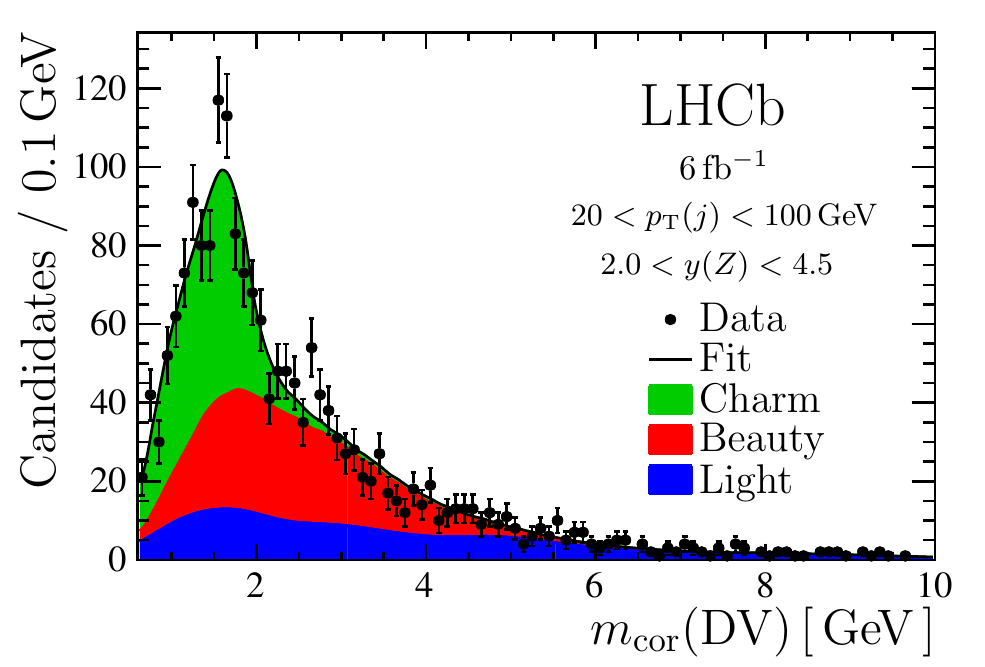}\hskip 0.01\textwidth
			\includegraphics[width=0.49\textwidth]{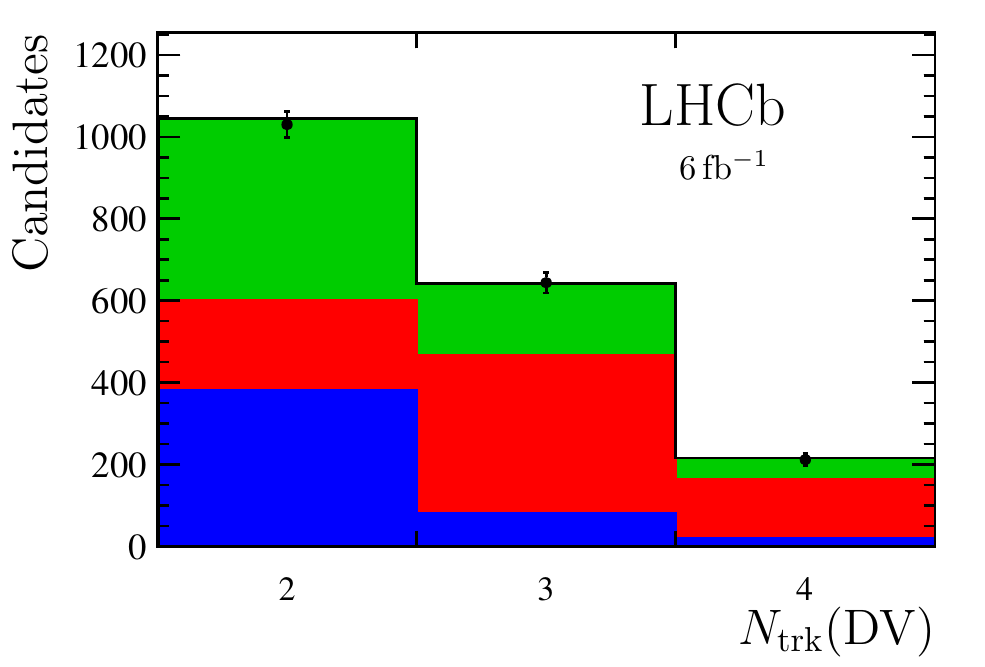}\\
		\end{tabular}
		\caption{Result of the two-dimensional fit for the corrected mass ($m_\mathrm{cor}$) and number of tracks ($N_\mathrm{trk}$) in the dispaced-vertex (DV). The contributions for charm, beauty and light jets are shown.}
		\label{fig:fitIC}
	\end{center}
\end{figure}

An integrated luminosity of 6 fb$^{-1}$ corresponding to the full proton-proton collision LHCb dataset at $\sqrt{s} = 13 \ \mathrm{TeV}$ is used \cite{lhcbIC}. Events with $Z \rightarrow \mu^+ \mu^-$ and at least one jet with transverse momentum  \mbox{$p_\mathrm{T} > 20 \ \mathrm{GeV/}c$} are selected. Charm jets are identified by using a displaced-vertex (DV) tagger in bins of $p_\mathrm{T}(\mathrm{jet}), y(Z)$. A two-dimensional fit to the corrected DV mass and the number of tracks in the DV is performed to identify the flavour of each jet in the selected events. The result of the fit is shown in Fig.~\ref{fig:fitIC}. The efficiency of tagging a jet as a charm jet is estimated in simulation and calibrated in data. The $Z_c$ and $Z_j$ yields in each $y(Z)$ bin are corrected for their selection efficiency and resolution effects at detection.

\begin{figure}[ht]
	\begin{center}
		\includegraphics[width=0.49\textwidth]{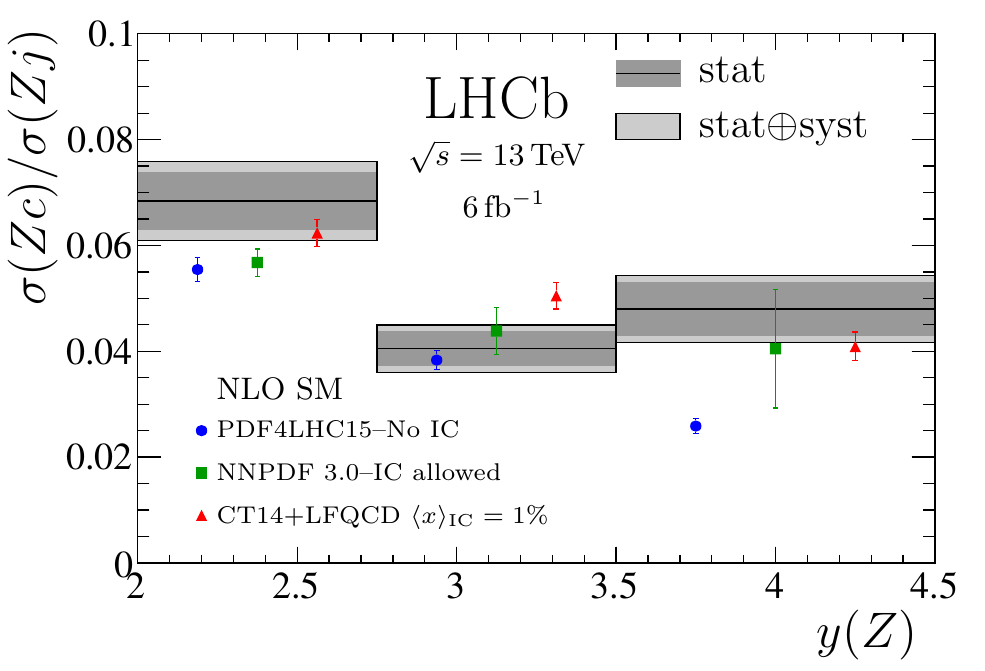}
		\caption{Results for $\sigma(Zc)/\sigma(Zj)$ compared to theory predictions allowing and excluding intrinsic charm content in the proton.}
		\label{fig:resultIC}
	\end{center}
\end{figure}

The measurement of a ratio results in most of the systematic uncertainties cancelling out. The dominant systematic uncertainty is related to the efficiency of identifying charm jets. A comparison of the measured $\sigma(Zc)/\sigma(Zj)$ values to different theory predictions is shown in Fig.~\ref{fig:resultIC}. The first two bins are compatible with both no IC and IC allowed content. The bin at higher $Z$ boson rapidity is consistent with proton IC models, and is about three standard deviations away from the prediction of no intrinsic charm content. The measurement is statistically limited and more data is needed to draw further conclusions. Moreover, these results need to be added to global PDF analyses for interpretation.

\section{Conclusions}
The LHCb detector can be used to perform precision QCD measurements, both in the low- and high-$x$ regions. The central exclusive production cross-section of $J/\psi$ allows to probe the region $x \sim 10^{-6}$. This measurement at $\sqrt{s} = 13 \ \mathrm{TeV}$ is in agreement with the JMRT NLO description, and further data is needed to observe if the same behaviour is present for $\psi(2S)$. The high-$x$ region provides access to large values $x > 0.1$, where the intrinsic charm of the content can be probed. While statistically limited, the first measurement of the proton intrinsic charm content in the forward region in proton-proton collisions has been performed.

\FloatBarrier

\nocite{*}
\bibliographystyle{auto_generated}
\bibliography{proceedings-CSanchezGras/sanchez_cristina/Sanchez}

%% file: proceedings_elba2021_ragoni/proceedings_elba2021_ragoni.tex
\vspace*{1.2cm}

\thispagestyle{empty}
\begin{center}
{\LARGE \bf Recent ALICE results on vector meson photoproduction}

\par\vspace*{7mm}\par

{

\bigskip

\large \bf Simone Ragoni on behalf of the ALICE Collaboration}

\bigskip

{\large \bf  E-Mail: \url{simone.ragoni@cern.ch}}

\bigskip

{The University of Birmingham, B15 2TT, Birmingham, United Kingdom (UK)}

\bigskip

{\it Presented at the Low-$x$ Workshop, Elba Island, Italy, September 27--October 1 2021}

\vspace*{15mm}

\end{center}
\vspace*{1mm}

\begin{abstract}
	
Ultra-peripheral collisions (UPC) are events characterised by large impact parameters between the two projectiles, larger than the sum of their radii. As a consequence, the protons and ions accelerated by the LHC are beyond the reach of the strong interaction and they can be considered as photon sources.

Vector mesons produced in UPC, e.g. \rhozero, \jpsi, and \psip, are of particular interest: vector meson photoproduction in UPC at the LHC is sensitive to the low Bjorken-$x$ gluon parton density.
As the photons involved in the interactions are \textit{quasireal}, the vector mesons should retain the polarisation of the photon, as postulated by the s-channel helicity conservation hypothesis.

ALICE has provided measurements of the production cross section at forward rapidity for \jpsi and at midrapidity for coherent \jpsi, \psip and \rhozero. The collaboration has also measured the $t$-dependence of coherent \jpsi production and compared it with models incorporating nuclear shadowing effects, thus providing a new tool to investigate gluon structure at low Bjorken-$x$. The measurement of photoproduction accompanied by neutron emission allows us to use a new technique to resolve the ambiguity in Bjorken-$x$ which arises in symmetric A--A UPC collisions.
\end{abstract}
  \part[Recent ALICE results on vector meson photoproductions\\ \phantom{x}\hspace{4ex}\it{Simone Ragoni on behalf of the ALICE Collaboration}]{}
 \section{Introduction}
Vector meson photoproduction is being investigated in ultra-peripheral collisions (UPCs) at the LHC. In this type of events, the two interacting objects lie at impact parameters larger than the sum of their radii. A photon from one of them interacts with a colourless object from the target, i.e. a gluon ladder, and vector mesons can be formed in the final state. The ALICE Collaboration has previously measured \rhozero \cite{Adam:2015gsa}, \jpsi \cite{Abelev:2012ba, Abbas:2013oua} and \psip \cite{Adam:2015sia} photoproduction at a centre-of-mass energy of \twosevensixnn in \PbPb, and exclusive \jpsi photoproduction in \pPb at \fivenn \cite{TheALICE:2014dwa, ALICE:2018oyo}. 

LHC Run 2 provided a large dataset of UPC events, which in turn allowed for more differential measurements of vector meson photoproduction processes, at \fivenn in \PbPb, such as the first measurement of the $t$-dependence of coherent \jpsi production. However, symmetric collision systems like \PbPb, have an ambiguity in the sign of the rapidity $y$ of the vector meson in the final state. This will be further explained below. There are two viable techniques to disentangle this ambiguity, and both will be presented.
\section{Exclusive \jpsi in \pPb}
\begin{figure}[ht!]
		\centering
		\includegraphics[width=0.8\textwidth]{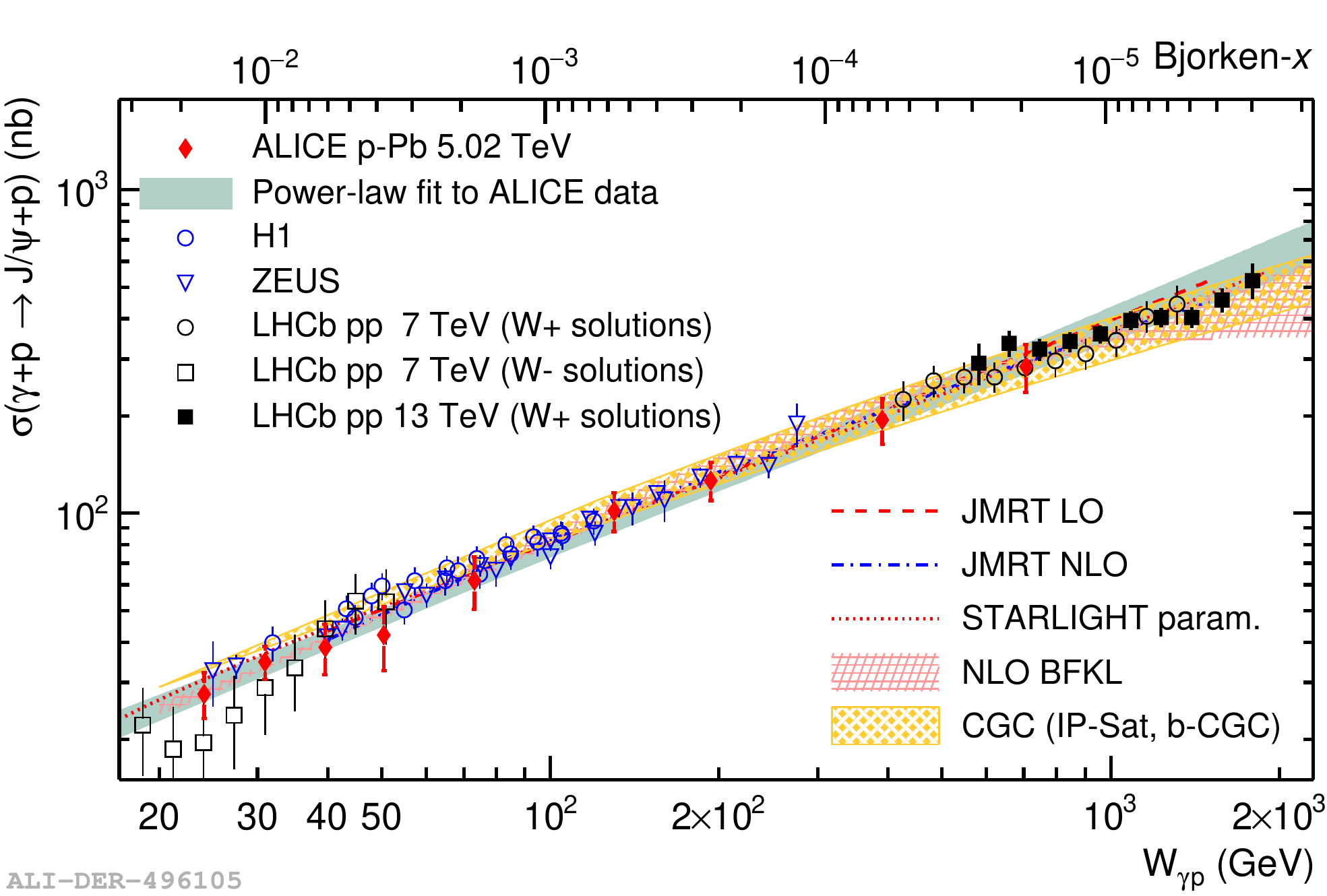}
		\caption{
			Exclusive \jpsi production cross section as a function of the centre-of-mass energy of the $\gamma$p system \cite{Aaij:2018arx, TheALICE:2014dwa, ALICE:2018oyo}. 
		} \label{fig:pPb}
\end{figure}

The UPC dataset collected with \pPb beams is quite valuable as it provides direct access to the proton gluon distributions down to a low Bjorken-$x$ of about $x \sim 10^{-5}$ with  ALICE's current kinematic reach. 
The rapidity $y$ of the vector meson in the final state, may be directly related to the probed Bjorken-$x$ as follows in Eq.~\ref{eq:bjorken-x} \cite{Contreras:2015dqa}:
\begin{equation}
\label{eq:bjorken-x}	
 x = \frac{M_{\rm VM}}{2 E_{\rm p}}\times \exp(-y) \text{ ,}
\end{equation}
where $M_{\rm VM}$ is the mass of the vector meson, and $ E_{\rm p}$ is the energy of the proton beam. 
ALICE results 
are presented in Fig.~\ref{fig:pPb} \cite{Aaij:2018arx, TheALICE:2014dwa, ALICE:2018oyo} together with LHCb's own \pp results \cite{Aaij:2013jxj, Aaij:2014iea, Aaij:2018arx}. ALICE points are obtained with three different configurations:
\begin{itemize}
	\item[\color{red}{$\blacktriangleright$}] \textbf{forward:} two muons\footnote{At pseudorapidities belonging to $-4 < \eta < -2.5$ ALICE can only detect muons.} satisfying the requirement of pseudorapidity $-4 < \eta < -2.5$;
	\item[\color{red}{$\blacktriangleright$}] \textbf{semiforward:} one muon satisfying the requirement of pseudorapidity $-4 < \eta < -2.5$, with the other in $-0.9 < \eta < 0.9$;	
	\item[\color{red}{$\blacktriangleright$}] \textbf{central:}  both muons lie in $-0.9 < \eta < 0.9$.
\end{itemize}
As shown in Fig.~\ref{fig:pPb}, the three configurations provide almost continuous coverage from about 25 \GeVns up to about 700 \GeVns in centre-of-mass energy of the photon--proton system, ${\rm W}_{\gamma p}$, which is determined starting from the vector meson rapidity as ${\rm W}_{\gamma p}^2 = 2E_{\rm p} M_{\jpsi}\times e^{-y}$.
LHCb \pp data extending to almost 2 \TeVns\ are also shown. The power-law growth of the cross sections can then be related to a power-law growth of gluon distributions down to $x\sim 10^{-6}$. The lack of a clear deviation from the power-law trend indicates a lack of clear signs of gluon saturation. Gluon saturation is considered to be the most straightforward mechanism by which the growth of the cross sections could be tamed. However, neither ALICE nor LHCb have observed compelling evidence for such an effect so far. Gluon saturation at low-$x$ would also have implications for the early stages of ultra-relativistic heavy-ion collisions, thus becoming a key investigation topic for current experiments.
\section{Coherent \jpsi in \PbPb}
\begin{figure}[ht!]
	\centering
	\includegraphics[width=0.6\textwidth]{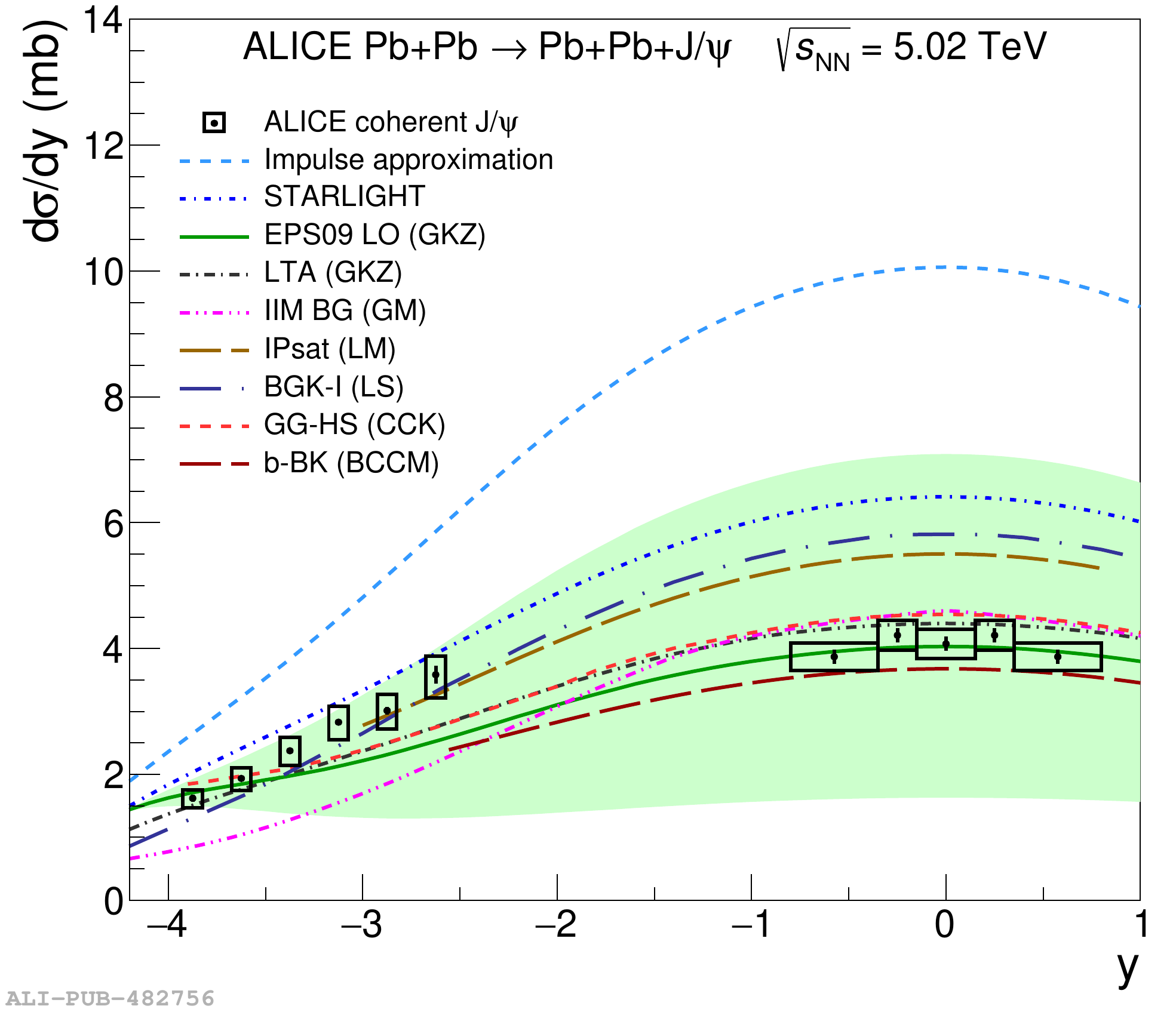}
	\caption{
		Coherent \jpsi photoproduction cross sections as a function of rapidity \cite{Acharya:2019vlb, Acharya:2021ugn}. 
	} \label{fig:coherent-xsec}
\end{figure}

The ALICE Collaboration has measured the production cross section of coherent \jpsi at forward and midrapidity. The results are shown in Fig.~\ref{fig:coherent-xsec}  \cite{Acharya:2019vlb, Acharya:2021ugn}. ALICE results can then be directly compared to e.g. the Impulse Approximation, which is a model without nuclear effects, apart from coherence.
This is particularly useful to measure the nuclear suppression factor $S_{{\rm Pb}}(x)$, which is defined as follows in Eq.~\ref{eq:nuclear-suppression-factor} \cite{Guzey:2020ntc}:
\begin{equation}
	\label{eq:nuclear-suppression-factor}
	S_{{\rm Pb}}(x) = \sqrt{\frac{\sigma(\gamma A \longrightarrow \jpsi A)_{\rm measured}}{\sigma(\gamma A \longrightarrow \jpsi A)_{\rm IA}}} \text{ ,}
\end{equation}
where $\sigma(\gamma A \longrightarrow \jpsi A)_{\rm measured}$ is the measured cross section for coherent \jpsi photoproduction, while $\sigma(\gamma A \longrightarrow \jpsi A)_{\rm IA}$ is the cross section computed with the Impulse Approximation model. 
The Impulse Approximation model is derived starting from photoproduction data on protons without including nuclear effects except for coherence. The nuclear suppression factor thus provides a way to test the agreement of ALICE data with the existing datasets. ALICE data are consistent with a $S_{{\rm Pb}}(x) = 0.65 \pm 0.03$ \cite{Acharya:2021ugn} at midrapidity, signalling strong nuclear effects which are unaccounted for in the available nuclear parton distribution function (PDF) sets. \starlight \cite{Klein:2016yzr}, a Glauber-like model which considers the interaction as a single dipole moving through the nucleus, also overpredicts the data. This would imply that a Glauber-like description only would not have been enough to describe the suppression of the coherent \jpsi. Guzey, Kryshen and Zhalov (GKZ) \cite{Guzey:2016piu} provide two models, one based on EPS09 nPDF parametrisation and the other on leading twist approximation (LTA). Both models describe the data. This implies that the \jpsi data agrees with existing measurements of nuclear shadowing.

\begin{figure}[ht!]
	\begin{center}
		\begin{subfigure}{.5\textwidth}
  \centering
			\includegraphics[width=0.95\textwidth]{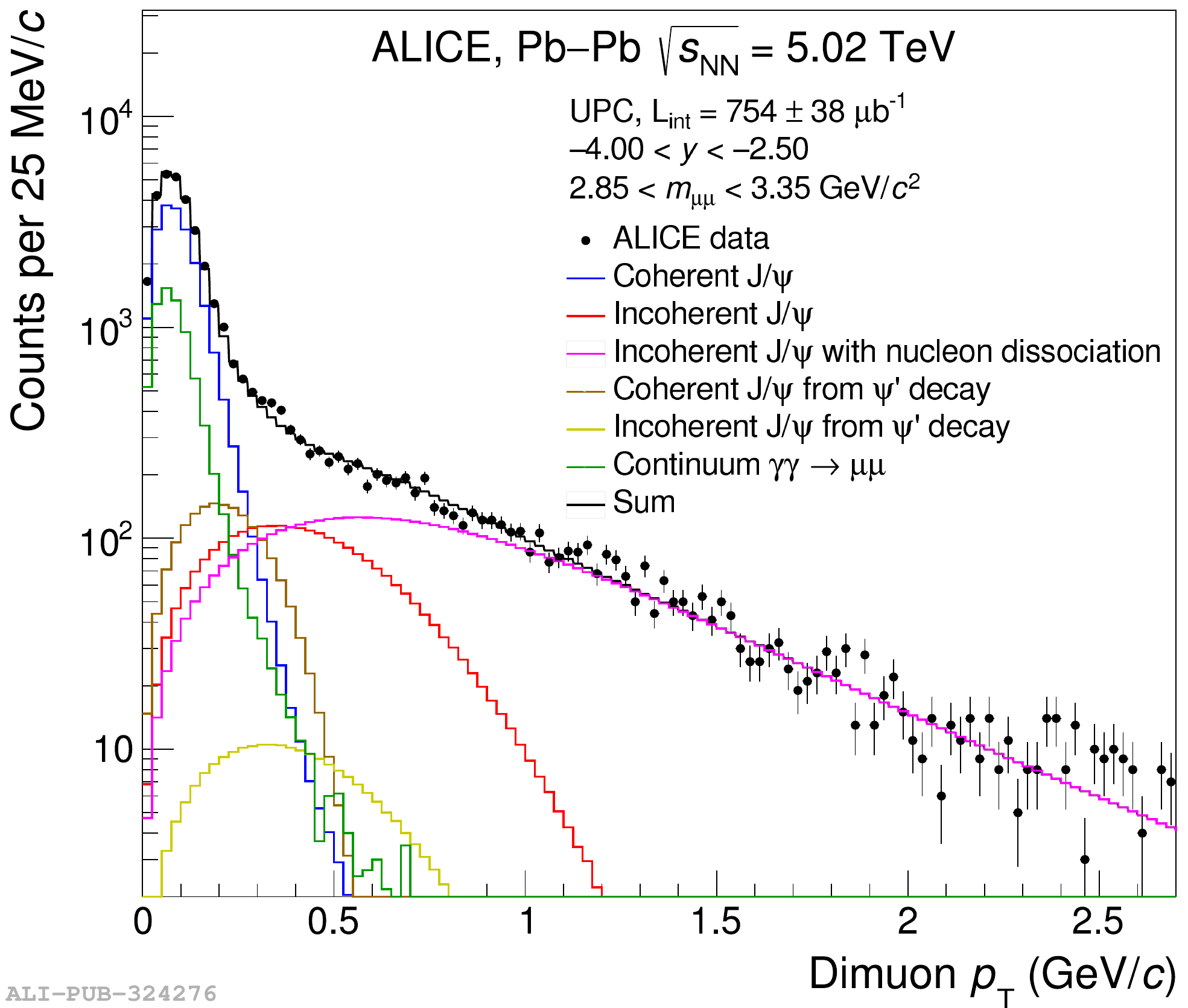}
			\caption{}
			\label{fig:pt-forward-midrapidity-1}
		\end{subfigure}%
		\begin{subfigure}{.5\textwidth}
  \centering
			\includegraphics[width=0.985\textwidth]{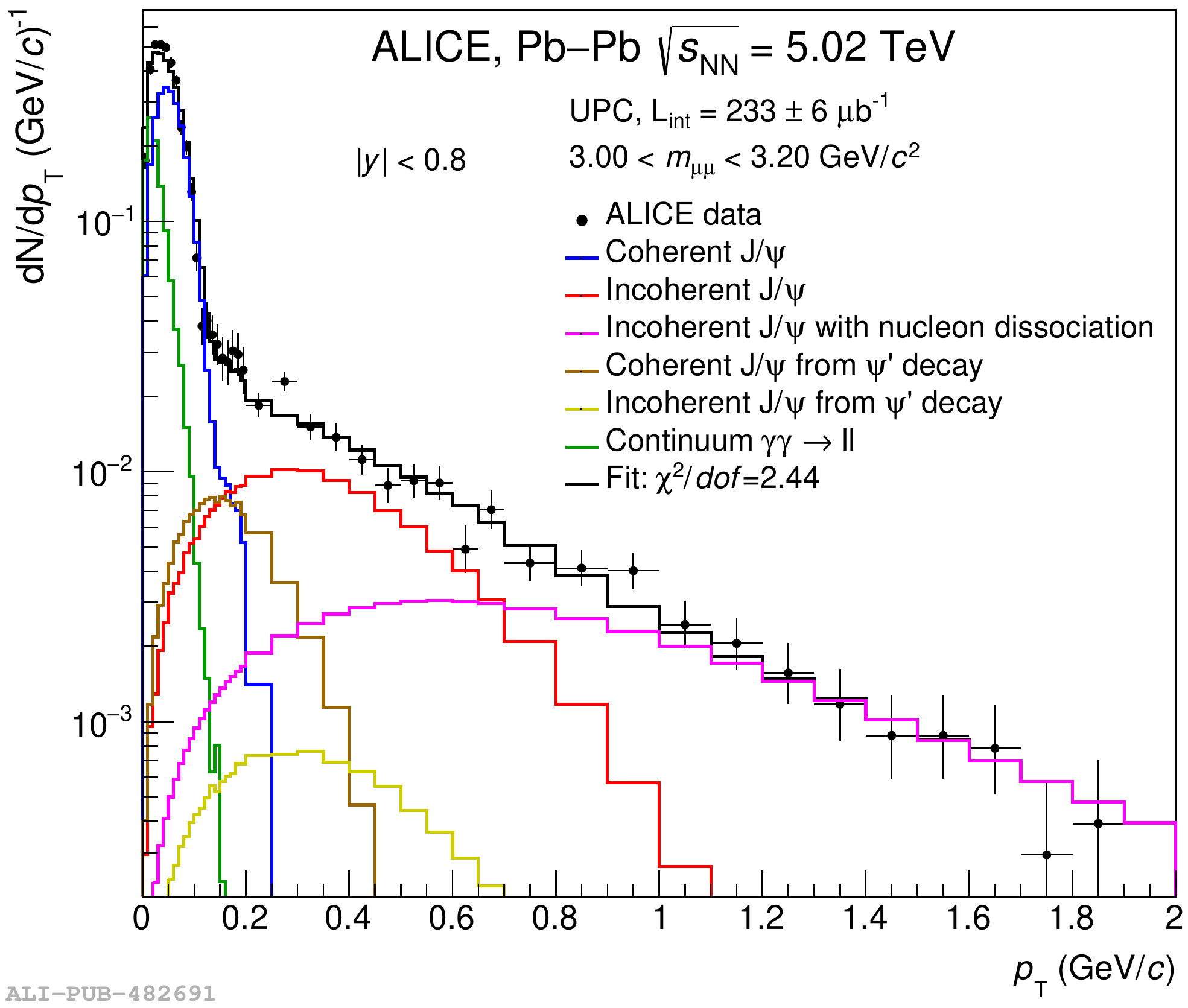}
			\caption{}
			\label{fig:pt-forward-midrapidity-2}
		\end{subfigure}\\
	\end{center}
	\caption{Dimuon \pt distribution for candidates with masses in the \jpsi mass peak region, i.e. $2.85 < M_{\mu\mu} < 3.35$ \GeVmass for Fig.~\ref{fig:pt-forward-midrapidity-1} at forward rapidity \cite{Acharya:2019vlb}, and $3.00 < M_{\mu\mu} < 3.20$ \GeVmass for Fig.~\ref{fig:pt-forward-midrapidity-2} at midrapidity \cite{Acharya:2021ugn}. 
	}
	\label{fig:pt-forward-midrapidity}
\end{figure}

The \pt distributions for those dimuons lying in the \jpsi mass peak region are shown in Fig.~\ref{fig:pt-forward-midrapidity-1}~\cite{Acharya:2019vlb} and in Fig.~\ref{fig:pt-forward-midrapidity-2}~\cite{Acharya:2021ugn}, i.e. for dimuon masses $2.85 < M_{\mu\mu} < 3.35$ \GeVmass and $3.00 < M_{\mu\mu} < 3.20$ \GeVmass at forward and midrapidity, respectively. Coherent \jpsi is characterised by a  lower average \pt compared to incoherent \jpsi, owing to the different size of the corresponding photon targets, i.e. the photon couples with the entire nucleus \textit{coherently} for coherent \jpsi, and with a single nucleon for incoherent \jpsi. It is also shown how incoherent \jpsi with dissociation has a lower impact at midrapidity compared to forward rapidity. 

\begin{figure}[ht!]
	\centering
	\includegraphics[width=0.6\textwidth]{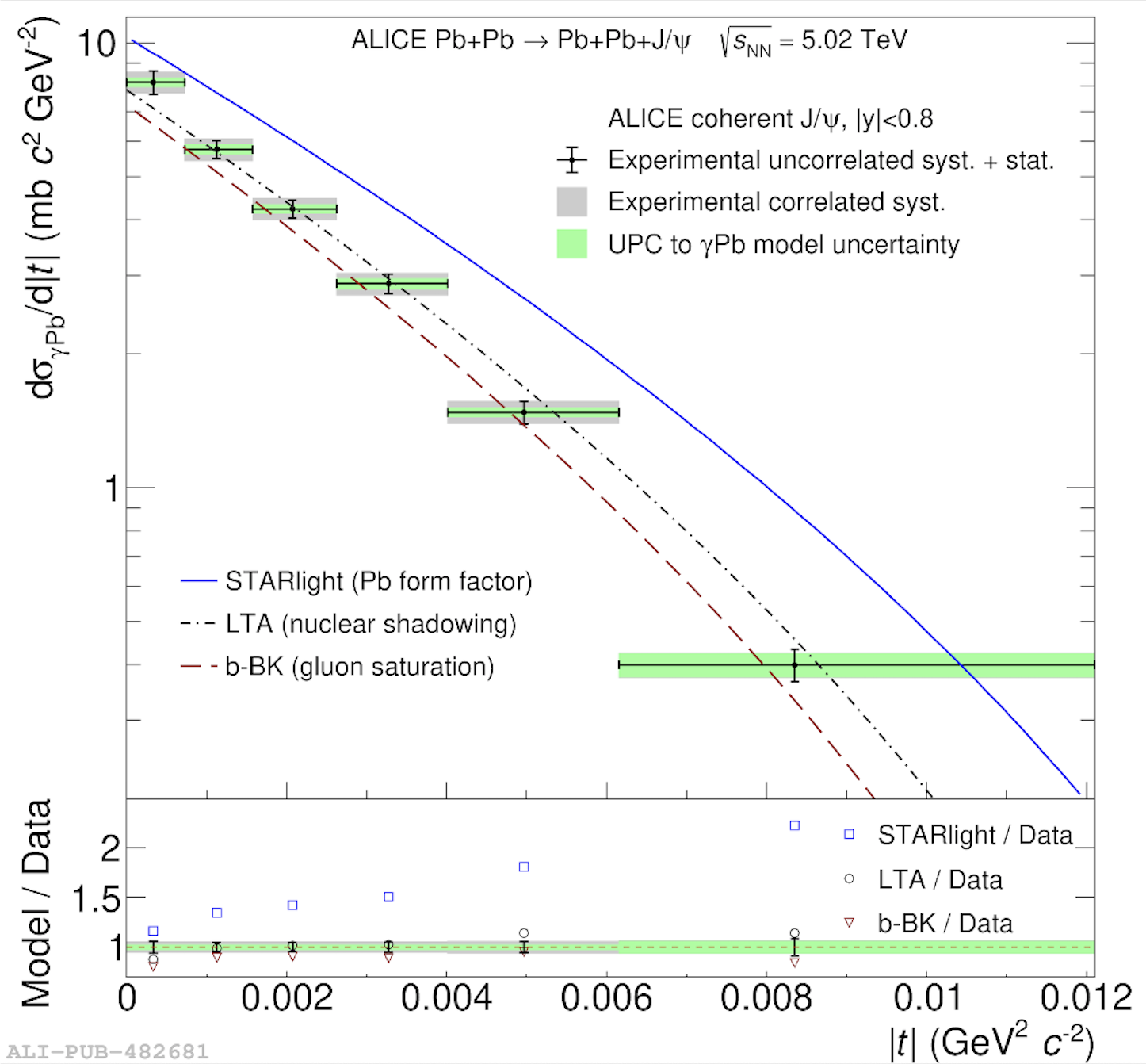}
	\caption{
		$|t|$-dependence of the photonuclear cross section for coherent \jpsi at midrapidity.
	} \label{fig:t-dependence}
\end{figure}

The ALICE Collaboration has also measured for the first time the $|t|$-dependence of coherent \jpsi in \PbPb \cite{ALICE:2021tyx}. This is shown in Fig.~\ref{fig:t-dependence}. Since the measured $|t|$-dependence shows a trend compatible with models incorporating QCD effects,  it then constitutes a valuable new observable to probe the transverse gluonic structure of the nucleus at low Bjorken-$x$. 
\section{Viable techniques to solve the Bjorken-$x$ ambiguity}
Ultra-peripheral collisions with symmetric systems such as \PbPb or \pp, have an ambiguity as either of the projectiles might have emitted the photon. The rapidity $y$ of the vector meson in the final state will then be related to the Bjorken-$x$ of the process by means of Eq.~\ref{eq:rapidity-x}:
\begin{equation}
	\label{eq:rapidity-x}
	x = \frac{M_{\rm VM}}{\sqrt{s_{\rm NN}}} \times e^{\pm |y|} \text{ ,}
\end{equation}
where $M_{\rm VM}$ is the mass of the vector meson and $\sqrt{s_{\rm NN}}$ is the centre-of-mass energy of the collision system. With ALICE kinematics this would imply that, at forward rapidity, $x\sim 10^{-2} $ or $10^{-5}$.
 
 Two techniques have currently been proposed to disentangle the ambiguity. They both leverage on the impact parameter:
 \begin{itemize}
 	\item[\color{red}{$\blacktriangleright$}] \textit{neutron emission:} generators such as n$^0_0$n  \cite{Broz:2019kpl} predict that different neutron emission classes are characterised by quite different impact parameters \cite{Baltz:2002pp}. If neutron emission is considered only in four cases, 0N0N no neutron emitted, 0NXN neutrons emitted only on one side with respect to the interaction vertex, XNXN neutrons emitted on both sides of the interaction vertex, 	n$^0_0$n predicts that there will be a hierarchy of average impact parameters, as shown in Eq.~\ref{eq:hierarchy-neutron}: 
	\begin{equation}
		\label{eq:hierarchy-neutron}
	 	\langle{b_{\rm XNXN}}\rangle < \langle{b_{\rm 0NXN}}\rangle < \langle{b_{\rm 0N0N}}\rangle \text{ ,}
	\end{equation}
    where $\langle{b_{\rm 0N0N}}\rangle $, $\langle{b_{\rm 0NXN}}\rangle $, and $\langle{b_{\rm XNXN}}\rangle $ are the average impact parameters for the 0N0N, 0NXN, and XNXN classes, respectively;
 	\item[\color{red}{$\blacktriangleright$}] \textit{peripheral photoproduction:} in peripheral collisions the impact parameter is less than the sum of the nuclear radii, as opposed to UPCs, where the impact parameter is larger instead. 
 \end{itemize}
Using either neutron emission classes \cite{Guzey:2013jaa} or peripheral photoproduction \cite{Contreras:2016pkc} would enable the separation of the low- and high-$x$ photonuclear cross sections, as shown in  Eq.~\ref{eq:neutron-emission-splitting}:
\begin{equation}
	\label{eq:neutron-emission-splitting}
	\begin{split}
		\frac{{\rm d}\sigma^{0{\rm N}0{\rm N}}_{{\rm Pb}{\rm Pb}}}{{\rm d}y} &= n_{0{\rm N}0{\rm N}}(\gamma, +y) \cdot {\color{red}\sigma_{\gamma {\rm Pb}}(+y)} +  n_{0{\rm N}0{\rm N}}(\gamma, -y) \cdot {\color{red}\sigma_{\gamma {\rm Pb}}(-y)}  \text{ ,}\\
		\frac{{\rm d}\sigma^{0{\rm NX}{\rm N}}_{{\rm Pb}{\rm Pb}}}{{\rm d}y}  &= n_{0{\rm NX}{\rm N}}(\gamma, +y) \cdot {\color{red}\sigma_{\gamma {\rm Pb}}(+y)} +  n_{0{\rm NX}{\rm N}}(\gamma, -y) \cdot {\color{red}\sigma_{\gamma {\rm Pb}}(-y)}  \text{ ,}\\
		\frac{{\rm d}\sigma^{{\rm XNX}{\rm N}}_{{\rm Pb}{\rm Pb}}}{{\rm d}y}  &= n_{{\rm XNX}{\rm N}}(\gamma, +y) \cdot {\color{red}\sigma_{\gamma {\rm Pb}}(+y)} +  n_{{\rm N}{\rm XNX}}(\gamma, -y) \cdot {\color{red}\sigma_{\gamma {\rm Pb}}(-y)} \text{ ,}
	\end{split}
\end{equation}
for neutron emission, where ${\rm d}\sigma^{0{\rm N}0{\rm N}}_{{\rm Pb}{\rm Pb}}/{\rm d}y$,  ${\rm d}\sigma^{0{\rm NX}{\rm N}}_{{\rm Pb}{\rm Pb}}/{\rm d}y$ and  ${\rm d}\sigma^{{\rm XNX}{\rm N}}_{{\rm Pb}{\rm Pb}}/{\rm d}y$ are the measured UPC cross sections for 0N0N, 0NXN, and XNXN, respectively, $n_{0{\rm N}0{\rm N}}$, $n_{0{\rm NX}{\rm N}}$, and $n_{{\rm XNX}{\rm N}}$ are the corresponding photon fluxes, at either positive or negative rapidities, and finally, $\sigma_{\gamma {\rm Pb}}(\pm y)$ are the photonuclear cross sections, at the two rapidities. It is also possible to simultaneously use peripheral and UPC results as shown in Eq.~\ref{eq:peripheral-photoproduction}:
\begin{equation}
	\label{eq:peripheral-photoproduction}
	\begin{split}
		\frac{{\rm d}\sigma_{{\rm Pb}{\rm Pb}}^{\rm P}}{{\rm d}y} &= n_{\rm P}(\gamma, +y)\cdot {\color{red} \sigma_{\gamma {\rm Pb}}(+y)} + n_{\rm P}(\gamma, -y)\cdot {\color{red}\sigma_{\gamma {\rm Pb}}(-y)} \text{ ,} \\
		\frac{{\rm d}\sigma_{{\rm Pb}{\rm Pb}}^{\rm U}}{{\rm d}y} &= n_{\rm U}(\gamma, +y)\cdot {\color{red}\sigma_{\gamma {\rm Pb}}(+y)} + n_{\rm U}(\gamma, -y)\cdot {\color{red}\sigma_{\gamma {\rm Pb}}(-y)}\text{ ,}
	\end{split}
\end{equation}
where ${\rm d}\sigma^{{\rm P}}_{{\rm Pb}{\rm Pb}}/{\rm d}y$ and ${\rm d}\sigma^{{\rm U}}_{{\rm Pb}{\rm Pb}}/{\rm d}y$ are the measured peripheral and UPC cross sections, respectively, while $n_{\rm P}$ and $n_{\rm U}$ are the corresponding fluxes, computed for the appropriate impact parameters.

Neutron emission was recently applied by the ALICE Collaboration for the measurement of coherent \rhozero photoproduction in \PbPb \cite{Acharya:2020sbc}, and overall agreement was found between the ALICE data points and the available models. This indicates a satisfactory knowledge of the photon fluxes.

\begin{figure}[ht!]
	\begin{center}
		\begin{subfigure}{.5\textwidth}
  \centering
			\includegraphics[width=0.95\textwidth]{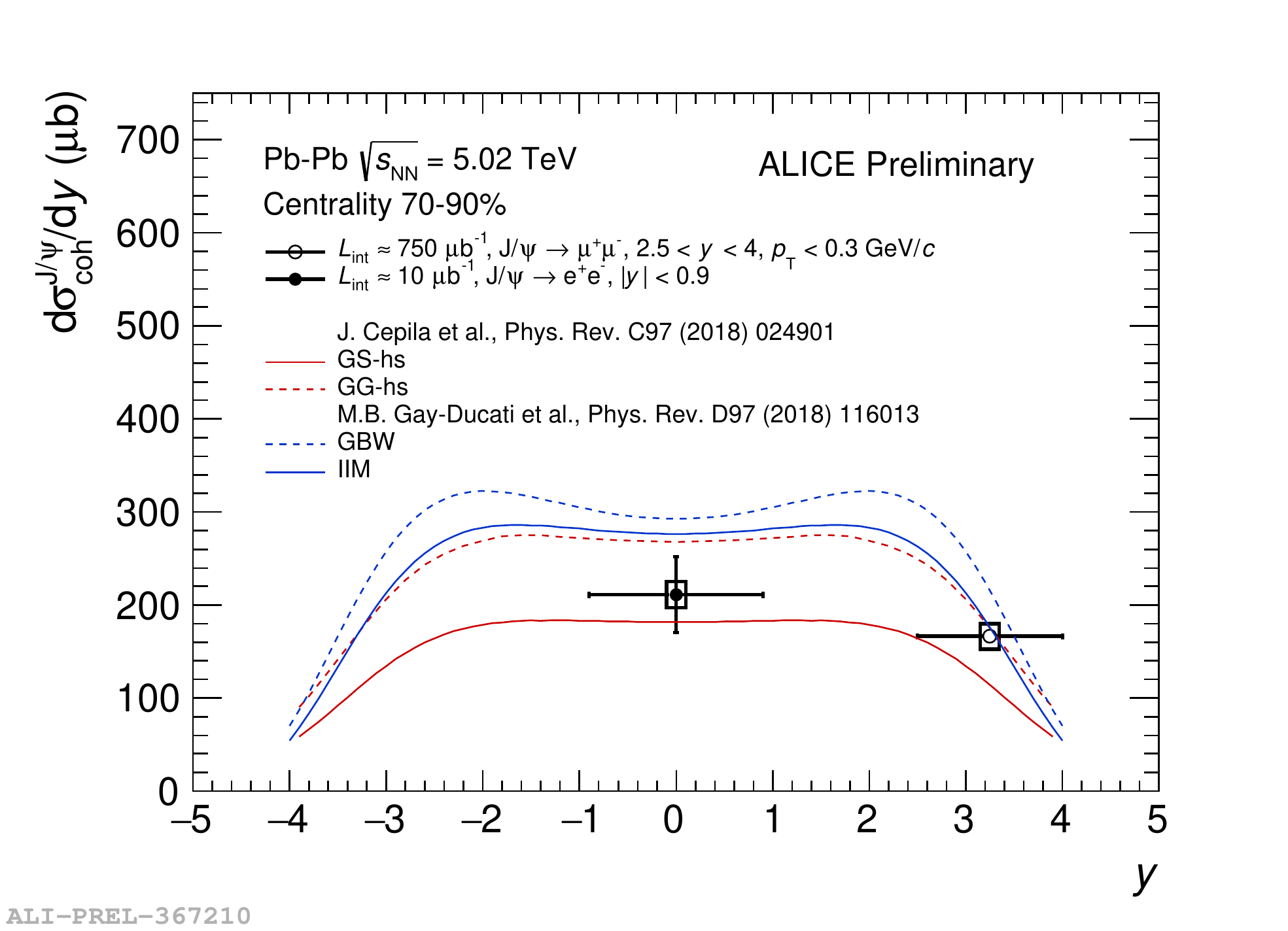}
			\caption{}
			\label{fig:peripheral-run2-7090}
		\end{subfigure}%
		\begin{subfigure}{.5\textwidth}
  \centering
			\includegraphics[width=0.95\textwidth]{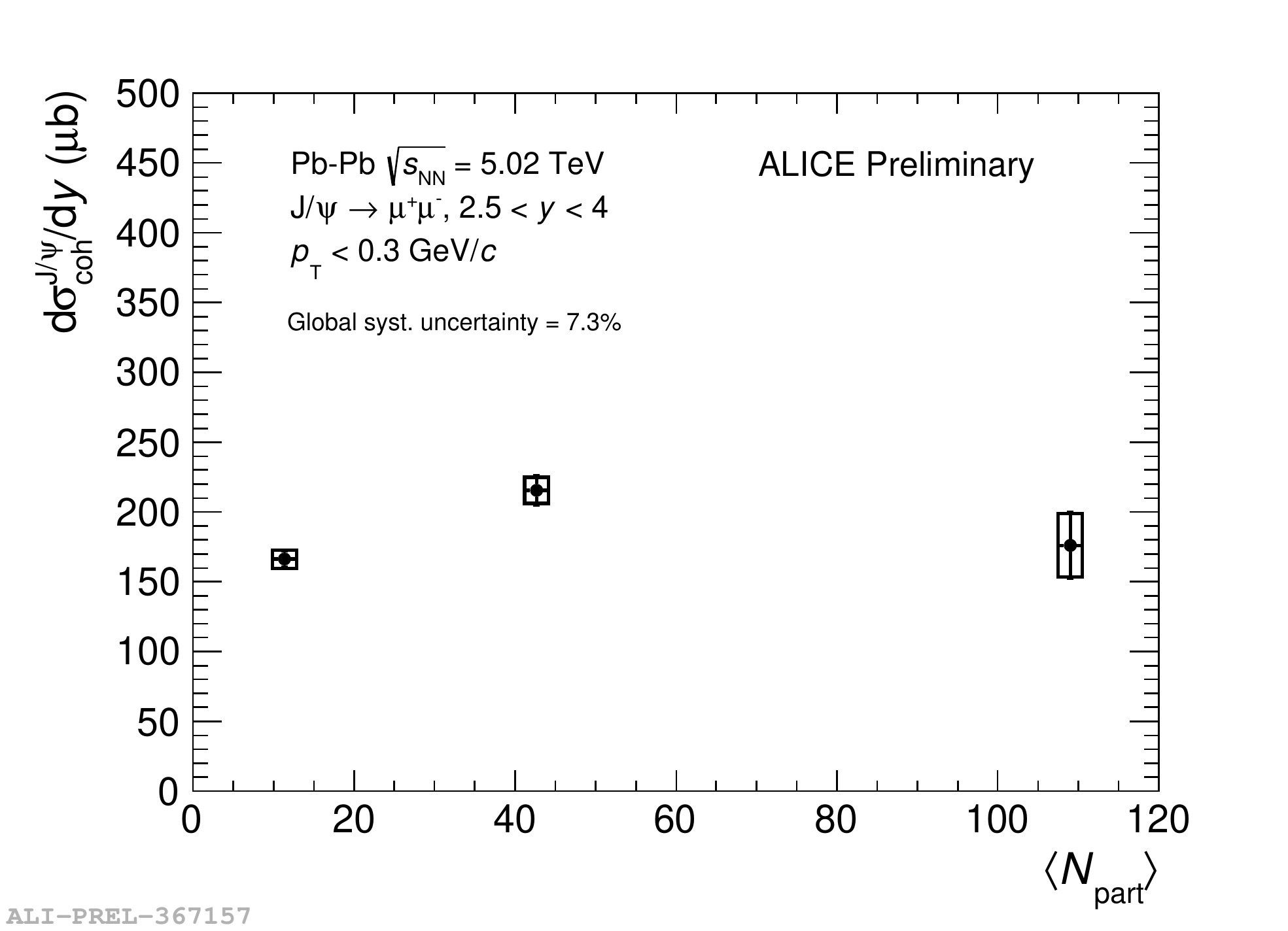}
			\caption{}
			\label{fig:peripheral-run2-centrality}
		\end{subfigure}\\
	\end{center}
	\caption{Coherent \jpsi photoproduction with the ALICE Run 2 peripheral sample. Cross sections for the most peripheral centrality class in Fig.~\ref{fig:peripheral-run2-7090}, and cross sections as a function of centrality in  Fig.~\ref{fig:peripheral-run2-centrality}  \cite{Bugnon:2020vti}.}
	\label{fig:peripheral-run2}
\end{figure}

Peripheral photoproduction of coherent \jpsi was first observed in ALICE Run 1 data \cite{Adam:2015gba}. The STAR  \cite{STAR:2019yox} and LHCb \cite{LHCb:2021hoq} collaborations also report the observation. ALICE has also measured coherent \jpsi photoproduction in the peripheral sample with Run 2 data, at forward rapidity \cite{Bugnon:2020vti}.  Fig.~\ref{fig:peripheral-run2-7090} and Fig.~\ref{fig:peripheral-run2-centrality} show the cross section for the most peripheral centrality class, and the cross sections as a function of centrality at forward rapidity, respectively\footnote{The more central a collision, the higher the number of participants.}. 

\section{Summary}
Exclusive \jpsi photoproduction in \pPb constitutes a valuable tool to probe for gluon saturation in protons. Coherent \jpsi photoproduction in \PbPb UPCs is particularly useful to measure nuclear shadowing in Pb nuclei. Finally, the ambiguity on the sign of the rapidity of the vector meson in the final state, with respect to the photon emitter -- which occurs in symmetric collision systems -- can be solved by analysing data samples characterised by different average impact parameters. Two techniques are presented, using neutron emission and peripheral photoproduction.
%
%
%

%


\nocite{*}
\bibliographystyle{auto_generated}
\bibliography{proceedings_elba2021_ragoni/proceedings_elba2021_ragoni}


%% file: Zhang_ATLAS/Zhang_ATLAS.tex
\vspace*{1.2cm}

\thispagestyle{empty}
\begin{center}
{\LARGE \bf Determination of Proton Parton Distribution}\\\vspace{2mm}
{\LARGE \bf  Functions using ATLAS Data}

\par\vspace*{7mm}\par

{

\bigskip

\large \bf Zhiqing Zhang, on behalf of the ATLAS Collaboration\footnote{Copyright 2021 CERN for the benefit of the ATLAS Collaboration. Reproduction of this article or parts of it is allowed as specified in the CC-BY-4.0 license.}}

\bigskip

{\large \bf  E-Mail: Zhiqing.Zhang@ijclab.in2p3.fr}

\bigskip

{Laboratoire de Physique des 2 infinis Ir\`ene Joliot-Curie $-$ IJCLab, Universit\'e Paris-Saclay, CNRS/IN2P3, 91405 Orsay, France}

\bigskip

{\it Presented at the Low-$x$ Workshop, Elba Island, Italy, September 27--October 1 2021}

\vspace*{15mm}

\end{center}
\vspace*{1mm}

\begin{abstract}

Using differential cross-section data of inclusive $W$ and $Z/\gamma^\ast$ boson production and of their production in association with jets, and differential cross-section data of top-quark pair production in the lepton + jets and dilepton channels in proton--proton collisions at the CERN Large Hadron Collider at center-of-mass energies of $\sqrt{s}=7$ and 8~TeV, together with deep inelastic scattering data from electron--proton collisions at the HERA collider, several determinations of parton distribution functions of the proton have been performed by the ATLAS experiment. The results are presented here. The emphasis is given to the determination of the strange sea quark distribution and its evolution over time.
\end{abstract}
  \part[Determination of Proton Parton Distribution Functions using ATLAS Data\\ \phantom{x}\hspace{4ex}\it{Zhiqing Zhang on behalf of the ATLAS Collaboration}]{}

 \section{Introduction}
Parton distribution functions (PDFs) of the proton cannot be predicted from first principle and they used to be determined in deep inelastic scattering (DIS) experiments using point-like beams of charged or neutral leptons to probe the proton or other nucleon targets, in earlier fixed target mode and later collision mode at HERA. Thanks to the factorization theorem, the PDFs determined from one process can be used as prediction for other processes. 

At hadron colliders such as the Large Hadron Collider (LHC), the uncertainty of the PDFs is now the dominant uncertainty source for precision measurements, one example is the recent determination of the $W$ boson mass~\cite{1701.07240} by ATLAS~\cite{detector}. It also becomes a limiting factor for searches beyond the Standard Model. It is therefore important to improve our knowledge on PDFs by using relevant measurements at the LHC.

Since 2012, several analyses at next-to-next-to-leading order (NNLO) accuracy in quantum chromodynamics (QCD) have been performed by the ATLAS experiment using Drell-Yan or top-quark pair cross-section data as outlined in Table~\ref{tab:fits}. In the following sections, these QCD analyses and their corresponding results will be briefly described and shown.
\begin{table}
\centering
\caption{Overview of ATLAS QCD analyses and the used data sets including the neutral current (NC) and charged current (CC) data from HERA-I and HERA-I+II, and various ATLAS measurements of the Drell-Yan processes $W\to\ell\nu$ and $Z(\gamma^\ast)\to\ell\ell$ and of top-quark pair production at different center-of-mass energies.}
        \label{tab:fits}
        \resizebox{\columnwidth}{!}
        { 
          \begin{tabular}{c|cccc}
          \hline
          Data sets & epWZ12~\cite{1203.4051} & epWZ16~\cite{1612.03016} & epWZtop18~\cite{ATL-PHYS-PUB-2018-017} & epWZVjet20~\cite{2101.05095} \\\hline
          HERA-I NC, CC~\cite{0911.0884} & \checkmark &  &  &  \\
          HERA-I+II NC, CC~\cite{1506.06042} &  & \checkmark & \checkmark & \checkmark \\\hline
          ATLAS $W$, $Z$, 7 TeV 35 pb$^{-1}$~\cite{1109.5141}  & \checkmark & & & \\
          ATLAS $W$, $Z/\gamma^\ast$, 7 TeV 4.2 fb$^{-1}$~\cite{1612.03016} & & \checkmark & \checkmark & \checkmark \\\hline
          ATLAS $t\bar{t}$ $(\ell+$ jets, dilepton$)$ 8 TeV~\cite{1511.04716,1607.07281} & & & \checkmark & \\\hline
          ATLAS $W/Z+$ jets 8 TeV~\cite{1711.03296,1907.06728} & & & & \checkmark \\\hline
          \end{tabular}
        }
\end{table}
 
\section{Analysis ATLAS-epWZ12}

In 2012, a NNLO QCD analysis, ATLAS-epWZ12~\cite{1203.4051}, was performed by the ATLAS experiment to assess the strange quark distribution, using the first differential cross-section measurements of inclusive $W^\pm$ and $Z$ boson production based on a $pp$ collision data sample corresponding to an integrated luminosity of 35 pb$^{-1}$ recorded in 2010 at a center-of-mass energy of $\sqrt{s}=7$~TeV~\cite{1109.5141}, together with the combined $e^\pm p$ neutral current (NC) and charged current (CC) cross-section measurements of H1 and ZEUS from HERA-I~\cite{0911.0884}. 

The HERA measurements were included since they are the primary source for constraining PDFs with their large kinematic coverage for $Q^2$, the absolute four-momentum transfer squared, from near 1~GeV$^2$ to above $10^4$~GeV$^2$ and of Bjorken-$x$ from $\sim 0.6$ down to $10^{-4}$. However, they do have limitations. They cannot distinguish quark flavor between the down-type sea quarks, $\bar{d}$ and $\bar{s}$. Also the measurement precision at high $x$ and large $Q^2$ are statistically limited.

The ATLAS data, measured as functions of the $W$ decay lepton pseudorapidity, $\eta_\ell$, and of the $Z$ boson rapidity, $y_Z$, with a typical precision of $(2-3)\%$, access a kinematic range prescribed by the boson masses, $M_{W, Z}$, and the proton beam energy of $E_p=3.5$~TeV, corresponding to $Q^2\simeq M^2_{W,Z}$ and an $x$ range $0.001\lesssim x\lesssim 0.1$, with a mean $x=M_Z/2E_p=0.013$ for the $Z$ boson. The data provide new constraints on the strange quark distribution at high scale, $Q^2\sim M^2_{W,Z}$, which imply constraints at low $Q^2$ through perturbative  QCD evolution.

The QCD analysis used the HERAFitter framework~\cite{HERAFitter}. The light quark coefficient functions were calculated to NNLO as implemented in QCDNUM~\cite{1005.1481}. The contributions of heavy quarks are calculated in the general-mass variable-flavor-number scheme of Refs.~\cite{9709442,0601245}. The electroweak (EW) parameters and corrections relevant for the $W$ and $Z$ boson production processes were determined following the procedure described in Ref.~\cite{1109.5141} and cross-checked between the FEWZ~\cite{0312266} and the DYNNLO~\cite{0903.2120} programs. The fit package used APPLGRID code~\cite{0911.2985} interfaced to the MCFM program~\cite{1007.3492} for fast calculation of the differential $W$ and $Z$ boson cross sections at next-to-leading order (NLO) and a $K$-factor technique to correct from NLO to NNLO predictions. The data were compared to the theory using the $\chi^2$ function defined in Ref.~\cite{0904.0929}. 

The quark and gluon distributions at an initial scale of $Q^2_0=1.9$~GeV$^2$, chosen such that it is below the charm mass threshold $m_c$, were parameterized by generic forms:
\begin{eqnarray}
&& xq_i(x)=A_ix^{B_i}(1-x)^{C_i}P_i(x)\,,\nonumber\\
&& xg(x)=A_gx^{B_g}(1-x)^{C_g}P_g(x)-A^\prime_gx^{B^\prime_g}(1-x)^{C^\prime_g}\,,\nonumber
\end{eqnarray}
where $P_{i, g}$ denotes polynomials in powers of $x$. The fitted quark distributions were the valence quark distributions $(xu_v, xd_v)$ and the light quark sea distributions $(x\bar{u}, x\bar{d}, x\bar{s})$. 
The parameters $A_{u_v}$ and $A_{d_v}$ were fixed using the quark counting rule, $A_g$ using the momentum sum rule, and $C^\prime_g$ was set to 25 to suppress negative contributions at high $x$.  The normalization and slope parameters, $A$ and $B$, of $\bar{u}$ and $\bar{d}$ were set equal such that $x\bar{u} = x\bar{d}$ at $x\to 0$. For the strange quark distribution results presented here, it was parametrized with $P_{\bar{s}}=1$ and $B_{\bar{s}}=B_{\bar{d}}$. Terms were
added in the polynomial expansion $P_i(x)$ only if required by the data with a significant decrease in $\chi^2$ of the fit results, following the procedure described in Ref.~\cite{0911.0884}. This led to one additional term, $P_{u_v}=1+E_{u_v}x^2$, for the $u$ valence quark distribution, giving in total 15 free parameters for the fit.
The parton distributions at other $Q^2$ values are obtained using the evolution equations. 

\begin{figure}
\begin{center}
\includegraphics[width=0.7\columnwidth]{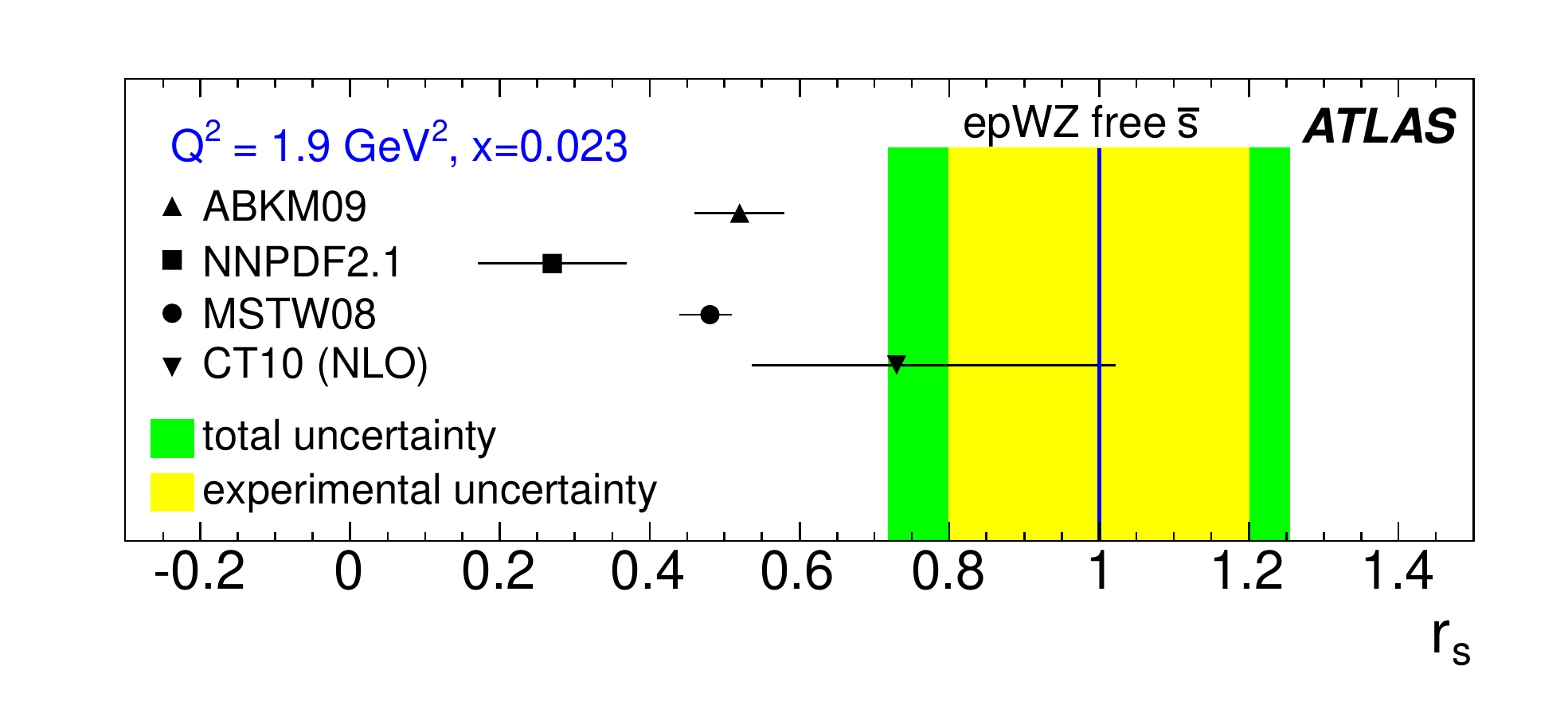}
\caption{Relative strange-to-down sea quark ratio for $Q^2_0=1.9$~GeV$^2$ and $x=0.023$ comparing the determination from ATLAS (shown in the vertical line with bands for experimental and total uncertainties) with predictions from different PDF sets (shown in closed symbols with horizontal error bars). The plot is taken from Ref.~\cite{1203.4051}.}
\label{fig:epwz12}
\end{center}
\end{figure}

The fit resulted in a good overall $\chi^2$ value per number of degree of freedom of 538.4/565 and determined the ratio of the strange quark distribution over the down quark one at $Q^2_0$ and $x=0.023$ (the $x$ value corresponds to 0.013 at $Q^2=M^2_Z$) to be:
\begin{equation}
r_s=\frac{s+\bar{s}}{2\bar{d}}=1.00\pm 0.20 (\textrm{exp})\pm 0.07 (\textrm{mod})^{+0.10}_{-0.15} (\textrm{par}) ^{+0.06}_{-0.07} (\alpha_s)\pm 0.08 (\textrm{th})\,,\nonumber
\end{equation}
where uncertainties are experimental, model, parametrization, $\alpha_s$ and theoretical. The determination, consistent with the prediction that the light quark sea at low $x$ is flavor symmetric, was compared in Figure~\ref{fig:epwz12} with predictions obtained from four global PDF determinations. The ABKM09~\cite{0908.2766} and MSTW08~\cite{0901.0002} determinations gave a value around 0.5 and the NNPDF2.1~\cite{1107.2652} determination was lower at about 0.25. On the other hand, the CT10 (NLO)~\cite{1007.2241} determination gave a large ratio consistent with the ATLAS determination.

\section{Analysis ATLAS-epWZ16}

The ATLAS-epWZ16 analysis~\cite{1612.03016} is very similar to ATLAS-epWZ12. The main differences are that the HERA-I $ep$ data were replaced with the final NC and CC DIS HERA-I+II data~\cite{1506.06042}. The ATLAS Drell-Yan measurements used~\cite{1612.03016} were based on the full 7~TeV  data sample of 4.6 fb$^{-1}$ with improved precision. In addition, the measurement of the $Z/\gamma^\ast$ process was extended in three mass ranges of $46<m_{\ell\ell}<66$~GeV, $66<m_{\ell\ell}<116$~GeV and $116<m_{\ell\ell}<150$~GeV for the central rapidity region of $|y_{\ell\ell}|<2.5$ and in two mass ranges $66<m_{\ell\ell}<116$~GeV and $116<m_{\ell\ell}<150$~GeV in the forward rapidity region up to 3.6. The NNLO QCD fit with the same PDF parameterization as the ATLAS-epWZ12 fit was performed using the xFitter program~\cite{HERAFitter}.

\begin{figure}
\begin{center}
\includegraphics[width=0.55\columnwidth]{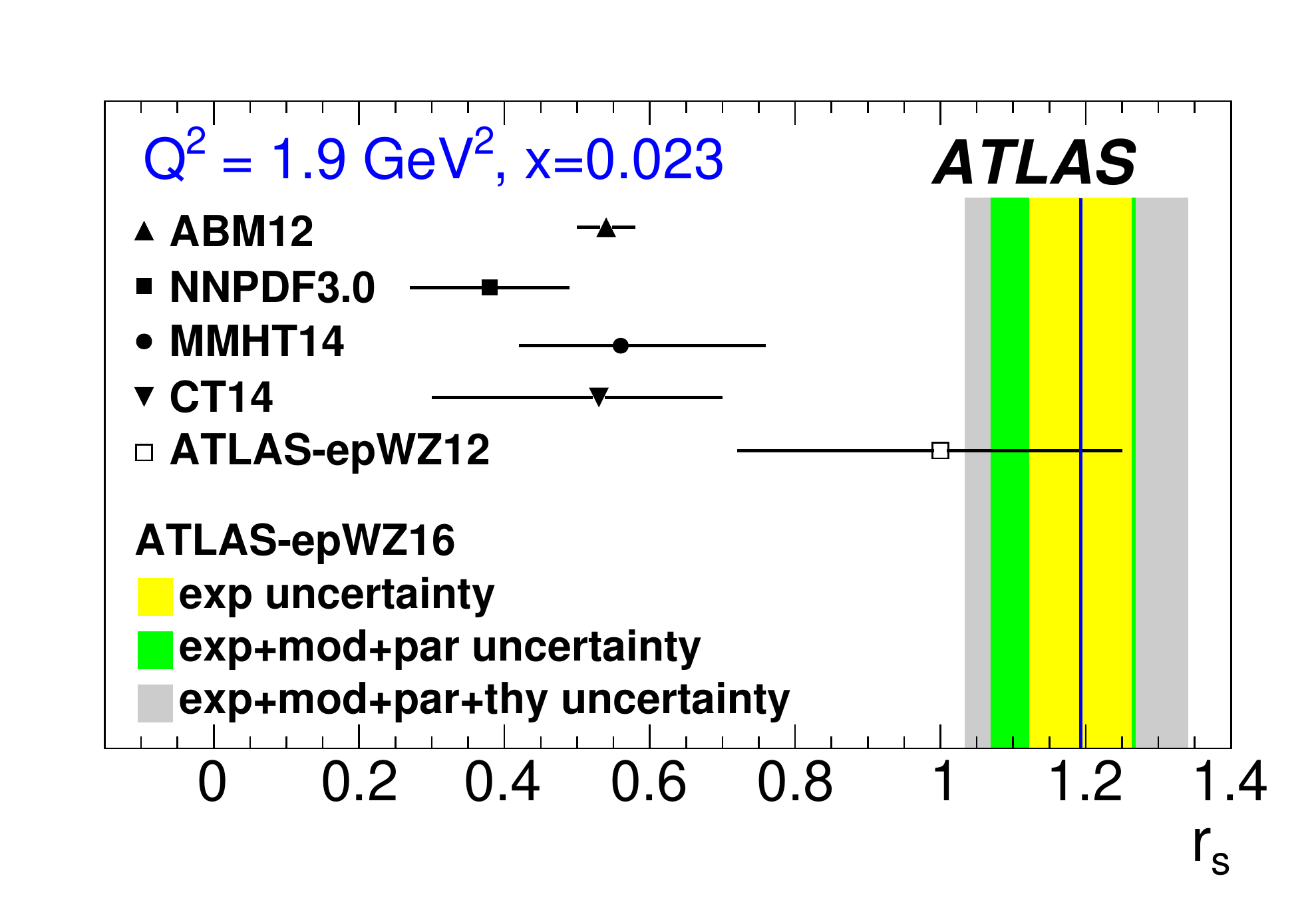}
\includegraphics[width=0.4\columnwidth]{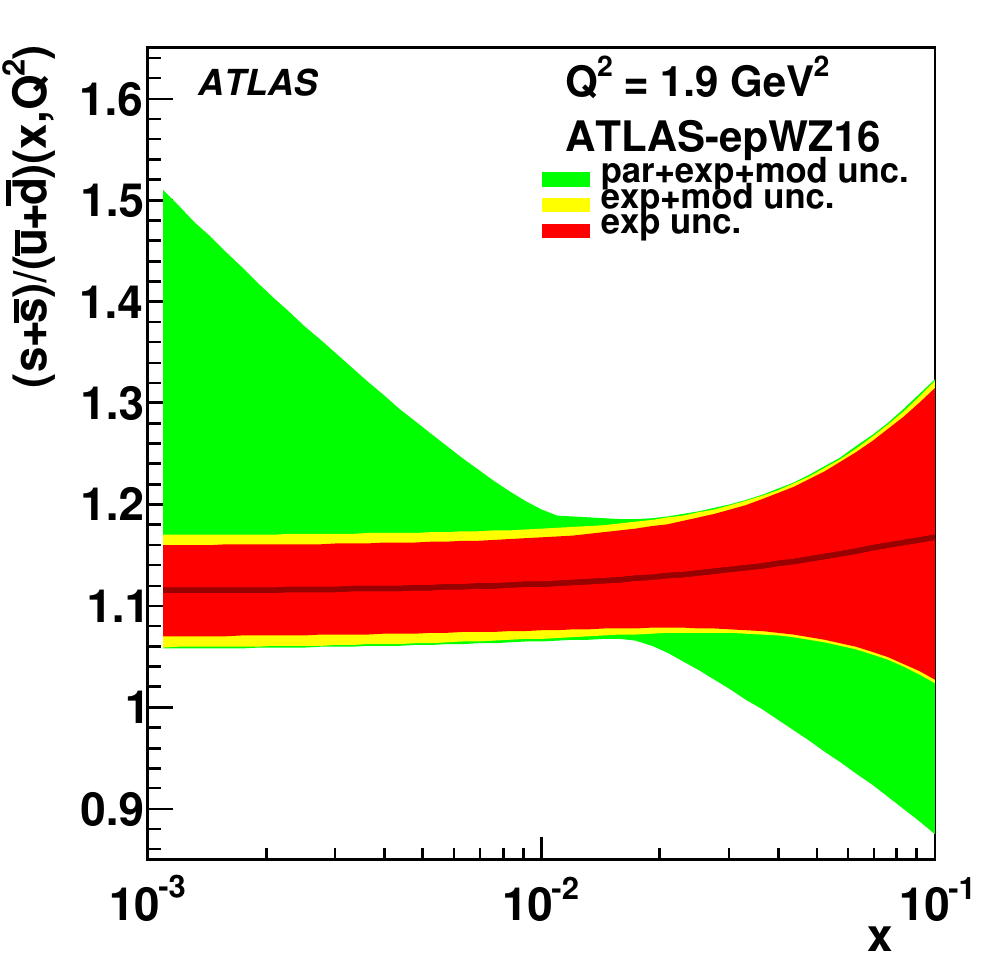}
\caption{Left: Relative strange-to-down sea quark ratio $r_s$ for $Q^2_0=1.9$~GeV$^2$ and $x=0.023$ comparing the determination from ATLAS (shown in the vertical line with bands for experimental data, QCD fit and theoretical uncertainties) with predictions from different NNLO PDF sets (shown in closed symbols with horizontal error bars) as well as previous ATLAS results (shown in open square). Right: relative strange quark ratio over the sum of up and down sea quarks as a function of Bjorken-$x$. The plots are taken from Ref.~\cite{1612.03016}.}
\label{fig:epwz16}
\end{center}
\end{figure}

\sloppy
The strange quark ratio $r_s$ was determined with improved precision at $Q^2_0$ and $x=0.023$:
\begin{equation}
r_s=1.19\pm 0.07 (\textrm{exp})^{+0.13}_{-0.14} (\textrm{mod}+\textrm{par}+\textrm{th})\,.\nonumber
\end{equation} 
The comparison with predictions from other global PDF sets ABM12~\cite{1310.3059}, NNPDF3.0~\cite{1410.8849}, MMHT14~\cite{1412.3989}, CT14~\cite{1506.07443}, as well as from ATLAS-epWZ12 is shown in Figure~\ref{fig:epwz16} (left). In addition, the ratio of strange quarks over the sum of up and down sea quarks $R_s=(s+\bar{s})/(\bar{u}+\bar{d})$ as a function of $x$ has also been obtained (Figure~\ref{fig:epwz16} (right)), though the uncertainty in particular the parametrization uncertainty was large.

\section{Analysis ATLAS-epWZtop18}

Another NNLO QCD analysis, ATLAS-epWZtop18~\cite{ATL-PHYS-PUB-2018-017}, included, in addition to those data used in the ATLAS-epWZ16 fit, top-quark pair production data measured in the lepton + jets~\cite{1511.04716} and dilepton~\cite{1607.07281} channels at 8~TeV corresponding to an integrated luminosity of 20.2~fb$^{-1}$. The top-quark pair production data are complementary to the other data sets in their PDF constraining power since they are expected to be sensitive to the high-$x$ gluon distribution~\cite{1611.08609}. The corresponding NNLO prediction~\cite{1704.08551} were supplied in the form of FastNLO grids~\cite{1208.3641} for the data in the lepton + jets channel. For the dilepton channel, APPLGRID was interfaced to MCFM to produce NLO grids and a $K$-factor technique was used to correct from NLO to NNLO predictions. 


The setup of the fit differs from that of ATLAS-epWZ16 in three aspects, apart from the addition of top-quark pair production data. Firstly, the low mass off-peak $Z/\gamma^\ast$ boson data was not used because it became clear that the lower $x$ region probed by the data was subject to further QCD~\cite{1710.05935} corrections which were not readily calculable, this had negligible impact on the fits. Secondly, the minimum $Q^2$ selection on the HERA data was raised from 7.5~GeV$^2$ to 10~GeV$^2$, this larger cut was already considered as one of the model variations of the ATLS-epWZ16 fit and was used as standard in the new analysis to avoid the region of the HERA data which may be subject to $\ln(1/x)$-resummation effects~\cite{1710.05935} that were not accounted for in the prediction. Thirdly, the ATLAS-epWZtop18 fit had one additional term, $P_{\bar{d}}=1+D_{\bar{d}}x$, for the down sea quark distribution than those of ATLAS-epWZ16 where it was considered as a parameterization variation.

\begin{figure}
\begin{center}
\includegraphics[width=0.6\columnwidth]{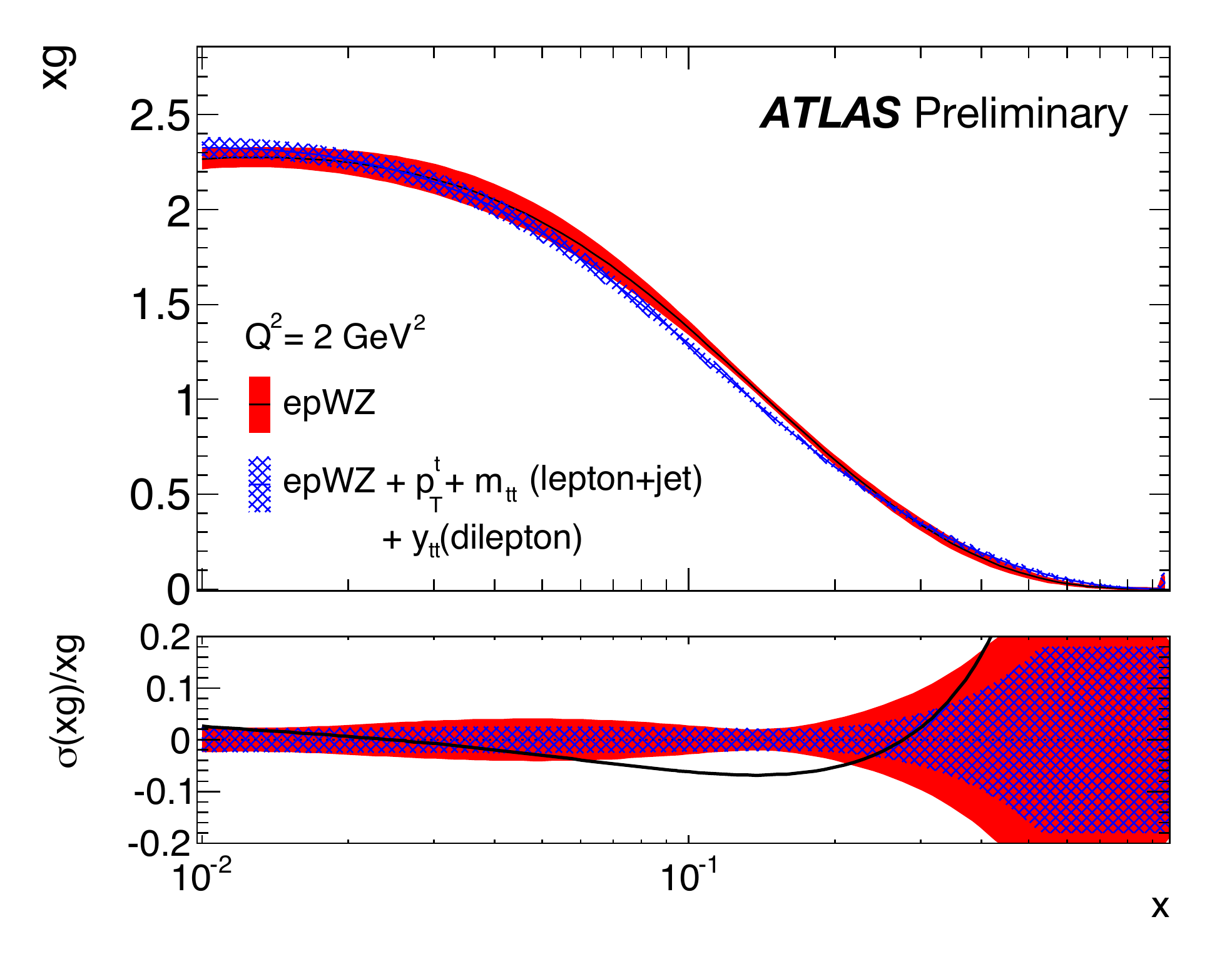}
\caption{Gluon distribution as a function of Bjorken-$x$ from fitting HERA data and ATLAS $W$ and $Z/\gamma^\ast$ boson data plus the $t\bar{t}$ lepton + jet $\pt^{t}$ and $m_{t\bar{t}}$ data and the $t\bar{t}$ dilepton $y_{t\bar{t}}$ data compared to the fit to HERA and ATLAS $W$ and $Z/\gamma^\ast$ boson data alone. The plot is taken from Ref.~\protect\cite{ATL-PHYS-PUB-2018-017}.}
\label{fig:top18}
\end{center}
\end{figure}

The main result of the fit is presented in Figure~\ref{fig:top18}, showing the impact of the top-quark pair production data on the gluon distribution; reducing its uncertainty in particular at high $x$ and making it softer and harder at medium $x$ and high $x$, respectively.

\section{Analysis ATLAS-epWZVjet20}

The analysis ATLAS-epWZVjet20~\cite{2101.05095} included new ATLAS data on the production of a vector ($W$ or $Z$) boson in association with at least on jet~\cite{1711.03296,1907.06728}. The measurements were based on the same data sample as for the top-quark pair production at 8~TeV. They provided a novel source of input to PDF determination that is sensitive to partons at higher $x$ and $Q^2$ than can be accessed by $W$ and $Z$ boson data alone. These data are thus complementary to the inclusive $W/Z$ boson measurements. 

Predictions for $W +$ jets and $Z +$ jets production~\cite{1511.08692} were obtained similarly to the $W$ and $Z$ predictions to NLO in QCD and leading order in EW couplings by using the APPLGRID code interfaced to the MCFM program. NNLO (NLO) corrections in QCD (EW) were implemented as $K$-factors. 

\begin{figure}
\begin{center}
\includegraphics[width=0.49\columnwidth]{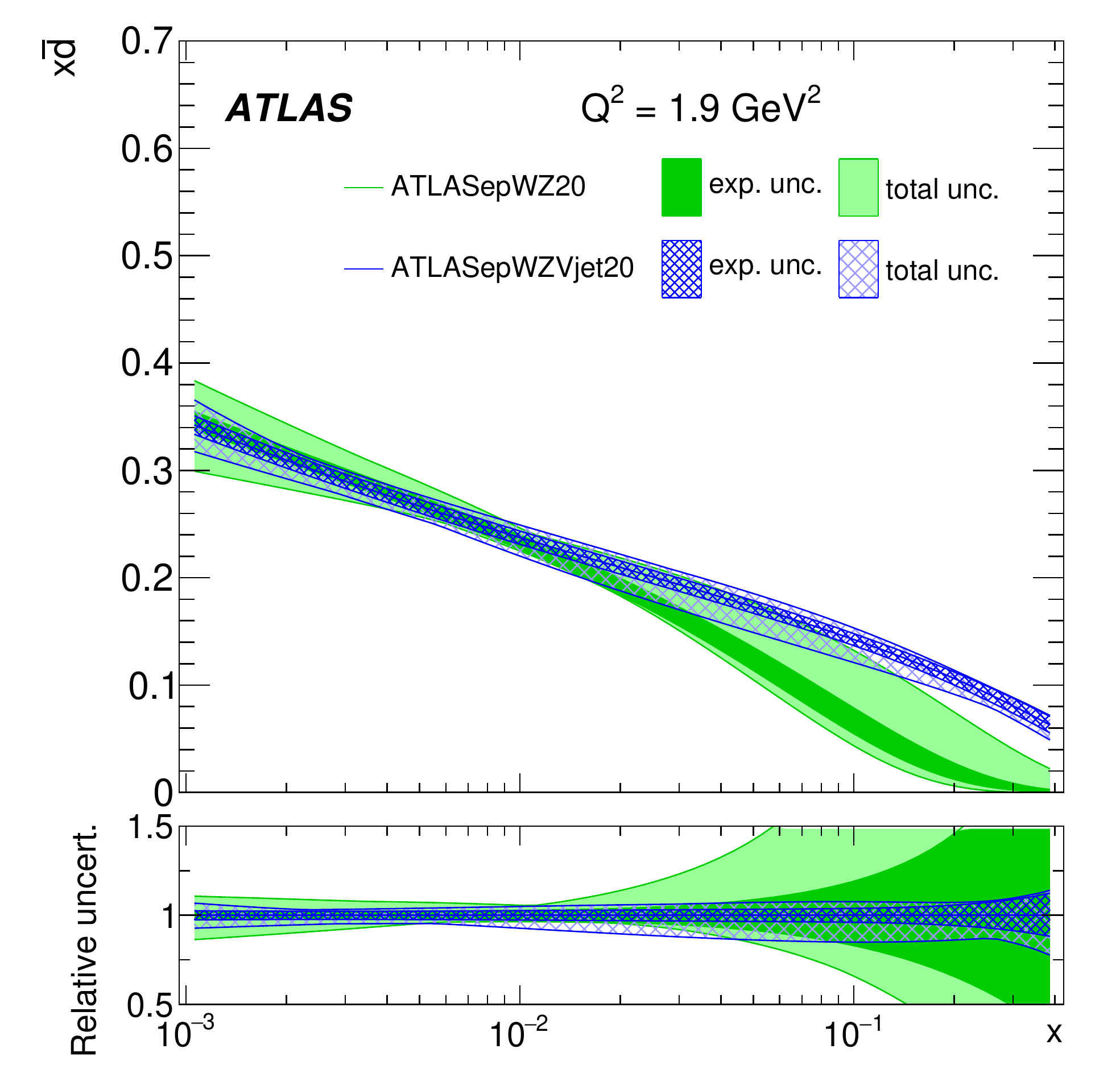}
\includegraphics[width=0.49\columnwidth]{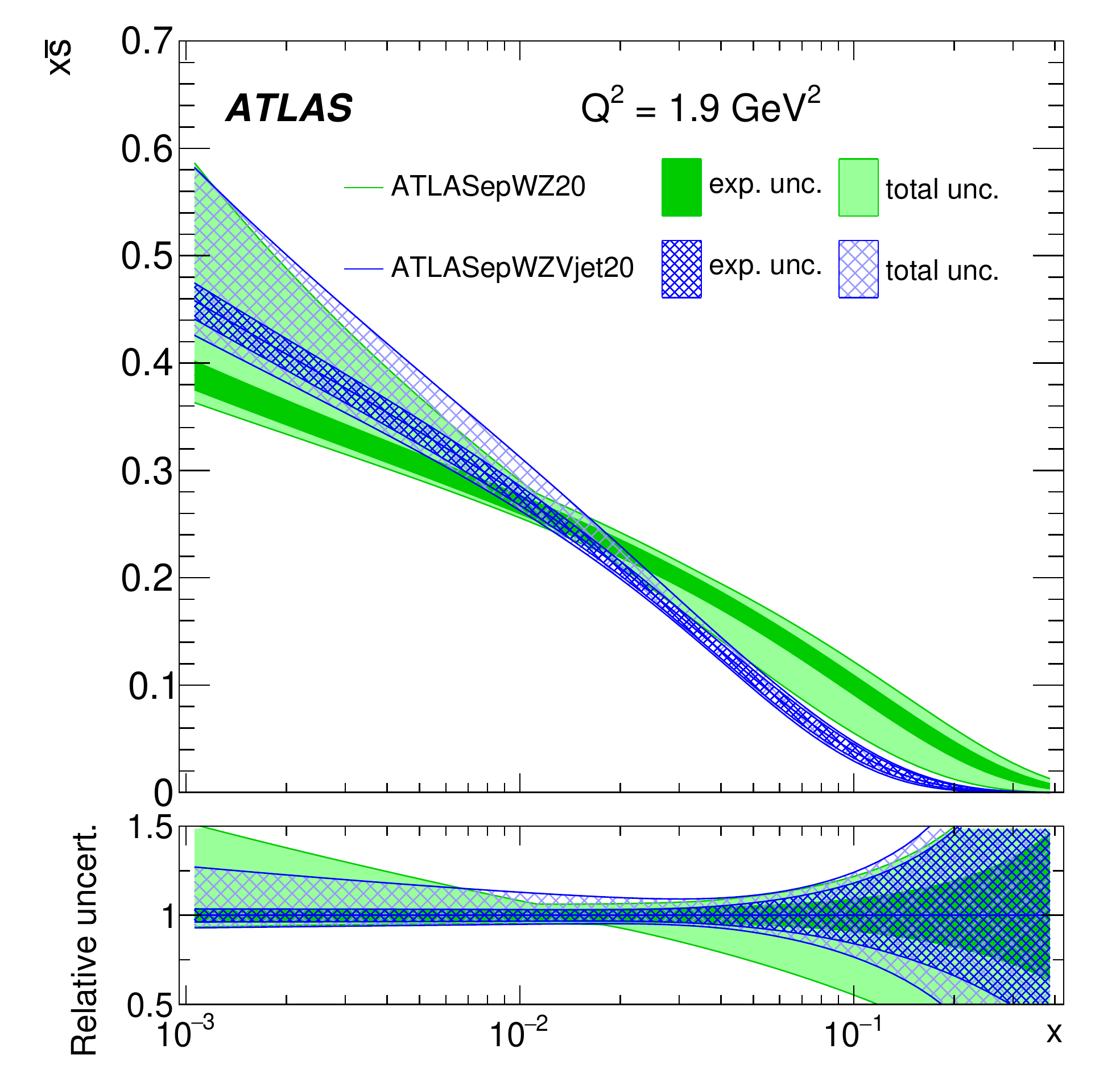}
\includegraphics[width=0.49\columnwidth]{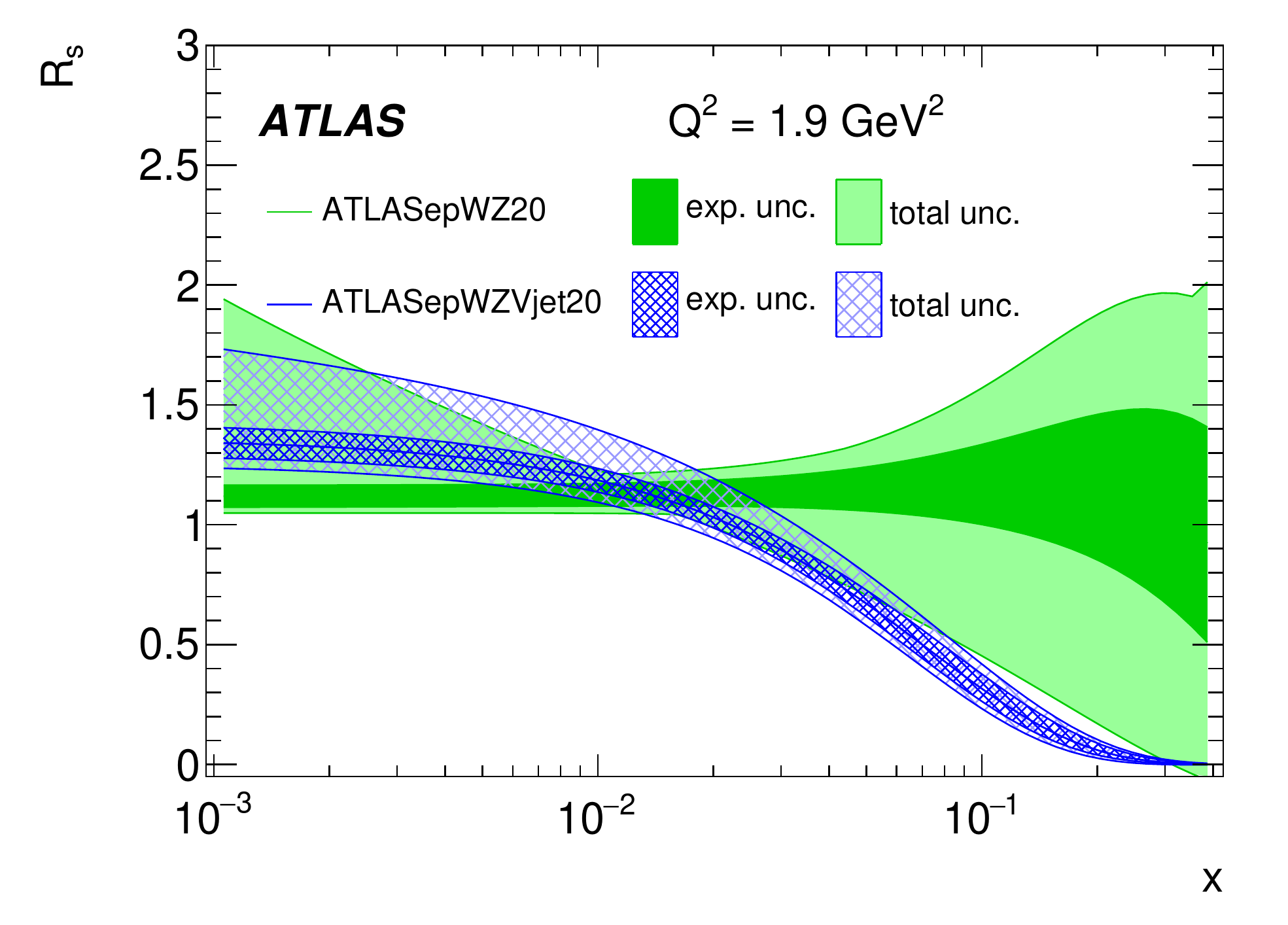}
\includegraphics[width=0.49\columnwidth]{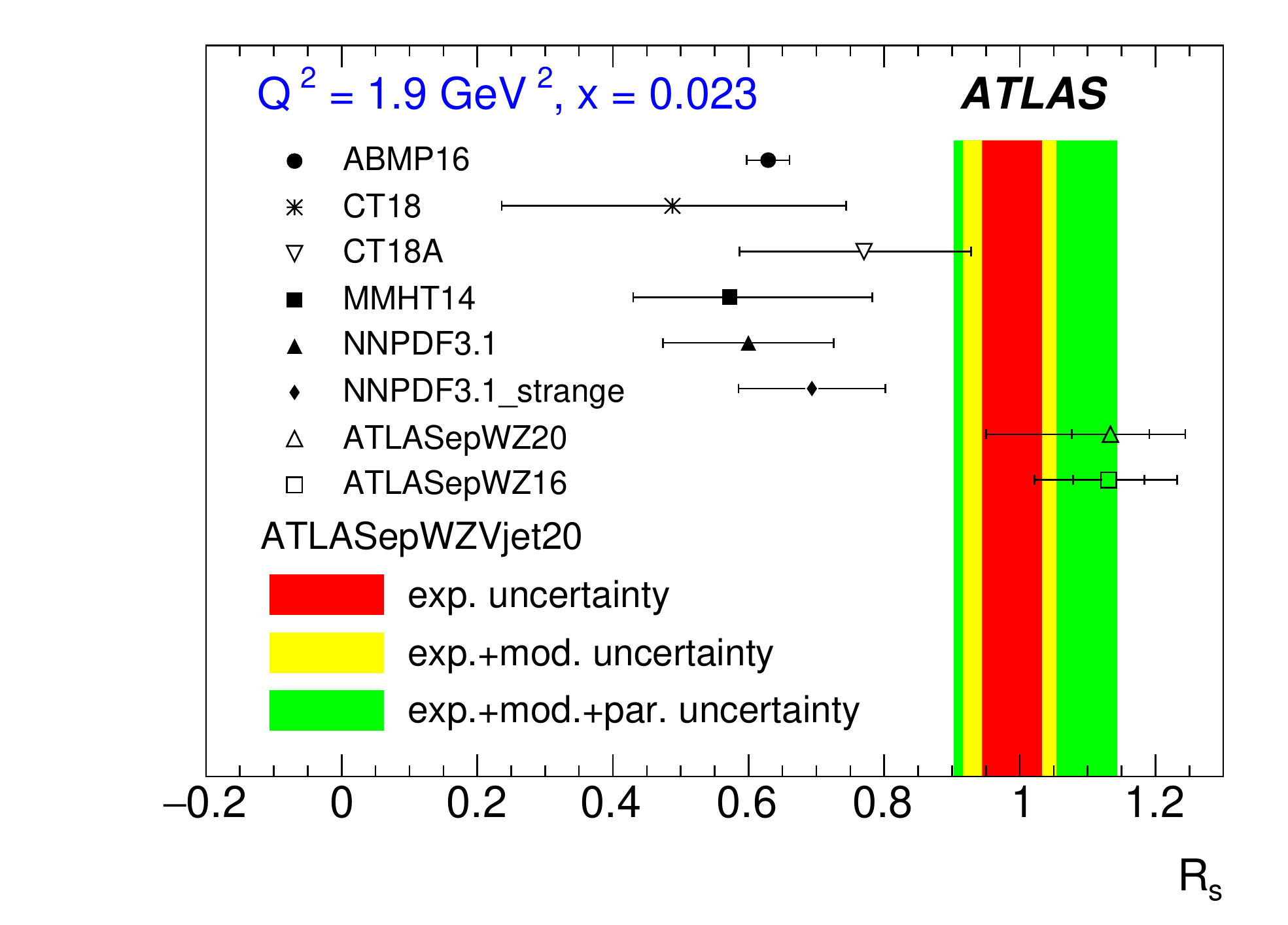}
\caption{Top-left: Down sea quark distribution as a function of Bjorken-$x$ when including (blue) or without including (green) the $V+$ jets data in the fits. Top-right: similar plot for strange sea quark distribution. Bottom-left: similar plot for relative strange quark ratio over the sum of up and down sea quarks $R_s$ as a function of Bjorken-$x$. Bottom-right: $R_s$ determination in comparison with other predictions for $Q^2_0=1.9$~GeV$^2$ and $x=0.023$. The plots are taken from Ref.~\cite{2101.05095}.}
\label{fig:ds_vjet20}
\end{center}
\end{figure}

The setup of the ATLAS-epWZVjet20 fit differs from that of ATLAS-epWZ16 in that there was one more term, $P_g=1+D_gx$, for the gluon distribution and a tighter selection of $Q^2>10$~GeV$^2$ instead of 7.5~GeV$^2$ on the DIS data. The resulting down and strange sea quark distributions of the fit and $R_s$ at the starting scale $Q^2_0$ are shown in Figure~\ref{fig:ds_vjet20} in comparison with the results of the ATLAS-epWZ20 fit when the $V+$ jets data were not included. The inclusion of the $V+$ jets data made the $x\bar{d}$ distribution notably higher in the range of $x\gtrsim 0.02$ and the $x\bar{s}$ distribution lower in the same range, with significantly smaller experimental and parameterization uncertainties at high $x$. The difference between the two $R_s$ determinations at high $x$ is essentially covered by the large uncertainty dominated by the parameterization uncertainty in the ATLAS-epWZ20 fit. The new $R_s$ determination at $x=0.023$ was again compared with the predictions from ABMP16~\cite{1701.05838}, CT18 and CT18A~\cite{1912.10053}, MMHT14~\cite{1412.3989}, NNPDF3.1 and NNPDF3.1\_strange~\cite{1706.00428} and the ATLAS determinations ATLAS-epWZ20 and ATLAS-epWZ16, the latter two being different mainly in their PDF parameterizations. Better agreement was observed with the CT18A PDF set, which included both the data used in the CT18 fit and the ATLAS 7 TeV data, although tension remains with the NNPDF3.1\_strange PDF set, which also used this data.

\begin{figure}
\begin{center}
\includegraphics[width=0.5\columnwidth]{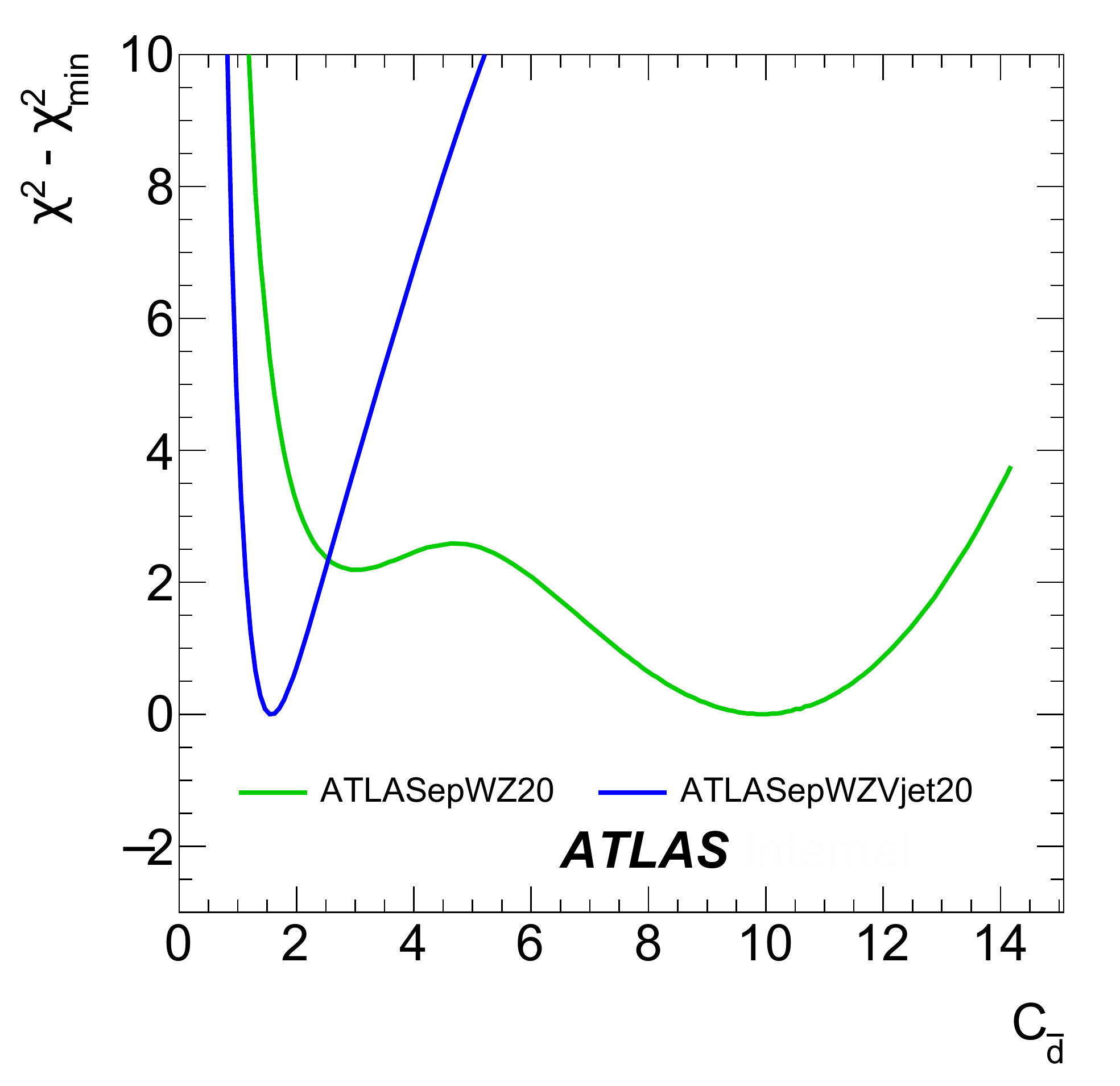}
\caption{Variation of $\chi^2$ subtracting the minimum value when including (blue) or without including (green) the $V+$ jets data in the fits as a function of parameter $C_{\bar{d}}$ . The plot is taken from Ref.~\cite{2101.05095}.}
\label{fig:dm_vjet20}
\end{center}
\end{figure}

The large change at high $x$ was investigated by performing a $\chi^2$ scan of the $C_{\bar{d}}$ parameter which controls the high-$x$ behavior of the $x\bar{d}$ distribution. The $\chi^2$ variation subtracting the minimum value is presented in Figure~\ref{fig:dm_vjet20}. The other data sets showed double minima in $\chi^2$ with a preferred solution for a $C_{\bar{d}}$ value around 10 corresponding to a soft $x\bar{d}$ distribution at high $x$. The inclusion of the $V+$ jet data allowed to resolve the ambiguity and resulted in a $C_{\bar{d}}$ value around 1.6 and thus a harder $x\bar{d}$ distribution at high $x$. Since the sum of $x\bar{d}+x\bar{s}$ is well constrained by e.g.\ the DIS data, the harder $x\bar{d}$ distribution gives thus the softer $x\bar{s}$ distribution and smaller $R_s$ at high $x$.

\FloatBarrier

\section{Summary and prospect}

The results of several NNLO QCD analyses performed using the Drell-Yan production and top-quark pair production cross sections measured by ATLAS with $pp$ collision data at the LHC taken at center-of-mass energies of 7 and 8~TeV, together with the neutral and charged current deep inelastic scattering cross sections from $ep$ collisions at HERA, have been presented. The analyses show the constraining power of the ATLAS measurements on parton distribution functions of the proton, complementary to the HERA data. In particular, the strange quark distribution at low $x$ is found unsuppressed contrary to what was assumed and obtained in the other PDF analyses. The shape of the strange quark ratio over the other light sea quarks at high $x$ has been better understood thanks to the measurement of vector-boson production in association with at least one jet.

Another valuable constraint for the strange quark distribution at the LHC concerns the associated production of $W$ bosons and charm quarks since it probes the strange quark content of the proton directly through the leading order processes $g + \bar{s} \to W^+ +\bar{c}$ and $g + s \to W^- +c$. Earlier analyses, not presented here, performed NLO fits using the measurements of ATLAS at 7~TeV~\cite{1402.6263} and of CMS at 7~TeV~\cite{1310.1138} and at 13~TeV~\cite{1811.10021}. With the recent available NNLO predictions~\cite{2011.01011}, there is a good prospect in including the $W+c$ data in the future NNLO analyses, though the interpretation of the data is sensitive to the modelling of $c$-quark fragmentation.

\nocite{*}
\bibliographystyle{auto_generated.bst} 
\bibliography{Zhang_ATLAS/Zhang_ATLAS}
